%% file: dynamicgelsphere.tex
\documentclass{jfm}
\usepackage{graphicx}
\usepackage{epstopdf, epsfig}

\usepackage[utf8]{inputenc}
\usepackage{amsmath}
\usepackage{amsfonts}
\usepackage{amssymb}

\usepackage[usenames,dvipsnames]{xcolor}
\usepackage{natbib}
\usepackage{appendix}
\usepackage{subcaption}
\usepackage[normalem]{ulem}

\usepackage{pgfplots}
\pgfplotsset{compat=newest}

\shorttitle{Swelling and shrinking of thermo-responsive hydrogels}
\shortauthor{M. Butler, T. Montenegro-Johnson}

\title{The swelling and shrinking of spherical thermo-responsive hydrogels}

\author{Matthew D. Butler 
  \corresp{\email{m.butler.3@bham.ac.uk}} \and
  Thomas D. Montenegro-Johnson}

\affiliation{School of Mathematics, University of Birmingham, Edgbaston, Birmingham, B15 2TT, UK}

\newcommand{\FreeEnergy}{\dimensional{W}}
\newcommand{\temp}{\dimensional{T}}
\newcommand{\elongation}{\lambda}
\newcommand{\osmotic}{\Pi} 
\newcommand{\mix}{_{\mathrm{mix}}}
\newcommand{\elastic}{_{\mathrm{elast}}}
\newcommand{\Boltzman}{\dimensional{k}_B}

\newcommand{\FloryHuggins}{\chi}
\newcommand{\solvol}{\dimensional{\Omega}_f}
\newcommand{\polyvol}{\dimensional{\Omega}_p}
\newcommand{\chempotential}{\mu}
\newcommand{\Param}{\Omega}
\newcommand{\shearmodulus}{\dimensional{G}}
\newcommand{\dryradius}{\dimensional{a}_d}
\newcommand{\externalchempot}{\chempotential_0}

\newcommand{\stress}{\sigma}
\newcommand{\porosity}{\phi}
\newcommand{\viscosity}{\dimensional{\eta}}
\newcommand{\permeability}{k}
\newcommand{\Terzaghi}{\stress'}
\newcommand{\fluidvelocity}{\dimensional{v}_f}
\newcommand{\polyvelocity}{\dimensional{v}_p}

\newcommand{\tempstart}{\temp_{\mathrm{start}}}
\newcommand{\tempend}{\temp_{\mathrm{end}}}

\newcommand{\dimensional}[1]{\tilde{#1}}
\newcommand{\timescale}{\dimensional{\tau}}

\newcommand{\Rfront}{R_f}

\newcommand{\dosage}{\dimensional{v}}

\newcommand{\zerothorder}[1]{{#1}^{(0)}}
\newcommand{\firstorder}[1]{{#1}^{(1)}}

\newcommand{\diff}[2]{\frac{\mathrm{d} #1}{\mathrm{d} #2}}
\newcommand{\partdiff}[2]{\frac{\partial #1}{\partial #2}}
\newcommand{\intstart}[2]{ \int_{#1}^{#2} }
\newcommand{\dd}{\mathrm{d}}

\input{parula}
\usepackage{tikz,pgfplots}
\usetikzlibrary{fadings}
\usetikzlibrary{decorations.pathmorphing}
\usetikzlibrary{decorations.markings}
\tikzset{snake it/.style={decorate, decoration=snake}}

\begin{document}

\maketitle

\begin{abstract}
Thermo-responsive hydrogels are a promising material for creating controllable actuators for use in micro-scale devices, since they expand and contract significantly (absorbing or expelling fluid) in response to relatively small temperature changes. Understanding such systems can be difficult because of the spatially- and temporally-varying properties of the gel, and the complex relationships between the fluid dynamics, elastic deformation of the gel and chemical interaction between the polymer and fluid. We address this using a poro-elastic model, considering the dynamics of a thermo-responsive spherical hydrogel after a sudden change in the temperature that should result in substantial swelling or shrinking. We focus on two model examples, with equilibrium parameters extracted from data in the literature. We find a range of qualitatively different behaviours when swelling and shrinking, including cases where swelling and shrinking happen smoothly from the edge, and other situations which result in the formation of an inwards-travelling spherical front that separates a core and shell with markedly different degrees of swelling. We then characterise when each of these scenarios is expected to occur. An approximate analytical form for the front dynamics is developed, with two levels of constant porosity, that well-approximates the numerical solutions. This system can be evolved forward in time, and is simpler to solve than the full numerics, allowing for more efficient predictions to be made, such as when deciding dosing strategies for drug-laden hydrogels.
\end{abstract}


\section{Introduction}

\begin{figure}
        \centering
     \subcaptionbox{}{    
    \begin{tikzpicture}
        \node[xshift = -0.1cm] at (-4.5,0) {\includegraphics*[width=3cm,viewport = 80 400 280 600,clip]{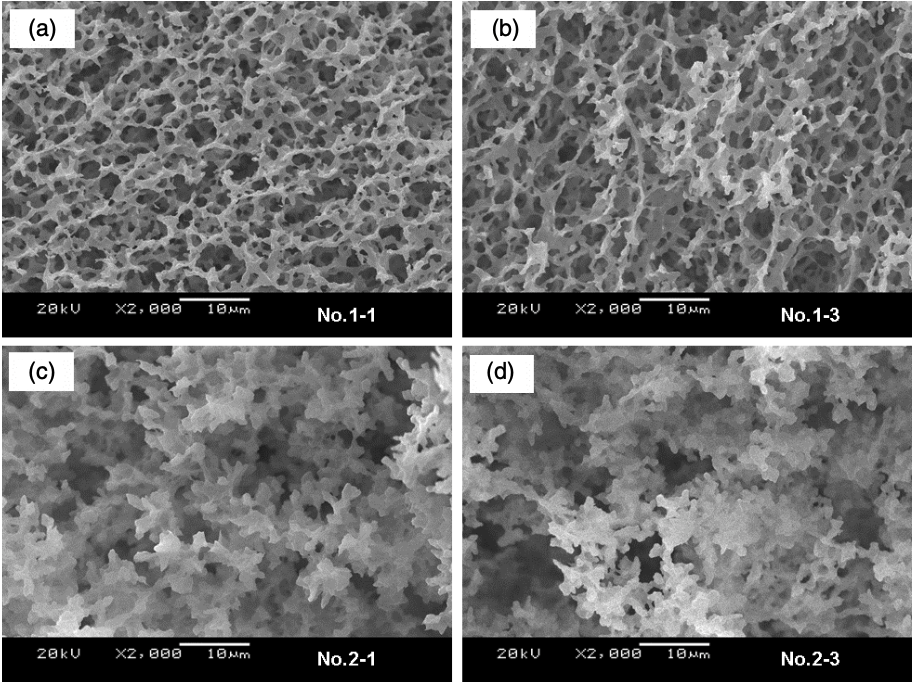}};
        \draw[thick, white,xshift = -0.1cm] (-4.45,-1.2) -- (-3.42,-1.2);
        \node[align=center,xshift = -0.1cm] at (-3.935,-1) {\color{white}$10\mu$m};
        \node at (-4.5,-2) {\footnotesize{\phantom{p} PNIPAM SEM \phantom{p}}};
     \end{tikzpicture}
     }
     \subcaptionbox{}{
     \begin{tikzpicture}
        \node at (0,0) {\input{Figs/pnipam_schematic_1}};
        \node at (0,-2) {\footnotesize{Swollen hydrogel}};
        \node at (2.5,0) {\input{Figs/arrow}};
        \node at (5,0) {\input{Figs/pnipaam_schematic_2}};
        \node at (5,-2) {\footnotesize{Deswollen hydrogel}};
    \end{tikzpicture}
     }
    \caption{(a) An SEM image of the micro-structure of PNIPAM, reprinted from \citet{Ju2006} with permission from IOP Publishing. (b) Schematic of a thermo-responsive hydrogel's structure  --- as the gel is heated, it loses its affinity for the fluid in the porespace, hence it contracts and expels fluid.} 
    \label{fig:schematic}
\end{figure}

Thermo-responsive hydrogels are polymers whose degree of swelling depends on the ambient temperature. Hydrogels absorb and retain significant amounts of fluid, resulting in swelling of the solid structure as the fluid fills the interstitial space between polymer chains (see fig.~\ref{fig:schematic}). This swelling can be substantial, with the volume of some hydrogels increasing several fold compared to their dry form \citep{Tanaka1978,Hirokawa1984}. Thermo-responsive hydrogels have a sharp decrease in their affinity for the fluid when heated above a certain temperature, called the volume phase transition temperature, resulting in expulsion of much of the interstitial fluid and significant shrinking of the gel. Many other responsive gels have also been developed in recent years, with a variety of different controlling external stimuli, such as pH, electric charge, and light \citep{Erol2019,Koetting2015}. However, responsive hydrogels actuated by temperature are of particular interest, especially gels such as poly(N-isopropylacrylamide) --- or PNIPAM --- because they have a sharp transition in their degree of swelling at temperatures that are easily obtainable experimentally, and often biomedically-relevant \citep{Haq2017}. 

Developing materials that change size in a controllable way gives a route to creating controllable shape change. For example, having two (or more) materials joined together that swell or shrink differently under a given stimulus can generate internal elastic stresses that are relaxed by deforming; differential swelling can cause out-of-plane buckling, and careful choice of the material properties and geometry during fabrication can allow for a whole host of possible programmed transformations \citep{Holmes2019}. 
Hydrogels with controllable swelling are therefore a promising avenue for developing shape-changing devices. This is particularly true at small-scales, where recent technological advances allow for micro-scale design and manufacture of gel structures through techniques such as 3D printing via two photon polymerisation, otherwise known as 2PP, \citep{Ji2021,Hippler2019} and halftone gel or stop flow lithography \citep{Kim2012,Sharan2021}. Such small devices have a wide range of physical applications \citep{Ionov2014}, for example as microfluidic valves \citep{Harmon2003,Richter2003}, as grippers that could be used by soft robots \citep{Li2017}, and as actuation for swimming in microbots or artificial active matter \citep{Nikolov2015,Mourran2017,Masoud2012,Wischnewski2020,MontenegroJohnson2018}.

There are many opportunities to exploit stimuli-responsive hydrogels in biomedical settings \citep{Sood2016,Klouda2015}. Controllable delivery of small quantities of drugs to a specific site could improve patient outcomes by reducing the required dosage necessary for a treatment; such dosing cannot be addressed by conventional drug delivery methods \citep{Li2016}. Responsive hydrogels are an appropriate and flexible means to achieve this targeted drug delivery \citep{Peppas1997,Qiu2001,Qureshi2019,Marques2021,Schmaljohann2006}. The drug can either be embedded within the microstructure of the gel itself and released as it contracts \citep{Bhattarai2010,Kulkarni2007}, or encapsulated within a microscopic delivery device that opens when actuated \citep{Fan2016,Stoychev2011}, with the stimuli either externally-imposed or dependent on the immediate environment. Additional control over the drug release can be obtained by combining multiple stimuli \citep{Fu2018} or through timed actuation, allowing multiple dose strategies to be developed.
Responsive hydrogels also have applicability in other medically-relevant scenarios, for example being used as scaffolds in tissue engineering, as an aid for wound healing and as components in microtools for use in microsurgery \citep{Erol2019,Sood2016}.

To fully understand the shape-changing properties of responsive hydrogel devices, a key first step is to understand how the size change occurs in a simple homogeneous gel. In particular, the dynamics of the swelling or shrinking of such a gel will be an important component in determining how a more complicated material behaves during the swelling and shrinking process. 
Experiments on these thermo-responsive gels have suggested that there is a difference between the behaviour of a homogeneous gel when swelling, compared to when shrinking: dynamic swelling has been observed to equilibrate exponentially, whereas the dynamics of a shrinking gel appears more complicated and can exhibit an instability that forms lobe-like structures \citep{Matsuo1988}. 
Temperature changes beyond transition have been observed to cause separation into distinct swollen and shrunken states along the length of cylinders, with a similar appearance to the fluid-dynamical Rayleigh-Plateau instability \citep{Matsuo1992}, as well as through the appearance of surface blisters that can cause deformation of rods and tori \citep{Shen2019,Chang2018}.

Since the size change of hydrogels is dependent on the movement of fluid into or out of the polymer structure, the dynamics of swelling and shrinking are fundamentally problems of fluid dynamics, coupled with solid mechanics and polymer physics.
Some of the earliest approaches to modelling such systems focused on the diffusion of the cross-linked polymer network, whilst ignoring the fluid motion \citep{Tanaka1973,Tanaka1979}. 
More recent work has incorporated all three physical building blocks into continuum mechanical models of hydrogels \citep[e.g.][]{Doi2009,Hong2008,Engelsberg2013,Chester2011,Bertrand2016,Drozdov2017}.
These models are typically poro-elastic, and are constructed via a model for the free energy density of swelling, based on Flory-Rehner theory \citep{FloryRehner1943I,FloryRehner1943II}, that has three main components: mixing of the fluid and polymer, deformation of the polymer structure, and work against the chemical potential or fluid pressure. The hydrogel can undergo large deformations when swelling and shrinking, and so the material is often taken to be hyperelastic, for example using a Neo-Hookean energy density. The fluid motion within the hydrogel mixture is then modelled diffusively \citep{Hong2008,Engelsberg2013} or using Darcy flow \citep{Doi2009,Bertrand2016}. 

These models have been applied to a range of scenarios involving generic hydrogels, including the evolution of gel filters \citep{Doi2009} and gel systems under load \citep{Hong2008}, and as a model system for plant transpiration \citep{Etzold2021}. Finite element simulations have been used to investigate the large deformation behaviour of swelling hydrogels \citep{Lucantonio2013}. Other studies modelling hydrogel spheres have also uncovered complex internal dynamics when swelling and shrinking, such as core-shell behaviour and transient surface instabilities \citep{Bertrand2016,Curatolo2017}. In addition, theoretical modelling of one-dimensional hydrogels has isolated the physics of the swelling from the geometry, from which it is possible to investigate when and how phase separation into states with different degree of swelling occurs \citep{Hennessy2020}.

These theoretical models have also been previously applied specifically to thermo-responsive hydrogels. Diffusive models of swelling and shrinking thermo-responsive gels suggest that the timescale for swelling or shrinking scales with the square of the gel size, agreeing with experimental observations \citep{Tanaka1985}, and have been used to predict the dynamic swelling of spherical shells \citep{Wahrmund2009}. Following predictions for the phase separation of thermo-responsive hydrogels based on their free energy density \citep{Sekimoto1993}, the dynamics of a one-dimensional hydrogel bar with regions of different swelling was explored using a mechanical model that balances internal hydrogel stresses with friction between polymer and fluid within the gel \citep{Tomari1994}. This modelling approach was further developed to investigate the dynamics of spherical thermo-responsive hydrogels close to the transition temperature \citep{Tomari1995}. This work theoretically reproduced the experimental observations of two-stage swelling and shrinking dynamics, and predicted the occurrence of transient phase separation into concentric swollen and collapsed states close to the volume phase transition, similar to the previously-mentioned core-shell behaviour observed in other hydrogel systems, which has since been enforced by other studies \citep{Doi2009}. In recent years, many several studies have focused on applying large deformation mechanics to thermo-responsive hydrogel problems, which has resulted in finite element simulations for the swelling of cubes, discs and more complex geometries \citep{Chester2011,Drozdov2017}. These methods are excellent for specific applications, however, there is still room for more fundamental studies of the generic swelling behaviour of hydrogels in simplified systems.

In this work, we apply the poro-elastic model of \citet{Bertrand2016}, which was used to study the swelling of a dry spherical gel bead immersed in water and the subsequent drying by evaporation, to investigate the dynamic swelling and shrinking of a thermo-responsive hydrogel sphere in a quiescent fluid bath. In particular, we consider the evolution of a thermo-responsive hydrogel sphere after a sudden temperature change for two different model systems that have been observed experimentally, and consider their behaviour during both swelling and shrinking. The temperature change is taken as instantaneous since diffusion of heat occurs much faster than hydrogel swelling.
We aim to extend the work of \citet{Tomari1995} in a few key ways. 
Firstly, we note that \citet{Tomari1995} focused on the dynamics for temperatures close to the transition between swollen and shrunken states, whereas we shall consider characterising the behaviours for more significant temperature changes.
In addition, we use a fully poro-elastic model that explicitly includes the fluid flow within the mixture, which \citet{Tomari1995} account for using a viscous dissipation term. This additional component makes an important contribution to the physics of the system, which can result in a significant difference to resultant behaviours \citep{Hui2005}.
Moreover, we shall investigate an approximate analytic solution that encapsulates the key physics of the core-shell shrinking dynamics, which is simple enough to be used in applications, such as designing drug-dosing strategies. We will comment on any similarities and differences between our results and those of \citet{Tomari1995} as they arise.

\begin{figure}
    \centering
    \begin{tikzpicture}
        \node at (-0.6,0) {\includegraphics[width=3cm,viewport = 600 320 1800 1520, clip]{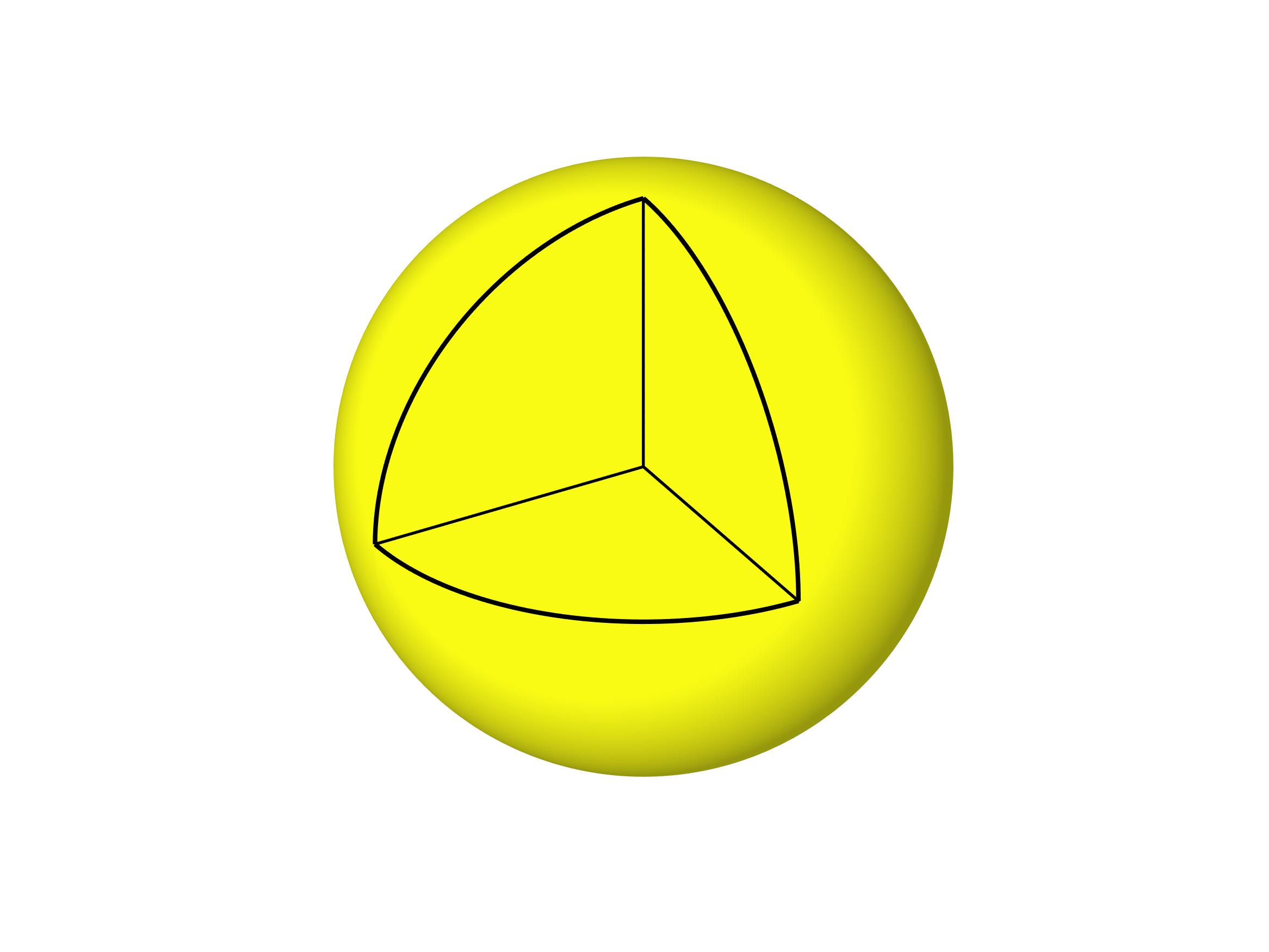}};
        \node at (3,1.5) {\includegraphics[width=3cm,viewport = 600 320 1800 1520, clip]{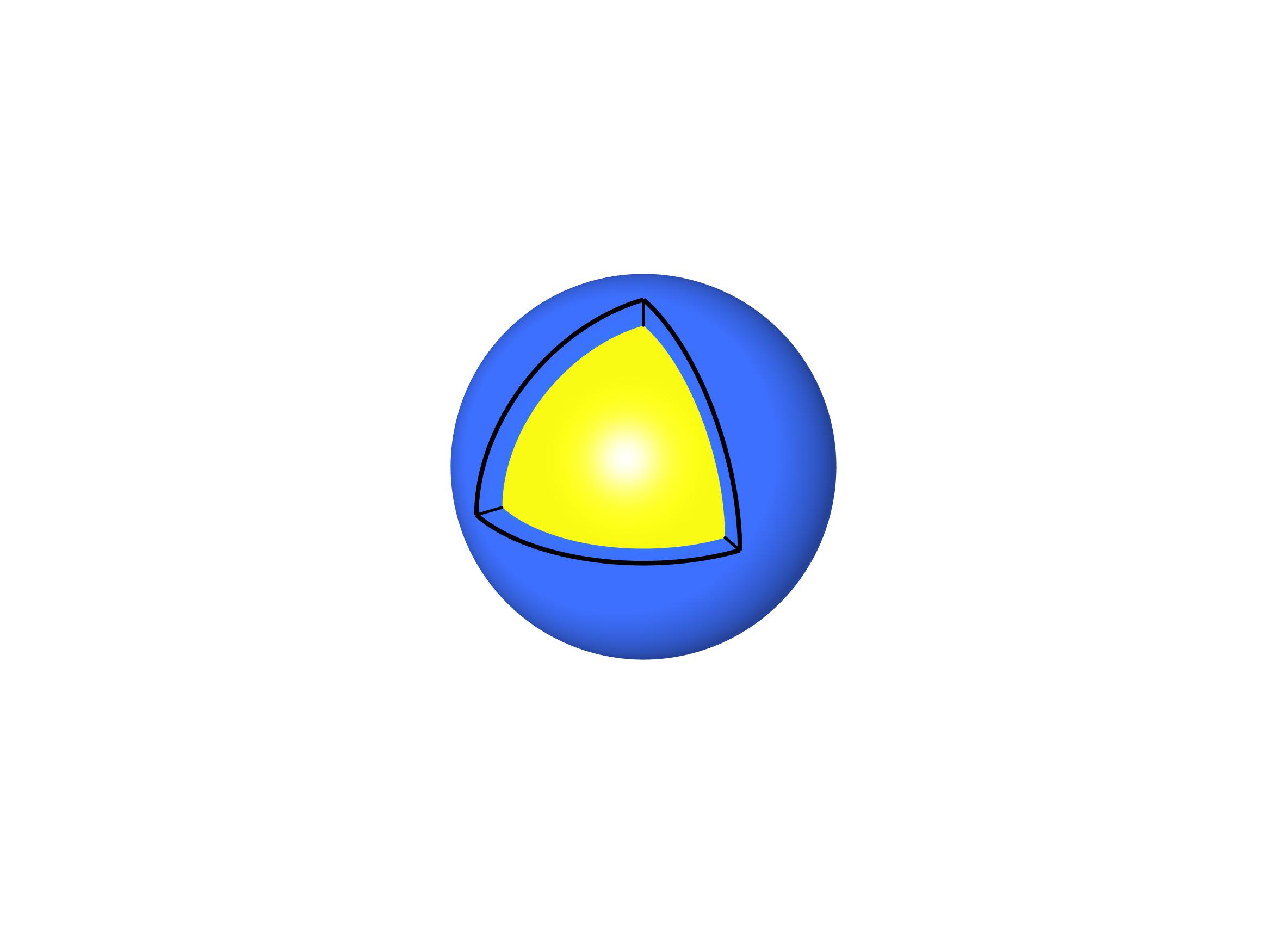}};
        \node at (6,1.5) {\includegraphics[width=3cm,viewport = 600 320 1800 1520, clip]{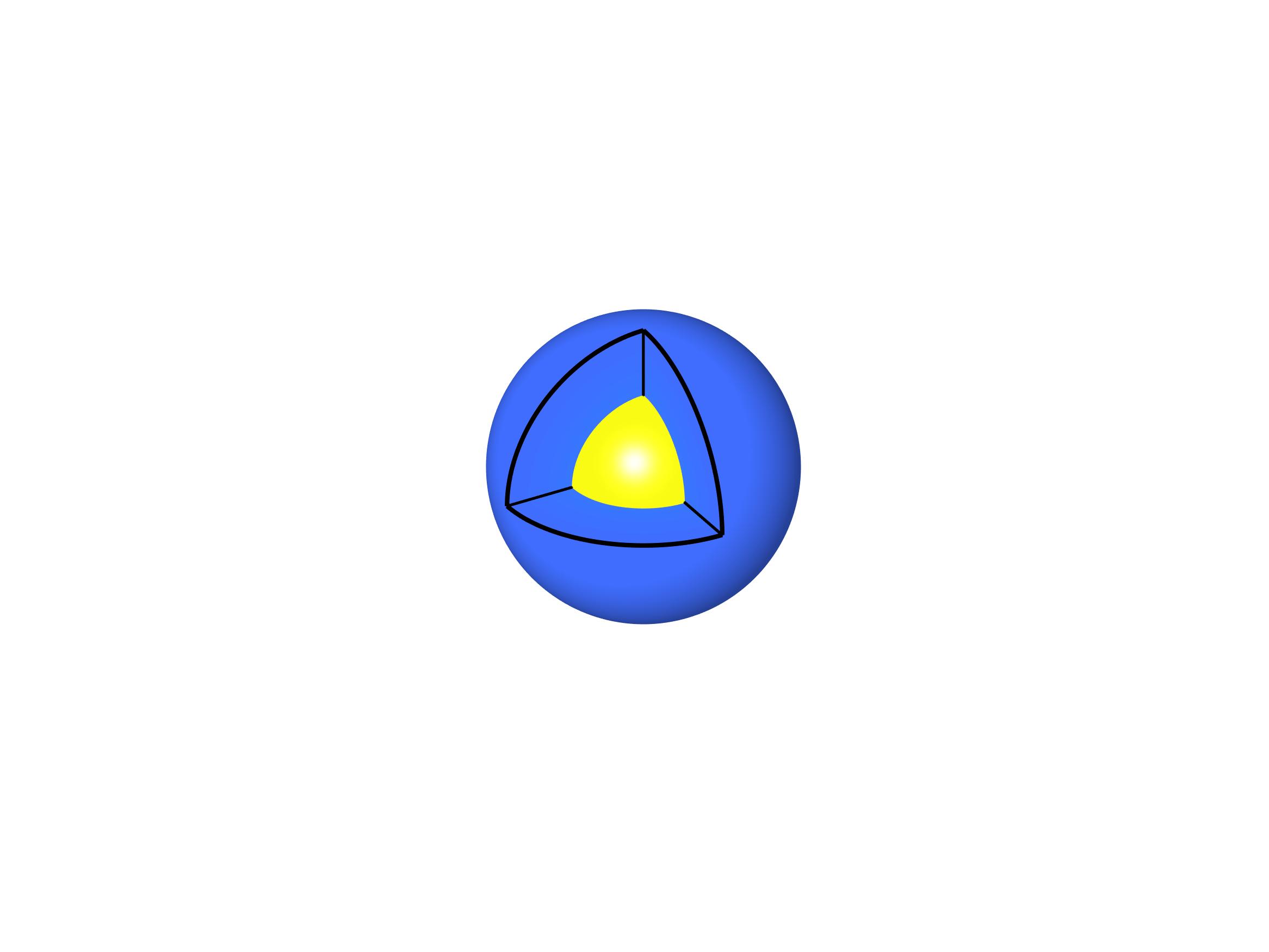}};
        \node at (9,0) {\includegraphics[width=3cm,viewport = 600 320 1800 1520, clip]{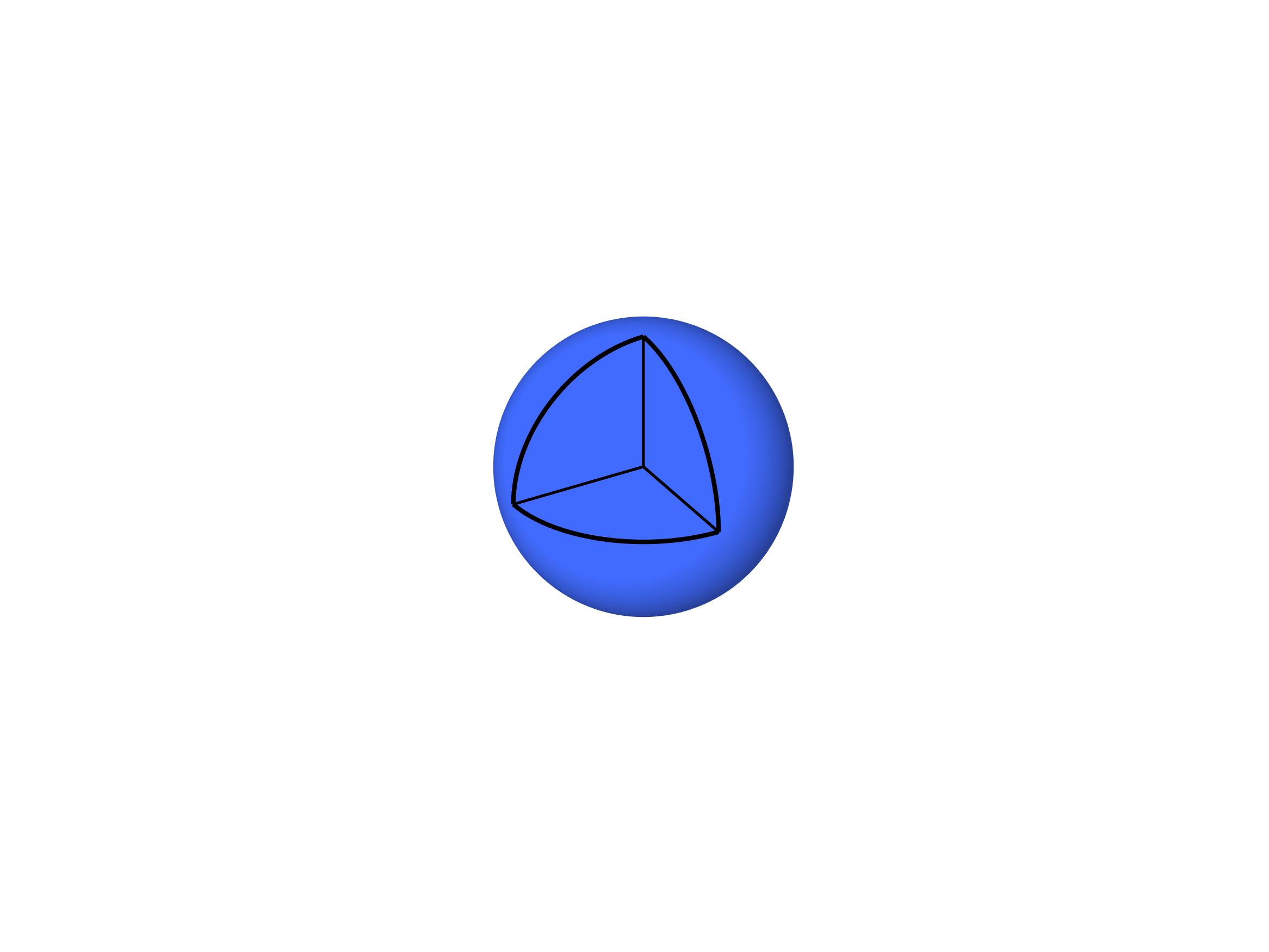}};
        \node at (3,-1.5) {\includegraphics[width=3cm,viewport = 600 320 1800 1520, clip]{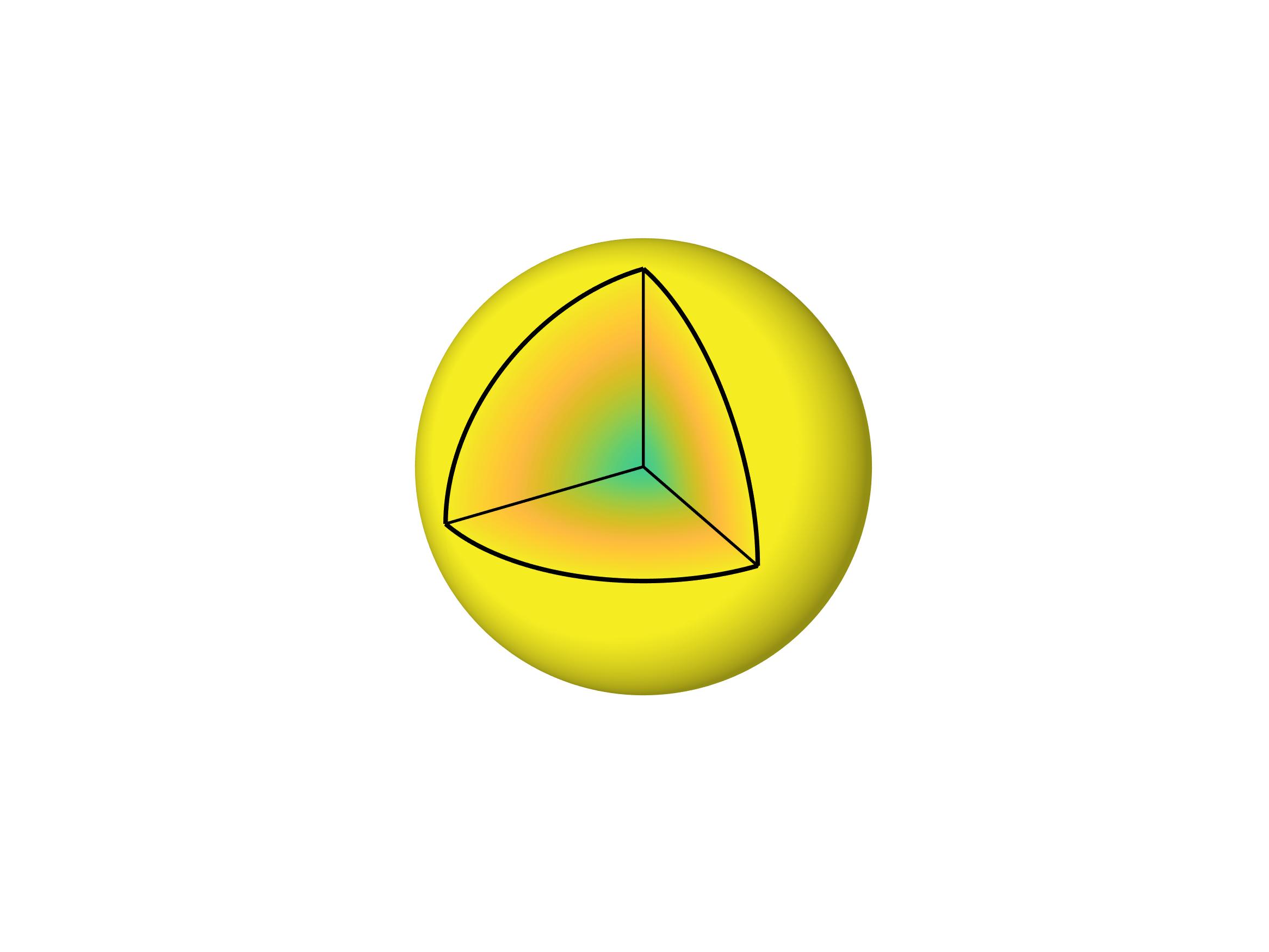}};
        \node at (6,-1.5) {\includegraphics[width=3cm,viewport = 600 320 1800 1520, clip]{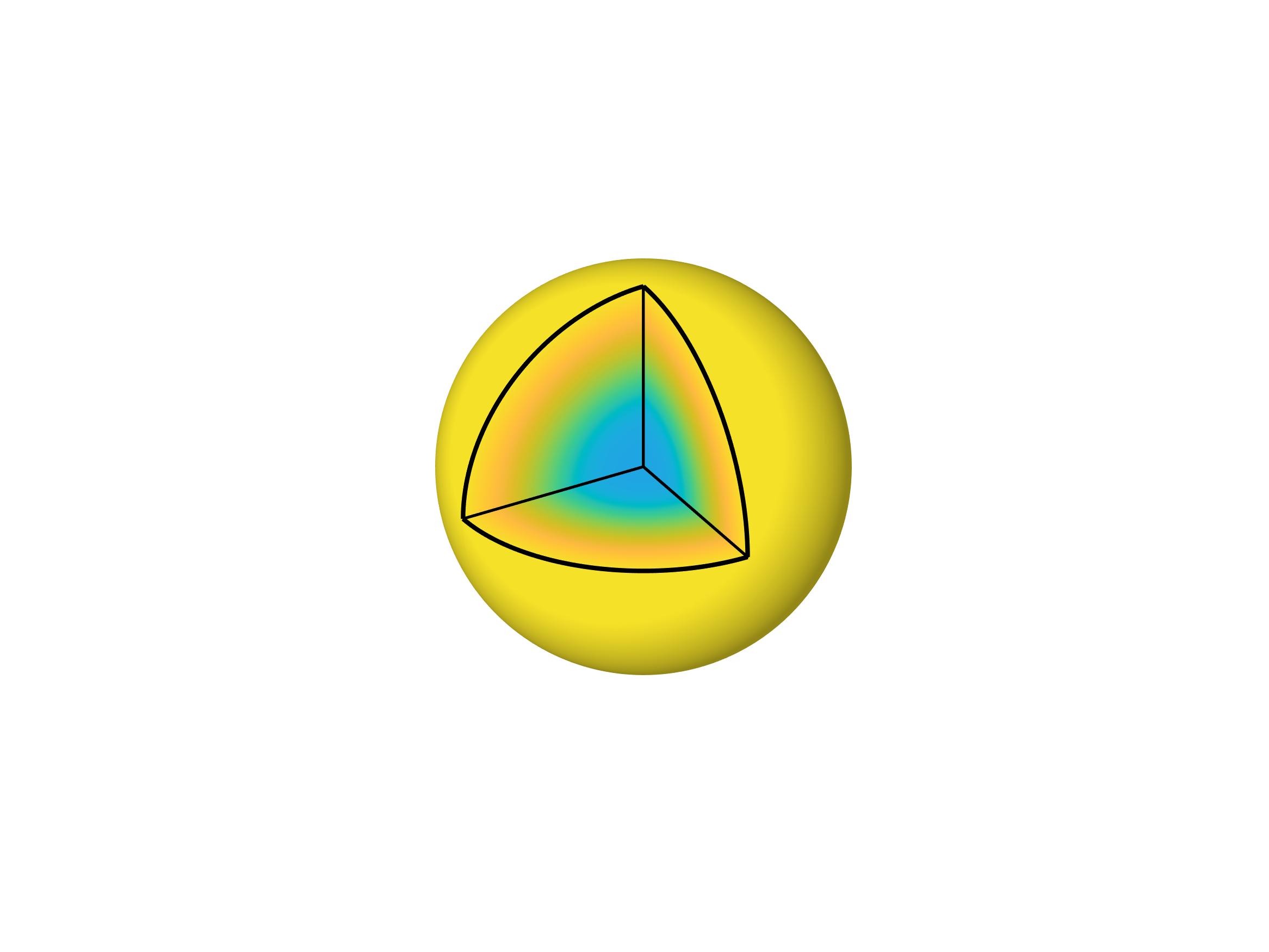}};
        %
        \draw[very thick,red,yshift = 2cm,xshift=-0.5cm,path fading=west] (0,0.2) -- (1.5,0.2) -- (1.5,0.5) -- (2,0) -- (1.5,-0.5) -- (1.5,-0.2) -- (0,-0.2);  
        \node at (-0,2) {\color{red}\it HEAT};
        \draw[very thick,blue,yshift = -1.5cm,xshift=9.5cm,rotate=180,path fading=east] (0,0.2) -- (1.5,0.2) -- (1.5,0.5) -- (2,0) -- (1.5,-0.5) -- (1.5,-0.2) -- (0,-0.2); 
        \node at (9,-1.5) {\color{blue}\it COOL};
    \end{tikzpicture}
    \caption{A thermo-responsive hydrogel sphere shrinks as it is heated and swells as it is cooled. Colour denotes porosity (or degree of swelling). Swelling occurs smoothly from the edge, whereas for shrinking, a sharp inward-propagating front is formed between two (nearly) uniform porosity regions. These illustrations are results from our study (see figs.~\ref{fig:Results_CaiSuo_SwellNoFront} \& \ref{fig:Results_CaiSuo_ShrinkFront}), presented here to orient the reader with the discussions that follow.}
    \label{fig:shrink_swell_schematic}
\end{figure}
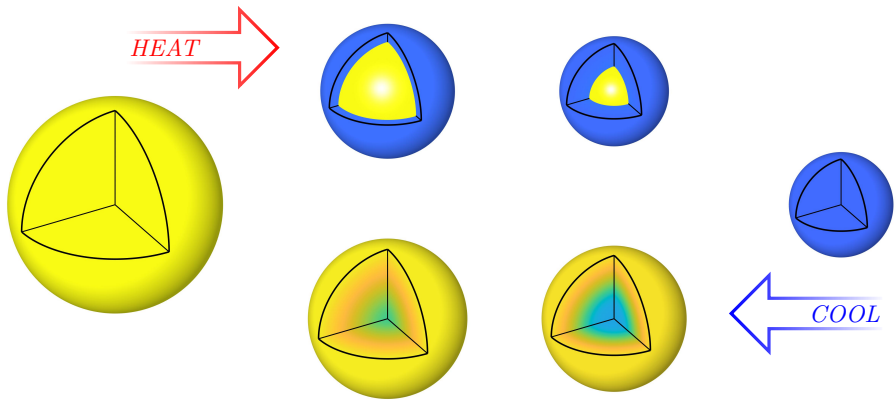

We begin in \S\ref{sec:Eql} by introducing the equilibrium model for the thermo-responsive hydrogel, before outlining the dynamic model in \S\ref{sec:DynamicModel}. The poro-elastic equations for the evolution of the hydrogel sphere are then solved numerically, with example results for swelling and shrinking of each of the two model systems presented in \S\ref{sec:Results}. We observe a range of different behaviours in the swelling and shrinking dynamics in both cases, such as swelling occurring smoothly inwards from the edge upon cooling, but reheating resulting in a sharp front between a swollen core and shrunken shell (see schematic in fig.~\ref{fig:shrink_swell_schematic}). Other distinct behaviours are also seen, and we provide an overview of when each is expected to occur. The dynamics of the shrinking gel is often characterised by an inwards-travelling front, which we investigate further in \S\ref{sec:FrontSoln} using an approximate analytic solution. We show that this solution can be evolved forward in time, and investigate a potential application in calculating drug-dosage strategies. Finally, we summarise our results in \S\ref{sec:Conclusion}. 

\section{Equilibrium swelling of a hydrogel} \label{sec:Eql}

\subsection{Free energy density of the hydrogel}

We consider a hydrogel in a quiescent bath of fluid that has absorbed fluid and deformed. Both of these processes have energy costs associated with them: a chemical contribution from the mixing of fluid molecules and polymer chains, and a physical contribution from the elastic deformation of the polymer. The nominal free energy density of the hydrogel in such a state can then be written as the sum of these two contributions \citep{FloryRehner1943I,FloryRehner1943II}
\begin{equation}     \label{eq:FreeEnergy}
\FreeEnergy = \FreeEnergy\elastic + \FreeEnergy\mix,
\end{equation}
where $\FreeEnergy\elastic$ represents the elastic energy density and $\FreeEnergy\mix$ is the mixing energy density. Here, $\FreeEnergy$ is the Helmholtz free energy of the mixture per unit volume of the undeformed dry polymer. Throughout this work we use tildes, $\dimensional{\cdot}$, to denote dimensional quantities.

We consider a gel deformation from a reference state, resulting in a local factor increase in volume of the body, $J$, relative to this reference state (at which $J=1$). The principal stretches, $\elongation_i$ for $i\in\{1,2,3\}$, are the the factor increase in lengths from the reference state in each direction; they are the
major and minor axes of an ellipsoid that results from the deformation (locally) of a unit sphere in the reference state. The local volume change, $J$, is therefore the product of the stretches, $J=\elongation_1\elongation_2\elongation_3$.

Such a deformation has an energetic cost to stretching (or compressing) the polymer chains. We consider a Neo-Hookean model for the elastic energy density 
\begin{equation} \label{eq:ElasticEnergy}
    \FreeEnergy\elastic = \frac{\shearmodulus}{2} \left[ \elongation_1^2 + \elongation_2^2 + \elongation_3^2 - 3 - 2 \log J \right],
\end{equation}
where $\shearmodulus$ is the shear modulus of the gel. Commonly, the shear modulus is found to follow a linear behaviour with temperature, $\shearmodulus=\Boltzman\temp/\polyvol$, where $\temp$ is the temperature (in Kelvin), $\Boltzman$ is the Boltzman constant and $\polyvol$ is the volume of polymer per polymer molecule in the reference state. (Note that often this is written as $\shearmodulus=\dimensional{N}\Boltzman\temp$ with $\dimensional{N}$ the number density of polymer chains in the reference state, but $\dimensional{N}$ and $\polyvol$ are simply related through $\dimensional{N}=1/\polyvol$.) 

For the mixing contribution to the free energy density, we use the Flory-Huggins polymer theory \citep{Flory1942,Huggins1942,FloryRehner1943II} to get the mixing energy density 
\begin{equation} \label{eq:MixingEnergy}
    \FreeEnergy\mix 
    = \frac{\Boltzman \temp}{\solvol} J \left[ \porosity\log\porosity + \FloryHuggins\porosity(1-\porosity) \right].
\end{equation}
Here, $\porosity$ is the porosity (i.e. the local proportion of fluid per unit mixture volume) and $\solvol$ is the volume of fluid per fluid molecule in the unmixed state. 
Note that in eqn.~\eqref{eq:MixingEnergy}, there is a factor $J$ to account for the local volume change relative to the reference state, due to deformation of the gel.

The term $\FloryHuggins$ in \eqref{eq:MixingEnergy} is the (Flory) mixing parameter, and gives the degree to which the polymer and fluid like to mix; in general, this mixing parameter can depend on both temperature and mixture composition, $\FloryHuggins = \FloryHuggins(\temp,\porosity)$. It has been suggested that both components are required to accurately reproduce experimental data \citep{LopezLeon2007}. There are many possible options for this function, but for simplicity, we focus on a relationship that is linear in both the temperature and the polymer fraction
\begin{equation} \label{eq:FloryParam}
\begin{aligned} 
    &\FloryHuggins(\temp,\porosity) = \FloryHuggins_0(\temp) + \FloryHuggins_1(\temp) (1-\porosity), \\
    \FloryHuggins_0(\temp)& = A_0 + B_0 \temp, \quad
    \FloryHuggins_1(\temp) = A_1 + B_1 \temp,
\end{aligned}
\end{equation}
where the $A,B$ are constants, as modelled by \citet{Cai2011}. We note, however, that many other models for this mixing parameter exist: constant \citep{Bertrand2016,Hong2008,Hennessy2020}, dependent only on the temperature \citep{Tomari1995,Chester2011}, or with more complicated behaviours, such as including terms inversely proportional to temperature \citep{QuesadaPerez2011,Tanaka1978} or quadratic powers \citep{Drozdov2017}.
We have chosen to use the linear relation \eqref{eq:FloryParam} because it is simple, yet still includes the important local composition-dependence. Despite this simplicity, different choices of the constants can give a range of different behaviours, both quantitatively and qualitatively. 

To relate the polymer fraction and $J$, we must define the reference state, where $J~=~1$ everywhere. For simplicity, and in line with many other works on hydrogel swelling \citep{Bertrand2016,Hennessy2020,Doi2009}, we consider the reference state to be the dry state, where the polymer is a solid block with $\porosity=0$; we note however that there is some debate as to what an appropriate reference state is, and whether such a completely dry state is indeed physical \citep{QuesadaPerez2011}. Treating the polymer chains as incompressible and the mixture as ideal, the relation between the volume change and solid fraction is then simply given by
\begin{equation} \label{eq:Jeqn}
J=\frac{1}{1-\porosity}.
\end{equation}

\subsection{Free-swelling equilibria}

We now turn to consider the equilibrium swelling of a hydrogel. Under free-swelling conditions, where no external stresses are applied to the hydrogel and the background fluid in the surrounding environment is static, the stretches in each direction must be equal, $\elongation_i=\elongation$ with $J=\elongation^3$. The  free energy density can therefore be rewritten as a function of $\elongation$ only
\begin{equation} \label{eq:FreeSwellEnergy}
    \FreeEnergy(\elongation) = \frac{3\shearmodulus}{2} \left[ \elongation^2 - 1 - 2 \log \elongation \right]
    + \frac{\Boltzman \temp}{\solvol} \left[ (\elongation^3-1)\log\left(1-\frac{1}{\elongation^3}\right) + \FloryHuggins\left(1-\frac{1}{\elongation^3}\right) \right],
\end{equation}
with $\FloryHuggins=\FloryHuggins_0(\temp) + \FloryHuggins_1(\temp)/\elongation^3$.

Equilibria are found by minimising the free energy density so that $\dd \FreeEnergy(\elongation)/\dd \elongation=0$; the equilibrium stretch, $\elongation$, must therefore satisfy the equation
\begin{equation} \label{eq:EqlEqn}
    \frac{\elongation^2 - 1}{\elongation^3}
    + \Param \left[ \log\left(1-\frac{1}{\elongation^3}\right) + \frac{1}{\elongation^3} + \frac{\FloryHuggins_0-\FloryHuggins_1}{\elongation^6} + \frac{2\FloryHuggins_1}{\elongation^9} \right] = 0.
\end{equation}
Given the mixing parameter, $\FloryHuggins(\temp,\porosity)$, the equilibrium solutions, $\elongation=\elongation(\temp;\Param)$, can then be calculated as a function of temperature with a single parameter, $\Param = \polyvol/\solvol$. This parameter encodes the relative importance of the mixing and elastic energies, but it is also the ratio of volume per polymer chain compared to a fluid molecule (or, alternatively, a ratio of their densities) and is in general expected to be large. The value of this parameter may vary depending on factors such as the gel composition and curing conditions (since these affect the stiffness of the gel), and can take values from tens \citep{Drozdov2014,QuesadaPerez2011} to hundreds \citep{Cai2011,Tanaka1978} and above \citep{Bertrand2016}.

To complete the calculation of the equilibrium swelling, it therefore remains to define the functions $\FloryHuggins_0$ and $\FloryHuggins_1$, i.e.~choose values for the constants $A$ and $B$ in \eqref{eq:FloryParam}. Even after restricting the equilibrium swelling to certain qualitative behaviour (such as being more swollen at low temperatures, with a swelling transition within a certain temperature range), there still remains a wide range of possible choices for these parameters; we consider two possibilities using equilibrium data extracted from the literature.

\begin{figure}
\centering
    \subcaptionbox{}{\input{Figs/Fig3a_Eql_CaiSuo}}
    \subcaptionbox{}{\input{Figs/Fig3b_Eql_Hirotsu}}
    \caption{Equilibrium swelling of thermo-responsive hydrogels under free-swelling conditions (solid curves) for (a) the ANB parameters \eqref{eq:Params_CaiSuo}, and (b) the HHT parameters \eqref{eq:Params_Fitted}. The stretch (or factor increase in lengths compared to the dry state), $\lambda$, is calculated as a function of temperature, $\dimensional{T}$, using eqn.~\eqref{eq:EqlEqn}. The dashed lines show the spinodal curves, where $\partial^2 \FreeEnergy/\partial \elongation_i^2=0$, with the spinodal regions being the darker shaded areas, $\partial^2 \FreeEnergy/\partial \elongation_i^2<0$. The lighter shaded regions are where the isotropically-stretched state can coexist alongside another with a discontinuous normal stretch across the interface. In (b), the fitted curve is compared to (rescaled) data from \citet{Hirotsu1987}, and the inset shows a zoom of the region close to the swelling transition.}
    \label{fig:EqlSwell}
\end{figure}
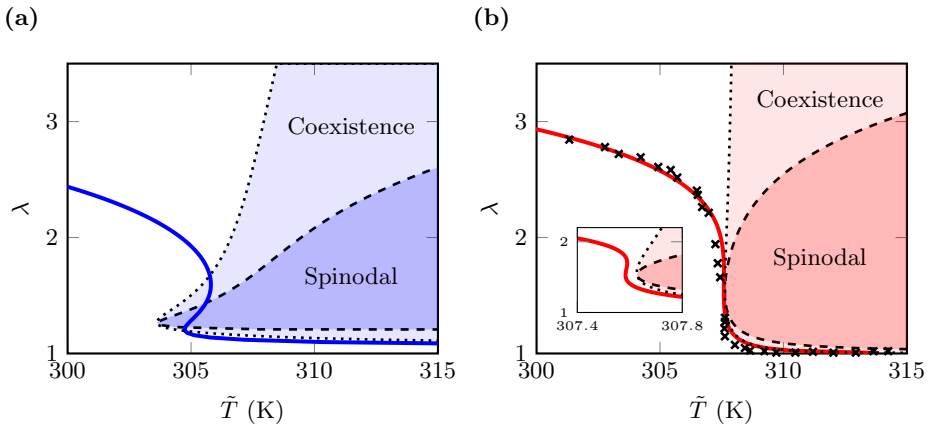

In the first case, we take the values previously used by \citet{Cai2011} in modelling the mechanics of a thermo-responsive hydrogel:
\begin{equation} \label{eq:Params_CaiSuo}
    \begin{aligned}
    \Param = 100&, \\
    A_0 = -12.947, \quad \qquad &A_1 = 17.92, \\
    B_0 = 0.04496~\text{K}^{-1}, \quad &B_1 = -0.0569~\text{K}^{-1}.
    \end{aligned}
\end{equation}
The values of $A$ and $B$ given here were determined by \citet{Afroze2000} from a prepared sample of PNIPAM. In this case, the equilibrium curve is multi-valued, taking a characteristic S-shape as shown in fig.~\ref{fig:EqlSwell}a. We shall refer to this as the Afroze-Nies-Berghmans (ANB) solution. Small changes in temperature can result in a large (discontinuous) jump in the level of equilibrium swelling. This equilibrium behaviour has been observed experimentally in some thermo-responsive hydrogels \citep[e.g.][]{Matsuo1988, Hirokawa1984,Tanaka1978}. The volume phase transition temperature for shrinking in this example is $\temp_c\approx305.8~$K, and we expect a large degree of shrinking when the temperature is increased above this. However, we note that there is a small range of temperatures over which there are multiple possible equilibrium solutions (the system exhibits hysteresis); when decreasing the temperature from an originally deswollen state, the hydrogel will only swell significantly once below the volume phase transition for swelling, $\temp_c\approx304.7~$K, which is lower than that for the shrinking transition. 

We also consider an example that is found by fitting eqn.~\eqref{eq:EqlEqn} to data captured from fig.~3 of \citet{Hirotsu1987} for another sample of PNIPAM. To do this, we rearrange \eqref{eq:EqlEqn} to $\temp=\temp(\elongation)$, and fit the appropriate 6 parameters, which are $\Param,A_0,B_0,A_1,B_1$ and the scaling between their gel radius and the stretch (i.e.~determine the radius at which $\lambda=1$, since this is unknown in the experiment). This fitting was implemented using the MATLAB least-squares fitting \texttt{nonlinlsq} from the Optimization Toolbox, running the fitting a large number of times to remove any local minima that were found. The best-fit parameters are found to be 
\begin{equation} \label{eq:Params_Fitted}
    \begin{aligned}
    \Param = 720&, \\
    A_0 = -62.22, \quad \qquad &A_1 = -58.28, \\
    B_0 = 0.20470~\text{K}^{-1}, \quad &B_1 = 0.19044~\text{K}^{-1},
    \end{aligned}
\end{equation}
with the scaling between the data and the equilibrium stretch found to be 0.4127. These parameters are noticeably different from those in \eqref{eq:Params_CaiSuo}, with each changing by at least several fold between the two examples, but this is not particularly unexpected considering $\Param$ has a wide range of reported values in the literature, depending on the choice of monomer and cross-linker and how the hydrogel is prepared.

The equilibrium curve that is calculated from these parameters is shown in fig.~\ref{fig:EqlSwell}b, with the comparison between the fitting and the data also illustrated. There is a sharp transition in the equilibrium degree of swelling close to $\temp_c\approx307.6~$K (both when swelling and shrinking). The fitted equilibrium solutions are again multi-valued around the transition, but over a much smaller region (approximately $0.01$K wide) that would certainly not be experimentally detecteable, as can be seen in the inset to fig.~\ref{fig:EqlSwell}b. This solution and parameter set shall be referred to as the `Hirotsu-Hirokawa-Tanaka' (HHT) solution, to distinguish it from the ANB case~\eqref{eq:Params_CaiSuo}.

\subsection{Coexistence and spinodal curves}

In addition to the equilibrium curves in fig.~\ref{fig:EqlSwell}, we have also plotted two shaded regions for each model hydrogel. To understand these, we first note that the  free energy density can be written as a function of the principal stretches only, $\FreeEnergy=\FreeEnergy(\elongation_1,\elongation_2,\elongation_3)$, since $J=\elongation_1\elongation_2\elongation_3=1/(1-\porosity)$. The darker shaded region is bounded by the dashed curve 
\begin{equation} \label{eq:Spinodal}
    \frac{\partial^2\FreeEnergy}{\partial \elongation_i^2} (\elongation_1=\elongation_2=\elongation_3=\elongation; \temp) = 0.
\end{equation}
Inside the dark shaded region, a uniformly swollen hydrogel is linearly unstable to uniaxial perturbations. We call this region the spinodal region, and the curve the spinodal curve.

We also consider when an isotropically swollen state, with stretch $\elongation$, can be in local equilibrium with an adjoining gel state with a different degree of swelling, with a sharp interface between them. The two stretches in directions tangential to the interface must vary continuously across the interface, $\elongation_t=\elongation$, but the stretch in the normal direction can be discontinuous, $\elongation_n=\elongation_{\mathrm{new}}\neq\elongation$, and can be either more or less swollen than the isotropic state. At the interface, the following balance condition must be satisfied \citep{Sekimoto1989,Tomari1995,Sekimoto1993}
\begin{equation} \label{eq:Coexistence}
        \partdiff{\FreeEnergy}{\elongation_n} (\elongation,\elongation;\temp) = \partdiff{\FreeEnergy}{\elongation_n} (\elongation_{\mathrm{new}}, \elongation;\temp) = \frac{\FreeEnergy(\elongation_{\mathrm{new}},\elongation;\temp)-\FreeEnergy(\elongation,\elongation;\temp)}{\elongation_{\mathrm{new}}-\elongation},
\end{equation}
where, for simplicity, we here write the arguments of the  free energy density as $\FreeEnergy=\FreeEnergy(\elongation_n,\elongation_t;\temp)$.
As we shall see later when defining the stress, the first equality here is enforcing a mechanical stress balance across the interface so that the normal stress is continuous. The second equality ensures that the two sides are in chemical equilibrium, and so the gradient, $\partial \FreeEnergy/\partial \elongation_n$, is well-defined at the interface. At a given temperature, $\temp$, eqn.~\eqref{eq:Coexistence} therefore gives two equations in two unknowns, $\elongation$ and $\elongation_{\mathrm{new}}$, which can be solved and plotted in $(\temp,\elongation)$-space; we instead choose values of $\elongation$ and solve for both $\temp$ and $\elongation_{\mathrm{new}}$. Since the  free energy density is symmetric in the principal stretches (i.e.~$\elongation_1,\elongation_2,\elongation_3$), we can solve eqn.~\eqref{eq:Coexistence} by taking derivatives with respect to any principal stretch as the normal direction, so that, for example, $\elongation_n=\elongation_1$, without loss of generality, and plot the solution as the dotted line in fig.~\ref{fig:EqlSwell}, called the coexistence curve. 
In the light shaded region, bounded by this curve, the described coexistence between the neighbouring isotropic and anisotropic states can still mechanically occur, but the latter is energetically favourable and will be expected to invade the domain. 

Considering a hydrogel that is initially isotropically swollen with a stretch $\elongation$ at a temperature $\temp$, we may expect to observe different dynamic behaviours depending upon which region of $(\temp,\elongation,)$-space the hydrogel starts in. In the spinodal region, the initial state is unstable and so we might see spontaneous swelling or shrinkage within the interior of the hydrogel due to the growth of any small perturbations --- spinodal decomposition. If, instead, the initial state is within the coexistence region, there is an energetically-favourable alternative state. However, the gel must be stimulated at some specific location to overcome the energy barrier required to generate this alternative state, which can then invade the domain --- this is nucleation and growth. Therefore, when starting in these two regions, we expect to see phase separation in the hydrogel. The behaviours of gels in the non-shaded regions of fig.~\ref{fig:EqlSwell} are slightly more complicated to predict, since as a gel swells or shrinks it is generally no longer isotropically swollen, and thus phase separation could occur after a delay, or not at all. We shall turn to investigate some of these behaviours in a spherical thermo-responsive hydrogel, but first we introduce a model for the dynamics of such a gel. 

\section{Poro-mechanics of a swelling spherical gel} \label{sec:DynamicModel}

We now turn to focus on the dynamics of a spherical hydrogel, based on the poro-elastic model of \citet{Bertrand2016} that was used to investigate gel swelling from a dry state and subsequent drying by evaporation. The main point of difference in this study is the inclusion of the effect of temperature- and composition-dependence in the mixing parameter, $\FloryHuggins(\temp,\porosity)$, to enable modelling of the thermo-responsive physics, as well as the removal of the external forcing due to the chemical potential of the surrounding fluid bath and a small difference in the form of the mixing energy density, eqn.~\eqref{eq:MixingEnergy}, since we do not include any term of the form $(1-\porosity)\log(1-\porosity)$ because its prefactor is negligible \citep{Engelsberg2013}. We repeat some key points of the mechanical model here for completeness. 

\subsection{Strains}

For a sphere undergoing a spherically-symmetric deformation with a radial displacement, $\dimensional{u}(\dimensional{r},\dimensional{t})$, the principal stretches are \citep{Bertrand2016}
\begin{equation}
    \elongation_{r} = \left( 1 - \partdiff{\dimensional{u}}{\dimensional{r}} \right)^{-1}, \qquad
    \elongation_\theta = \elongation_\varphi = \left( 1 - \frac{\dimensional{u}}{\dimensional{r}} \right)^{-1},
\end{equation}
where $\dimensional{r}$ is the radial coordinate from the centre of the sphere, $\theta$ and $\varphi$ are the spherical angle coordinates, and $\dimensional{t}$ is time. This can be seen by considering a small piece of the gel that is initially at a radial distance $\dimensional{r}_0$ in the undeformed reference state; once deformed, it is locally stretched by an amount $\partial \dimensional{r}/\partial \dimensional{r}_0$ in the radial direction and $\dimensional{r}/\dimensional{r}_0$ in the angular directions, and we have that $\dimensional{u}=\dimensional{r}-\dimensional{r}_0$.

We can then relate the stretches to the porosity through the local volume change, $J$, since we have $J = \elongation_r \elongation_\theta^2$, but also eqn.~\eqref{eq:Jeqn} holds (since the reference volume is taken to be the dry state with $\porosity=0$).  
Combining these relations for $J$, we find a first order differential equation in $\dimensional{r}$ for the displacement $\dimensional{u}$. This can be integrated once to determine the displacement in terms of the porosity
\begin{equation}
    \dimensional{u}(\dimensional{r},\dimensional{t}) = \dimensional{r} - \left( \dimensional{r}^3-3\intstart{0}{\dimensional{r}} x^2 \porosity(x,\dimensional{t}) ~\dd x \right)^{1/3},
\end{equation}
with $\dimensional{u}(0,\dimensional{t})=0$ at the centre of the sphere and $\dimensional{u}(\dimensional{a},\dimensional{t})=\dimensional{a}-\dryradius$ at the sphere edge.

Therefore, given the porosity, $\phi(\dimensional{r},\dimensional{t})$, in the gel sphere, we can calculate the radial displacements, $\dimensional{u}(\dimensional{r},\dimensional{t})$, and hence the stretches, $\lambda_i(\dimensional{r},\dimensional{t})$. 

\subsection{Stresses}

Having determined the strains in the sphere, it remains to calculate the stresses. The mechanical stresses, $\dimensional{\stress}$, acting on the gel can be written in terms of gradients of the  free energy density, $\FreeEnergy$, given in eqns.~\eqref{eq:FreeEnergy}--\eqref{eq:MixingEnergy}. We decompose the stress components into contributions from the elastic deformation and the chemical mixing 
\begin{equation}
    \dimensional{\stress}_i = \dimensional{\stress}_i' - \dimensional{p}, 
\end{equation}
where $\dimensional{\stress}_i$ are the principal Cauchy stresses within the mixture. The Terzaghi effective stress (which is the elastic contribution) is defined by 
\begin{equation}
    \dimensional{\stress}_i' = \frac{\elongation_i}{J} \partdiff{\FreeEnergy\elastic}{\elongation_i} = G \frac{\elongation_i^2-1}{J}.
\end{equation}
The pressure is split into an osmotic pressure, $\dimensional{\osmotic}$, due to the interaction of the fluid and polymer, and the chemical potential, $\dimensional{\chempotential}$,
\begin{equation} \label{eq:pressure}
    \dimensional{p} = \dimensional{\osmotic} + \frac{\dimensional{\chempotential}}{\solvol},
\end{equation}
where the osmotic pressure is given by
\begin{equation}
    \dimensional{\osmotic} = - \diff{\FreeEnergy\mix}{J} = -\frac{\Boltzman \temp}{\solvol} \left[ \log\left(1-\frac{1}{J}\right) + \frac{1}{J} + \frac{\FloryHuggins_0-\FloryHuggins_1}{J^2} + \frac{2\FloryHuggins_1}{J^3} \right].
\end{equation}
The justification for the elastic stress and osmotic pressure being written in terms of derivatives of the  free energy density can be found by considering the work done when changing the hydrogel mixture by a small volume locally \citep[see eqn.~4 \& 5 of][]{Bertrand2016}. More generally, the stress tensor can be written in terms of derivatives of the  free energy density with respect to the deformation tensor \citep[e.g.][]{Doi2009,Tadmor2012}, but we stick to the simpler spherically symmetric model outlined above.

The term $\dimensional{\chempotential}/\solvol$ in eqn.~\eqref{eq:pressure} arises to account for the work required to move fluid in to the mixture. 
An alternative view of this term is that it is the pervadic pressure of the fluid in the mixture --- a connected fluid-only bath in local equilibrium with the fluid in the mixture would be at this pressure \citep{Peppin2005,Etzold2021}.

Note that the precise value of the chemical potential does not affect the equilibrium calculations, since it must be uniform and equal to the ambient value in the surrounding fluid. As we shall see, only gradients in the chemical potential alter the dynamics. 
In reality, the value of the ambient chemical potential should differ at each temperature, but this shall turn out to be unimportant for our simulations as we only consider dynamics at a fixed temperature. More care may be needed in more complex scenarios, but note that adding a uniform amount to the chemical potential only results in a uniform increase in the stress, with otherwise no effect on our results. 

\subsection{Poro-elastic flow}

To resolve the dynamic behaviour of the swelling and shrinking hydrogel sphere, we need to understand the fluid flow within the gel pore space. The fluid velocity, $\fluidvelocity$, relative to the motion of the gel, $\polyvelocity$, is driven by the chemical potential gradient (Darcy flow): 
\begin{equation} \label{eq:Darcy}
    \porosity(\fluidvelocity-\polyvelocity) = - \frac{\dimensional{\permeability}(\porosity)}{\viscosity} \partdiff{}{\dimensional{r}}\left( \frac{\dimensional{\chempotential}}{\solvol} \right),
\end{equation}
where $\viscosity$ is the fluid viscosity and $\dimensional{\permeability}$ is the permeability of the hydrogel. Note that the Darcy flow here is driven only by the chemical potential because, as previously mentioned, it is equivalent to the pressure of the fluid within the mixture by itself \citep{Peppin2005}. This Darcy flow is what makes the dynamic modelling different from that of \citet{Tomari1995}, since here we explicitly account for the interstitial fluid flow relative to the solid matrix, rather than accounting for it through an energy dissipation via friction.

We take the permeability to be isotropic and a function of the porosity, $\dimensional{\permeability}(\porosity)$. For simplicity, we shall present results with a constant permeability, $\dimensional{\permeability}_0$, but keep a general form in our equations. We discuss an alternative permeability model used by \citet{Bertrand2016} in App.~\ref{app:VaryPermeability}.

As the hydrogel swells or shrinks, any deformation of the solid structures must be accompanied by a corresponding displacement of fluid; in this manner, conservation of solid volume dictates that the porosity must obey
\begin{equation} \label{eq:MassConservation}
    \partdiff{}{\dimensional{t}}(1-\porosity) + \frac{1}{\dimensional{r}^2} \partdiff{}{\dimensional{r}} [\dimensional{r}^2 (1-\porosity)\polyvelocity] = 0,
\end{equation}
and a similar equation holds for the fluid velocity $\fluidvelocity$ with $\porosity$ replacing $1-\porosity$. Combining these two relations gives a simple relation of no net flux in any cross-section
\begin{equation} \label{eq:VolConservation}
    \porosity \fluidvelocity + (1-\porosity) \polyvelocity = 0.
\end{equation}

Combining the equations \eqref{eq:Darcy}--\eqref{eq:VolConservation}, we find a partial differential equation (PDE) for the evolution of the porosity in terms of the gradient of the chemical potential 
\begin{equation} \label{eq:PDE}
    \partdiff{\porosity}{\dimensional{t}} = \frac{1}{\dimensional{r}^2} \partdiff{}{\dimensional{r}} \left[ \dimensional{r}^2(1-\porosity)\frac{\dimensional{k}(\porosity)}{\viscosity \solvol} \partdiff{\dimensional{\chempotential}}{\dimensional{r}} \right].
\end{equation}
Note that this is a Reynolds' equation of the form $\partial\porosity/\partial \dimensional{t} + \nabla \cdot \mathbf{Q} = 0$, with a radial flux $Q = -[(1-\porosity)\dimensional{k}(\porosity) /(\viscosity\solvol)] \times \partial\dimensional{\chempotential}/\partial \dimensional{r}$. Conservation of fluid volume then means that the sphere radius changes due to the outward radial flux at the sphere edge
\begin{equation}
    \diff{\dimensional{a}}{\dimensional{t}} = \frac{\dimensional{\permeability}(\porosity)}{\viscosity\solvol} \partdiff{\dimensional{\chempotential}}{\dimensional{r}} \qquad \text{at }\dimensional{r}=\dimensional{a}.
\end{equation}

A stress balance in the solid enforces that $\nabla\cdot\dimensional{\stress}=0$; considering the radial component of this, we can determine the chemical potential gradient in terms of the Terzaghi stresses and osmotic pressure via
\begin{equation}
    \solvol^{-1} \partdiff{\dimensional{\chempotential}}{\dimensional{r}} = \partdiff{\dimensional{\stress}_r'}{\dimensional{r}} + 2 \frac{\dimensional{\stress}_r'-\dimensional{\stress}_\theta'}{\dimensional{r}} - \partdiff{\dimensional{\osmotic}}{\dimensional{r}}.
\end{equation}

\subsection{Non-dimensionalisation}

We rescale the governing equations using the dry sphere radius $\dryradius$, the permeability scale $\dimensional{\permeability}_0$, the stress scale $\Boltzman \tempend/\polyvol$ (for a final temperature $\tempend$), the chemical potential scale $\Boltzman \tempend\solvol/\polyvol$ and the timescale 
\begin{equation}
\timescale = \frac{\viscosity \polyvol \dryradius^2}{\dimensional{\permeability}_0\Boltzman \tempend}.
\label{eq:timescale}
\end{equation}
Note that the timescale here is proportional to the square of the size of the gel sphere, $\timescale\propto \dryradius^2$, just as was found for the swelling of gels by \citet{Tanaka1979}; here, the effective diffusivity is $\dimensional{D} \sim (\dimensional{\permeability}_0\Boltzman \tempend) / (\viscosity \polyvol)$. In general, we expect this timescale to be much longer than that for thermal diffusion: \citet{Tanaka1979} reported a diffusivity of $3\times10^{-7}~\text{cm}^2/\text{s}$ which is $10^4$ times smaller than a typical thermal diffusivity for water, suggesting that swelling typically takes $100$ times longer than the diffusion of heat. We obtain a similar effective diffusivity for our swelling gel model when $\viscosity=10^{-3}~\text{Pa s}$, $\polyvol=3\times10^{-25}~\text{m}^3$ and $\dimensional{\permeability}_0=8\times10^{-20}~\text{m}^3$ \citep{Bertrand2016}. We therefore model temperature changes as being effectively instantaneous and uniform over the whole system. 

We also note that, in general, the fluid viscosity, $\viscosity$, may be temperature-dependent. Since this viscosity only appears in the dimensionless model through the timescale, $\timescale$, defined in eqn.~\eqref{eq:timescale}, and all simulations are to be conducted at fixed temperature, this does not affect any of our dimensionless results. However, there will be a quantitative effect that must be accounted for when considering the real dimensional dynamic swelling and shrinking times.

From this point onwards, we write all equations and results in dimensionless terms, removing the tildes used in earlier equations that denote the dimensional forms of variables. The only remaining dimensional quantity is the temperature, $\temp$, since its dimensional value matters, although in all of our equations it shall be non-dimensionalised by multiplying by $B_0$ or $B_1$ in $\FloryHuggins$. 

The dimensionless PDE for the evolution of the porosity is
\begin{equation} \label{eq:PDE_Dimless}
    \partdiff{\porosity}{t} = \frac{1}{r^2} \partdiff{}{r} \left[ r^2(1-\porosity)k(\porosity) \partdiff{\chempotential}{r} \right],
\end{equation}
for a dimensionless permeability $k(\porosity)$ (which we take to be $\permeability=1$). The chemical potential gradient is given in terms of the stresses
\begin{equation} \label{eq:ChemPotential_Dimless}
    \partdiff{\chempotential}{r} = \partdiff{\Terzaghi_r}{r} + 2 \frac{\Terzaghi_r-\Terzaghi_\theta}{r} - \partdiff{\osmotic}{r}.
\end{equation}
These stresses are defined in terms of the strains through
\begin{equation} \label{eq:Stress_Dimless}
    \Terzaghi_i = \frac{\elongation_i^2-1}{J},
\end{equation}
and similarly for the osmotic pressure
\begin{equation} \label{eq:Osmotic_Dimless}
    \osmotic = - \Param \left[ \log\left(1-\frac{1}{J}\right) + \frac{1}{J} + \frac{\FloryHuggins_0-\FloryHuggins_1}{J^2} + \frac{2\FloryHuggins_1}{J^3} \right],
\end{equation}
where $\Param=\polyvol/\solvol$ is the previously mentioned parameter that relates the relative importance of the mixing and elastic energies.

The strains can be calculated directly from the porosity, $\porosity(r,t)$, by first determining the radial displacement
\begin{equation} \label{eq:Displacement_Dimless}
    u = r - \left( r^3 - 3\intstart{0}{r} x^2 \porosity ~\dd x \right)^{1/3}.
\end{equation}
The stretches and volume change are then
\begin{equation} \label{eq:Strain_Dimless}
    J = \frac{1}{1-\porosity}, \quad \elongation_\theta = \frac{1}{1-u/r}, \quad \elongation_r = \frac{J}{\elongation_\theta^2}.
\end{equation}

The sphere edge moves according to the chemical potential gradient there 
\begin{equation} \label{eq:SphereEvolve_Dimless}
    \diff{a}{t} =  \permeability(\porosity) \partdiff{\chempotential}{r} \qquad \text{at }r=a.
\end{equation}

Note that at the sphere edge, the displacement is $u=a-1$ and $\lambda_\theta=a$. We also enforce chemical equilibrium, $\chempotential=\externalchempot$, and a radial stress balance, $\stress_r=-\externalchempot$, at the sphere boundary, where $\externalchempot$ is the chemical potential of the surrounding background fluid. Note that the radial stress just inside the sphere is not zero because it must balance the external fluid pressure (c.f. pervadic pressure). Combining these two conditions enforces that $\Terzaghi_r=\osmotic$, which gives a nonlinear equation for $J$ at the sphere edge 
\begin{equation} \label{eq:Boundary_Dimless}
    \frac{J}{a^4} - \frac{1}{J} + \Param \left[\log\left(1-\frac{1}{J}\right) + \frac{1}{J} + \frac{\FloryHuggins_0-\FloryHuggins_1}{J^2} + \frac{2 \FloryHuggins_1}{J^3} \right] = 0.
\end{equation}
Note that these boundary conditions mean that the hydrogel dynamics have no dependence on the external chemical potential, $\externalchempot$. This differs from the model of \citet{Bertrand2016}, where there is a stress discontinuity at the sphere edge, but does align with other models \citep[e.g.][]{Hennessy2020}.

Further, we are neglecting any mechanical feedback from the subsequent external fluid flow. We expect the outside flow to be unimportant to the swelling dynamics because the polymer's incompressibility means that any fluid flux through the boundary must be exactly compensated for by the boundary motion itself (except for, perhaps, some inhomogeneity at the pore-scale). Some recent studies have shown responsive hydrogels are able to generate substantial bulk fluid motions, such as cyclic swelling of bilayer ribbons causing net swimming \citep{Tanasijevic2022}, but in this reported case this was possible because of some significant asymmetry in the material composition of the particle due to impermeable sections of the boundary, that caused shape change and asymmetric fluid flow across the boundary.

In summary, our model comprises a Neo-Hookean free energy density for deformation \eqref{eq:ElasticEnergy}, the Flory-Huggins free energy density of mixing for a non-ionic gel \eqref{eq:MixingEnergy}, a particular form for the temperature- and composition-dependence of the mixing parameter \eqref{eq:FloryParam}, and poro-elastic dynamics with fluid motion given by Darcy flow \eqref{eq:Darcy}. This combination results in the dimensionless governing equations for the evolution of the swelling or shrinking thermo-responsive gel, given in eqns.~\eqref{eq:PDE_Dimless}--\eqref{eq:Boundary_Dimless}.

\subsection{Numerical scheme}

We numerically solve these equations using a spatial grid in the range of $N=200-400$ points and a timestep of $\Delta t = 10^{-7}-10^{-8}$, with results recorded every $t=10^{-4}$. Such a small timestep was used to maintain numerical stability with high spatial resolution for this simple numerical scheme. Details of the scheme are given in Appendix \ref{app:NumericalScheme}. 

\section{Dynamics of a swelling or shrinking sphere} \label{sec:Results}

To investigate the dynamics of a thermo-responsive hydrogel sphere, we consider scenarios where the sphere is initially in equilibrium at temperature $\temp=\tempstart$, before applying an instantaneous change in the temperature to $\temp=\tempend$. In particular, we consider temperature changes across the volume phase transition temperature. These are considered for both of the example gels introduced in \S\ref{sec:Eql}; we begin by considering the results for the \citet{Cai2011} parameters (ANB), before looking at the results from the parameter fitting to the \citet{Hirotsu1987} data (HHT) in \S\ref{sec:Hirotsu_Dynamics}.

\begin{figure}
\centering
    \subcaptionbox{}{\input{Figs/Fig4a_Trajectory}}
    \subcaptionbox{}{\input{Figs/Fig4b_Trajectory}}
    \caption{An illustration of the dynamic trajectories that are simulated for the two model hydrogels: (a) ANB and (b) HHT. The paths labelled (i)--(iii) correspond to the results presented in the main text (in the order they occur). Immediately after the instantaneous temperature change, the hydrogels are at the points labelled by crosses, and subsequently evolve towards equilibrium at a fixed temperature. The inset shows a zoom close to the volume phase transition, along with path (iii). Note that during the dynamics, the stretches in the radial and angular directions will vary, and so the true paths do not remain on this diagram of isotropic stretch, which is merely illustrative.}
    \label{fig:Trajectories}
\end{figure}
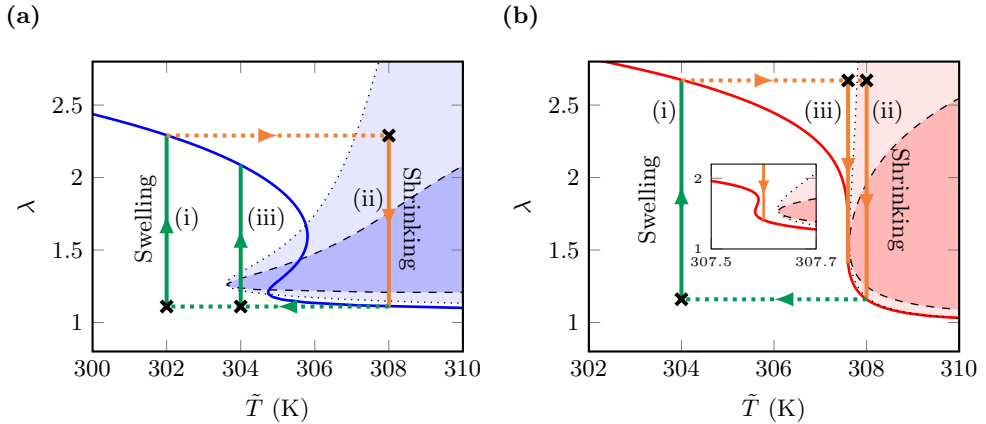

\subsection{Results for the ANB parameter set} \label{sec:CaiSuo_Dynamics}

Taking the ANB parameters \eqref{eq:Params_CaiSuo}, we obtain the multi-valued equilibrium curve shown in blue in fig.~\ref{fig:Trajectories}a with a characteristic S-shape. We will consider the three dynamic trajectories illustrated in fig.~\ref{fig:Trajectories}a.
We first consider the swelling dynamics: starting at higher temperatures, with a gel at equilibrium on the lower branch of the curve, the temperature is instantaneously decreased below $\temp_c\approx304.7~$K (the volume phase transition temperature for swelling). The gel sphere swells significantly, evolving according to the equations presented in \S\ref{sec:DynamicModel}.

\begin{figure}
\centering
    \subcaptionbox{}{\input{Figs/Fig5a_ColourMap_CaiSuo_Swell_NoFront}}
    \subcaptionbox{}{\input{Figs/Fig5b_Porosity_CaiSuo_Swell_NoFront}}
    \caption{Swelling of a hydrogel for the ANB equilibrium curve. The temperature is decreased from $\tempstart=308~$K to $\tempend=302~$K. In (b), porosity profiles are plotted at fixed times from $t=0$ to $t=0.2$ in intervals of $t=0.02$, alongside the expected final equilibrium profile (dashed line). The dotted profiles are to illustrate the  late-time dynamics, and are plotted at times $t=0.4,0.6,0.8,1$.
    } 
    \label{fig:Results_CaiSuo_SwellNoFront}
\end{figure}
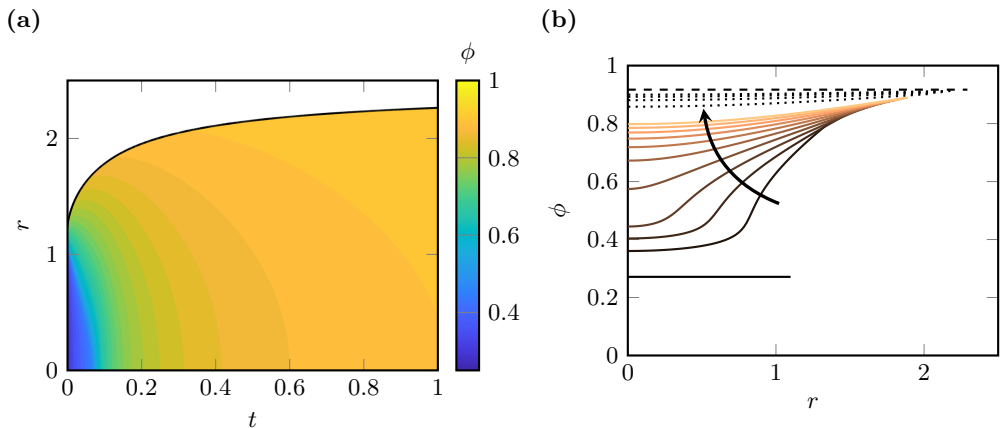

Results for the dynamic swelling of a gel from an initial temperature $\tempstart=308~$K to a final temperature $\tempend=302~$K (trajectory (i) in fig.~\ref{fig:Trajectories}), are shown in fig.~\ref{fig:Results_CaiSuo_SwellNoFront}. The sphere swells from an initial state with radius $a=1.11$ and uniform porosity $\porosity=0.27$ towards an equilibrium that is over twice as big, with radius $a=2.29$ and porosity $\porosity=0.92$. The swelling initiates at the edge of the sphere and propagates smoothly inwards, towards the centre, with the system getting close to equilibrium after a dimensionless time $t=1$.

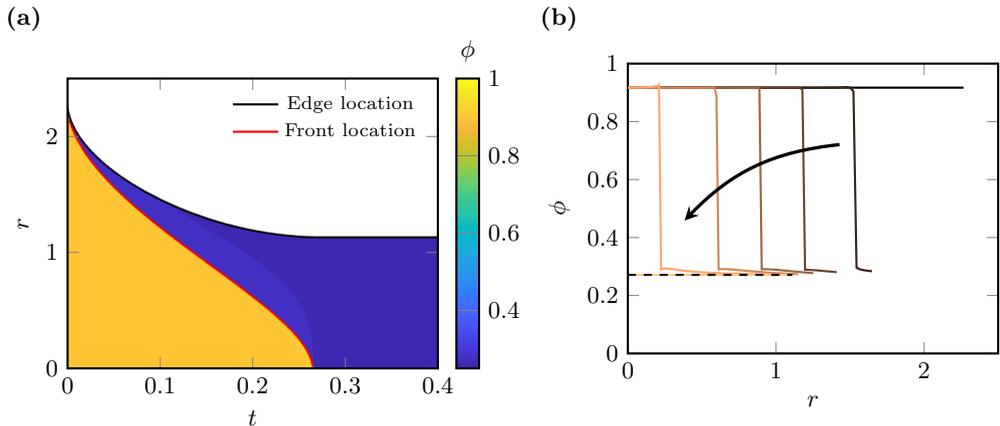
\begin{figure}
\centering
    \subcaptionbox{}{\input{Figs/Fig6a_ColourMap_CaiSuo_Shrink_Front}}
    \subcaptionbox{}{\input{Figs/Fig6b_Porosity_CaiSuo_Shrink_Front}}
    \caption{Shrinking of a hydrogel for the ANB equilibrium curve. The temperature is increased from $\tempstart=302~$K to $\tempend=308~$K. In (b), porosity profiles are plotted at fixed times from $t=0$ to $t=0.3$ in intervals of $t=0.05$, alongside the expected final equilibrium profile (dashed line).} 
    \label{fig:Results_CaiSuo_ShrinkFront}
\end{figure}

When the opposite scenario is considered, with the temperature increasing from $\tempstart=302~$K to $\tempend=308~$K to promote shrinking of the gel (see the trajectory (ii) in fig.~\ref{fig:Trajectories}), the dynamics is starkly different. In this case, the temperature is increased above $\temp_c\approx305.8~$K (the volume phase transition temperature for shrinking), resulting in significant shrinking. In fig.~\ref{fig:Results_CaiSuo_ShrinkFront}, we show the evolution of the sphere after this instant heating. 

Similar to the swelling dynamics, the shrinking begins from the edge and propagates inwards. However, in this case, the sphere separates into a swollen core and shrunken shell with a sharp boundary between them; we call this sharp transition in porosity/swelling a `front'. This front (whose location is illustrated in red in fig.~\ref{fig:Results_CaiSuo_ShrinkFront}a) forms close to the edge, and moves inwards towards the centre, driving the dynamics of the shrinking. 

Looking at the paths in $(\temp,\elongation)$-space shown in fig.~\ref{fig:Trajectories}a, we see that a key difference between the swelling and shrinking is that, after the instantaneous temperature change when shrinking, the initial homogeneous state is transported to the coexistence region in the equilibrium diagram, illustrated by the lighter shaded area. After this, a shrunken state quickly forms at the sphere edge that is separated from the swollen core by a sharp front; this shrunken phase is energetically favourable and invades the domain.

We can also see in fig.~\ref{fig:EqlSwell}a, that the coexistence and spinodal regions extend a small amount to the left of the S-bends of the equilibrium curve. From an initially deswollen equilibrium state, although it is not possible to move directly to the coexistence region, it may be possible that a decrease in the temperature just past the volume phase transition for swelling could result in the gel being affected in a similar manner as it swells; the question is then: is this phase separation behaviour seen for these swelling cases?

\begin{figure}
    \centering
    \subcaptionbox{}{\input{Figs/Fig7a_ColourMap_CaiSuo_Swell_Front}}
    \subcaptionbox{}{\input{Figs/Fig7b_Porosity_CaiSuo_Swell_Front}}
    \caption{Swelling of a hydrogel for the ANB equilibrium curve. The temperature is decreased from $\tempstart=308~$K to $\tempend=304~$K. In (b), porosity profiles are plotted {at fixed times} from $t=0$ to $t={0.5}$ in intervals of $t={0.05}$, alongside the expected final equilibrium profile (dashed line). The dotted profiles are to illustrate the early- and late-time dynamics, and are plotted at times $t=0.001,0.01$ and $t=1$. 
    }
    \label{fig:Results_CaiSuo_SwellFront}
\end{figure}
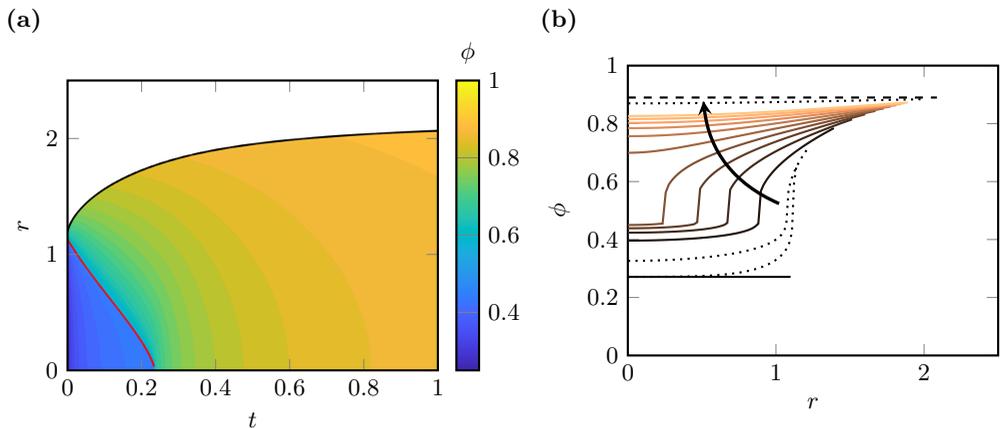

To test this, we calculated the evolution of the hydrogel sphere from an initial temperature $\tempstart=308~$K to a final temperature $\tempend=304~$K (just below the $\temp_c\approx304.7~$K, shown by path (iii) in fig.~\ref{fig:Trajectories}). The results are shown in fig.~\ref{fig:Results_CaiSuo_SwellFront}. Largely, the profiles look similar to those in fig.~\ref{fig:Results_CaiSuo_SwellNoFront}, but the profiles of $\phi$ at given times shown in fig.~\ref{fig:Results_CaiSuo_SwellFront}b {now do have a sharp jump in porosity.} {A} sharp front has formed in {the interior of the gel}, joining two smoothly varying profiles together with a porosity jump of around $\Delta \phi \approx 0.1-0.2$, although it is {noticeably} less {pronounced} than the front in fig.~\ref{fig:Results_CaiSuo_ShrinkFront}.

\subsection{Results for the {HHT parameter set}} \label{sec:Hirotsu_Dynamics}

We also consider the swelling and shrinking of the hydrogel from the HHT data. As discussed in \S\ref{sec:Eql}, we take the parameters \eqref{eq:Params_Fitted} that are fitted from the data of \citet{Hirotsu1987}, which results in the red solid curve plotted in fig.~\ref{fig:Trajectories}b, alongside the dynamic paths that we consider. 

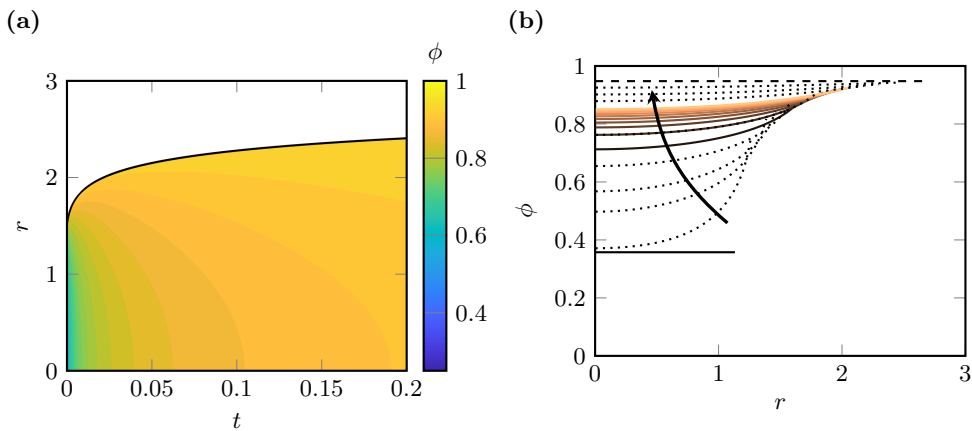
\begin{figure}
\centering
    \subcaptionbox{}{\input{Figs/Fig8a_ColourMap_Hirotsu_Swell_NoFront}}
    \subcaptionbox{}{\input{Figs/Fig8b_Porosity_Hirotsu_Swell_NoFront}}
    \caption{Swelling of a hydrogel for the HHT equilibrium curve. The temperature is decreased from $\tempstart=308~$K to $\tempend=304~$K. In (b), porosity profiles are plotted at fixed times from $t=0$ to $t={0.05}$ in intervals of $t=0.005$, alongside the expected final equilibrium profile (dashed line). The dotted profiles are to illustrate the early- and late-time dynamics, and are plotted at times $t=0.0001,0.0005,0.001,0.0025$ and $t=0.1,0.2,0.5$. 
    } 
    \label{fig:Results_Hirotsu_SwellNoFront}
\end{figure}

For these parameters, we again first consider the swelling dynamics. In fig.~\ref{fig:Results_Hirotsu_SwellNoFront}, we plot the porosity as a function of space and time following a temperature decrease from $\tempstart=308~$K to $\tempend=304~$K, where we expect to see a swelling of the hydrogel from an initial porosity $\porosity=0.36$ and radius $a=1.16$ to a final porosity $\porosity=0.95$ and radius $a=2.67$ (trajectory (i) in fig.~\ref{fig:Trajectories}b). The dynamics of swelling appears qualitatively similar to the swelling seen in \S\ref{sec:CaiSuo_Dynamics}, although occurs much faster. Note, however, that in contrast to the ANB case, the spinodal and coexistence region{s} do not cross the equilibrium curve. As such, there is no opportunity to find a swelling solution in which a front develops.

\begin{figure}
    \centering
    \subcaptionbox[width=0.495\textwidth]{}{\input{Figs/Fig9a_ColourMap_Hirotsu_Shrink_Front}}
    \subcaptionbox{}{\input{Figs/Fig9b_Porosity_Hirotsu_Shrink_Front}}
    \caption{Shrinking of a hydrogel for the HHT equilibrium curve. The temperature is increased from $\tempstart=304~$K to $\tempend=308~$K. In (b), porosity profiles are plotted at fixed times from $t=0$ to $t={0.4}$ in intervals of $t={0.05}$, alongside the expected final equilibrium profile (dashed line).}
    \label{fig:Results_Hirotsu_ShrinkFront}
\end{figure}
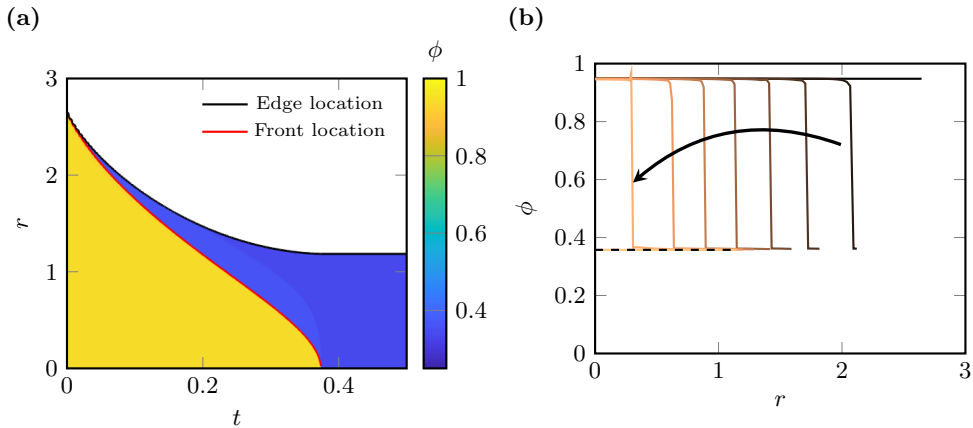

Reversing the temperature change, so that $\temp$ increases from $304~$K to $308~$K (path (ii) in fig.~\ref{fig:Trajectories}), we obtain the results shown in fig.~\ref{fig:Results_Hirotsu_ShrinkFront}. The dynamics of shrinking appears to again be dominated by the formation of a sharp front that travels radially inwards. 

\begin{figure}
    \centering
    \subcaptionbox{}{\input{Figs/Fig10a_ColourMap_Hirotsu_Shrink_NoFront}}
    \subcaptionbox{}{\input{Figs/Fig10b_Porosity_Hirotsu_Shrink_NoFront}}
    \caption{Shrinking of a hydrogel for the HHT equilibrium curve. The temperature is increased from $\tempstart=304~$K to $\tempend={307.6}~$K. In (b), porosity profiles are plotted at fixed times from $t=0$ to $t={6}$ in intervals of $t={0.5}$, alongside the expected final equilibrium profile (dashed line). The dotted profiles are to illustrate the faster dynamics at later times, and are plotted at times $t=5.25,5.75$.} 
    \label{fig:Results_Hirotsu_ShrinkNoFront}
\end{figure}
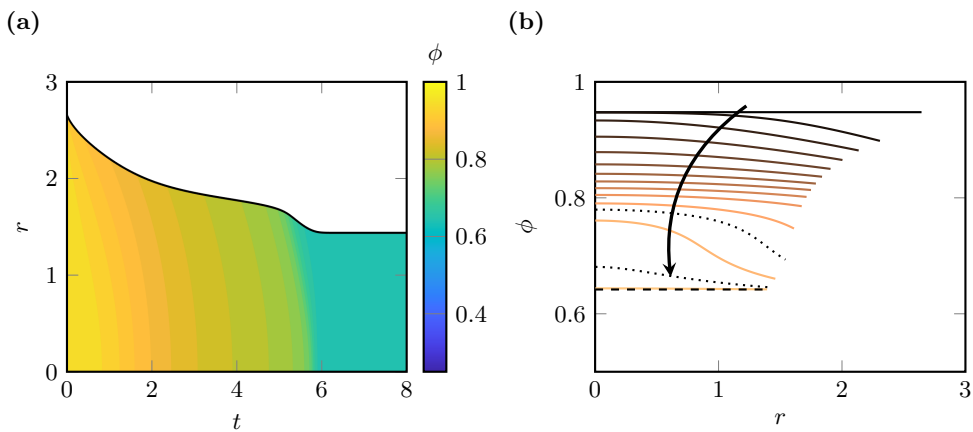

For this gel, however, we note that it is now possible to shrink by increasing the temperature past the volume phase transition temperature without entering, or at least lingering, in either the coexistence or spinodal regions (see path  (iii) and inset to fig.~\ref{fig:Trajectories}). When considering a smaller temperature increase from $\tempstart=304~$K to $\tempend=307.6~$K, shown in fig.~\ref{fig:Results_Hirotsu_ShrinkNoFront}, we no longer find that a front is formed; the gel shrinks smoothly over a dimensionless time of order $t=6$. In this case, although the gel still undergoes a significant shrinkage from $a=2.67$ to $a=1.41$, its path in the $(\temp,\elongation)$-space of fig.~\ref{fig:EqlSwell}b does not pass through the coexistence or spinodal regions, and phase separation does not occur. 

We note, however, that the shrinking here occurs in two stages with different timescales. We believe that, in this specific case, it is because the temperature is only just above the volume phase temperature: the dynamics are `close to equilibrium' as the dynamic path passes close to the fold in the equilibrium curve, since it is near a swollen equilibrium that that exists at marginally smaller temperatures. This slowing down behaviour near a bifurcation is well-known, often referred to as `critical slowing down' or a `bottleneck' due to the `ghost' of a nearby equilibrium \citep{Gomez2017,Virgin1986,Tredicce2004}.

\subsection{Comparison of {results}}

In both model systems, we have seen a range of different dynamic behaviours.
Both cases exhibited swelling that progressed smoothly inwards, and shrinking that resulted in a sharp travelling front between a swollen core and shrunken shell. In each model gel, we were also able to find an example of the opposite behaviour.

The occurrence of front formation during shrinking seems to align with whether the initial homogeneous state is within the bounds of the coexistence regions that are illustrated in fig.~\ref{fig:EqlSwell}. When starting there, the gel initially phase separates close to the boundary, forming two regions with different degrees of swelling; the newly created shrunken state then spreads inwards, replacing the swollen core. 

Similar behaviour for shrinking gels was also found by \citet{Tomari1995}. They showed that (for their model hydrogel) there are equilibria that appear linearly stable but exist within the coexistence region and so are thermodynamically unstable, and calculated the intersection between the coexistence and equilibrium curves. Their dynamic simulations revealed an inward-propagating front, similar to ours, but with an an overshoot in the amount of shrinking behind (outside) the front that we do not observe. It is not clear whether this difference is a numerical artifact or related to the change in the dynamic model to incorporate Darcy flow.

We also found that phase separation, although apparently common in the phase-space of our shrinking hydrogel examples, does not always occur when heating a thermo-responsive hydrogel. For small temperature changes above the transition, one of our model hydrogels (see fig.~\ref{fig:Results_Hirotsu_ShrinkNoFront}) shrank smoothly with no front formation.

Swelling of the hydrogel generally occurs smoothly towards the ultimate equilibrium state, since the dynamics are often not affected by the phase separation mechanisms seen in the shrinking dynamics; the coexistence and spinodal regions are mostly at higher temperatures. However, we did find that phase separation could occur when swelling in some cases, with the formation of a small but sharp front in one of our examples (fig.~\ref{fig:Results_CaiSuo_SwellFront}). Here, the initial condition was not in the coexistence region, and so no front immediately formed, but as the solution evolved, some (now anisotropic) part of the hydrogel passed its coexistence limit and a small inward-travelling front formed between a swollen shell and a shrunken core. (Recall that the coexistence curve given by eqn.~\ref{eq:Coexistence} was calculated for an isotropic hydrogel; a similar equation holds for hydrogels in an anisotropic state but with normal and tangential stretches that differ from one another.) Similar behaviour may explain the two-stage dynamics observed in the experiments of \citet{Matsuo1988} for a swelling thermo-responsive hydrogel just above the transition temperature.

Although the shaded coexistence and spinodal regions illustrated in fig.~\ref{fig:EqlSwell} are only valid for an isotropically swollen hydrogel, and as the hydrogel swells or shrinks it will in general not remain isotropically swollen, they are still useful in determining when this delayed front formation may occur. This is because, in this enforced spherical symmetry, the centre of the sphere must always remain isotropically-stretched --- the radial and angular stretches converge near the centre, $\elongation_r-\elongation_\theta\to0$ as $r\to0$, since the stretches are given by \eqref{eq:Strain_Dimless} and $u/r\to\partial u/\partial r$ as $r\to0$ (consider a Taylor expansion close to $r=0$ with $u(r=0)=0$). As such, any trajectory (see fig.~\ref{fig:Trajectories}) at fixed temperature that begins outside the coexistence region and passes through the coexistence curve will be expected to undergo (delayed) phase separation. For example, shrinking the ANB hydrogel from $302$K to $306$K resulted in delayed front formation (see App.~\ref{app:OtherDynamics}).

Beyond this, we have not presented any results that started in the spinodal region. Simulations in this case exhibit spontaneous phase separation, with shrunken regions appearing in the interior of the hydrogel (see App.~\ref{app:OtherDynamics}). Similar dynamics have been observed in one-dimensional hydrogels \citep{Hennessy2020}. However, we do not explore this more fully because the spherically-symmetric constraint of our model causes unrealistic results: the internal phase separation observed is not likely to occur at a given radius all at once to form shrunken concentric spherical shells, but instead in small localised regions across the whole hydrogel that break the spherical symmetry. 

We have therefore characterised and classified a range of different dynamic behaviours at fixed temperature using the coexistence and spinodal regions shown in fig.~\ref{fig:EqlSwell}. The importance of the coexistence and spinodal conditions have been previously highlighted in thermo-responsive hydrogels \citep[e.g.][]{Sekimoto1993}, and our results are an extension of the predictions of \citet{Tomari1995}, who found phase separation behaviour inside the coexistence region that made a small range of apparently stable equilibrium states close to the transition to be effectively unstable in practice. We have demonstrated that this dynamic behaviour extends beyond the transition region and found where it occurs. 

The results from our two model hydrogels suggest that phase separation is much common for shrinking hydrogels than those that are swelling. In many scenarios, such as during the expulsion phase of a drug-laden hydrogel, it is particularly important to understand the shrinking behaviour and dynamics of these thermo-responsive hydrogels, and we have seen that front propagation is a key feature of shrinking gels. We therefore turn to investigate the dynamics of the front in more detail now.

\section{Core-shell shrinking front dynamics} \label{sec:FrontSoln}

When the gel is heated to shrink significantly, simulations suggest a travelling front can form, which separates a swollen core from a shrunken shell. This front travels inwards, and its speed appears to have a dependence on its position {since it does not follow a straight line in either fig.~\ref{fig:Results_CaiSuo_ShrinkFront}a or fig.~\ref{fig:Results_Hirotsu_ShrinkFront}a}. This behaviour may be difficult to observe experimentally, since it occurs within the gel body, and so a theoretical solution would be helpful in understanding the key features. 
Therefore, we turn to investigate an approximate solution for the front shape, with the aim of understanding its behaviour and motion.

\subsection{Step-function approximation}

\begin{figure}
\centering
    \subcaptionbox{}{\input{Figs/Fig11a_Front_Porosity}}
    \subcaptionbox{}{\input{Figs/Fig11b_Front_PorosityPert}}
    \caption{(a) Porosity for the ANB shrinking front of fig.~\ref{fig:Results_CaiSuo_ShrinkFront} at time $t={0.2}$, compared to a step-function approximation that has levels matched to the {initial equilibrium} value at the centre and {so that the leading order osmotic pressure is zero} towards the edge. (b) The perturbation in porosity from the step{-}function in the numerical solution (solid curves) compared to the calculated values of ${\firstorder{\porosity}_\pm}$ from solving eqn.~\eqref{eq:BVP_Front} (dashed){, also at time $t=0.2$}. The position of the front is denoted by a vertical dotted line.} 
    \label{fig:Front_Porosity}
\end{figure}
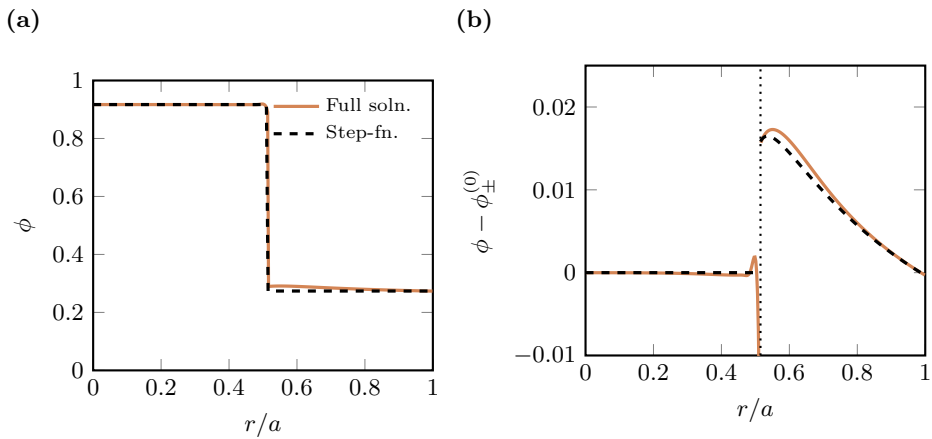

{Solutions that begin in the coexistence region form a sharp front on short timescales.} Once the front has formed, profiles of the porosity suggest that it is very close to being uniform in space on either side of the front (see figs.~\ref{fig:Results_CaiSuo_ShrinkFront} \& \ref{fig:Results_Hirotsu_ShrinkFront}). {In fig.~\ref{fig:Front_Porosity}a, we plot the porosity as a function of radius at an arbitrarily chosen time, $t=0.2$, from the simulation for fig.~\ref{fig:Results_CaiSuo_ShrinkFront}. We see that this solution looks like a step-function, and similar profiles are seen at other times and in the other shrinking front solutions.} 

{To justify the formation of this step-function porosity profile, we note that in our system, the parameter relating the relative importance of the mixing and elastic contributions is large: $\Param\gg1$. This means that in general the osmotic pressure, given by eqn.~\eqref{eq:Osmotic_Dimless}, dominates the total stress in the mixture. However, we note that the osmotic pressure is simply a function of the porosity multiplied by the large factor $\Param$, i.e. we can write $\osmotic=\Param P(\porosity)$ for a function $P$. Therefore, any $O(1)$ gradients in the porosity will be expected to result in $O(\Param)$ gradients in the osmotic pressure, and hence also in the total radial stress and chemical potential gradients due to the stress balance, eqn.~\eqref{eq:ChemPotential_Dimless}. These large stresses will be relaxed on a fast timescale of $O(\Param^{-1})$, and so on an $O(1)$ timescale we expect to see solutions with constant osmotic pressure and porosity.} 

Based on this observation, we look for solutions to the poro-elastic model with a front at $r=\Rfront$ using an {expansion in terms of the small parameter $\Param^{-1}\ll1$:}
\begin{equation}
    \porosity(r,t) = 
    \begin{cases}
    \zerothorder{\porosity}_- + \Param^{-1} \firstorder{\porosity}_-(r,t) + O(\Param^{-2}) &\text{for }0\leq r<\Rfront,\\
    \zerothorder{\porosity}_+ + \Param^{-1} \firstorder{\porosity}_+(r,t) + O(\Param^{-2}) &\text{for }\Rfront<r\leq a, 
    \end{cases}
\end{equation}
where the $-$ {terms} refer to the solution inside the front and the $+$ terms to solutions outside the front.

For the leading order porosity, we take the inner value as the initial equilibrium, $\zerothorder{\porosity}_-=\porosity(r=0,t=0)$, as we assume that the inner region has not been able to expel any fluid during the initial stages of front formation. For the porosity in the outer shell, we expect a uniform value that satisfies the no radial stress boundary condition at the edge. Since the osmotic pressure is dominant, this would mean that the porosity is such that the leading-order osmotic pressure vanishes in the entire region, $P(\zerothorder{\porosity}_+)=0$. We therefore have a shrunken shell where the leading-order osmotic pressure vanishes. A step-function with these two levels is plotted in fig.~\ref{fig:Front_Porosity}a against the numerical solution, and we see good agreement. The solid line in fig.~\ref{fig:Front_Porosity}b shows that the difference between the numerical result and this step-function approximation is only a few percent.

We {now} calculate the leading order contribution to the strains and stresses in the hydrogel. In each region, the local volume change (relative to the dry reference state) is found from expanding eqn.~\eqref{eq:Jeqn} to give
\begin{equation}
    J = \frac{1}{1-\zerothorder{\porosity}_\pm} + O(\Param^{-1}), 
\end{equation}
and the radial displacement, calculated using eqn.~\eqref{eq:Displacement_Dimless}, is
\begin{equation}
    u = 
    \begin{cases}
        \left[ 1-\left(1-\zerothorder{\porosity}_-\right)^{1/3} \right]r + O({\Param^{-1}}) &\text{for }0\leq r<\Rfront,\\
        \left[ 1-f \right]r + O({\Param^{-1}}) &\text{for }\Rfront<r\leq a.   
    \end{cases}
\end{equation}
Here, we have defined the function
\begin{equation}
    f(r) = \left[\left(1-\zerothorder{\porosity}_+\right)+\left(\zerothorder{\porosity}_+-\zerothorder{\porosity}_-\right)\left( \frac{R_f}{r} \right)^3 \right]^{1/3},
\end{equation}
for $\Rfront\leq r\leq a$, which takes the values $f(\Rfront)=(1-\zerothorder{\porosity}_-)^{1/3}$ and $f(a)=1/a$. Note that $(4\pi r^3/3)\times f(r)^3$ is equal to the total volume of solid material inside $r$ (when $r>\Rfront$), and so $f^3$ could be viewed as the average solid fraction within a sphere of radius $r$ once the front has formed. 

We then use the displacements to calculate the stretches, $\elongation$, from eqn.~\eqref{eq:Strain_Dimless}, which are
\begin{equation} \label{eq:Front_Strain_theta}
    \elongation_\theta = 
    \begin{cases}
        \left(1-\zerothorder{\porosity}_-\right)^{-1/3} + O({\Param^{-1}}) &\text{for }0\leq r<\Rfront,\\
        f^{-1} + O({\Param^{-1}}) &\text{for }\Rfront<r\leq a,
    \end{cases}
\end{equation}
\begin{equation}
    \elongation_r = 
    \begin{cases}
        \left(1-\zerothorder{\porosity}_-\right)^{-1/3} + O({\Param^{-1}}) &\text{for }0\leq r<\Rfront,\\
        \frac{f^2}{1-\zerothorder{\porosity}_+} + O({\Param^{-1}}) &\text{for }\Rfront<r\leq a.
    \end{cases}
\end{equation}

The Terzaghi stresses are calculated from the strains using eqn.~\eqref{eq:Stress_Dimless}, giving
\begin{equation}
    \Terzaghi_\theta = 
    \begin{cases}
        \left(1-\zerothorder{\porosity}_-\right)^{1/3} - (1-\zerothorder{\porosity}_-) + O({\Param^{-1}}) &\text{for }0\leq r<\Rfront,\\
        \left(1-\zerothorder{\porosity}_+\right)\left(f^{-2}-1\right) + O({\Param^{-1}}) &\text{for }\Rfront<r\leq a,
    \end{cases}
\end{equation}
\begin{equation} \label{eq:Front_Stress_r}
    \Terzaghi_r = 
    \begin{cases}
        \left(1-\zerothorder{\porosity}_-\right)^{1/3} - (1-\zerothorder{\porosity}_-) + O({\Param^{-1}}) &\text{for }0\leq r<\Rfront,\\
        \frac{f^4-\left(1-\zerothorder{\porosity}_+\right)^2}{1-\zerothorder{\porosity}_+} + O({\Param^{-1}}) &\text{for }\Rfront<r\leq a.
    \end{cases}
\end{equation}

The (dominant) osmotic pressure, {given in} eqn.~\eqref{eq:Osmotic_Dimless}, {can be expanded in powers of $\Param^{-1}$} to get 
\begin{equation}
    \osmotic = {\Param P\left(\zerothorder{\porosity}_\pm\right) + P'\left(\zerothorder{\porosity}_\pm\right) \firstorder{\porosity}_\pm + O(\Param^{-1})}, 
\end{equation}
{where we recall that $\zerothorder{\porosity}_+$ satisfies $P(\zerothorder{\porosity}_+)=0$, and note that the function $P$ and its first derivative are given by
\begin{align}
    P(\porosity) &= - \log\porosity - (1-\porosity) - (\FloryHuggins_0-\FloryHuggins_1)(1-\porosity)^2 - 2\FloryHuggins_1(1-\porosity)^3 ,\\
    P'(\porosity) &=  -\frac{1}{\porosity} + 1 + 2(\FloryHuggins_0-\FloryHuggins_1)(1-\porosity) + 6\FloryHuggins_1(1-\porosity)^2 .
\end{align}
}

The leading order contribution to the chemical potential is then calculated using eqn.~\eqref{eq:ChemPotential_Dimless} --- noting that $f'(r)=-(\zerothorder{\porosity}_+-\zerothorder{\porosity}_-)\Rfront^3/r^4f^2$ and simplifying --- to obtain
\begin{equation} \label{eq:ChemPotential_Front}
    \partdiff{\chempotential}{r} = 
    \begin{cases}
    - P'\left(\zerothorder{\porosity}_-\right) \partdiff{\firstorder{\porosity}_-}{r} + O(\Param^{-1}) &\text{for }0\leq r<\Rfront,\\
    \frac{-2\left(\zerothorder{\porosity}_+-\zerothorder{\porosity}_-\right)^2 \Rfront^6}{\left(1-\zerothorder{\porosity}_+\right)f^2 r^7} - P'\left(\zerothorder{\porosity}_+\right) \partdiff{\firstorder{\porosity}_+}{r} + O(\Param^{-1}) &\text{for }\Rfront<r\leq a.
    \end{cases}
\end{equation}

To sustain a profile which has uniform porosity on either side of a travelling front, we expect the total fluid flux to be uniform in space at any given time, so that the fluid pushed outwards by the front moving inwards (causing the gel to locally contract as the front passes by) is equal to the amount of fluid expelled out from the boundary. For this to be the case, $r^2\partial\mu/\partial r$ must be uniform in each region. Since $\partial\mu/\partial r=0$ at the centre of the sphere, the section of swollen core in our solution must therefore have no flux and the chemical potential should be of the form
\begin{equation} \label{eq:ChemPotential_Const}
    \partdiff{\chempotential}{r} = 
    \begin{cases}
    0 &\text{for }0\leq r<\Rfront,\\
    A/r^2 &\text{for }\Rfront<r\leq a,
    \end{cases}
\end{equation}
for some $A$ which we wish to determine --- its value quantifies the total (inward) fluid flux from the gel sphere. Note that $A$ is independent of $r$ but may vary in time, $A=A(t)$. 

Equating eqns.~\eqref{eq:ChemPotential_Front} \& \eqref{eq:ChemPotential_Const}, we find a differential equation for the porosity perturbation in the shrunken shell of the gel, ${\firstorder{\porosity}_+}$, {which is valid for $\Rfront<r<a$:}
\begin{equation} \label{eq:BVP_Front}
    {P'\left(\zerothorder{\porosity}_+\right) \partdiff{\firstorder{\porosity}_+}{r}} = \frac{-2\left(\zerothorder{\porosity}_+-\zerothorder{\porosity}_-\right)^2 \Rfront^6}{\left(1-\zerothorder{\porosity}_+\right)f(r)^2 r^7}  - \frac{A}{r^2}.
\end{equation}
At each point in time, this is a first order differential equation in $r$ for ${\firstorder{\porosity}_+}(r,t)$ with one unknown constant, and so we require two boundary conditions. 

The first of these is simply the radial stress balance at the outer boundary: $\stress_r=-\externalchempot$ (i.e.~$\Terzaghi_r=\osmotic$) at $r=a$. This gives one condition on ${\firstorder{\porosity}_+}$ at the sphere edge, namely
\begin{equation} \label{eq:BVP_Front_BCEdge}
    {P'\left(\zerothorder{\porosity}_+\right) \firstorder{\porosity}_+} = \frac{1}{a^4\left(1-\zerothorder{\porosity}_+\right)} - \left(1-\zerothorder{\porosity}_+\right) \qquad \text{at } r=a.
\end{equation}

The second condition imposes that the radial stress is continuous across $r=\Rfront$: $[\sigma_r]_-^+=0$. Enforcing that the chemical potential is continuous across the front (i.e.~$[\chempotential]_-^+=0$), we then find a condition for the porosity perturbation at the front 
\begin{multline} \label{eq:BVP_Front_BCFront}
    P'\left(\zerothorder{\porosity}_+\right) \firstorder{\porosity}_+ = \left(1-\zerothorder{\porosity}_-\right)^{4/3} \left[\frac{1}{1-\zerothorder{\porosity}_+} - \frac{1}{1-\zerothorder{\porosity}_-} \right] \\
    +  \left(\zerothorder{\porosity}_+-\zerothorder{\porosity}_-\right) + P\left(\zerothorder{\porosity}_- \right)  \quad \text{at } r=\Rfront.
\end{multline}

Given a sphere radius $a$, front position $\Rfront$ and uniform porosities $\zerothorder{\porosity}_\pm$, at any point in time, we can therefore solve the boundary value problem given by the differential equation \eqref{eq:BVP_Front} and boundary conditions \eqref{eq:BVP_Front_BCEdge} \& \eqref{eq:BVP_Front_BCFront}, to determine the porosity perturbation ${\firstorder{\porosity}_+}$ and (perhaps more importantly) the value of $A$; hence, we can calculate the radial fluid flux. Note that the differential equation \eqref{eq:BVP_Front} could be integrated to determine ${\firstorder{\porosity}_+}$ analytically and give a transcendental equation for $A$, but we choose instead to solve \eqref{eq:BVP_Front} numerically using MATLAB's in-built boundary value problem solver \texttt{bvp4c}.

\subsection{Comparison to the numerical solution}

\begin{figure}
\centering
    \subcaptionbox[width=0.33\textwidth]{}{\input{Figs/Fig12a_Front_Strain}}
    \subcaptionbox[width=0.33\textwidth]{}{\input{Figs/Fig12b_Front_Stress}}
    \subcaptionbox[width=0.33\textwidth]{}{\input{Figs/Fig12c_Front_ChemGrad}}
    \caption{Comparison of the step-function approximation and the numerical solutions for (a) the {stretches}, {(b) the Terzaghi (elastic) }stresses, and {(c) the} chemical potential gradient after a front has formed. Here, solid curves show the full numerical results for the multi-valued shrinking front, as presented in fig.~\ref{fig:Results_CaiSuo_ShrinkFront}, at a dimensionless time $t={0.2}$. The approximate solutions --- calculated after measuring the front position and the sphere radius --- are shown as dashed curves. In (a) and (b), the red-orange (lower) solid curves denote the radial stretches and Terzaghi stresses, whereas the blue (upper) solid curves represent the angular stretches and Terzaghi stresses.}
    \label{fig:Front_StrainStress}
\end{figure}
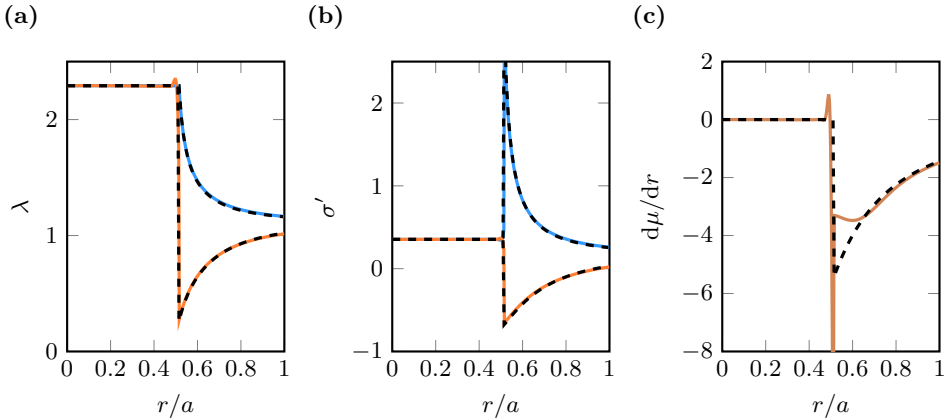

Taking the values for the porosity and sphere radius used in fig.~\ref{fig:Front_Porosity}a at time $t={0.2}$, we now plot the leading order strains and stresses, calculated from eqns.~\eqref{eq:Front_Strain_theta}--\eqref{eq:Front_Stress_r}, in figs.~\ref{fig:Front_StrainStress}a,b. Again, we find good agreement between the step-function approximation and the numerical solutions.

We can also consider calculating the chemical potential gradient, $\partial\mu/\partial r$, given in $\eqref{eq:ChemPotential_Const}$ by solving the boundary value problem \eqref{eq:BVP_Front} for $A$ and ${\firstorder{\porosity}_+}$. Using the same parameters as in figs.~\ref{fig:Front_Porosity}a \& \ref{fig:Front_StrainStress}a,b, we determine $A={-3.25}$, and obtain the chemical potential gradient shown in fig.~\ref{fig:Front_StrainStress}c. We see here that the corrected solution for the chemical potential with a constant fluid flux approximates the numerical solution well. 

In solving \eqref{eq:BVP_Front}, we also determine ${\firstorder{\porosity}_+}$, and in fig.~\ref{fig:Front_Porosity}b we plot the calculated ${\firstorder{\porosity}_\pm}$ alongside the values of $\porosity(r)-\zerothorder{\porosity}_+$ obtained from the full numerical solution. We find good agreement between the two, suggesting that our analytic solution is indeed capturing the key features of this shrinking gel. {(There is a visible discrepancy just ahead of the front in figs.~\ref{fig:Front_Porosity}b \& \ref{fig:Front_StrainStress}c, but we attribute this to errors arising from numerically calculating gradients close to a sharp front.)}

\subsection{Dynamics of the front}

We have seen that at a given time we can well-reproduce the stresses, strains and chemical potential gradient after observing the porosity at the centre, the sphere radius and the front position. We should therefore be able to numerically evolve the front solution forwards in time, since we can calculate the fluid flux at each time point after solving the boundary value problem \eqref{eq:BVP_Front}{--\eqref{eq:BVP_Front_BCFront}}. The sphere radius evolves according to \eqref{eq:SphereEvolve_Dimless}, and conservation of solid in the sphere gives the evolution of the front by
\begin{equation} \label{eq:FrontEvolve}
    \dot{\Rfront} = - \frac{\left(1-\zerothorder{\porosity}_+\right)a^2 \dot{a}}{\left(\zerothorder{\porosity}_+-\zerothorder{\porosity}_-\right)\Rfront^2}  = - \frac{\left(1-\zerothorder{\porosity}_+\right)k\left(\zerothorder{\porosity}_+\right)}{\left(\zerothorder{\porosity}_+-\zerothorder{\porosity}_-\right)} \frac{A(t)}{\Rfront(t)^2}.
\end{equation}

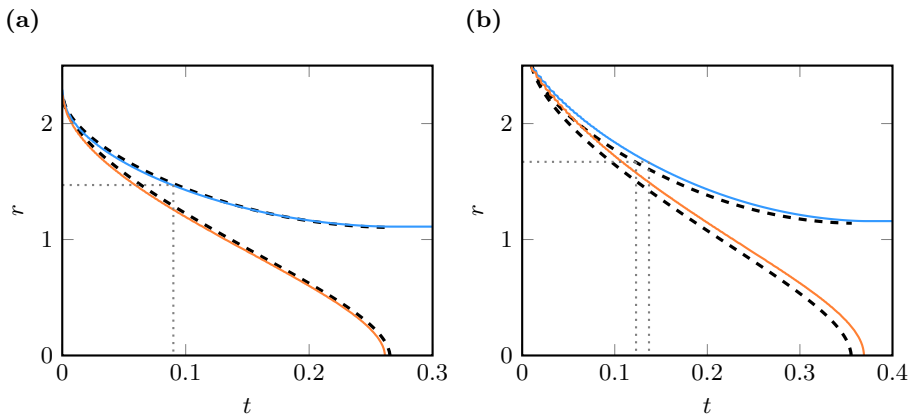
\begin{figure}
\centering
    \subcaptionbox{}{\input{Figs/Fig13a_Front_Dynamics_CaiSuo}}
    \subcaptionbox{}{\input{Figs/Fig13b_Front_Dynamics_Hirotsu}}
    \caption{Dynamics of a shrinking hydrogel once a front has formed, comparing the full numerical solution (solid curves) with the step-function approximation (dashed curves) for (a) the ANB shrinking front of fig.~\ref{fig:Results_CaiSuo_ShrinkFront}, and (b) the HHT shrinking front of fig.~\ref{fig:Results_Hirotsu_ShrinkFront}. For the numerical solution, the hydrogel sphere radius is shown in blue, whilst the front position is plotted in red-orange. In (a) the step-function solution is initiated from the numerics at a time $t=0.001$, in (b) it is at $t=0.01$. The dotted lines illustrate the radius and predicted time at which $80\%$ of the fluid has been expelled, as described in the drug delivery application of \S\ref{sec:Application}.}  
    \label{fig:Front_Dynamics}
\end{figure}

In fig.~\ref{fig:Front_Dynamics}{a}, we show the full numerical results for the evolution of the front position and sphere radius against the evolved step-function solution, using the same parameters as in fig.~\ref{fig:Results_CaiSuo_ShrinkFront}. The evolution of this solution was {initiated after observing the} sphere radius and front position {from} the numerical solution at time $t={0.001}$; the equations are only valid once the front has formed and so it is not possible to initiate the simulation from $t=0$ (we require $\Rfront<a$ to solve the boundary value problem \eqref{eq:BVP_Front}). {Similarly, fig.~\ref{fig:Front_Dynamics}b shows the evolution of the analytic solutions compared to the results of fig.~\ref{fig:Results_Hirotsu_ShrinkFront}, started at time $t=0.01$.}

The evolution of the front and sphere edge appears to be reasonably well-approximated by the step-function solution. The shapes of the curves are matched well, suggesting that this observed core-shell behaviour in the porosity is indeed dominating the dynamics. However, we note that the step-function approximation slightly {miscalculates} the timescale for the front to reach the centre of the sphere{, particularly in fig.~\ref{fig:Front_Dynamics}b}; calculating the next order correction to the solution using the solution for ${\firstorder{\porosity}_+}$ may improve the accuracy of this approximation and reduce the discrepancy. 

\subsection{Application: Predicting multi-dose strategies for targeted drug delivery} \label{sec:Application}

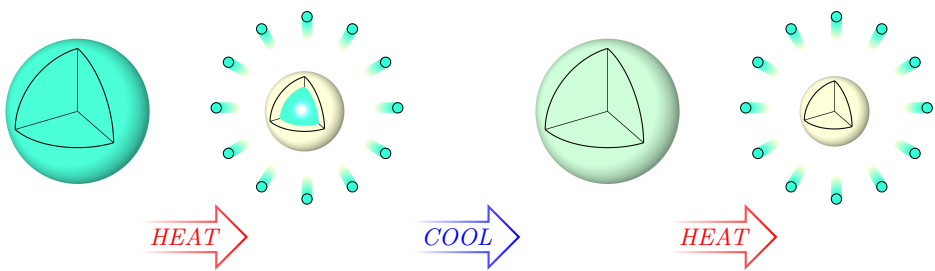
\begin{figure}
    \centering
    \begin{tikzpicture}
        \node at (0,0) {\includegraphics[width=2cm,viewport = 600 320 1800 1520, clip]{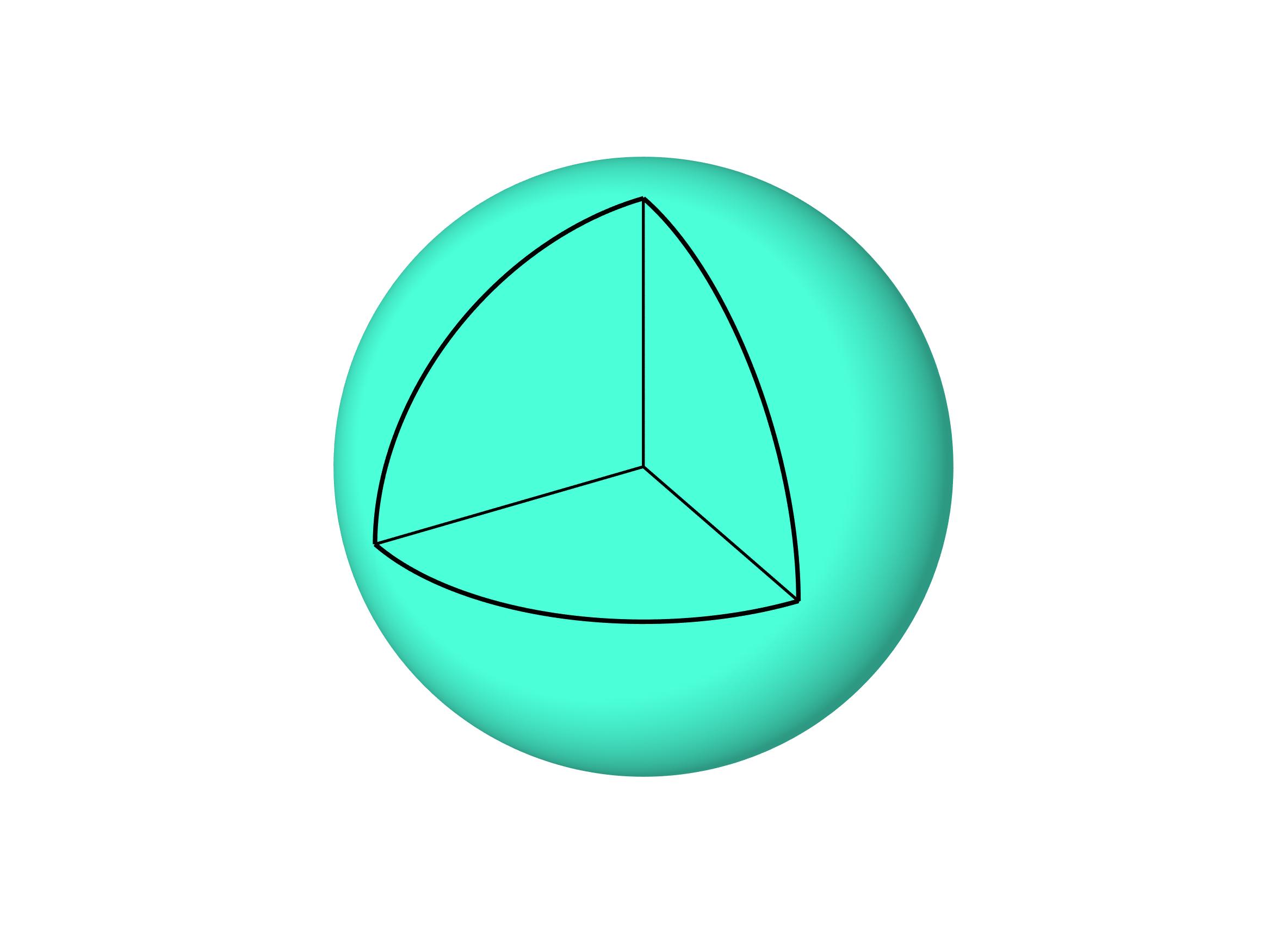}};
        \node at (3,0) {\includegraphics[width=2cm,viewport = 600 320 1800 1520, clip]{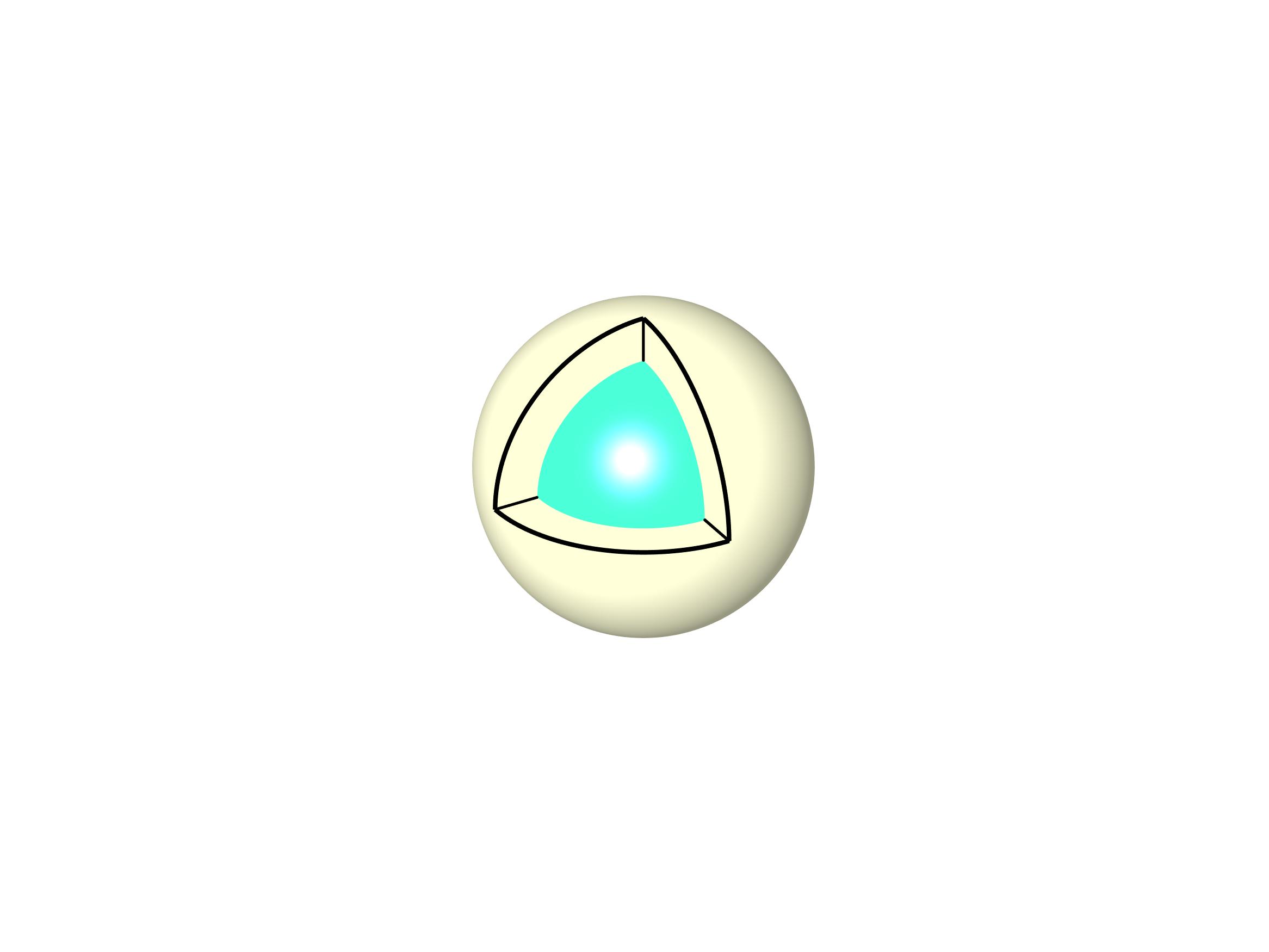}};
        \node at (7,0) {\includegraphics[width=2cm,viewport = 600 320 1800 1520, clip]{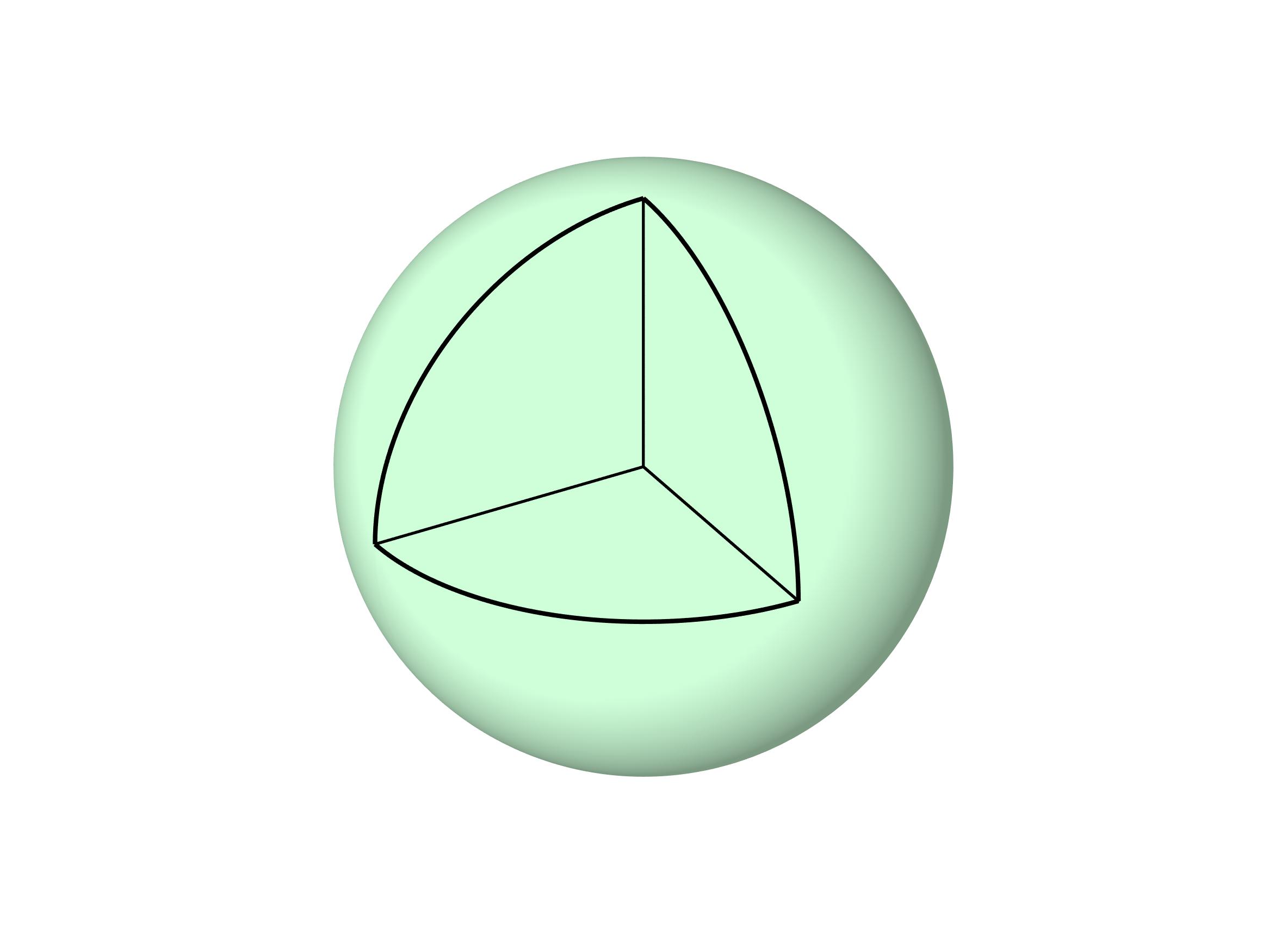}};
        \node at (10,0) {\includegraphics[width=2cm,viewport = 600 320 1800 1520, clip]{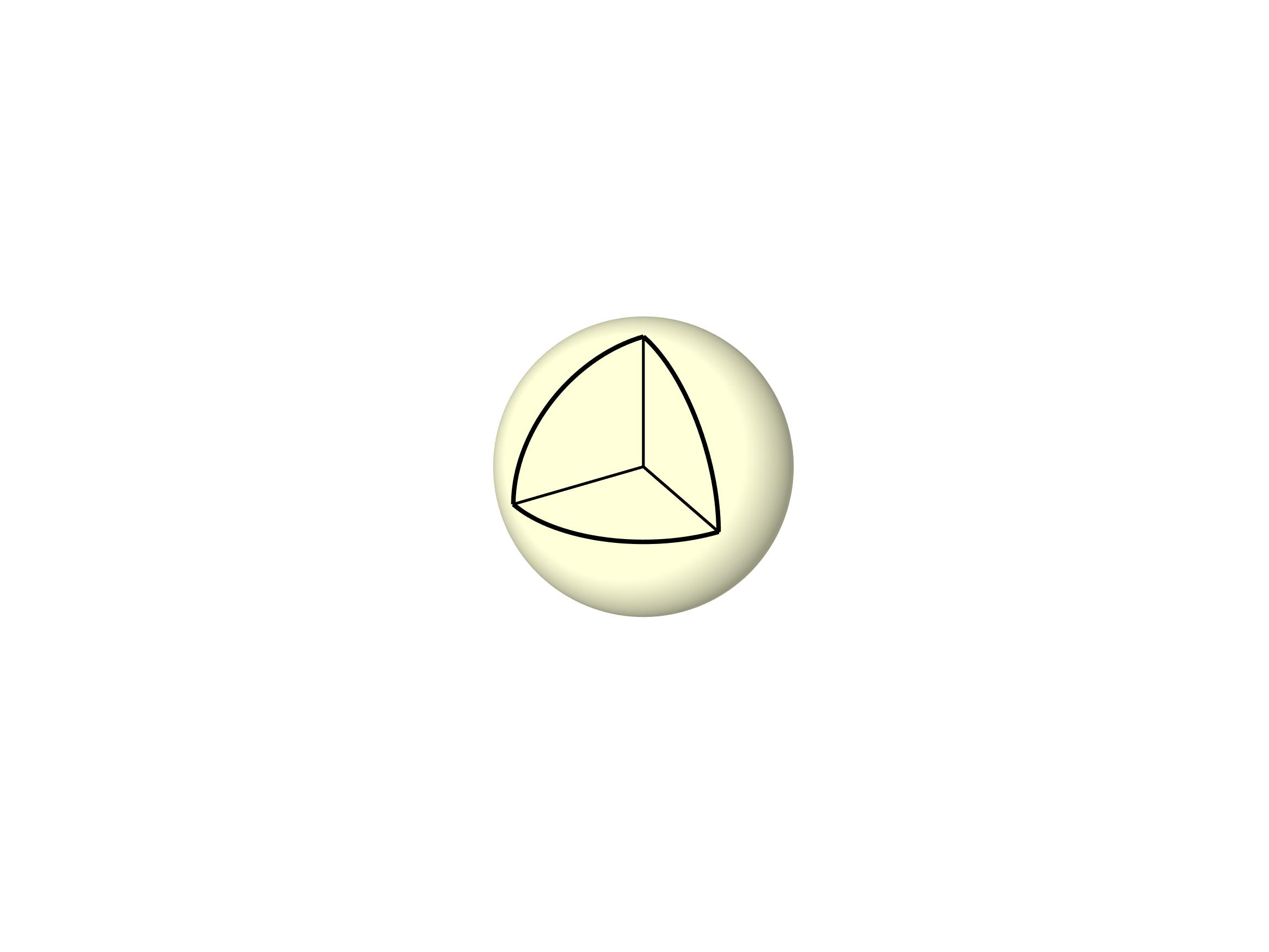}};
        
        \node at (1.5,2) {\textbf{\textit{FIRST DOSE}}};
        \node at (8.5,2) {\textbf{\textit{SECOND DOSE}}};
        
        \draw[thick,red,yshift = -1.7cm,xshift=0.8cm,path fading=west] (0,0.2) -- (1,0.2) -- (1,0.4) -- (1.4,0) -- (1,-0.4) -- (1,-0.2) -- (0,-0.2);  
        \node[anchor=west] at (0.8,-1.7) {\color{red}\it HEAT};
        \draw[thick,red,yshift = -1.7cm,xshift=7.8cm,path fading=west] (0,0.2) -- (1,0.2) -- (1,0.4) -- (1.4,0) -- (1,-0.4) -- (1,-0.2) -- (0,-0.2);  
        \node[anchor=west] at (7.8,-1.7) {\color{red}\it HEAT};
        \draw[thick,blue,yshift = -1.7cm,xshift=4.4cm,path fading=west] (0,0.2) -- (1,0.2) -- (1,0.4) -- (1.4,0) -- (1,-0.4) -- (1,-0.2) -- (0,-0.2);  
        \node[anchor=west] at (4.4,-1.7) {\color{blue}\it COOL};
        
        \node at (3,0) {\input{Figs/expulsion_particles}};
        \node at (10,0) {\input{Figs/expulsion_particles}};
    \end{tikzpicture}
    \caption{Illustration of a multi-dose drug strategy. A swollen drug-laden hydrogel is heated, causing it to shrink and expel its load. The heat source is removed before all of the drug is released, allowing for a second dose to be administered at a later time. To release a given dosage at each stage, the actuation time of the heating must be carefully controlled.}
    \label{fig:multidose_schematic}
\end{figure}  

A potential use for this simplified step-function solution is in determining dosage strategies for targeted drug delivery. In this scenario, we consider a swollen hydrogel sphere that contains a given quantity of a drug within its interstitial pore space. Upon heating, the hydrogel contracts and expels its previously-absorbed fluid, thereby releasing its load of drugs. As we have seen, in many situations we can expect this temperature change to result in front formation and propagation that dominates the dynamics. If we know the equilibrium behaviour of our hydrogel, and therefore the parameters $\Param, A_i$ and $B_i$, we can then use our simplified step-function front solution to quickly evolve our model system to calculate the (dimensionless) time required for expulsion of a dosage. If any of the inputs change (e.g.~starting or final temperature, or gel sphere size), this model can be updated and new results found in realistic timescales for clinical practice.

This could be extended to develop multi-dose strategies: a drug-laden hydrogel sphere could be actuated on more than one occasion to release specific doses at given times, as illustrated in fig.~\ref{fig:multidose_schematic}. Upon each actuation, we want to release a specified amount of the drug. The question is then: how long do we have to apply the temperature stimulus to achieve the required dosage? 

For an incompressible solid (as was assumed for the polymer in our model), the volume of polymer within the boundary of the sphere remains constant, and any change in total volume must be due to expulsion or absorption of fluid into the pore space. The expelled volume of fluid when the sphere changes from an initial radius $a(0)$ to the current radius $a(t)$, regardless of the spatial arrangement of fluid and solid within the hydrogel sphere in each state, can then simply be calculated as
\begin{equation} \label{eq:FluidVolExpelled}
    \Delta V = \frac{4\pi}{3} \left[a(0)^3-a(t)^3\right],
\end{equation}
where the dimensionless expelled volume here is $\Delta V = \Delta \dimensional{V}/\dryradius^3$, for a dimensional expelled volume $\Delta \dimensional{V}$.

We consider a situation where we are given a hydrogel with known properties (i.e.~we know its equilibrium parameters, permeability, etc.) and are changing between two set temperatures, $\tempstart$ and $\tempend$, and we wish to expel a dosage volume $\dosage$ that is less than the total fluid volume within the swollen hydrogel. Using the following procedure, we can calculate the required actuation time:
\begin{enumerate}
    \item Find the initial equilibrium stretch, $\elongation$, by solving the energy minimisation, eqn.~\eqref{eq:EqlEqn}, at the initial temperature, $\tempstart$. This is the initial dimensionless sphere radius $a(0)=\elongation$.
    \item Evolve the analytic step-function solution forward in time until the front reaches the sphere centre.
    \item Calculate the appropriate final radius using eqn.~\eqref{eq:FluidVolExpelled}
    \begin{equation}
        a(t) = \left[ a(0)^3 - \frac{\dosage}{4\pi\dryradius^3/3} \right]^{1/3}.
    \end{equation}
    \item From the equivalent to fig.~\ref{fig:Front_Dynamics}, read off the dimensionless time at which this dimensionless sphere radius is attained.
\end{enumerate}
The dimensional time can then be simply calculated by multiplying by the timescale, $\timescale$, defined in eqn.~\eqref{eq:timescale}. 

For example, if we were considering the ANB hydrogel with a temperature change from $\tempstart=302$ to $\tempend=308$, we would obtain fig.~\ref{fig:Front_Dynamics}a. To expel $80\%$ of the stored fluid volume, which is equivalent to a dosage volume of $\dosage=(4\pi\dryradius^3/3)\times(0.2a(0)^3+0.8)$, we find that we must shrink from an initial radius $a=2.29$ to a final radius $a=1.47$ (larger than the final equilibrium radius $a=1.11$). From fig.~\ref{fig:Front_Dynamics}a, we then see that we require a dimensionless time $t=0.092$ from our approximate front solution; our numerics suggest this should instead be $t=0.089$, which results in a difference in the expelled fluid volume of approximately $1\%$. For the same volumes with the HHT hydrogel, the shrinking must occur from an initial radius $a=2.67$ to a radius $a=1.67$; the approximate solution suggests a time $t=0.123$ compared to $t=0.137$ for the numerics. While this error in the timing appears significant, the resulting dosage volume is only out by less than $5\%$.

\section{Conclusions} \label{sec:Conclusion}

In this paper, we have considered the evolution of a thermo-responsive hydrogel after an instantaneous temperature change that causes it to significantly swell or shrink. By adapting the poro-elastic model of \citet{Bertrand2016} for a spherically-symmetric hydrogel bead, we considered the swelling and shrinking of two model systems, whose equilibrium parameters were extracted from the literature. 

{Through our numerical simulations, we observed a range of different dynamic behaviours. In some cases, the evolution occurred smoothly, whereas in others there were sharp gradients and phase separation. The occurrence of these qualitative dynamic behaviours appears to be well-captured by two key regions in the equilibrium curve space: the coexistence and spinodal regions (see e.g.~fig.~\ref{fig:EqlSwell}). When starting in these regions, the hydrogel soon undergoes phase separation. In the coexistence region, a sharp front forms that propagates inwards and invades the domain (akin to nucleation and growth). This core-shell behaviour has also been observed in other similar hydrogel systems \cite[e.g.][]{Bertrand2016,Doi2009}. Starting close to this coexistence region can also result in delayed front formation in the interior to the hydrogel. At a fixed temperature, we expect this to occur if the coexistence region lies between the start and end of the trajectories in $(\temp,\elongation)$-space, such as those shown in fig.~\ref{fig:Trajectories}. Meanwhile, dynamics starting in the spinodal region exhibit spontaneous localised phase separation (spinodal decomposition), but our model is not well suited to investigating these dynamics properly due to the enforced spherical symmetry, and so this is left for further future study.}

In general, we found that the dynamics of swelling was qualitatively different from the dynamics of shrinking in our two example hydrogels. Phase separation seems to be common when shrinking past the volume phase transition, and relatively rare for swelling, as can be seen from the location of the shaded regions in fig.~\ref{fig:EqlSwell}. However, only one of our theoretical gels exhibited smooth shrinking, and the other had a small range of temperatures just below the volume phase transition that showed front formation when shrinking (which could help explain the two-stage dynamics of swelling seen around these temperatures in the experiments of \citet{Matsuo1988}). A more thorough future investigation of different thermo-responsive hydrogels may reveal which of these behaviours is more prevalent, and if any others are seen in different thermo-responsive hydrogels. In particular, we would be interested to see future experimental observations of the internal structure of swelling and shrinking hydrogels, perhaps using MRI to image the phase separation within small hydrogel beads.

Our results extend upon those of \citet{Tomari1995}, who investigated the dynamics and phase separation using a theoretical model of a particular thermo-responsive hydrogel close to the volume phase transition. Despite some differences in our modelling approach, we saw many similar results, such as phase separation and front formation, as well as two-stage dynamics with a delayed front formation. Their results showed that this core-shell dynamics can occur in both swelling and shrinking hydrogels. However, our results suggest that the (qualitative) symmetry between these breaks down as temperatures are changed beyond the volume phase transition: phase separation appears to be more common amongst shrinking gels in our example systems.

We investigated the propagation of the shrinking front in more detail by considering a step-function approximation for the porosity that is based upon having two regions of uniform porosity on either side of the front. The formation of this step-function is explained by the dominance of the osmotic pressure in the stress, which must be relaxed in the outer region due to stress-free boundary conditions. We determined the leading-order solutions for the stresses and strains and showed that these matched well to the solutions found by numerically integrating the full poro-elastic equations. More careful calculation was required for the approximated chemical potential gradient and fluid flux, because the osmotic pressure is sensitive to small changes in the porosity due to the presence of a large parameter. Evolving this step-function approximation forwards in time, we were able to well-reproduce the front propagation and evolution of the sphere size. Being able to solve this simpler approximated system is useful because evolving the full numerical system required high temporal resolution, and hence took much longer to run: the approximate system can be solved in tens of seconds on a laptop, compared to approximately half a day for the full numerics. We then demonstrated how this step-function solution could be used to determine actuation times for expelling a specified volume of fluid,for use in applications such as targeted drug delivery applications.

From this asymptotic solution for a shrinking front we are able to gleam some insight into the front evolution: in a gel with a dominant osmotic pressure (i.e.~large parameter $\Param$), once phase separation has been instigated at the edge by the sudden change in temperature, the outer shrunken shell must relax the osmotic pressure and maintain a constant outward fluid flux which constrains the resulting dynamics. Applying a full Maxwell construction at the interface may help to expand this analytic solution to more examples, such as the delayed and swelling fronts observed in the numerics. In addition, we note that although this solution was found for thermo-responsive hydrogels, the key physics should also work for other swelling and shrinking hydrogel systems, and it would be interesting to see how well this analysis carries across to other response modes.

Our study of the swelling and shrinking of a spherical thermo-responsive hydrogel has highlighted many interesting features that may be observed in such a system. However, there are some details omitted from our work that could be important in real systems and deserve further study. For example, the model enforced a spherical symmetry on the gel that may not be realistic, and precludes dynamics such as spinodal decomposition. These spinodal dynamics may be the cause of the experimentally-observed blistering instabilities that have been found to cause shape change in gel tori and rods \citep{Chang2018,Shen2019}.
In addition, spherical (non-thermo-responsive) gel beads have been observed to form lobes or wrinkles as they swell \citep{Bertrand2016,Doi2009}, while thermo-responsive gels can exhibit a similar instability as they shrink \citep{Matsuo1988}.
While this wrinkling cannot be reproduced with a spherically-symmetric system, we note that our step-function solution has a discontinuous jump in the hoop stresses at the front, which could be relaxed by deforming this interface in a non-radially symmetric manner. Therefore, the front formation observed in our solution could be the origin of the lobe-like instability seen by \citet{Matsuo1988}; further study is required to investigate this.

For our results, we have also assumed that the background fluid is quiescent and the boundary of the hydrogel only experiences stresses from the external fluid pressure. This is often not true in practice: in real-life applications, a hydrogel structure will often be present in a background fluid flow, with a range of other external forces applied. Fluid injection into a gel has been shown to induce phase separation behaviour in swelling and shrinking hydrogels in other scenarios \citep{Hennessy2020} and so we should expect these aspects to modify the results presented here. Our results should remain valid provided that any external stresses are small compared to the stress scale $\Boltzman\tempend/\polyvol$ and the background fluid flow is small compared to the velocity scale $\dryradius/\timescale$.

As the biomedical and engineering potential of hydrogels becomes increasingly achievable with the development of new manufacturing techniques, we hope that further theoretical studies will provide opportunities to analyse and understand the wealth of rich dynamics of thermo-responsive hydrogels, as well as help to inspire new applications, such as novel treatment strategies.

\section*{Acknowledgements} 
This work was supported by the Leverhulme Trust Research Leadership Award “Shape-Transforming Active Microfluidics”. We would like to thank Prof.~C.~MacMinn and Dr.~M.~Hennessy for helpful discussions on this subject.

\section*{Declaration of interests} 
The authors report no conflict of interest.

\section*{Author ORCID} 
M. Butler, https://orcid.org/0000-0002-7110-163X; 

T. Montenegro-Johnson, https://orcid.org/0000-0002-9370-7720

\bibliographystyle{jfm}
\bibliography{dynamicgelsphere}

\begin{appendix}

\section{Numerical scheme} \label{app:NumericalScheme}

To numerically evolve the free-boundary problem defined in eqns.~\eqref{eq:PDE_Dimless}--\eqref{eq:Boundary_Dimless}, we first rescale the equations onto a fixed domain using the radial coordinate scaling $R=r/a(t)$, so that the moving sphere boundary remains fixed at $R=1$ throughout. Taking account of the chain rule, we find that the porosity satisfies a (different) Reynolds' equation of the form
\begin{equation} \label{eq:NumericalPDE}
    \partdiff{(a^3 \porosity)}{t}  = \frac{1}{R^2} \partdiff{(R^2 Q)}{R},
\end{equation}
where the (inward) flux $Q$ is given by 
\begin{equation} \label{eq:NumericalFlux}
    Q = a (1-\porosity) \permeability(\porosity) \partdiff{\chempotential}{R} + \frac{R \porosity}{a} \diff{a}{t}.
\end{equation}

The system is discretised onto a staggered radial grid with gridpoints $R_i=(i-1/2)/n$ for $i=1,2,\dots,n$, designed to conserve volumes accurately. At each timestep, the porosity at each gridpoint, $\porosity_i$, and the sphere radius, $a$, are both initially known, and the following procedure is followed: 
\begin{enumerate}
    \item the stresses and strains (i.e. the $\Terzaghi,\osmotic,\elongation$ and $J$) are calculated at each gridpoint, $R_i$, from the known porosities, $\porosity_i$, using eqns.~\eqref{eq:Stress_Dimless}--\eqref{eq:Strain_Dimless},
    \item $J$ is calculated at the boundary, $R=1$, by solving the non-linear eqn.~\eqref{eq:Boundary_Dimless}, and the result is used to find the boundary porosity, strains, and stresses,
    \item the chemical potential, $(\partial\chempotential/\partial r)_{i+1/2}$, is calculated at midpoints ($R_{i+1/2}=i/n$) using \eqref{eq:ChemPotential_Dimless} with central differences for the derivatives and averages for stresses evaluated at midpoints,
    \item the chemical potential is also calculated at the boundary, $R=1$, using a one-sided (inward) derivative, 
    \item the flux, $Q_{i+1/2}$, is calculated at midpoints and the boundary using eqn.~\eqref{eq:NumericalFlux}, with $Q=0$ enforced at $R=0$ (no flux at the centre of the sphere),
    \item the porosities, $\porosity_i$, and sphere radius, $a$, are then evolved by one timestep using a Forward Euler method, for simplicity, on eqns.~\eqref{eq:NumericalPDE} \& \eqref{eq:SphereEvolve_Dimless}, so that, for example
    \begin{equation}
        a(t+\Delta t)^3 \porosity_i(t+\Delta t) = a(t)^3 \porosity_i(t) + \Delta t \left[ 3\frac{R_{i+1/2}^2 Q_{i+1/2} - R_{i-1/2}^2 Q_{i-1/2}}{R_{i+1/2}^3 - R_{i-1/2}^3} \right],
    \end{equation}
    where the discretised spatial derivative has here been calculated by integrating eqn.~\eqref{eq:NumericalPDE} over a spherical shell between $R_{i-1/2}$ and $R_{i+1/2}$, and
    \item the procedure is then repeated from step (i) for the next timestep.
\end{enumerate}
 
\section{Other numerical simulations} \label{app:OtherDynamics}

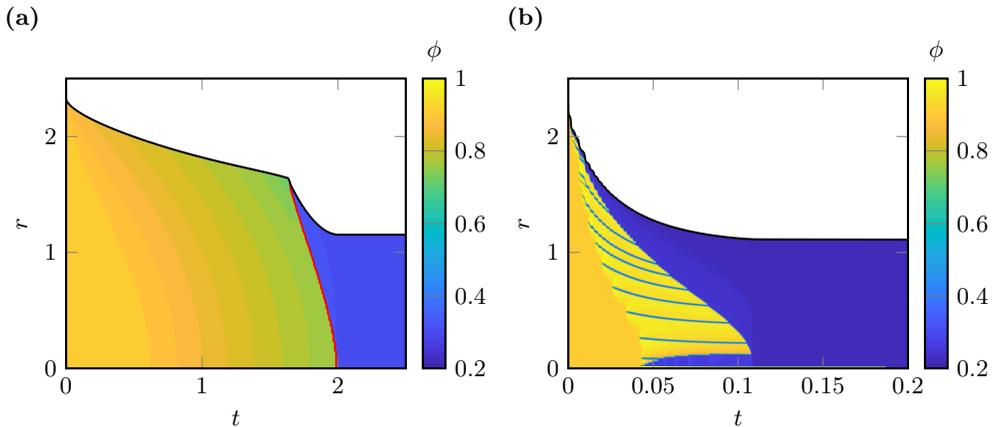
\begin{figure}
\centering
    \subcaptionbox{}{\input{Figs/FigA1a_ColourMap_CaiSuo_Shrink_DelayedFront}}
    \subcaptionbox{}{\input{Figs/FigA1b_ColourMap_CaiSuo_Shrink_Spinodal}}
    \caption{Porosity colour maps for the ANB parameters showing (a) delayed bifurcation when heated from $\tempstart=302~$K to $\tempend=306~$K, and (b) spinodal decomposition when heated from $\tempstart=302~$K to $\tempend=312~$K. The front is shown in red in (a), but this is omitted on the sharp jump in (b) so that the detail of the solution can be clearly seen.}
    \label{fig:App_Dynamics}
\end{figure}

Here, we show two more results that demonstrate qualitatively different behaviour than that seen in the main text. These results are for the ANB parameters, \eqref{eq:Params_CaiSuo}, when increasing the temperature from $\tempstart=302~$K to $\tempend=306~$K (fig.~\ref{fig:App_Dynamics}a) or $\tempend=312~$K (fig.~\ref{fig:App_Dynamics}b). 

In the first instance, following the temperature change, the hydrogel is initially outside the coexistence region, but will pass through it. We observe front formation but only after a delay where the hydrogel initially shrinks smoothly.

In the second case, the starting point is just inside the spinodal region. We see points in the interior of the hydrogel collapsing to a shrunken state, surrounded by swollen regions of hydrogel. This is spinodal decomposition. We note that it is perhaps surprising that we have been able to capture this behaviour without using a full phase-field model. It is not clear to us what is selecting the wavelength of the observed pattern here, whether it is physical in origin or just due to numerical instability. However, we cannot make any strong conclusions about the dynamic behaviour from this solution: due to the enforced spherical symmetry, we are observing concentric shells of collapsed and swollen hydrogel, which we do not expect to be realistic. Instead, we would expect the spinodal decomposition to result in local pockets of collapsed gel. 

\section{Varying the permeability} \label{app:VaryPermeability}

For simplicity, we focused on presenting results where the permeability had a constant value, $\permeability=1$. However, in general, the permeability may depend on the porosity, $\permeability=\permeability(\porosity)$ (or even be anisotropic in which case it would be a tensor rather than a scalar). A common form taken for the permeability of a hydrogel is 
\begin{equation} \label{eq:Permeability}
    \permeability(\porosity) = \frac{\porosity}{(1-\porosity)^\beta},
\end{equation}
where $\beta$ is a constant typically found to be in the range 1.5--2 \citep{Tokita1991,Grattoni2001,Engelsberg2013}. With this form of the permeability, the hydrogel is slightly less permeable at low porosities, when the gel is shrunken, but much more permeable at the high porosities for a swollen hydrogel.

\begin{figure}
\centering
    \subcaptionbox{}{\input{Figs/FigA2a_ColourMap_Swell_VaryPerm}}
    \subcaptionbox{}{\input{Figs/FigA2b_ColourMap_ShrinkFront_VaryPerm}}
    \caption{Porosity colour maps for the dynamics of the ANB hydrogel with permeability given by \eqref{eq:Permeability} with $\beta=1.5$. In (a), the remaining parameters are the same as for fig.~\ref{fig:Results_CaiSuo_SwellNoFront}. In (b), they are the same as fig.~\ref{fig:Results_CaiSuo_ShrinkFront}.}
    \label{fig:App_VaryPermeability}
\end{figure}
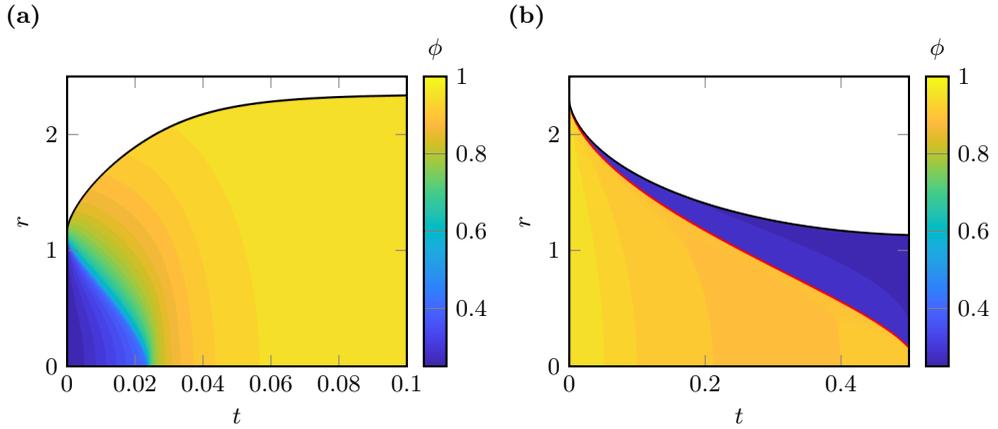

Taking a value $\beta=1.5$, as in \citet{Bertrand2016}, the results are qualitatively the same although the details of the dynamics, such as the timescale of swelling or shrinking, do vary. We show some example simulations of the ANB hydrogel in fig.~\ref{fig:App_VaryPermeability} for the same parameters as figs.~\ref{fig:Results_CaiSuo_SwellNoFront} \& \ref{fig:Results_CaiSuo_ShrinkFront}. 

We see the same smooth swelling and front-dominated shrinking as in the constant permeability case. The qualitative behaviour also remains the same for the other temperature changes and with the HHT parameters. However, the swelling timescale is noticeably faster, and the shrinking takes slightly more time. This can be simply explained by increased permeability in the swollen state, and decrease in the shrunken state. Changing the constant $\beta$ accentuates the difference in the swelling and shrinking dynamics, but still keeps the same qualitative behaviour.

One other noticeable difference between the shrinking results of figs.~\ref{fig:Results_CaiSuo_ShrinkFront}a \& \ref{fig:App_VaryPermeability}b for the two different permeability functions is that in this non-constant permeability case, the porosity of the swollen core decreases slightly over time, which was not seen for constant permeability. We believe this variation is due to the high permeability in the swollen core accentuating any tiny porosity gradients present to drive a small but significant fluid flow out through the front. However, at any given time, the porosity is still close-to-uniform and a step-function approximates the solution well. Its dynamic evolution approximates the numerics well, provided that the initial condition is started at a late enough time, and inputting the current porosity from the numerical solution rather than using the initial condition for the inner swollen region.

\end{appendix}

\end{document}

%% file: parula.tex
\pgfplotsset{
	colormap={parula}{
        rgb=(0.242200, 0.150400, 0.660300)
        rgb=(0.246416, 0.156930, 0.684790)
        rgb=(0.250330, 0.164865, 0.707270)
        rgb=(0.254043, 0.173297, 0.728853)
        rgb=(0.257657, 0.181526, 0.750443)
        rgb=(0.261167, 0.189454, 0.772329)
        rgb=(0.264576, 0.197384, 0.794201)
        rgb=(0.267683, 0.205431, 0.815357)
        rgb=(0.270482, 0.214090, 0.835186)
        rgb=(0.272978, 0.223447, 0.853395)
        rgb=(0.274971, 0.233409, 0.869791)
        rgb=(0.276669, 0.243968, 0.884589)
        rgb=(0.278166, 0.254820, 0.897886)
        rgb=(0.279441, 0.265863, 0.909973)
        rgb=(0.280244, 0.277006, 0.920962)
        rgb=(0.280924, 0.288150, 0.931055)
        rgb=(0.281313, 0.299193, 0.940254)
        rgb=(0.281387, 0.310236, 0.948849)
        rgb=(0.281072, 0.321180, 0.956824)
        rgb=(0.280640, 0.332108, 0.964109)
        rgb=(0.279690, 0.342950, 0.970688)
        rgb=(0.278268, 0.353793, 0.976763)
        rgb=(0.276392, 0.364753, 0.982098)
        rgb=(0.273828, 0.375814, 0.986544)
        rgb=(0.270641, 0.387058, 0.990065)
        rgb=(0.266547, 0.398422, 0.992617)
        rgb=(0.261486, 0.409967, 0.994784)
        rgb=(0.254771, 0.421533, 0.996691)
        rgb=(0.246220, 0.433201, 0.998476)
        rgb=(0.235576, 0.445070, 0.999577)
        rgb=(0.223559, 0.457341, 0.998098)
        rgb=(0.211287, 0.469689, 0.994263)
        rgb=(0.199189, 0.481761, 0.990121)
        rgb=(0.189417, 0.493355, 0.986102)
        rgb=(0.184272, 0.504446, 0.981204)
        rgb=(0.181469, 0.515261, 0.975828)
        rgb=(0.179217, 0.525874, 0.969993)
        rgb=(0.177654, 0.536415, 0.962964)
        rgb=(0.176680, 0.546756, 0.954683)
        rgb=(0.175039, 0.556966, 0.946041)
        rgb=(0.170655, 0.567075, 0.938166)
        rgb=(0.163641, 0.577114, 0.931202)
        rgb=(0.156209, 0.586987, 0.924209)
        rgb=(0.150589, 0.596591, 0.917015)
        rgb=(0.147015, 0.605894, 0.910157)
        rgb=(0.143998, 0.615094, 0.904008)
        rgb=(0.139894, 0.624266, 0.898967)
        rgb=(0.134179, 0.633402, 0.894523)
        rgb=(0.127477, 0.642501, 0.890131)
        rgb=(0.120588, 0.651398, 0.885120)
        rgb=(0.113901, 0.660093, 0.879061)
        rgb=(0.107014, 0.668546, 0.871834)
        rgb=(0.099057, 0.676597, 0.863498)
        rgb=(0.088938, 0.684344, 0.854155)
        rgb=(0.075669, 0.691731, 0.843946)
        rgb=(0.058772, 0.698672, 0.833075)
        rgb=(0.038670, 0.705311, 0.821598)
        rgb=(0.019854, 0.711646, 0.809707)
        rgb=(0.007045, 0.717624, 0.797469)
        rgb=(0.001435, 0.723301, 0.784873)
        rgb=(0.003123, 0.728728, 0.772176)
        rgb=(0.012666, 0.733901, 0.759178)
        rgb=(0.030963, 0.738772, 0.745978)
        rgb=(0.057047, 0.743441, 0.732576)
        rgb=(0.083501, 0.747858, 0.718973)
        rgb=(0.108339, 0.752075, 0.705269)
        rgb=(0.130501, 0.756239, 0.691414)
        rgb=(0.149231, 0.760408, 0.677454)
        rgb=(0.164594, 0.764624, 0.663245)
        rgb=(0.177251, 0.768941, 0.648734)
        rgb=(0.188166, 0.773313, 0.633711)
        rgb=(0.198139, 0.777730, 0.618083)
        rgb=(0.209331, 0.781934, 0.601851)
        rgb=(0.221594, 0.786050, 0.584957)
        rgb=(0.236636, 0.789749, 0.567314)
        rgb=(0.254887, 0.793086, 0.549167)
        rgb=(0.276322, 0.795878, 0.530375)
        rgb=(0.300357, 0.798106, 0.511180)
        rgb=(0.325277, 0.799891, 0.491420)
        rgb=(0.350113, 0.801335, 0.470869)
        rgb=(0.374911, 0.802415, 0.449434)
        rgb=(0.400328, 0.803028, 0.427320)
        rgb=(0.427310, 0.802871, 0.405162)
        rgb=(0.455776, 0.802008, 0.383407)
        rgb=(0.484957, 0.800505, 0.362155)
        rgb=(0.513971, 0.798564, 0.340906)
        rgb=(0.542448, 0.796120, 0.319387)
        rgb=(0.570422, 0.793341, 0.297533)
        rgb=(0.598162, 0.790122, 0.275886)
        rgb=(0.625631, 0.786469, 0.255122)
        rgb=(0.652796, 0.782412, 0.235639)
        rgb=(0.679558, 0.777952, 0.217881)
        rgb=(0.705572, 0.773334, 0.201922)
        rgb=(0.731027, 0.768543, 0.187128)
        rgb=(0.755830, 0.763550, 0.173389)
        rgb=(0.779902, 0.758530, 0.162060)
        rgb=(0.803193, 0.753510, 0.155186)
        rgb=(0.825525, 0.748667, 0.153576)
        rgb=(0.847080, 0.744002, 0.155989)
        rgb=(0.868029, 0.739663, 0.160883)
        rgb=(0.888195, 0.735704, 0.168543)
        rgb=(0.907396, 0.732407, 0.180149)
        rgb=(0.925468, 0.729997, 0.195666)
        rgb=(0.942431, 0.728594, 0.213399)
        rgb=(0.959097, 0.728482, 0.229842)
        rgb=(0.975597, 0.730192, 0.241020)
        rgb=(0.989041, 0.735921, 0.243181)
        rgb=(0.996285, 0.744997, 0.235777)
        rgb=(0.997185, 0.756062, 0.227328)
        rgb=(0.996928, 0.767492, 0.218709)
        rgb=(0.996254, 0.779024, 0.210076)
        rgb=(0.995001, 0.790757, 0.201717)
        rgb=(0.992477, 0.802691, 0.194313)
        rgb=(0.988731, 0.814928, 0.187842)
        rgb=(0.983766, 0.827365, 0.181997)
        rgb=(0.978074, 0.839814, 0.175983)
        rgb=(0.972243, 0.852363, 0.169769)
        rgb=(0.967281, 0.864904, 0.163828)
        rgb=(0.963520, 0.877361, 0.158384)
        rgb=(0.960869, 0.889809, 0.153357)
        rgb=(0.959628, 0.901969, 0.148149)
        rgb=(0.959691, 0.914021, 0.141951)
        rgb=(0.960772, 0.925968, 0.134550)
        rgb=(0.962865, 0.937714, 0.126235)
        rgb=(0.965665, 0.949265, 0.116913)
        rgb=(0.969172, 0.960810, 0.106187)
        rgb=(0.972985, 0.972355, 0.093850)
        rgb=(0.976900, 0.983900, 0.080500)
	}
}

%% file: Figs/pnipam_schematic_1.tex
\begin{tikzpicture}[thick,scale = 1]
    \draw[blue,snake it] (0,0) .. controls (1,1) and (2,-1) .. (3.3,0);
    \draw[blue,snake it,rotate=-5,xshift = 0cm,yshift = -0.7cm] (0,0) .. controls (1,0) and (2,0.5) .. (3.4,0);
    \draw[blue,snake it,rotate=-5,xshift = 0cm,yshift = 0.8cm] (0,0) .. controls (1,0.5) and (2,-0.5) .. (3.1,0.3);
    \draw[blue,snake it,rotate=0,xshift = 0cm,yshift = 1.6cm] (0,0) .. controls (1,0) and (2,-0.5) .. (3,0.5);
    \draw[blue,snake it,rotate=90,xshift = -1cm,yshift = -0.3cm] (0,0) .. controls (1,-1) and (2,0.5) .. (3,0);
    \draw[blue,snake it,rotate=95,xshift = -1cm,yshift = -1.4cm] (0,0) .. controls (1,0) and (2,0.5) .. (3,0);
    \draw[blue,snake it,rotate=95,xshift = -1cm,yshift = -2.4cm] (-0.3,0) .. controls (1,1) and (2,-0.5) .. (3,0.5);
    \draw[blue,snake it,rotate=90,xshift = -1cm,yshift = -2.8cm] (0,0) .. controls (1,-1) and (2,0.5) .. (3,0);
    %
    \draw[thin,fill = teal] (0.1,-0.4) circle (.5ex);
    \draw[thin,fill = teal] (0.3,0) circle (.5ex);
    \draw[thin,fill = teal] (0.4,-0.3) circle (.5ex);
    \draw[thin,fill = teal] (0.95,-0.4) circle (.5ex);
    \draw[thin,fill = teal] (0.85,0.1) circle (.5ex);
    \draw[thin,fill = teal] (1.1,0) circle (.5ex);
    \draw[thin,fill = teal] (1.6,-0.2) circle (.5ex);
    \draw[thin,fill = teal] (1.8,-0.4) circle (.5ex);
    \draw[thin,fill = teal] (2.3,-0.45) circle (.5ex);
    \draw[thin,fill = teal] (2.85,-0.35) circle (.5ex);
    \draw[thin,fill = teal] (0.25,0.6) circle (.5ex);
    \draw[thin,fill = teal] (0.15,0.35) circle (.5ex);
    \draw[thin,fill = teal] (0.7,0.6) circle (.5ex);
    \draw[thin,fill = teal] (1,0.45) circle (.5ex);
    \draw[thin,fill = teal] (1.5,0.4) circle (.5ex);
    \draw[thin,fill = teal] (1.7,0.5) circle (.5ex);
    \draw[thin,fill = teal] (1.8,0.2) circle (.5ex);
    \draw[thin,fill = teal] (2.25,0) circle (.5ex);
    \draw[thin,fill = teal] (2.9,0.1) circle (.5ex);
    \draw[thin,fill = teal] (2.4,0.4) circle (.5ex);
    \draw[thin,fill = teal] (2.65,0.3) circle (.5ex);
    \draw[thin,fill = teal] (0.7,1.1) circle (.5ex);
    \draw[thin,fill = teal] (0.5,1.4) circle (.5ex);
    \draw[thin,fill = teal] (0.9,1.3) circle (.5ex);
    \draw[thin,fill = teal] (1.2,1.32) circle (.5ex);
    \draw[thin,fill = teal] (1.3,1.05) circle (.5ex);
    \draw[thin,fill = teal] (1.7,1.3) circle (.5ex);
    \draw[thin,fill = teal] (1.9,0.85) circle (.5ex);
    \draw[thin,fill = teal] (2.3,1.3) circle (.5ex);
    \draw[thin,fill = teal] (2.27,0.85) circle (.5ex);
    \draw[thin,fill = teal] (2.6,0.9) circle (.5ex);
    \draw[thin,fill = teal] (2.55,1.5) circle (.5ex);
    \draw[thin,fill = teal] (0.6,1.8) circle (.5ex);
    \draw[thin,fill = teal] (1.3,1.65) circle (.5ex);
    \draw[thin,fill = teal] (1.8,1.7) circle (.5ex);
    \draw[thin,fill = teal] (2.3,1.9) circle (.5ex);
    \end{tikzpicture}

%% file: Figs/arrow.tex
\begin{tikzpicture}[xscale=1]
    \draw[very thick,red,yshift = 2cm,xshift=-0.5cm,path fading=west] (0.6,0.2) -- (1.5,0.2) -- (1.5,0.5) -- (2,0) -- (1.5,-0.5) -- (1.5,-0.2) -- (0.6,-0.2);  
    \node at (0.6,2) {\color{red}\it HEAT};
\end{tikzpicture}

%% file: Figs/pnipaam_schematic_2.tex
\begin{tikzpicture}[thick,scale = 0.6]

\tikzset{test/.style n args={3}{
    draw = none,
    postaction={
    decorate,
    decoration={
    markings,
    mark=between positions 0 and \pgfdecoratedpathlength step 0.2pt with {
    \pgfmathsetmacro\myval{multiply(
        divide(
        \pgfkeysvalueof{/pgf/decoration/mark info/distance from start}, \pgfdecoratedpathlength
        ),
        100
    )};
    \pgfsetfillcolor{#3!\myval!#2};
    \pgfpathcircle{\pgfpointorigin}{#1};
    \pgfusepath{fill};}
}}}}

     \foreach \angle / \label in
    {0/3, 30/2, 60/1, 90/12, 120/11, 150/10, 180/9,
     210/8, 240/7, 270/6, 300/5, 330/4}
  {
  
    \draw[test={0.5ex}{white}{teal},xshift = 1.7cm, yshift = 0.5cm] plot coordinates {(\angle:2cm) (\angle:2.4cm)};
    \draw[thin,fill = teal,xshift = 1.7cm, yshift = 0.5cm] (\angle:2.4cm) circle (.833ex);

  }

    \draw[red,snake it] (0,0) .. controls (1,1) and (2,-1) .. (3.3,0);
    \draw[red,snake it,rotate=-5,xshift = 0cm,yshift = -0.7cm] (0,0) .. controls (1,0) and (2,0.5) .. (3.4,0);
    \draw[red,snake it,rotate=-5,xshift = 0cm,yshift = 0.8cm] (0,0) .. controls (1,0.5) and (2,-0.5) .. (3.1,0.3);
    \draw[red,snake it,rotate=0,xshift = 0cm,yshift = 1.6cm] (0,0) .. controls (1,0) and (2,-0.5) .. (3,0.5);
    \draw[red,snake it,rotate=90,xshift = -1cm,yshift = -0.3cm] (0,0) .. controls (1,-1) and (2,0.5) .. (3,0);
    \draw[red,snake it,rotate=95,xshift = -1cm,yshift = -1.4cm] (0,0) .. controls (1,0) and (2,0.5) .. (3,0);
    \draw[red,snake it,rotate=95,xshift = -1cm,yshift = -2.4cm] (-0.3,0) .. controls (1,1) and (2,-0.5) .. (3,0.5);
    \draw[red,snake it,rotate=90,xshift = -1cm,yshift = -2.8cm] (0,0) .. controls (1,-1) and (2,0.5) .. (3,0);
    %
    
    %

    
    \end{tikzpicture}

%% file: Figs/Fig3a_Eql_CaiSuo.tex
%
%
\begin{tikzpicture}

\begin{axis}[%
thick, 
width=0.48\textwidth, 
height=0.4\textwidth,
xmin=300,
xmax=315,
ymin=1,
ymax=3.5,
xlabel = {$\temp$  (K)},
ylabel = {$\elongation$},
yticklabel style = overlay]

\addplot[area legend, dashed, line width=1pt, draw=black, fill=blue, fill opacity=0.2, forget plot]
table[row sep=crcr] {%
x	y\\
452.703178586498	1.18898769637776\\
307.825060062335	1.20813851197385\\
304.406584755581	1.22728932756995\\
303.706450500737	1.24644014316604\\
303.638469225084	1.26559095876213\\
303.782147335624	1.28474177435822\\
304.004291613861	1.30389258995431\\
304.25309882987	1.3230434055504\\
304.506225116804	1.3421942211465\\
304.753650154162	1.36134503674259\\
304.990999410707	1.38049585233868\\
305.216633424736	1.39964666793477\\
305.430275404835	1.41879748353086\\
305.632327194351	1.43794829912695\\
305.823515148984	1.45709911472305\\
306.004702604642	1.47624993031914\\
306.176789858059	1.49540074591523\\
306.340661475454	1.51455156151132\\
306.497159703924	1.53370237710741\\
306.647072425229	1.5528531927035\\
306.791129204043	1.5720040082996\\
306.930001771	1.59115482389569\\
307.064306840241	1.61030563949178\\
307.194610051163	1.62945645508787\\
307.321430340032	1.64860727068396\\
307.445244349667	1.66775808628005\\
307.56649066397	1.68690890187615\\
307.685573759503	1.70605971747224\\
307.802867628269	1.72521053306833\\
307.918719061593	1.74436134866442\\
308.033450605017	1.76351216426051\\
308.147363204661	1.7826629798566\\
308.260738570309	1.8018137954527\\
308.37384128202	1.82096461104879\\
308.486920666624	1.84011542664488\\
308.600212469018	1.85926624224097\\
308.713940341226	1.87841705783706\\
308.828317170027	1.89756787343315\\
308.943546261794	1.91671868902925\\
309.059822401149	1.93586950462534\\
309.177332798111	1.95502032022143\\
309.296257936702	1.97417113581752\\
309.416772336417	1.99332195141361\\
309.539045236569	2.0124727670097\\
309.663241212347	2.0316235826058\\
309.789520730287	2.05077439820189\\
309.918040649982	2.06992521379798\\
310.04895467799	2.08907602939407\\
310.18241377919	2.10822684499016\\
310.318566550232	2.12737766058625\\
310.45755955913	2.14652847618235\\
310.599537654618	2.16567929177844\\
310.744644248433	2.18483010737453\\
310.893021573339	2.20398092297062\\
311.044810919378	2.22313173856671\\
311.200152850559	2.2422825541628\\
311.359187403932	2.26143336975889\\
311.52205427281	2.28058418535499\\
311.688892975663	2.29973500095108\\
311.85984301209	2.31888581654717\\
312.035044007087	2.33803663214326\\
312.214635844725	2.35718744773935\\
312.398758792214	2.37633826333544\\
312.58755361525	2.39548907893154\\
312.781161685429	2.41463989452763\\
312.979725080448	2.43379071012372\\
313.183386677729	2.45294152571981\\
313.392290242048	2.4720923413159\\
313.606580507694	2.49124315691199\\
313.826403255613	2.51039397250809\\
314.051905385987	2.52954478810418\\
314.283234986605	2.54869560370027\\
314.520541397402	2.56784641929636\\
314.76397527146	2.58699723489245\\
315.013688632769	2.60614805048854\\
315.269834931013	2.62529886608464\\
315.532569093603	2.64444968168073\\
315.802047575196	2.66360049727682\\
316.078428404869	2.68275131287291\\
316.361871231159	2.701902128469\\
316.652537365105	2.72105294406509\\
316.95058982146	2.74020375966119\\
317.256193358211	2.75935457525728\\
317.569514514513	2.77850539085337\\
317.890721647186	2.79765620644946\\
318.219984965842	2.81680702204555\\
318.557476566779	2.83595783764164\\
318.903370465692	2.85510865323774\\
319.257842629322	2.87425946883383\\
319.621071006084	2.89341028442992\\
319.993235555773	2.91256110002601\\
320.374518278386	2.9317119156221\\
320.765103242144	2.95086273121819\\
321.165176610752	2.97001354681429\\
321.57492666995	2.98916436241038\\
321.994543853413	3.00831517800647\\
322.424220768029	3.02746599360256\\
322.864152218602	3.04661680919865\\
323.314535232021	3.06576762479474\\
323.775569080921	3.08491844039084\\
324.247455306875	3.10406925598693\\
324.730397743141	3.12322007158302\\
325.224602537001	3.14237088717911\\
325.730278171704	3.1615217027752\\
326.247635488051	3.18067251837129\\
326.776887705634	3.19982333396739\\
327.318250443755	3.21897414956348\\
327.871941742044	3.23812496515957\\
328.438182080787	3.25727578075566\\
329.017194400999	3.27642659635175\\
329.60920412423	3.29557741194784\\
330.214439172148	3.31472822754394\\
330.83312998589	3.33387904314003\\
331.465509545208	3.35302985873612\\
332.111813387412	3.37218067433221\\
332.772279626131	3.3913314899283\\
333.447148969893	3.41048230552439\\
334.136664740541	3.42963312112048\\
334.841072891493	3.44878393671658\\
335.560622025847	3.46793475231267\\
336.295563414354	3.48708556790876\\
337.046151013251	3.50623638350485\\
337.812641481972	3.52538719910094\\
338.595294200739	3.54453801469703\\
339.394371288038	3.56368883029313\\
340.210137617996	3.58283964588922\\
341.042860837642	3.60199046148531\\
341.89281138409	3.6211412770814\\
342.760262501619	3.64029209267749\\
343.645490258669	3.65944290827359\\
344.548773564758	3.67859372386968\\
345.470394187319	3.69774453946577\\
346.410636768465	3.71689535506186\\
347.369788841682	3.73604617065795\\
348.348140848457	3.75519698625404\\
349.345986154838	3.77434780185014\\
350.363621067947	3.79349861744623\\
351.401344852418	3.81264943304232\\
352.459459746793	3.83180024863841\\
353.538270979863	3.8509510642345\\
354.638086786958	3.87010187983059\\
355.759218426192	3.88925269542668\\
356.901980194662	3.90840351102278\\
358.066689444608	3.92755432661887\\
359.253666599524	3.94670514221496\\
360.46323517024	3.96585595781105\\
361.695721770963	3.98500677340714\\
362.951456135278	4.00415758900323\\
364.230771132127	4.02330840459933\\
365.534002781742	4.04245922019542\\
366.86149027156	4.06161003579151\\
368.2135759721	4.0807608513876\\
369.590605452817	4.09991166698369\\
370.992927497924	4.11906248257978\\
372.420894122199	4.13821329817588\\
373.874860586752	4.15736411377197\\
375.355185414786	4.17651492936806\\
376.862230407322	4.19566574496415\\
378.396360658907	4.21481656056024\\
379.957944573309	4.23396737615633\\
381.547353879177	4.25311819175243\\
383.164963645699	4.27226900734852\\
384.811152298231	4.29141982294461\\
386.486301633913	4.3105706385407\\
388.190796837267	4.32972145413679\\
389.925026495781	4.34887226973288\\
391.689382615477	4.36802308532897\\
393.484260636467	4.38717390092507\\
395.310059448488	4.40632471652116\\
397.167181406433	4.42547553211725\\
399.056032345864	4.44462634771334\\
400.97702159851	4.46377716330943\\
402.930562007763	4.48292797890553\\
404.917069944153	4.50207879450162\\
406.936965320815	4.52122961009771\\
408.99067160895	4.5403804256938\\
411.078615853271	4.55953124128989\\
413.201228687438	4.57868205688598\\
415.358944349493	4.59783287248207\\
417.552200697275	4.61698368807817\\
419.781439223834	4.63613450367426\\
422.047105072833	4.65528531927035\\
424.349647053945	4.67443613486644\\
426.68951765824	4.69358695046253\\
429.067173073566	4.71273776605863\\
431.483073199922	4.73188858165472\\
433.937681664828	4.75103939725081\\
436.431465838683	4.7701902128469\\
438.964896850118	4.78934102844299\\
441.53844960135	4.80849184403908\\
444.152602783518	4.82764265963517\\
446.807838892026	4.84679347523127\\
449.50464424187	4.86594429082736\\
452.243508982969	4.88509510642345\\
455.024927115483	4.90424592201954\\
457.849396505131	4.92339673761563\\
460.717418898506	4.94254755321172\\
463.62949993838	4.96169836880782\\
466.586149179006	4.98084918440391\\
469.587880101424	5\\
}--cycle;
\addplot [color=blue, line width=1.5pt, forget plot]
  table[row sep=crcr]{%
349.535952845841	1.0585825729887\\
318.655804650909	1.07848872160997\\
310.349061046778	1.09839487023124\\
307.165052829233	1.1183010188525\\
305.764273731907	1.13820716747377\\
305.122152084437	1.15811331609504\\
304.841728050359	1.17801946471631\\
304.745734636546	1.19792561333758\\
304.747110269917	1.21783176195885\\
304.800137377127	1.23773791058012\\
304.879631885414	1.25764405920139\\
304.971246034342	1.27755020782266\\
305.066624945173	1.29745635644393\\
305.16085063507	1.3173625050652\\
305.251031895602	1.33726865368647\\
305.335497603347	1.35717480230774\\
305.413321298016	1.377080950929\\
305.484034086001	1.39698709955027\\
305.547447823303	1.41689324817154\\
305.603544521203	1.43679939679281\\
305.652406378146	1.45670554541408\\
305.694171189502	1.47661169403535\\
305.729003851143	1.49651784265662\\
305.757078195519	1.51642399127789\\
305.778565525274	1.53633013989916\\
305.793627517913	1.55623628852043\\
305.802411994064	1.5761424371417\\
305.805050562467	1.59604858576297\\
305.801657490193	1.61595473438423\\
305.792329365198	1.6358608830055\\
305.777145262357	1.65576703162677\\
305.756167219799	1.67567318024804\\
305.729440896498	1.69557932886931\\
305.696996325219	1.71548547749058\\
305.658848704155	1.73539162611185\\
305.614999190425	1.75529777473312\\
305.565435672062	1.77520392335439\\
305.510133504281	1.79511007197566\\
305.449056201975	1.81501622059693\\
305.382156084547	1.8349223692182\\
305.309374871918	1.85482851783947\\
305.230644232346	1.87473466646073\\
305.145886283778	1.894640815082\\
305.055014051179	1.91454696370327\\
304.957931882573	1.93445311232454\\
304.854535826737	1.95435926094581\\
304.744713975476	1.97426540956708\\
304.628346773332	1.99417155818835\\
304.505307297468	2.01407770680962\\
304.375461510285	2.03398385543089\\
304.23866848718	2.05389000405216\\
304.09478062164	2.07379615267343\\
303.943643809731	2.0937023012947\\
303.785097615816	2.11360844991597\\
303.618975421222	2.13351459853723\\
303.445104557394	2.1534207471585\\
303.263306424935	2.17332689577977\\
303.073396599817	2.19323304440104\\
302.875184927909	2.21313919302231\\
302.668475608872	2.23304534164358\\
302.45306727036	2.25295149026485\\
302.228753033407	2.27285763888612\\
301.995320569745	2.29276378750739\\
301.752552151782	2.31266993612866\\
301.500224695869	2.33257608474993\\
301.238109799425	2.3524822333712\\
300.965973772462	2.37238838199246\\
300.683577663963	2.39229453061373\\
300.390677283566	2.412200679235\\
300.08702321893	2.43210682785627\\
299.772360849144	2.45201297647754\\
299.446430354507	2.47191912509881\\
299.108966722967	2.49182527372008\\
298.759699753492	2.51173142234135\\
298.398354056615	2.53163757096262\\
298.024649052378	2.55154371958389\\
297.638298965878	2.57144986820516\\
297.239012820602	2.59135601682643\\
296.826494429721	2.6112621654477\\
296.400442385491	2.63116831406896\\
295.960550046925	2.65107446269023\\
295.506505525836	2.6709806113115\\
295.037991671396	2.69088675993277\\
294.554686053316	2.71079290855404\\
294.05626094373	2.73069905717531\\
293.542383297904	2.75060520579658\\
293.01271473383	2.77051135441785\\
292.466911510804	2.79041750303912\\
291.904624507043	2.81032365166039\\
291.325499196417	2.83022980028166\\
290.72917562436	2.85013594890292\\
290.115288383008	2.87004209752419\\
289.483466585624	2.88994824614546\\
288.833333840352	2.90985439476673\\
288.164508223349	2.929760543388\\
287.476602251343	2.94966669200927\\
286.769222853635	2.96957284063054\\
286.041971343604	2.98947898925181\\
285.294443389729	3.00938513787308\\
284.526228986181	3.02929128649435\\
283.736912422979	3.04919743511562\\
282.926072255776	3.06910358373689\\
282.093281275267	3.08900973235816\\
281.238106476263	3.10891588097943\\
280.360109026443	3.12882202960069\\
279.458844234797	3.14872817822196\\
278.533861519799	3.16863432684323\\
277.584704377306	3.1885404754645\\
276.610910348209	3.20844662408577\\
275.612010985851	3.22835277270704\\
274.587531823219	3.24825892132831\\
273.536992339942	3.26816506994958\\
272.459905929075	3.28807121857085\\
271.355779863708	3.30797736719212\\
270.224115263405	3.32788351581339\\
269.064407060477	3.34778966443466\\
267.876143966089	3.36769581305592\\
266.65880843624	3.38760196167719\\
265.411876637595	3.40750811029846\\
264.134818413183	3.42741425891973\\
262.827097247984	3.447320407541\\
261.488170234394	3.46722655616227\\
260.117488037574	3.48713270478354\\
258.714494860708	3.50703885340481\\
257.278628410151	3.52694500202608\\
255.809319860493	3.54685115064735\\
254.305993819529	3.56675729926862\\
252.768068293154	3.58666344788989\\
251.194954650164	3.60656959651116\\
249.586057586996	3.62647574513242\\
247.940775092394	3.64638189375369\\
246.258498412003	3.66628804237496\\
244.538612012907	3.68619419099623\\
242.780493548092	3.7061003396175\\
240.983513820868	3.72600648823877\\
239.147036749234	3.74591263686004\\
237.270419330176	3.76581878548131\\
235.353011603928	3.78572493410258\\
233.39415661819	3.80563108272385\\
231.393190392294	3.82553723134512\\
229.349441881323	3.84544337996639\\
227.262232940198	3.86534952858766\\
225.130878287732	3.88525567720892\\
222.954685470628	3.90516182583019\\
220.732954827458	3.92506797445146\\
218.464979452606	3.94497412307273\\
216.150045160178	3.964880271694\\
213.787430447881	3.98478642031527\\
211.376406460863	4.00469256893654\\
208.916236955556	4.02459871755781\\
206.406178263464	4.04450486617908\\
203.845479254928	4.06441101480035\\
201.233381302877	4.08431716342162\\
198.569118246571	4.10422331204289\\
195.851916355279	4.12412946066415\\
193.080994291979	4.14403560928542\\
190.255563077009	4.16394175790669\\
187.374826051723	4.18384790652796\\
184.437978842113	4.20375405514923\\
181.444209322411	4.2236602037705\\
178.392697578697	4.24356635239177\\
175.282615872452	4.26347250101304\\
172.113128604147	4.28337864963431\\
168.883392276752	4.30328479825558\\
165.592555459314	4.32319094687685\\
162.239758750442	4.34309709549812\\
158.824134741834	4.36300324411939\\
155.344807981748	4.38290939274065\\
151.800894938506	4.40281554136192\\
148.191503963971	4.42272168998319\\
144.515735256972	4.44262783860446\\
140.772680826802	4.46253398722573\\
136.961424456613	4.482440135847\\
133.081041666883	4.50234628446827\\
129.130599678809	4.52225243308954\\
125.109157377759	4.54215858171081\\
121.015765276626	4.56206473033208\\
116.849465479255	4.58197087895335\\
112.60929164385	4.60187702757462\\
108.2942689463	4.62178317619589\\
103.903414043633	4.64168932481715\\
99.4357350372876	4.66159547343842\\
94.8902314365471	4.68150162205969\\
90.2658941218965	4.70140777068096\\
85.5617053082983	4.72131391930223\\
80.7766385086208	4.7412200679235\\
75.9096584969154	4.76112621654477\\
70.9597212717904	4.78103236516604\\
65.9257740197095	4.80093851378731\\
60.8067550783217	4.82084466240858\\
55.6015938997906	4.84075081102985\\
50.3092110140663	4.86065695965111\\
44.9285179922508	4.88056310827239\\
39.4584174098755	4.90046925689365\\
33.8978028101568	4.92037540551492\\
28.2455586673923	4.94028155413619\\
22.5005603501463	4.96018770275746\\
16.6616740846112	4.98009385137873\\
10.7277569178498	5\\
};

\addplot[area legend, dotted, line width=1pt, draw=black, fill=blue, fill opacity=0.1, forget plot]
table[row sep=crcr] {%
x	y\\
319.572915057938	1.1\\
315.347400514719	1.11\\
312.473818617123	1.12\\
310.429778917417	1.13\\
308.91724341931	1.14\\
307.756129483131	1.15\\
306.83467262228	1.16\\
306.084953682245	1.17\\
305.468393599867	1.18\\
304.964176452394	1.19\\
304.559757289958	1.2\\
304.244742299345	1.21\\
304.008268408261	1.22\\
303.838772175501	1.23\\
303.724692904636	1.24\\
303.655229583098	1.25\\
303.621049852416	1.26\\
303.614956769053	1.27\\
303.629762120972	1.28\\
303.660693237033	1.29\\
303.703762041387	1.3\\
303.755843108327	1.31\\
303.814501844071	1.32\\
303.877846468355	1.33\\
303.944409084276	1.34\\
304.013051946797	1.35\\
304.082894339494	1.36\\
304.153255601313	1.37\\
304.223612090687	1.38\\
304.293561576847	1.39\\
304.362797927622	1.4\\
304.43109020767	1.41\\
304.498266619521	1.42\\
304.564201935875	1.43\\
304.628807633496	1.44\\
304.692024109662	1.45\\
304.753814539909	1.46\\
304.81416000685	1.47\\
304.873055628097	1.48\\
304.930507474731	1.49\\
304.986530110424	1.5\\
305.041144626088	1.51\\
305.094377066003	1.52\\
305.146257168934	1.53\\
305.196817359254	1.54\\
305.246091940231	1.55\\
305.294116448818	1.56\\
305.34092714131	1.57\\
305.386560584599	1.58\\
305.43105333229	1.59\\
305.474441670446	1.6\\
305.516761418767	1.61\\
305.558047777355	1.62\\
305.598335210358	1.63\\
305.637657359224	1.64\\
305.676046980189	1.65\\
305.713535901236	1.66\\
305.750154994802	1.67\\
305.785934163006	1.68\\
305.820902333185	1.69\\
305.855087461282	1.7\\
305.888516541738	1.71\\
305.921215622242	1.72\\
305.953209822399	1.73\\
305.984523355246	1.74\\
306.015179550977	1.75\\
306.045200882164	1.76\\
306.074608990009	1.77\\
306.103424711289	1.78\\
306.131668105499	1.79\\
306.159358482112	1.8\\
306.186514427629	1.81\\
306.213153832307	1.82\\
306.239293916362	1.83\\
306.264951255685	1.84\\
306.290141806844	1.85\\
306.314880931381	1.86\\
306.339183419434	1.87\\
306.363063512497	1.88\\
306.386534925462	1.89\\
306.409610867864	1.9\\
306.432304064304	1.91\\
306.454626774137	1.92\\
306.47659081035	1.93\\
306.498207557722	1.94\\
306.519487990226	1.95\\
306.540442687715	1.96\\
306.561081851932	1.97\\
306.581415321828	1.98\\
306.601452588241	1.99\\
306.621202807956	2\\
306.640674817139	2.01\\
306.659877144228	2.02\\
306.678818022235	2.03\\
306.697505400535	2.04\\
306.715946956144	2.05\\
306.734150104492	2.06\\
306.752122009755	2.07\\
306.769869594708	2.08\\
306.787399550184	2.09\\
306.804718344093	2.1\\
306.821832230072	2.11\\
306.838747255754	2.12\\
306.855469270681	2.13\\
306.872003933879	2.14\\
306.888356721106	2.15\\
306.904532931793	2.16\\
306.920537695689	2.17\\
306.936375979222	2.18\\
306.952052591592	2.19\\
306.967572190612	2.2\\
306.982939288293	2.21\\
306.998158256204	2.22\\
307.013233330609	2.23\\
307.028168617383	2.24\\
307.042968096735	2.25\\
307.057635627729	2.26\\
307.072174952621	2.27\\
307.086589701024	2.28\\
307.100883393896	2.29\\
307.115059447376	2.3\\
307.129121176453	2.31\\
307.143071798504	2.32\\
307.156914436676	2.33\\
307.170652123144	2.34\\
307.184287802233	2.35\\
307.197824333423	2.36\\
307.211264494231	2.37\\
307.224610982989	2.38\\
307.237866421501	2.39\\
307.251033357613	2.4\\
307.26411426767	2.41\\
307.27711155889	2.42\\
307.290027571642	2.43\\
307.302864581639	2.44\\
307.315624802046	2.45\\
307.328310385515	2.46\\
307.34092342614	2.47\\
307.353465961336	2.48\\
307.365939973657	2.49\\
307.378347392538	2.5\\
307.390690095982	2.51\\
307.402969912176	2.52\\
307.415188621061	2.53\\
307.427347955828	2.54\\
307.439449604378	2.55\\
307.451495210721	2.56\\
307.46348637632	2.57\\
307.475424661401	2.58\\
307.487311586207	2.59\\
307.499148632206	2.6\\
307.510937243268	2.61\\
307.522678826789	2.62\\
307.534374754784	2.63\\
307.546026364939	2.64\\
307.557634961628	2.65\\
307.569201816894	2.66\\
307.580728171398	2.67\\
307.592215235338	2.68\\
307.603664189328	2.69\\
307.615076185264	2.7\\
307.626452347141	2.71\\
307.637793771861	2.72\\
307.649101530002	2.73\\
307.660376666569	2.74\\
307.671620201715	2.75\\
307.682833131443	2.76\\
307.694016428281	2.77\\
307.705171041941	2.78\\
307.71629789995	2.79\\
307.727397908264	2.8\\
307.738471951866	2.81\\
307.749520895337	2.82\\
307.760545583416	2.83\\
307.771546841539	2.84\\
307.782525476362	2.85\\
307.793482276266	2.86\\
307.804418011852	2.87\\
307.81533343641	2.88\\
307.826229286387	2.89\\
307.837106281827	2.9\\
307.847965126811	2.91\\
307.85880650987	2.92\\
307.869631104398	2.93\\
307.880439569042	2.94\\
307.891232548089	2.95\\
307.902010671836	2.96\\
307.91277455695	2.97\\
307.923524806817	2.98\\
307.934262011884	2.99\\
307.944986749986	3\\
307.955699586669	3.01\\
307.966401075494	3.02\\
307.977091758346	3.03\\
307.987772165721	3.04\\
307.998442817014	3.05\\
308.009104220791	3.06\\
308.019756875061	3.07\\
308.030401267535	3.08\\
308.041037875878	3.09\\
308.051667167955	3.1\\
308.062289602072	3.11\\
308.072905627206	3.12\\
308.083515683232	3.13\\
308.094120201141	3.14\\
308.104719603255	3.15\\
308.115314303434	3.16\\
308.125904707276	3.17\\
308.136491212317	3.18\\
308.147074208216	3.19\\
308.157654076946	3.2\\
308.168231192972	3.21\\
308.178805923428	3.22\\
308.189378628285	3.23\\
308.19994966052	3.24\\
308.210519366278	3.25\\
308.221088085028	3.26\\
308.231656149717	3.27\\
308.242223886919	3.28\\
308.252791616983	3.29\\
308.263359654169	3.3\\
308.273928306791	3.31\\
308.284497877349	3.32\\
308.29506866266	3.33\\
308.305640953985	3.34\\
308.316215037154	3.35\\
308.326791192687	3.36\\
308.33736969591	3.37\\
308.347950817069	3.38\\
308.358534821449	3.39\\
308.369121969471	3.4\\
308.379712516811	3.41\\
308.390306714493	3.42\\
308.400904808996	3.43\\
308.411507042352	3.44\\
308.422113652239	3.45\\
308.432724872076	3.46\\
308.443340931116	3.47\\
308.453962054532	3.48\\
308.464588463509	3.49\\
308.47522037532	3.5\\
320	3.5\\
}--cycle;
\end{axis}

%
\node at (3.75,3) {Coexistence};
\node at (3.75, 1) {Spinodal};
\end{tikzpicture}%

%% file: Figs/Fig3b_Eql_Hirotsu.tex
%
%
\begin{tikzpicture}

\begin{axis}[%
thick, 
width=0.48\textwidth, 
height=0.4\textwidth,
xmin=300,
xmax=315,
ymin=1,
ymax=3.5,
xlabel = {$\temp$  (K)},
ylabel = {$\elongation$},
yticklabel style = overlay]

\addplot[area legend, dashed, line width=1pt, draw=black, fill=red, fill opacity=0.2, forget plot]
table[row sep=crcr] {%
x	y\\
372.509525841867	1.00432560582563\\
318.384372451929	1.0244043716255\\
313.131031720635	1.04448313742537\\
311.152278496081	1.06456190322524\\
310.11785151261	1.08464066902512\\
309.48470177788	1.10471943482499\\
309.059271101255	1.12479820062486\\
308.755309453041	1.14487696642473\\
308.528566287007	1.1649557322246\\
308.35400317162	1.18503449802447\\
308.21639052314	1.20511326382434\\
308.105934887657	1.22519202962421\\
308.016054038673	1.24527079542409\\
307.942162227865	1.26534956122396\\
307.880968167373	1.28542832702383\\
307.830049857315	1.3055070928237\\
307.787586748414	1.32558585862357\\
307.752185265861	1.34566462442344\\
307.722761817114	1.36574339022331\\
307.698462343156	1.38582215602318\\
307.678605759673	1.40590092182306\\
307.662643406024	1.42597968762293\\
307.650129458302	1.4460584534228\\
307.640699000827	1.46613721922267\\
307.634051542455	1.48621598502254\\
307.629938466413	1.50629475082241\\
307.628153363499	1.52637351662228\\
307.628524507214	1.54645228242215\\
307.630908939561	1.56653104822202\\
307.635187781715	1.5866098140219\\
307.641262485917	1.60668857982177\\
307.649051817581	1.62676734562164\\
307.658489409046	1.64684611142151\\
307.669521764516	1.66692487722138\\
307.68210662394	1.68700364302125\\
307.696211614482	1.70708240882112\\
307.711813134	1.72716117462099\\
307.728895422888	1.74723994042087\\
307.74744978974	1.76731870622074\\
307.767473963357	1.78739747202061\\
307.788971549035	1.80747623782048\\
307.811951571389	1.82755500362035\\
307.836428089267	1.84763376942022\\
307.862419871016	1.86771253522009\\
307.889950120452	1.88779130101996\\
307.919046245583	1.90787006681984\\
307.949739663513	1.92794883261971\\
307.982065636051	1.94802759841958\\
308.01606313145	1.96810636421945\\
308.051774708443	1.98818513001932\\
308.089246419336	2.00826389581919\\
308.128527729435	2.02834266161906\\
308.169671450469	2.04842142741893\\
308.212733686059	2.0685001932188\\
308.25777378751	2.08857895901868\\
308.304854318502	2.10865772481855\\
308.354041027426	2.12873649061842\\
308.405402826293	2.14881525641829\\
308.459011775274	2.16889402221816\\
308.514943072086	2.18897278801803\\
308.573275045512	2.2090515538179\\
308.634089152441	2.22913031961777\\
308.697469977917	2.24920908541765\\
308.763505237706	2.26928785121752\\
308.832285783	2.28936661701739\\
308.903905606891	2.30944538281726\\
308.9784618523	2.32952414861713\\
309.056054821106	2.349602914417\\
309.136787984217	2.36968168021687\\
309.220767992387	2.38976044601674\\
309.308104687594	2.40983921181662\\
309.398911114816	2.42991797761649\\
309.493303534074	2.44999674341636\\
309.59140143261	2.47007550921623\\
309.693327537111	2.4901542750161\\
309.799207825877	2.51023304081597\\
309.909171540859	2.53031180661584\\
310.02335119951	2.55039057241571\\
310.14188260638	2.57046933821558\\
310.264904864424	2.59054810401546\\
310.392560385976	2.61062686981533\\
310.524994903365	2.6307056356152\\
310.662357479149	2.65078440141507\\
310.804800515943	2.67086316721494\\
310.952479765845	2.69094193301481\\
311.105554339438	2.71102069881468\\
311.264186714376	2.73109946461455\\
311.428542743548	2.75117823041443\\
311.598791662833	2.7712569962143\\
311.775106098446	2.79133576201417\\
311.957662073891	2.81141452781404\\
312.146639016538	2.83149329361391\\
312.342219763823	2.85157205941378\\
312.544590569109	2.87165082521365\\
312.753941107209	2.89172959101352\\
312.970464479606	2.91180835681339\\
313.194357219372	2.93188712261327\\
313.42581929583	2.95196588841314\\
313.665054118958	2.97204465421301\\
313.912268543579	2.99212342001288\\
314.167672873344	3.01220218581275\\
314.431480864541	3.03228095161262\\
314.703909729751	3.05235971741249\\
314.985180141367	3.07243848321236\\
315.275516235017	3.09251724901224\\
315.575145612889	3.11259601481211\\
315.88429934701	3.13267478061198\\
316.20321198247	3.15275354641185\\
316.532121540637	3.17283231221172\\
316.871269522378	3.19291107801159\\
317.220900911295	3.21298984381146\\
317.581264177012	3.23306860961133\\
317.952611278525	3.25314737541121\\
318.335197667626	3.27322614121108\\
318.729282292439	3.29330490701095\\
319.135127601062	3.31338367281082\\
319.552999545344	3.33346243861069\\
319.983167584814	3.35354120441056\\
320.425904690769	3.37361997021043\\
320.881487350538	3.3936987360103\\
321.350195571934	3.41377750181017\\
321.832312887911	3.43385626761005\\
322.328126361431	3.45393503340992\\
322.83792659055	3.47401379920979\\
323.362007713744	3.49409256500966\\
323.900667415466	3.51417133080953\\
324.454206931966	3.5342500966094\\
325.022931057345	3.55432886240927\\
325.6071481499	3.57440762820914\\
326.20717013871	3.59448639400902\\
326.823312530514	3.61456515980889\\
327.45589441686	3.63464392560876\\
328.105238481534	3.65472269140863\\
328.771671008274	3.6748014572085\\
329.455521888777	3.69488022300837\\
330.157124630984	3.71495898880824\\
330.876816367665	3.73503775460811\\
331.614937865291	3.75511652040799\\
332.371833533193	3.77519528620786\\
333.147851433019	3.79527405200773\\
333.943343288476	3.8153528178076\\
334.758664495366	3.83543158360747\\
335.594174131899	3.85551034940734\\
336.450234969308	3.87558911520721\\
337.327213482729	3.89566788100708\\
338.225479862376	3.91574664680695\\
339.145408024989	3.93582541260683\\
340.087375625551	3.9559041784067\\
341.051764069282	3.97598294420657\\
342.038958523899	3.99606171000644\\
343.049347932141	4.01614047580631\\
344.083325024548	4.03621924160618\\
345.141286332497	4.05629800740605\\
346.223632201493	4.07637677320592\\
347.330766804696	4.0964555390058\\
348.463098156696	4.11653430480567\\
349.621038127524	4.13661307060554\\
350.805002456885	4.15669183640541\\
352.015410768631	4.17677060220528\\
353.252686585439	4.19684936800515\\
354.517257343717	4.21692813380502\\
355.80955440871	4.23700689960489\\
357.130013089819	4.25708566540477\\
358.479072656114	4.27716443120464\\
359.857176352039	4.29724319700451\\
361.264771413316	4.31732196280438\\
362.702309083019	4.33740072860425\\
364.170244627835	4.35747949440412\\
365.669037354496	4.37755826020399\\
367.199150626375	4.39763702600386\\
368.761051880251	4.41771579180373\\
370.355212643221	4.43779455760361\\
371.982108549777	4.45787332340348\\
373.642219359017	4.47795208920335\\
375.33602897201	4.49803085500322\\
377.064025449289	4.51810962080309\\
378.826701028481	4.53818838660296\\
380.624552142062	4.55826715240283\\
382.458079435241	4.5783459182027\\
384.327787783952	4.59842468400258\\
386.234186312973	4.61850344980245\\
388.177788414141	4.63858221560232\\
390.159111764683	4.65866098140219\\
392.178678345644	4.67873974720206\\
394.237014460407	4.69881851300193\\
396.33465075332	4.7188972788018\\
398.472122228391	4.73897604460167\\
400.649968268094	4.75905481040154\\
402.868732652229	4.77913357620142\\
405.128963576882	4.79921234200129\\
407.43121367345	4.81929110780116\\
409.776040027734	4.83936987360103\\
412.164004199106	4.8594486394009\\
414.595672239737	4.87952740520077\\
417.071614713886	4.89960617100064\\
419.592406717247	4.91968493680051\\
422.158627896358	4.93976370260039\\
424.770862468047	4.95984246840026\\
427.429699238943	4.97992123420013\\
430.135731625027	5\\
}--cycle;
\addplot [color=red, line width=1.5pt, forget plot]
  table[row sep=crcr]{%
313.793680480592	1.00432560582563\\
310.309050195154	1.0244043716255\\
309.366954761083	1.04448313742537\\
308.873600836652	1.06456190322524\\
308.562325001658	1.08464066902512\\
308.346621417456	1.10471943482499\\
308.188428203112	1.12479820062486\\
308.067967117797	1.14487696642473\\
307.973797349906	1.1649557322246\\
307.898780541925	1.18503449802447\\
307.838203006093	1.20511326382434\\
307.788811062892	1.22519202962421\\
307.748276856096	1.24527079542409\\
307.714884523929	1.26534956122396\\
307.68733677262	1.28542832702383\\
307.664630826657	1.3055070928237\\
307.645976159492	1.32558585862357\\
307.630738343136	1.34566462442344\\
307.618399757392	1.36574339022331\\
307.608531488675	1.38582215602318\\
307.600772838464	1.40590092182306\\
307.594816119475	1.42597968762293\\
307.590395197362	1.4460584534228\\
307.587276731633	1.46613721922267\\
307.585253392229	1.48621598502254\\
307.584138542751	1.50629475082241\\
307.583762026557	1.52637351662228\\
307.583966792055	1.54645228242215\\
307.584606163551	1.56653104822202\\
307.585541613732	1.5866098140219\\
307.586640929606	1.60668857982177\\
307.587776689807	1.62676734562164\\
307.588824990302	1.64684611142151\\
307.589664369853	1.66692487722138\\
307.590174897282	1.68700364302125\\
307.590237390739	1.70708240882112\\
307.589732745382	1.72716117462099\\
307.588541350692	1.74723994042087\\
307.586542582352	1.76731870622074\\
307.583614356547	1.78739747202061\\
307.579632736824	1.80747623782048\\
307.574471585494	1.82755500362035\\
307.568002252967	1.84763376942022\\
307.560093299612	1.86771253522009\\
307.55061024564	1.88779130101996\\
307.539415345274	1.90787006681984\\
307.5263673821	1.92794883261971\\
307.511321482951	1.94802759841958\\
307.494128948145	1.96810636421945\\
307.474637096191	1.98818513001932\\
307.452689121373	2.00826389581919\\
307.428123962859	2.02834266161906\\
307.400776184159	2.04842142741893\\
307.370475861927	2.0685001932188\\
307.33704848324	2.08857895901868\\
307.300314850579	2.10865772481855\\
307.260090993849	2.12873649061842\\
307.216188088857	2.14881525641829\\
307.168412381712	2.16889402221816\\
307.116565118687	2.18897278801803\\
307.060442481132	2.2090515538179\\
306.999835525046	2.22913031961777\\
306.934530124982	2.24920908541765\\
306.86430692196	2.26928785121752\\
306.788941275108	2.28936661701739\\
306.708203216767	2.30944538281726\\
306.621857410806	2.32952414861713\\
306.529663113932	2.349602914417\\
306.431374139771	2.36968168021687\\
306.326738825521	2.38976044601674\\
306.215500000991	2.40983921181662\\
306.097394959845	2.42991797761649\\
305.972155432877	2.44999674341636\\
305.839507563163	2.47007550921623\\
305.699171882937	2.4901542750161\\
305.550863292033	2.51023304081597\\
305.394291037767	2.53031180661584\\
305.229158696121	2.55039057241571\\
305.055164154096	2.57046933821558\\
304.871999593124	2.59054810401546\\
304.679351473416	2.61062686981533\\
304.476900519129	2.6307056356152\\
304.264321704265	2.65078440141507\\
304.041284239177	2.67086316721494\\
303.807451557608	2.69094193301481\\
303.562481304165	2.71102069881468\\
303.306025322127	2.73109946461455\\
303.037729641525	2.75117823041443\\
302.757234467419	2.7712569962143\\
302.464174168264	2.79133576201417\\
302.158177264341	2.81141452781404\\
301.838866416166	2.83149329361391\\
301.505858412794	2.85157205941378\\
301.158764160029	2.87165082521365\\
300.797188668393	2.89172959101352\\
300.420731040894	2.91180835681339\\
300.02898446048	2.93188712261327\\
299.621536177183	2.95196588841314\\
299.197967494889	2.97204465421301\\
298.757853757694	2.99212342001288\\
298.300764335849	3.01220218581275\\
297.826262611229	3.03228095161262\\
297.333905962296	3.05235971741249\\
296.82324574858	3.07243848321236\\
296.2938272946	3.09251724901224\\
295.745189873246	3.11259601481211\\
295.176866688601	3.13267478061198\\
294.588384858153	3.15275354641185\\
293.97926539446	3.17283231221172\\
293.349023186165	3.19291107801159\\
292.697166978463	3.21298984381146\\
292.023199352893	3.23306860961133\\
291.326616706577	3.25314737541121\\
290.606909230804	3.27322614121108\\
289.863560888992	3.29330490701095\\
289.096049394103	3.31338367281082\\
288.303846185354	3.33346243861069\\
287.486416404398	3.35354120441056\\
286.643218870865	3.37361997021043\\
285.773706057314	3.3936987360103\\
284.877324063646	3.41377750181017\\
283.95351259085	3.43385626761005\\
283.001704914259	3.45393503340992\\
282.021327856231	3.47401379920979\\
281.011801758231	3.49409256500966\\
279.972540452483	3.51417133080953\\
278.902951232973	3.5342500966094\\
277.802434826039	3.55432886240927\\
276.670385360395	3.57440762820914\\
275.506190336713	3.59448639400902\\
274.309230596737	3.61456515980889\\
273.078880291856	3.63464392560876\\
271.814506851321	3.65472269140863\\
270.515470950009	3.6748014572085\\
269.181126475742	3.69488022300837\\
267.810820496192	3.71495898880824\\
266.40389322542	3.73503775460811\\
264.959677990063	3.75511652040799\\
263.477501195096	3.77519528620786\\
261.956682289205	3.79527405200773\\
260.396533730034	3.8153528178076\\
258.796360948815	3.83543158360747\\
257.155462314869	3.85551034940734\\
255.473129099788	3.87558911520721\\
253.748645441292	3.89566788100708\\
251.981288306826	3.91574664680695\\
250.170327456897	3.93582541260683\\
248.315025408094	3.9559041784067\\
246.414637396067	3.97598294420657\\
244.468411337995	3.99606171000644\\
242.475587795054	4.01614047580631\\
240.435399934652	4.03621924160618\\
238.347073492294	4.05629800740605\\
236.20982673354	4.07637677320592\\
234.022870415485	4.0964555390058\\
231.785407748369	4.11653430480567\\
229.49663435678	4.13661307060554\\
227.15573824088	4.15669183640541\\
224.761899737365	4.17677060220528\\
222.314291480398	4.19684936800515\\
219.812078362399	4.21692813380502\\
217.254417494584	4.23700689960489\\
214.640458167668	4.25708566540477\\
211.969341812168	4.27716443120464\\
209.240201958806	4.29724319700451\\
206.452164198781	4.31732196280438\\
203.604346143959	4.33740072860425\\
200.695857386908	4.35747949440412\\
197.725799461025	4.37755826020399\\
194.693265800479	4.39763702600386\\
191.597341700051	4.41771579180373\\
188.437104275171	4.43779455760361\\
185.211622421482	4.45787332340348\\
181.919956774953	4.47795208920335\\
178.561159671272	4.49803085500322\\
175.134275105754	4.51810962080309\\
171.638338693058	4.53818838660296\\
168.072377626492	4.55826715240283\\
164.435410638136	4.5783459182027\\
160.726447958129	4.59842468400258\\
156.944491274164	4.61850344980245\\
153.088533691466	4.63858221560232\\
149.157559691909	4.65866098140219\\
145.150545093925	4.67873974720206\\
141.066457011937	4.69881851300193\\
136.904253815697	4.7188972788018\\
132.66288509027	4.73897604460167\\
128.34129159533	4.75905481040154\\
123.93840522467	4.77913357620142\\
119.453148965862	4.79921234200129\\
114.884436860058	4.81929110780116\\
110.231173961082	4.83936987360103\\
105.49225629545	4.8594486394009\\
100.666570821941	4.87952740520077\\
95.7529953909418	4.89960617100064\\
90.7503987044713	4.91968493680051\\
85.6576402756422	4.93976370260039\\
80.4735703883136	4.95984246840026\\
75.197030056833	4.97992123420013\\
69.8268509859278	5\\
};

\addplot[area legend, dotted, line width=1pt, draw=black, fill=red, fill opacity=0.1, forget plot]
table[row sep=crcr] {%
x	y\\
317.425520359657	1.001\\
311.917014451707	1.011\\
310.674153661165	1.021\\
310.007567857547	1.031\\
309.573023776336	1.041\\
309.261413269661	1.051\\
309.024736982539	1.061\\
308.837884769422	1.071\\
308.686197454723	1.081\\
308.560437257317	1.091\\
308.454443823547	1.101\\
308.363925392889	1.111\\
308.285785874921	1.121\\
308.217727126144	1.131\\
308.15800223866	1.141\\
308.105256275137	1.151\\
308.058419951167	1.161\\
308.016636605667	1.171\\
307.979210784195	1.181\\
307.945571251571	1.191\\
307.915243875595	1.201\\
307.887831410549	1.211\\
307.862998196591	1.221\\
307.840458422163	1.231\\
307.819967008945	1.241\\
307.801312454125	1.251\\
307.784311152389	1.261\\
307.768802849318	1.271\\
307.754646969346	1.281\\
307.741719626283	1.291\\
307.729911171433	1.301\\
307.719124168674	1.311\\
307.709271711389	1.321\\
307.700276014946	1.331\\
307.692067233092	1.341\\
307.684582457044	1.351\\
307.677764865105	1.361\\
307.671562996674	1.371\\
307.665930129166	1.381\\
307.660823742247	1.391\\
307.656205053661	1.401\\
307.65203861661	1.411\\
307.648291969211	1.421\\
307.644935328594	1.431\\
307.641941319976	1.441\\
307.639284742773	1.451\\
307.63694394069	1.461\\
307.634892717367	1.471\\
307.633115971858	1.481\\
307.631593753839	1.491\\
307.629408856794	1.51\\
307.628475377902	1.52\\
307.627784932017	1.53\\
307.627285558432	1.54\\
307.626955884586	1.55\\
307.626784974668	1.56\\
307.626762450513	1.57\\
307.626878605672	1.58\\
307.627124364396	1.59\\
307.62749123706	1.6\\
307.627971268688	1.61\\
307.628557032875	1.62\\
307.629241545193	1.63\\
307.630018262059	1.64\\
307.630881047092	1.65\\
307.631824141679	1.66\\
307.6328421368	1.67\\
307.633929952233	1.68\\
307.635082814645	1.69\\
307.636296236063	1.7\\
307.637565996865	1.71\\
307.638888128505	1.72\\
307.640258896882	1.73\\
307.641674788239	1.74\\
307.643132495772	1.75\\
307.644628906107	1.76\\
307.646161088089	1.77\\
307.647726281724	1.78\\
307.64932188774	1.79\\
307.65094545805	1.8\\
307.652594686965	1.81\\
307.654267402715	1.82\\
307.655961559666	1.83\\
307.657675231032	1.84\\
307.659406602033	1.85\\
307.66115396349	1.86\\
307.662915705767	1.87\\
307.664690313228	1.88\\
307.666476358846	1.89\\
307.668272499258	1.9\\
307.67007747014	1.91\\
307.671890081737	1.92\\
307.673709214785	1.93\\
307.675533816603	1.94\\
307.677362897425	1.95\\
307.679195526934	1.96\\
307.681030831054	1.97\\
307.682867988824	1.98\\
307.684706229561	1.99\\
307.686544830092	2\\
307.688383112195	2.01\\
307.690220440161	2.02\\
307.692056218497	2.03\\
307.693889889762	2.04\\
307.695720932508	2.05\\
307.697548859347	2.06\\
307.699373215111	2.07\\
307.701193575128	2.08\\
307.70300954357	2.09\\
307.704820751909	2.1\\
307.706626857444	2.11\\
307.708427541915	2.12\\
307.71022251018	2.13\\
307.712011488972	2.14\\
307.713794225717	2.15\\
307.715570487414	2.16\\
307.717340059576	2.17\\
307.719102745227	2.18\\
307.720858363943	2.19\\
307.722606750957	2.2\\
307.724347756302	2.21\\
307.726081243993	2.22\\
307.727807091266	2.23\\
307.729525187845	2.24\\
307.731235435247	2.25\\
307.732937746133	2.26\\
307.734632043677	2.27\\
307.736318260981	2.28\\
307.737996340514	2.29\\
307.739666233578	2.3\\
307.741327899806	2.31\\
307.742981306683	2.32\\
307.74462642909	2.33\\
307.746263248875	2.34\\
307.747891754446	2.35\\
307.749511940377	2.36\\
307.751123807047	2.37\\
307.752727360286	2.38\\
307.754322611044	2.39\\
307.755909575079	2.4\\
307.757488272655	2.41\\
307.759058728262	2.42\\
307.760620970347	2.43\\
307.762175031056	2.44\\
307.763720945995	2.45\\
307.765258754001	2.46\\
307.766788496922	2.47\\
307.768310219415	2.48\\
307.769823968745	2.49\\
307.771329794601	2.5\\
307.772827748926	2.51\\
307.774317885739	2.52\\
307.77580026099	2.53\\
307.777274932397	2.54\\
307.778741959316	2.55\\
307.780201402598	2.56\\
307.781653324466	2.57\\
307.783097788394	2.58\\
307.78453485899	2.59\\
307.785964601892	2.6\\
307.787387083662	2.61\\
307.788802371692	2.62\\
307.790210534111	2.63\\
307.791611639699	2.64\\
307.793005757805	2.65\\
307.794392958268	2.66\\
307.795773311349	2.67\\
307.797146887657	2.68\\
307.798513758084	2.69\\
307.799873993749	2.7\\
307.801227665933	2.71\\
307.802574846029	2.72\\
307.803915605491	2.73\\
307.805250015781	2.74\\
307.80657814833	2.75\\
307.807900074488	2.76\\
307.809215865491	2.77\\
307.810525592418	2.78\\
307.811829326157	2.79\\
307.813127137372	2.8\\
307.814419096473	2.81\\
307.815705273585	2.82\\
307.816985738521	2.83\\
307.818260560758	2.84\\
307.819529809412	2.85\\
307.820793553218	2.86\\
307.822051860504	2.87\\
307.823304799179	2.88\\
307.824552436712	2.89\\
307.825794840114	2.9\\
307.827032075925	2.91\\
307.828264210201	2.92\\
307.829491308498	2.93\\
307.830713435864	2.94\\
307.831930656826	2.95\\
307.833143035379	2.96\\
307.834350634982	2.97\\
307.835553518544	2.98\\
307.836751748422	2.99\\
307.837945386411	3\\
307.83913449374	3.01\\
307.840319131068	3.02\\
307.841499358478	3.03\\
307.842675235472	3.04\\
307.843846820972	3.05\\
307.845014173315	3.06\\
307.84617735025	3.07\\
307.847336408939	3.08\\
307.848491405954	3.09\\
307.849642397276	3.1\\
307.850789438298	3.11\\
307.851932583821	3.12\\
307.853071888058	3.13\\
307.854207404632	3.14\\
307.85533918658	3.15\\
307.856467286352	3.16\\
307.857591755814	3.17\\
307.858712646251	3.18\\
307.859830008368	3.19\\
307.860943892292	3.2\\
307.862054347577	3.21\\
307.863161423204	3.22\\
307.864265167587	3.23\\
307.865365628575	3.24\\
307.866462853456	3.25\\
307.86755688896	3.26\\
307.868647781261	3.27\\
307.869735575987	3.28\\
307.870820318218	3.29\\
307.871902052491	3.3\\
307.872980822809	3.31\\
307.874056672638	3.32\\
307.875129644918	3.33\\
307.876199782063	3.34\\
307.87726712597	3.35\\
307.878331718019	3.36\\
307.879393599081	3.37\\
307.88045280952	3.38\\
307.8815093892	3.39\\
307.882563377491	3.4\\
307.88361481327	3.41\\
307.884663734929	3.42\\
307.885710180378	3.43\\
307.886754187052	3.44\\
307.887795791914	3.45\\
307.888835031461	3.46\\
307.889871941729	3.47\\
307.890906558298	3.48\\
307.891938916295	3.49\\
307.892969050403	3.5\\
320	4\\
}--cycle;
\addplot [color=black, line width=1pt, only marks, mark size=2.0pt, mark=x, mark options={solid, black}, forget plot]
  table[row sep=crcr]{%
318.137	1.00084202081454\\
315.457	1.01476512022757\\
314.243	1.02023181963183\\
313.645	1.01944906320437\\
312.931	1.0077795262758\\
312.025	1.00700632366981\\
311.196	1.01476512022757\\
310.483	1.0077795262758\\
309.712	1.0077795262758\\
309.23	1.02023181963183\\
308.651	1.02572796900611\\
308.439	1.04720618310162\\
308.034	1.07406903888056\\
307.629	1.14823874663726\\
307.61	1.21627616976842\\
307.629	1.27262035251482\\
307.629	1.30827446371269\\
307.437	1.65716415624286\\
307.34	1.77841125327102\\
307.244	1.94252014390036\\
306.974	2.21325879043445\\
306.704	2.26307456967358\\
306.53	2.36791185111658\\
306.492	2.4045410616237\\
305.701	2.51786359058679\\
305.431	2.58245164172363\\
304.93	2.60834798553389\\
304.217	2.69173435673389\\
303.33	2.72081406800565\\
302.771	2.77991935819674\\
301.325	2.84467199834834\\
298.876	2.98331330132783\\
297.43	3.02249458260951\\
295.425	3.09764894466446\\
293.324	3.21883610785232\\
};
\end{axis}

%
%
\begin{axis}[xshift=.04\textwidth,yshift=.04\textwidth,
     axis on top,  
    width=0.22\textwidth,
    height = 0.2\textwidth,
    enlargelimits=false, 
    xmin=307.4,xmax=307.8,ymin=1,ymax=2.2, 
    xtick={307.4,307.8}, 
    ytick={1,2},
    label style={font=\tiny}, 
    tick label style={font=\tiny},
  ]
 \addplot[area legend, dashed, line width=1pt, draw=black, fill=red, fill opacity=0.2, forget plot]
table[row sep=crcr] {%
x	y\\
308.016054038673	1.24527079542409\\
307.942162227865	1.26534956122396\\
307.880968167373	1.28542832702383\\
307.830049857315	1.3055070928237\\
307.787586748414	1.32558585862357\\
307.752185265861	1.34566462442344\\
307.722761817114	1.36574339022331\\
307.698462343156	1.38582215602318\\
307.678605759673	1.40590092182306\\
307.662643406024	1.42597968762293\\
307.650129458302	1.4460584534228\\
307.640699000827	1.46613721922267\\
307.634051542455	1.48621598502254\\
307.629938466413	1.50629475082241\\
307.628153363499	1.52637351662228\\
307.628524507214	1.54645228242215\\
307.630908939561	1.56653104822202\\
307.635187781715	1.5866098140219\\
307.641262485917	1.60668857982177\\
307.649051817581	1.62676734562164\\
307.658489409046	1.64684611142151\\
307.669521764516	1.66692487722138\\
307.68210662394	1.68700364302125\\
307.696211614482	1.70708240882112\\
307.711813134	1.72716117462099\\
307.728895422888	1.74723994042087\\
307.74744978974	1.76731870622074\\
307.767473963357	1.78739747202061\\
307.788971549035	1.80747623782048\\
307.811951571389	1.82755500362035\\
307.836428089267	1.84763376942022\\
307.862419871016	1.86771253522009\\
307.889950120452	1.88779130101996\\
307.919046245583	1.90787006681984\\
307.949739663513	1.92794883261971\\
307.982065636051	1.94802759841958\\
308.01606313145	1.96810636421945\\
}--cycle;
\addplot [color=red, line width=1.5pt, forget plot]
  table[row sep=crcr]{%
308.067967117797	1.14487696642473\\
307.973797349906	1.1649557322246\\
307.898780541925	1.18503449802447\\
307.838203006093	1.20511326382434\\
307.788811062892	1.22519202962421\\
307.748276856096	1.24527079542409\\
307.714884523929	1.26534956122396\\
307.68733677262	1.28542832702383\\
307.664630826657	1.3055070928237\\
307.645976159492	1.32558585862357\\
307.630738343136	1.34566462442344\\
307.618399757392	1.36574339022331\\
307.608531488675	1.38582215602318\\
307.600772838464	1.40590092182306\\
307.594816119475	1.42597968762293\\
307.590395197362	1.4460584534228\\
307.587276731633	1.46613721922267\\
307.585253392229	1.48621598502254\\
307.584138542751	1.50629475082241\\
307.583762026557	1.52637351662228\\
307.583966792055	1.54645228242215\\
307.584606163551	1.56653104822202\\
307.585541613732	1.5866098140219\\
307.586640929606	1.60668857982177\\
307.587776689807	1.62676734562164\\
307.588824990302	1.64684611142151\\
307.589664369853	1.66692487722138\\
307.590174897282	1.68700364302125\\
307.590237390739	1.70708240882112\\
307.589732745382	1.72716117462099\\
307.588541350692	1.74723994042087\\
307.586542582352	1.76731870622074\\
307.583614356547	1.78739747202061\\
307.579632736824	1.80747623782048\\
307.574471585494	1.82755500362035\\
307.568002252967	1.84763376942022\\
307.560093299612	1.86771253522009\\
307.55061024564	1.88779130101996\\
307.539415345274	1.90787006681984\\
307.5263673821	1.92794883261971\\
307.511321482951	1.94802759841958\\
307.494128948145	1.96810636421945\\
307.474637096191	1.98818513001932\\
307.452689121373	2.00826389581919\\
307.428123962859	2.02834266161906\\
307.400776184159	2.04842142741893\\
307.370475861927	2.0685001932188\\
307.33704848324	2.08857895901868\\
307.300314850579	2.10865772481855\\
307.260090993849	2.12873649061842\\
307.216188088857	2.14881525641829\\
307.168412381712	2.16889402221816\\
307.116565118687	2.18897278801803\\
307.060442481132	2.2090515538179\\
306.999835525046	2.22913031961777\\
};

\addplot[area legend, dotted, line width=1pt, draw=black, fill=red, fill opacity=0.1, forget plot]
table[row sep=crcr] {%
x	y\\
308.016636605667	1.171\\
307.979210784195	1.181\\
307.945571251571	1.191\\
307.915243875595	1.201\\
307.887831410549	1.211\\
307.862998196591	1.221\\
307.840458422163	1.231\\
307.819967008945	1.241\\
307.801312454125	1.251\\
307.784311152389	1.261\\
307.768802849318	1.271\\
307.754646969346	1.281\\
307.741719626283	1.291\\
307.729911171433	1.301\\
307.719124168674	1.311\\
307.709271711389	1.321\\
307.700276014946	1.331\\
307.692067233092	1.341\\
307.684582457044	1.351\\
307.677764865105	1.361\\
307.671562996674	1.371\\
307.665930129166	1.381\\
307.660823742247	1.391\\
307.656205053661	1.401\\
307.65203861661	1.411\\
307.648291969211	1.421\\
307.644935328594	1.431\\
307.641941319976	1.441\\
307.639284742773	1.451\\
307.63694394069	1.461\\
307.634892717367	1.471\\
307.633115971858	1.481\\
307.631593753839	1.491\\
307.629408856794	1.51\\
307.628475377902	1.52\\
307.627784932017	1.53\\
307.627285558432	1.54\\
307.626955884586	1.55\\
307.626784974668	1.56\\
307.626762450513	1.57\\
307.626878605672	1.58\\
307.627124364396	1.59\\
307.62749123706	1.6\\
307.627971268688	1.61\\
307.628557032875	1.62\\
307.629241545193	1.63\\
307.630018262059	1.64\\
307.630881047092	1.65\\
307.631824141679	1.66\\
307.6328421368	1.67\\
307.633929952233	1.68\\
307.635082814645	1.69\\
307.636296236063	1.7\\
307.637565996865	1.71\\
307.638888128505	1.72\\
307.640258896882	1.73\\
307.641674788239	1.74\\
307.643132495772	1.75\\
307.644628906107	1.76\\
307.646161088089	1.77\\
307.647726281724	1.78\\
307.64932188774	1.79\\
307.65094545805	1.8\\
307.652594686965	1.81\\
307.654267402715	1.82\\
307.655961559666	1.83\\
307.657675231032	1.84\\
307.659406602033	1.85\\
307.66115396349	1.86\\
307.662915705767	1.87\\
307.664690313228	1.88\\
307.666476358846	1.89\\
307.668272499258	1.9\\
307.67007747014	1.91\\
307.671890081737	1.92\\
307.673709214785	1.93\\
307.675533816603	1.94\\
307.677362897425	1.95\\
307.679195526934	1.96\\
307.681030831054	1.97\\
307.682867988824	1.98\\
307.684706229561	1.99\\
307.686544830092	2\\
307.688383112195	2.01\\
307.690220440161	2.02\\
307.692056218497	2.03\\
307.693889889762	2.04\\
307.695720932508	2.05\\
307.697548859347	2.06\\
307.699373215111	2.07\\
307.701193575128	2.08\\
307.70300954357	2.09\\
307.704820751909	2.1\\
307.706626857444	2.11\\
307.708427541915	2.12\\
307.71022251018	2.13\\
307.712011488972	2.14\\
307.713794225717	2.15\\
307.715570487414	2.16\\
307.717340059576	2.17\\
307.719102745227	2.18\\
307.720858363943	2.19\\
307.722606750957	2.2\\
307.724347756302	2.21\\
307.726081243993	2.22\\
307.727807091266	2.23\\
307.729525187845	2.24\\
307.731235435247	2.25\\
307.732937746133	2.26\\
307.734632043677	2.27\\
307.736318260981	2.28\\
307.737996340514	2.29\\
307.739666233578	2.3\\
307.741327899806	2.31\\
307.742981306683	2.32\\
307.74462642909	2.33\\
307.746263248875	2.34\\
307.747891754446	2.35\\
307.749511940377	2.36\\
307.751123807047	2.37\\
307.752727360286	2.38\\
307.754322611044	2.39\\
307.755909575079	2.4\\
307.757488272655	2.41\\
307.759058728262	2.42\\
307.760620970347	2.43\\
307.762175031056	2.44\\
307.763720945995	2.45\\
307.765258754001	2.46\\
307.766788496922	2.47\\
307.768310219415	2.48\\
307.769823968745	2.49\\
307.771329794601	2.5\\
307.772827748926	2.51\\
307.774317885739	2.52\\
307.77580026099	2.53\\
307.777274932397	2.54\\
307.778741959316	2.55\\
307.780201402598	2.56\\
307.781653324466	2.57\\
307.783097788394	2.58\\
307.78453485899	2.59\\
307.785964601892	2.6\\
307.787387083662	2.61\\
307.788802371692	2.62\\
307.790210534111	2.63\\
307.791611639699	2.64\\
307.793005757805	2.65\\
307.794392958268	2.66\\
307.795773311349	2.67\\
307.797146887657	2.68\\
307.798513758084	2.69\\
307.799873993749	2.7\\
307.801227665933	2.71\\
307.802574846029	2.72\\
307.803915605491	2.73\\
307.805250015781	2.74\\
307.80657814833	2.75\\
307.807900074488	2.76\\
307.809215865491	2.77\\
307.810525592418	2.78\\
307.811829326157	2.79\\
307.813127137372	2.8\\
307.814419096473	2.81\\
307.815705273585	2.82\\
307.816985738521	2.83\\
307.818260560758	2.84\\
307.819529809412	2.85\\
307.820793553218	2.86\\
307.822051860504	2.87\\
307.823304799179	2.88\\
307.824552436712	2.89\\
307.825794840114	2.9\\
307.827032075925	2.91\\
307.828264210201	2.92\\
307.829491308498	2.93\\
307.830713435864	2.94\\
307.831930656826	2.95\\
307.833143035379	2.96\\
307.834350634982	2.97\\
307.835553518544	2.98\\
307.836751748422	2.99\\
307.837945386411	3\\
307.83913449374	3.01\\
307.840319131068	3.02\\
307.841499358478	3.03\\
307.842675235472	3.04\\
307.843846820972	3.05\\
307.845014173315	3.06\\
307.84617735025	3.07\\
307.847336408939	3.08\\
307.848491405954	3.09\\
307.849642397276	3.1\\
307.850789438298	3.11\\
307.851932583821	3.12\\
307.853071888058	3.13\\
307.854207404632	3.14\\
307.85533918658	3.15\\
307.856467286352	3.16\\
307.857591755814	3.17\\
307.858712646251	3.18\\
307.859830008368	3.19\\
307.860943892292	3.2\\
307.862054347577	3.21\\
307.863161423204	3.22\\
307.864265167587	3.23\\
307.865365628575	3.24\\
307.866462853456	3.25\\
307.86755688896	3.26\\
307.868647781261	3.27\\
307.869735575987	3.28\\
307.870820318218	3.29\\
307.871902052491	3.3\\
307.872980822809	3.31\\
307.874056672638	3.32\\
307.875129644918	3.33\\
307.876199782063	3.34\\
307.87726712597	3.35\\
307.878331718019	3.36\\
307.879393599081	3.37\\
307.88045280952	3.38\\
307.8815093892	3.39\\
307.882563377491	3.4\\
307.88361481327	3.41\\
307.884663734929	3.42\\
307.885710180378	3.43\\
307.886754187052	3.44\\
307.887795791914	3.45\\
307.888835031461	3.46\\
307.889871941729	3.47\\
307.890906558298	3.48\\
307.891938916295	3.49\\
307.892969050403	3.5\\
320	4\\
}--cycle;
\end{axis}
%
\node at (3.75,3.35) {Coexistence};
\node at (3.75, 1.25) {Spinodal};

\end{tikzpicture}%

%% file: Figs/Fig4a_Trajectory.tex

\newcommand{\swellcolor}{ForestGreen}%
\newcommand{\shrinkcolor}{Orange}%

\begin{tikzpicture}[scale=1]
    \begin{axis}[axis on top,  
thick, 
width=0.48\textwidth, 
height=0.4\textwidth,
enlargelimits=false, 
xmin=300,xmax=310,ymin=0.8,ymax=2.8, 
xlabel = {$\temp$ (K)},
ylabel = {$\lambda$},
yticklabel style = overlay,
]
\addplot[area legend, dashed, line width=0.5pt, draw=black, fill=blue, fill opacity=0.2, forget plot]
table[row sep=crcr] {%
x	y\\
452.703178586498	1.18898769637776\\
307.825060062335	1.20813851197385\\
304.406584755581	1.22728932756995\\
303.706450500737	1.24644014316604\\
303.638469225084	1.26559095876213\\
303.782147335624	1.28474177435822\\
304.004291613861	1.30389258995431\\
304.25309882987	1.3230434055504\\
304.506225116804	1.3421942211465\\
304.753650154162	1.36134503674259\\
304.990999410707	1.38049585233868\\
305.216633424736	1.39964666793477\\
305.430275404835	1.41879748353086\\
305.632327194351	1.43794829912695\\
305.823515148984	1.45709911472305\\
306.004702604642	1.47624993031914\\
306.176789858059	1.49540074591523\\
306.340661475454	1.51455156151132\\
306.497159703924	1.53370237710741\\
306.647072425229	1.5528531927035\\
306.791129204043	1.5720040082996\\
306.930001771	1.59115482389569\\
307.064306840241	1.61030563949178\\
307.194610051163	1.62945645508787\\
307.321430340032	1.64860727068396\\
307.445244349667	1.66775808628005\\
307.56649066397	1.68690890187615\\
307.685573759503	1.70605971747224\\
307.802867628269	1.72521053306833\\
307.918719061593	1.74436134866442\\
308.033450605017	1.76351216426051\\
308.147363204661	1.7826629798566\\
308.260738570309	1.8018137954527\\
308.37384128202	1.82096461104879\\
308.486920666624	1.84011542664488\\
308.600212469018	1.85926624224097\\
308.713940341226	1.87841705783706\\
308.828317170027	1.89756787343315\\
308.943546261794	1.91671868902925\\
309.059822401149	1.93586950462534\\
309.177332798111	1.95502032022143\\
309.296257936702	1.97417113581752\\
309.416772336417	1.99332195141361\\
309.539045236569	2.0124727670097\\
309.663241212347	2.0316235826058\\
309.789520730287	2.05077439820189\\
309.918040649982	2.06992521379798\\
310.04895467799	2.08907602939407\\
310.18241377919	2.10822684499016\\
310.318566550232	2.12737766058625\\
310.45755955913	2.14652847618235\\
310.599537654618	2.16567929177844\\
310.744644248433	2.18483010737453\\
310.893021573339	2.20398092297062\\
311.044810919378	2.22313173856671\\
311.200152850559	2.2422825541628\\
311.359187403932	2.26143336975889\\
311.52205427281	2.28058418535499\\
311.688892975663	2.29973500095108\\
311.85984301209	2.31888581654717\\
312.035044007087	2.33803663214326\\
312.214635844725	2.35718744773935\\
312.398758792214	2.37633826333544\\
312.58755361525	2.39548907893154\\
312.781161685429	2.41463989452763\\
312.979725080448	2.43379071012372\\
313.183386677729	2.45294152571981\\
313.392290242048	2.4720923413159\\
313.606580507694	2.49124315691199\\
313.826403255613	2.51039397250809\\
314.051905385987	2.52954478810418\\
314.283234986605	2.54869560370027\\
314.520541397402	2.56784641929636\\
314.76397527146	2.58699723489245\\
315.013688632769	2.60614805048854\\
315.269834931013	2.62529886608464\\
315.532569093603	2.64444968168073\\
315.802047575196	2.66360049727682\\
316.078428404869	2.68275131287291\\
316.361871231159	2.701902128469\\
316.652537365105	2.72105294406509\\
316.95058982146	2.74020375966119\\
317.256193358211	2.75935457525728\\
317.569514514513	2.77850539085337\\
317.890721647186	2.79765620644946\\
318.219984965842	2.81680702204555\\
318.557476566779	2.83595783764164\\
318.903370465692	2.85510865323774\\
319.257842629322	2.87425946883383\\
319.621071006084	2.89341028442992\\
319.993235555773	2.91256110002601\\
320.374518278386	2.9317119156221\\
320.765103242144	2.95086273121819\\
321.165176610752	2.97001354681429\\
321.57492666995	2.98916436241038\\
321.994543853413	3.00831517800647\\
322.424220768029	3.02746599360256\\
322.864152218602	3.04661680919865\\
323.314535232021	3.06576762479474\\
323.775569080921	3.08491844039084\\
324.247455306875	3.10406925598693\\
324.730397743141	3.12322007158302\\
325.224602537001	3.14237088717911\\
325.730278171704	3.1615217027752\\
326.247635488051	3.18067251837129\\
326.776887705634	3.19982333396739\\
327.318250443755	3.21897414956348\\
327.871941742044	3.23812496515957\\
328.438182080787	3.25727578075566\\
329.017194400999	3.27642659635175\\
329.60920412423	3.29557741194784\\
330.214439172148	3.31472822754394\\
330.83312998589	3.33387904314003\\
331.465509545208	3.35302985873612\\
332.111813387412	3.37218067433221\\
332.772279626131	3.3913314899283\\
333.447148969893	3.41048230552439\\
334.136664740541	3.42963312112048\\
334.841072891493	3.44878393671658\\
335.560622025847	3.46793475231267\\
336.295563414354	3.48708556790876\\
337.046151013251	3.50623638350485\\
337.812641481972	3.52538719910094\\
338.595294200739	3.54453801469703\\
339.394371288038	3.56368883029313\\
340.210137617996	3.58283964588922\\
341.042860837642	3.60199046148531\\
341.89281138409	3.6211412770814\\
342.760262501619	3.64029209267749\\
343.645490258669	3.65944290827359\\
344.548773564758	3.67859372386968\\
345.470394187319	3.69774453946577\\
346.410636768465	3.71689535506186\\
347.369788841682	3.73604617065795\\
348.348140848457	3.75519698625404\\
349.345986154838	3.77434780185014\\
350.363621067947	3.79349861744623\\
351.401344852418	3.81264943304232\\
352.459459746793	3.83180024863841\\
353.538270979863	3.8509510642345\\
354.638086786958	3.87010187983059\\
355.759218426192	3.88925269542668\\
356.901980194662	3.90840351102278\\
358.066689444608	3.92755432661887\\
359.253666599524	3.94670514221496\\
360.46323517024	3.96585595781105\\
361.695721770963	3.98500677340714\\
362.951456135278	4.00415758900323\\
364.230771132127	4.02330840459933\\
365.534002781742	4.04245922019542\\
366.86149027156	4.06161003579151\\
368.2135759721	4.0807608513876\\
369.590605452817	4.09991166698369\\
370.992927497924	4.11906248257978\\
372.420894122199	4.13821329817588\\
373.874860586752	4.15736411377197\\
375.355185414786	4.17651492936806\\
376.862230407322	4.19566574496415\\
378.396360658907	4.21481656056024\\
379.957944573309	4.23396737615633\\
381.547353879177	4.25311819175243\\
383.164963645699	4.27226900734852\\
384.811152298231	4.29141982294461\\
386.486301633913	4.3105706385407\\
388.190796837267	4.32972145413679\\
389.925026495781	4.34887226973288\\
391.689382615477	4.36802308532897\\
393.484260636467	4.38717390092507\\
395.310059448488	4.40632471652116\\
397.167181406433	4.42547553211725\\
399.056032345864	4.44462634771334\\
400.97702159851	4.46377716330943\\
402.930562007763	4.48292797890553\\
404.917069944153	4.50207879450162\\
406.936965320815	4.52122961009771\\
408.99067160895	4.5403804256938\\
411.078615853271	4.55953124128989\\
413.201228687438	4.57868205688598\\
415.358944349493	4.59783287248207\\
417.552200697275	4.61698368807817\\
419.781439223834	4.63613450367426\\
422.047105072833	4.65528531927035\\
424.349647053945	4.67443613486644\\
426.68951765824	4.69358695046253\\
429.067173073566	4.71273776605863\\
431.483073199922	4.73188858165472\\
433.937681664828	4.75103939725081\\
436.431465838683	4.7701902128469\\
438.964896850118	4.78934102844299\\
441.53844960135	4.80849184403908\\
444.152602783518	4.82764265963517\\
446.807838892026	4.84679347523127\\
449.50464424187	4.86594429082736\\
452.243508982969	4.88509510642345\\
455.024927115483	4.90424592201954\\
457.849396505131	4.92339673761563\\
460.717418898506	4.94254755321172\\
463.62949993838	4.96169836880782\\
466.586149179006	4.98084918440391\\
469.587880101424	5\\
}--cycle;
\addplot [color=blue, line width=1pt, forget plot]
  table[row sep=crcr]{%
349.535952845841	1.0585825729887\\
318.655804650909	1.07848872160997\\
310.349061046778	1.09839487023124\\
307.165052829233	1.1183010188525\\
305.764273731907	1.13820716747377\\
305.122152084437	1.15811331609504\\
304.841728050359	1.17801946471631\\
304.745734636546	1.19792561333758\\
304.747110269917	1.21783176195885\\
304.800137377127	1.23773791058012\\
304.879631885414	1.25764405920139\\
304.971246034342	1.27755020782266\\
305.066624945173	1.29745635644393\\
305.16085063507	1.3173625050652\\
305.251031895602	1.33726865368647\\
305.335497603347	1.35717480230774\\
305.413321298016	1.377080950929\\
305.484034086001	1.39698709955027\\
305.547447823303	1.41689324817154\\
305.603544521203	1.43679939679281\\
305.652406378146	1.45670554541408\\
305.694171189502	1.47661169403535\\
305.729003851143	1.49651784265662\\
305.757078195519	1.51642399127789\\
305.778565525274	1.53633013989916\\
305.793627517913	1.55623628852043\\
305.802411994064	1.5761424371417\\
305.805050562467	1.59604858576297\\
305.801657490193	1.61595473438423\\
305.792329365198	1.6358608830055\\
305.777145262357	1.65576703162677\\
305.756167219799	1.67567318024804\\
305.729440896498	1.69557932886931\\
305.696996325219	1.71548547749058\\
305.658848704155	1.73539162611185\\
305.614999190425	1.75529777473312\\
305.565435672062	1.77520392335439\\
305.510133504281	1.79511007197566\\
305.449056201975	1.81501622059693\\
305.382156084547	1.8349223692182\\
305.309374871918	1.85482851783947\\
305.230644232346	1.87473466646073\\
305.145886283778	1.894640815082\\
305.055014051179	1.91454696370327\\
304.957931882573	1.93445311232454\\
304.854535826737	1.95435926094581\\
304.744713975476	1.97426540956708\\
304.628346773332	1.99417155818835\\
304.505307297468	2.01407770680962\\
304.375461510285	2.03398385543089\\
304.23866848718	2.05389000405216\\
304.09478062164	2.07379615267343\\
303.943643809731	2.0937023012947\\
303.785097615816	2.11360844991597\\
303.618975421222	2.13351459853723\\
303.445104557394	2.1534207471585\\
303.263306424935	2.17332689577977\\
303.073396599817	2.19323304440104\\
302.875184927909	2.21313919302231\\
302.668475608872	2.23304534164358\\
302.45306727036	2.25295149026485\\
302.228753033407	2.27285763888612\\
301.995320569745	2.29276378750739\\
301.752552151782	2.31266993612866\\
301.500224695869	2.33257608474993\\
301.238109799425	2.3524822333712\\
300.965973772462	2.37238838199246\\
300.683577663963	2.39229453061373\\
300.390677283566	2.412200679235\\
300.08702321893	2.43210682785627\\
299.772360849144	2.45201297647754\\
299.446430354507	2.47191912509881\\
299.108966722967	2.49182527372008\\
298.759699753492	2.51173142234135\\
298.398354056615	2.53163757096262\\
298.024649052378	2.55154371958389\\
297.638298965878	2.57144986820516\\
297.239012820602	2.59135601682643\\
296.826494429721	2.6112621654477\\
296.400442385491	2.63116831406896\\
295.960550046925	2.65107446269023\\
295.506505525836	2.6709806113115\\
295.037991671396	2.69088675993277\\
294.554686053316	2.71079290855404\\
294.05626094373	2.73069905717531\\
293.542383297904	2.75060520579658\\
293.01271473383	2.77051135441785\\
292.466911510804	2.79041750303912\\
291.904624507043	2.81032365166039\\
291.325499196417	2.83022980028166\\
290.72917562436	2.85013594890292\\
290.115288383008	2.87004209752419\\
289.483466585624	2.88994824614546\\
288.833333840352	2.90985439476673\\
288.164508223349	2.929760543388\\
287.476602251343	2.94966669200927\\
286.769222853635	2.96957284063054\\
286.041971343604	2.98947898925181\\
285.294443389729	3.00938513787308\\
284.526228986181	3.02929128649435\\
283.736912422979	3.04919743511562\\
282.926072255776	3.06910358373689\\
282.093281275267	3.08900973235816\\
281.238106476263	3.10891588097943\\
280.360109026443	3.12882202960069\\
279.458844234797	3.14872817822196\\
278.533861519799	3.16863432684323\\
277.584704377306	3.1885404754645\\
276.610910348209	3.20844662408577\\
275.612010985851	3.22835277270704\\
274.587531823219	3.24825892132831\\
273.536992339942	3.26816506994958\\
272.459905929075	3.28807121857085\\
271.355779863708	3.30797736719212\\
270.224115263405	3.32788351581339\\
269.064407060477	3.34778966443466\\
267.876143966089	3.36769581305592\\
266.65880843624	3.38760196167719\\
265.411876637595	3.40750811029846\\
264.134818413183	3.42741425891973\\
262.827097247984	3.447320407541\\
261.488170234394	3.46722655616227\\
260.117488037574	3.48713270478354\\
258.714494860708	3.50703885340481\\
257.278628410151	3.52694500202608\\
255.809319860493	3.54685115064735\\
254.305993819529	3.56675729926862\\
252.768068293154	3.58666344788989\\
251.194954650164	3.60656959651116\\
249.586057586996	3.62647574513242\\
247.940775092394	3.64638189375369\\
246.258498412003	3.66628804237496\\
244.538612012907	3.68619419099623\\
242.780493548092	3.7061003396175\\
240.983513820868	3.72600648823877\\
239.147036749234	3.74591263686004\\
237.270419330176	3.76581878548131\\
235.353011603928	3.78572493410258\\
233.39415661819	3.80563108272385\\
231.393190392294	3.82553723134512\\
229.349441881323	3.84544337996639\\
227.262232940198	3.86534952858766\\
225.130878287732	3.88525567720892\\
222.954685470628	3.90516182583019\\
220.732954827458	3.92506797445146\\
218.464979452606	3.94497412307273\\
216.150045160178	3.964880271694\\
213.787430447881	3.98478642031527\\
211.376406460863	4.00469256893654\\
208.916236955556	4.02459871755781\\
206.406178263464	4.04450486617908\\
203.845479254928	4.06441101480035\\
201.233381302877	4.08431716342162\\
198.569118246571	4.10422331204289\\
195.851916355279	4.12412946066415\\
193.080994291979	4.14403560928542\\
190.255563077009	4.16394175790669\\
187.374826051723	4.18384790652796\\
184.437978842113	4.20375405514923\\
181.444209322411	4.2236602037705\\
178.392697578697	4.24356635239177\\
175.282615872452	4.26347250101304\\
172.113128604147	4.28337864963431\\
168.883392276752	4.30328479825558\\
165.592555459314	4.32319094687685\\
162.239758750442	4.34309709549812\\
158.824134741834	4.36300324411939\\
155.344807981748	4.38290939274065\\
151.800894938506	4.40281554136192\\
148.191503963971	4.42272168998319\\
144.515735256972	4.44262783860446\\
140.772680826802	4.46253398722573\\
136.961424456613	4.482440135847\\
133.081041666883	4.50234628446827\\
129.130599678809	4.52225243308954\\
125.109157377759	4.54215858171081\\
121.015765276626	4.56206473033208\\
116.849465479255	4.58197087895335\\
112.60929164385	4.60187702757462\\
108.2942689463	4.62178317619589\\
103.903414043633	4.64168932481715\\
99.4357350372876	4.66159547343842\\
94.8902314365471	4.68150162205969\\
90.2658941218965	4.70140777068096\\
85.5617053082983	4.72131391930223\\
80.7766385086208	4.7412200679235\\
75.9096584969154	4.76112621654477\\
70.9597212717904	4.78103236516604\\
65.9257740197095	4.80093851378731\\
60.8067550783217	4.82084466240858\\
55.6015938997906	4.84075081102985\\
50.3092110140663	4.86065695965111\\
44.9285179922508	4.88056310827239\\
39.4584174098755	4.90046925689365\\
33.8978028101568	4.92037540551492\\
28.2455586673923	4.94028155413619\\
22.5005603501463	4.96018770275746\\
16.6616740846112	4.98009385137873\\
10.7277569178498	5\\
};

\addplot[area legend, dotted, line width=0.5pt, draw=black, fill=blue, fill opacity=0.1, forget plot]
table[row sep=crcr] {%
x	y\\
319.572915057938	1.1\\
315.347400514719	1.11\\
312.473818617123	1.12\\
310.429778917417	1.13\\
308.91724341931	1.14\\
307.756129483131	1.15\\
306.83467262228	1.16\\
306.084953682245	1.17\\
305.468393599867	1.18\\
304.964176452394	1.19\\
304.559757289958	1.2\\
304.244742299345	1.21\\
304.008268408261	1.22\\
303.838772175501	1.23\\
303.724692904636	1.24\\
303.655229583098	1.25\\
303.621049852416	1.26\\
303.614956769053	1.27\\
303.629762120972	1.28\\
303.660693237033	1.29\\
303.703762041387	1.3\\
303.755843108327	1.31\\
303.814501844071	1.32\\
303.877846468355	1.33\\
303.944409084276	1.34\\
304.013051946797	1.35\\
304.082894339494	1.36\\
304.153255601313	1.37\\
304.223612090687	1.38\\
304.293561576847	1.39\\
304.362797927622	1.4\\
304.43109020767	1.41\\
304.498266619521	1.42\\
304.564201935875	1.43\\
304.628807633496	1.44\\
304.692024109662	1.45\\
304.753814539909	1.46\\
304.81416000685	1.47\\
304.873055628097	1.48\\
304.930507474731	1.49\\
304.986530110424	1.5\\
305.041144626088	1.51\\
305.094377066003	1.52\\
305.146257168934	1.53\\
305.196817359254	1.54\\
305.246091940231	1.55\\
305.294116448818	1.56\\
305.34092714131	1.57\\
305.386560584599	1.58\\
305.43105333229	1.59\\
305.474441670446	1.6\\
305.516761418767	1.61\\
305.558047777355	1.62\\
305.598335210358	1.63\\
305.637657359224	1.64\\
305.676046980189	1.65\\
305.713535901236	1.66\\
305.750154994802	1.67\\
305.785934163006	1.68\\
305.820902333185	1.69\\
305.855087461282	1.7\\
305.888516541738	1.71\\
305.921215622242	1.72\\
305.953209822399	1.73\\
305.984523355246	1.74\\
306.015179550977	1.75\\
306.045200882164	1.76\\
306.074608990009	1.77\\
306.103424711289	1.78\\
306.131668105499	1.79\\
306.159358482112	1.8\\
306.186514427629	1.81\\
306.213153832307	1.82\\
306.239293916362	1.83\\
306.264951255685	1.84\\
306.290141806844	1.85\\
306.314880931381	1.86\\
306.339183419434	1.87\\
306.363063512497	1.88\\
306.386534925462	1.89\\
306.409610867864	1.9\\
306.432304064304	1.91\\
306.454626774137	1.92\\
306.47659081035	1.93\\
306.498207557722	1.94\\
306.519487990226	1.95\\
306.540442687715	1.96\\
306.561081851932	1.97\\
306.581415321828	1.98\\
306.601452588241	1.99\\
306.621202807956	2\\
306.640674817139	2.01\\
306.659877144228	2.02\\
306.678818022235	2.03\\
306.697505400535	2.04\\
306.715946956144	2.05\\
306.734150104492	2.06\\
306.752122009755	2.07\\
306.769869594708	2.08\\
306.787399550184	2.09\\
306.804718344093	2.1\\
306.821832230072	2.11\\
306.838747255754	2.12\\
306.855469270681	2.13\\
306.872003933879	2.14\\
306.888356721106	2.15\\
306.904532931793	2.16\\
306.920537695689	2.17\\
306.936375979222	2.18\\
306.952052591592	2.19\\
306.967572190612	2.2\\
306.982939288293	2.21\\
306.998158256204	2.22\\
307.013233330609	2.23\\
307.028168617383	2.24\\
307.042968096735	2.25\\
307.057635627729	2.26\\
307.072174952621	2.27\\
307.086589701024	2.28\\
307.100883393896	2.29\\
307.115059447376	2.3\\
307.129121176453	2.31\\
307.143071798504	2.32\\
307.156914436676	2.33\\
307.170652123144	2.34\\
307.184287802233	2.35\\
307.197824333423	2.36\\
307.211264494231	2.37\\
307.224610982989	2.38\\
307.237866421501	2.39\\
307.251033357613	2.4\\
307.26411426767	2.41\\
307.27711155889	2.42\\
307.290027571642	2.43\\
307.302864581639	2.44\\
307.315624802046	2.45\\
307.328310385515	2.46\\
307.34092342614	2.47\\
307.353465961336	2.48\\
307.365939973657	2.49\\
307.378347392538	2.5\\
307.390690095982	2.51\\
307.402969912176	2.52\\
307.415188621061	2.53\\
307.427347955828	2.54\\
307.439449604378	2.55\\
307.451495210721	2.56\\
307.46348637632	2.57\\
307.475424661401	2.58\\
307.487311586207	2.59\\
307.499148632206	2.6\\
307.510937243268	2.61\\
307.522678826789	2.62\\
307.534374754784	2.63\\
307.546026364939	2.64\\
307.557634961628	2.65\\
307.569201816894	2.66\\
307.580728171398	2.67\\
307.592215235338	2.68\\
307.603664189328	2.69\\
307.615076185264	2.7\\
307.626452347141	2.71\\
307.637793771861	2.72\\
307.649101530002	2.73\\
307.660376666569	2.74\\
307.671620201715	2.75\\
307.682833131443	2.76\\
307.694016428281	2.77\\
307.705171041941	2.78\\
307.71629789995	2.79\\
307.727397908264	2.8\\
307.738471951866	2.81\\
307.749520895337	2.82\\
307.760545583416	2.83\\
307.771546841539	2.84\\
307.782525476362	2.85\\
307.793482276266	2.86\\
307.804418011852	2.87\\
307.81533343641	2.88\\
307.826229286387	2.89\\
307.837106281827	2.9\\
307.847965126811	2.91\\
307.85880650987	2.92\\
307.869631104398	2.93\\
307.880439569042	2.94\\
307.891232548089	2.95\\
307.902010671836	2.96\\
307.91277455695	2.97\\
307.923524806817	2.98\\
307.934262011884	2.99\\
307.944986749986	3\\
307.955699586669	3.01\\
307.966401075494	3.02\\
307.977091758346	3.03\\
307.987772165721	3.04\\
307.998442817014	3.05\\
308.009104220791	3.06\\
308.019756875061	3.07\\
308.030401267535	3.08\\
308.041037875878	3.09\\
308.051667167955	3.1\\
308.062289602072	3.11\\
308.072905627206	3.12\\
308.083515683232	3.13\\
308.094120201141	3.14\\
308.104719603255	3.15\\
308.115314303434	3.16\\
308.125904707276	3.17\\
308.136491212317	3.18\\
308.147074208216	3.19\\
308.157654076946	3.2\\
308.168231192972	3.21\\
308.178805923428	3.22\\
308.189378628285	3.23\\
308.19994966052	3.24\\
308.210519366278	3.25\\
308.221088085028	3.26\\
308.231656149717	3.27\\
308.242223886919	3.28\\
308.252791616983	3.29\\
308.263359654169	3.3\\
308.273928306791	3.31\\
308.284497877349	3.32\\
308.29506866266	3.33\\
308.305640953985	3.34\\
308.316215037154	3.35\\
308.326791192687	3.36\\
308.33736969591	3.37\\
308.347950817069	3.38\\
308.358534821449	3.39\\
308.369121969471	3.4\\
308.379712516811	3.41\\
308.390306714493	3.42\\
308.400904808996	3.43\\
308.411507042352	3.44\\
308.422113652239	3.45\\
308.432724872076	3.46\\
308.443340931116	3.47\\
308.453962054532	3.48\\
308.464588463509	3.49\\
308.47522037532	3.5\\
320	3.5\\
}--cycle;
%
\addplot[color=\swellcolor, mark=none, -latex, line width=1.5pt]
  table[row sep=crcr]{%
302 1.11\\
302	1.75\\
};
\addplot[color=\swellcolor, mark=none, line width=1.5pt]
  table[row sep=crcr]{%
302 1.70\\
302	2.29\\
};
%
\addplot[color=\shrinkcolor, mark=none, -latex, line width=1.5pt]
  table[row sep=crcr]{%
308 2.29\\
308	1.65\\
};
\addplot[color=\shrinkcolor, mark=none, line width=1.5pt]
  table[row sep=crcr]{%
308 1.70\\
308	1.11\\
};
%
\addplot[color=\swellcolor, mark=none, -latex, line width=1.5pt]
  table[row sep=crcr]{%
304 1.11\\
304	1.65\\
};
\addplot[color=\swellcolor, mark=none, line width=1.5pt]
  table[row sep=crcr]{%
304 2.09\\
304	1.60\\
};
%
\addplot[color=\shrinkcolor, mark=none, -latex, dotted, line width=1.5pt]
  table[row sep=crcr]{%
302 2.29\\
305	2.29\\
};
\addplot[color=\shrinkcolor, mark=none, dotted, line width=1.5pt]
  table[row sep=crcr]{%
305 2.29\\
308	2.29\\
};
\addplot[color=\swellcolor, mark=none, -latex, dotted, line width=1.5pt]
  table[row sep=crcr]{%
308 1.11\\
305	1.11\\
};
\addplot[color=\swellcolor, mark=none, dotted, line width=1.5pt]
  table[row sep=crcr]{%
305 1.11\\
302	1.11\\
};
%
\addplot[color=black, mark=x, only marks, mark size=3.0pt, line width=1.5pt]
    table[row sep=crcr]{%
302 1.11\\
308 2.29\\
304 1.11\\
};
%
\end{axis}

\node[rotate=90,color=black] at (0.7,1.75) {Swelling};
\node[rotate=-90,color=black] at (4.2, 1.75) {Shrinking};

\node[color=black] at (1.25,1.75) {(i)};
\node[color=black] at (3.65,2) {(ii)};
\node[color=black] at (2.3,1.75) {(iii)};

\end{tikzpicture}

%% file: Figs/Fig4b_Trajectory.tex

\newcommand{\swellcolor}{ForestGreen}%
\newcommand{\shrinkcolor}{Orange}%

\begin{tikzpicture}[scale=1]
    \begin{axis}[axis on top,  
thick, 
width=0.48\textwidth, 
height=0.4\textwidth,
enlargelimits=false, 
xmin=302,xmax=310,ymin=0.8,ymax=2.8, 
xlabel = {$\temp$ (K)},
ylabel = {$\lambda$},
yticklabel style = overlay,
]
\addplot[area legend, dashed, line width=0.5pt, draw=black, fill=red, fill opacity=0.2, forget plot]
table[row sep=crcr] {%
x	y\\
372.509525841867	1.00432560582563\\
318.384372451929	1.0244043716255\\
313.131031720635	1.04448313742537\\
311.152278496081	1.06456190322524\\
310.11785151261	1.08464066902512\\
309.48470177788	1.10471943482499\\
309.059271101255	1.12479820062486\\
308.755309453041	1.14487696642473\\
308.528566287007	1.1649557322246\\
308.35400317162	1.18503449802447\\
308.21639052314	1.20511326382434\\
308.105934887657	1.22519202962421\\
308.016054038673	1.24527079542409\\
307.942162227865	1.26534956122396\\
307.880968167373	1.28542832702383\\
307.830049857315	1.3055070928237\\
307.787586748414	1.32558585862357\\
307.752185265861	1.34566462442344\\
307.722761817114	1.36574339022331\\
307.698462343156	1.38582215602318\\
307.678605759673	1.40590092182306\\
307.662643406024	1.42597968762293\\
307.650129458302	1.4460584534228\\
307.640699000827	1.46613721922267\\
307.634051542455	1.48621598502254\\
307.629938466413	1.50629475082241\\
307.628153363499	1.52637351662228\\
307.628524507214	1.54645228242215\\
307.630908939561	1.56653104822202\\
307.635187781715	1.5866098140219\\
307.641262485917	1.60668857982177\\
307.649051817581	1.62676734562164\\
307.658489409046	1.64684611142151\\
307.669521764516	1.66692487722138\\
307.68210662394	1.68700364302125\\
307.696211614482	1.70708240882112\\
307.711813134	1.72716117462099\\
307.728895422888	1.74723994042087\\
307.74744978974	1.76731870622074\\
307.767473963357	1.78739747202061\\
307.788971549035	1.80747623782048\\
307.811951571389	1.82755500362035\\
307.836428089267	1.84763376942022\\
307.862419871016	1.86771253522009\\
307.889950120452	1.88779130101996\\
307.919046245583	1.90787006681984\\
307.949739663513	1.92794883261971\\
307.982065636051	1.94802759841958\\
308.01606313145	1.96810636421945\\
308.051774708443	1.98818513001932\\
308.089246419336	2.00826389581919\\
308.128527729435	2.02834266161906\\
308.169671450469	2.04842142741893\\
308.212733686059	2.0685001932188\\
308.25777378751	2.08857895901868\\
308.304854318502	2.10865772481855\\
308.354041027426	2.12873649061842\\
308.405402826293	2.14881525641829\\
308.459011775274	2.16889402221816\\
308.514943072086	2.18897278801803\\
308.573275045512	2.2090515538179\\
308.634089152441	2.22913031961777\\
308.697469977917	2.24920908541765\\
308.763505237706	2.26928785121752\\
308.832285783	2.28936661701739\\
308.903905606891	2.30944538281726\\
308.9784618523	2.32952414861713\\
309.056054821106	2.349602914417\\
309.136787984217	2.36968168021687\\
309.220767992387	2.38976044601674\\
309.308104687594	2.40983921181662\\
309.398911114816	2.42991797761649\\
309.493303534074	2.44999674341636\\
309.59140143261	2.47007550921623\\
309.693327537111	2.4901542750161\\
309.799207825877	2.51023304081597\\
309.909171540859	2.53031180661584\\
310.02335119951	2.55039057241571\\
310.14188260638	2.57046933821558\\
310.264904864424	2.59054810401546\\
310.392560385976	2.61062686981533\\
310.524994903365	2.6307056356152\\
310.662357479149	2.65078440141507\\
310.804800515943	2.67086316721494\\
310.952479765845	2.69094193301481\\
311.105554339438	2.71102069881468\\
311.264186714376	2.73109946461455\\
311.428542743548	2.75117823041443\\
311.598791662833	2.7712569962143\\
311.775106098446	2.79133576201417\\
311.957662073891	2.81141452781404\\
312.146639016538	2.83149329361391\\
312.342219763823	2.85157205941378\\
312.544590569109	2.87165082521365\\
312.753941107209	2.89172959101352\\
312.970464479606	2.91180835681339\\
313.194357219372	2.93188712261327\\
313.42581929583	2.95196588841314\\
313.665054118958	2.97204465421301\\
313.912268543579	2.99212342001288\\
314.167672873344	3.01220218581275\\
314.431480864541	3.03228095161262\\
314.703909729751	3.05235971741249\\
314.985180141367	3.07243848321236\\
315.275516235017	3.09251724901224\\
315.575145612889	3.11259601481211\\
315.88429934701	3.13267478061198\\
316.20321198247	3.15275354641185\\
316.532121540637	3.17283231221172\\
316.871269522378	3.19291107801159\\
317.220900911295	3.21298984381146\\
317.581264177012	3.23306860961133\\
317.952611278525	3.25314737541121\\
318.335197667626	3.27322614121108\\
318.729282292439	3.29330490701095\\
319.135127601062	3.31338367281082\\
319.552999545344	3.33346243861069\\
319.983167584814	3.35354120441056\\
320.425904690769	3.37361997021043\\
320.881487350538	3.3936987360103\\
321.350195571934	3.41377750181017\\
321.832312887911	3.43385626761005\\
322.328126361431	3.45393503340992\\
322.83792659055	3.47401379920979\\
323.362007713744	3.49409256500966\\
323.900667415466	3.51417133080953\\
324.454206931966	3.5342500966094\\
325.022931057345	3.55432886240927\\
325.6071481499	3.57440762820914\\
326.20717013871	3.59448639400902\\
326.823312530514	3.61456515980889\\
327.45589441686	3.63464392560876\\
328.105238481534	3.65472269140863\\
328.771671008274	3.6748014572085\\
329.455521888777	3.69488022300837\\
330.157124630984	3.71495898880824\\
330.876816367665	3.73503775460811\\
331.614937865291	3.75511652040799\\
332.371833533193	3.77519528620786\\
333.147851433019	3.79527405200773\\
333.943343288476	3.8153528178076\\
334.758664495366	3.83543158360747\\
335.594174131899	3.85551034940734\\
336.450234969308	3.87558911520721\\
337.327213482729	3.89566788100708\\
338.225479862376	3.91574664680695\\
339.145408024989	3.93582541260683\\
340.087375625551	3.9559041784067\\
341.051764069282	3.97598294420657\\
342.038958523899	3.99606171000644\\
343.049347932141	4.01614047580631\\
344.083325024548	4.03621924160618\\
345.141286332497	4.05629800740605\\
346.223632201493	4.07637677320592\\
347.330766804696	4.0964555390058\\
348.463098156696	4.11653430480567\\
349.621038127524	4.13661307060554\\
350.805002456885	4.15669183640541\\
352.015410768631	4.17677060220528\\
353.252686585439	4.19684936800515\\
354.517257343717	4.21692813380502\\
355.80955440871	4.23700689960489\\
357.130013089819	4.25708566540477\\
358.479072656114	4.27716443120464\\
359.857176352039	4.29724319700451\\
361.264771413316	4.31732196280438\\
362.702309083019	4.33740072860425\\
364.170244627835	4.35747949440412\\
365.669037354496	4.37755826020399\\
367.199150626375	4.39763702600386\\
368.761051880251	4.41771579180373\\
370.355212643221	4.43779455760361\\
371.982108549777	4.45787332340348\\
373.642219359017	4.47795208920335\\
375.33602897201	4.49803085500322\\
377.064025449289	4.51810962080309\\
378.826701028481	4.53818838660296\\
380.624552142062	4.55826715240283\\
382.458079435241	4.5783459182027\\
384.327787783952	4.59842468400258\\
386.234186312973	4.61850344980245\\
388.177788414141	4.63858221560232\\
390.159111764683	4.65866098140219\\
392.178678345644	4.67873974720206\\
394.237014460407	4.69881851300193\\
396.33465075332	4.7188972788018\\
398.472122228391	4.73897604460167\\
400.649968268094	4.75905481040154\\
402.868732652229	4.77913357620142\\
405.128963576882	4.79921234200129\\
407.43121367345	4.81929110780116\\
409.776040027734	4.83936987360103\\
412.164004199106	4.8594486394009\\
414.595672239737	4.87952740520077\\
417.071614713886	4.89960617100064\\
419.592406717247	4.91968493680051\\
422.158627896358	4.93976370260039\\
424.770862468047	4.95984246840026\\
427.429699238943	4.97992123420013\\
430.135731625027	5\\
}--cycle;
\addplot [color=red, line width=1pt, forget plot]
  table[row sep=crcr]{%
313.793680480592	1.00432560582563\\
310.309050195154	1.0244043716255\\
309.366954761083	1.04448313742537\\
308.873600836652	1.06456190322524\\
308.562325001658	1.08464066902512\\
308.346621417456	1.10471943482499\\
308.188428203112	1.12479820062486\\
308.067967117797	1.14487696642473\\
307.973797349906	1.1649557322246\\
307.898780541925	1.18503449802447\\
307.838203006093	1.20511326382434\\
307.788811062892	1.22519202962421\\
307.748276856096	1.24527079542409\\
307.714884523929	1.26534956122396\\
307.68733677262	1.28542832702383\\
307.664630826657	1.3055070928237\\
307.645976159492	1.32558585862357\\
307.630738343136	1.34566462442344\\
307.618399757392	1.36574339022331\\
307.608531488675	1.38582215602318\\
307.600772838464	1.40590092182306\\
307.594816119475	1.42597968762293\\
307.590395197362	1.4460584534228\\
307.587276731633	1.46613721922267\\
307.585253392229	1.48621598502254\\
307.584138542751	1.50629475082241\\
307.583762026557	1.52637351662228\\
307.583966792055	1.54645228242215\\
307.584606163551	1.56653104822202\\
307.585541613732	1.5866098140219\\
307.586640929606	1.60668857982177\\
307.587776689807	1.62676734562164\\
307.588824990302	1.64684611142151\\
307.589664369853	1.66692487722138\\
307.590174897282	1.68700364302125\\
307.590237390739	1.70708240882112\\
307.589732745382	1.72716117462099\\
307.588541350692	1.74723994042087\\
307.586542582352	1.76731870622074\\
307.583614356547	1.78739747202061\\
307.579632736824	1.80747623782048\\
307.574471585494	1.82755500362035\\
307.568002252967	1.84763376942022\\
307.560093299612	1.86771253522009\\
307.55061024564	1.88779130101996\\
307.539415345274	1.90787006681984\\
307.5263673821	1.92794883261971\\
307.511321482951	1.94802759841958\\
307.494128948145	1.96810636421945\\
307.474637096191	1.98818513001932\\
307.452689121373	2.00826389581919\\
307.428123962859	2.02834266161906\\
307.400776184159	2.04842142741893\\
307.370475861927	2.0685001932188\\
307.33704848324	2.08857895901868\\
307.300314850579	2.10865772481855\\
307.260090993849	2.12873649061842\\
307.216188088857	2.14881525641829\\
307.168412381712	2.16889402221816\\
307.116565118687	2.18897278801803\\
307.060442481132	2.2090515538179\\
306.999835525046	2.22913031961777\\
306.934530124982	2.24920908541765\\
306.86430692196	2.26928785121752\\
306.788941275108	2.28936661701739\\
306.708203216767	2.30944538281726\\
306.621857410806	2.32952414861713\\
306.529663113932	2.349602914417\\
306.431374139771	2.36968168021687\\
306.326738825521	2.38976044601674\\
306.215500000991	2.40983921181662\\
306.097394959845	2.42991797761649\\
305.972155432877	2.44999674341636\\
305.839507563163	2.47007550921623\\
305.699171882937	2.4901542750161\\
305.550863292033	2.51023304081597\\
305.394291037767	2.53031180661584\\
305.229158696121	2.55039057241571\\
305.055164154096	2.57046933821558\\
304.871999593124	2.59054810401546\\
304.679351473416	2.61062686981533\\
304.476900519129	2.6307056356152\\
304.264321704265	2.65078440141507\\
304.041284239177	2.67086316721494\\
303.807451557608	2.69094193301481\\
303.562481304165	2.71102069881468\\
303.306025322127	2.73109946461455\\
303.037729641525	2.75117823041443\\
302.757234467419	2.7712569962143\\
302.464174168264	2.79133576201417\\
302.158177264341	2.81141452781404\\
301.838866416166	2.83149329361391\\
301.505858412794	2.85157205941378\\
301.158764160029	2.87165082521365\\
300.797188668393	2.89172959101352\\
300.420731040894	2.91180835681339\\
300.02898446048	2.93188712261327\\
299.621536177183	2.95196588841314\\
299.197967494889	2.97204465421301\\
298.757853757694	2.99212342001288\\
298.300764335849	3.01220218581275\\
297.826262611229	3.03228095161262\\
297.333905962296	3.05235971741249\\
296.82324574858	3.07243848321236\\
296.2938272946	3.09251724901224\\
295.745189873246	3.11259601481211\\
295.176866688601	3.13267478061198\\
294.588384858153	3.15275354641185\\
293.97926539446	3.17283231221172\\
293.349023186165	3.19291107801159\\
292.697166978463	3.21298984381146\\
292.023199352893	3.23306860961133\\
291.326616706577	3.25314737541121\\
290.606909230804	3.27322614121108\\
289.863560888992	3.29330490701095\\
289.096049394103	3.31338367281082\\
288.303846185354	3.33346243861069\\
287.486416404398	3.35354120441056\\
286.643218870865	3.37361997021043\\
285.773706057314	3.3936987360103\\
284.877324063646	3.41377750181017\\
283.95351259085	3.43385626761005\\
283.001704914259	3.45393503340992\\
282.021327856231	3.47401379920979\\
281.011801758231	3.49409256500966\\
279.972540452483	3.51417133080953\\
278.902951232973	3.5342500966094\\
277.802434826039	3.55432886240927\\
276.670385360395	3.57440762820914\\
275.506190336713	3.59448639400902\\
274.309230596737	3.61456515980889\\
273.078880291856	3.63464392560876\\
271.814506851321	3.65472269140863\\
270.515470950009	3.6748014572085\\
269.181126475742	3.69488022300837\\
267.810820496192	3.71495898880824\\
266.40389322542	3.73503775460811\\
264.959677990063	3.75511652040799\\
263.477501195096	3.77519528620786\\
261.956682289205	3.79527405200773\\
260.396533730034	3.8153528178076\\
258.796360948815	3.83543158360747\\
257.155462314869	3.85551034940734\\
255.473129099788	3.87558911520721\\
253.748645441292	3.89566788100708\\
251.981288306826	3.91574664680695\\
250.170327456897	3.93582541260683\\
248.315025408094	3.9559041784067\\
246.414637396067	3.97598294420657\\
244.468411337995	3.99606171000644\\
242.475587795054	4.01614047580631\\
240.435399934652	4.03621924160618\\
238.347073492294	4.05629800740605\\
236.20982673354	4.07637677320592\\
234.022870415485	4.0964555390058\\
231.785407748369	4.11653430480567\\
229.49663435678	4.13661307060554\\
227.15573824088	4.15669183640541\\
224.761899737365	4.17677060220528\\
222.314291480398	4.19684936800515\\
219.812078362399	4.21692813380502\\
217.254417494584	4.23700689960489\\
214.640458167668	4.25708566540477\\
211.969341812168	4.27716443120464\\
209.240201958806	4.29724319700451\\
206.452164198781	4.31732196280438\\
203.604346143959	4.33740072860425\\
200.695857386908	4.35747949440412\\
197.725799461025	4.37755826020399\\
194.693265800479	4.39763702600386\\
191.597341700051	4.41771579180373\\
188.437104275171	4.43779455760361\\
185.211622421482	4.45787332340348\\
181.919956774953	4.47795208920335\\
178.561159671272	4.49803085500322\\
175.134275105754	4.51810962080309\\
171.638338693058	4.53818838660296\\
168.072377626492	4.55826715240283\\
164.435410638136	4.5783459182027\\
160.726447958129	4.59842468400258\\
156.944491274164	4.61850344980245\\
153.088533691466	4.63858221560232\\
149.157559691909	4.65866098140219\\
145.150545093925	4.67873974720206\\
141.066457011937	4.69881851300193\\
136.904253815697	4.7188972788018\\
132.66288509027	4.73897604460167\\
128.34129159533	4.75905481040154\\
123.93840522467	4.77913357620142\\
119.453148965862	4.79921234200129\\
114.884436860058	4.81929110780116\\
110.231173961082	4.83936987360103\\
105.49225629545	4.8594486394009\\
100.666570821941	4.87952740520077\\
95.7529953909418	4.89960617100064\\
90.7503987044713	4.91968493680051\\
85.6576402756422	4.93976370260039\\
80.4735703883136	4.95984246840026\\
75.197030056833	4.97992123420013\\
69.8268509859278	5\\
};

\addplot[area legend, dotted, line width=0.5pt, draw=black, fill=red, fill opacity=0.1, forget plot]
table[row sep=crcr] {%
x	y\\
317.425520359657	1.001\\
311.917014451707	1.011\\
310.674153661165	1.021\\
310.007567857547	1.031\\
309.573023776336	1.041\\
309.261413269661	1.051\\
309.024736982539	1.061\\
308.837884769422	1.071\\
308.686197454723	1.081\\
308.560437257317	1.091\\
308.454443823547	1.101\\
308.363925392889	1.111\\
308.285785874921	1.121\\
308.217727126144	1.131\\
308.15800223866	1.141\\
308.105256275137	1.151\\
308.058419951167	1.161\\
308.016636605667	1.171\\
307.979210784195	1.181\\
307.945571251571	1.191\\
307.915243875595	1.201\\
307.887831410549	1.211\\
307.862998196591	1.221\\
307.840458422163	1.231\\
307.819967008945	1.241\\
307.801312454125	1.251\\
307.784311152389	1.261\\
307.768802849318	1.271\\
307.754646969346	1.281\\
307.741719626283	1.291\\
307.729911171433	1.301\\
307.719124168674	1.311\\
307.709271711389	1.321\\
307.700276014946	1.331\\
307.692067233092	1.341\\
307.684582457044	1.351\\
307.677764865105	1.361\\
307.671562996674	1.371\\
307.665930129166	1.381\\
307.660823742247	1.391\\
307.656205053661	1.401\\
307.65203861661	1.411\\
307.648291969211	1.421\\
307.644935328594	1.431\\
307.641941319976	1.441\\
307.639284742773	1.451\\
307.63694394069	1.461\\
307.634892717367	1.471\\
307.633115971858	1.481\\
307.631593753839	1.491\\
307.629408856794	1.51\\
307.628475377902	1.52\\
307.627784932017	1.53\\
307.627285558432	1.54\\
307.626955884586	1.55\\
307.626784974668	1.56\\
307.626762450513	1.57\\
307.626878605672	1.58\\
307.627124364396	1.59\\
307.62749123706	1.6\\
307.627971268688	1.61\\
307.628557032875	1.62\\
307.629241545193	1.63\\
307.630018262059	1.64\\
307.630881047092	1.65\\
307.631824141679	1.66\\
307.6328421368	1.67\\
307.633929952233	1.68\\
307.635082814645	1.69\\
307.636296236063	1.7\\
307.637565996865	1.71\\
307.638888128505	1.72\\
307.640258896882	1.73\\
307.641674788239	1.74\\
307.643132495772	1.75\\
307.644628906107	1.76\\
307.646161088089	1.77\\
307.647726281724	1.78\\
307.64932188774	1.79\\
307.65094545805	1.8\\
307.652594686965	1.81\\
307.654267402715	1.82\\
307.655961559666	1.83\\
307.657675231032	1.84\\
307.659406602033	1.85\\
307.66115396349	1.86\\
307.662915705767	1.87\\
307.664690313228	1.88\\
307.666476358846	1.89\\
307.668272499258	1.9\\
307.67007747014	1.91\\
307.671890081737	1.92\\
307.673709214785	1.93\\
307.675533816603	1.94\\
307.677362897425	1.95\\
307.679195526934	1.96\\
307.681030831054	1.97\\
307.682867988824	1.98\\
307.684706229561	1.99\\
307.686544830092	2\\
307.688383112195	2.01\\
307.690220440161	2.02\\
307.692056218497	2.03\\
307.693889889762	2.04\\
307.695720932508	2.05\\
307.697548859347	2.06\\
307.699373215111	2.07\\
307.701193575128	2.08\\
307.70300954357	2.09\\
307.704820751909	2.1\\
307.706626857444	2.11\\
307.708427541915	2.12\\
307.71022251018	2.13\\
307.712011488972	2.14\\
307.713794225717	2.15\\
307.715570487414	2.16\\
307.717340059576	2.17\\
307.719102745227	2.18\\
307.720858363943	2.19\\
307.722606750957	2.2\\
307.724347756302	2.21\\
307.726081243993	2.22\\
307.727807091266	2.23\\
307.729525187845	2.24\\
307.731235435247	2.25\\
307.732937746133	2.26\\
307.734632043677	2.27\\
307.736318260981	2.28\\
307.737996340514	2.29\\
307.739666233578	2.3\\
307.741327899806	2.31\\
307.742981306683	2.32\\
307.74462642909	2.33\\
307.746263248875	2.34\\
307.747891754446	2.35\\
307.749511940377	2.36\\
307.751123807047	2.37\\
307.752727360286	2.38\\
307.754322611044	2.39\\
307.755909575079	2.4\\
307.757488272655	2.41\\
307.759058728262	2.42\\
307.760620970347	2.43\\
307.762175031056	2.44\\
307.763720945995	2.45\\
307.765258754001	2.46\\
307.766788496922	2.47\\
307.768310219415	2.48\\
307.769823968745	2.49\\
307.771329794601	2.5\\
307.772827748926	2.51\\
307.774317885739	2.52\\
307.77580026099	2.53\\
307.777274932397	2.54\\
307.778741959316	2.55\\
307.780201402598	2.56\\
307.781653324466	2.57\\
307.783097788394	2.58\\
307.78453485899	2.59\\
307.785964601892	2.6\\
307.787387083662	2.61\\
307.788802371692	2.62\\
307.790210534111	2.63\\
307.791611639699	2.64\\
307.793005757805	2.65\\
307.794392958268	2.66\\
307.795773311349	2.67\\
307.797146887657	2.68\\
307.798513758084	2.69\\
307.799873993749	2.7\\
307.801227665933	2.71\\
307.802574846029	2.72\\
307.803915605491	2.73\\
307.805250015781	2.74\\
307.80657814833	2.75\\
307.807900074488	2.76\\
307.809215865491	2.77\\
307.810525592418	2.78\\
307.811829326157	2.79\\
307.813127137372	2.8\\
307.814419096473	2.81\\
307.815705273585	2.82\\
307.816985738521	2.83\\
307.818260560758	2.84\\
307.819529809412	2.85\\
307.820793553218	2.86\\
307.822051860504	2.87\\
307.823304799179	2.88\\
307.824552436712	2.89\\
307.825794840114	2.9\\
307.827032075925	2.91\\
307.828264210201	2.92\\
307.829491308498	2.93\\
307.830713435864	2.94\\
307.831930656826	2.95\\
307.833143035379	2.96\\
307.834350634982	2.97\\
307.835553518544	2.98\\
307.836751748422	2.99\\
307.837945386411	3\\
307.83913449374	3.01\\
307.840319131068	3.02\\
307.841499358478	3.03\\
307.842675235472	3.04\\
307.843846820972	3.05\\
307.845014173315	3.06\\
307.84617735025	3.07\\
307.847336408939	3.08\\
307.848491405954	3.09\\
307.849642397276	3.1\\
307.850789438298	3.11\\
307.851932583821	3.12\\
307.853071888058	3.13\\
307.854207404632	3.14\\
307.85533918658	3.15\\
307.856467286352	3.16\\
307.857591755814	3.17\\
307.858712646251	3.18\\
307.859830008368	3.19\\
307.860943892292	3.2\\
307.862054347577	3.21\\
307.863161423204	3.22\\
307.864265167587	3.23\\
307.865365628575	3.24\\
307.866462853456	3.25\\
307.86755688896	3.26\\
307.868647781261	3.27\\
307.869735575987	3.28\\
307.870820318218	3.29\\
307.871902052491	3.3\\
307.872980822809	3.31\\
307.874056672638	3.32\\
307.875129644918	3.33\\
307.876199782063	3.34\\
307.87726712597	3.35\\
307.878331718019	3.36\\
307.879393599081	3.37\\
307.88045280952	3.38\\
307.8815093892	3.39\\
307.882563377491	3.4\\
307.88361481327	3.41\\
307.884663734929	3.42\\
307.885710180378	3.43\\
307.886754187052	3.44\\
307.887795791914	3.45\\
307.888835031461	3.46\\
307.889871941729	3.47\\
307.890906558298	3.48\\
307.891938916295	3.49\\
307.892969050403	3.5\\
320	4\\
}--cycle;
%
\addplot[color=\swellcolor, mark=none, -latex, line width=1.5pt]
  table[row sep=crcr]{%
304 1.16\\
304	1.95\\
};
\addplot[color=\swellcolor, mark=none, line width=1.5pt]
  table[row sep=crcr]{%
304 1.80\\
304	2.67\\
};
%
\addplot[color=\shrinkcolor, mark=none, -latex, line width=1.5pt]
  table[row sep=crcr]{%
308 2.67\\
308	1.75\\
};
\addplot[color=\shrinkcolor, mark=none, line width=1.5pt]
  table[row sep=crcr]{%
308 1.80\\
308	1.16\\
};
\addplot[color=\shrinkcolor, mark=none, -latex, line width=1.5pt]
  table[row sep=crcr]{%
307.6 2.67\\
307.6 2.0\\
};
\addplot[color=\shrinkcolor, mark=none, line width=1.5pt]
  table[row sep=crcr]{%
307.6 2.05\\
307.6 1.41\\
};
%
\addplot[color=\shrinkcolor, mark=none, -latex, dotted, line width=1.5pt]
  table[row sep=crcr]{%
304 2.67\\
306	2.67\\
};
\addplot[color=\shrinkcolor, mark=none, dotted, line width=1.5pt]
  table[row sep=crcr]{%
306 2.67\\
308	2.67\\
};
\addplot[color=\swellcolor, mark=none, -latex, dotted, line width=1.5pt]
  table[row sep=crcr]{%
308 1.16\\
306	1.16\\
};
\addplot[color=\swellcolor, mark=none, dotted, line width=1.5pt]
  table[row sep=crcr]{%
306 1.16\\
304	1.16\\
};
%
\addplot[color=black, mark=x, only marks, mark size=3.0pt, line width=1.5pt]
    table[row sep=crcr]{%
304 1.16\\
308 2.67\\
307.6 2.67\\
};
%
\end{axis}
%
%
\begin{axis}[xshift=.12\textwidth,yshift=.1\textwidth,
     axis on top,  
    width=0.22\textwidth,
    height = 0.2\textwidth,
    enlargelimits=false, 
    xmin=307.5,xmax=307.7,ymin=1,ymax=2.2, 
    xtick={307.5,307.7}, 
    ytick={1,2},
    label style={font=\tiny}, 
    tick label style={font=\tiny},
  ]
 \addplot[area legend, dashed, line width=0.5pt, draw=black, fill=red, fill opacity=0.2, forget plot]
table[row sep=crcr] {%
x	y\\
308.016054038673	1.24527079542409\\
307.942162227865	1.26534956122396\\
307.880968167373	1.28542832702383\\
307.830049857315	1.3055070928237\\
307.787586748414	1.32558585862357\\
307.752185265861	1.34566462442344\\
307.722761817114	1.36574339022331\\
307.698462343156	1.38582215602318\\
307.678605759673	1.40590092182306\\
307.662643406024	1.42597968762293\\
307.650129458302	1.4460584534228\\
307.640699000827	1.46613721922267\\
307.634051542455	1.48621598502254\\
307.629938466413	1.50629475082241\\
307.628153363499	1.52637351662228\\
307.628524507214	1.54645228242215\\
307.630908939561	1.56653104822202\\
307.635187781715	1.5866098140219\\
307.641262485917	1.60668857982177\\
307.649051817581	1.62676734562164\\
307.658489409046	1.64684611142151\\
307.669521764516	1.66692487722138\\
307.68210662394	1.68700364302125\\
307.696211614482	1.70708240882112\\
307.711813134	1.72716117462099\\
307.728895422888	1.74723994042087\\
307.74744978974	1.76731870622074\\
307.767473963357	1.78739747202061\\
307.788971549035	1.80747623782048\\
307.811951571389	1.82755500362035\\
307.836428089267	1.84763376942022\\
307.862419871016	1.86771253522009\\
307.889950120452	1.88779130101996\\
307.919046245583	1.90787006681984\\
307.949739663513	1.92794883261971\\
307.982065636051	1.94802759841958\\
308.01606313145	1.96810636421945\\
}--cycle;
\addplot [color=red, line width=1pt, forget plot]
  table[row sep=crcr]{%
308.067967117797	1.14487696642473\\
307.973797349906	1.1649557322246\\
307.898780541925	1.18503449802447\\
307.838203006093	1.20511326382434\\
307.788811062892	1.22519202962421\\
307.748276856096	1.24527079542409\\
307.714884523929	1.26534956122396\\
307.68733677262	1.28542832702383\\
307.664630826657	1.3055070928237\\
307.645976159492	1.32558585862357\\
307.630738343136	1.34566462442344\\
307.618399757392	1.36574339022331\\
307.608531488675	1.38582215602318\\
307.600772838464	1.40590092182306\\
307.594816119475	1.42597968762293\\
307.590395197362	1.4460584534228\\
307.587276731633	1.46613721922267\\
307.585253392229	1.48621598502254\\
307.584138542751	1.50629475082241\\
307.583762026557	1.52637351662228\\
307.583966792055	1.54645228242215\\
307.584606163551	1.56653104822202\\
307.585541613732	1.5866098140219\\
307.586640929606	1.60668857982177\\
307.587776689807	1.62676734562164\\
307.588824990302	1.64684611142151\\
307.589664369853	1.66692487722138\\
307.590174897282	1.68700364302125\\
307.590237390739	1.70708240882112\\
307.589732745382	1.72716117462099\\
307.588541350692	1.74723994042087\\
307.586542582352	1.76731870622074\\
307.583614356547	1.78739747202061\\
307.579632736824	1.80747623782048\\
307.574471585494	1.82755500362035\\
307.568002252967	1.84763376942022\\
307.560093299612	1.86771253522009\\
307.55061024564	1.88779130101996\\
307.539415345274	1.90787006681984\\
307.5263673821	1.92794883261971\\
307.511321482951	1.94802759841958\\
307.494128948145	1.96810636421945\\
307.474637096191	1.98818513001932\\
307.452689121373	2.00826389581919\\
307.428123962859	2.02834266161906\\
307.400776184159	2.04842142741893\\
307.370475861927	2.0685001932188\\
307.33704848324	2.08857895901868\\
307.300314850579	2.10865772481855\\
307.260090993849	2.12873649061842\\
307.216188088857	2.14881525641829\\
307.168412381712	2.16889402221816\\
307.116565118687	2.18897278801803\\
307.060442481132	2.2090515538179\\
306.999835525046	2.22913031961777\\
};

\addplot[area legend, dotted, line width=0.5pt, draw=black, fill=red, fill opacity=0.1, forget plot]
table[row sep=crcr] {%
x	y\\
308.016636605667	1.171\\
307.979210784195	1.181\\
307.945571251571	1.191\\
307.915243875595	1.201\\
307.887831410549	1.211\\
307.862998196591	1.221\\
307.840458422163	1.231\\
307.819967008945	1.241\\
307.801312454125	1.251\\
307.784311152389	1.261\\
307.768802849318	1.271\\
307.754646969346	1.281\\
307.741719626283	1.291\\
307.729911171433	1.301\\
307.719124168674	1.311\\
307.709271711389	1.321\\
307.700276014946	1.331\\
307.692067233092	1.341\\
307.684582457044	1.351\\
307.677764865105	1.361\\
307.671562996674	1.371\\
307.665930129166	1.381\\
307.660823742247	1.391\\
307.656205053661	1.401\\
307.65203861661	1.411\\
307.648291969211	1.421\\
307.644935328594	1.431\\
307.641941319976	1.441\\
307.639284742773	1.451\\
307.63694394069	1.461\\
307.634892717367	1.471\\
307.633115971858	1.481\\
307.631593753839	1.491\\
307.629408856794	1.51\\
307.628475377902	1.52\\
307.627784932017	1.53\\
307.627285558432	1.54\\
307.626955884586	1.55\\
307.626784974668	1.56\\
307.626762450513	1.57\\
307.626878605672	1.58\\
307.627124364396	1.59\\
307.62749123706	1.6\\
307.627971268688	1.61\\
307.628557032875	1.62\\
307.629241545193	1.63\\
307.630018262059	1.64\\
307.630881047092	1.65\\
307.631824141679	1.66\\
307.6328421368	1.67\\
307.633929952233	1.68\\
307.635082814645	1.69\\
307.636296236063	1.7\\
307.637565996865	1.71\\
307.638888128505	1.72\\
307.640258896882	1.73\\
307.641674788239	1.74\\
307.643132495772	1.75\\
307.644628906107	1.76\\
307.646161088089	1.77\\
307.647726281724	1.78\\
307.64932188774	1.79\\
307.65094545805	1.8\\
307.652594686965	1.81\\
307.654267402715	1.82\\
307.655961559666	1.83\\
307.657675231032	1.84\\
307.659406602033	1.85\\
307.66115396349	1.86\\
307.662915705767	1.87\\
307.664690313228	1.88\\
307.666476358846	1.89\\
307.668272499258	1.9\\
307.67007747014	1.91\\
307.671890081737	1.92\\
307.673709214785	1.93\\
307.675533816603	1.94\\
307.677362897425	1.95\\
307.679195526934	1.96\\
307.681030831054	1.97\\
307.682867988824	1.98\\
307.684706229561	1.99\\
307.686544830092	2\\
307.688383112195	2.01\\
307.690220440161	2.02\\
307.692056218497	2.03\\
307.693889889762	2.04\\
307.695720932508	2.05\\
307.697548859347	2.06\\
307.699373215111	2.07\\
307.701193575128	2.08\\
307.70300954357	2.09\\
307.704820751909	2.1\\
307.706626857444	2.11\\
307.708427541915	2.12\\
307.71022251018	2.13\\
307.712011488972	2.14\\
307.713794225717	2.15\\
307.715570487414	2.16\\
307.717340059576	2.17\\
307.719102745227	2.18\\
307.720858363943	2.19\\
307.722606750957	2.2\\
307.724347756302	2.21\\
307.726081243993	2.22\\
307.727807091266	2.23\\
307.729525187845	2.24\\
307.731235435247	2.25\\
307.732937746133	2.26\\
307.734632043677	2.27\\
307.736318260981	2.28\\
307.737996340514	2.29\\
307.739666233578	2.3\\
307.741327899806	2.31\\
307.742981306683	2.32\\
307.74462642909	2.33\\
307.746263248875	2.34\\
307.747891754446	2.35\\
307.749511940377	2.36\\
307.751123807047	2.37\\
307.752727360286	2.38\\
307.754322611044	2.39\\
307.755909575079	2.4\\
307.757488272655	2.41\\
307.759058728262	2.42\\
307.760620970347	2.43\\
307.762175031056	2.44\\
307.763720945995	2.45\\
307.765258754001	2.46\\
307.766788496922	2.47\\
307.768310219415	2.48\\
307.769823968745	2.49\\
307.771329794601	2.5\\
307.772827748926	2.51\\
307.774317885739	2.52\\
307.77580026099	2.53\\
307.777274932397	2.54\\
307.778741959316	2.55\\
307.780201402598	2.56\\
307.781653324466	2.57\\
307.783097788394	2.58\\
307.78453485899	2.59\\
307.785964601892	2.6\\
307.787387083662	2.61\\
307.788802371692	2.62\\
307.790210534111	2.63\\
307.791611639699	2.64\\
307.793005757805	2.65\\
307.794392958268	2.66\\
307.795773311349	2.67\\
307.797146887657	2.68\\
307.798513758084	2.69\\
307.799873993749	2.7\\
307.801227665933	2.71\\
307.802574846029	2.72\\
307.803915605491	2.73\\
307.805250015781	2.74\\
307.80657814833	2.75\\
307.807900074488	2.76\\
307.809215865491	2.77\\
307.810525592418	2.78\\
307.811829326157	2.79\\
307.813127137372	2.8\\
307.814419096473	2.81\\
307.815705273585	2.82\\
307.816985738521	2.83\\
307.818260560758	2.84\\
307.819529809412	2.85\\
307.820793553218	2.86\\
307.822051860504	2.87\\
307.823304799179	2.88\\
307.824552436712	2.89\\
307.825794840114	2.9\\
307.827032075925	2.91\\
307.828264210201	2.92\\
307.829491308498	2.93\\
307.830713435864	2.94\\
307.831930656826	2.95\\
307.833143035379	2.96\\
307.834350634982	2.97\\
307.835553518544	2.98\\
307.836751748422	2.99\\
307.837945386411	3\\
307.83913449374	3.01\\
307.840319131068	3.02\\
307.841499358478	3.03\\
307.842675235472	3.04\\
307.843846820972	3.05\\
307.845014173315	3.06\\
307.84617735025	3.07\\
307.847336408939	3.08\\
307.848491405954	3.09\\
307.849642397276	3.1\\
307.850789438298	3.11\\
307.851932583821	3.12\\
307.853071888058	3.13\\
307.854207404632	3.14\\
307.85533918658	3.15\\
307.856467286352	3.16\\
307.857591755814	3.17\\
307.858712646251	3.18\\
307.859830008368	3.19\\
307.860943892292	3.2\\
307.862054347577	3.21\\
307.863161423204	3.22\\
307.864265167587	3.23\\
307.865365628575	3.24\\
307.866462853456	3.25\\
307.86755688896	3.26\\
307.868647781261	3.27\\
307.869735575987	3.28\\
307.870820318218	3.29\\
307.871902052491	3.3\\
307.872980822809	3.31\\
307.874056672638	3.32\\
307.875129644918	3.33\\
307.876199782063	3.34\\
307.87726712597	3.35\\
307.878331718019	3.36\\
307.879393599081	3.37\\
307.88045280952	3.38\\
307.8815093892	3.39\\
307.882563377491	3.4\\
307.88361481327	3.41\\
307.884663734929	3.42\\
307.885710180378	3.43\\
307.886754187052	3.44\\
307.887795791914	3.45\\
307.888835031461	3.46\\
307.889871941729	3.47\\
307.890906558298	3.48\\
307.891938916295	3.49\\
307.892969050403	3.5\\
320	4\\
}--cycle;
%
\addplot[color=\shrinkcolor, mark=none, line width=1pt]
  table[row sep=crcr]{%
307.6 2.67\\
307.6 1.41\\
};
\addplot[color=\shrinkcolor, mark=none, -latex, line width = 1pt]
  table[row sep=crcr]{%
307.6 2.67\\
307.6 1.75\\
};
%
\end{axis}

\node[rotate=90,color=black] at (0.8,2) {Swelling};
\node[rotate=-90,color=black] at (4.1, 2) {Shrinking};

\node[color=black] at (1,3.15) {(i)};
\node[color=black] at (3.95,3.15) {(ii)};
\node[color=black] at (3.1,3.15) {(iii)};

\end{tikzpicture}

%% file: Figs/Fig5a_ColourMap_CaiSuo_Swell_NoFront.tex
\begin{tikzpicture}
\begin{axis}[axis on top,  
thick, 
/pgf/number format/.cd,
                fixed,
width=0.48\textwidth, 
height=0.4\textwidth,
enlargelimits=false, 
colorbar, 
point meta min = 0.25, 
point meta max = 1,
xmin=0,xmax=1,ymin=0,ymax=2.5, 
colorbar style = {
width = 0.3cm,
thick,
black,
title = {$\phi$},
title style = {overlay,yshift = -3pt},
at={(1.05,1)}},
xlabel = {$t$},
ylabel = {$r$},
legend style = {
draw=none,
fill=none,
font=\scriptsize},]
\addplot[forget plot] graphics [xmin=0,xmax=1,ymin=0,ymax=2.5] {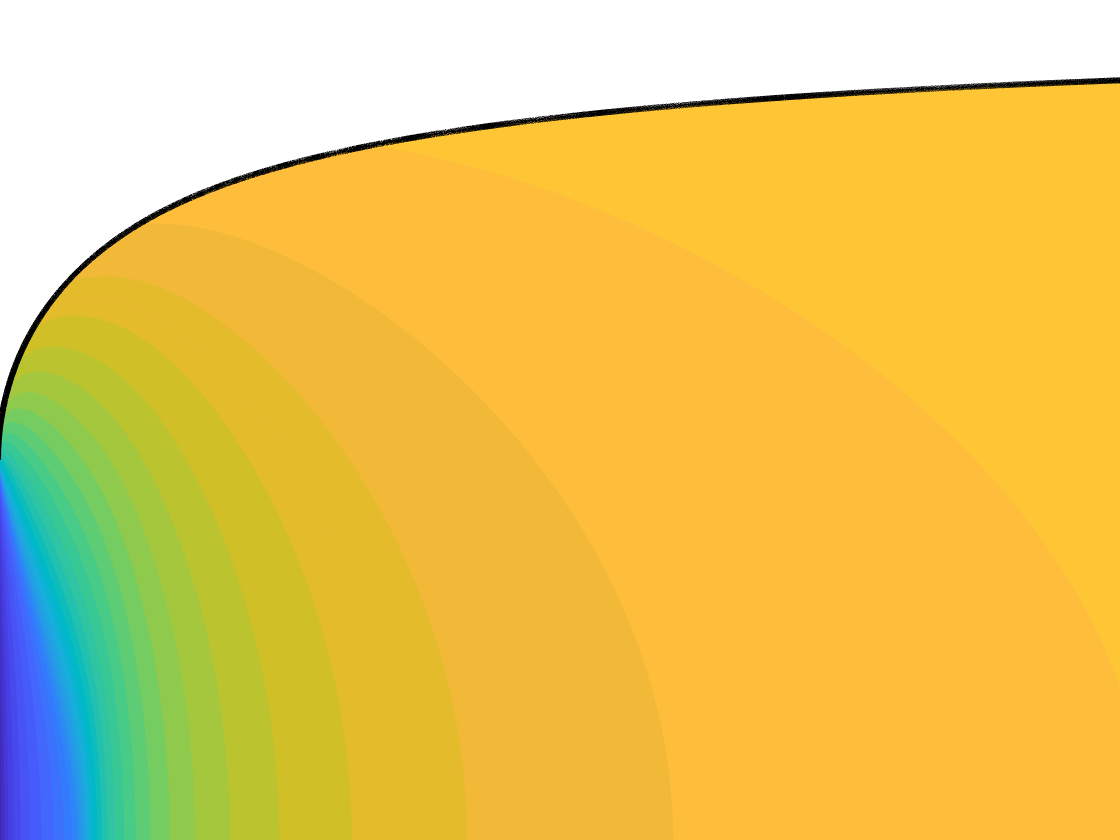};
%
%
\end{axis}
%
%
\end{tikzpicture}

%% file: Figs/Fig5b_Porosity_CaiSuo_Swell_NoFront.tex
%
%
\definecolor{mycolor1}{rgb}{0.12500,0.07812,0.04975}%
\definecolor{mycolor2}{rgb}{0.25000,0.15624,0.09950}%
\definecolor{mycolor3}{rgb}{0.37500,0.23436,0.14925}%
\definecolor{mycolor4}{rgb}{0.50000,0.31248,0.19900}%
\definecolor{mycolor5}{rgb}{0.62500,0.39060,0.24875}%
\definecolor{mycolor6}{rgb}{0.75000,0.46872,0.29850}%
\definecolor{mycolor7}{rgb}{0.87500,0.54684,0.34825}%
\definecolor{mycolor8}{rgb}{1.00000,0.62496,0.39800}%
\definecolor{mycolor9}{rgb}{1.00000,0.70308,0.44775}%
\definecolor{mycolor10}{rgb}{1.00000,0.78120,0.49750}%
\begin{tikzpicture}

\begin{axis}[%
thick, 
width=0.48\textwidth, 
height=0.4\textwidth,
xmin=0,xmax=2.5,ymin=0,ymax=1, 
xlabel = {$r$},
ylabel = {$\phi$},
yticklabel style = overlay]
\addplot [color=black, forget plot]
  table[row sep=crcr]{%
0.00138904704000746	0.271248974374373\\
0.0152795174400821	0.271248974374373\\
0.0291699878401568	0.271248974374373\\
0.0430604582402314	0.271248974374373\\
0.056950928640306	0.271248974374373\\
0.0708413990403807	0.271248974374373\\
0.0847318694404553	0.271248974374373\\
0.09862233984053	0.271248974374373\\
0.112512810240605	0.271248974374373\\
0.126403280640679	0.271248974374373\\
0.140293751040754	0.271248974374373\\
0.154184221440829	0.271248974374373\\
0.168074691840903	0.271248974374373\\
0.181965162240978	0.271248974374373\\
0.195855632641052	0.271248974374373\\
0.209746103041127	0.271248974374373\\
0.223636573441202	0.271248974374373\\
0.237527043841276	0.271248974374373\\
0.251417514241351	0.271248974374373\\
0.265307984641426	0.271248974374373\\
0.2791984550415	0.271248974374373\\
0.293088925441575	0.271248974374373\\
0.30697939584165	0.271248974374373\\
0.320869866241724	0.271248974374373\\
0.334760336641799	0.271248974374373\\
0.348650807041874	0.271248974374373\\
0.362541277441948	0.271248974374373\\
0.376431747842023	0.271248974374373\\
0.390322218242098	0.271248974374373\\
0.404212688642172	0.271248974374373\\
0.418103159042247	0.271248974374373\\
0.431993629442321	0.271248974374373\\
0.445884099842396	0.271248974374373\\
0.459774570242471	0.271248974374373\\
0.473665040642545	0.271248974374373\\
0.48755551104262	0.271248974374373\\
0.501445981442695	0.271248974374373\\
0.515336451842769	0.271248974374373\\
0.529226922242844	0.271248974374373\\
0.543117392642919	0.271248974374373\\
0.557007863042993	0.271248974374373\\
0.570898333443068	0.271248974374373\\
0.584788803843143	0.271248974374373\\
0.598679274243217	0.271248974374373\\
0.612569744643292	0.271248974374373\\
0.626460215043366	0.271248974374373\\
0.640350685443441	0.271248974374373\\
0.654241155843516	0.271248974374373\\
0.66813162624359	0.271248974374373\\
0.682022096643665	0.271248974374373\\
0.69591256704374	0.271248974374373\\
0.709803037443814	0.271248974374373\\
0.723693507843889	0.271248974374373\\
0.737583978243964	0.271248974374373\\
0.751474448644038	0.271248974374373\\
0.765364919044113	0.271248974374373\\
0.779255389444188	0.271248974374373\\
0.793145859844262	0.271248974374373\\
0.807036330244337	0.271248974374373\\
0.820926800644411	0.271248974374373\\
0.834817271044486	0.271248974374373\\
0.848707741444561	0.271248974374373\\
0.862598211844635	0.271248974374373\\
0.87648868224471	0.271248974374373\\
0.890379152644785	0.271248974374373\\
0.904269623044859	0.271248974374373\\
0.918160093444934	0.271248974374373\\
0.932050563845009	0.271248974374373\\
0.945941034245083	0.271248974374373\\
0.959831504645158	0.271248974374373\\
0.973721975045233	0.271248974374373\\
0.987612445445307	0.271248974374373\\
1.00150291584538	0.271248974374373\\
1.01539338624546	0.271248974374373\\
1.02928385664553	0.271248974374373\\
1.04317432704561	0.271248974374373\\
1.05706479744568	0.271248974374373\\
1.07095526784576	0.271248974374373\\
1.08484573824583	0.271248974374373\\
1.0987362086459	0.271248974374373\\
};
\addplot [color=mycolor1, forget plot]
  table[row sep=crcr]{%
0.00177774426667866	0.360428877378754\\
0.0195551869334653	0.360450577759885\\
0.0373326296002519	0.360508573850327\\
0.0551100722670385	0.360602994484901\\
0.0728875149338251	0.360734071883809\\
0.0906649576006117	0.360902133881758\\
0.108442400267398	0.361107605627739\\
0.126219842934185	0.361351013379336\\
0.143997285600971	0.361632989588008\\
0.161774728267758	0.361954279231797\\
0.179552170934545	0.362315747495595\\
0.197329613601331	0.362718388961108\\
0.215107056268118	0.363163338518797\\
0.232884498934904	0.363651884268085\\
0.250661941601691	0.364185482736118\\
0.268439384268478	0.364765776823962\\
0.286216826935264	0.365394616987251\\
0.303994269602051	0.366074086282296\\
0.321771712268837	0.366806530066669\\
0.339549154935624	0.367594591346602\\
0.357326597602411	0.368441253027359\\
0.375104040269197	0.369349888668254\\
0.392881482935984	0.370324323800846\\
0.41065892560277	0.371368910478855\\
0.428436368269557	0.372488618551685\\
0.446213810936344	0.37368914827733\\
0.46399125360313	0.374977070442979\\
0.481768696269917	0.376360002334699\\
0.499546138936703	0.377846830981371\\
0.51732358160349	0.379447999540756\\
0.535101024270277	0.381175879200383\\
0.552878466937063	0.383045258658132\\
0.57065590960385	0.385073997964767\\
0.588433352270636	0.387283916319589\\
0.606210794937423	0.389702019545498\\
0.62398823760421	0.392362231576087\\
0.641765680270996	0.395307891714392\\
0.659543122937783	0.398595445507032\\
0.677320565604569	0.402300047018761\\
0.695098008271356	0.406524305780155\\
0.712875450938143	0.411412330155485\\
0.730652893604929	0.417172783791953\\
0.748430336271716	0.424116811027329\\
0.766207778938503	0.432716362188675\\
0.783985221605289	0.443665631563678\\
0.801762664272076	0.45779815838997\\
0.819540106938862	0.475402009085155\\
0.837317549605649	0.495094664910998\\
0.855094992272436	0.514608815134987\\
0.872872434939222	0.5327171693061\\
0.890649877606009	0.549249094557851\\
0.908427320272795	0.564403778160839\\
0.926204762939582	0.57842726223659\\
0.943982205606369	0.591529577618074\\
0.961759648273155	0.603875883932996\\
0.979537090939942	0.615594024341042\\
0.997314533606728	0.626783478912365\\
1.01509197627351	0.637522605111073\\
1.0328694189403	0.647873986370898\\
1.05064686160709	0.657888290067655\\
1.06842430427387	0.667607047552257\\
1.08620174694066	0.67706467298376\\
1.10397918960745	0.686289945283021\\
1.12175663227423	0.69530710822634\\
1.13953407494102	0.704136695429754\\
1.15731151760781	0.712796154137592\\
1.17508896027459	0.721300319477333\\
1.19286640294138	0.729661775695258\\
1.21064384560817	0.737891130494412\\
1.22842128827495	0.74599722140945\\
1.24619873094174	0.753987268158942\\
1.26397617360853	0.761866981454593\\
1.28175361627531	0.769640636389114\\
1.2995310589421	0.777311116993928\\
1.31730850160889	0.784879937678473\\
1.33508594427567	0.792347246924419\\
1.35286338694246	0.799711818738695\\
1.37064082960925	0.806971037913606\\
1.38841827227603	0.814120886037842\\
1.40619571494282	0.821155936356188\\
};
\addplot [color=mycolor2, forget plot]
  table[row sep=crcr]{%
0.00192577036689919	0.403689832694128\\
0.0211834740358911	0.403731041290236\\
0.0404411777048831	0.403841532451902\\
0.059698881373875	0.404021968796616\\
0.0789565850428669	0.404273540608429\\
0.0982142887118589	0.40459795010103\\
0.117471992380851	0.404997446109733\\
0.136729696049843	0.405474878575203\\
0.155987399718835	0.40603377172912\\
0.175245103387827	0.406678419252972\\
0.194502807056819	0.407414006638907\\
0.213760510725811	0.408246768067547\\
0.233018214394802	0.409184187866978\\
0.252275918063794	0.410235260432471\\
0.271533621732786	0.411410827894407\\
0.290791325401778	0.412724022598192\\
0.31004902907077	0.414190852776578\\
0.329306732739762	0.415830986428643\\
0.348564436408754	0.417668813041928\\
0.367822140077746	0.419734899304477\\
0.387079843746738	0.422068008633633\\
0.40633754741573	0.42471793095815\\
0.425595251084722	0.42774947005043\\
0.444852954753714	0.431248039394432\\
0.464110658422706	0.435327319879129\\
0.483368362091698	0.440138965247142\\
0.502626065760689	0.445882286886235\\
0.521883769429681	0.452805399588393\\
0.541141473098673	0.461174182386908\\
0.560399176767665	0.471167839406483\\
0.579656880436657	0.482696008502387\\
0.598914584105649	0.495291483003501\\
0.618172287774641	0.50828454032851\\
0.637429991443633	0.521115241992846\\
0.656687695112625	0.53347008949794\\
0.675945398781617	0.545232857584571\\
0.695203102450609	0.556395250611703\\
0.714460806119601	0.566995666501623\\
0.733718509788593	0.577087792089386\\
0.752976213457585	0.586726685796931\\
0.772233917126577	0.595963351169375\\
0.791491620795569	0.60484308770882\\
0.81074932446456	0.613405412705685\\
0.830007028133553	0.621684571323801\\
0.849264731802544	0.629710212370879\\
0.868522435471536	0.637508057170129\\
0.887780139140528	0.645100498665557\\
0.90703784280952	0.652507114836318\\
0.926295546478512	0.659745099489298\\
0.945553250147504	0.666829620183985\\
0.964810953816496	0.6737741144579\\
0.984068657485488	0.680590534826053\\
1.00332636115448	0.687289551624434\\
1.02258406482347	0.69388072125532\\
1.04184176849246	0.70037262600629\\
1.06109947216146	0.706772990429444\\
1.08035717583045	0.713088778287092\\
1.09961487949944	0.719326273273545\\
1.11887258316843	0.725491146080875\\
1.13813028683742	0.73158850986108\\
1.15738799050642	0.737622965723221\\
1.17664569417541	0.743598639571636\\
1.1959033978444	0.749519211323623\\
1.21516110151339	0.755387937329624\\
1.23441880518238	0.761207666646305\\
1.25367650885137	0.766980851676098\\
1.27293421252037	0.772709553581097\\
1.29219191618936	0.77839544280185\\
1.31144961985835	0.784039794961786\\
1.33070732352734	0.789643482416104\\
1.34996502719633	0.795206961712387\\
1.36922273086533	0.800730257272182\\
1.38848043453432	0.806212941683292\\
1.40773813820331	0.811654113116711\\
1.4269958418723	0.817052370556112\\
1.44625354554129	0.822405787756436\\
1.46551124921029	0.827711887134223\\
1.48476895287928	0.832967615135153\\
1.50402665654827	0.838169321015358\\
1.52328436021726	0.843312741394747\\
};
\addplot [color=mycolor3, forget plot]
  table[row sep=crcr]{%
0.00203065110902722	0.444905097554615\\
0.0223371621992995	0.445045837242017\\
0.0426436732895717	0.445426840031525\\
0.0629501843798439	0.446056922940147\\
0.0832566954701162	0.44695169505628\\
0.103563206560388	0.44813417145273\\
0.123869717650661	0.449636065412679\\
0.144176228740933	0.451499557100849\\
0.164482739831205	0.453779454873008\\
0.184789250921477	0.456545465899865\\
0.20509576201175	0.459883780051609\\
0.225402273102022	0.463896140741121\\
0.245708784192294	0.468692919823247\\
0.266015295282566	0.474375161073826\\
0.286321806372838	0.48100213822509\\
0.306628317463111	0.48855083487544\\
0.326934828553383	0.496890588040163\\
0.347241339643655	0.505799724588469\\
0.367547850733927	0.515022209208769\\
0.3878543618242	0.52432841360008\\
0.408160872914472	0.533548134850649\\
0.428467384004744	0.542573393485746\\
0.448773895095016	0.551345538495042\\
0.469080406185289	0.559839503847279\\
0.489386917275561	0.568050966200917\\
0.509693428365833	0.575987464319515\\
0.529999939456105	0.583662764975536\\
0.550306450546378	0.591093469427123\\
0.57061296163665	0.598297047169299\\
0.590919472726922	0.60529074132453\\
0.611225983817194	0.612090993198426\\
0.631532494907466	0.618713170367982\\
0.651839005997739	0.625171468813375\\
0.672145517088011	0.631478912138068\\
0.692452028178283	0.637647402489911\\
0.712758539268555	0.643687796645648\\
0.733065050358828	0.649609991947967\\
0.7533715614491	0.655423013460065\\
0.773678072539372	0.661135097661258\\
0.793984583629644	0.666753770340213\\
0.814291094719917	0.672285917701427\\
0.834597605810189	0.677737850475303\\
0.854904116900461	0.683115361247135\\
0.875210627990733	0.68842377543691\\
0.895517139081005	0.693667996455594\\
0.915823650171278	0.698852545586452\\
0.93613016126155	0.703981597123197\\
0.956436672351822	0.709059009259262\\
0.976743183442095	0.714088351175413\\
0.997049694532367	0.719072926722851\\
1.01735620562264	0.724015795049424\\
1.03766271671291	0.72891978846973\\
1.05796922780318	0.733787527836419\\
1.07827573889346	0.738621435630345\\
1.09858224998373	0.743423746951392\\
1.118888761074	0.748196518559351\\
1.13919527216427	0.7529416360851\\
1.15950178325454	0.757660819506291\\
1.17980829434482	0.762355626958242\\
1.20011480543509	0.767027456929859\\
1.22042131652536	0.771677548876103\\
1.24072782761563	0.776306982262444\\
1.26103433870591	0.780916674043627\\
1.28134084979618	0.78550737456865\\
1.30164736088645	0.790079661897108\\
1.32195387197672	0.794633934509317\\
1.34226038306699	0.799170402395234\\
1.36256689415727	0.803689076516028\\
1.38287340524754	0.808189756649021\\
1.40317991633781	0.812672017653413\\
1.42348642742808	0.817135194232773\\
1.44379293851836	0.82157836432339\\
1.46409944960863	0.826000331307646\\
1.4844059606989	0.830399605341216\\
1.50471247178917	0.834774384194374\\
1.52501898287944	0.839122534142415\\
1.54532549396972	0.843441571598267\\
1.56563200505999	0.847728646359533\\
1.58593851615026	0.851980527537095\\
1.60624502724053	0.856193593433223\\
};
\addplot [color=mycolor4, forget plot]
  table[row sep=crcr]{%
0.0021125390300004	0.574401703768678\\
0.0232379293300044	0.574689203370155\\
0.0443633196300083	0.575452258854129\\
0.0654887099300123	0.576673211022077\\
0.0866141002300163	0.578325766049176\\
0.10773949053002	0.58037664541448\\
0.128864880830024	0.582787848892771\\
0.149990271130028	0.585518950418183\\
0.171115661430032	0.588529150062965\\
0.192241051730036	0.591778926636469\\
0.21336644203004	0.595231233777103\\
0.234491832330044	0.59885225743967\\
0.255617222630048	0.602611799000176\\
0.276742612930052	0.606483368176877\\
0.297868003230056	0.610444070658822\\
0.31899339353006	0.614474364700069\\
0.340118783830064	0.618557745530354\\
0.361244174130068	0.622680400618391\\
0.382369564430072	0.62683086495946\\
0.403494954730076	0.630999694550896\\
0.42462034503008	0.635179168101496\\
0.445745735330084	0.639363021401242\\
0.466871125630088	0.643546215122593\\
0.487996515930092	0.647724734613834\\
0.509121906230096	0.65189541904037\\
0.530247296530099	0.656055816691822\\
0.551372686830103	0.660204063153907\\
0.572498077130107	0.664338779172055\\
0.593623467430111	0.668458985293407\\
0.614748857730115	0.672564030691538\\
0.635874248030119	0.676653533908747\\
0.656999638330123	0.680727333568085\\
0.678125028630127	0.684785447397885\\
0.699250418930131	0.688828038169628\\
0.720375809230135	0.69285538537441\\
0.741501199530139	0.696867861655512\\
0.762626589830143	0.700865913177506\\
0.783751980130147	0.704850043249293\\
0.804877370430151	0.708820798632873\\
0.826002760730155	0.71277875806481\\
0.847128151030159	0.716724522596202\\
0.868253541330163	0.720658707422181\\
0.889378931630167	0.724581934925566\\
0.910504321930171	0.728494828703583\\
0.931629712230175	0.732398008382656\\
0.952755102530179	0.736292085056067\\
0.973880492830183	0.740177657203288\\
0.995005883130187	0.744055306969555\\
1.01613127343019	0.747925596699958\\
1.03725666373019	0.751789065635146\\
1.0583820540302	0.755646226685649\\
1.0795074443302	0.759497563209796\\
1.10063283463021	0.763343525726155\\
1.12175822493021	0.767184528495967\\
1.14288361523021	0.771020945914191\\
1.16400900553022	0.774853108650072\\
1.18513439583022	0.778681299479493\\
1.20625978613023	0.782505748752169\\
1.22738517643023	0.786326629437205\\
1.24851056673023	0.790144051690831\\
1.26963595703024	0.79395805689056\\
1.29076134733024	0.797768611080864\\
1.31188673763025	0.801575597777324\\
1.33301212793025	0.805378810078873\\
1.35413751823025	0.809177942042783\\
1.37526290853026	0.812972579284005\\
1.39638829883026	0.8167621887709\\
1.41751368913027	0.820546107803733\\
1.43863907943027	0.824323532181713\\
1.45976446973027	0.828093503590061\\
1.48088986003028	0.831854896271552\\
1.50201525033028	0.835606403088791\\
1.52314064063029	0.839346521135165\\
1.54426603093029	0.843073537115091\\
1.56539142123029	0.846785512788662\\
1.5865168115303	0.850480270861963\\
1.6076422018303	0.854155381801531\\
1.62876759213031	0.85780815215728\\
1.64989298243031	0.86143561508876\\
1.67101837273031	0.865034523898499\\
};
\addplot [color=mycolor5, forget plot]
  table[row sep=crcr]{%
0.00217940179665804	0.672214747385296\\
0.0239734197632384	0.67229299810535\\
0.0457674377298189	0.672501862703293\\
0.0675614556963993	0.672840491164385\\
0.0893554736629797	0.673307648164011\\
0.11114949162956	0.67390166947364\\
0.13294350959614	0.674620477747791\\
0.154737527562721	0.675461610724828\\
0.176531545529301	0.676422255352258\\
0.198325563495882	0.677499285848924\\
0.220119581462462	0.678689304387924\\
0.241913599429043	0.679988683258368\\
0.263707617395623	0.681393607462508\\
0.285501635362203	0.682900116805754\\
0.307295653328784	0.684504146654183\\
0.329089671295364	0.68620156666521\\
0.350883689261945	0.687988216935887\\
0.372677707228525	0.689859941152836\\
0.394471725195105	0.691812616461828\\
0.416265743161686	0.693842179898414\\
0.438059761128266	0.695944651330343\\
0.459853779094847	0.698116152955468\\
0.481647797061427	0.700352925474665\\
0.503441815028007	0.702651341118073\\
0.525235832994588	0.705007913745782\\
0.547029850961168	0.707419306272419\\
0.568823868927749	0.709882335680835\\
0.590617886894329	0.71239397589527\\
0.612411904860909	0.7149513587811\\
0.63420592282749	0.717551773528365\\
0.65599994079407	0.720192664661561\\
0.677793958760651	0.722871628900121\\
0.699587976727231	0.725586411073975\\
0.721381994693811	0.728334899277548\\
0.743176012660392	0.731115119424273\\
0.764970030626972	0.733925229343134\\
0.786764048593553	0.736763512538749\\
0.808558066560133	0.73962837171817\\
0.830352084526713	0.742518322170197\\
0.852146102493294	0.745431985067519\\
0.873940120459874	0.748368080747732\\
0.895734138426455	0.751325422016823\\
0.917528156393035	0.754302907507227\\
0.939322174359616	0.757299515112884\\
0.961116192326196	0.760314295514862\\
0.982910210292776	0.763346365803446\\
1.00470422825936	0.766394903196033\\
1.02649824622594	0.769459138844199\\
1.04829226419252	0.772538351718109\\
1.0700862821591	0.775631862552107\\
1.09188030012568	0.778739027831162\\
1.11367431809226	0.781859233794382\\
1.13546833605884	0.784991890428659\\
1.15726235402542	0.788136425422729\\
1.179056371992	0.791292278049351\\
1.20085038995858	0.794458892941295\\
1.22264440792516	0.797635713724929\\
1.24443842589174	0.800822176473515\\
1.26623244385832	0.804017702941587\\
1.2880264618249	0.807221693541041\\
1.30982047979148	0.81043352001973\\
1.33161449775806	0.813652517804402\\
1.35340851572464	0.816877977971781\\
1.37520253369122	0.820109138815083\\
1.3969965516578	0.823345176978317\\
1.41879056962438	0.826585198137999\\
1.44058458759096	0.829828227221693\\
1.46237860555755	0.833073198165824\\
1.48417262352413	0.836318943231898\\
1.50596664149071	0.839564181921418\\
1.52776065945729	0.84280750955597\\
1.54955467742387	0.846047385620884\\
1.57134869539045	0.849282122008989\\
1.59314271335703	0.852509871345705\\
1.61493673132361	0.855728615628089\\
1.63673074929019	0.858936155467919\\
1.65852476725677	0.862130100291542\\
1.68031878522335	0.865307859914987\\
1.70211280318993	0.868466637978661\\
1.72390682115651	0.87160342778763\\
};
\addplot [color=mycolor6, forget plot]
  table[row sep=crcr]{%
0.00223538575392717	0.718582985779825\\
0.0245892432931988	0.718629973008918\\
0.0469431008324705	0.718755466205682\\
0.0692969583717422	0.718959217907251\\
0.0916508159110139	0.71924090193459\\
0.114004673450286	0.719600080879835\\
0.136358530989557	0.720036204186947\\
0.158712388528829	0.72054861089998\\
0.181066246068101	0.721136533904788\\
0.203420103607372	0.721799105051692\\
0.225773961146644	0.722535360939169\\
0.248127818685916	0.723344249227872\\
0.270481676225187	0.724224635379313\\
0.292835533764459	0.725175309721983\\
0.315189391303731	0.726194994751799\\
0.337543248843002	0.727282352577147\\
0.359897106382274	0.728435992422929\\
0.382250963921546	0.7296544781129\\
0.404604821460817	0.730936335455613\\
0.426958679000089	0.73228005946583\\
0.449312536539361	0.733684121360727\\
0.471666394078632	0.735146975277669\\
0.494020251617904	0.736667064668513\\
0.516374109157176	0.7382428283329\\
0.538727966696447	0.739872706060854\\
0.561081824235719	0.741555143862245\\
0.583435681774991	0.743288598767454\\
0.605789539314262	0.745071543189916\\
0.628143396853534	0.746902468846768\\
0.650497254392806	0.748779890238848\\
0.672851111932077	0.75070234769555\\
0.695204969471349	0.752668409993546\\
0.717558827010621	0.754676676561335\\
0.739912684549892	0.756725779283891\\
0.762266542089164	0.758814383923137\\
0.784620399628436	0.760941191171211\\
0.806974257167707	0.763104937354008\\
0.829328114706979	0.765304394802493\\
0.851681972246251	0.767538371909104\\
0.874035829785523	0.769805712885792\\
0.896389687324794	0.772105297239327\\
0.918743544864066	0.774436038978195\\
0.941097402403338	0.776796885564035\\
0.963451259942609	0.779186816618811\\
0.985805117481881	0.781604842397244\\
1.00815897502115	0.784050002032026\\
1.03051283256042	0.786521361557332\\
1.0528666900997	0.789018011714148\\
1.07522054763897	0.791539065538759\\
1.09757440517824	0.794083655733652\\
1.11992826271751	0.796650931817977\\
1.14228212025678	0.799240057052496\\
1.16463597779605	0.801850205132083\\
1.18698983533533	0.804480556636785\\
1.2093436928746	0.807130295230534\\
1.23169755041387	0.809798603595226\\
1.25405140795314	0.812484659086079\\
1.27640526549241	0.815187629093343\\
1.29875912303168	0.817906666094575\\
1.32111298057096	0.820640902381489\\
1.34346683811023	0.823389444445653\\
1.3658206956495	0.826151367008798\\
1.38817455318877	0.828925706685281\\
1.41052841072804	0.831711455267753\\
1.43288226826731	0.834507552632008\\
1.45523612580659	0.837312879263345\\
1.47758998334586	0.840126248415516\\
1.49994384088513	0.842946397924346\\
1.5222976984244	0.845771981711934\\
1.54465155596367	0.848601561034412\\
1.56700541350294	0.85143359554665\\
1.58935927104222	0.854266434281763\\
1.61171312858149	0.857098306671372\\
1.63406698612076	0.859927313764948\\
1.65642084366003	0.862751419842428\\
1.6787747011993	0.865568444653391\\
1.70112855873857	0.868376056557173\\
1.72348241627785	0.871171766879852\\
1.74583627381712	0.873952925843807\\
1.76819013135639	0.876716720460527\\
};
\addplot [color=mycolor7, forget plot]
  table[row sep=crcr]{%
0.0022830331877134	0.747994314831557\\
0.0251133650648474	0.748029174675099\\
0.0479436969419814	0.748122300117117\\
0.0707740288191154	0.748273567035443\\
0.0936043606962495	0.74848282936966\\
0.116434692573383	0.748749893919919\\
0.139265024450517	0.749074517478234\\
0.162095356327652	0.749456407072272\\
0.184925688204786	0.749895220997871\\
0.20775602008192	0.750390570223579\\
0.230586351959054	0.750942020045879\\
0.253416683836188	0.75154909194292\\
0.276247015713322	0.752211265594887\\
0.299077347590456	0.752927981045838\\
0.32190767946759	0.753698640984295\\
0.344738011344724	0.754522613120781\\
0.367568343221858	0.755399232640997\\
0.390398675098992	0.756327804713689\\
0.413229006976126	0.757307607032699\\
0.43605933885326	0.758337892373339\\
0.458889670730394	0.75941789114392\\
0.481720002607528	0.760546813914225\\
0.504550334484662	0.761723853903605\\
0.527380666361796	0.762948189412653\\
0.55021099823893	0.764218986183523\\
0.573041330116064	0.765535399675136\\
0.595871661993198	0.766896577240948\\
0.618701993870332	0.768301660198122\\
0.641532325747466	0.76974978577819\\
0.6643626576246	0.771240088950566\\
0.687192989501734	0.772771704111386\\
0.710023321378868	0.774343766631199\\
0.732853653256002	0.775955414256032\\
0.755683985133136	0.777605788357244\\
0.77851431701027	0.779294035026317\\
0.801344648887404	0.7810193060114\\
0.824174980764538	0.782780759493016\\
0.847005312641672	0.784577560696524\\
0.869835644518806	0.786408882339541\\
0.89266597639594	0.788273904912403\\
0.915496308273074	0.790171816789925\\
0.938326640150208	0.792101814172617\\
0.961156972027342	0.794063100855244\\
0.983987303904476	0.796054887820524\\
1.00681763578161	0.798076392655077\\
1.02964796765874	0.80012683878449\\
1.05247829953588	0.802205454523698\\
1.07530863141301	0.804311471938161\\
1.09813896329015	0.80644412551089\\
1.12096929516728	0.808602650609279\\
1.14379962704441	0.81078628174524\\
1.16662995892155	0.812994250621218\\
1.18946029079868	0.81522578395398\\
1.21229062267582	0.817480101067496\\
1.23512095455295	0.819756411245404\\
1.25795128643008	0.822053910833379\\
1.28078161830722	0.824371780081387\\
1.30361195018435	0.826709179715822\\
1.32644228206149	0.829065247231936\\
1.34927261393862	0.831439092897794\\
1.37210294581575	0.833829795462308\\
1.39493327769289	0.836236397562058\\
1.41776360957002	0.838657900824319\\
1.44059394144716	0.841093260667709\\
1.46342427332429	0.84354138080677\\
1.48625460520142	0.846001107473337\\
1.50908493707856	0.848471223375307\\
1.53191526895569	0.850950441423368\\
1.55474560083283	0.853437398267918\\
1.57757593270996	0.855930647702329\\
1.60040626458709	0.858428654005057\\
1.62323659646423	0.860929785311758\\
1.64606692834136	0.86343230712961\\
1.6688972602185	0.865934376129367\\
1.69172759209563	0.868434034375738\\
1.71455792397276	0.870929204183021\\
1.7373882558499	0.873417683809559\\
1.76021858772703	0.875897144230237\\
1.78304891960417	0.878365127249255\\
1.8058792514813	0.880819045233843\\
};
\addplot [color=mycolor8, forget plot]
  table[row sep=crcr]{%
0.00232412306167334	0.769030871186859\\
0.0255653536784067	0.769059163851922\\
0.0488065842951401	0.769134757049891\\
0.0720478149118735	0.769257571170624\\
0.0952890455286069	0.769427522786466\\
0.11853027614534	0.769644503482028\\
0.141771506762074	0.769908377096779\\
0.165012737378807	0.770218979452007\\
0.188253967995541	0.770576118640862\\
0.211495198612274	0.770979575542575\\
0.234736429229007	0.771429104467869\\
0.257977659845741	0.771924433900674\\
0.281218890462474	0.772465267318636\\
0.304460121079207	0.773051284080923\\
0.327701351695941	0.773682140374113\\
0.350942582312674	0.774357470207789\\
0.374183812929408	0.775076886451835\\
0.397425043546141	0.775839981907443\\
0.420666274162874	0.776646330404024\\
0.443907504779608	0.777495487914165\\
0.467148735396341	0.778386993678739\\
0.490389966013075	0.779320371334572\\
0.513631196629808	0.780295130036988\\
0.536872427246541	0.781310765569853\\
0.560113657863275	0.782366761435807\\
0.583354888480008	0.783462589919693\\
0.606596119096741	0.784597713118343\\
0.629837349713475	0.785771583930165\\
0.653078580330208	0.786983646998232\\
0.676319810946942	0.78823333960081\\
0.699561041563675	0.789520092483455\\
0.722802272180408	0.790843330627357\\
0.746043502797142	0.792202473948207\\
0.769284733413875	0.793596937920771\\
0.792525964030609	0.79502613412403\\
0.815767194647342	0.796489470702127\\
0.839008425264075	0.797986352736363\\
0.862249655880809	0.799516182523552\\
0.885490886497542	0.801078359756244\\
0.908732117114276	0.802672281600015\\
0.931973347731009	0.804297342663218\\
0.955214578347743	0.805952934854384\\
0.978455808964476	0.807638447122329\\
1.00169703958121	0.8093532650738\\
1.02493827019794	0.811096770463491\\
1.04817950081468	0.812868340550575\\
1.07142073143141	0.814667347316089\\
1.09466196204814	0.816493156534911\\
1.11790319266488	0.818345126695865\\
1.14114442328161	0.820222607763172\\
1.16438565389834	0.822124939772144\\
1.18762688451508	0.824051451251831\\
1.21086811513181	0.826001457467094\\
1.23410934574854	0.82797425847244\\
1.25735057636528	0.829969136970057\\
1.28059180698201	0.831985355964601\\
1.30383303759874	0.834022156207684\\
1.32707426821548	0.836078753425639\\
1.35031549883221	0.838154335325151\\
1.37355672944894	0.840248058372651\\
1.39679796006568	0.842359044345254\\
1.42003919068241	0.844486376653384\\
1.44328042129914	0.846629096438395\\
1.46652165191588	0.848786198452434\\
1.48976288253261	0.850956626732291\\
1.51300411314934	0.853139270085186\\
1.53624534376608	0.855332957411159\\
1.55948657438281	0.857536452895114\\
1.58272780499954	0.859748451111106\\
1.60596903561628	0.861967572092601\\
1.62921026623301	0.864192356434864\\
1.65245149684974	0.866421260509701\\
1.67569272746648	0.868652651887935\\
1.69893395808321	0.870884805081373\\
1.72217518869994	0.873115897733053\\
1.74541641931668	0.875344007401662\\
1.76865764993341	0.877567109102749\\
1.79189888055014	0.879783073784315\\
1.81514011116688	0.88198966792685\\
1.83838134178361	0.884184554466307\\
};
\addplot [color=mycolor9, forget plot]
  table[row sep=crcr]{%
0.00235998100743682	0.78510943998452\\
0.025959791081805	0.78513353311212\\
0.0495596011561732	0.785197913558485\\
0.0731594112305414	0.785302522791109\\
0.0967592213049096	0.785447305864786\\
0.120359031379278	0.78563219277702\\
0.143958841453646	0.785857095911689\\
0.167558651528014	0.786121909622869\\
0.191158461602382	0.786426510276928\\
0.21475827167675	0.786770756457786\\
0.238358081751119	0.787154489256001\\
0.261957891825487	0.787577532613429\\
0.285557701899855	0.788039693710699\\
0.309157511974223	0.788540763390143\\
0.332757322048591	0.789080516608961\\
0.35635713212296	0.789658712918204\\
0.379956942197328	0.790275096963565\\
0.403556752271696	0.790929399004002\\
0.427156562346064	0.791621335444286\\
0.450756372420432	0.792350609377519\\
0.474356182494801	0.793116911133604\\
0.497955992569169	0.793919918829701\\
0.521555802643537	0.7947592989185\\
0.545155612717905	0.795634706730294\\
0.568755422792273	0.796545787004696\\
0.592355232866641	0.797492174407894\\
0.61595504294101	0.798473494031297\\
0.639554853015378	0.799489361867474\\
0.663154663089746	0.800539385259239\\
0.686754473164114	0.801623163317787\\
0.710354283238482	0.802740287305775\\
0.73395409331285	0.803890340981164\\
0.757553903387219	0.805072900897908\\
0.781153713461587	0.806287536659134\\
0.804753523535955	0.807533811118796\\
0.828353333610323	0.808811280527584\\
0.851953143684691	0.810119494618872\\
0.87555295375906	0.811457996630379\\
0.899152763833428	0.812826323257222\\
0.922752573907796	0.814224004531833\\
0.946352383982164	0.8156505636262\\
0.969952194056532	0.817105516571744\\
0.993552004130901	0.818588371891869\\
1.01715181420527	0.82009863014244\\
1.04075162427964	0.821635783354731\\
1.06435143435401	0.823199314375769\\
1.08795124442837	0.82478869610048\\
1.11155105450274	0.826403390590039\\
1.13515086457711	0.828042848070658\\
1.15875067465148	0.82970650580697\\
1.18235048472585	0.831393786844054\\
1.20595029480021	0.833104098612201\\
1.22955010487458	0.834836831388571\\
1.25314991494895	0.836591356609994\\
1.27674972502332	0.838367025031811\\
1.30034953509769	0.840163164727579\\
1.32394934517206	0.84197907892559\\
1.34754915524642	0.843814043678921\\
1.37114896532079	0.845667305366905\\
1.39474877539516	0.847538078027506\\
1.41834858546953	0.849425540522002\\
1.4419483955439	0.851328833535845\\
1.46554820561826	0.853247056422469\\
1.48914801569263	0.855179263900441\\
1.512747825767	0.857124462618513\\
1.53634763584137	0.859081607608261\\
1.55994744591574	0.86104959864967\\
1.58354725599011	0.863027276581781\\
1.60714706606447	0.865013419598126\\
1.63074687613884	0.867006739575067\\
1.65434668621321	0.869005878490602\\
1.67794649628758	0.871009405001234\\
1.70154630636195	0.873015811255462\\
1.72514611643631	0.875023510033351\\
1.74874592651068	0.877030832313127\\
1.77234573658505	0.879036025376584\\
1.79594554665942	0.88103725157508\\
1.81954535673379	0.883032587886461\\
1.84314516680816	0.885020026399331\\
1.86674497688252	0.886997475864076\\
};
\addplot [color=mycolor10, forget plot]
  table[row sep=crcr]{%
0.00239160909272102	0.797935108070376\\
0.0263077000199312	0.797956236084106\\
0.0502237909471414	0.798012699105749\\
0.0741398818743516	0.798104450311917\\
0.0980559728015618	0.798231449887791\\
0.121972063728772	0.798393648131579\\
0.145888154655982	0.798590983079188\\
0.169804245583192	0.798823380046726\\
0.193720336510403	0.799090751570877\\
0.217636427437613	0.799392997481016\\
0.241552518364823	0.799730005032066\\
0.265468609292033	0.800101649073459\\
0.289384700219243	0.800507792243514\\
0.313300791146454	0.80094828518358\\
0.337216882073664	0.801422966768657\\
0.361132973000874	0.80193166435099\\
0.385049063928084	0.802474194014629\\
0.408965154855294	0.803050360838393\\
0.432881245782505	0.803659959164881\\
0.456797336709715	0.804302772873171\\
0.480713427636925	0.80497857565276\\
0.504629518564135	0.805687131276259\\
0.528545609491345	0.806428193868289\\
0.552461700418556	0.807201508167982\\
0.576377791345766	0.80800680978238\\
0.600293882272976	0.808843825428043\\
0.624209973200186	0.809712273158033\\
0.648126064127396	0.810611862571431\\
0.672042155054607	0.811542295002504\\
0.695958245981817	0.812503263686504\\
0.719874336909027	0.813494453899108\\
0.743790427836237	0.814515543066379\\
0.767706518763447	0.815566200842064\\
0.791622609690658	0.816646089149068\\
0.815538700617868	0.817754862181704\\
0.839454791545078	0.818892166365367\\
0.863370882472288	0.820057640270152\\
0.887286973399498	0.821250914474821\\
0.911203064326709	0.822471611377459\\
0.935119155253919	0.823719344949043\\
0.959035246181129	0.824993720426019\\
0.982951337108339	0.826294333937911\\
1.00686742803555	0.827620772065847\\
1.03078351896276	0.828972611327789\\
1.05469960988997	0.83034941758615\\
1.07861570081718	0.831750745373393\\
1.10253179174439	0.833176137131141\\
1.1264478826716	0.834625122358278\\
1.15036397359881	0.836097216663332\\
1.17428006452602	0.837591920717042\\
1.19819615545323	0.839108719100125\\
1.22211224638044	0.840647079042306\\
1.24602833730765	0.842206449048415\\
1.26994442823486	0.843786257407843\\
1.29386051916207	0.845385910584102\\
1.31777661008928	0.847004791481804\\
1.34169270101649	0.848642257589235\\
1.3656087919437	0.850297638995652\\
1.38952488287091	0.851970236283724\\
1.41344097379812	0.853659318299041\\
1.43735706472533	0.855364119800507\\
1.46127315565254	0.857083838997643\\
1.48518924657975	0.858817634983521\\
1.50910533750696	0.860564625075068\\
1.53302142843417	0.862323882076138\\
1.55693751936138	0.864094431482867\\
1.58085361028859	0.865875248655509\\
1.6047697012158	0.867665255986186\\
1.62868579214301	0.869463320097849\\
1.65260188307022	0.871268249116104\\
1.67651797399743	0.873078790062445\\
1.70043406492465	0.87489362642458\\
1.72435015585186	0.876711375967132\\
1.74826624677907	0.878530588853512\\
1.77218233770628	0.880349746157049\\
1.79609842863349	0.882167258846344\\
1.8200145195607	0.883981467335655\\
1.84393061048791	0.885790641695645\\
1.86784670141512	0.887592982622532\\
1.89176279234233	0.889386623263767\\
};
\addplot [color=black, dotted, forget plot]
  table[row sep=crcr]{%
0.00258317418845549	0.858183944175907\\
0.0284149160730104	0.858193855823716\\
0.0542466579575653	0.858220362272829\\
0.0800783998421202	0.858263442456255\\
0.105910141726675	0.858323083783744\\
0.13174188361123	0.858399272008236\\
0.157573625495785	0.85849198973662\\
0.18340536738034	0.85860121607226\\
0.209237109264895	0.858726926478862\\
0.23506885114945	0.858869092706661\\
0.260900593034005	0.859027682738844\\
0.28673233491856	0.859202660744104\\
0.312564076803114	0.859393987029657\\
0.338395818687669	0.859601617992127\\
0.364227560572224	0.859825506064941\\
0.390059302456779	0.860065599661463\\
0.415891044341334	0.860321843113273\\
0.441722786225889	0.860594176603237\\
0.467554528110444	0.860882536092966\\
0.493386269994999	0.861186853244387\\
0.519218011879554	0.861507055335014\\
0.545049753764109	0.861843065166803\\
0.570881495648664	0.862194800968074\\
0.596713237533218	0.862562176288309\\
0.622544979417773	0.862945099885456\\
0.648376721302328	0.863343475605431\\
0.674208463186883	0.863757202253466\\
0.700040205071438	0.864186173456969\\
0.725871946955993	0.864630277519564\\
0.751703688840548	0.865089397265935\\
0.777535430725103	0.86556340987716\\
0.803367172609658	0.86605218671616\\
0.829198914494213	0.86655559314296\\
0.855030656378768	0.867073488319394\\
0.880862398263322	0.867605725002967\\
0.906694140147877	0.86815214932956\\
0.932525882032432	0.86871260058471\\
0.958357623916987	0.869286910963189\\
0.984189365801542	0.869874905316712\\
1.0100211076861	0.870476400889569\\
1.03585284957065	0.871091207042042\\
1.06168459145521	0.871719124961574\\
1.08751633333976	0.872359947361642\\
1.11334807522432	0.873013458168453\\
1.13917981710887	0.873679432195582\\
1.16501155899343	0.874357634806811\\
1.19084330087798	0.8750478215676\\
1.21667504276254	0.875749737885587\\
1.24250678464709	0.876463118640841\\
1.26833852653165	0.877187687806615\\
1.2941702684162	0.877923158061594\\
1.32000201030076	0.878669230394774\\
1.34583375218531	0.879425593704384\\
1.37166549406987	0.88019192439241\\
1.39749723595442	0.880967885956608\\
1.42332897783898	0.881753128582109\\
1.44916071972353	0.882547288735046\\
1.47499246160809	0.883349988760904\\
1.50082420349264	0.88416083649067\\
1.5266559453772	0.884979424858147\\
1.55248768726175	0.885805331532207\\
1.5783194291463	0.88663811856812\\
1.60415117103086	0.887477332082428\\
1.62998291291541	0.888322501956307\\
1.65581465479997	0.889173141572672\\
1.68164639668452	0.89002874759269\\
1.70747813856908	0.890888799777745\\
1.73330988045363	0.891752760863227\\
1.75914162233819	0.892620076490831\\
1.78497336422274	0.893490175206345\\
1.8108051061073	0.894362468530119\\
1.83663684799185	0.895236351107554\\
1.86246858987641	0.896111200947066\\
1.88830033176096	0.896986379752942\\
1.91413207364552	0.897861233360432\\
1.93996381553007	0.898735092280139\\
1.96579555741463	0.899607272358084\\
1.99162729929918	0.900477075558905\\
2.01745904118374	0.901343790876133\\
2.04329078306829	0.902206695375457\\
};
\addplot [color=black, dotted, forget plot]
  table[row sep=crcr]{%
0.00267710709187037	0.880605164924273\\
0.0294481780105741	0.880611411131624\\
0.0562192489292778	0.880628122249546\\
0.0829903198479815	0.880655282680613\\
0.109761390766685	0.88069288275597\\
0.136532461685389	0.88074091139975\\
0.163303532604093	0.88079935504737\\
0.190074603522796	0.880868197385092\\
0.2168456744415	0.880947419250216\\
0.243616745360204	0.881036998576469\\
0.270387816278908	0.881136910354128\\
0.297158887197611	0.881247126594697\\
0.323929958116315	0.881367616296101\\
0.350701029035019	0.881498345406616\\
0.377472099953722	0.881639276786669\\
0.404243170872426	0.881790370168029\\
0.43101424179113	0.881951582110148\\
0.457785312709834	0.88212286595347\\
0.484556383628537	0.882304171769606\\
0.511327454547241	0.882495446308355\\
0.538098525465945	0.882696632941447\\
0.564869596384648	0.88290767160303\\
0.591640667303352	0.883128498726863\\
0.618411738222056	0.883359047180206\\
0.64518280914076	0.883599246194384\\
0.671953880059463	0.883849021292072\\
0.698724950978167	0.884108294211266\\
0.725496021896871	0.884376982825999\\
0.752267092815574	0.884655001063805\\
0.779038163734278	0.884942258820019\\
0.805809234652982	0.88523866186892\\
0.832580305571685	0.885544111771832\\
0.859351376490389	0.885858505782246\\
0.886122447409093	0.886181736748067\\
0.912893518327797	0.886513693011147\\
0.9396645892465	0.886854258304188\\
0.966435660165204	0.887203311645232\\
0.993206731083908	0.887560727229919\\
1.01997780200261	0.887926374321716\\
1.04674887292132	0.888300117140388\\
1.07351994384002	0.888681814748967\\
1.10029101475872	0.889071320939558\\
1.12706208567743	0.889468484118311\\
1.15383315659613	0.889873147189951\\
1.18060422751483	0.890285147442307\\
1.20737529843354	0.890704316431284\\
1.23414636935224	0.89113047986682\\
1.26091744027094	0.891563457500385\\
1.28768851118965	0.892003063014622\\
1.31445958210835	0.892449103915818\\
1.34123065302706	0.8929013814299\\
1.36800172394576	0.893359690402761\\
1.39477279486446	0.893823819205729\\
1.42154386578317	0.894293549647242\\
1.44831493670187	0.894768656890571\\
1.47508600762057	0.895248909380985\\
1.50185707853928	0.895734068781094\\
1.52862814945798	0.896223889916406\\
1.55539922037669	0.896718120732303\\
1.58217029129539	0.897216502263567\\
1.60894136221409	0.897718768617755\\
1.6357124331328	0.898224646973706\\
1.6624835040515	0.898733857596576\\
1.6892545749702	0.899246113870772\\
1.71602564588891	0.899761122352211\\
1.74279671680761	0.900278582841332\\
1.76956778772632	0.90079818847832\\
1.79633885864502	0.901319625861963\\
1.82310992956372	0.901842575193579\\
1.84988100048243	0.902366710447404\\
1.87665207140113	0.902891699568782\\
1.90342314231983	0.903417204701464\\
1.93019421323854	0.903942882445217\\
1.95696528415724	0.904468384144852\\
1.98373635507595	0.904993356211713\\
2.01050742599465	0.905517440478438\\
2.03727849691335	0.906040274587758\\
2.06404956783206	0.906561492415871\\
2.09082063875076	0.907080724530698\\
2.11759170966946	0.907597598685222\\
};
\addplot [color=black, dotted, forget plot]
  table[row sep=crcr]{%
0.00273333427907367	0.892502534287542\\
0.0300666770698104	0.89250682277455\\
0.0574000198605471	0.892518299494999\\
0.0847333626512838	0.892536952357994\\
0.11206670544202	0.892562773456955\\
0.139400048232757	0.892595753580539\\
0.166733391023494	0.892635881403665\\
0.194066733814231	0.892683143295099\\
0.221400076604967	0.892737523246186\\
0.248733419395704	0.892799002834188\\
0.276066762186441	0.892867561197389\\
0.303400104977177	0.892943175014356\\
0.330733447767914	0.893025818484358\\
0.358066790558651	0.893115463307649\\
0.385400133349387	0.893212078664979\\
0.412733476140124	0.893315631196045\\
0.440066818930861	0.89342608497671\\
0.467400161721598	0.89354340149492\\
0.494733504512334	0.893667539625305\\
0.522066847303071	0.893798455602442\\
0.549400190093808	0.893936102992813\\
0.576733532884544	0.894080432665487\\
0.604066875675281	0.894231392761553\\
0.631400218466018	0.894388928662391\\
0.658733561256754	0.894552982956794\\
0.686066904047491	0.89472349540704\\
0.713400246838228	0.894900402913965\\
0.740733589628965	0.895083639481153\\
0.768066932419701	0.895273136178289\\
0.795400275210438	0.895468821103797\\
0.822733618001175	0.895670619346878\\
0.850066960791911	0.895878452949043\\
0.877400303582648	0.896092240865259\\
0.904733646373385	0.896311898924879\\
0.932066989164121	0.896537339792485\\
0.959400331954858	0.896768472928765\\
0.986733674745595	0.897005204551644\\
1.01406701753633	0.897247437597818\\
1.04140036032707	0.897495071684886\\
1.06873370311781	0.897748003074275\\
1.09606704590854	0.898006124635214\\
1.12340038869928	0.89826932580992\\
1.15073373149002	0.89853749258031\\
1.17806707428075	0.898810507436439\\
1.20540041707149	0.899088249346966\\
1.23273375986222	0.899370593731927\\
1.26006710265296	0.899657412438115\\
1.2874004454437	0.899948573717366\\
1.31473378823444	0.90024394220811\\
1.34206713102517	0.900543378920496\\
1.36940047381591	0.900846741225479\\
1.39673381660665	0.90115388284806\\
1.42406715939738	0.901464653865533\\
1.45140050218812	0.901778900710533\\
1.47873384497886	0.902096466179657\\
1.50606718776959	0.902417189447963\\
1.53340053056033	0.902740906089784\\
1.56073387335107	0.903067448106258\\
1.5880672161418	0.903396643960028\\
1.61540055893254	0.903728318617517\\
1.64273390172328	0.904062293599232\\
1.67006724451401	0.90439838703851\\
1.69740058730475	0.904736413749139\\
1.72473393009549	0.905076185302279\\
1.75206727288622	0.905417510113027\\
1.77940061567696	0.905760193537157\\
1.8067339584677	0.906104037978244\\
1.83406730125843	0.906448843005634\\
1.86140064404917	0.906794405483557\\
1.88873398683991	0.907140519711679\\
1.91606732963064	0.907486977577366\\
1.94340067242138	0.907833568719842\\
1.97073401521212	0.908180080706581\\
1.99806735800285	0.908526299221855\\
2.02540070079359	0.908872008267687\\
2.05273404358433	0.909216990377199\\
2.08006738637506	0.909561026840324\\
2.1074007291658	0.909903897941814\\
2.13473407195654	0.910245383211388\\
2.16206741474727	0.910585261685779\\
};
\addplot [color=black, dotted, forget plot]
  table[row sep=crcr]{%
0.002770162258495	0.89977617132707\\
0.030471784843445	0.899779235814214\\
0.058173407428395	0.899787438568615\\
0.085875030013345	0.899800770176157\\
0.113576652598295	0.899819224230253\\
0.141278275183245	0.89984279317839\\
0.168979897768195	0.89987146771089\\
0.196681520353145	0.899905236617541\\
0.224383142938095	0.899944086736643\\
0.252084765523045	0.899988002930845\\
0.279786388107995	0.900036968072468\\
0.307488010692945	0.900090963032564\\
0.335189633277895	0.900149966671465\\
0.362891255862845	0.900213955829684\\
0.390592878447795	0.900282905319023\\
0.418294501032745	0.900356787913382\\
0.445996123617695	0.900435574339261\\
0.473697746202645	0.900519233265926\\
0.501399368787595	0.900607731295207\\
0.529100991372545	0.900701032950947\\
0.556802613957495	0.900799100668119\\
0.584504236542445	0.900901894781657\\
0.612205859127395	0.90100937351503\\
0.639907481712345	0.901121492968622\\
0.667609104297295	0.901238207107957\\
0.695310726882245	0.901359467751844\\
0.723012349467195	0.901485224560491\\
0.750713972052145	0.901615425023665\\
0.778415594637095	0.901750014448977\\
0.806117217222045	0.90188893595033\\
0.833818839806995	0.902032130436693\\
0.861520462391945	0.902179536601195\\
0.889222084976895	0.902331090910692\\
0.916923707561845	0.902486727595883\\
0.944625330146795	0.902646378642052\\
0.972326952731745	0.902809973780606\\
1.00002857531669	0.902977440481439\\
1.02773019790164	0.903148703946294\\
1.05543182048659	0.903323687103211\\
1.08313344307154	0.903502310602205\\
1.11083506565649	0.90368449281229\\
1.13853668824144	0.903870149819989\\
1.16623831082639	0.904059195429427\\
1.19393993341134	0.904251541164195\\
1.22164155599629	0.904447096271206\\
1.24934317858124	0.904645767726478\\
1.27704480116619	0.904847460243174\\
1.30474642375114	0.905052076282017\\
1.33244804633609	0.905259516064189\\
1.36014966892104	0.905469677586936\\
1.38785129150599	0.905682456641996\\
1.41555291409094	0.905897746837037\\
1.44325453667589	0.906115439620263\\
1.47095615926084	0.906335424308347\\
1.49865778184579	0.906557588117875\\
1.52635940443074	0.906781816200433\\
1.55406102701569	0.907007991681529\\
1.58176264960064	0.907235995703494\\
1.60946427218559	0.907465707472478\\
1.63716589477054	0.907697004309817\\
1.66486751735549	0.90792976170774\\
1.69256913994044	0.9081638533897\\
1.72027076252539	0.908399151375364\\
1.74797238511034	0.908635526050456\\
1.77567400769529	0.908872846241492\\
1.80337563028024	0.909110979295581\\
1.83107725286519	0.909349791165314\\
1.85877887545014	0.90958914649886\\
1.88648049803509	0.909828908735293\\
1.91418212062004	0.910068940205239\\
1.94188374320499	0.910309102236806\\
1.96958536578994	0.910549255266856\\
1.99728698837489	0.910789258957585\\
2.02498861095984	0.911028972318368\\
2.05269023354479	0.911268253832827\\
2.08039185612974	0.91150696159104\\
2.10809347871469	0.911744953426777\\
2.13579510129964	0.911982087059632\\
2.16349672388459	0.912218220241924\\
2.19119834646954	0.912453210910152\\
};
\addplot [color=black, dashed, forget plot]
  table[row sep=crcr]{%
0	0.91698730536472\\
2.29237235717068	0.91698730536472\\
};
\end{axis}
%
%
\draw[->,-stealth, line width=1.25pt] (2,2) to [bend right=-30] (1,3.25);

\end{tikzpicture}%

%% file: Figs/Fig6a_ColourMap_CaiSuo_Shrink_Front.tex
\begin{tikzpicture}
\begin{axis}[axis on top,  
thick, 
width=0.48\textwidth, 
height=0.4\textwidth,
enlargelimits=false, 
colorbar, 
point meta min = 0.25, 
point meta max = 1,
xmin=0,xmax=0.4,ymin=0,ymax=2.5, 
colorbar style = { 
width = 0.3cm,
thick,
black,
title = {$\phi$},
title style = {overlay,yshift = -3pt},
at={(1.05,1)}},
xlabel = {$t$},
ylabel = {$r$},
legend style = {
draw=none,
fill=none,
font=\scriptsize},]
\addplot[forget plot] graphics [xmin=0,xmax=0.4,ymin=0,ymax=2.5] {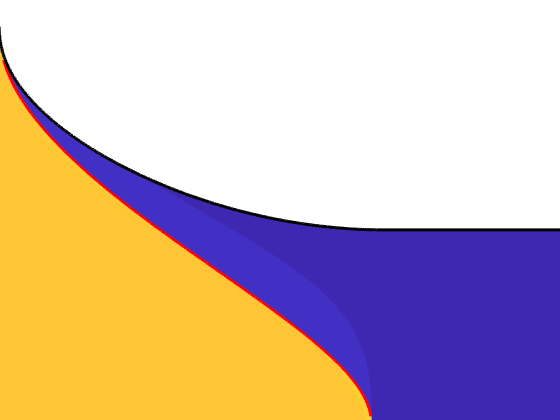};
%
%
\addplot[black, mark=none, domain=-1:-0.1] {x};
    \addlegendentry{Edge location};

\addplot[red, mark=none, domain=-1:-0.1] {x};
    \addlegendentry{Front location}
\end{axis}
\end{tikzpicture}

%% file: Figs/Fig6b_Porosity_CaiSuo_Shrink_Front.tex
%
%
\definecolor{mycolor1}{rgb}{0.20833,0.13020,0.08292}%
\definecolor{mycolor2}{rgb}{0.41667,0.26040,0.16583}%
\definecolor{mycolor3}{rgb}{0.62500,0.39060,0.24875}%
\definecolor{mycolor4}{rgb}{0.83333,0.52080,0.33167}%
\definecolor{mycolor5}{rgb}{1.00000,0.65100,0.41458}%
\definecolor{mycolor6}{rgb}{1.00000,0.78120,0.49750}%
\begin{tikzpicture}

\begin{axis}[%
thick, 
width=0.48\textwidth, 
height=0.4\textwidth,
xmin=0,xmax=2.5,ymin=0,ymax=1, 
xlabel = {$r$},
ylabel = {$\phi$},
yticklabel style = overlay]
\addplot [color=black, forget plot]
  table[row sep=crcr]{%
0.00286546544646336	0.91698730536472\\
0.0315201199110969	0.91698730536472\\
0.0601747743757305	0.91698730536472\\
0.088829428840364	0.91698730536472\\
0.117484083304998	0.91698730536472\\
0.146138737769631	0.91698730536472\\
0.174793392234265	0.91698730536472\\
0.203448046698898	0.91698730536472\\
0.232102701163532	0.91698730536472\\
0.260757355628165	0.91698730536472\\
0.289412010092799	0.91698730536472\\
0.318066664557432	0.91698730536472\\
0.346721319022066	0.91698730536472\\
0.3753759734867	0.91698730536472\\
0.404030627951333	0.91698730536472\\
0.432685282415967	0.91698730536472\\
0.4613399368806	0.91698730536472\\
0.489994591345234	0.91698730536472\\
0.518649245809867	0.91698730536472\\
0.547303900274501	0.91698730536472\\
0.575958554739134	0.91698730536472\\
0.604613209203768	0.91698730536472\\
0.633267863668401	0.91698730536472\\
0.661922518133035	0.91698730536472\\
0.690577172597669	0.91698730536472\\
0.719231827062302	0.91698730536472\\
0.747886481526936	0.91698730536472\\
0.776541135991569	0.91698730536472\\
0.805195790456203	0.91698730536472\\
0.833850444920836	0.91698730536472\\
0.86250509938547	0.91698730536472\\
0.891159753850103	0.91698730536472\\
0.919814408314737	0.91698730536472\\
0.94846906277937	0.91698730536472\\
0.977123717244004	0.91698730536472\\
1.00577837170864	0.91698730536472\\
1.03443302617327	0.91698730536472\\
1.0630876806379	0.91698730536472\\
1.09174233510254	0.91698730536472\\
1.12039698956717	0.91698730536472\\
1.14905164403181	0.91698730536472\\
1.17770629849644	0.91698730536472\\
1.20636095296107	0.91698730536472\\
1.23501560742571	0.91698730536472\\
1.26367026189034	0.91698730536472\\
1.29232491635497	0.91698730536472\\
1.32097957081961	0.91698730536472\\
1.34963422528424	0.91698730536472\\
1.37828887974887	0.91698730536472\\
1.40694353421351	0.91698730536472\\
1.43559818867814	0.91698730536472\\
1.46425284314277	0.91698730536472\\
1.49290749760741	0.91698730536472\\
1.52156215207204	0.91698730536472\\
1.55021680653668	0.91698730536472\\
1.57887146100131	0.91698730536472\\
1.60752611546594	0.91698730536472\\
1.63618076993058	0.91698730536472\\
1.66483542439521	0.91698730536472\\
1.69349007885984	0.91698730536472\\
1.72214473332448	0.91698730536472\\
1.75079938778911	0.91698730536472\\
1.77945404225374	0.91698730536472\\
1.80810869671838	0.91698730536472\\
1.83676335118301	0.91698730536472\\
1.86541800564764	0.91698730536472\\
1.89407266011228	0.91698730536472\\
1.92272731457691	0.91698730536472\\
1.95138196904154	0.91698730536472\\
1.98003662350618	0.91698730536472\\
2.00869127797081	0.91698730536472\\
2.03734593243545	0.91698730536472\\
2.06600058690008	0.91698730536472\\
2.09465524136471	0.91698730536472\\
2.12330989582935	0.91698730536472\\
2.15196455029398	0.91698730536472\\
2.18061920475861	0.91698730536472\\
2.20927385922325	0.91698730536472\\
2.23792851368788	0.91698730536472\\
2.26658316815251	0.91698730536472\\
};
\addplot [color=mycolor1, forget plot]
  table[row sep=crcr]{%
0.00208375760323911	0.916987052247548\\
0.0229213336356302	0.916987052247548\\
0.0437589096680212	0.916987052247547\\
0.0645964857004123	0.916987052247547\\
0.0854340617328034	0.916987052247548\\
0.106271637765194	0.916987052247548\\
0.127109213797585	0.916987052247549\\
0.147946789829977	0.916987052247549\\
0.168784365862368	0.916987052247549\\
0.189621941894759	0.916987052247549\\
0.21045951792715	0.91698705224755\\
0.231297093959541	0.916987052247549\\
0.252134669991932	0.91698705224755\\
0.272972246024323	0.916987052247549\\
0.293809822056714	0.91698705224755\\
0.314647398089105	0.91698705224755\\
0.335484974121496	0.916987052247549\\
0.356322550153887	0.91698705224755\\
0.377160126186278	0.91698705224755\\
0.397997702218669	0.91698705224755\\
0.41883527825106	0.91698705224755\\
0.439672854283451	0.916987052247551\\
0.460510430315843	0.916987052247551\\
0.481348006348234	0.916987052247552\\
0.502185582380625	0.916987052247553\\
0.523023158413016	0.916987052247554\\
0.543860734445407	0.916987052247559\\
0.564698310477798	0.916987052247568\\
0.585535886510189	0.916987052247585\\
0.60637346254258	0.916987052247623\\
0.627211038574971	0.916987052247704\\
0.648048614607362	0.91698705224788\\
0.668886190639753	0.916987052248251\\
0.689723766672144	0.916987052249031\\
0.710561342704535	0.916987052250631\\
0.731398918736927	0.916987052253883\\
0.752236494769318	0.916987052260418\\
0.773074070801709	0.916987052273492\\
0.7939116468341	0.916987052299318\\
0.814749222866491	0.916987052349855\\
0.835586798898882	0.916987052447858\\
0.856424374931273	0.916987052636105\\
0.877261950963664	0.916987052994334\\
0.898099526996055	0.916987053669583\\
0.918937103028446	0.916987054930408\\
0.939774679060837	0.916987057262227\\
0.960612255093228	0.916987061533688\\
0.981449831125619	0.916987069283459\\
1.00228740715801	0.91698708320911\\
1.0231249831904	0.916987107991539\\
1.04396255922279	0.916987151669301\\
1.06480013525518	0.916987227903695\\
1.08563771128757	0.916987359668972\\
1.10647528731997	0.916987585194373\\
1.12731286335236	0.91698796741846\\
1.14815043938475	0.916988608851068\\
1.16898801541714	0.916989674648825\\
1.18982559144953	0.916991427992452\\
1.21066316748192	0.916994283623716\\
1.23150074351431	0.916998887783726\\
1.2523383195467	0.917006235891106\\
1.27317589557909	0.917017843049571\\
1.29401347161149	0.917035986952108\\
1.31485104764388	0.917064050597761\\
1.33568862367627	0.917107006890328\\
1.35652619970866	0.917172055376161\\
1.37736377574105	0.917269140523859\\
1.39820135177344	0.917411192329852\\
1.41903892780583	0.917619727282921\\
1.43987650383822	0.917950882758348\\
1.46071407987061	0.918473182789277\\
1.481551655903	0.918772789777386\\
1.5023892319354	0.916393585718714\\
1.52322680796779	0.904390214770775\\
1.54406438400018	0.295675989249191\\
1.56490196003257	0.290577670997698\\
1.58573953606496	0.28802733929544\\
1.60657711209735	0.286269158661054\\
1.62741468812974	0.284800557585761\\
1.64825226416213	0.283447517787163\\
};
\addplot [color=mycolor2, forget plot]
  table[row sep=crcr]{%
0.00178312219389466	0.916987038159545\\
0.0196143441328412	0.916987038160344\\
0.0374455660717878	0.916987038162588\\
0.0552767880107343	0.916987038166537\\
0.0731080099496809	0.916987038172664\\
0.0909392318886275	0.916987038181686\\
0.108770453827574	0.916987038194677\\
0.126601675766521	0.916987038213193\\
0.144432897705467	0.916987038239478\\
0.162264119644414	0.916987038276765\\
0.18009534158336	0.916987038329592\\
0.197926563522307	0.916987038404463\\
0.215757785461253	0.916987038510559\\
0.2335890074002	0.916987038660887\\
0.251420229339146	0.916987038873784\\
0.269251451278093	0.916987039175113\\
0.28708267321704	0.916987039601191\\
0.304913895155986	0.916987040202971\\
0.322745117094933	0.916987041051736\\
0.340576339033879	0.916987042246925\\
0.358407560972826	0.916987043926903\\
0.376238782911772	0.916987046283629\\
0.394070004850719	0.916987049582635\\
0.411901226789666	0.916987054190065\\
0.429732448728612	0.916987060609207\\
0.447563670667559	0.916987069529509\\
0.465394892606505	0.916987081892188\\
0.483226114545452	0.916987098977595\\
0.501057336484398	0.916987122521012\\
0.518888558423345	0.916987154865283\\
0.536719780362292	0.916987199161134\\
0.554551002301238	0.916987259628518\\
0.572382224240185	0.916987341895614\\
0.590213446179131	0.916987453436026\\
0.608044668118078	0.916987604128841\\
0.625875890057024	0.916987806971183\\
0.643707111995971	0.91698807897813\\
0.661538333934917	0.916988442309823\\
0.679369555873864	0.916988925670694\\
0.697200777812811	0.916989566029156\\
0.715031999751757	0.916990410707365\\
0.732863221690704	0.916991519887697\\
0.75069444362965	0.916992969573273\\
0.768525665568597	0.91699485502017\\
0.786356887507543	0.916997294624793\\
0.80418810944649	0.917000434194187\\
0.822019331385437	0.917004451441819\\
0.839850553324383	0.917009560424328\\
0.85768177526333	0.917016015452505\\
0.875512997202276	0.917024113752243\\
0.893344219141223	0.917034195795735\\
0.911175441080169	0.917046641738202\\
0.929006663019116	0.917061861749111\\
0.946837884958063	0.917080277207304\\
0.964669106897009	0.917102288681958\\
0.982500328835956	0.917128224682843\\
1.0003315507749	0.917158261245387\\
1.01816277271385	0.917192306325882\\
1.0359939946528	0.917229892590164\\
1.05382521659174	0.917270091319365\\
1.07165643853069	0.917310363065442\\
1.08948766046964	0.917341957856746\\
1.10731888240858	0.917358792996482\\
1.12515010434753	0.91744826275171\\
1.14298132628647	0.917775200580864\\
1.16081254822542	0.917146138201551\\
1.17864377016437	0.908434016112864\\
1.19647499210331	0.288655497458955\\
1.21430621404226	0.288553738575991\\
1.23213743598121	0.288183266377009\\
1.24996865792015	0.287643462529074\\
1.2677998798591	0.286991638161351\\
1.28563110179805	0.286264302288157\\
1.30346232373699	0.28548631638685\\
1.32129354567594	0.284675433862283\\
1.33912476761489	0.283844783292581\\
1.35695598955383	0.283004328851078\\
1.37478721149278	0.282161778507052\\
1.39261843343173	0.281323174620716\\
1.41044965537067	0.280493292464902\\
};
\addplot [color=mycolor3, forget plot]
  table[row sep=crcr]{%
0.00158325402802487	0.916987213494181\\
0.0174157943082736	0.916987215751599\\
0.0332483345885224	0.916987221908934\\
0.0490808748687711	0.916987232134385\\
0.0649134151490198	0.916987246725195\\
0.0807459554292686	0.916987266104566\\
0.0965784957095173	0.916987290828226\\
0.112411035989766	0.916987321593898\\
0.128243576270015	0.916987359252363\\
0.144076116550264	0.916987404819402\\
0.159908656830512	0.916987459487664\\
0.175741197110761	0.916987524637308\\
0.19157373739101	0.916987601843605\\
0.207406277671259	0.916987692879253\\
0.223238817951507	0.916987799708125\\
0.239071358231756	0.916987924466287\\
0.254903898512005	0.916988069424532\\
0.270736438792253	0.916988236925099\\
0.286568979072502	0.916988429282907\\
0.302401519352751	0.916988648638876\\
0.318234059633	0.916988896749424\\
0.334066599913248	0.91698917469203\\
0.349899140193497	0.916989482461416\\
0.365731680473746	0.916989818424584\\
0.381564220753995	0.916990178595284\\
0.397396761034243	0.916990555679232\\
0.413229301314492	0.916990937830629\\
0.429061841594741	0.916991307047529\\
0.44489438187499	0.916991637118934\\
0.460726922155238	0.916991891019171\\
0.476559462435487	0.916992017625753\\
0.492392002715736	0.916991947615152\\
0.508224542995985	0.91699158836686\\
0.524057083276233	0.916990817680484\\
0.539889623556482	0.916989476083377\\
0.555722163836731	0.916987357478989\\
0.57155470411698	0.916984197859577\\
0.587387244397228	0.916979661783264\\
0.603219784677477	0.91697332629707\\
0.619052324957726	0.916964661977666\\
0.634884865237974	0.91695301076458\\
0.650717405518223	0.916937560282241\\
0.666549945798472	0.916917314395638\\
0.682382486078721	0.916891059814422\\
0.698215026358969	0.916857328620984\\
0.714047566639218	0.916814356959735\\
0.729880106919467	0.916760042035202\\
0.745712647199716	0.916691896963096\\
0.761545187479965	0.916606961321616\\
0.777377727760213	0.916501661935145\\
0.793210268040462	0.916372672977028\\
0.809042808320711	0.916219973879025\\
0.824875348600959	0.916026286389225\\
0.840707888881208	0.915655156812158\\
0.856540429161457	0.915433066191139\\
0.872372969441706	0.918213722553519\\
0.888205509721954	0.916249295628991\\
0.904038050002203	0.290451326914492\\
0.919870590282452	0.290783756666016\\
0.935703130562701	0.290767408356583\\
0.951535670842949	0.290513940179225\\
0.967368211123198	0.290090393738884\\
0.983200751403447	0.289542087292659\\
0.999033291683696	0.288902167160525\\
1.01486583196394	0.288196145669926\\
1.03069837224419	0.287444259129364\\
1.04653091252444	0.286662789490665\\
1.06236345280469	0.285864867026924\\
1.07819599308494	0.285061005557618\\
1.09402853336519	0.284259497243423\\
1.10986107364544	0.283466731603657\\
1.12569361392569	0.282687470985063\\
1.14152615420593	0.281925097718025\\
1.15735869448618	0.281181839492196\\
1.17319123476643	0.280458975300251\\
1.18902377504668	0.279757022552854\\
1.20485631532693	0.279075905479623\\
1.22068885560718	0.278415105028566\\
1.23652139588743	0.277773790785489\\
1.25235393616768	0.277150935755186\\
};
\addplot [color=mycolor4, forget plot]
  table[row sep=crcr]{%
0.0014558625361943	0.91698135659561\\
0.0160144878981373	0.916981271407596\\
0.0305731132600804	0.916981039356784\\
0.0451317386220234	0.91698065495945\\
0.0596903639839664	0.916980108455333\\
0.0742489893459094	0.916979385997337\\
0.0888076147078525	0.916978469499417\\
0.103366240069796	0.916977336393891\\
0.117924865431739	0.916975959336602\\
0.132483490793682	0.916974305874125\\
0.147042116155625	0.91697233808653\\
0.161600741517568	0.916970012222308\\
0.176159366879511	0.916967278347491\\
0.190717992241454	0.916964080038055\\
0.205276617603397	0.916960354153775\\
0.21983524296534	0.916956030743454\\
0.234393868327283	0.916951033146206\\
0.248952493689226	0.916945278372116\\
0.263511119051169	0.91693867786903\\
0.278069744413112	0.916931138811893\\
0.292628369775055	0.916922566087684\\
0.307186995136998	0.916912865195658\\
0.321745620498941	0.91690194634054\\
0.336304245860884	0.91688973006924\\
0.350862871222827	0.916876154892999\\
0.36542149658477	0.916861187451766\\
0.379980121946713	0.916844835921744\\
0.394538747308656	0.91682716754898\\
0.409097372670599	0.916808331420945\\
0.423655998032542	0.916788587848212\\
0.438214623394485	0.916768346076575\\
0.452773248756428	0.916748213389446\\
0.467331874118371	0.916729058577935\\
0.481890499480314	0.916712063584696\\
0.496449124842257	0.916698792615206\\
0.5110077502042	0.916692288433357\\
0.525566375566143	0.91669722039591\\
0.540125000928086	0.916684974993762\\
0.554683626290029	0.916651225693298\\
0.569242251651972	0.917665344020813\\
0.583800877013915	0.918366473536542\\
0.598359502375859	0.89020160881572\\
0.612918127737802	0.290370543324517\\
0.627476753099745	0.290912871316439\\
0.642035378461688	0.291049634103499\\
0.656594003823631	0.290923981858431\\
0.671152629185574	0.290617617548275\\
0.685711254547517	0.290183085307962\\
0.70026987990946	0.289656957516504\\
0.714828505271403	0.289066094646639\\
0.729387130633346	0.288430935414588\\
0.743945755995289	0.287767367288888\\
0.758504381357232	0.287087866399602\\
0.773063006719175	0.286402241220942\\
0.787621632081118	0.285718152222381\\
0.802180257443061	0.285041499591268\\
0.816738882805004	0.284376729370229\\
0.831297508166947	0.283727085812163\\
0.84585613352889	0.283094825377127\\
0.860414758890833	0.282481400989838\\
0.874973384252776	0.281887621491897\\
0.889532009614719	0.281313789297501\\
0.904090634976662	0.280759818319228\\
0.918649260338605	0.280225333823092\\
0.933207885700548	0.279709755739255\\
0.947766511062491	0.279212366946762\\
0.962325136424434	0.278732368082821\\
0.976883761786377	0.278268920453828\\
0.99144238714832	0.277821178624097\\
1.00600101251026	0.277388314220817\\
1.02055963787221	0.276969532420006\\
1.03511826323415	0.276564082473009\\
1.04967688859609	0.276171263503855\\
1.06423551395804	0.275790426662839\\
1.07879413931998	0.275420974569675\\
1.09335276468192	0.275062358828056\\
1.10791139004386	0.274714076249059\\
1.12247001540581	0.274375664288743\\
1.13702864076775	0.274046696089\\
1.15158726612969	0.273726775412439\\
};
\addplot [color=mycolor5, forget plot]
  table[row sep=crcr]{%
0.00139455760741318	0.917516031139654\\
0.015340133681545	0.917527345941074\\
0.0292857097556768	0.91755853911176\\
0.0432312858298086	0.917611413464058\\
0.0571768619039404	0.917689171651141\\
0.0711224379780721	0.91779663830123\\
0.0850680140522039	0.917940662859154\\
0.0990135901263357	0.918130713798311\\
0.112959166200468	0.918379723032572\\
0.126904742274599	0.918705273968813\\
0.140850318348731	0.91913121211084\\
0.154795894422863	0.919689800144606\\
0.168741470496995	0.920430286961397\\
0.182687046571126	0.921357536871921\\
0.196632622645258	0.922713439555845\\
0.21057819871939	0.927950670411343\\
0.224523774793522	0.289963124062719\\
0.238469350867654	0.292592704921454\\
0.252414926941785	0.293136608362163\\
0.266360503015917	0.292864743318834\\
0.280306079090049	0.292191863140775\\
0.294251655164181	0.291312692188402\\
0.308197231238313	0.290335403541809\\
0.322142807312444	0.289325639420193\\
0.336088383386576	0.288324644921725\\
0.350033959460708	0.287358083834347\\
0.36397953553484	0.286441018793295\\
0.377925111608972	0.285581116067737\\
0.391870687683103	0.284780909839931\\
0.405816263757235	0.284039485655289\\
0.419761839831367	0.283353751740526\\
0.433707415905499	0.282719389623777\\
0.447652991979631	0.282131544544711\\
0.461598568053762	0.2815853036719\\
0.475544144127894	0.281076004410224\\
0.489489720202026	0.280599411165584\\
0.503435296276158	0.280151794921913\\
0.51738087235029	0.279729945336696\\
0.531326448424421	0.279331139874491\\
0.545272024498553	0.27895308913619\\
0.559217600572685	0.278593872419722\\
0.573163176646817	0.27825187300991\\
0.587108752720949	0.277925718958306\\
0.60105432879508	0.277614232257286\\
0.614999904869212	0.27731638730027\\
0.628945480943344	0.277031278237241\\
0.642891057017476	0.276758094128133\\
0.656836633091607	0.276496100506594\\
0.670782209165739	0.276244625949318\\
0.684727785239871	0.276003052385935\\
0.698673361314003	0.275770808096297\\
0.712618937388135	0.275547362569927\\
0.726564513462266	0.275332222613206\\
0.740510089536398	0.275124929267805\\
0.75445566561053	0.274925055244183\\
0.768401241684662	0.274732202678701\\
0.782346817758794	0.27454600109731\\
0.796292393832925	0.274366105519059\\
0.810237969907057	0.274192194665023\\
0.824183545981189	0.274023969257683\\
0.838129122055321	0.273861150406573\\
0.852074698129453	0.273703478081078\\
0.866020274203584	0.273550709672965\\
0.879965850277716	0.273402618650915\\
0.893911426351848	0.273258993308135\\
0.90785700242598	0.273119635602536\\
0.921802578500112	0.27298436008745\\
0.935748154574243	0.272852992929522\\
0.949693730648375	0.272725371009375\\
0.963639306722507	0.272601341099951\\
0.977584882796639	0.272480759116987\\
0.991530458870771	0.2723634894359\\
1.0054760349449	0.2722494042694\\
1.01942161101903	0.272138383100303\\
1.03336718709317	0.272030312164314\\
1.0473127631673	0.271925083977895\\
1.06125833924143	0.271822596906736\\
1.07520391531556	0.27172275477073\\
1.08914949138969	0.271625466481782\\
1.10309506746382	0.271530645711124\\
};
\addplot [color=mycolor6, forget plot]
  table[row sep=crcr]{%
0.00138904841259715	0.27124897474171\\
0.0152795325385686	0.271248974741542\\
0.0291700166645401	0.271248974741094\\
0.0430605007905116	0.271248974740366\\
0.0569509849164831	0.27124897473936\\
0.0708414690424545	0.271248974738076\\
0.084731953168426	0.271248974736514\\
0.0986224372943975	0.271248974734678\\
0.112512921420369	0.271248974732568\\
0.12640340554634	0.271248974730186\\
0.140293889672312	0.271248974727535\\
0.154184373798283	0.271248974724616\\
0.168074857924255	0.271248974721433\\
0.181965342050226	0.271248974717988\\
0.195855826176198	0.271248974714284\\
0.209746310302169	0.271248974710326\\
0.223636794428141	0.271248974706115\\
0.237527278554112	0.271248974701657\\
0.251417762680084	0.271248974696955\\
0.265308246806055	0.271248974692014\\
0.279198730932027	0.271248974686838\\
0.293089215057998	0.271248974681431\\
0.30697969918397	0.271248974675799\\
0.320870183309941	0.271248974669946\\
0.334760667435913	0.271248974663878\\
0.348651151561884	0.2712489746576\\
0.362541635687856	0.271248974651118\\
0.376432119813827	0.271248974644437\\
0.390322603939799	0.271248974637564\\
0.40421308806577	0.271248974630505\\
0.418103572191741	0.271248974623266\\
0.431994056317713	0.271248974615853\\
0.445884540443684	0.271248974608273\\
0.459775024569656	0.271248974600533\\
0.473665508695627	0.271248974592639\\
0.487555992821599	0.271248974584599\\
0.50144647694757	0.271248974576419\\
0.515336961073542	0.271248974568107\\
0.529227445199513	0.27124897455967\\
0.543117929325485	0.271248974551116\\
0.557008413451456	0.27124897454245\\
0.570898897577428	0.271248974533682\\
0.584789381703399	0.271248974524819\\
0.598679865829371	0.271248974515868\\
0.612570349955342	0.271248974506838\\
0.626460834081314	0.271248974497735\\
0.640351318207285	0.271248974488567\\
0.654241802333257	0.271248974479343\\
0.668132286459228	0.27124897447007\\
0.6820227705852	0.271248974460755\\
0.695913254711171	0.271248974451407\\
0.709803738837143	0.271248974442033\\
0.723694222963114	0.27124897443264\\
0.737584707089086	0.271248974423238\\
0.751475191215057	0.271248974413832\\
0.765365675341028	0.271248974404431\\
0.779256159467	0.271248974395043\\
0.793146643592971	0.271248974385674\\
0.807037127718943	0.271248974376332\\
0.820927611844914	0.271248974367024\\
0.834818095970886	0.271248974357757\\
0.848708580096857	0.271248974348539\\
0.862599064222829	0.271248974339375\\
0.8764895483488	0.271248974330274\\
0.890380032474772	0.271248974321241\\
0.904270516600743	0.271248974312284\\
0.918161000726715	0.271248974303407\\
0.932051484852686	0.271248974294618\\
0.945941968978658	0.271248974285923\\
0.959832453104629	0.271248974277327\\
0.973722937230601	0.271248974268835\\
0.987613421356572	0.271248974260453\\
1.00150390548254	0.271248974252187\\
1.01539438960852	0.271248974244041\\
1.02928487373449	0.271248974236021\\
1.04317535786046	0.271248974228129\\
1.05706584198643	0.271248974220372\\
1.0709563261124	0.271248974212752\\
1.08484681023837	0.271248974205274\\
1.09873729436434	0.271248974197941\\
};
\addplot [color=black, dashed, forget plot]
  table[row sep=crcr]{%
0	0.271248974374373\\
1.11123763200597	0.271248974374373\\
};
\end{axis}
%
%
\draw[->,-stealth, line width=1.25pt] (2.8,2.75) to [bend right=20] (0.75,1.75);

\end{tikzpicture}%

%% file: Figs/Fig7a_ColourMap_CaiSuo_Swell_Front.tex
\begin{tikzpicture}
\begin{axis}[axis on top,  
thick, 
/pgf/number format/.cd,
                fixed,
width=0.48\textwidth, 
height=0.4\textwidth,
enlargelimits=false, 
colorbar, 
point meta min = 0.25, 
point meta max = 1,
xmin=0,xmax=1,ymin=0,ymax=2.5, 
colorbar style = { 
width = 0.3cm,
thick,
black,
title = {$\phi$},
title style = {overlay,yshift = -3pt},
at={(1.05,1)}},
xlabel = {$t$},
ylabel = {$r$},
legend style = {
draw=none,
fill=none,
font=\scriptsize},]
\addplot[forget plot] graphics [xmin=0,xmax=1,ymin=0,ymax=2.5] {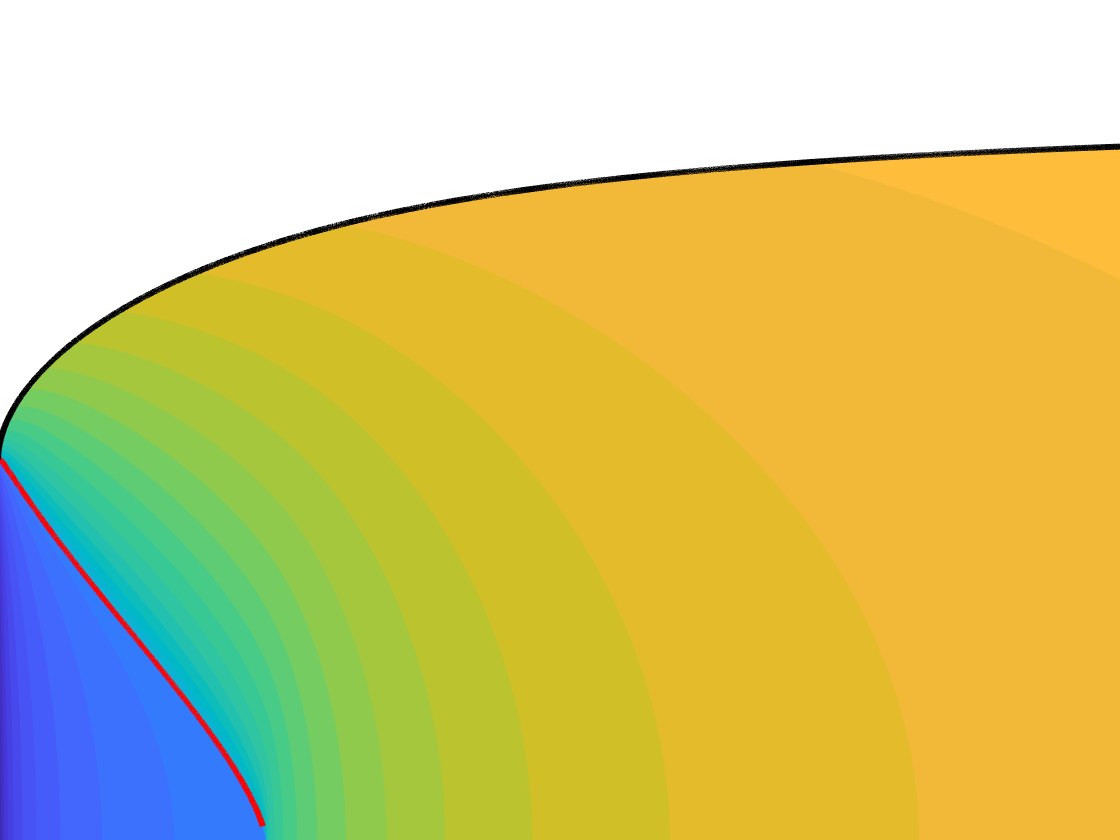};
%
\end{axis}
\end{tikzpicture}

%% file: Figs/Fig7b_Porosity_CaiSuo_Swell_Front.tex
%
%
\definecolor{mycolor1}{rgb}{0.12500,0.07812,0.04975}%
\definecolor{mycolor2}{rgb}{0.25000,0.15624,0.09950}%
\definecolor{mycolor3}{rgb}{0.37500,0.23436,0.14925}%
\definecolor{mycolor4}{rgb}{0.50000,0.31248,0.19900}%
\definecolor{mycolor5}{rgb}{0.62500,0.39060,0.24875}%
\definecolor{mycolor6}{rgb}{0.75000,0.46872,0.29850}%
\definecolor{mycolor7}{rgb}{0.87500,0.54684,0.34825}%
\definecolor{mycolor8}{rgb}{1.00000,0.62496,0.39800}%
\definecolor{mycolor9}{rgb}{1.00000,0.70308,0.44775}%
\definecolor{mycolor10}{rgb}{1.00000,0.78120,0.49750}%
\begin{tikzpicture}

\begin{axis}[%
thick, 
width=0.48\textwidth, 
height=0.4\textwidth,
xmin=0,xmax=2.5,ymin=0,ymax=1, 
xlabel = {$r$},
ylabel = {$\phi$},
yticklabel style = overlay]
\addplot [color=black, forget plot]
  table[row sep=crcr]{%
0.00138904704000746	0.271248974374373\\
0.0152795174400821	0.271248974374373\\
0.0291699878401568	0.271248974374373\\
0.0430604582402314	0.271248974374373\\
0.056950928640306	0.271248974374373\\
0.0708413990403807	0.271248974374373\\
0.0847318694404553	0.271248974374373\\
0.09862233984053	0.271248974374373\\
0.112512810240605	0.271248974374373\\
0.126403280640679	0.271248974374373\\
0.140293751040754	0.271248974374373\\
0.154184221440829	0.271248974374373\\
0.168074691840903	0.271248974374373\\
0.181965162240978	0.271248974374373\\
0.195855632641052	0.271248974374373\\
0.209746103041127	0.271248974374373\\
0.223636573441202	0.271248974374373\\
0.237527043841276	0.271248974374373\\
0.251417514241351	0.271248974374373\\
0.265307984641426	0.271248974374373\\
0.2791984550415	0.271248974374373\\
0.293088925441575	0.271248974374373\\
0.30697939584165	0.271248974374373\\
0.320869866241724	0.271248974374373\\
0.334760336641799	0.271248974374373\\
0.348650807041874	0.271248974374373\\
0.362541277441948	0.271248974374373\\
0.376431747842023	0.271248974374373\\
0.390322218242098	0.271248974374373\\
0.404212688642172	0.271248974374373\\
0.418103159042247	0.271248974374373\\
0.431993629442321	0.271248974374373\\
0.445884099842396	0.271248974374373\\
0.459774570242471	0.271248974374373\\
0.473665040642545	0.271248974374373\\
0.48755551104262	0.271248974374373\\
0.501445981442695	0.271248974374373\\
0.515336451842769	0.271248974374373\\
0.529226922242844	0.271248974374373\\
0.543117392642919	0.271248974374373\\
0.557007863042993	0.271248974374373\\
0.570898333443068	0.271248974374373\\
0.584788803843143	0.271248974374373\\
0.598679274243217	0.271248974374373\\
0.612569744643292	0.271248974374373\\
0.626460215043366	0.271248974374373\\
0.640350685443441	0.271248974374373\\
0.654241155843516	0.271248974374373\\
0.66813162624359	0.271248974374373\\
0.682022096643665	0.271248974374373\\
0.69591256704374	0.271248974374373\\
0.709803037443814	0.271248974374373\\
0.723693507843889	0.271248974374373\\
0.737583978243964	0.271248974374373\\
0.751474448644038	0.271248974374373\\
0.765364919044113	0.271248974374373\\
0.779255389444188	0.271248974374373\\
0.793145859844262	0.271248974374373\\
0.807036330244337	0.271248974374373\\
0.820926800644411	0.271248974374373\\
0.834817271044486	0.271248974374373\\
0.848707741444561	0.271248974374373\\
0.862598211844635	0.271248974374373\\
0.87648868224471	0.271248974374373\\
0.890379152644785	0.271248974374373\\
0.904269623044859	0.271248974374373\\
0.918160093444934	0.271248974374373\\
0.932050563845009	0.271248974374373\\
0.945941034245083	0.271248974374373\\
0.959831504645158	0.271248974374373\\
0.973721975045233	0.271248974374373\\
0.987612445445307	0.271248974374373\\
1.00150291584538	0.271248974374373\\
1.01539338624546	0.271248974374373\\
1.02928385664553	0.271248974374373\\
1.04317432704561	0.271248974374373\\
1.05706479744568	0.271248974374373\\
1.07095526784576	0.271248974374373\\
1.08484573824583	0.271248974374373\\
1.0987362086459	0.271248974374373\\
};
\addplot [color=mycolor1, forget plot]
  table[row sep=crcr]{%
0.00175834720137486	0.396291304404258\\
0.0193418192151234	0.396303721743788\\
0.036925291228872	0.396336937622697\\
0.0545087632426206	0.396390987735893\\
0.0720922352563692	0.396465953967655\\
0.0896757072701177	0.396561953980871\\
0.107259179283866	0.396679140424993\\
0.124842651297615	0.396817701618421\\
0.142426123311363	0.396977862736286\\
0.160009595325112	0.397159887352386\\
0.177593067338861	0.397364079315766\\
0.195176539352609	0.397590784978494\\
0.212760011366358	0.397840395807109\\
0.230343483380106	0.398113351421579\\
0.247926955393855	0.398410143116415\\
0.265510427407603	0.398731317930619\\
0.283093899421352	0.399077483347146\\
0.300677371435101	0.399449312719518\\
0.318260843448849	0.399847551543849\\
0.335844315462598	0.400273024719977\\
0.353427787476346	0.400726644977099\\
0.371011259490095	0.401209422678912\\
0.388594731503843	0.401722477273386\\
0.406178203517592	0.402267050715966\\
0.423761675531341	0.40284452327682\\
0.441345147545089	0.40345643224853\\
0.458928619558838	0.40410449420885\\
0.476512091572586	0.404790631675364\\
0.494095563586335	0.405517005231765\\
0.511679035600084	0.406286052532736\\
0.529262507613832	0.407100536040832\\
0.546845979627581	0.407963601965855\\
0.564429451641329	0.408878853742546\\
0.582012923655078	0.409850444614945\\
0.599596395668826	0.410883195680797\\
0.617179867682575	0.411982748381592\\
0.634763339696323	0.413155764383171\\
0.652346811710072	0.414410191888727\\
0.669930283723821	0.415755627097266\\
0.687513755737569	0.417203815449005\\
0.705097227751318	0.418769364676662\\
0.722680699765066	0.420470790486333\\
0.740264171778815	0.422332103994944\\
0.757847643792564	0.424385308162908\\
0.775431115806312	0.426674456603041\\
0.793014587820061	0.42926255204306\\
0.810598059833809	0.432244628407073\\
0.828181531847558	0.435779279517155\\
0.845765003861306	0.440189176554565\\
0.863348475875055	0.44631817960896\\
0.880931947888804	0.454728900340306\\
0.898515419902552	0.567914042487759\\
0.916098891916301	0.588144356107451\\
0.933682363930049	0.603129811170388\\
0.951265835943798	0.615899683181679\\
0.968849307957546	0.62721794752061\\
0.986432779971295	0.637496347368411\\
1.00401625198504	0.646990859475354\\
1.02159972399879	0.655868862033019\\
1.03918319601254	0.664245152978063\\
1.05676666802629	0.672202295097665\\
1.07435014004004	0.679802119955214\\
1.09193361205379	0.687092412112796\\
1.10951708406754	0.694110999712518\\
1.12710055608128	0.700888403040088\\
1.14468402809503	0.70744963100848\\
1.16226750010878	0.713815442194851\\
1.17985097212253	0.720003253314683\\
1.19743444413628	0.726027807947111\\
1.21501791615003	0.731901678257581\\
1.23260138816377	0.737635647964344\\
1.25018486017752	0.743239009230626\\
1.26776833219127	0.748719796060512\\
1.28535180420502	0.754084970104923\\
1.30293527621877	0.759340570299011\\
1.32051874823252	0.764491834678252\\
1.33810222024627	0.769543300573467\\
1.35568569226002	0.774498887858643\\
1.37326916427376	0.779361968823574\\
1.39085263628751	0.784135427436613\\
};
\addplot [color=mycolor2, forget plot]
  table[row sep=crcr]{%
0.00191378283703752	0.42356349033692\\
0.0210516112074127	0.423578916649453\\
0.0401894395777879	0.423620332837007\\
0.0593272679481631	0.423687822412329\\
0.0784650963185383	0.423781587563206\\
0.0976029246889135	0.42390192207142\\
0.116740753059289	0.424049211025408\\
0.135878581429664	0.424223935336578\\
0.155016409800039	0.424426678499117\\
0.174154238170414	0.424658135353404\\
0.193292066540789	0.424919123041025\\
0.212429894911165	0.425210594518515\\
0.23156772328154	0.425533655145335\\
0.250705551651915	0.425889583032356\\
0.26984338002229	0.426279854056379\\
0.288981208392665	0.426706172739685\\
0.308119036763041	0.427170510595136\\
0.327256865133416	0.42767515409649\\
0.346394693503791	0.428222765225325\\
0.365532521874166	0.428816458686672\\
0.384670350244541	0.429459901559187\\
0.403808178614916	0.430157443649018\\
0.422946006985292	0.430914290643735\\
0.442083835355667	0.431736738185553\\
0.461221663726042	0.432632494841214\\
0.480359492096417	0.433611138760197\\
0.499497320466792	0.434684781877339\\
0.518635148837168	0.4358690636469\\
0.537772977207543	0.437184673045366\\
0.556910805577918	0.438659765068078\\
0.576048633948293	0.440334313630277\\
0.595186462318668	0.442270127353841\\
0.614324290689044	0.444573701703057\\
0.633462119059419	0.447393837346537\\
0.652599947429794	0.450748299378073\\
0.671737775800169	0.457538911919145\\
0.690875604170544	0.57167327517682\\
0.71001343254092	0.588356667958202\\
0.729151260911295	0.601656612566355\\
0.74828908928167	0.612861119345159\\
0.767426917652045	0.622783601620402\\
0.78656474602242	0.631809274579115\\
0.805702574392795	0.640146879054491\\
0.824840402763171	0.647932184052383\\
0.843978231133546	0.655262263473663\\
0.863116059503921	0.662209475752561\\
0.882253887874296	0.668829242004457\\
0.901391716244671	0.675165039390191\\
0.920529544615047	0.681251717342739\\
0.939667372985422	0.687117730578055\\
0.958805201355797	0.692786673773978\\
0.977943029726172	0.698278365073469\\
0.997080858096547	0.703609632425079\\
1.01621868646692	0.708794898291728\\
1.0353565148373	0.713846623268734\\
1.05449434320767	0.71877564833351\\
1.07363217157805	0.723591462709954\\
1.09276999994842	0.728302416196682\\
1.1119078283188	0.73291588940656\\
1.13104565668917	0.737438431673153\\
1.15018348505955	0.741875873805988\\
1.16932131342992	0.746233421054985\\
1.1884591418003	0.750515730337545\\
1.20759697017067	0.75472697483089\\
1.22673479854105	0.758870898332029\\
1.24587262691142	0.76295086126479\\
1.2650104552818	0.766969879818904\\
1.28414828365218	0.770930659405395\\
1.30328611202255	0.774835623380831\\
1.32242394039293	0.778686937813557\\
1.3415617687633	0.782486532924178\\
1.36069959713368	0.786236121722155\\
1.37983742550405	0.789937216272545\\
1.39897525387443	0.793591141957061\\
1.4181130822448	0.797199050037666\\
1.43725091061518	0.800761928785896\\
1.45638873898555	0.804280613404671\\
1.47552656735593	0.807755794939894\\
1.4946643957263	0.811188028354946\\
1.51380222409668	0.814577739921485\\
};
\addplot [color=mycolor3, forget plot]
  table[row sep=crcr]{%
0.00202621173447636	0.438822408135596\\
0.02228832907924	0.438843915875044\\
0.0425504464240036	0.438902010834573\\
0.0628125637687673	0.438996980232053\\
0.0830746811135309	0.439129467694308\\
0.103336798458295	0.439300418667295\\
0.123598915803058	0.439511099468977\\
0.143861033147822	0.43976313734913\\
0.164123150492585	0.440058577801537\\
0.184385267837349	0.440399962761546\\
0.204647385182113	0.44079043656263\\
0.224909502526876	0.441233890302068\\
0.24517161987164	0.44173516093817\\
0.265433737216404	0.442300310734795\\
0.285695854561167	0.442937029108604\\
0.305957971905931	0.443655227932133\\
0.326220089250695	0.444467941273723\\
0.346482206595458	0.445392687556337\\
0.366744323940222	0.446453814348987\\
0.387006441284985	0.447688608970591\\
0.407268558629749	0.449158834546828\\
0.427530675974513	0.450900423965803\\
0.447792793319276	0.45292165031392\\
0.46805491066404	0.459539794655863\\
0.488317028008804	0.575737995769225\\
0.508579145353567	0.591157447965634\\
0.528841262698331	0.603938477001336\\
0.549103380043094	0.614762001862323\\
0.569365497387858	0.624266576560167\\
0.589627614732622	0.632836347549325\\
0.609889732077385	0.640693772992287\\
0.630151849422149	0.647981326266161\\
0.650413966766913	0.654799812451522\\
0.670676084111676	0.661224688823972\\
0.69093820145644	0.667314187252053\\
0.711200318801204	0.673114233939411\\
0.731462436145967	0.678661737790791\\
0.751724553490731	0.683986859438316\\
0.771986670835494	0.689114602871705\\
0.792248788180258	0.694065953528141\\
0.812510905525022	0.698858710461514\\
0.832773022869785	0.70350810832491\\
0.853035140214549	0.708027291317971\\
0.873297257559313	0.712427680270872\\
0.893559374904076	0.71671926093658\\
0.91382149224884	0.720910813170277\\
0.934083609593604	0.72501009510558\\
0.954345726938367	0.729023992624241\\
0.974607844283131	0.732958641743447\\
0.994869961627894	0.736819529640237\\
1.01513207897266	0.740611578655846\\
1.03539419631742	0.744339216615193\\
1.05565631366219	0.748006436050082\\
1.07591843100695	0.751616844354714\\
1.09618054835171	0.755173706477783\\
1.11644266569648	0.758679981430073\\
1.13670478304124	0.762138353635372\\
1.156966900386	0.765551259956479\\
1.17722901773077	0.768920913074093\\
1.19749113507553	0.772249321774532\\
1.21775325242029	0.775538308605051\\
1.23801536976506	0.778789525277663\\
1.25827748710982	0.782004466139541\\
1.27853960445459	0.785184479977171\\
1.29880172179935	0.788330780379918\\
1.31906383914411	0.791444454854761\\
1.33932595648888	0.794526472856135\\
1.35958807383364	0.797577692871887\\
1.3798501911784	0.800598868687461\\
1.40011230852317	0.803590654934811\\
1.42037442586793	0.806553612019604\\
1.44063654321269	0.809488210509588\\
1.46089866055746	0.812394835058252\\
1.48116077790222	0.815273787930407\\
1.50142289524699	0.818125292190463\\
1.52168501259175	0.820949494609111\\
1.54194712993651	0.823746468339999\\
1.56220924728128	0.826516215414607\\
1.58247136462604	0.829258669100627\\
1.6027334819708	0.831973696166739\\
};
\addplot [color=mycolor4, forget plot]
  table[row sep=crcr]{%
0.00211403843273189	0.450156471191134\\
0.0232544227600508	0.450192833949903\\
0.0443948070873698	0.450292346122226\\
0.0655351914146887	0.450456547338179\\
0.0866755757420077	0.450688607697416\\
0.107815960069327	0.450993360722075\\
0.128956344396646	0.451377716901275\\
0.150096728723964	0.451851634329942\\
0.171237113051283	0.452427346034161\\
0.192377497378602	0.453114484189744\\
0.213517881705921	0.454113738694907\\
0.23465826603324	0.454541914822157\\
0.255798650360559	0.562505427038716\\
0.276939034687878	0.589855418817942\\
0.298079419015197	0.604901815704474\\
0.319219803342516	0.616745401064194\\
0.340360187669835	0.626798045059274\\
0.361500571997154	0.635574748232472\\
0.382640956324473	0.643400639291851\\
0.403781340651792	0.65049666593778\\
0.424921724979111	0.657015939290609\\
0.44606210930643	0.663067804008703\\
0.467202493633749	0.66873311748789\\
0.488342877961068	0.674073535689462\\
0.509483262288386	0.679137227578253\\
0.530623646615705	0.683962556629418\\
0.551764030943024	0.688580569340994\\
0.572904415270343	0.693016739125501\\
0.594044799597662	0.69729221999059\\
0.615185183924981	0.701424766184864\\
0.6363255682523	0.705429418958797\\
0.657465952579619	0.709319027849636\\
0.678606336906938	0.713104652251639\\
0.699746721234257	0.716795874886389\\
0.720887105561576	0.720401049417231\\
0.742027489888895	0.723927498140206\\
0.763167874216214	0.727381671350071\\
0.784308258543533	0.730769276948008\\
0.805448642870852	0.734095386700587\\
0.826589027198171	0.737364524002311\\
0.847729411525489	0.740580736855248\\
0.868869795852808	0.743747658936237\\
0.890010180180127	0.746868560991278\\
0.911150564507446	0.749946394319595\\
0.932290948834765	0.752983827745547\\
0.953431333162084	0.755983279195901\\
0.974571717489403	0.758946942781932\\
0.995712101816722	0.761876812115014\\
1.01685248614404	0.76477470044979\\
1.03799287047136	0.767642258141913\\
1.05913325479868	0.770480987821984\\
1.080273639126	0.773292257618479\\
1.10141402345332	0.776077312706899\\
1.12255440778064	0.778837285417156\\
1.14369479210795	0.781573204094305\\
1.16483517643527	0.784286000877408\\
1.18597556076259	0.786976518536344\\
1.20711594508991	0.78964551648573\\
1.22825632941723	0.792293676077948\\
1.24939671374455	0.794921605263014\\
1.27053709807187	0.797529842691109\\
1.29167748239919	0.800118861323764\\
1.31281786672651	0.80268907161043\\
1.33395825105383	0.805240824282467\\
1.35509863538114	0.807774412808052\\
1.37623901970846	0.810290075547503\\
1.39737940403578	0.812787997645974\\
1.4185197883631	0.815268312694966\\
1.43966017269042	0.817731104192177\\
1.46080055701774	0.820176406827064\\
1.48194094134506	0.822604207617357\\
1.50308132567238	0.825014446920282\\
1.5242217099997	0.827407019341053\\
1.54536209432701	0.829781774560247\\
1.56650247865433	0.832138518100856\\
1.58764286298165	0.83447701205536\\
1.60878324730897	0.836796975792611\\
1.62992363163629	0.839098086663981\\
1.65106401596361	0.841379980727903\\
1.67220440029093	0.84364225351165\\
};
\addplot [color=mycolor5, forget plot]
  table[row sep=crcr]{%
0.00218487625078874	0.699274426829742\\
0.0240336387586762	0.699376226107167\\
0.0458824012665636	0.699647835315542\\
0.067731163774451	0.70008638271155\\
0.0895799262823384	0.700687621300196\\
0.111428688790226	0.70144592772309\\
0.133277451298113	0.702354466468099\\
0.155126213806001	0.703405398937307\\
0.176974976313888	0.704590111045694\\
0.198823738821776	0.705899444426256\\
0.220672501329663	0.707323919532247\\
0.24252126383755	0.708853941532107\\
0.264370026345438	0.710479982540607\\
0.286218788853325	0.712192736300517\\
0.308067551361213	0.713983243730241\\
0.3299163138691	0.715842989647756\\
0.351765076376988	0.717763972409041\\
0.373613838884875	0.719738749164484\\
0.395462601392762	0.72176045998943\\
0.41731136390065	0.723822834359565\\
0.439160126408537	0.725920183400682\\
0.461008888916425	0.728047381124009\\
0.482857651424312	0.730199837530463\\
0.5047064139322	0.732373466083882\\
0.526555176440087	0.734564647654772\\
0.548403938947974	0.736770192650075\\
0.570252701455862	0.738987302688261\\
0.592101463963749	0.741213532863159\\
0.613950226471637	0.743446755366578\\
0.635798988979524	0.745685125011169\\
0.657647751487412	0.747927047006729\\
0.679496513995299	0.750171147192613\\
0.701345276503186	0.752416244811012\\
0.723194039011074	0.754661327816226\\
0.745042801518961	0.756905530648763\\
0.766891564026849	0.759148114356686\\
0.788740326534736	0.761388448915441\\
0.810589089042623	0.763625997579075\\
0.832437851550511	0.765860303086916\\
0.854286614058398	0.768090975548326\\
0.876135376566286	0.770317681832077\\
0.897984139074173	0.772540136294635\\
0.919832901582061	0.774758092691915\\
0.941681664089948	0.776971337130881\\
0.963530426597835	0.779179681929846\\
0.985379189105723	0.781382960269177\\
1.00722795161361	0.783581021526401\\
1.0290767141215	0.785773727201765\\
1.05092547662939	0.78796094735129\\
1.07277423913727	0.790142557454793\\
1.09462300164516	0.79231843565558\\
1.11647176415305	0.794488460316961\\
1.13832052666093	0.796652507848301\\
1.16016928916882	0.798810450759927\\
1.18201805167671	0.800962155912046\\
1.2038668141846	0.803107482928041\\
1.22571557669248	0.805246282746972\\
1.24756433920037	0.807378396293974\\
1.26941310170826	0.809503653250693\\
1.29126186421615	0.811621870910879\\
1.31311062672403	0.813732853108805\\
1.33495938923192	0.815836389210504\\
1.35680815173981	0.81793225315976\\
1.3786569142477	0.820020202572601\\
1.40050567675558	0.822099977875593\\
1.42235443926347	0.824171301484675\\
1.44420320177136	0.826233877022523\\
1.46605196427925	0.828287388573681\\
1.48790072678713	0.830331499977712\\
1.50974948929502	0.832365854161477\\
1.53159825180291	0.834390072513109\\
1.5534470143108	0.836403754300438\\
1.57529577681868	0.838406476137923\\
1.59714453932657	0.840397791506697\\
1.61899330183446	0.842377230333109\\
1.64084206434235	0.844344298631768\\
1.66269082685023	0.846298478219619\\
1.68453958935812	0.848239226508131\\
1.70638835186601	0.850165976381028\\
1.7282371143739	0.8520781361653\\
};
\addplot [color=mycolor6, forget plot]
  table[row sep=crcr]{%
0.00224279363538354	0.756480022535545\\
0.0246707299892189	0.756513265221838\\
0.0470986663430543	0.756602169816705\\
0.0695266026968897	0.756746519206151\\
0.0919545390507251	0.756946049386478\\
0.11438247540456	0.757200410243951\\
0.136810411758396	0.757509161899529\\
0.159238348112231	0.757871776248264\\
0.181666284466067	0.758287639957254\\
0.204094220819902	0.758756058227289\\
0.226522157173737	0.759276259092816\\
0.248950093527573	0.759847398143624\\
0.271378029881408	0.760468563583042\\
0.293805966235244	0.761138781548363\\
0.316233902589079	0.76185702162368\\
0.338661838942914	0.762622202478627\\
0.36108977529675	0.763433197569557\\
0.383517711650585	0.76428884084322\\
0.40594564800442	0.765187932387196\\
0.428373584358256	0.766129243975951\\
0.450801520712091	0.767111524466595\\
0.473229457065927	0.768133505003846\\
0.495657393419762	0.769193903999164\\
0.518085329773597	0.770291431855104\\
0.540513266127433	0.771424795410899\\
0.562941202481268	0.772592702091172\\
0.585369138835104	0.773793863744547\\
0.607797075188939	0.775027000163705\\
0.630225011542774	0.776290842282781\\
0.65265294789661	0.777584135051908\\
0.675080884250445	0.778905639992136\\
0.69750882060428	0.780254137437046\\
0.719936756958116	0.781628428469725\\
0.742364693311951	0.783027336565988\\
0.764792629665787	0.784449708956402\\
0.787220566019622	0.785894417720929\\
0.809648502373457	0.787360360630913\\
0.832076438727293	0.788846461753849\\
0.854504375081128	0.790351671836587\\
0.876932311434964	0.791874968482892\\
0.899360247788799	0.793415356141002\\
0.921788184142634	0.794971865916709\\
0.94421612049647	0.796543555226953\\
0.966644056850305	0.798129507308442\\
0.989071993204141	0.799728830595242\\
1.01149992955798	0.801340657978637\\
1.03392786591181	0.8029641459616\\
1.05635580226565	0.804598473719834\\
1.07878373861948	0.806242842080431\\
1.10121167497332	0.807896472428425\\
1.12363961132715	0.809558605550932\\
1.14606754768099	0.811228500427768\\
1.16849548403482	0.812905432976937\\
1.19092342038866	0.814588694762677\\
1.21335135674249	0.816277591673017\\
1.23577929309633	0.817971442573768\\
1.25820722945017	0.81966957794491\\
1.280635165804	0.821371338505292\\
1.30306310215784	0.823076073830946\\
1.32549103851167	0.824783140972194\\
1.34791897486551	0.82649190307445\\
1.37034691121934	0.828201728007392\\
1.39277484757318	0.829911987007038\\
1.41520278392701	0.831622053335254\\
1.43763072028085	0.833331300961079\\
1.46005865663468	0.835039103268363\\
1.48248659298852	0.836744831794189\\
1.50491452934235	0.838447855002686\\
1.52734246569619	0.840147537098981\\
1.54977040205003	0.841843236888173\\
1.57219833840386	0.843534306684435\\
1.5946262747577	0.845220091275547\\
1.61705421111153	0.846899926948426\\
1.63948214746537	0.848573140581422\\
1.6619100838192	0.850239048809455\\
1.68433802017304	0.851896957268261\\
1.70676595652687	0.853546159924264\\
1.72919389288071	0.855185938496833\\
1.75162182923454	0.856815561979784\\
1.77404976558838	0.858434286269199\\
};
\addplot [color=mycolor7, forget plot]
  table[row sep=crcr]{%
0.00229033277724533	0.784134322940281\\
0.0251936605496986	0.784156200607742\\
0.0480969883221519	0.784214725875638\\
0.0710003160946051	0.784309804248977\\
0.0939036438670584	0.784441339442822\\
0.116806971639512	0.78460920702676\\
0.139710299411965	0.784813250941046\\
0.162613627184418	0.785053283122741\\
0.185516954956871	0.785329083832213\\
0.208420282729325	0.785640402254536\\
0.231323610501778	0.785986957258963\\
0.254226938274231	0.786368438272714\\
0.277130266046685	0.786784506247251\\
0.300033593819138	0.787234794702786\\
0.322936921591591	0.787718910839848\\
0.345840249364044	0.788236436707686\\
0.368743577136498	0.788786930420065\\
0.391646904908951	0.789369927409158\\
0.414550232681404	0.789984941708448\\
0.437453560453857	0.790631467255762\\
0.460356888226311	0.791308979207864\\
0.483260215998764	0.792016935258174\\
0.506163543771217	0.792754776949666\\
0.52906687154367	0.793521930975307\\
0.551970199316124	0.794317810458894\\
0.574873527088577	0.795141816209524\\
0.59777685486103	0.795993337943565\\
0.620680182633483	0.796871755468423\\
0.643583510405937	0.797776439823003\\
0.66648683817839	0.798706754370292\\
0.689390165950843	0.799662055837941\\
0.712293493723297	0.8006416953037\\
0.73519682149575	0.801645019122395\\
0.758100149268203	0.802671369792178\\
0.781003477040656	0.803720086758231\\
0.80390680481311	0.804790507152355\\
0.826810132585563	0.805881966467522\\
0.849713460358016	0.806993799166813\\
0.87261678813047	0.808125339226537\\
0.895520115902923	0.809275920613747\\
0.918423443675376	0.810444877698653\\
0.941326771447829	0.811631545602753\\
0.964230099220282	0.81283526048377\\
0.987133426992736	0.814055359758713\\
1.01003675476519	0.815291182266699\\
1.03294008253764	0.816542068373191\\
1.0558434103101	0.817807360017657\\
1.07874673808255	0.81908640070675\\
1.101650065855	0.820378535455221\\
1.12455339362746	0.821683110677002\\
1.14745672139991	0.822999474028822\\
1.17036004917236	0.82432697420923\\
1.19326337694482	0.825664960715488\\
1.21616670471727	0.827012783561299\\
1.23907003248972	0.828369792958305\\
1.26197336026217	0.829735338964425\\
1.28487668803463	0.831108771102154\\
1.30778001580708	0.832489437950157\\
1.33068334357953	0.833876686711609\\
1.35358667135199	0.835269862762772\\
1.37648999912444	0.836668309185588\\
1.39939332689689	0.838071366288177\\
1.42229665466935	0.839478371117158\\
1.4451999824418	0.840888656966275\\
1.46810331021425	0.842301552885541\\
1.49100663798671	0.843716383195604\\
1.51390996575916	0.845132467012229\\
1.53681329353161	0.846549117785928\\
1.55971662130407	0.847965642862033\\
1.58261994907652	0.849381343066766\\
1.60552327684897	0.850795512324982\\
1.62842660462143	0.852207437315587\\
1.65132993239388	0.853616397170755\\
1.67423326016633	0.855021663225246\\
1.69713658793879	0.856422498822297\\
1.72003991571124	0.857818159182682\\
1.74294324348369	0.859207891343545\\
1.76584657125615	0.860590934173655\\
1.7887498990286	0.861966518471764\\
1.81165322680105	0.863333867154566\\
};
\addplot [color=mycolor8, forget plot]
  table[row sep=crcr]{%
0.0023295688688329	0.802440638441851\\
0.0256252575571619	0.802457353354112\\
0.0489209462454909	0.802502074340234\\
0.07221663493382	0.802574741084621\\
0.095512323622149	0.802675299836488\\
0.118808012310478	0.802803682861005\\
0.142103700998807	0.802959805551739\\
0.165399389687136	0.803143565903411\\
0.188695078375465	0.803354844480602\\
0.211990767063794	0.803593504555514\\
0.235286455752123	0.803859392327432\\
0.258582144440452	0.804152337193489\\
0.281877833128781	0.804472152057559\\
0.30517352181711	0.804818633670281\\
0.328469210505439	0.805191562995663\\
0.351764899193768	0.805590705600711\\
0.375060587882097	0.806015812065052\\
0.398356276570426	0.806466618407601\\
0.421651965258755	0.806942846527802\\
0.444947653947084	0.807444204658374\\
0.468243342635413	0.807970387827136\\
0.491539031323742	0.808521078325219\\
0.514834720012071	0.809095946179075\\
0.5381304087004	0.809694649623708\\
0.56142609738873	0.810316835574689\\
0.584721786077058	0.810962140096504\\
0.608017474765387	0.811630188864908\\
0.631313163453716	0.812320597621075\\
0.654608852142046	0.813032972615394\\
0.677904540830375	0.813766911038904\\
0.701200229518704	0.814522001440478\\
0.724495918207033	0.815297824127988\\
0.747791606895362	0.816093951551872\\
0.771087295583691	0.816909948669572\\
0.79438298427202	0.817745373289524\\
0.817678672960349	0.818599776393505\\
0.840974361648678	0.819472702436347\\
0.864270050337007	0.820363689622048\\
0.887565739025336	0.821272270155637\\
0.910861427713665	0.822197970470135\\
0.934157116401994	0.82314031142821\\
0.957452805090323	0.82409880849825\\
0.980748493778652	0.825072971904721\\
1.00404418246698	0.826062306752814\\
1.02733987115531	0.827066313127541\\
1.05063555984364	0.828084486167634\\
1.07393124853197	0.829116316114651\\
1.0972269372203	0.830161288337931\\
1.12052262590863	0.831218883336141\\
1.14381831459695	0.832288576716319\\
1.16711400328528	0.833369839151448\\
1.19040969197361	0.83446213631778\\
1.21370538066194	0.835564928813249\\
1.23700106935027	0.836677672058475\\
1.2602967580386	0.837799816182025\\
1.28359244672693	0.83893080589175\\
1.30688813541526	0.840070080334176\\
1.33018382410359	0.841217072944093\\
1.35347951279192	0.842371211286669\\
1.37677520148025	0.843531916894599\\
1.40007089016857	0.844698605102532\\
1.4233665788569	0.845870684882757\\
1.44666226754523	0.847047558683763\\
1.46995795623356	0.848228622275555\\
1.49325364492189	0.849413264605063\\
1.51654933361022	0.850600867665038\\
1.53984502229855	0.851790806380137\\
1.56314071098688	0.852982448514021\\
1.58643639967521	0.854175154601416\\
1.60973208836354	0.855368277909246\\
1.63302777705186	0.856561164431014\\
1.65632346574019	0.857753152918704\\
1.67961915442852	0.858943574956568\\
1.70291484311685	0.860131755081112\\
1.72621053180518	0.861317010951689\\
1.74950622049351	0.862498653575963\\
1.77280190918184	0.863675987594471\\
1.79609759787017	0.86484831162839\\
1.8193932865585	0.866014918694328\\
1.84268897524683	0.867175096689852\\
};
\addplot [color=mycolor9, forget plot]
  table[row sep=crcr]{%
0.00236230274124863	0.815858121770611\\
0.0259853301537349	0.815871723425032\\
0.0496083575662212	0.815908119152662\\
0.0732313849787075	0.815967263818406\\
0.0968544123911938	0.816049120804964\\
0.12047743980368	0.816153644909744\\
0.144100467216166	0.816280779889367\\
0.167723494628653	0.8164304579498\\
0.191346522041139	0.816602599633956\\
0.214969549453625	0.816797113835818\\
0.238592576866111	0.817013897869191\\
0.262215604278598	0.817252837566618\\
0.285838631691084	0.817513807398385\\
0.30946165910357	0.817796670606703\\
0.333084686516057	0.818101279352277\\
0.356707713928543	0.818427474871354\\
0.380330741341029	0.818775087641762\\
0.403953768753515	0.819143937556684\\
0.427576796166002	0.819533834104974\\
0.451199823578488	0.819944576556897\\
0.474822850990974	0.820375954154189\\
0.498445878403461	0.820827746303349\\
0.522068905815947	0.821299722771078\\
0.545691933228433	0.821791643880812\\
0.56931496064092	0.822303260709285\\
0.592937988053406	0.822834315282096\\
0.616561015465892	0.82338454076726\\
0.640184042878378	0.82395366166577\\
0.663807070290865	0.824541393998216\\
0.687430097703351	0.825147445486529\\
0.711053125115837	0.825771515730011\\
0.734676152528323	0.826413296374803\\
0.75829917994081	0.82707247127602\\
0.781922207353296	0.827748716651847\\
0.805545234765782	0.828441701228932\\
0.829168262178269	0.829151086378491\\
0.852791289590755	0.829876526242585\\
0.876414317003241	0.830617667850139\\
0.900037344415727	0.831374151222303\\
0.923660371828214	0.83214560946688\\
0.9472833992407	0.832931668861562\\
0.970906426653186	0.833731948925897\\
0.994529454065673	0.834546062481903\\
1.01815248147816	0.835373615703363\\
1.04177550889065	0.836214208153988\\
1.06539853630313	0.837067432814617\\
1.08902156371562	0.837932876099859\\
1.1126445911281	0.838810117864519\\
1.13626761854059	0.839698731400474\\
1.15989064595308	0.840598283424522\\
1.18351367336556	0.841508334058023\\
1.20713670077805	0.842428436799199\\
1.23075972819054	0.843358138489058\\
1.25438275560302	0.844296979272068\\
1.27800578301551	0.84524449255277\\
1.30162881042799	0.846200204949705\\
1.32525183784048	0.847163636248086\\
1.34887486525297	0.848134299352827\\
1.37249789266545	0.849111700243628\\
1.39612092007794	0.850095337933898\\
1.41974394749043	0.851084704435621\\
1.44336697490291	0.85207928473204\\
1.4669900023154	0.853078556760464\\
1.49061302972788	0.854081991407463\\
1.51423605714037	0.855089052518854\\
1.53785908455286	0.856099196926975\\
1.56148211196534	0.857111874497805\\
1.58510513937783	0.85812652820063\\
1.60872816679032	0.85914259420291\\
1.6323511942028	0.860159501993164\\
1.65597422161529	0.861176674534615\\
1.67959724902778	0.862193528452399\\
1.70322027644026	0.863209474257141\\
1.72684330385275	0.86422391660756\\
1.75046633126523	0.865236254614814\\
1.77408935867772	0.866245882191118\\
1.79771238609021	0.867252188445054\\
1.82133541350269	0.86825455812583\\
1.84495844091518	0.869252372118532\\
1.86858146832767	0.870245007992159\\
};
\addplot [color=mycolor10, forget plot]
  table[row sep=crcr]{%
0.00238994171149858	0.826235710680707\\
0.0262893588264844	0.826247157750254\\
0.0501887759414702	0.826277791313225\\
0.0740881930564561	0.826327574777016\\
0.0979876101714419	0.826396480253813\\
0.121887027286428	0.826484473837388\\
0.145786444401414	0.826591513468291\\
0.169685861516399	0.826717548466991\\
0.193585278631385	0.82686251940323\\
0.217484695746371	0.827026358068662\\
0.241384112861357	0.827208987491143\\
0.265283529976343	0.827410321969903\\
0.289182947091328	0.827630267123155\\
0.313082364206314	0.827868719944323\\
0.3369817813213	0.828125568864714\\
0.360881198436286	0.828400693821424\\
0.384780615551272	0.82869396632953\\
0.408680032666258	0.829005249557857\\
0.432579449781243	0.829334398407765\\
0.456478866896229	0.829681259594305\\
0.480378284011215	0.830045671729253\\
0.504277701126201	0.830427465405516\\
0.528177118241187	0.830826463282345\\
0.552076535356173	0.831242480170925\\
0.575975952471158	0.831675323119789\\
0.599875369586144	0.832124791499589\\
0.62377478670113	0.832590677086767\\
0.647674203816116	0.833072764145662\\
0.671573620931102	0.833570829508589\\
0.695473038046088	0.834084642653492\\
0.719372455161073	0.83461396577877\\
0.743271872276059	0.835158553874879\\
0.767171289391045	0.8357181547924\\
0.791070706506031	0.836292509306213\\
0.814970123621017	0.836881351175567\\
0.838869540736003	0.837484407199591\\
0.862768957850988	0.838101397268439\\
0.886668374965974	0.838732034409574\\
0.91056779208096	0.839376024829126\\
0.934467209195946	0.840033067948351\\
0.958366626310932	0.840702856435137\\
0.982266043425918	0.841385076230582\\
1.0061654605409	0.842079406570776\\
1.03006487765589	0.842785520003888\\
1.05396429477087	0.843503082402851\\
1.07786371188586	0.844231752973857\\
1.10176312900085	0.844971184261077\\
1.12566254611583	0.845721022147971\\
1.14956196323082	0.846480905855748\\
1.1734613803458	0.847250467939475\\
1.19736079746079	0.848029334282528\\
1.22126021457578	0.848817124090095\\
1.24515963169076	0.849613449882526\\
1.26905904880575	0.850417917489419\\
1.29295846592073	0.851230126045401\\
1.31685788303572	0.852049667988636\\
1.3407573001507	0.852876129063217\\
1.36465671726569	0.853709088326625\\
1.38855613438068	0.854548118163534\\
1.41245555149566	0.855392784307335\\
1.43635496861065	0.85624264587081\\
1.46025438572563	0.857097255387462\\
1.48415380284062	0.857956158865087\\
1.50805321995561	0.858818895853184\\
1.53195263707059	0.859684999525933\\
1.55585205418558	0.86055399678243\\
1.57975147130056	0.861425408365987\\
1.60365088841555	0.862298749004252\\
1.62755030553053	0.863173527571995\\
1.65144972264552	0.864049247278324\\
1.67534913976051	0.864925405880118\\
1.69924855687549	0.86580149592361\\
1.72314797399048	0.86667700501555\\
1.74704739110546	0.867551416125825\\
1.77094680822045	0.868424207923022\\
1.79484622533544	0.869294855144406\\
1.81874564245042	0.87016282900165\\
1.84264505956541	0.871027597623485\\
1.86654447668039	0.871888626536329\\
1.89044389379538	0.872745379183631\\
};
\addplot [color=black, dotted, forget plot]
  table[row sep=crcr]{%
0.0014369690159404	0.271412091488527\\
0.0158066591753444	0.271413160668384\\
0.0301763493347484	0.271416026824873\\
0.0445460394941523	0.271420728848129\\
0.0589157296535563	0.27142733048575\\
0.0732854198129603	0.271435920732016\\
0.0876551099723643	0.271446614484203\\
0.102024800131768	0.271459553389496\\
0.116394490291172	0.271474906869103\\
0.130764180450576	0.271492873312369\\
0.14513387060998	0.27151368143382\\
0.159503560769384	0.271537591785463\\
0.173873250928788	0.271564898415916\\
0.188242941088192	0.271595930667229\\
0.202612631247596	0.271631055099707\\
0.216982321407	0.27167067753458\\
0.231352011566404	0.271715245204189\\
0.245721701725808	0.271765248999348\\
0.260091391885212	0.271821225803885\\
0.274461082044616	0.271883760906996\\
0.28883077220402	0.271953490485098\\
0.303200462363424	0.272031104146311\\
0.317570152522828	0.272117347532636\\
0.331939842682232	0.272213024977301\\
0.346309532841636	0.272319002217761\\
0.36067922300104	0.272436209168329\\
0.375048913160444	0.27256564276062\\
0.389418603319848	0.272708369864735\\
0.403788293479252	0.272865530309594\\
0.418157983638656	0.273038340026971\\
0.43252767379806	0.273228094350702\\
0.446897363957464	0.273436171510219\\
0.461267054116868	0.27366403636617\\
0.475636744276272	0.273913244445421\\
0.490006434435676	0.274185446343434\\
0.50437612459508	0.274482392573944\\
0.518745814754484	0.274805938959398\\
0.533115504913888	0.275158052670963\\
0.547485195073292	0.275540819044521\\
0.561854885232696	0.275956449319545\\
0.5762245753921	0.276407289471631\\
0.590594265551504	0.276895830337809\\
0.604963955710908	0.277424719267597\\
0.619333645870312	0.277996773573585\\
0.633703336029716	0.278614996105096\\
0.64807302618912	0.279282593329524\\
0.662442716348524	0.280002996381601\\
0.676812406507928	0.280779885635216\\
0.691182096667332	0.281617219471067\\
0.705551786826736	0.282519268063788\\
0.71992147698614	0.283490653204177\\
0.734291167145544	0.284536395419247\\
0.748660857304948	0.285661969973605\\
0.763030547464352	0.286873373755798\\
0.777400237623756	0.288177205608984\\
0.791769927783159	0.289580763407821\\
0.806139617942564	0.291092162187167\\
0.820509308101967	0.292720479001314\\
0.834878998261371	0.294475932096437\\
0.849248688420775	0.296370104657383\\
0.863618378580179	0.298416227218296\\
0.877988068739583	0.300629538394601\\
0.892357758898987	0.303027751848872\\
0.906727449058391	0.305631669904414\\
0.921097139217795	0.308466003606003\\
0.935466829377199	0.311560489900134\\
0.949836519536603	0.314951447286709\\
0.964206209696007	0.318683997438171\\
0.978575899855411	0.322815332778936\\
0.992945590014815	0.327419693256771\\
1.00731528017422	0.332596272811805\\
1.02168497033362	0.338482452317151\\
1.03605466049303	0.34527746536013\\
1.05042435065243	0.35328858204674\\
1.06479404081184	0.363032746508498\\
1.07916373097124	0.375503532560922\\
1.09353342113064	0.393111922784879\\
1.10790311129005	0.426201018387755\\
1.12227280144945	0.594672344794295\\
1.13664249160886	0.648358551471024\\
};
\addplot [color=black, dotted, forget plot]
  table[row sep=crcr]{%
0.00154772010087882	0.326330144123744\\
0.017024921109667	0.326341934577983\\
0.0325021221184552	0.3263734058186\\
0.0479793231272434	0.326424583346118\\
0.0634565241360315	0.326495514147186\\
0.0789337251448197	0.326586264314246\\
0.0944109261536079	0.326696918914126\\
0.109888127162396	0.326827582207003\\
0.125365328171184	0.326978377992078\\
0.140842529179972	0.327149450043605\\
0.156319730188761	0.327340962630575\\
0.171796931197549	0.327553101120944\\
0.187274132206337	0.3277860726742\\
0.202751333215125	0.328040107027632\\
0.218228534223913	0.328315457382935\\
0.233705735232702	0.328612401400959\\
0.24918293624149	0.328931242313653\\
0.264660137250278	0.329272310163637\\
0.280137338259066	0.329635963183335\\
0.295614539267854	0.330022589327367\\
0.311091740276642	0.330432607973859\\
0.326568941285431	0.330866471812629\\
0.342046142294219	0.331324668940817\\
0.357523343303007	0.331807725189596\\
0.373000544311795	0.332316206709145\\
0.388477745320583	0.332850722843222\\
0.403954946329371	0.333411929329531\\
0.41943214733816	0.334000531867829\\
0.434909348346948	0.334617290104499\\
0.450386549355736	0.335263022090338\\
0.465863750364524	0.335938609277937\\
0.481340951373312	0.336645002136493\\
0.496818152382101	0.337383226475721\\
0.512295353390889	0.338154390587188\\
0.527772554399677	0.338959693331699\\
0.543249755408465	0.339800433326019\\
0.558726956417253	0.3406780194126\\
0.574204157426041	0.341593982633323\\
0.58968135843483	0.342549989974754\\
0.605158559443618	0.343547860210335\\
0.620635760452406	0.344589582237856\\
0.636112961461194	0.345677336402834\\
0.651590162469982	0.346813519416141\\
0.667067363478771	0.34800077362563\\
0.682544564487559	0.349242021597825\\
0.698021765496347	0.350540507222691\\
0.713498966505135	0.351899844894103\\
0.728976167513923	0.353324078772126\\
0.744453368522711	0.354817754745815\\
0.7599305695315	0.356386008552904\\
0.775407770540288	0.358034674673595\\
0.790884971549076	0.359770422248155\\
0.806362172557864	0.361600926601386\\
0.821839373566652	0.363535088351902\\
0.837316574575441	0.365583317122745\\
0.852793775584229	0.36775790451281\\
0.868270976593017	0.370073522865428\\
0.883748177601805	0.372547905321821\\
0.899225378610593	0.375202793783322\\
0.914702579619382	0.37806529425684\\
0.93017978062817	0.38116987219626\\
0.945656981636958	0.384561392177379\\
0.961134182645746	0.388299941503098\\
0.976611383654534	0.392468882696729\\
0.992088584663322	0.397189216346412\\
1.00756578567211	0.402647636078686\\
1.0230429866809	0.409158860830326\\
1.03852018768969	0.417332568689767\\
1.05399738869848	0.428672994233205\\
1.06947458970726	0.449704146993473\\
1.08495179071605	0.570334854019233\\
1.10042899172484	0.601872703426357\\
1.11590619273363	0.624233675234316\\
1.13138339374242	0.642592016155631\\
1.1468605947512	0.658572150621701\\
1.16233779575999	0.672934844849823\\
1.17781499676878	0.686099206299196\\
1.19329219777757	0.698321136334021\\
1.20876939878636	0.709768877225193\\
1.22424659979515	0.72055901537714\\
};
\addplot [color=black, dotted, forget plot]
  table[row sep=crcr]{%
0.00252923911170456	0.869772166357725\\
0.0278216302287501	0.869775707634935\\
0.0531140213457957	0.869785190305831\\
0.0784064124628412	0.869800602188821\\
0.103698803579887	0.8698219348326\\
0.128991194696932	0.869849178224591\\
0.154283585813978	0.869882320013939\\
0.179575976931023	0.869921345333584\\
0.204868368048069	0.869966236741625\\
0.230160759165115	0.870016974198082\\
0.25545315028216	0.870073535054945\\
0.280745541399206	0.870135894052098\\
0.306037932516251	0.870204023316245\\
0.331330323633297	0.870277892361538\\
0.356622714750342	0.870357468091344\\
0.381915105867388	0.870442714800808\\
0.407207496984434	0.870533594180092\\
0.432499888101479	0.870630065318192\\
0.457792279218525	0.870732084707276\\
0.48308467033557	0.870839606247607\\
0.508377061452616	0.870952581252975\\
0.533669452569661	0.871070958456691\\
0.558961843686707	0.871194684018173\\
0.584254234803752	0.871323701530146\\
0.609546625920798	0.871457952026457\\
0.634839017037844	0.871597373990619\\
0.660131408154889	0.871741903365042\\
0.685423799271935	0.871891473561041\\
0.71071619038898	0.872046015469656\\
0.736008581506026	0.872205457473336\\
0.761300972623071	0.872369725458529\\
0.786593363740117	0.872538742829234\\
0.811885754857162	0.872712430521589\\
0.837178145974208	0.87289070701952\\
0.862470537091254	0.873073488371541\\
0.887762928208299	0.873260688208754\\
0.913055319325345	0.873452217764111\\
0.93834771044239	0.873647985893013\\
0.963640101559436	0.87384789909531\\
0.988932492676481	0.874051861538767\\
1.01422488379353	0.874259775084082\\
1.03951727491057	0.874471539311507\\
1.06480966602762	0.874687051549173\\
1.09010205714466	0.874906206903174\\
1.11539444826171	0.875128898289491\\
1.14068683937875	0.875355016467853\\
1.1659792304958	0.875584450077627\\
1.19127162161285	0.875817085675415\\
1.21656401272989	0.876052807775574\\
1.24185640384694	0.876291498892306\\
1.26714879496398	0.876533039584277\\
1.29244118608103	0.876777308501612\\
1.31773357719807	0.877024182435346\\
1.34302596831512	0.877273536369412\\
1.36831835943216	0.877525243535218\\
1.39361075054921	0.877779175468909\\
1.41890314166626	0.878035202071352\\
1.4441955327833	0.878293191670926\\
1.46948792390035	0.878553011089171\\
1.49478031501739	0.878814525709338\\
1.52007270613444	0.879077599547905\\
1.54536509725148	0.879342095329094\\
1.57065748836853	0.879607874562427\\
1.59594987948557	0.879874797623355\\
1.62124227060262	0.880142723836979\\
1.64653466171967	0.880411511564885\\
1.67182705283671	0.880681018295103\\
1.69711944395376	0.880951100735181\\
1.7224118350708	0.881221614908379\\
1.74770422618785	0.88149241625295\\
1.77299661730489	0.88176335972448\\
1.79828900842194	0.882034299901276\\
1.82358139953898	0.8823050910927\\
1.84887379065603	0.882575587450445\\
1.87416618177308	0.88284564308263\\
1.89945857289012	0.88311511217065\\
1.92475096400717	0.883383849088704\\
1.95004335512421	0.883651708525844\\
1.97533574624126	0.883918545610462\\
2.0006281373583	0.884184216037045\\
};
\addplot [color=black, dashed, forget plot]
  table[row sep=crcr]{%
0	0.889893658725525\\
2.0863919069197	0.889893658725525\\
};
\end{axis}
%
%
\draw[->,-stealth, line width=1.25pt] (2,2) to [bend right=-30] (1,3.35);

\end{tikzpicture}%

%% file: Figs/Fig8a_ColourMap_Hirotsu_Swell_NoFront.tex
\begin{tikzpicture}
\begin{axis}[axis on top,  
thick, 
width=0.45\textwidth, 
height=0.4\textwidth,
/pgf/number format/.cd,
                fixed,
enlargelimits=false, 
colorbar, 
point meta min = 0.25, 
point meta max = 1,
xmin=0,xmax=0.2,ymin=0,ymax=3, 
colorbar style = { 
width = 0.3cm,
thick,
black,
title = {$\phi$},
title style = {overlay,yshift = -3pt},
at={(1.05,1)}},
xlabel = {$t$},
ylabel = {$r$},
legend style = {
draw=none,
fill=none,
font=\scriptsize},]
\addplot[forget plot] graphics [xmin=0,xmax=0.2,ymin=0,ymax=3] {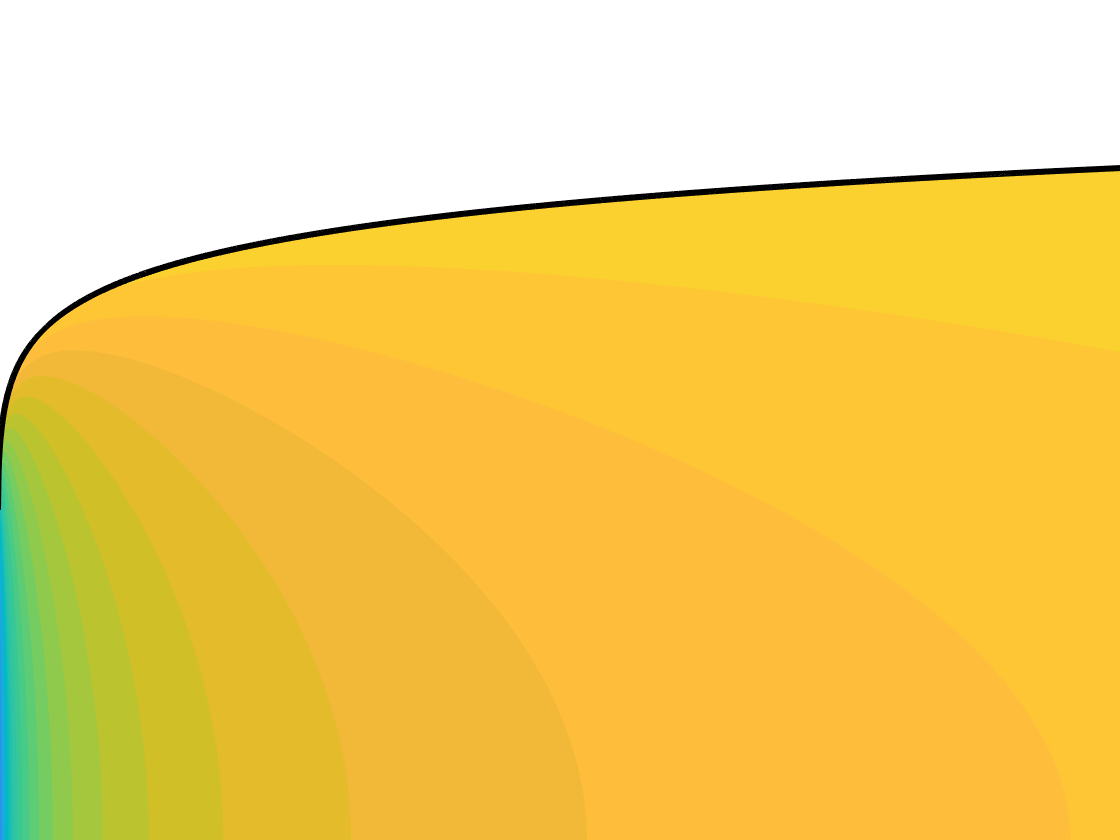};
%
\end{axis}
\end{tikzpicture}

%% file: Figs/Fig8b_Porosity_Hirotsu_Swell_NoFront.tex
%
%
\definecolor{mycolor1}{rgb}{0.12500,0.07812,0.04975}%
\definecolor{mycolor2}{rgb}{0.25000,0.15624,0.09950}%
\definecolor{mycolor3}{rgb}{0.37500,0.23436,0.14925}%
\definecolor{mycolor4}{rgb}{0.50000,0.31248,0.19900}%
\definecolor{mycolor5}{rgb}{0.62500,0.39060,0.24875}%
\definecolor{mycolor6}{rgb}{0.75000,0.46872,0.29850}%
\definecolor{mycolor7}{rgb}{0.87500,0.54684,0.34825}%
\definecolor{mycolor8}{rgb}{1.00000,0.62496,0.39800}%
\definecolor{mycolor9}{rgb}{1.00000,0.70308,0.44775}%
\definecolor{mycolor10}{rgb}{1.00000,0.78120,0.49750}%
\begin{tikzpicture}

\begin{axis}[%
thick, 
width=0.48\textwidth, 
height=0.4\textwidth,
xmin=0,xmax=3,ymin=0,ymax=1, 
xlabel = {$r$},
ylabel = {$\phi$},
yticklabel style = overlay]
\addplot [color=black, forget plot]
  table[row sep=crcr]{%
0.00289720200258154	0.357484371523868\\
0.0318692220283969	0.357484371523868\\
0.0608412420542123	0.357484371523868\\
0.0898132620800277	0.357484371523868\\
0.118785282105843	0.357484371523868\\
0.147757302131658	0.357484371523868\\
0.176729322157474	0.357484371523868\\
0.205701342183289	0.357484371523868\\
0.234673362209105	0.357484371523868\\
0.26364538223492	0.357484371523868\\
0.292617402260735	0.357484371523868\\
0.321589422286551	0.357484371523868\\
0.350561442312366	0.357484371523868\\
0.379533462338181	0.357484371523868\\
0.408505482363997	0.357484371523868\\
0.437477502389812	0.357484371523868\\
0.466449522415628	0.357484371523868\\
0.495421542441443	0.357484371523868\\
0.524393562467258	0.357484371523868\\
0.553365582493074	0.357484371523868\\
0.582337602518889	0.357484371523868\\
0.611309622544704	0.357484371523868\\
0.64028164257052	0.357484371523868\\
0.669253662596335	0.357484371523868\\
0.698225682622151	0.357484371523868\\
0.727197702647966	0.357484371523868\\
0.756169722673781	0.357484371523868\\
0.785141742699597	0.357484371523868\\
0.814113762725412	0.357484371523868\\
0.843085782751227	0.357484371523868\\
0.872057802777043	0.357484371523868\\
0.901029822802858	0.357484371523868\\
0.930001842828674	0.357484371523868\\
0.958973862854489	0.357484371523868\\
0.987945882880304	0.357484371523868\\
1.01691790290612	0.357484371523868\\
1.04588992293193	0.357484371523868\\
1.07486194295775	0.357484371523868\\
1.10383396298357	0.357484371523868\\
1.13280598300938	0.357484371523868\\
};
\addplot [color=mycolor1, forget plot]
  table[row sep=crcr]{%
0.00428514551592536	0.712547259782213\\
0.047136600675179	0.712621257847925\\
0.0899880558344326	0.712818793005106\\
0.132839510993686	0.713140370993309\\
0.17569096615294	0.713586825381872\\
0.218542421312193	0.714159323381303\\
0.261393876471447	0.714859377577944\\
0.304245331630701	0.715688861825074\\
0.347096786789954	0.716650031429406\\
0.389948241949208	0.717745548087504\\
0.432799697108461	0.718978510194744\\
0.475651152267715	0.720352489323004\\
0.518502607426969	0.721871573870174\\
0.561354062586222	0.723540421140897\\
0.604205517745476	0.725364319441039\\
0.647056972904729	0.727349262180897\\
0.689908428063983	0.729502036514522\\
0.732759883223237	0.731830329737172\\
0.77561133838249	0.734342857578821\\
0.818462793541744	0.737049519753042\\
0.861314248700998	0.739961589768153\\
0.904165703860251	0.743091948257659\\
0.947017159019505	0.746455372201004\\
0.989868614178758	0.750068896776634\\
1.03272006933801	0.753952272820071\\
1.07557152449727	0.758128551891228\\
1.11842297965652	0.762624844288116\\
1.16127443481577	0.76747331542536\\
1.20412588997503	0.772712516915005\\
1.24697734513428	0.778389197463182\\
1.28982880029353	0.784560817733903\\
1.33268025545279	0.791299125249704\\
1.37553171061204	0.798695372844419\\
1.41838316577129	0.80686817034554\\
1.46123462093055	0.815975711485199\\
1.5040860760898	0.826235559926638\\
1.54693753124905	0.837957995858149\\
1.58978898640831	0.851604207544857\\
1.63264044156756	0.867887299758694\\
1.67549189672682	0.887908239794328\\
};
\addplot [color=mycolor2, forget plot]
  table[row sep=crcr]{%
0.00455779977438598	0.762443498736815\\
0.0501357975182458	0.762505651193054\\
0.0957137952621055	0.762671571003272\\
0.141291793005965	0.762941667446941\\
0.186869790749825	0.763316618272847\\
0.232447788493685	0.763797372723416\\
0.278025786237545	0.764385160596833\\
0.323603783981404	0.765081504761052\\
0.369181781725264	0.765888237078917\\
0.414759779469124	0.766807518064643\\
0.460337777212984	0.767841860737356\\
0.505915774956844	0.768994159271468\\
0.551493772700703	0.770267723198317\\
0.597071770444563	0.771666318102668\\
0.642649768188423	0.773194213993817\\
0.688227765932283	0.774856242829981\\
0.733805763676142	0.776657867057054\\
0.779383761420002	0.778605261517113\\
0.824961759163862	0.780705411727111\\
0.870539756907722	0.782966232378673\\
0.916117754651582	0.78539671104303\\
0.961695752395441	0.788007083591114\\
1.0072737501393	0.790809049918012\\
1.05285174788316	0.79381604142838\\
1.09842974562702	0.797043555745605\\
1.14400774337088	0.800509579782332\\
1.18958574111474	0.804235130466886\\
1.2351637388586	0.808244954327578\\
1.28074173660246	0.81256844480319\\
1.32631973434632	0.817240862798747\\
1.37189773209018	0.82230498688233\\
1.41747572983404	0.82781338318154\\
1.4630537275779	0.833831585338952\\
1.50863172532176	0.84044263324886\\
1.55420972306562	0.847753664037643\\
1.59978772080948	0.855905594537417\\
1.64536571855334	0.865087263148394\\
1.6909437162972	0.875554874166336\\
1.73652171404106	0.887651848888846\\
1.78209971178492	0.901792116369156\\
};
\addplot [color=mycolor3, forget plot]
  table[row sep=crcr]{%
0.00472593040033928	0.788121970848947\\
0.0519852344037321	0.788177801294114\\
0.099244538407125	0.788326849470802\\
0.146503842410518	0.788569473052731\\
0.193763146413911	0.788906266813999\\
0.241022450417304	0.789338063952492\\
0.288281754420696	0.789865943664526\\
0.335541058424089	0.790491241804455\\
0.382800362427482	0.791215564473441\\
0.430059666430875	0.79204080478025\\
0.477318970434268	0.792969163151512\\
0.524578274437661	0.794003171681323\\
0.571837578441053	0.795145723135669\\
0.619096882444446	0.796400105378705\\
0.666356186447839	0.797770042175429\\
0.713615490451232	0.799259741560674\\
0.760874794454625	0.80087395326292\\
0.808134098458018	0.802618037053763\\
0.855393402461411	0.804498044387962\\
0.902652706464803	0.806520816343204\\
0.949912010468196	0.808694101716538\\
0.997171314471589	0.811026700260691\\
1.04443061847498	0.813528637554136\\
1.09168992247837	0.816211380045761\\
1.13894922648177	0.819088101617051\\
1.18620853048516	0.822174016880791\\
1.23346783448855	0.825486801853502\\
1.28072713849195	0.829047130288309\\
1.32798644249534	0.832879364851466\\
1.37524574649873	0.837012457948835\\
1.42250505050212	0.841481139448951\\
1.46976435450552	0.846327500529511\\
1.51702365850891	0.851603127160639\\
1.5642829625123	0.857371993295824\\
1.6115422665157	0.863714379266671\\
1.65880157051909	0.870732071786442\\
1.70606087452248	0.878554789510842\\
1.75332017852587	0.887346314630206\\
1.80057948252927	0.897303253884637\\
1.84783878653266	0.908620050515785\\
};
\addplot [color=mycolor4, forget plot]
  table[row sep=crcr]{%
0.00484866641629309	0.804864847651788\\
0.053335330579224	0.804916457576029\\
0.101821994742155	0.805054243036241\\
0.150308658905086	0.805278526823775\\
0.198795323068017	0.805589847816716\\
0.247281987230948	0.805988961051724\\
0.295768651393879	0.806476844265781\\
0.34425531555681	0.807054707280174\\
0.392741979719741	0.807724003984127\\
0.441228643882672	0.808486447106142\\
0.489715308045602	0.809344026089093\\
0.538201972208533	0.810299028482804\\
0.586688636371464	0.811354065373523\\
0.635175300534395	0.812512101495254\\
0.683661964697326	0.813776490822027\\
0.732148628860257	0.815151018631924\\
0.780635293023188	0.81663995127472\\
0.829121957186119	0.818248095180941\\
0.87760862134905	0.819980867041447\\
0.926095285511981	0.821844377590897\\
0.974581949674912	0.823845532084075\\
1.02306861383784	0.82599215141204\\
1.07155527800077	0.82829311893761\\
1.1200419421637	0.830758559635103\\
1.16852860632664	0.833400060134864\\
1.21701527048957	0.836230940988951\\
1.2655019346525	0.839266596151664\\
1.31398859881543	0.842524919660203\\
1.36247526297836	0.846026846263252\\
1.41096192714129	0.849797041815455\\
1.45944859130422	0.853864791108095\\
1.50793525546715	0.858265145368213\\
1.55642191963008	0.863040406863084\\
1.60490858379301	0.868242035724034\\
1.65339524795595	0.873933038412959\\
1.70188191211888	0.880190761974016\\
1.75036857628181	0.88710955338864\\
1.79885524044474	0.894801320275649\\
1.84734190460767	0.903387900862457\\
1.8958285687706	0.912967995939753\\
};
\addplot [color=mycolor5, forget plot]
  table[row sep=crcr]{%
0.0049456446392817	0.817049937494701\\
0.0544020910320987	0.817098417454953\\
0.103858537424916	0.817227850767067\\
0.153314983817733	0.81743853398723\\
0.20277143021055	0.817730964063876\\
0.252227876603367	0.818105837409682\\
0.301684322996184	0.818564055651093\\
0.351140769389001	0.819106734040853\\
0.400597215781818	0.819735212220595\\
0.450053662174635	0.820451067479026\\
0.499510108567452	0.821256130774828\\
0.548966554960269	0.822152505880015\\
0.598423001353086	0.823142592090058\\
0.647879447745903	0.82422911105314\\
0.69733589413872	0.825415138399739\\
0.746792340531537	0.826704141012669\\
0.796248786924354	0.828100020975758\\
0.845705233317171	0.829607167488186\\
0.895161679709988	0.831230518346329\\
0.944618126102805	0.832975632996048\\
0.994074572495622	0.834848779672382\\
1.04353101888844	0.836857039806178\\
1.09298746528126	0.839008433735898\\
1.14244391167407	0.841312072880184\\
1.19190035806689	0.843778344985163\\
1.24135680445971	0.846419140965119\\
1.29081325085252	0.849248134335003\\
1.34026969724534	0.852281127431763\\
1.38972614363816	0.855536482660505\\
1.43918259003097	0.859035661879458\\
1.48863903642379	0.862803902363745\\
1.53809548281661	0.866871062119117\\
1.58755192920943	0.871272666524508\\
1.63700837560224	0.876051171514824\\
1.68646482199506	0.881257398150851\\
1.73592126838788	0.886951919870442\\
1.78537771478069	0.893205722476252\\
1.83483416117351	0.900098277258195\\
1.88429060756633	0.907708267614989\\
1.93374705395914	0.916085694566434\\
};
\addplot [color=mycolor6, forget plot]
  table[row sep=crcr]{%
0.00502592217314662	0.826507415861141\\
0.0552851439046128	0.826553426579218\\
0.105544365636079	0.826676271059408\\
0.155803587367545	0.82687622487217\\
0.206062809099011	0.827153751495733\\
0.256322030830477	0.827509500551614\\
0.306581252561944	0.827944312911169\\
0.35684047429341	0.828459228324085\\
0.407099696024876	0.829055495196539\\
0.457358917756342	0.829734582629742\\
0.507618139487808	0.830498194950241\\
0.557877361219274	0.831348289041089\\
0.608136582950741	0.832287094861547\\
0.658395804682207	0.833317139633628\\
0.708655026413673	0.834441276282419\\
0.758914248145139	0.835662716850286\\
0.809173469876605	0.836985071769145\\
0.859432691608071	0.838412396079047\\
0.909691913339538	0.839949243936641\\
0.959951135071004	0.841600733077907\\
1.01021035680247	0.843372621304655\\
1.06046957853394	0.845271397577382\\
1.1107288002654	0.847304390948186\\
1.16098802199687	0.849479901393788\\
1.21124724372833	0.851807357654153\\
1.2615064654598	0.854297508494458\\
1.31176568719127	0.856962655428287\\
1.36202490892273	0.859816936877721\\
1.4122841306542	0.862876675918544\\
1.46254335238567	0.866160805854316\\
1.51280257411713	0.869691389033287\\
1.5630617958486	0.873494242492888\\
1.61332101758006	0.877599674339928\\
1.66358023931153	0.882043306114057\\
1.713839461043	0.886866883303132\\
1.76409868277446	0.892118800728894\\
1.81435790450593	0.897853660862868\\
1.86461712623739	0.904129255650234\\
1.91487634796886	0.910997353710309\\
1.96513556970033	0.918480812897068\\
};
\addplot [color=mycolor7, forget plot]
  table[row sep=crcr]{%
0.0050944530572284	0.834166039719723\\
0.0560389836295124	0.834210021208585\\
0.106983514201796	0.83432745110934\\
0.15792804477408	0.834518587519027\\
0.208872575346364	0.83478386609409\\
0.259817105918648	0.835123897565234\\
0.310761636490932	0.835539472302809\\
0.361706167063216	0.836031567288491\\
0.4126506976355	0.836601355069436\\
0.463595228207784	0.83725021477644\\
0.514539758780068	0.837979745405849\\
0.565484289352352	0.838791781635617\\
0.616428819924636	0.839688412514593\\
0.66737335049692	0.84067200344186\\
0.718317881069204	0.841745221945434\\
0.769262411641488	0.842911067881401\\
0.820206942213772	0.844172908811206\\
0.871151472786056	0.845534521482675\\
0.92209600335834	0.847000140547749\\
0.973040533930624	0.848574515906562\\
1.02398506450291	0.850262980386177\\
1.07492959507519	0.85207152985766\\
1.12587412564748	0.854006918384347\\
1.17681865621976	0.856076771596124\\
1.22776318679204	0.858289722215364\\
1.27870771736433	0.860655572528281\\
1.32965224793661	0.863185489582536\\
1.3805967785089	0.865892239922236\\
1.43154130908118	0.8687904715373\\
1.48248583965346	0.871897050920593\\
1.53343037022575	0.875231461621702\\
1.58437490079803	0.878816265156388\\
1.63531943137032	0.882677610678441\\
1.6862639619426	0.886845746068764\\
1.73720849251488	0.891355408085066\\
1.78815302308717	0.896245806723108\\
1.83909755365945	0.901559573631743\\
1.89004208423174	0.907339330740667\\
1.94098661480402	0.913619147007753\\
1.9919311453763	0.920405838927759\\
};
\addplot [color=mycolor8, forget plot]
  table[row sep=crcr]{%
0.00515425281787746	0.840557712523945\\
0.056696780996652	0.840599977242544\\
0.108239309175427	0.840712826389102\\
0.159781837354201	0.84089650309688\\
0.211324365532976	0.841151419228795\\
0.26286689371175	0.8414781522587\\
0.314409421890525	0.841877449369687\\
0.365951950069299	0.842350233861071\\
0.417494478248074	0.842897613394706\\
0.469037006426849	0.843520890137043\\
0.520579534605623	0.844221572969645\\
0.572122062784398	0.845001392005627\\
0.623664590963172	0.845862315709808\\
0.675207119141947	0.846806570987437\\
0.726749647320721	0.847836666685015\\
0.778292175499496	0.84895542104097\\
0.829834703678271	0.850165993737661\\
0.881377231857045	0.85147192334413\\
0.93291976003582	0.852877171107075\\
0.984462288214594	0.85438617225188\\
1.03600481639337	0.85600389620416\\
1.08754734457214	0.857735917443085\\
1.13908987275092	0.859588499058364\\
1.19063240092969	0.86156869150815\\
1.24217492910847	0.86368444956244\\
1.29371745728724	0.865944770945664\\
1.34525998546602	0.868359860706445\\
1.39680251364479	0.870941325710556\\
1.44834504182357	0.873702403595606\\
1.49988757000234	0.876658229468359\\
1.55143009818111	0.879826140411351\\
1.60297262635989	0.883226010180661\\
1.65451515453866	0.886880589703814\\
1.70605768271744	0.89081579391497\\
1.75760021089621	0.895060803778347\\
1.80914273907499	0.899647708086495\\
1.86068526725376	0.904610124594229\\
1.91222779543254	0.909979696108144\\
1.96377032361131	0.915778396909861\\
2.01531285179009	0.922003197833935\\
};
\addplot [color=mycolor9, forget plot]
  table[row sep=crcr]{%
0.00520729787902474	0.846013562356719\\
0.0572802766692722	0.846054342764057\\
0.10935325545952	0.846163231476213\\
0.161426234249767	0.846340458543141\\
0.213499213040015	0.846586415047226\\
0.265572191830262	0.846901649419272\\
0.317645170620509	0.847286871126223\\
0.369718149410757	0.847742956583326\\
0.421791128201004	0.848270956780783\\
0.473864106991252	0.84887210665918\\
0.525937085781499	0.849547836383037\\
0.578010064571747	0.850299784721392\\
0.630083043361994	0.851129814797577\\
0.682156022152242	0.852040032528234\\
0.734229000942489	0.853032808138664\\
0.786301979732737	0.854110801220833\\
0.838374958522984	0.855276989894802\\
0.890447937313231	0.856534704747297\\
0.942520916103479	0.857887668356241\\
0.994593894893726	0.859340041371255\\
1.04666687368397	0.860896476311193\\
1.09873985247422	0.862562180463639\\
1.15081283126447	0.864342989528289\\
1.20288581005472	0.866245453931218\\
1.25495878884496	0.868276940033248\\
1.30703176763521	0.870445748723884\\
1.35910474642546	0.872761254049971\\
1.41117772521571	0.875234064410999\\
1.46325070400595	0.877876208142232\\
1.5153236827962	0.88070134339275\\
1.56739666158645	0.883724987927856\\
1.6194696403767	0.886964755639427\\
1.67154261916694	0.890440568944679\\
1.72361559795719	0.89417478194836\\
1.77568857674744	0.898192083284709\\
1.82776155553769	0.902518922428291\\
1.87983453432793	0.907181970973285\\
1.93190751311818	0.912204717699604\\
1.98398049190843	0.917600631744849\\
2.03605347069868	0.923360503905012\\
};
\addplot [color=mycolor10, forget plot]
  table[row sep=crcr]{%
0.00525495604775517	0.850752757933336\\
0.0578045165253069	0.850792233200693\\
0.110354077002859	0.850897639586176\\
0.16290363748041	0.851069195517132\\
0.215453197957962	0.851307273631918\\
0.268002758435514	0.851612396585858\\
0.320552318913065	0.85198524037298\\
0.373101879390617	0.85242663979835\\
0.425651439868169	0.852937595554291\\
0.478201000345721	0.853519282915327\\
0.530750560823272	0.854173062180493\\
0.583300121300824	0.854900491046996\\
0.635849681778376	0.855703339146134\\
0.688399242255927	0.85658360502255\\
0.740948802733479	0.857543535894694\\
0.793498363211031	0.858585650601016\\
0.846047923688583	0.859712766214123\\
0.898597484166134	0.860928028897205\\
0.951147044643686	0.862234949684153\\
1.00369660512124	0.863637445989868\\
1.05624616559879	0.865139889800778\\
1.10879572607634	0.866747163655972\\
1.16134528655389	0.868464725702286\\
1.21389484703144	0.870298685279176\\
1.266444407509	0.872255890634718\\
1.31899396798655	0.874344030440011\\
1.3715435284641	0.876571750657779\\
1.42409308894165	0.878948787848001\\
1.4766426494192	0.88148611882477\\
1.52919220989675	0.884196124101589\\
1.58174177037431	0.887092757661624\\
1.63429133085186	0.890191706229236\\
1.68684089132941	0.89351050369086\\
1.73939045180696	0.897068533891053\\
1.79194001228451	0.900886795693319\\
1.84448957276206	0.9049871971775\\
1.89703913323962	0.909390957943323\\
1.94958869371717	0.914115386208748\\
2.00213825419472	0.919167838312849\\
2.05468781467227	0.924535191088803\\
};
\addplot [color=black, dotted, forget plot]
  table[row sep=crcr]{%
0.00322611110747884	0.371233417167123\\
0.0354872221822672	0.37131623084484\\
0.0677483332570556	0.371537771191375\\
0.100009444331844	0.371899953541568\\
0.132270555406632	0.372405898148915\\
0.164531666481421	0.373059940829452\\
0.196792777556209	0.373867649257197\\
0.229053888630997	0.374835845887806\\
0.261314999705786	0.375972639027065\\
0.293576110780574	0.377287463971015\\
0.325837221855363	0.378791136580179\\
0.358098332930151	0.380495922148519\\
0.390359444004939	0.382415623019182\\
0.422620555079728	0.384565689122998\\
0.454881666154516	0.386963356525295\\
0.487142777229305	0.389627820234027\\
0.519403888304093	0.392580449047586\\
0.551664999378881	0.395845052242876\\
0.58392611045367	0.399448210620567\\
0.616187221528458	0.40341968811786\\
0.648448332603247	0.40779294528346\\
0.680709443678035	0.412605782999949\\
0.712970554752823	0.417901154867763\\
0.745231665827612	0.423728201073745\\
0.7774927769024	0.430143577632384\\
0.809753887977188	0.437213186291691\\
0.842014999051977	0.445014458250621\\
0.874276110126765	0.453639419585848\\
0.906537221201553	0.463198886368963\\
0.938798332276342	0.473828336632081\\
0.97105944335113	0.485696349087497\\
1.00332055442592	0.499017114087597\\
1.03558166550071	0.514069685011352\\
1.0678427765755	0.531228970748175\\
1.10010388765028	0.551018506190583\\
1.13236499872507	0.574206960517421\\
1.16462610979986	0.602002145794268\\
1.19688722087465	0.636496063784519\\
1.22914833194944	0.681911722070176\\
1.26140944302423	0.749564172278627\\
};
\addplot [color=black, dotted, forget plot]
  table[row sep=crcr]{%
0.00354690315652309	0.497359277539671\\
0.039015934721754	0.497475962583303\\
0.0744849662869849	0.497787469017776\\
0.109953997852216	0.498294724466126\\
0.145423029417447	0.498999249254933\\
0.180892060982678	0.499903171441984\\
0.216361092547909	0.501009249306405\\
0.251830124113139	0.502320901160777\\
0.28729915567837	0.503842243143796\\
0.322768187243601	0.505578135944988\\
0.358237218808832	0.507534241693801\\
0.393706250374063	0.509717092578111\\
0.429175281939294	0.512134173167636\\
0.464644313504525	0.514794018933388\\
0.500113345069756	0.517706334110902\\
0.535582376634987	0.520882132900837\\
0.571051408200218	0.524333909102971\\
0.606520439765449	0.528075840732524\\
0.64198947133068	0.532124038105159\\
0.67745850289591	0.536496846491899\\
0.712927534461141	0.541215218020334\\
0.748396566026372	0.546303172454248\\
0.783865597591603	0.551788373456826\\
0.819334629156834	0.557702856914649\\
0.854803660722065	0.564083962417201\\
0.890272692287296	0.570975540540809\\
0.925741723852527	0.57842954129371\\
0.961210755417758	0.586508139936223\\
0.996679786982989	0.595286637667387\\
1.03214881854822	0.604857508667378\\
1.06761785011345	0.61533619391042\\
1.10308688167868	0.62686964971259\\
1.13855591324391	0.639649420385396\\
1.17402494480914	0.653932510785368\\
1.20949397637437	0.670076528925356\\
1.24496300793961	0.688602948957405\\
1.28043203950484	0.710321360983768\\
1.31590107107007	0.736604165423864\\
1.3513701026353	0.770106837507761\\
1.38683913420053	0.817228050466163\\
};
\addplot [color=black, dotted, forget plot]
  table[row sep=crcr]{%
0.00373787559968907	0.567649652247075\\
0.0411166315965798	0.567754626041578\\
0.0784953875934706	0.568034852210428\\
0.115874143590361	0.568491112650449\\
0.153252899587252	0.569124689952482\\
0.190631655584143	0.569937379573336\\
0.228010411581033	0.570931508741886\\
0.265389167577924	0.572109961597559\\
0.302767923574815	0.573476211069351\\
0.340146679571706	0.57503435828687\\
0.377525435568596	0.576789180556266\\
0.414904191565487	0.578746189213941\\
0.452282947562378	0.580911699014887\\
0.489661703559269	0.583292911143737\\
0.527040459556159	0.585898012484934\\
0.56441921555305	0.588736294494027\\
0.601797971549941	0.591818295930395\\
0.639176727546832	0.595155974920334\\
0.676555483543722	0.598762917428596\\
0.713934239540613	0.602654591384752\\
0.751312995537504	0.606848658669667\\
0.788691751534395	0.611365361260677\\
0.826070507531285	0.616228003580254\\
0.863449263528176	0.621463561289423\\
0.900828019525067	0.627103458665119\\
0.938206775521958	0.633184574307227\\
0.975585531518848	0.63975056153727\\
1.01296428751574	0.646853611055574\\
1.05034304351263	0.654556848939785\\
1.08772179950952	0.662937670432837\\
1.12510055550641	0.672092492081533\\
1.1624793115033	0.682143726189266\\
1.19985806750019	0.69325037559781\\
1.23723682349708	0.70562480632119\\
1.27461557949397	0.719560671497624\\
1.31199433549086	0.735482411428197\\
1.34937309148776	0.754040338423869\\
1.38675184748465	0.776313729566299\\
1.42413060348154	0.80431221672809\\
1.46150935947843	0.842478488446478\\
};
\addplot [color=black, dotted, forget plot]
  table[row sep=crcr]{%
0.00403345525117133	0.654806005078289\\
0.0443680077628847	0.65489302940968\\
0.084702560274598	0.655125335027558\\
0.125037112786311	0.655503536839182\\
0.165371665298025	0.656028645553068\\
0.205706217809738	0.656702076212625\\
0.246040770321451	0.657525662832604\\
0.286375322833165	0.658501678007521\\
0.326709875344878	0.659632857795568\\
0.367044427856591	0.660922432472151\\
0.407378980368304	0.662374163943412\\
0.447713532880018	0.663992390826748\\
0.488048085391731	0.665782082468288\\
0.528382637903445	0.667748903495293\\
0.568717190415158	0.669899290917133\\
0.609051742926871	0.672240546321552\\
0.649386295438585	0.674780946404457\\
0.689720847950298	0.67752987597828\\
0.730055400462011	0.680497988806548\\
0.770389952973724	0.683697403225884\\
0.810724505485438	0.687141941708151\\
0.851059057997151	0.690847426531362\\
0.891393610508864	0.694832047936975\\
0.931728163020578	0.699116827115924\\
0.972062715532291	0.703726204960124\\
1.012397268044	0.708688800127839\\
1.05273182055572	0.714038398853792\\
1.09306637306743	0.719815267843542\\
1.13340092557914	0.726067926963585\\
1.17373547809086	0.732855591661451\\
1.21407003060257	0.740251616995003\\
1.25440458311428	0.748348485626548\\
1.294739135626	0.757265260589106\\
1.33507368813771	0.767159136972246\\
1.37540824064942	0.778244146767742\\
1.41574279316114	0.79082308313779\\
1.45607734567285	0.805345578488508\\
1.49641189818456	0.822522182979521\\
1.53674645069628	0.843568142551764\\
1.57708100320799	0.870742522380374\\
};
\addplot [color=black, dotted, forget plot]
  table[row sep=crcr]{%
0.00455779977438598	0.762443498736815\\
0.0501357975182458	0.762505651193054\\
0.0957137952621055	0.762671571003272\\
0.141291793005965	0.762941667446941\\
0.186869790749825	0.763316618272847\\
0.232447788493685	0.763797372723416\\
0.278025786237545	0.764385160596833\\
0.323603783981404	0.765081504761052\\
0.369181781725264	0.765888237078917\\
0.414759779469124	0.766807518064643\\
0.460337777212984	0.767841860737356\\
0.505915774956844	0.768994159271468\\
0.551493772700703	0.770267723198317\\
0.597071770444563	0.771666318102668\\
0.642649768188423	0.773194213993817\\
0.688227765932283	0.774856242829981\\
0.733805763676142	0.776657867057054\\
0.779383761420002	0.778605261517113\\
0.824961759163862	0.780705411727111\\
0.870539756907722	0.782966232378673\\
0.916117754651582	0.78539671104303\\
0.961695752395441	0.788007083591114\\
1.0072737501393	0.790809049918012\\
1.05285174788316	0.79381604142838\\
1.09842974562702	0.797043555745605\\
1.14400774337088	0.800509579782332\\
1.18958574111474	0.804235130466886\\
1.2351637388586	0.808244954327578\\
1.28074173660246	0.81256844480319\\
1.32631973434632	0.817240862798747\\
1.37189773209018	0.82230498688233\\
1.41747572983404	0.82781338318154\\
1.4630537275779	0.833831585338952\\
1.50863172532176	0.84044263324886\\
1.55420972306562	0.847753664037643\\
1.59978772080948	0.855905594537417\\
1.64536571855334	0.865087263148394\\
1.6909437162972	0.875554874166336\\
1.73652171404106	0.887651848888846\\
1.78209971178492	0.901792116369156\\
};
\addplot [color=black, dotted, forget plot]
  table[row sep=crcr]{%
0.00557125455888319	0.878794384271811\\
0.0612838001477151	0.878825727228896\\
0.116996345736547	0.878909435683583\\
0.172708891325379	0.879045653030364\\
0.228421436914211	0.879234633061947\\
0.284133982503043	0.879476732405249\\
0.339846528091875	0.879772411497066\\
0.395559073680707	0.880122237260724\\
0.451271619269538	0.880526886696756\\
0.50698416485837	0.880987151282693\\
0.562696710447202	0.88150394218865\\
0.618409256036034	0.88207829634702\\
0.674121801624866	0.882711383427879\\
0.729834347213698	0.883404513779296\\
0.78554689280253	0.884159147395652\\
0.841259438391362	0.884976903977599\\
0.896971983980194	0.885859574142607\\
0.952684529569026	0.886809131833194\\
1.00839707515786	0.887827747946912\\
1.06410962074669	0.888917805172672\\
1.11982216633552	0.890081913953889\\
1.17553471192435	0.891322929398141\\
1.23124725751319	0.892643968798334\\
1.28695980310202	0.894048429195439\\
1.34267234869085	0.895540004060103\\
1.39838489427968	0.897122697643352\\
1.45409743986851	0.898800834763228\\
1.50980998545734	0.90057906263344\\
1.56552253104618	0.902462339627015\\
1.62123507663501	0.904455903350169\\
1.67694762222384	0.906565206725406\\
1.73266016781267	0.90879580546728\\
1.7883727134015	0.911153172769847\\
1.84408525899034	0.913642406533529\\
1.89979780457917	0.916267780529416\\
1.955510350168	0.919032073797433\\
2.01122289575683	0.921935594738075\\
2.06693544134566	0.924974805152606\\
2.1226479869345	0.928140461884712\\
2.17836053252333	0.931415262155431\\
};
\addplot [color=black, dotted, forget plot]
  table[row sep=crcr]{%
0.00588423048109367	0.901841475565186\\
0.0647265352920304	0.901865236438433\\
0.123568840102967	0.901928713068749\\
0.182411144913904	0.902031976648684\\
0.241253449724841	0.902175166261288\\
0.300095754535777	0.902358478426275\\
0.358938059346714	0.902582166057045\\
0.417780364157651	0.902846538749246\\
0.476622668968587	0.903151963421608\\
0.535464973779524	0.903498865121465\\
0.594307278590461	0.903887727924673\\
0.653149583401398	0.904319095880775\\
0.711991888212334	0.904793573952492\\
0.770834193023271	0.905311828887864\\
0.829676497834208	0.905874589946842\\
0.888518802645144	0.906482649381482\\
0.947361107456081	0.907136862539614\\
1.00620341226702	0.907838147423827\\
1.06504571707795	0.908587483488854\\
1.12388802188889	0.909385909397893\\
1.18273032669983	0.910234519378568\\
1.24157263151076	0.911134457717323\\
1.3004149363217	0.912086910801731\\
1.35925724113264	0.913093095956793\\
1.41809954594358	0.914154246115871\\
1.47694185075451	0.915271589111118\\
1.53578415556545	0.91644632005337\\
1.59462646037639	0.91767956489016\\
1.65346876518732	0.918972332779343\\
1.71231106999826	0.920325454400556\\
1.7711533748092	0.921739502769352\\
1.82999567962013	0.923214692571211\\
1.88883798443107	0.924750753594731\\
1.94768028924201	0.926346773690464\\
2.00652259405294	0.928001007098326\\
2.06536489886388	0.929710645402634\\
2.12420720367482	0.931471551388133\\
2.18304950848575	0.9332779614289\\
2.24189181329669	0.935122170510756\\
2.30073411810763	0.93699422604444\\
};
\addplot [color=black, dotted, forget plot]
  table[row sep=crcr]{%
0.00626339791321697	0.925292851297236\\
0.0688973770453866	0.925306969909335\\
0.131531356177556	0.925344705096994\\
0.194165335309726	0.925406049877254\\
0.256799314441896	0.925491015603705\\
0.319433293574065	0.925599620753197\\
0.382067272706235	0.925731888737064\\
0.444701251838404	0.925887846720254\\
0.507335230970574	0.926067524469001\\
0.569969210102744	0.926270953039544\\
0.632603189234913	0.926498163240436\\
0.695237168367083	0.926749183826229\\
0.757871147499253	0.927024039384902\\
0.820505126631422	0.92732274788001\\
0.883139105763592	0.927645317804802\\
0.945773084895762	0.92799174490113\\
1.00840706402793	0.928362008391038\\
1.0710410431601	0.928756066663872\\
1.13367502229227	0.929173852356996\\
1.19630900142444	0.929615266763885\\
1.25894298055661	0.930080173499996\\
1.32157695968878	0.930568391355011\\
1.38421093882095	0.931079686260344\\
1.44684491795312	0.93161376230437\\
1.50947889708529	0.932170251735466\\
1.57211287621746	0.932748703906569\\
1.63474685534963	0.933348573135638\\
1.6973808344818	0.933969205486641\\
1.76001481361397	0.934609824517435\\
1.82264879274614	0.935269516096414\\
1.88528277187831	0.93594721246161\\
1.94791675101048	0.936641675785857\\
2.01055073014265	0.937351481620846\\
2.07318470927482	0.938075002721698\\
2.13581868840699	0.938810393899487\\
2.19845266753915	0.939555578708086\\
2.26108664667132	0.940308238933316\\
2.32372062580349	0.941065808006371\\
2.38635460493566	0.941825469588972\\
2.44898858406783	0.942584162653855\\
};
\addplot [color=black, dashed, forget plot]
  table[row sep=crcr]{%
0	0.947726305996732\\
2.67447742393749	0.947726305996732\\
};
\end{axis}
%
%
\draw[->,-stealth, line width=1.25pt] (1.75,1.75) to [bend right=-20] (0.75,3.5);

\end{tikzpicture}%

%% file: Figs/Fig9a_ColourMap_Hirotsu_Shrink_Front.tex
\begin{tikzpicture}
\begin{axis}[axis on top,  
thick, 
width=0.45\textwidth, 
height=0.4\textwidth,
/pgf/number format/.cd,
                fixed,
enlargelimits=false, 
colorbar, 
point meta min = 0.25, 
point meta max = 1,
xmin=0,xmax=0.5,ymin=0,ymax=3, 
colorbar style = { 
width = 0.3cm,
thick,
black,
title = {$\phi$},
title style = {overlay,yshift = -3pt},
at={(1.05,1)}},
xlabel = {$t$},
ylabel = {$r$},
legend style = {
draw=none,
fill=none,
font=\scriptsize},]
\addplot[forget plot] graphics [xmin=0,xmax=0.5,ymin=0,ymax=3] {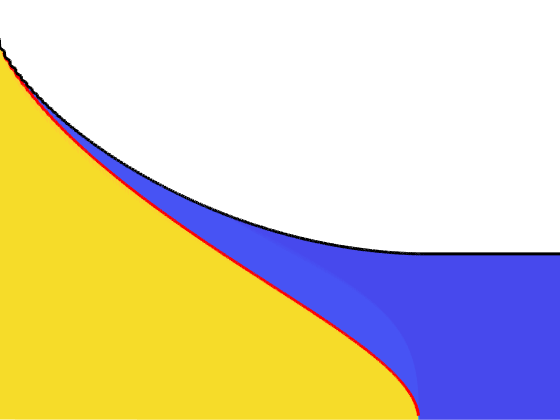};
%
%
\addplot[black, mark=none, domain=-1:-0.1] {x};
    \addlegendentry{Edge location};

\addplot[red, mark=none, domain=-1:-0.1] {x};
    \addlegendentry{Front location}
\end{axis}
\end{tikzpicture}

%% file: Figs/Fig9b_Porosity_Hirotsu_Shrink_Front.tex
%
%
\definecolor{mycolor1}{rgb}{0.15625,0.09765,0.06219}%
\definecolor{mycolor2}{rgb}{0.31250,0.19530,0.12437}%
\definecolor{mycolor3}{rgb}{0.46875,0.29295,0.18656}%
\definecolor{mycolor4}{rgb}{0.62500,0.39060,0.24875}%
\definecolor{mycolor5}{rgb}{0.78125,0.48825,0.31094}%
\definecolor{mycolor6}{rgb}{0.93750,0.58590,0.37312}%
\definecolor{mycolor7}{rgb}{1.00000,0.68355,0.43531}%
\definecolor{mycolor8}{rgb}{1.00000,0.78120,0.49750}%
\begin{tikzpicture}

\begin{axis}[%
thick, 
width=0.48\textwidth, 
height=0.4\textwidth,
xmin=0,xmax=3,ymin=0,ymax=1, 
xlabel = {$r$},
ylabel = {$\phi$},
yticklabel style = overlay]
\addplot [color=black, forget plot]
  table[row sep=crcr]{%
0.00334309677992186	0.947726305996732\\
0.0367740645791405	0.947726305996732\\
0.0702050323783591	0.947726305996732\\
0.103636000177578	0.947726305996732\\
0.137066967976796	0.947726305996732\\
0.170497935776015	0.947726305996732\\
0.203928903575234	0.947726305996732\\
0.237359871374452	0.947726305996732\\
0.270790839173671	0.947726305996732\\
0.304221806972889	0.947726305996732\\
0.337652774772108	0.947726305996732\\
0.371083742571327	0.947726305996732\\
0.404514710370545	0.947726305996732\\
0.437945678169764	0.947726305996732\\
0.471376645968983	0.947726305996732\\
0.504807613768201	0.947726305996732\\
0.53823858156742	0.947726305996732\\
0.571669549366638	0.947726305996732\\
0.605100517165857	0.947726305996732\\
0.638531484965076	0.947726305996732\\
0.671962452764294	0.947726305996732\\
0.705393420563513	0.947726305996732\\
0.738824388362732	0.947726305996732\\
0.77225535616195	0.947726305996732\\
0.805686323961169	0.947726305996732\\
0.839117291760388	0.947726305996732\\
0.872548259559606	0.947726305996732\\
0.905979227358825	0.947726305996732\\
0.939410195158043	0.947726305996732\\
0.972841162957262	0.947726305996732\\
1.00627213075648	0.947726305996732\\
1.0397030985557	0.947726305996732\\
1.07313406635492	0.947726305996732\\
1.10656503415414	0.947726305996732\\
1.13999600195336	0.947726305996732\\
1.17342696975257	0.947726305996732\\
1.20685793755179	0.947726305996732\\
1.24028890535101	0.947726305996732\\
1.27371987315023	0.947726305996732\\
1.30715084094945	0.947726305996732\\
1.34058180874867	0.947726305996732\\
1.37401277654789	0.947726305996732\\
1.4074437443471	0.947726305996732\\
1.44087471214632	0.947726305996732\\
1.47430567994554	0.947726305996732\\
1.50773664774476	0.947726305996732\\
1.54116761554398	0.947726305996732\\
1.5745985833432	0.947726305996732\\
1.60802955114242	0.947726305996732\\
1.64146051894163	0.947726305996732\\
1.67489148674085	0.947726305996732\\
1.70832245454007	0.947726305996732\\
1.74175342233929	0.947726305996732\\
1.77518439013851	0.947726305996732\\
1.80861535793773	0.947726305996732\\
1.84204632573695	0.947726305996732\\
1.87547729353616	0.947726305996732\\
1.90890826133538	0.947726305996732\\
1.9423392291346	0.947726305996732\\
1.97577019693382	0.947726305996732\\
2.00920116473304	0.947726305996732\\
2.04263213253226	0.947726305996732\\
2.07606310033148	0.947726305996732\\
2.1094940681307	0.947726305996732\\
2.14292503592991	0.947726305996732\\
2.17635600372913	0.947726305996732\\
2.20978697152835	0.947726305996732\\
2.24321793932757	0.947726305996732\\
2.27664890712679	0.947726305996732\\
2.31007987492601	0.947726305996732\\
2.34351084272523	0.947726305996732\\
2.37694181052444	0.947726305996732\\
2.41037277832366	0.947726305996732\\
2.44380374612288	0.947726305996732\\
2.4772347139221	0.947726305996732\\
2.51066568172132	0.947726305996732\\
2.54409664952054	0.947726305996732\\
2.57752761731976	0.947726305996732\\
2.61095858511897	0.947726305996732\\
2.64438955291819	0.947726305996732\\
};
\addplot [color=mycolor1, forget plot]
  table[row sep=crcr]{%
0.0026802302771697	0.947725729514236\\
0.0294825330488667	0.947725729514236\\
0.0562848358205638	0.947725729514236\\
0.0830871385922608	0.947725729514236\\
0.109889441363958	0.947725729514236\\
0.136691744135655	0.947725729514237\\
0.163494046907352	0.947725729514238\\
0.190296349679049	0.947725729514238\\
0.217098652450746	0.947725729514238\\
0.243900955222443	0.947725729514238\\
0.27070325799414	0.947725729514238\\
0.297505560765837	0.947725729514238\\
0.324307863537534	0.947725729514238\\
0.351110166309231	0.947725729514238\\
0.377912469080928	0.947725729514238\\
0.404714771852625	0.947725729514238\\
0.431517074624322	0.947725729514238\\
0.458319377396019	0.947725729514239\\
0.485121680167716	0.947725729514238\\
0.511923982939414	0.947725729514238\\
0.538726285711111	0.947725729514238\\
0.565528588482808	0.947725729514238\\
0.592330891254505	0.947725729514238\\
0.619133194026202	0.947725729514238\\
0.645935496797899	0.947725729514238\\
0.672737799569596	0.947725729514238\\
0.699540102341293	0.947725729514238\\
0.72634240511299	0.947725729514238\\
0.753144707884687	0.947725729514238\\
0.779947010656384	0.947725729514238\\
0.806749313428081	0.947725729514238\\
0.833551616199778	0.947725729514239\\
0.860353918971475	0.947725729514238\\
0.887156221743172	0.947725729514238\\
0.913958524514869	0.947725729514237\\
0.940760827286566	0.947725729514234\\
0.967563130058263	0.947725729514226\\
0.99436543282996	0.947725729514203\\
1.02116773560166	0.947725729514146\\
1.04797003837335	0.947725729514013\\
1.07477234114505	0.947725729513697\\
1.10157464391675	0.947725729512951\\
1.12837694668845	0.947725729511206\\
1.15517924946014	0.947725729507179\\
1.18198155223184	0.947725729498034\\
1.20878385500354	0.947725729477529\\
1.23558615777523	0.947725729432023\\
1.26238846054693	0.947725729332321\\
1.28919076331863	0.94772572911672\\
1.31599306609032	0.947725728656483\\
1.34279536886202	0.947725727686478\\
1.36959767163372	0.947725725668093\\
1.39639997440542	0.947725721521467\\
1.42320227717711	0.947725713110133\\
1.45000457994881	0.947725696262769\\
1.47680688272051	0.94772566294166\\
1.5036091854922	0.947725597861448\\
1.5304114882639	0.947725472332212\\
1.5572137910356	0.947725233203308\\
1.5840160938073	0.9477247832824\\
1.61081839657899	0.947723947134469\\
1.63762069935069	0.947722412164441\\
1.66442300212239	0.947719628495675\\
1.69122530489408	0.947714641083241\\
1.71802760766578	0.947705811923922\\
1.74482991043748	0.947690366969178\\
1.77163221320917	0.947663669111685\\
1.79843451598087	0.947618069172744\\
1.82523681875257	0.947541089292733\\
1.85203912152427	0.947412447393897\\
1.87884142429596	0.947198994001107\\
1.90564372706766	0.946847117059871\\
1.93244602983936	0.946277786673006\\
1.95924833261105	0.945397040558665\\
1.98605063538275	0.944080274283125\\
2.01285293815445	0.941846653434365\\
2.03965524092614	0.936722718045026\\
2.06645754369784	0.922788988384833\\
2.09325984646954	0.361749987198865\\
2.12006214924124	0.361599284934439\\
};
\addplot [color=mycolor2, forget plot]
  table[row sep=crcr]{%
0.0022980263826703	0.947725690442266\\
0.0252782902093733	0.947725690442266\\
0.0482585540360763	0.947725690442265\\
0.0712388178627793	0.947725690442264\\
0.0942190816894823	0.947725690442264\\
0.117199345516185	0.947725690442263\\
0.140179609342888	0.947725690442261\\
0.163159873169591	0.947725690442258\\
0.186140136996294	0.947725690442251\\
0.209120400822997	0.947725690442241\\
0.2321006646497	0.947725690442224\\
0.255080928476403	0.947725690442194\\
0.278061192303106	0.947725690442145\\
0.301041456129809	0.947725690442063\\
0.324021719956512	0.947725690441927\\
0.347001983783215	0.947725690441701\\
0.369982247609918	0.947725690441327\\
0.392962511436621	0.947725690440714\\
0.415942775263324	0.947725690439707\\
0.438923039090027	0.947725690438064\\
0.46190330291673	0.947725690435373\\
0.484883566743433	0.947725690430987\\
0.507863830570136	0.947725690423857\\
0.530844094396839	0.947725690412294\\
0.553824358223542	0.947725690393601\\
0.576804622050245	0.947725690363433\\
0.599784885876948	0.947725690314966\\
0.622765149703651	0.947725690237355\\
0.645745413530354	0.947725690113549\\
0.668725677357057	0.947725689916784\\
0.69170594118376	0.947725689605265\\
0.714686205010463	0.947725689113997\\
0.737666468837166	0.947725688342282\\
0.760646732663869	0.947725687134889\\
0.783626996490572	0.947725685253502\\
0.806607260317275	0.947725682333857\\
0.829587524143978	0.947725677821682\\
0.852567787970681	0.947725670877392\\
0.875548051797384	0.947725660234976\\
0.898528315624087	0.947725643994086\\
0.92150857945079	0.947725619315384\\
0.944488843277493	0.947725581976175\\
0.967469107104196	0.947725525725723\\
0.990449370930899	0.947725441354643\\
1.0134296347576	0.947725315358692\\
1.03640989858431	0.947725128030062\\
1.05939016241101	0.947724850745347\\
1.08237042623771	0.947724442132777\\
1.10535069006441	0.9477238426846\\
1.12833095389112	0.94772296722519\\
1.15131121771782	0.947721694439059\\
1.17429148154452	0.947719852391887\\
1.19727174537123	0.947717198622638\\
1.22025200919793	0.947713392924629\\
1.24323227302463	0.947707960339517\\
1.26621253685134	0.947700241126527\\
1.28919280067804	0.947689323497445\\
1.31217306450474	0.947673953673183\\
1.33515332833144	0.947652416263255\\
1.35813359215815	0.947622376021627\\
1.38111385598485	0.947580669500405\\
1.40409411981155	0.94752303140351\\
1.42707438363826	0.94744373462573\\
1.45005464746496	0.947335117351249\\
1.47303491129166	0.947186980516209\\
1.49601517511837	0.946985884247339\\
1.51899543894507	0.946714332523719\\
1.54197570277177	0.946349032802799\\
1.56495596659847	0.945854908671685\\
1.58793623042518	0.945173651283749\\
1.61091649425188	0.944249212474576\\
1.63389675807858	0.943222593021295\\
1.65687702190529	0.942531927924281\\
1.67985728573199	0.941079927288184\\
1.70283754955869	0.931528469643821\\
1.72581781338539	0.361942038864637\\
1.7487980772121	0.361876421948965\\
1.7717783410388	0.361728358005752\\
1.7947586048655	0.361531017411007\\
1.81773886869221	0.361302370988868\\
};
\addplot [color=mycolor3, forget plot]
  table[row sep=crcr]{%
0.00201118484128905	0.9477256747269\\
0.0221230332541795	0.947725674716121\\
0.0422348816670699	0.947725674686266\\
0.0623467300799604	0.947725674634991\\
0.0824585784928508	0.947725674558158\\
0.102570426905741	0.947725674449535\\
0.122682275318632	0.947725674300238\\
0.142794123731522	0.947725674097963\\
0.162905972144413	0.947725673825921\\
0.183017820557303	0.947725673461375\\
0.203129668970194	0.947725672973699\\
0.223241517383084	0.947725672321753\\
0.243353365795974	0.947725671450364\\
0.263465214208865	0.947725670285673\\
0.283577062621755	0.947725668728943\\
0.303688911034646	0.947725666648344\\
0.323800759447536	0.94772566386807\\
0.343912607860427	0.947725660154016\\
0.364024456273317	0.947725655194882\\
0.384136304686208	0.947725648577294\\
0.404248153099098	0.94772563975323\\
0.424360001511988	0.947725627997319\\
0.444471849924879	0.947725612351059\\
0.464583698337769	0.947725591550153\\
0.48469554675066	0.947725563929959\\
0.50480739516355	0.947725527302879\\
0.524919243576441	0.947725478799619\\
0.545031091989331	0.947725414664286\\
0.565142940402222	0.947725329990467\\
0.585254788815112	0.947725218382249\\
0.605366637228003	0.947725071519865\\
0.625478485640893	0.947724878604677\\
0.645590334053783	0.947724625651824\\
0.665702182466674	0.947724294591216\\
0.685814030879564	0.947723862128085\\
0.705925879292455	0.947723298302814\\
0.726037727705345	0.947722564675945\\
0.746149576118236	0.947721612047355\\
0.766261424531126	0.947720377598406\\
0.786373272944017	0.947718781321829\\
0.806485121356907	0.94771672157504\\
0.826596969769797	0.947714069558557\\
0.846708818182688	0.94771066248059\\
0.866820666595578	0.947706295121154\\
0.886932515008469	0.947700709453101\\
0.907044363421359	0.947693581911943\\
0.92715621183425	0.947684507829684\\
0.94726806024714	0.94767298245874\\
0.967379908660031	0.947658377907915\\
0.987491757072921	0.947639915191049\\
1.00760360548581	0.947616630446826\\
1.0277154538987	0.947587334221249\\
1.04782730231159	0.947550562505088\\
1.06793915072448	0.947504517975674\\
1.08805099913737	0.947446999593185\\
1.10816284755026	0.947375318375354\\
1.12827469596315	0.947286196922428\\
1.14838654437604	0.947175649846734\\
1.16849839278894	0.947038839474024\\
1.18861024120183	0.946869892402001\\
1.20872208961472	0.94666166446405\\
1.22883393802761	0.946405547165924\\
1.2489457864405	0.94609167781634\\
1.26905763485339	0.945709339918047\\
1.28916948326628	0.945241849466719\\
1.30928133167917	0.944640706562883\\
1.32939318009206	0.943822384513409\\
1.34950502850495	0.943050431711355\\
1.36961687691784	0.943660375136524\\
1.38972872533073	0.944542925708001\\
1.40984057374362	0.931762086492535\\
1.42995242215651	0.362224615649428\\
1.4500642705694	0.362190367042554\\
1.47017611898229	0.362079410706873\\
1.49028796739518	0.361920493117663\\
1.51039981580807	0.361729863806781\\
1.53051166422096	0.361517745501477\\
1.55062351263385	0.361291042788165\\
1.57073536104674	0.361054659557774\\
1.59084720945963	0.36081221006216\\
};
\addplot [color=mycolor4, forget plot]
  table[row sep=crcr]{%
0.00179059112390935	0.947725394060881\\
0.0196965023630029	0.947725390566459\\
0.0376024136020964	0.947725381042551\\
0.0555083248411899	0.947725365171852\\
0.0734142360802834	0.947725342401204\\
0.0913201473193769	0.947725311932178\\
0.10922605855847	0.94772527269359\\
0.127131969797564	0.947725223303023\\
0.145037881036658	0.947725162017033\\
0.162943792275751	0.947725086668523\\
0.180849703514845	0.947724994589372\\
0.198755614753938	0.947724882515792\\
0.216661525993032	0.94772474647339\\
0.234567437232125	0.947724581638293\\
0.252473348471219	0.947724382170124\\
0.270379259710312	0.947724141011513\\
0.288285170949406	0.947723849648318\\
0.306191082188499	0.947723497823234\\
0.324096993427593	0.947723073194536\\
0.342002904666686	0.947722560930072\\
0.35990881590578	0.947721943225097\\
0.377814727144873	0.947721198730436\\
0.395720638383967	0.947720301875468\\
0.41362654962306	0.947719222067733\\
0.431532460862154	0.947717922748074\\
0.449438372101247	0.947716360276847\\
0.467344283340341	0.947714482622931\\
0.485250194579434	0.947712227822783\\
0.503156105818528	0.947709522171675\\
0.521062017057621	0.947706278103504\\
0.538967928296715	0.947702391708734\\
0.556873839535808	0.947697739832435\\
0.574779750774902	0.947692176685339\\
0.592685662013996	0.947685529890702\\
0.610591573253089	0.947677595877644\\
0.628497484492183	0.947668134517892\\
0.646403395731276	0.947656862886488\\
0.66430930697037	0.947643448007895\\
0.682215218209463	0.947627498426327\\
0.700121129448557	0.947608554412075\\
0.71802704068765	0.947586076583344\\
0.735932951926744	0.947559432683705\\
0.753838863165837	0.947527882208079\\
0.771744774404931	0.947490558511201\\
0.789650685644024	0.947446447960762\\
0.807556596883118	0.947394365607147\\
0.825462508122211	0.947332926728137\\
0.843368419361305	0.947260513463695\\
0.861274330600398	0.947175235585002\\
0.879180241839492	0.94707488424782\\
0.897086153078585	0.946956877226179\\
0.914992064317679	0.946818193063483\\
0.932897975556772	0.946655290281922\\
0.950803886795866	0.94646401507127\\
0.968709798034959	0.946239530821182\\
0.986615709274053	0.945976256943695\\
1.00452162051315	0.945667230853103\\
1.02242753175224	0.945301658657262\\
1.04033344299133	0.944866590303434\\
1.05823935423043	0.944383349718066\\
1.07614526546952	0.943960866362663\\
1.09405117670861	0.943438918198262\\
1.11195708794771	0.940747188033725\\
1.1298629991868	0.928802262550019\\
1.14776891042589	0.361765268614546\\
1.16567482166499	0.361912297850504\\
1.18358073290408	0.361941648748936\\
1.20148664414318	0.361902649902981\\
1.21939255538227	0.361819813978542\\
1.23729846662136	0.36170730891434\\
1.25520437786046	0.361574115430146\\
1.27311028909955	0.361426290563848\\
1.29101620033864	0.361268104329619\\
1.30892211157774	0.36110266751241\\
1.32682802281683	0.360932303938755\\
1.34473393405592	0.360758783710864\\
1.36263984529502	0.360583475867687\\
1.38054575653411	0.360407451920445\\
1.3984516677732	0.360231558157827\\
1.4163575790123	0.360056467390802\\
};
\addplot [color=mycolor5, forget plot]
  table[row sep=crcr]{%
0.00162639781759672	0.947711413912726\\
0.0178903759935639	0.947711317575825\\
0.0341543541695311	0.947711056611363\\
0.0504183323454983	0.947710626747747\\
0.0666823105214655	0.947710020293407\\
0.0829462886974328	0.947709226288993\\
0.0992102668734	0.94770823034529\\
0.115474245049367	0.947707014381723\\
0.131738223225334	0.947705556288365\\
0.148002201401302	0.947703829509755\\
0.164266179577269	0.947701802543996\\
0.180530157753236	0.947699438347863\\
0.196794135929203	0.947696693636754\\
0.21305811410517	0.947693518065819\\
0.229322092281138	0.947689853276411\\
0.245586070457105	0.94768563178891\\
0.261850048633072	0.947680775719982\\
0.278114026809039	0.947675195298215\\
0.294378004985006	0.947668787148063\\
0.310641983160974	0.947661432306377\\
0.326905961336941	0.947652993930157\\
0.343169939512908	0.947643314646704\\
0.359433917688875	0.947632213489143\\
0.375697895864842	0.947619482350146\\
0.39196187404081	0.947604881875221\\
0.408225852216777	0.947588136702645\\
0.424489830392744	0.947568929941048\\
0.440753808568711	0.947546896755988\\
0.457017786744678	0.947521616913609\\
0.473281764920646	0.947492606102071\\
0.489545743096613	0.947459305818858\\
0.50580972127258	0.947421071572732\\
0.522073699448547	0.947377159103233\\
0.538337677624515	0.947326708264618\\
0.554601655800482	0.947268724155406\\
0.570865633976449	0.947202054994725\\
0.587129612152416	0.947125366151628\\
0.603393590328383	0.947037109618032\\
0.61965756850435	0.946935488075428\\
0.635921546680318	0.946818412533579\\
0.652185524856285	0.946683452329789\\
0.668449503032252	0.946527776088962\\
0.684713481208219	0.946348081723894\\
0.700977459384187	0.946140511596276\\
0.717241437560154	0.945900549350428\\
0.733505415736121	0.945622924044281\\
0.749769393912088	0.945301581941414\\
0.766033372088055	0.944929296648082\\
0.782297350264023	0.944495019833052\\
0.79856132843999	0.943983790903195\\
0.814825306615957	0.943419981585181\\
0.831089284791924	0.942933282907398\\
0.847353262967891	0.942190051122366\\
0.863617241143859	0.937887785674404\\
0.879881219319826	0.91825176806503\\
0.896145197495793	0.361913495170385\\
0.91240917567176	0.362081390667496\\
0.928673153847727	0.362134682918134\\
0.944937132023695	0.362120056351595\\
0.961201110199662	0.362061318442936\\
0.977465088375629	0.361972408249331\\
0.993729066551596	0.361862217277543\\
1.00999304472756	0.36173675783286\\
1.02625702290353	0.361600270438624\\
1.0425210010795	0.361455842906118\\
1.05878497925547	0.361305780601824\\
1.07504895743143	0.361151839796029\\
1.0913129356074	0.360995380954798\\
1.10757691378337	0.36083747287766\\
1.12384089195933	0.360678965417717\\
1.1401048701353	0.360520541436366\\
1.15636884831127	0.36036275463937\\
1.17263282648724	0.360206057579271\\
1.1888968046632	0.360050822665486\\
1.20516078283917	0.35989735811285\\
1.22142476101514	0.359745920169097\\
1.2376887391911	0.359596722569734\\
1.25395271736707	0.359449943902681\\
1.27021669554304	0.359305733381038\\
1.28648067371901	0.359164215392735\\
};
\addplot [color=mycolor6, forget plot]
  table[row sep=crcr]{%
0.00151599036252619	0.947498098878024\\
0.0166758939877881	0.947497029818187\\
0.0318357976130499	0.947494138508164\\
0.0469957012383118	0.947489390397615\\
0.0621556048635737	0.947482721456296\\
0.0773155084888355	0.947474040484732\\
0.0924754121140974	0.947463228203037\\
0.107635315739359	0.947450135524942\\
0.122795219364621	0.947434581301367\\
0.137955122989883	0.947416349557075\\
0.153115026615145	0.947395186198097\\
0.168274930240407	0.947370795151866\\
0.183434833865669	0.947342833893243\\
0.19859473749093	0.947310908302299\\
0.213754641116192	0.947274566793314\\
0.228914544741454	0.947233293648318\\
0.244074448366716	0.94718650148258\\
0.259234351991978	0.947133522764099\\
0.27439425561724	0.947073600304206\\
0.289554159242502	0.947005876632394\\
0.304714062867764	0.946929382165369\\
0.319873966493025	0.946843022078739\\
0.335033870118287	0.946745561789837\\
0.350193773743549	0.946635610962641\\
0.365353677368811	0.946511605950871\\
0.380513580994073	0.946371790604048\\
0.395673484619335	0.946214195372643\\
0.410833388244597	0.946036614663863\\
0.425993291869858	0.94583658245471\\
0.44115319549512	0.945611346263592\\
0.456313099120382	0.945357838895777\\
0.471473002745644	0.945072645467876\\
0.486632906370906	0.944751979993605\\
0.501792809996168	0.944391742559171\\
0.51695271362143	0.943987337775307\\
0.532112617246692	0.943531364792654\\
0.547272520871953	0.943016400316011\\
0.562432424497215	0.942490542528567\\
0.577592328122477	0.942027864111739\\
0.592752231747739	0.940449742250493\\
0.607912135373001	0.931515052300898\\
0.623072038998263	0.87098175023396\\
0.638231942623525	0.362199690159015\\
0.653391846248787	0.362364036588955\\
0.668551749874048	0.362413377798036\\
0.68371165349931	0.36239650948729\\
0.698871557124572	0.362337967718441\\
0.714031460749834	0.362251835107798\\
0.729191364375096	0.362146889492128\\
0.744351268000358	0.3620289184718\\
0.75951117162562	0.361901895852018\\
0.774671075250881	0.361768634322653\\
0.789830978876143	0.361631172139854\\
0.804990882501405	0.361491014198711\\
0.820150786126667	0.361349288545535\\
0.835310689751929	0.361206851379669\\
0.850470593377191	0.361064359418497\\
0.865630497002453	0.36092232089392\\
0.880790400627715	0.360781132165636\\
0.895950304252976	0.360641104422949\\
0.911110207878238	0.360502483418329\\
0.9262701115035	0.360365464216993\\
0.941430015128762	0.360230202328809\\
0.956589918754024	0.360096822180665\\
0.971749822379286	0.359965423612214\\
0.986909726004548	0.359836086888646\\
1.00206962962981	0.359708876591774\\
1.01722953325507	0.359583844656748\\
1.03238943688033	0.359461032754012\\
1.0475493405056	0.359340474166823\\
1.06270924413086	0.359222195278242\\
1.07786914775612	0.359106216754475\\
1.09302905138138	0.358992554491057\\
1.10818895500664	0.358881220373018\\
1.1233488586319	0.358772222888388\\
1.13850876225717	0.358665567625431\\
1.15366866588243	0.358561257677043\\
1.16882856950769	0.35845929397039\\
1.18398847313295	0.358359675535731\\
1.19914837675821	0.358262399725091\\
};
\addplot [color=mycolor7, forget plot]
  table[row sep=crcr]{%
0.00145707394956215	0.94548761215745\\
0.0160278134451836	0.945481128116897\\
0.0305985529408051	0.945463679849798\\
0.0451692924364266	0.945435303711883\\
0.0597400319320481	0.945396026531669\\
0.0743107714276696	0.945345902522838\\
0.0888815109232911	0.945285041652975\\
0.103452250418913	0.945213646298882\\
0.118022989914534	0.945132061145943\\
0.132593729410156	0.94504084097974\\
0.147164468905777	0.94494084248915\\
0.161735208401399	0.944833348186257\\
0.17630594789702	0.944720231783588\\
0.190876687392641	0.944604199413189\\
0.205447426888263	0.944489142441323\\
0.220018166383884	0.944379507345403\\
0.234588905879506	0.944283661013\\
0.249159645375127	0.94426068013428\\
0.263730384870749	0.944036513791676\\
0.27830112436637	0.941517026681265\\
0.292871863861992	0.968946902057111\\
0.307442603357613	0.365884350854397\\
0.322013342853235	0.366317925015508\\
0.336584082348856	0.366326510710054\\
0.351154821844478	0.36609112620707\\
0.365725561340099	0.365703850461187\\
0.380296300835721	0.365226128791598\\
0.394867040331342	0.364703347169197\\
0.409437779826964	0.364169523723038\\
0.424008519322585	0.363649329664636\\
0.438579258818207	0.363159466050304\\
0.453149998313828	0.362709967380776\\
0.46772073780945	0.362305529478857\\
0.482291477305071	0.361946812920847\\
0.496862216800693	0.361631646993288\\
0.511432956296314	0.36135607452573\\
0.526003695791936	0.361115203874291\\
0.540574435287557	0.3609038589232\\
0.555145174783179	0.360717037316601\\
0.5697159142788	0.360550200408963\\
0.584286653774422	0.360399425939981\\
0.598857393270043	0.360261456972875\\
0.613428132765665	0.360133679223222\\
0.627998872261286	0.360014054749961\\
0.642569611756908	0.359901034308592\\
0.657140351252529	0.359793464577672\\
0.671711090748151	0.359690500807517\\
0.686281830243772	0.35959153073313\\
0.700852569739394	0.359496112065608\\
0.715423309235015	0.359403923511261\\
0.729994048730636	0.359314727894832\\
0.744564788226258	0.35922834534289\\
0.75913552772188	0.359144634370561\\
0.773706267217501	0.359063478899312\\
0.788277006713123	0.358984779557097\\
0.802847746208744	0.358908447969603\\
0.817418485704365	0.358834403081753\\
0.831989225199987	0.358762568824089\\
0.846559964695608	0.358692872652212\\
0.86113070419123	0.358625244644212\\
0.875701443686851	0.358559616950873\\
0.890272183182473	0.358495923467633\\
0.904842922678094	0.358434099645847\\
0.919413662173716	0.358374082391763\\
0.933984401669337	0.358315810020946\\
0.948555141164959	0.358259222247735\\
0.96312588066058	0.358204260196599\\
0.977696620156202	0.358150866426695\\
0.992267359651823	0.358098984963787\\
1.00683809914744	0.358048561335444\\
1.02140883864307	0.357999542606692\\
1.03597957813869	0.357951877414153\\
1.05055031763431	0.357905515997347\\
1.06512105712993	0.357860410226326\\
1.07969179662555	0.357816513625205\\
1.09426253612117	0.35777378139145\\
1.1088332756168	0.357732170411029\\
1.12340401511242	0.35769163926972\\
1.13797475460804	0.357652148261002\\
1.15254549410366	0.357613659391043\\
};
\addplot [color=mycolor8, forget plot]
  table[row sep=crcr]{%
0.00144860608644158	0.357484371306451\\
0.0159346669508574	0.357484371306451\\
0.0304207278152732	0.357484371306451\\
0.0449067886796891	0.357484371306451\\
0.0593928495441049	0.357484371306451\\
0.0738789104085207	0.357484371306451\\
0.0883649712729366	0.357484371306451\\
0.102851032137352	0.357484371306451\\
0.117337093001768	0.357484371306451\\
0.131823153866184	0.357484371306451\\
0.1463092147306	0.357484371306451\\
0.160795275595016	0.357484371306451\\
0.175281336459432	0.357484371306451\\
0.189767397323847	0.357484371306448\\
0.204253458188263	0.357484371306444\\
0.218739519052679	0.357484371306439\\
0.233225579917095	0.357484371306434\\
0.247711640781511	0.357484371306429\\
0.262197701645927	0.357484371306423\\
0.276683762510342	0.357484371306418\\
0.291169823374758	0.357484371306412\\
0.305655884239174	0.357484371306406\\
0.32014194510359	0.357484371306401\\
0.334628005968006	0.357484371306396\\
0.349114066832422	0.35748437130639\\
0.363600127696837	0.357484371306385\\
0.378086188561253	0.357484371306378\\
0.392572249425669	0.357484371306372\\
0.407058310290085	0.357484371306366\\
0.421544371154501	0.35748437130636\\
0.436030432018917	0.357484371306353\\
0.450516492883332	0.357484371306348\\
0.465002553747748	0.357484371306341\\
0.479488614612164	0.357484371306334\\
0.49397467547658	0.357484371306327\\
0.508460736340996	0.357484371306321\\
0.522946797205411	0.357484371306315\\
0.537432858069827	0.357484371306308\\
0.551918918934243	0.357484371306301\\
0.566404979798659	0.357484371306294\\
0.580891040663075	0.357484371306288\\
0.595377101527491	0.357484371306282\\
0.609863162391907	0.357484371306275\\
0.624349223256322	0.357484371306268\\
0.638835284120738	0.357484371306261\\
0.653321344985154	0.357484371306253\\
0.66780740584957	0.357484371306244\\
0.682293466713986	0.357484371306234\\
0.696779527578402	0.357484371306224\\
0.711265588442817	0.357484371306214\\
0.725751649307233	0.357484371306205\\
0.740237710171649	0.357484371306195\\
0.754723771036065	0.357484371306185\\
0.769209831900481	0.357484371306177\\
0.783695892764896	0.357484371306168\\
0.798181953629312	0.357484371306159\\
0.812668014493728	0.357484371306149\\
0.827154075358144	0.35748437130614\\
0.84164013622256	0.35748437130613\\
0.856126197086976	0.35748437130612\\
0.870612257951391	0.357484371306111\\
0.885098318815807	0.357484371306102\\
0.899584379680223	0.357484371306092\\
0.914070440544639	0.357484371306082\\
0.928556501409055	0.357484371306071\\
0.943042562273471	0.357484371306059\\
0.957528623137886	0.357484371306049\\
0.972014684002302	0.357484371306037\\
0.986500744866718	0.357484371306025\\
1.00098680573113	0.357484371306014\\
1.01547286659555	0.357484371306002\\
1.02995892745997	0.357484371305989\\
1.04444498832438	0.357484371305977\\
1.0589310491888	0.357484371305965\\
1.07341711005321	0.357484371305953\\
1.08790317091763	0.357484371305941\\
1.10238923178204	0.357484371305928\\
1.11687529264646	0.357484371305917\\
1.13136135351088	0.357484371305905\\
1.14584741437529	0.357484371305893\\
};
\addplot [color=black, dashed, forget plot]
  table[row sep=crcr]{%
0	0.357484371523868\\
1.15888080103262	0.357484371523868\\
};
\end{axis}
%
%
\draw[->,-stealth, line width=1.25pt] (3.25,2.75) to [bend right=30] (0.5,2.25);

\end{tikzpicture}%

%% file: Figs/Fig10a_ColourMap_Hirotsu_Shrink_NoFront.tex
\begin{tikzpicture}
\begin{axis}[axis on top,  
thick, 
width=0.45\textwidth, 
height=0.4\textwidth,
/pgf/number format/.cd,
                fixed,
enlargelimits=false, 
colorbar, 
point meta min = 0.25, 
point meta max = 1,
xmin=0,xmax=8,ymin=0,ymax=3, 
colorbar style = { 
width = 0.3cm,
thick,
black,
title = {$\phi$},
title style = {overlay,yshift = -3pt},
at={(1.05,1)}},
xlabel = {$t$},
ylabel = {$r$},
legend style = {
draw=none,
fill=none,
font=\scriptsize},]
\addplot[forget plot] graphics [xmin=0,xmax=8,ymin=0,ymax=3] {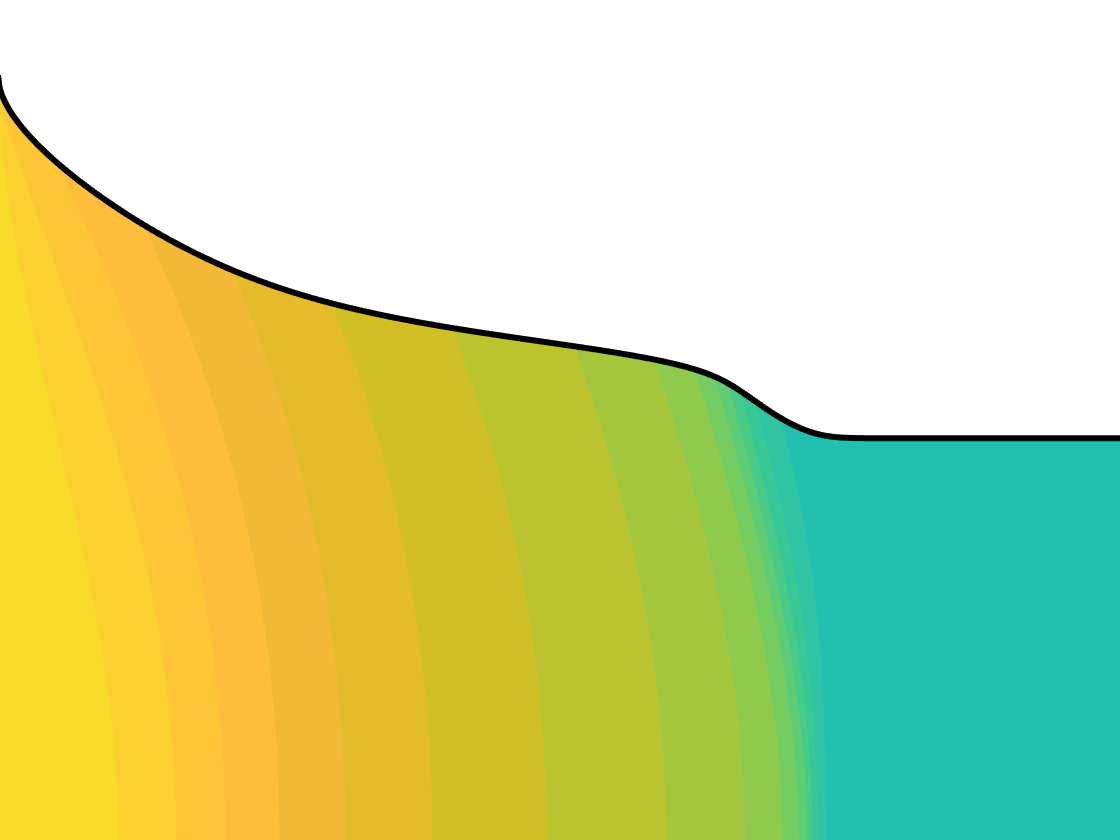};
%
\end{axis}
\end{tikzpicture}

%% file: Figs/Fig10b_Porosity_Hirotsu_Shrink_NoFront.tex
%
%
\definecolor{mycolor1}{rgb}{0.10417,0.06510,0.04146}%
\definecolor{mycolor2}{rgb}{0.20833,0.13020,0.08292}%
\definecolor{mycolor3}{rgb}{0.31250,0.19530,0.12437}%
\definecolor{mycolor4}{rgb}{0.41667,0.26040,0.16583}%
\definecolor{mycolor5}{rgb}{0.52083,0.32550,0.20729}%
\definecolor{mycolor6}{rgb}{0.62500,0.39060,0.24875}%
\definecolor{mycolor7}{rgb}{0.72917,0.45570,0.29021}%
\definecolor{mycolor8}{rgb}{0.83333,0.52080,0.33167}%
\definecolor{mycolor9}{rgb}{0.93750,0.58590,0.37312}%
\definecolor{mycolor10}{rgb}{1.00000,0.65100,0.41458}%
\definecolor{mycolor11}{rgb}{1.00000,0.71610,0.45604}%
\definecolor{mycolor12}{rgb}{1.00000,0.78120,0.49750}%
\begin{tikzpicture}

\begin{axis}[%
thick, 
width=0.48\textwidth, 
height=0.4\textwidth,
xmin=0,xmax=3,ymin=0.5,ymax=1, 
xlabel = {$r$},
ylabel = {$\phi$}]
\addplot [color=black, forget plot]
  table[row sep=crcr]{%
0.00334309677992186	0.947726305996732\\
0.0367740645791405	0.947726305996732\\
0.0702050323783591	0.947726305996732\\
0.103636000177578	0.947726305996732\\
0.137066967976796	0.947726305996732\\
0.170497935776015	0.947726305996732\\
0.203928903575234	0.947726305996732\\
0.237359871374452	0.947726305996732\\
0.270790839173671	0.947726305996732\\
0.304221806972889	0.947726305996732\\
0.337652774772108	0.947726305996732\\
0.371083742571327	0.947726305996732\\
0.404514710370545	0.947726305996732\\
0.437945678169764	0.947726305996732\\
0.471376645968983	0.947726305996732\\
0.504807613768201	0.947726305996732\\
0.53823858156742	0.947726305996732\\
0.571669549366638	0.947726305996732\\
0.605100517165857	0.947726305996732\\
0.638531484965076	0.947726305996732\\
0.671962452764294	0.947726305996732\\
0.705393420563513	0.947726305996732\\
0.738824388362732	0.947726305996732\\
0.77225535616195	0.947726305996732\\
0.805686323961169	0.947726305996732\\
0.839117291760388	0.947726305996732\\
0.872548259559606	0.947726305996732\\
0.905979227358825	0.947726305996732\\
0.939410195158043	0.947726305996732\\
0.972841162957262	0.947726305996732\\
1.00627213075648	0.947726305996732\\
1.0397030985557	0.947726305996732\\
1.07313406635492	0.947726305996732\\
1.10656503415414	0.947726305996732\\
1.13999600195336	0.947726305996732\\
1.17342696975257	0.947726305996732\\
1.20685793755179	0.947726305996732\\
1.24028890535101	0.947726305996732\\
1.27371987315023	0.947726305996732\\
1.30715084094945	0.947726305996732\\
1.34058180874867	0.947726305996732\\
1.37401277654789	0.947726305996732\\
1.4074437443471	0.947726305996732\\
1.44087471214632	0.947726305996732\\
1.47430567994554	0.947726305996732\\
1.50773664774476	0.947726305996732\\
1.54116761554398	0.947726305996732\\
1.5745985833432	0.947726305996732\\
1.60802955114242	0.947726305996732\\
1.64146051894163	0.947726305996732\\
1.67489148674085	0.947726305996732\\
1.70832245454007	0.947726305996732\\
1.74175342233929	0.947726305996732\\
1.77518439013851	0.947726305996732\\
1.80861535793773	0.947726305996732\\
1.84204632573695	0.947726305996732\\
1.87547729353616	0.947726305996732\\
1.90890826133538	0.947726305996732\\
1.9423392291346	0.947726305996732\\
1.97577019693382	0.947726305996732\\
2.00920116473304	0.947726305996732\\
2.04263213253226	0.947726305996732\\
2.07606310033148	0.947726305996732\\
2.1094940681307	0.947726305996732\\
2.14292503592991	0.947726305996732\\
2.17635600372913	0.947726305996732\\
2.20978697152835	0.947726305996732\\
2.24321793932757	0.947726305996732\\
2.27664890712679	0.947726305996732\\
2.31007987492601	0.947726305996732\\
2.34351084272523	0.947726305996732\\
2.37694181052444	0.947726305996732\\
2.41037277832366	0.947726305996732\\
2.44380374612288	0.947726305996732\\
2.4772347139221	0.947726305996732\\
2.51066568172132	0.947726305996732\\
2.54409664952054	0.947726305996732\\
2.57752761731976	0.947726305996732\\
2.61095858511897	0.947726305996732\\
2.64438955291819	0.947726305996732\\
};
\addplot [color=mycolor1, forget plot]
  table[row sep=crcr]{%
0.00291712313970204	0.94717996887806\\
0.0320883545367225	0.947177906314884\\
0.0612595859337429	0.947172346234437\\
0.0904308173307633	0.947163253881983\\
0.119602048727784	0.947150560436873\\
0.148773280124804	0.947134168384165\\
0.177944511521825	0.947113952008621\\
0.207115742918845	0.947089757144748\\
0.236286974315865	0.94706140070985\\
0.265458205712886	0.947028670118226\\
0.294629437109906	0.946991322608333\\
0.323800668506927	0.946949084500078\\
0.352971899903947	0.946901650395766\\
0.382143131300967	0.946848682337992\\
0.411314362697988	0.946789808938582\\
0.440485594095008	0.946724624493982\\
0.469656825492029	0.946652688104143\\
0.498828056889049	0.946573522813593\\
0.527999288286069	0.946486614795259\\
0.55717051968309	0.946391412599535\\
0.58634175108011	0.946287326493007\\
0.615512982477131	0.946173727913263\\
0.644684213874151	0.946049949068203\\
0.673855445271172	0.945915282710193\\
0.703026676668192	0.94576898211723\\
0.732197908065212	0.945610261315031\\
0.761369139462233	0.945438295575411\\
0.790540370859253	0.945252222227543\\
0.819711602256274	0.945051141819473\\
0.848882833653294	0.944834119667574\\
0.878054065050315	0.944600187831389\\
0.907225296447335	0.944348347550156\\
0.936396527844355	0.944077572175437\\
0.965567759241376	0.943786810631198\\
0.994738990638396	0.943474991428392\\
1.02391022203542	0.943141027255555\\
1.05308145343244	0.942783820159509\\
1.08225268482946	0.942402267321522\\
1.11142391622648	0.941995267423502\\
1.1405951476235	0.941561727586262\\
1.16976637902052	0.941100570847675\\
1.19893761041754	0.940610744132527\\
1.22810884181456	0.940091226648545\\
1.25728007321158	0.939541038624801\\
1.2864513046086	0.938959250289879\\
1.31562253600562	0.938344990968534\\
1.34479376740264	0.937697458158199\\
1.37396499879966	0.937015926430927\\
1.40313623019668	0.93629975599393\\
1.4323074615937	0.935548400733198\\
1.46147869299072	0.934761415561098\\
1.49064992438774	0.933938462891008\\
1.51982115578476	0.933079318070699\\
1.54899238718178	0.93218387362141\\
1.5781636185788	0.931252142151652\\
1.60733484997582	0.930284257842855\\
1.63650608137285	0.929280476437493\\
1.66567731276987	0.928241173697838\\
1.69484854416689	0.927166842343301\\
1.72401977556391	0.926058087515098\\
1.75319100696093	0.92491562085569\\
1.78236223835795	0.923740253326918\\
1.81153346975497	0.922532886921451\\
1.84070470115199	0.921294505447111\\
1.86987593254901	0.920026164581154\\
1.89904716394603	0.91872898140129\\
1.92821839534305	0.917404123602466\\
1.95738962674007	0.916052798602505\\
1.98656085813709	0.914676242728335\\
2.01573208953411	0.913275710656683\\
2.04490332093113	0.911852465261526\\
2.07407455232815	0.910407767996111\\
2.10324578372517	0.908942869911323\\
2.13241701512219	0.907459003385965\\
2.16158824651921	0.905957374619052\\
2.19075947791623	0.904439156910594\\
2.21993070931325	0.902905484735994\\
2.24910194071027	0.901357448600624\\
2.27827317210729	0.899796090645765\\
2.30744440350431	0.89822240096453\\
};
\addplot [color=mycolor2, forget plot]
  table[row sep=crcr]{%
0.0026994406635824	0.933420411839003\\
0.0296938472994064	0.933409902378997\\
0.0566882539352304	0.933381666950322\\
0.0836826605710545	0.933335740474224\\
0.110677067206879	0.933272122497267\\
0.137671473842703	0.93319080395405\\
0.164665880478527	0.933091771378273\\
0.191660287114351	0.932975008104222\\
0.218654693750175	0.932840494912851\\
0.245649100385999	0.9326882105508\\
0.272643507021823	0.932518132233014\\
0.299637913657647	0.932330236163421\\
0.326632320293471	0.932124498084657\\
0.353626726929295	0.931900893858952\\
0.380621133565119	0.93165940007865\\
0.407615540200943	0.931399994702613\\
0.434609946836767	0.931122657714267\\
0.461604353472591	0.930827371795659\\
0.488598760108415	0.930514123011992\\
0.515593166744239	0.930182901500615\\
0.542587573380063	0.929833702158185\\
0.569581980015887	0.929466525319634\\
0.596576386651711	0.929081377422551\\
0.623570793287535	0.928678271650587\\
0.650565199923359	0.928257228549562\\
0.677559606559183	0.927818276610208\\
0.704554013195007	0.927361452811609\\
0.731548419830831	0.926886803119811\\
0.758542826466655	0.92639438293648\\
0.785537233102479	0.925884257492749\\
0.812531639738303	0.925356502184126\\
0.839526046374127	0.924811202842811\\
0.866520453009951	0.924248455944332\\
0.893514859645775	0.923668368746186\\
0.920509266281599	0.92307105935671\\
0.947503672917423	0.922456656733256\\
0.974498079553247	0.921825300609355\\
1.00149248618907	0.921177141351274\\
1.0284868928249	0.920512339745244\\
1.05548129946072	0.919831066717073\\
1.08247570609654	0.919133502986559\\
1.10947011273237	0.91841983865993\\
1.13646451936819	0.91769027276384\\
1.16345892600402	0.916945012725026\\
1.19045333263984	0.916184273800222\\
1.21744773927566	0.915408278461142\\
1.24444214591149	0.914617255739741\\
1.27143655254731	0.913811440539069\\
1.29843095918314	0.912991072915268\\
1.32542536581896	0.912156397336106\\
1.35241977245478	0.911307661921644\\
1.37941417909061	0.910445117672366\\
1.40640858572643	0.90956901769002\\
1.43340299236226	0.908679616396138\\
1.46039739899808	0.907777168752989\\
1.4873918056339	0.906861929491344\\
1.51438621226973	0.905934152349132\\
1.54138061890555	0.904994089324631\\
1.56837502554138	0.904041989947466\\
1.5953694321772	0.903078100570253\\
1.62236383881302	0.902102663683327\\
1.64935824544885	0.901115917254555\\
1.67635265208467	0.900118094095737\\
1.7033470587205	0.899109421256882\\
1.73034146535632	0.898090119449054\\
1.75733587199214	0.897060402496208\\
1.78433027862797	0.896020476816078\\
1.81132468526379	0.894970540929818\\
1.83831909189962	0.893910784999833\\
1.86531349853544	0.89284139039493\\
1.89230790517126	0.891762529281658\\
1.91930231180709	0.89067436424059\\
1.94629671844291	0.889577047905803\\
1.97329112507874	0.88847072262608\\
2.00028553171456	0.887355520145835\\
2.02727993835038	0.886231561303755\\
2.05427434498621	0.88509895574707\\
2.08126875162203	0.883957801659216\\
2.10826315825786	0.882808185498578\\
2.13525756489368	0.881650181745918\\
};
\addplot [color=mycolor3, forget plot]
  table[row sep=crcr]{%
0.00253321150652861	0.905438600071511\\
0.0278653265718147	0.905429641761736\\
0.0531974416371008	0.905405595491283\\
0.0785295567023869	0.905366502014079\\
0.103861671767673	0.905312383146802\\
0.129193786832959	0.905243262369302\\
0.154525901898245	0.905159167865982\\
0.179858016963531	0.9050601331878\\
0.205190132028817	0.904946197432\\
0.230522247094103	0.904817405272281\\
0.255854362159389	0.904673806927337\\
0.281186477224676	0.90451545809732\\
0.306518592289962	0.904342419879973\\
0.331850707355248	0.904154758671712\\
0.357182822420534	0.903952546056397\\
0.38251493748582	0.903735858683265\\
0.407847052551106	0.903504778135057\\
0.433179167616392	0.903259390787075\\
0.458511282681678	0.90299978765751\\
0.483843397746964	0.902726064249867\\
0.50917551281225	0.902438320387729\\
0.534507627877536	0.902136660042383\\
0.559839742942822	0.901821191153761\\
0.585171858008108	0.901492025445146\\
0.610503973073395	0.901149278232133\\
0.635836088138681	0.90079306822631\\
0.661168203203967	0.900423517334176\\
0.686500318269253	0.900040750451782\\
0.711832433334539	0.899644895255631\\
0.737164548399825	0.899236081990336\\
0.762496663465111	0.898814443253605\\
0.787828778530397	0.89838011377905\\
0.813160893595683	0.897933230217367\\
0.838493008660969	0.897473930916434\\
0.863825123726255	0.89700235570085\\
0.889157238791542	0.896518645651414\\
0.914489353856828	0.896022942885097\\
0.939821468922114	0.895515390335982\\
0.9651535839874	0.894996131537664\\
0.990485699052686	0.894465310407592\\
1.01581781411797	0.89392307103377\\
1.04114992918326	0.893369557464296\\
1.06648204424854	0.892804913500099\\
1.09181415931383	0.892229282491279\\
1.11714627437912	0.891642807137386\\
1.1424783894444	0.89104562929202\\
1.16781050450969	0.890437889771924\\
1.19314261957497	0.88981972817098\\
1.21847473464026	0.889191282679265\\
1.24380684970555	0.888552689907384\\
1.26913896477083	0.88790408471627\\
1.29447107983612	0.887245600052572\\
1.3198031949014	0.886577366789769\\
1.34513530996669	0.885899513575082\\
1.37046742503198	0.88521216668223\\
1.39579954009726	0.884515449870078\\
1.42113165516255	0.88380948424715\\
1.44646377022784	0.883094388142003\\
1.47179588529312	0.882370276979379\\
1.49712800035841	0.881637263162095\\
1.52246011542369	0.880895455958532\\
1.54779223048898	0.880144961395632\\
1.57312434555427	0.879385882157252\\
1.59845646061955	0.878618317487699\\
1.62378857568484	0.877842363100296\\
1.64912069075012	0.87705811109078\\
1.67445280581541	0.876265649855318\\
1.6997849208807	0.875465064012942\\
1.72511703594598	0.874656434332158\\
1.75044915101127	0.873839837661501\\
1.77578126607655	0.873015346863803\\
1.80111338114184	0.872183030753893\\
1.82644549620713	0.871342954039509\\
1.85177761127241	0.870495177265143\\
1.8771097263377	0.869639756758559\\
1.90244184140298	0.868776744579716\\
1.92777395646827	0.867906188471846\\
1.95310607153356	0.867028131814405\\
1.97843818659884	0.866142613577658\\
2.00377030166413	0.865249668278579\\
};
\addplot [color=mycolor4, forget plot]
  table[row sep=crcr]{%
0.00241097792442463	0.878831670229204\\
0.0265207571686709	0.878825585555404\\
0.0506305364129172	0.87880925382959\\
0.0747403156571635	0.878782699746665\\
0.0988500949014098	0.87874593380699\\
0.122959874145656	0.87869896607567\\
0.147069653389902	0.878641808224296\\
0.171179432634149	0.878574473989402\\
0.195289211878395	0.878496979312889\\
0.219398991122641	0.8784093423863\\
0.243508770366887	0.878311583657518\\
0.267618549611134	0.878203725819559\\
0.29172832885538	0.878085793789234\\
0.315838108099626	0.877957814679121\\
0.339947887343873	0.877819817764606\\
0.364057666588119	0.877671834446898\\
0.388167445832365	0.877513898212573\\
0.412277225076611	0.877346044590003\\
0.436387004320858	0.877168311102948\\
0.460496783565104	0.87698073722149\\
0.48460656280935	0.876783364310472\\
0.508716342053597	0.876576235575607\\
0.532826121297843	0.876359396007387\\
0.556935900542089	0.876132892322935\\
0.581045679786335	0.87589677290591\\
0.605155459030582	0.87565108774461\\
0.629265238274828	0.8753958883684\\
0.653375017519074	0.875131227782593\\
0.677484796763321	0.874857160401901\\
0.701594576007567	0.874573741982607\\
0.725704355251813	0.874281029553562\\
0.749814134496059	0.873979081346151\\
0.773923913740306	0.873667956723351\\
0.798033692984552	0.873347716108006\\
0.822143472228798	0.87301842091045\\
0.846253251473045	0.872680133455628\\
0.870363030717291	0.872332916909746\\
0.894472809961537	0.871976835206676\\
0.918582589205784	0.871611952974191\\
0.94269236845003	0.871238335460152\\
0.966802147694276	0.870856048458738\\
0.990911926938522	0.870465158236864\\
1.01502170618277	0.870065731460861\\
1.03913148542702	0.869657835123544\\
1.06324126467126	0.869241536471752\\
1.08735104391551	0.868816902934457\\
1.11146082315975	0.868384002051535\\
1.135570602404	0.867942901403283\\
1.15968038164825	0.867493668540763\\
1.18379016089249	0.867036370917043\\
1.20789994013674	0.866571075819403\\
1.23200971938099	0.866097850302581\\
1.25611949862523	0.865616761123126\\
1.28022927786948	0.865127874674881\\
1.30433905711372	0.86463125692568\\
1.32844883635797	0.864126973355279\\
1.35255861560222	0.863615088894582\\
1.37666839484646	0.863095667866164\\
1.40077817409071	0.862568773926153\\
1.42488795333496	0.862034470007478\\
1.4489977325792	0.861492818264496\\
1.47310751182345	0.860943880019042\\
1.49721729106769	0.860387715707819\\
1.52132707031194	0.85982438483127\\
1.54543684955619	0.859253945903824\\
1.56954662880043	0.858676456405551\\
1.59365640804468	0.858091972735198\\
1.61776618728893	0.857500550164611\\
1.64187596653317	0.856902242794509\\
1.66598574577742	0.856297103511594\\
1.69009552502166	0.855685183946952\\
1.71420530426591	0.855066534435754\\
1.73831508351016	0.854441203978165\\
1.7624248627544	0.853809240201476\\
1.78653464199865	0.853170689323388\\
1.8106444212429	0.852525596116401\\
1.83475420048714	0.851874003873278\\
1.85886397973139	0.851215954373513\\
1.88297375897564	0.850551487850761\\
1.90708353821988	0.849880642961164\\
};
\addplot [color=mycolor5, forget plot]
  table[row sep=crcr]{%
0.00232428484060043	0.857728763130268\\
0.0255671332466048	0.857724465029535\\
0.0488099816526091	0.857712925253171\\
0.0720528300586134	0.857694160608223\\
0.0952956784646178	0.857668176691821\\
0.118538526870622	0.857634977975448\\
0.141781375276626	0.857594569308435\\
0.165024223682631	0.857546956261055\\
0.188267072088635	0.85749214523558\\
0.211509920494639	0.857430143508109\\
0.234752768900644	0.857360959244149\\
0.257995617306648	0.857284601502385\\
0.281238465712652	0.857201080232247\\
0.304481314118657	0.857110406267847\\
0.327724162524661	0.857012591319468\\
0.350967010930665	0.85690764796326\\
0.37420985933667	0.856795589629495\\
0.397452707742674	0.856676430589614\\
0.420695556148678	0.856550185942158\\
0.443938404554683	0.856416871597731\\
0.467181252960687	0.856276504263007\\
0.490424101366691	0.856129101423872\\
0.513666949772696	0.855974681327713\\
0.5369097981787	0.855813262964895\\
0.560152646584704	0.855644866049467\\
0.583395494990709	0.855469510999109\\
0.606638343396713	0.855287218914349\\
0.629881191802717	0.855098011557089\\
0.653124040208722	0.854901911328439\\
0.676366888614726	0.854698941245903\\
0.69960973702073	0.854489124919934\\
0.722852585426735	0.854272486529873\\
0.746095433832739	0.854049050799302\\
0.769338282238743	0.853818842970839\\
0.792581130644748	0.853581888780374\\
0.815823979050752	0.853338214430776\\
0.839066827456756	0.853087846565199\\
0.862309675862761	0.852830812239714\\
0.885552524268765	0.852567138895631\\
0.908795372674769	0.852296854331318\\
0.932038221080774	0.852019986673632\\
0.955281069486778	0.851736564348904\\
0.978523917892782	0.851446616053563\\
1.00176676629879	0.851150170724368\\
1.02500961470479	0.850847257508294\\
1.0482524631108	0.850537905732065\\
1.0714953115168	0.850222144871358\\
1.0947381599228	0.849900004519685\\
1.11798100832881	0.84957151435697\\
1.14122385673481	0.84923670411781\\
1.16446670514082	0.848895603559466\\
1.18770955354682	0.848548242429531\\
1.21095240195283	0.848194650433349\\
1.23419525035883	0.847834857201124\\
1.25743809876483	0.847468892254755\\
1.28068094717084	0.847096784974393\\
1.30392379557684	0.8467185645647\\
1.32716664398285	0.846334260020815\\
1.35040949238885	0.845943900094018\\
1.37365234079486	0.845547513257089\\
1.39689518920086	0.845145127669318\\
1.42013803760686	0.844736771141194\\
1.44338088601287	0.844322471098715\\
1.46662373441887	0.843902254547322\\
1.48986658282488	0.843476148035426\\
1.51310943123088	0.843044177617491\\
1.53635227963689	0.842606368816648\\
1.55959512804289	0.842162746586809\\
1.5828379764489	0.841713335274231\\
1.6060808248549	0.84125815857851\\
1.6293236732609	0.84079723951291\\
1.65256652166691	0.840330600364044\\
1.67580937007291	0.839858262650783\\
1.69905221847892	0.839380247082382\\
1.72229506688492	0.838896573515722\\
1.74553791529093	0.838407260911619\\
1.76878076369693	0.837912327290123\\
1.79202361210293	0.837411789684723\\
1.81526646050894	0.836905664095286\\
1.83850930891494	0.836393965439795\\
};
\addplot [color=mycolor6, forget plot]
  table[row sep=crcr]{%
0.00226104983002143	0.84141120640566\\
0.0248715481302357	0.841407938770623\\
0.0474820464304499	0.841399161625947\\
0.0700925447306642	0.841384887814766\\
0.0927030430308785	0.841365120680389\\
0.115313541331093	0.841339862200804\\
0.137924039631307	0.841309114205858\\
0.160534537931521	0.841272878656234\\
0.183145036231735	0.841231157734949\\
0.20575553453195	0.841183953883149\\
0.228366032832164	0.841131269814912\\
0.250976531132378	0.841073108522744\\
0.273587029432592	0.841009473278343\\
0.296197527732807	0.840940367630648\\
0.318808026033021	0.840865795402131\\
0.341418524333235	0.840785760683932\\
0.364029022633449	0.840700267830093\\
0.386639520933664	0.840609321450718\\
0.409250019233878	0.84051292640474\\
0.431860517534092	0.840411087791959\\
0.454471015834307	0.840303810944471\\
0.477081514134521	0.840191101417542\\
0.499692012434735	0.840072964979925\\
0.522302510734949	0.839949407603638\\
0.544913009035164	0.839820435453179\\
0.567523507335378	0.839686054874235\\
0.590134005635592	0.839546272381845\\
0.612744503935806	0.839401094648021\\
0.635355002236021	0.839250528488842\\
0.657965500536235	0.839094580851011\\
0.680575998836449	0.838933258797862\\
0.703186497136663	0.838766569494823\\
0.725796995436878	0.838594520194331\\
0.748407493737092	0.838417118220186\\
0.771017992037306	0.838234370951341\\
0.79362849033752	0.838046285805122\\
0.816238988637735	0.837852870219859\\
0.838849486937949	0.837654131636941\\
0.861459985238163	0.837450077482254\\
0.884070483538377	0.837240715147021\\
0.906680981838592	0.837026051968006\\
0.929291480138806	0.836806095207079\\
0.95190197843902	0.83658085203015\\
0.974512476739234	0.836350329485398\\
0.997122975039449	0.836114534480836\\
1.01973347333966	0.835873473761166\\
1.04234397163988	0.835627153883903\\
1.06495446994009	0.835375581194747\\
1.08756496824031	0.835118761802192\\
1.11017546654052	0.834856701551324\\
1.13278596484073	0.834589405996794\\
1.15539646314095	0.83431688037494\\
1.17800696144116	0.834039129575012\\
1.20061745974138	0.833756158109471\\
1.22322795804159	0.833467970083333\\
1.24583845634181	0.833174569162499\\
1.26844895464202	0.832875958541051\\
1.29105945294223	0.832572140907445\\
1.31366995124245	0.832263118409573\\
1.33628044954266	0.83194889261862\\
1.35889094784288	0.831629464491677\\
1.38150144614309	0.831304834333031\\
1.40411194444331	0.830975001754089\\
1.42672244274352	0.83063996563186\\
1.44933294104373	0.830299724065877\\
1.47194343934395	0.829954274333558\\
1.49455393764416	0.829603612843858\\
1.51716443594438	0.829247735089143\\
1.53977493424459	0.828886635595185\\
1.5623854325448	0.828520307869173\\
1.58499593084502	0.828148744345625\\
1.60760642914523	0.827771936330096\\
1.63021692744545	0.82738987394051\\
1.65282742574566	0.82700254604604\\
1.67543792404588	0.826609940203327\\
1.69804842234609	0.82621204258993\\
1.7206589206463	0.825808837934791\\
1.74326941894652	0.825400309445568\\
1.76587991724673	0.824986438732632\\
1.78849041554695	0.824567205729485\\
};
\addplot [color=mycolor7, forget plot]
  table[row sep=crcr]{%
0.00221087956651205	0.828255677451845\\
0.0243196752316325	0.828252941383589\\
0.046428470896753	0.828245587987267\\
0.0685372665618735	0.828233628154053\\
0.090646062226994	0.828217063867312\\
0.112754857892115	0.828195895489277\\
0.134863653557235	0.82817012285488\\
0.156972449222355	0.828139745522311\\
0.179081244887476	0.828104762854927\\
0.201190040552596	0.828065174052956\\
0.223298836217717	0.828020978166253\\
0.245407631882837	0.827972174098597\\
0.267516427547958	0.827918760607621\\
0.289625223213078	0.827860736302219\\
0.311734018878199	0.827798099638281\\
0.333842814543319	0.827730848913211\\
0.35595161020844	0.827658982259476\\
0.37806040587356	0.827582497637316\\
0.400169201538681	0.827501392826703\\
0.422277997203801	0.82741566541857\\
0.444386792868922	0.827325312805371\\
0.466495588534042	0.827230332170936\\
0.488604384199163	0.827130720479659\\
0.510713179864283	0.82702647446501\\
0.532821975529404	0.826917590617346\\
0.554930771194524	0.826804065171025\\
0.577039566859645	0.826685894090805\\
0.599148362524765	0.826563073057509\\
0.621257158189886	0.826435597452941\\
0.643365953855006	0.826303462344044\\
0.665474749520127	0.826166662466242\\
0.687583545185247	0.826025192205996\\
0.709692340850368	0.825879045582498\\
0.731801136515488	0.825728216228509\\
0.753909932180609	0.825572697370288\\
0.776018727845729	0.825412481806616\\
0.79812752351085	0.825247561886828\\
0.82023631917597	0.825077929487871\\
0.842345114841091	0.82490357599032\\
0.864453910506211	0.824724492253315\\
0.886562706171332	0.824540668588377\\
0.908671501836452	0.824352094732071\\
0.930780297501573	0.824158759817444\\
0.952889093166693	0.823960652344202\\
0.974997888831814	0.823757760147568\\
0.997106684496934	0.823550070365752\\
1.01921548016205	0.823337569405983\\
1.04132427582718	0.823120242909058\\
1.0634330714923	0.822898075712242\\
1.08554186715742	0.822671051810579\\
1.10765066282254	0.822439154316435\\
1.12975945848766	0.82220236541722\\
1.15186825415278	0.821960666331197\\
1.1739770498179	0.821714037261272\\
1.19608584548302	0.821462457346666\\
1.21819464114814	0.821205904612362\\
1.24030343681326	0.820944355916193\\
1.26241223247838	0.820677786893459\\
1.2845210281435	0.820406171898914\\
1.30662982380862	0.820129483946015\\
1.32873861947374	0.819847694643242\\
1.35084741513886	0.81956077412733\\
1.37295621080398	0.819268690993241\\
1.3950650064691	0.818971412220695\\
1.41717380213422	0.818668903097012\\
1.43928259779934	0.818361127136102\\
1.46139139346446	0.818048045993323\\
1.48350018912959	0.817729619375965\\
1.50560898479471	0.8174058049491\\
1.52771778045983	0.817076558236483\\
1.54982657612495	0.816741832516205\\
1.57193537179007	0.816401578710742\\
1.59404416745519	0.816055745271059\\
1.61615296312031	0.815704278054344\\
1.63826175878543	0.815347120194959\\
1.66037055445055	0.814984211968155\\
1.68247935011567	0.814615490646034\\
1.70458814578079	0.81424089034525\\
1.72669694144591	0.813860341865825\\
1.74880573711103	0.813473772520491\\
};
\addplot [color=mycolor8, forget plot]
  table[row sep=crcr]{%
0.00216559726306248	0.816661149928332\\
0.0238215698936872	0.81665855360217\\
0.045477542524312	0.816651571144794\\
0.0671335151549367	0.816640212921603\\
0.0887894877855615	0.816624479701548\\
0.110445460416186	0.816604370101293\\
0.132101433046811	0.816579881705955\\
0.153757405677436	0.816551011324425\\
0.175413378308061	0.816517755071171\\
0.197069350938685	0.816480108396081\\
0.21872532356931	0.816438066094394\\
0.240381296199935	0.816391622307527\\
0.26203726883056	0.816340770518951\\
0.283693241461184	0.816285503546995\\
0.305349214091809	0.816225813535476\\
0.327005186722434	0.816161691942564\\
0.348661159353059	0.816093129528155\\
0.370317131983683	0.816020116339835\\
0.391973104614308	0.815942641697531\\
0.413629077244933	0.815860694176853\\
0.435285049875558	0.815774261591117\\
0.456941022506182	0.815683330972056\\
0.478596995136807	0.815587888549188\\
0.500252967767432	0.815487919727798\\
0.521908940398057	0.815383409065498\\
0.543564913028681	0.815274340247327\\
0.565220885659306	0.815160696059336\\
0.586876858289931	0.815042458360582\\
0.608532830920556	0.814919608053504\\
0.630188803551181	0.814792125052592\\
0.651844776181805	0.814659988251285\\
0.67350074881243	0.814523175487004\\
0.695156721443055	0.814381663504274\\
0.71681269407368	0.814235427915807\\
0.738468666704304	0.814084443161475\\
0.760124639334929	0.813928682465059\\
0.781780611965554	0.813768117788676\\
0.803436584596179	0.813602719784762\\
0.825092557226803	0.81343245774547\\
0.846748529857428	0.813257299549379\\
0.868404502488053	0.813077211605352\\
0.890060475118678	0.812892158793385\\
0.911716447749302	0.812702104402312\\
0.933372420379927	0.812507010064133\\
0.955028393010552	0.81230683568484\\
0.976684365641177	0.812101539371494\\
0.998340338271801	0.811891077355356\\
1.01999631090243	0.811675403910819\\
1.04165228353305	0.811454471269924\\
1.06330825616368	0.811228229532148\\
1.0849642287943	0.8109966265692\\
1.10662020142493	0.810759607924508\\
1.12827617405555	0.81051711670705\\
1.14993214668617	0.81026909347919\\
1.1715881193168	0.810015476138086\\
1.19324409194742	0.8097561997903\\
1.21490006457805	0.809491196619104\\
1.23655603720867	0.809220395744031\\
1.2582120098393	0.808943723072096\\
1.27986798246992	0.808661101140164\\
1.30152395510055	0.808372448947777\\
1.32317992773117	0.808077681779813\\
1.3448359003618	0.807776711018224\\
1.36649187299242	0.80746944394204\\
1.38814784562305	0.807155783514795\\
1.40980381825367	0.80683562815841\\
1.4314597908843	0.806508871512531\\
1.45311576351492	0.806175402178161\\
1.47477173614555	0.805835103444408\\
1.49642770877617	0.805487852996962\\
1.5180836814068	0.805133522606885\\
1.53973965403742	0.804771977798084\\
1.56139562666805	0.804403077491734\\
1.58305159929867	0.804026673625694\\
1.60470757192929	0.80364261074684\\
1.62636354455992	0.803250725573966\\
1.64801951719054	0.802850846528699\\
1.66967548982117	0.802442793231617\\
1.69133146245179	0.802026375960409\\
1.71298743508242	0.801601395066711\\
};
\addplot [color=mycolor9, forget plot]
  table[row sep=crcr]{%
0.00211607129819279	0.804966607782068\\
0.0232767842801207	0.804963680918137\\
0.0444374972620486	0.804955802655849\\
0.0655982102439765	0.804942984376068\\
0.0867589232259044	0.80492522476309\\
0.107919636207832	0.804902518960258\\
0.12908034918976	0.804874859950801\\
0.150241062171688	0.804842238866937\\
0.171401775153616	0.804804645082628\\
0.192562488135544	0.804762066240395\\
0.213723201117472	0.804714488251873\\
0.2348839140994	0.804661895285358\\
0.256044627081328	0.804604269745524\\
0.277205340063256	0.804541592247543\\
0.298366053045184	0.804473841586636\\
0.319526766027111	0.804400994703537\\
0.340687479009039	0.804323026646071\\
0.361848191990967	0.804239910526925\\
0.383008904972895	0.804151617477564\\
0.404169617954823	0.804058116598243\\
0.425330330936751	0.803959374903965\\
0.446491043918679	0.803855357266267\\
0.467651756900607	0.803746026350641\\
0.488812469882535	0.803631342549409\\
0.509973182864463	0.803511263909833\\
0.531133895846391	0.803385746057205\\
0.552294608828318	0.803254742112695\\
0.573455321810246	0.803118202605611\\
0.594616034792174	0.802976075379831\\
0.615776747774102	0.802828305494016\\
0.63693746075603	0.802674835115268\\
0.658098173737958	0.80251560340582\\
0.679258886719886	0.80235054640234\\
0.700419599701814	0.802179596887386\\
0.721580312683742	0.80200268425248\\
0.74274102566567	0.801819734352295\\
0.763901738647598	0.801630669349318\\
0.785062451629525	0.801435407548364\\
0.806223164611453	0.801233863220245\\
0.827383877593381	0.801025946413802\\
0.848544590575309	0.800811562755493\\
0.869705303557237	0.800590613235629\\
0.890866016539165	0.800362993980253\\
0.912026729521093	0.800128596007613\\
0.933187442503021	0.799887304968019\\
0.954348155484949	0.799639000865844\\
0.975508868466877	0.799383557762218\\
0.996669581448805	0.799120843456927\\
1.01783029443073	0.798850719147781\\
1.03899100741266	0.798573039065652\\
1.06015172039459	0.798287650083116\\
1.08131243337652	0.797994391294498\\
1.10247314635844	0.797693093564845\\
1.12363385934037	0.797383579045123\\
1.1447945723223	0.797065660650676\\
1.16595528530423	0.796739141499615\\
1.18711599828616	0.796403814307532\\
1.20827671126808	0.796059460734455\\
1.22943742425001	0.795705850679606\\
1.25059813723194	0.795342741518964\\
1.27175885021387	0.7949698772801\\
1.2929195631958	0.794586987748143\\
1.31408027617772	0.794193787495973\\
1.33524098915965	0.793789974830978\\
1.35640170214158	0.793375230649794\\
1.37756241512351	0.792949217191366\\
1.39872312810544	0.792511576677548\\
1.41988384108736	0.792061929829065\\
1.44104455406929	0.79159987424316\\
1.46220526705122	0.791124982617465\\
1.48336598003315	0.790636800802636\\
1.50452669301507	0.790134845663979\\
1.525687405997	0.78961860272964\\
1.54684811897893	0.789087523599848\\
1.56800883196086	0.7885410230882\\
1.58916954494279	0.787978476061882\\
1.61033025792471	0.787399213942963\\
1.63149097090664	0.786802520827442\\
1.65265168388857	0.786187629172361\\
1.6738123968705	0.785553714993847\\
};
\addplot [color=mycolor10, forget plot]
  table[row sep=crcr]{%
0.00203677303428152	0.790389078569081\\
0.0224045033770967	0.790384729901797\\
0.0427722337199119	0.79037300777453\\
0.0631399640627271	0.790353926990769\\
0.0835076944055423	0.790327478184142\\
0.103875424748358	0.790293643195993\\
0.124243155091173	0.790252397382357\\
0.144610885433988	0.790203710092785\\
0.164978615776803	0.790147544771148\\
0.185346346119618	0.790083858933361\\
0.205714076462433	0.79001260408904\\
0.226081806805249	0.789933725629151\\
0.246449537148064	0.789847162687826\\
0.266817267490879	0.789752847981497\\
0.287184997833694	0.789650707626341\\
0.307552728176509	0.789540660933985\\
0.327920458519325	0.789422620184878\\
0.34828818886214	0.789296490378334\\
0.368655919204955	0.789162168957991\\
0.38902364954777	0.789019545511216\\
0.409391379890585	0.788868501440723\\
0.429759110233401	0.78870890960647\\
0.450126840576216	0.788540633935672\\
0.470494570919031	0.788363528998498\\
0.490862301261846	0.788177439546756\\
0.511230031604662	0.78798220001255\\
0.531597761947477	0.787777633963599\\
0.551965492290292	0.787563553511467\\
0.572333222633107	0.787339758668619\\
0.592700952975922	0.78710603664968\\
0.613068683318737	0.78686216111179\\
0.633436413661553	0.786607891328372\\
0.653804144004368	0.786342971289924\\
0.674171874347183	0.786067128724777\\
0.694539604689998	0.785780074031829\\
0.714907335032813	0.785481499116403\\
0.735275065375629	0.785171076119241\\
0.755642795718444	0.784848456027447\\
0.776010526061259	0.784513267154809\\
0.796378256404074	0.784165113477309\\
0.816745986746889	0.783803572807875\\
0.837113717089705	0.783428194792298\\
0.85748144743252	0.783038498705952\\
0.877849177775335	0.78263397102813\\
0.89821690811815	0.782214062767822\\
0.918584638460965	0.781778186511039\\
0.938952368803781	0.781325713155782\\
0.959320099146596	0.780855968295894\\
0.979687829489411	0.780368228209588\\
1.00005555983223	0.779861715402077\\
1.02042329017504	0.779335593644369\\
1.04079102051786	0.778788962441778\\
1.06115875086067	0.778220850855921\\
1.08152648120349	0.777630210592496\\
1.1018942115463	0.777015908254059\\
1.12226194188912	0.776376716641825\\
1.14262967223193	0.775711304973077\\
1.16299740257475	0.775018227860857\\
1.18336513291756	0.774295912879854\\
1.20373286326038	0.773542646516961\\
1.22410059360319	0.772756558276557\\
1.24446832394601	0.7719356026798\\
1.26483605428882	0.771077538864823\\
1.28520378463164	0.770179907462521\\
1.30557151497445	0.76924000439363\\
1.32593924531727	0.768254851212558\\
1.34630697566008	0.767221161620853\\
1.3666747060029	0.766135303802695\\
1.38704243634572	0.764993258319993\\
1.40741016668853	0.76379057148243\\
1.42777789703135	0.762522304436136\\
1.44814562737416	0.761182978783953\\
1.46851335771698	0.759766520498554\\
1.48888108805979	0.75826620542629\\
1.50924881840261	0.756674612113021\\
1.52961654874542	0.754983591453873\\
1.54998427908824	0.753184268370587\\
1.57035200943105	0.751267099072983\\
1.59071973977387	0.749222019151719\\
1.61108747011668	0.747038732962413\\
};
\addplot [color=mycolor11, forget plot]
  table[row sep=crcr]{%
0.00184580324797661	0.760796753311598\\
0.0203038357277428	0.760782404777158\\
0.0387618682075089	0.760743569911095\\
0.057219900687275	0.760680243768906\\
0.0756779331670412	0.760592273504329\\
0.0941359656468073	0.760479423167819\\
0.112593998126573	0.760341382314526\\
0.13105203060634	0.760177766166493\\
0.149510063086106	0.759988113647078\\
0.167968095565872	0.759771884380312\\
0.186426128045638	0.759528454918733\\
0.204884160525404	0.759257114260803\\
0.22334219300517	0.758957058653867\\
0.241800225484936	0.758627385652518\\
0.260258257964703	0.758267087389806\\
0.278716290444469	0.757875043012768\\
0.297174322924235	0.757450010232025\\
0.315632355404001	0.756990615937867\\
0.334090387883767	0.756495345843176\\
0.352548420363533	0.755962533129156\\
0.3710064528433	0.755390346095814\\
0.389464485323066	0.754776774860146\\
0.407922517802832	0.754119617206766\\
0.426380550282598	0.753416463786797\\
0.444838582762364	0.752664682992266\\
0.46329661524213	0.75186140602011\\
0.481754647721896	0.751003512901899\\
0.500212680201662	0.750087620637995\\
0.518670712681429	0.749110075069493\\
0.537128745161195	0.748066948785446\\
0.555586777640961	0.746954048236558\\
0.574044810120727	0.74576693434598\\
0.592502842600493	0.74450096229066\\
0.610960875080259	0.743151347743785\\
0.629418907560025	0.741713268598698\\
0.647876940039792	0.740182012750767\\
0.666334972519558	0.738553183346855\\
0.684793004999324	0.736822972102624\\
0.70325103747909	0.734988507483595\\
0.721709069958856	0.733048276058488\\
0.740167102438622	0.731002600591154\\
0.758625134918389	0.728854136990826\\
0.777083167398155	0.726608326347898\\
0.795541199877921	0.724273714523013\\
0.813999232357687	0.721862041252525\\
0.832457264837453	0.719388016058622\\
0.850915297317219	0.716868746353059\\
0.869373329796985	0.714322857211547\\
0.887831362276752	0.71176941943531\\
0.906289394756518	0.709226852312646\\
0.924747427236284	0.706711968810116\\
0.94320545971605	0.704239284771385\\
0.961663492195816	0.701820641902553\\
0.980121524675582	0.699465125484278\\
0.998579557155348	0.697179212100033\\
1.01703758963511	0.694967065664864\\
1.03549562211488	0.692830905577701\\
1.05395365459465	0.690771388411125\\
1.07241168707441	0.688787965014531\\
1.09086971955418	0.686879192691648\\
1.10932775203395	0.685042995080797\\
1.12778578451371	0.683276870546379\\
1.14624381699348	0.681578054285897\\
1.16470184947324	0.679943641165209\\
1.18315988195301	0.678370676537218\\
1.20161791443278	0.67685622171302\\
1.22007594691254	0.675397399814823\\
1.23853397939231	0.673991426722741\\
1.25699201187207	0.672635630876235\\
1.27545004435184	0.671327464866846\\
1.29390807683161	0.670064511077834\\
1.31236610931137	0.668844483080697\\
1.33082414179114	0.667665224070833\\
1.34928217427091	0.666524703294606\\
1.36774020675067	0.665421011168307\\
1.38619823923044	0.664352353599454\\
1.4046562717102	0.663317045878142\\
1.42311430418997	0.662313506399937\\
1.44157233666974	0.661340250403033\\
1.4600303691495	0.660395883844438\\
};
\addplot [color=mycolor12, forget plot]
  table[row sep=crcr]{%
0.00176156650546193	0.643856921232316\\
0.0193772315600812	0.643856433529108\\
0.0369928966147004	0.643855129582061\\
0.0546085616693197	0.643853011228771\\
0.072224226723939	0.643850080391498\\
0.0898398917785582	0.643846339571532\\
0.107455556833177	0.643841791918305\\
0.125071221887797	0.643836441241875\\
0.142686886942416	0.64383029201348\\
0.160302551997035	0.64382334936195\\
0.177918217051655	0.643815619068072\\
0.195533882106274	0.64380710755758\\
0.213149547160893	0.643797821893092\\
0.230765212215512	0.643787769765105\\
0.248380877270132	0.643776959482139\\
0.265996542324751	0.643765399960086\\
0.28361220737937	0.643753100710799\\
0.301227872433989	0.643740071829957\\
0.318843537488609	0.643726323984237\\
0.336459202543228	0.643711868397834\\
0.354074867597847	0.643696716838347\\
0.371690532652466	0.64368088160208\\
0.389306197707086	0.643664375498769\\
0.406921862761705	0.643647211835797\\
0.424537527816324	0.643629404401895\\
0.442153192870943	0.643610967450402\\
0.459768857925563	0.643591915682081\\
0.477384522980182	0.643572264227554\\
0.495000188034801	0.643552028629374\\
0.51261585308942	0.643531224823785\\
0.53023151814404	0.643509869122179\\
0.547847183198659	0.643487978192317\\
0.565462848253278	0.643465569039319\\
0.583078513307898	0.643442658986483\\
0.600694178362517	0.643419265655945\\
0.618309843417136	0.64339540694923\\
0.635925508471755	0.643371101027714\\
0.653541173526375	0.64334636629304\\
0.671156838580994	0.643321221367515\\
0.688772503635613	0.643295685074512\\
0.706388168690232	0.643269776418923\\
0.724003833744852	0.643243514567674\\
0.741619498799471	0.643216918830339\\
0.75923516385409	0.643190008639887\\
0.776850828908709	0.643162803533564\\
0.794466493963329	0.643135323133966\\
0.812082159017948	0.643107587130302\\
0.829697824072567	0.643079615259879\\
0.847313489127186	0.643051427289833\\
0.864929154181806	0.643023042999111\\
0.882544819236425	0.64299448216075\\
0.900160484291044	0.642965764524444\\
0.917776149345663	0.642936909799424\\
0.935391814400283	0.64290793763767\\
0.953007479454902	0.642878867617473\\
0.970623144509521	0.642849719227334\\
0.98823880956414	0.642820511850255\\
1.00585447461876	0.642791264748386\\
1.02347013967338	0.642761997048067\\
1.041085804728	0.642732727725272\\
1.05870146978262	0.642703475591434\\
1.07631713483724	0.642674259279696\\
1.09393279989186	0.642645097231563\\
1.11154846494648	0.64261600768398\\
1.12916413000109	0.642587008656812\\
1.14677979505571	0.642558117940756\\
1.16439546011033	0.642529353085679\\
1.18201112516495	0.642500731389358\\
1.19962679021957	0.642472269886664\\
1.21724245527419	0.642443985339147\\
1.23485812032881	0.642415894225051\\
1.25247378538343	0.64238801272974\\
1.27008945043805	0.642360356736531\\
1.28770511549267	0.642332941817941\\
1.30532078054729	0.64230578322734\\
1.32293644560191	0.642278895890991\\
1.34055211065653	0.6422522944005\\
1.35816777571114	0.642225993005639\\
1.37578344076576	0.642200005607561\\
1.39339910582038	0.642174345752384\\
};
\addplot [color=black, dotted, forget plot]
  table[row sep=crcr]{%
0.00194943766238754	0.779640317461798\\
0.021443814286263	0.779633920730969\\
0.0409381909101384	0.779616655061822\\
0.0604325675340139	0.779588537188963\\
0.0799269441578893	0.779549539911865\\
0.0994213207817647	0.779499616965787\\
0.11891569740564	0.779438706683606\\
0.138410074029516	0.77936673263763\\
0.157904450653391	0.779283603629143\\
0.177398827277266	0.779189213441055\\
0.196893203901142	0.779083440461285\\
0.216387580525017	0.778966147210463\\
0.235881957148893	0.7788371797845\\
0.255376333772768	0.778696367213758\\
0.274870710396644	0.778543520736579\\
0.294365087020519	0.778378432982605\\
0.313859463644395	0.778200877059779\\
0.33335384026827	0.778010605537574\\
0.352848216892145	0.777807349317714\\
0.372342593516021	0.777590816382366\\
0.391836970139896	0.777360690408362\\
0.411331346763772	0.777116629234444\\
0.430825723387647	0.776858263166824\\
0.450320100011523	0.776585193106346\\
0.469814476635398	0.776296988478382\\
0.489308853259274	0.775993184944067\\
0.508803229883149	0.775673281868616\\
0.528297606507024	0.775336739519248\\
0.5477919831309	0.774982975961541\\
0.567286359754775	0.774611363618795\\
0.586780736378651	0.77422122545418\\
0.606275113002526	0.773811830729954\\
0.625769489626402	0.773382390291699\\
0.645263866250277	0.772932051318398\\
0.664758242874152	0.772459891470987\\
0.684252619498028	0.771964912362751\\
0.703746996121903	0.771446032264402\\
0.723241372745779	0.770902077944936\\
0.742735749369654	0.770331775536111\\
0.76223012599353	0.769733740293831\\
0.781724502617405	0.769106465113958\\
0.801218879241281	0.768448307643116\\
0.820713255865156	0.767757475807962\\
0.840207632489031	0.767032011569557\\
0.859702009112907	0.76626977269524\\
0.879196385736782	0.765468412330962\\
0.898690762360658	0.7646253561571\\
0.918185138984533	0.763737776926901\\
0.937679515608409	0.762802566229458\\
0.957173892232284	0.761816303404284\\
0.976668268856159	0.760775221685917\\
0.996162645480035	0.759675171910608\\
1.01565702210391	0.758511584525699\\
1.03515139872779	0.757279431283435\\
1.05464577535166	0.755973188986309\\
1.07414015197554	0.754586809136687\\
1.09363452859941	0.753113699541949\\
1.11312890522329	0.751546727106827\\
1.13262328184716	0.749878255515236\\
1.15211765847104	0.748100237534832\\
1.17161203509491	0.746204389314406\\
1.19110641171879	0.744182482711333\\
1.21060078834266	0.742026799464817\\
1.23009516496654	0.739730793551271\\
1.24958954159042	0.737289997259342\\
1.26908391821429	0.734703170358571\\
1.28857829483817	0.731973617367774\\
1.30807267146204	0.729110482019481\\
1.32756704808592	0.726129695028616\\
1.34706142470979	0.723054169195515\\
1.36655580133367	0.719912901785434\\
1.38605017795754	0.716738915553869\\
1.40554455458142	0.713566374435501\\
1.42503893120529	0.710427538795665\\
1.44453330782917	0.707350273557016\\
1.46402768445305	0.704356566180365\\
1.48352206107692	0.701462124091789\\
1.5030164377008	0.698676816638054\\
1.52251081432467	0.696005606150099\\
1.54200519094855	0.693449646802249\\
};
\addplot [color=black, dotted, forget plot]
  table[row sep=crcr]{%
0.00177914352208738	0.680869219746381\\
0.0195705787429612	0.680843938896704\\
0.0373620139638349	0.680776225907504\\
0.0551534491847087	0.680666620605173\\
0.0729448844055825	0.68051586301872\\
0.0907363196264563	0.680324938829878\\
0.10852775484733	0.680095069885625\\
0.126319190068204	0.679827694008528\\
0.144110625289078	0.679524440327599\\
0.161902060509951	0.679187101929413\\
0.179693495730825	0.678817606954419\\
0.197484930951699	0.678417989081759\\
0.215276366172573	0.677990358242563\\
0.233067801393447	0.677536872299917\\
0.25085923661432	0.677059710319348\\
0.268650671835194	0.676561047928932\\
0.286442107056068	0.676043035138878\\
0.304233542276942	0.675507776863482\\
0.322024977497815	0.674957316269655\\
0.339816412718689	0.674393620970425\\
0.357607847939563	0.673818571992164\\
0.375399283160437	0.673233955372174\\
0.393190718381311	0.672641456189184\\
0.410982153602184	0.672042654792172\\
0.428773588823058	0.671439024971312\\
0.446565024043932	0.670831933806381\\
0.464356459264806	0.670222642930459\\
0.482147894485679	0.669612310957559\\
0.499939329706553	0.669001996839858\\
0.517730764927427	0.668392663941163\\
0.535522200148301	0.667785184636637\\
0.553313635369175	0.667180345272909\\
0.571105070590048	0.666578851346694\\
0.588896505810922	0.665981332782807\\
0.606687941031796	0.665388349213723\\
0.62447937625267	0.664800395182024\\
0.642270811473543	0.66421790520417\\
0.660062246694417	0.663641258648899\\
0.677853681915291	0.663070784396314\\
0.695645117136165	0.662506765254475\\
0.713436552357039	0.661949442119204\\
0.731227987577912	0.661399017870078\\
0.749019422798786	0.660855661001383\\
0.76681085801966	0.660319508991375\\
0.784602293240534	0.659790671416603\\
0.802393728461407	0.659269232820721\\
0.820185163682281	0.658755255348938\\
0.837976598903155	0.658248781160533\\
0.855768034124029	0.657749834632558\\
0.873559469344903	0.657258424368153\\
0.891350904565776	0.656774545022966\\
0.90914233978665	0.656298178962908\\
0.926933775007524	0.655829297766098\\
0.944725210228398	0.655367863581349\\
0.962516645449272	0.654913830354938\\
0.980308080670145	0.654467144936728\\
0.998099515891019	0.654027748076068\\
1.01589095111189	0.653595575317157\\
1.03368238633277	0.653170557802908\\
1.05147382155364	0.652752622995632\\
1.06926525677451	0.652341695322263\\
1.08705669199539	0.651937696751194\\
1.10484812721626	0.651540547307192\\
1.12263956243714	0.651150165530358\\
1.14043099765801	0.650766468884528\\
1.15822243287888	0.650389374120082\\
1.17601386809976	0.650018797595632\\
1.19380530332063	0.649654655562695\\
1.2115967385415	0.649296864417056\\
1.22938817376238	0.648945340920184\\
1.24717960898325	0.648600002393748\\
1.26497104420413	0.648260766889995\\
1.282762479425	0.647927553340481\\
1.30055391464587	0.647600281685412\\
1.31834534986675	0.647278872985621\\
1.33613678508762	0.646963249519046\\
1.35392822030849	0.64665333486332\\
1.37171965552937	0.646349053966018\\
1.38951109075024	0.646050333203882\\
1.40730252597112	0.645757100432223\\
};
\addplot [color=black, dashed, forget plot]
  table[row sep=crcr]{%
0	0.641908868402166\\
1.40821449896341	0.641908868402166\\
};
\end{axis}
%
%
\draw[->,-stealth, line width=1.25pt] (2,3.5) to [bend right=30] (1,1.25);

\end{tikzpicture}%

%% file: Figs/Fig11a_Front_Porosity.tex
%
%
\begin{tikzpicture}

\definecolor{mycolor}{rgb}{0.8333,0.5208,0.3317}%

\begin{axis}[%
thick, 
width=0.45\textwidth, 
height=0.4\textwidth,
xmin=0,xmax=1,ymin=0,ymax=1, 
xlabel = {$r/a$},
ylabel = {$\phi$},
legend style = {
draw=none,
fill=none,
font=\scriptsize},
legend cell align={left}]
\addplot [color=mycolor, very thick]
  table[row sep=crcr]{%
0.00125	0.91698135659561\\
0.00375	0.916981351151196\\
0.00625	0.916981339979871\\
0.00875	0.916981322958521\\
0.01125	0.916981300102592\\
0.01375	0.916981271407596\\
0.01625	0.916981236854599\\
0.01875	0.916981196414329\\
0.02125	0.916981150049143\\
0.02375	0.916981097713963\\
0.02625	0.916981039356784\\
0.02875	0.916980974918923\\
0.03125	0.916980904335162\\
0.03375	0.916980827533817\\
0.03625	0.916980744436765\\
0.03875	0.91698065495945\\
0.04125	0.91698055901085\\
0.04375	0.916980456493471\\
0.04625	0.91698034730327\\
0.04875	0.916980231329641\\
0.05125	0.916980108455333\\
0.05375	0.916979978556404\\
0.05625	0.916979841502143\\
0.05875	0.916979697155011\\
0.06125	0.916979545370557\\
0.06375	0.916979385997337\\
0.06625	0.916979218876842\\
0.06875	0.916979043843397\\
0.07125	0.916978860724088\\
0.07375	0.916978669338659\\
0.07625	0.916978469499417\\
0.07875	0.916978261011137\\
0.08125	0.916978043670968\\
0.08375	0.916977817268315\\
0.08625	0.91697758158475\\
0.08875	0.916977336393891\\
0.09125	0.916977081461301\\
0.09375	0.916976816544373\\
0.09625	0.91697654139222\\
0.09875	0.916976255745562\\
0.10125	0.916975959336602\\
0.10375	0.916975651888925\\
0.10625	0.916975333117368\\
0.10875	0.916975002727903\\
0.11125	0.916974660417524\\
0.11375	0.916974305874125\\
0.11625	0.916973938776389\\
0.11875	0.916973558793661\\
0.12125	0.916973165585837\\
0.12375	0.916972758803243\\
0.12625	0.91697233808653\\
0.12875	0.916971903066553\\
0.13125	0.916971453364267\\
0.13375	0.916970988590615\\
0.13625	0.916970508346423\\
0.13875	0.916970012222308\\
0.14125	0.916969499798571\\
0.14375	0.916968970645109\\
0.14625	0.91696842432133\\
0.14875	0.916967860376064\\
0.15125	0.916967278347491\\
0.15375	0.916966677763081\\
0.15625	0.916966058139519\\
0.15875	0.916965418982662\\
0.16125	0.916964759787483\\
0.16375	0.916964080038055\\
0.16625	0.916963379207507\\
0.16875	0.916962656758031\\
0.17125	0.916961912140865\\
0.17375	0.916961144796314\\
0.17625	0.916960354153775\\
0.17875	0.916959539631781\\
0.18125	0.916958700638061\\
0.18375	0.91695783656961\\
0.18625	0.916956946812794\\
0.18875	0.916956030743454\\
0.19125	0.916955087727052\\
0.19375	0.916954117118821\\
0.19625	0.91695311826396\\
0.19875	0.916952090497831\\
0.20125	0.916951033146206\\
0.20375	0.916949945525525\\
0.20625	0.916948826943198\\
0.20875	0.916947676697929\\
0.21125	0.916946494080089\\
0.21375	0.916945278372116\\
0.21625	0.916944028848945\\
0.21875	0.916942744778506\\
0.22125	0.916941425422241\\
0.22375	0.916940070035671\\
0.22625	0.91693867786903\\
0.22875	0.916937248167932\\
0.23125	0.916935780174096\\
0.23375	0.916934273126138\\
0.23625	0.916932726260401\\
0.23875	0.916931138811893\\
0.24125	0.916929510015245\\
0.24375	0.916927839105759\\
0.24625	0.916926125320543\\
0.24875	0.916924367899715\\
0.25125	0.916922566087684\\
0.25375	0.916920719134529\\
0.25625	0.916918826297469\\
0.25875	0.916916886842418\\
0.26125	0.916914900045669\\
0.26375	0.916912865195658\\
0.26625	0.916910781594859\\
0.26875	0.91690864856179\\
0.27125	0.916906465433146\\
0.27375	0.916904231566067\\
0.27625	0.91690194634054\\
0.27875	0.91689960916194\\
0.28125	0.916897219463753\\
0.28375	0.916894776710417\\
0.28625	0.916892280400368\\
0.28875	0.91688973006924\\
0.29125	0.91688712529326\\
0.29375	0.91688446569284\\
0.29625	0.916881750936369\\
0.29875	0.916878980744221\\
0.30125	0.916876154892999\\
0.30375	0.916873273219989\\
0.30625	0.91687033562789\\
0.30875	0.916867342089797\\
0.31125	0.916864292654446\\
0.31375	0.916861187451766\\
0.31625	0.916858026698706\\
0.31875	0.916854810705396\\
0.32125	0.916851539881643\\
0.32375	0.916848214743755\\
0.32625	0.916844835921744\\
0.32875	0.916841404166907\\
0.33125	0.916837920359805\\
0.33375	0.916834385518663\\
0.33625	0.91683080080821\\
0.33875	0.91682716754898\\
0.34125	0.916823487227086\\
0.34375	0.916819761504501\\
0.34625	0.916815992229865\\
0.34875	0.916812181449825\\
0.35125	0.916808331420945\\
0.35375	0.916804444622226\\
0.35625	0.916800523768222\\
0.35875	0.916796571822843\\
0.36125	0.916792592013863\\
0.36375	0.916788587848212\\
0.36625	0.916784563128125\\
0.36875	0.916780521968305\\
0.37125	0.916776468814163\\
0.37375	0.91677240846134\\
0.37625	0.916768346076575\\
0.37875	0.916764287219996\\
0.38125	0.916760237868724\\
0.38375	0.916756204441448\\
0.38625	0.916752193823245\\
0.38875	0.916748213389446\\
0.39125	0.916744271026754\\
0.39375	0.916740375149233\\
0.39625	0.916736534706497\\
0.39875	0.916732759181666\\
0.40125	0.916729058577935\\
0.40375	0.916725443395802\\
0.40625	0.916721924608511\\
0.40875	0.916718513652262\\
0.41125	0.916715222460161\\
0.41375	0.916712063584696\\
0.41625	0.91670905047046\\
0.41875	0.916706197952851\\
0.42125	0.916703523061613\\
0.42375	0.916701046188152\\
0.42625	0.916698792615206\\
0.42875	0.916696794284302\\
0.43125	0.916695091465804\\
0.43375	0.91669373367611\\
0.43625	0.916692778748579\\
0.43875	0.916692288433357\\
0.44125	0.916692318362552\\
0.44375	0.916692899860114\\
0.44625	0.916694011244364\\
0.44875	0.916695537513528\\
0.45125	0.91669722039591\\
0.45375	0.916698606627334\\
0.45625	0.91669901187182\\
0.45875	0.916697531256488\\
0.46125	0.916693143965161\\
0.46375	0.916684974993762\\
0.46625	0.916672784175192\\
0.46875	0.916657737780608\\
0.47125	0.916643462920635\\
0.47375	0.916637268680683\\
0.47625	0.916651225693298\\
0.47875	0.916702537304431\\
0.48125	0.916812373830653\\
0.48375	0.917002222625198\\
0.48625	0.917287049255353\\
0.48875	0.917665344020813\\
0.49125	0.918107331099301\\
0.49375	0.918543635474729\\
0.49625	0.918856598548274\\
0.49875	0.918874603174562\\
0.50125	0.918366473536542\\
0.50375	0.91702880385777\\
0.50625	0.914452614227363\\
0.50875	0.910038273805193\\
0.51125	0.90275600966492\\
0.51375	0.89020160881572\\
0.51625	0.289451354147737\\
0.51875	0.28973756249834\\
0.52125	0.28998243872581\\
0.52375	0.290191881197565\\
0.52625	0.290370543324517\\
0.52875	0.290522175209974\\
0.53125	0.290649853313653\\
0.53375	0.290756140000659\\
0.53625	0.29084319750811\\
0.53875	0.290912871316439\\
0.54125	0.290966752412292\\
0.54375	0.291006224635502\\
0.54625	0.291032501261906\\
0.54875	0.291046653672033\\
0.55125	0.291049634103499\\
0.55375	0.291042293914104\\
0.55625	0.291025398392194\\
0.55875	0.290999638878875\\
0.56125	0.290965642773904\\
0.56375	0.290923981858431\\
0.56625	0.290875179266545\\
0.56875	0.290819715362775\\
0.57125	0.290758032726741\\
0.57375	0.290690540403808\\
0.57625	0.290617617548275\\
0.57875	0.290539616560633\\
0.58125	0.290456865801068\\
0.58375	0.290369671946076\\
0.58625	0.290278322043035\\
0.58875	0.290183085307962\\
0.59125	0.290084214703968\\
0.59375	0.289981948331689\\
0.59625	0.289876510657925\\
0.59875	0.289768113604569\\
0.60125	0.289656957516504\\
0.60375	0.289543232024337\\
0.60625	0.289427116815516\\
0.60875	0.289308782325398\\
0.61125	0.289188390358235\\
0.61375	0.289066094646639\\
0.61625	0.288942041356952\\
0.61875	0.28881636954693\\
0.62125	0.28868921158133\\
0.62375	0.288560693510269\\
0.62625	0.288430935414588\\
0.62875	0.288300051721944\\
0.63125	0.288168151496883\\
0.63375	0.288035338707754\\
0.63625	0.287901712472978\\
0.63875	0.287767367288888\\
0.64125	0.287632393241084\\
0.64375	0.28749687620105\\
0.64625	0.287360898009528\\
0.64875	0.28722453664802\\
0.65125	0.287087866399602\\
0.65375	0.286950958000119\\
0.65625	0.286813878780675\\
0.65875	0.286676692802284\\
0.66125	0.286539460983391\\
0.66375	0.286402241220942\\
0.66625	0.286265088505563\\
0.66875	0.28612805503139\\
0.67125	0.285991190300991\\
0.67375	0.285854541225792\\
0.67625	0.285718152222381\\
0.67875	0.285582065305\\
0.68125	0.285446320174515\\
0.68375	0.285310954304126\\
0.68625	0.285176003022034\\
0.68875	0.285041499591268\\
0.69125	0.284907475286863\\
0.69375	0.284773959470526\\
0.69625	0.284640979662955\\
0.69875	0.284508561613921\\
0.70125	0.284376729370229\\
0.70375	0.284245505341665\\
0.70625	0.284114910365007\\
0.70875	0.283984963766189\\
0.71125	0.283855683420682\\
0.71375	0.283727085812163\\
0.71625	0.283599186089522\\
0.71875	0.283471998122263\\
0.72125	0.283345534554348\\
0.72375	0.283219806856515\\
0.72625	0.283094825377127\\
0.72875	0.282970599391566\\
0.73125	0.282847137150225\\
0.73375	0.28272444592511\\
0.73625	0.282602532055094\\
0.73875	0.282481400989838\\
0.74125	0.282361057332414\\
0.74375	0.282241504880646\\
0.74625	0.282122746667191\\
0.74875	0.282004784998389\\
0.75125	0.281887621491897\\
0.75375	0.281771257113127\\
0.75625	0.281655692210508\\
0.75875	0.281540926549597\\
0.76125	0.281426959346056\\
0.76375	0.281313789297501\\
0.76625	0.281201414614272\\
0.76875	0.281089833049107\\
0.77125	0.28097904192577\\
0.77375	0.28086903816663\\
0.77625	0.280759818319228\\
0.77875	0.280651378581837\\
0.78125	0.28054371482804\\
0.78375	0.280436822630348\\
0.78625	0.280330697282875\\
0.78875	0.280225333823092\\
0.79125	0.28012072705267\\
0.79375	0.280016871557451\\
0.79625	0.27991376172655\\
0.79875	0.279811391770613\\
0.80125	0.279709755739255\\
0.80375	0.2796088475377\\
0.80625	0.279508660942626\\
0.80875	0.279409189617268\\
0.81125	0.279310427125757\\
0.81375	0.279212366946762\\
0.81625	0.279115002486409\\
0.81875	0.27901832709054\\
0.82125	0.278922334056295\\
0.82375	0.278827016643064\\
0.82625	0.278732368082821\\
0.82875	0.278638381589845\\
0.83125	0.278545050369875\\
0.83375	0.278452367628696\\
0.83625	0.278360326580183\\
0.83875	0.278268920453828\\
0.84125	0.278178142501759\\
0.84375	0.27808798600527\\
0.84625	0.277998444280896\\
0.84875	0.277909510686027\\
0.85125	0.277821178624097\\
0.85375	0.27773344154936\\
0.85625	0.277646292971269\\
0.85875	0.277559726458473\\
0.86125	0.277473735642454\\
0.86375	0.277388314220817\\
0.86625	0.277303455960251\\
0.86875	0.277219154699166\\
0.87125	0.277135404350043\\
0.87375	0.277052198901485\\
0.87625	0.276969532420006\\
0.87875	0.276887399051558\\
0.88125	0.276805793022816\\
0.88375	0.276724708642229\\
0.88625	0.27664414030086\\
0.88875	0.276564082473009\\
0.89125	0.276484529716653\\
0.89375	0.276405476673698\\
0.89625	0.276326918070062\\
0.89875	0.276248848715596\\
0.90125	0.276171263503855\\
0.90375	0.276094157411733\\
0.90625	0.276017525498961\\
0.90875	0.275941362907488\\
0.91125	0.275865664860752\\
0.91375	0.275790426662839\\
0.91625	0.275715643697554\\
0.91875	0.275641311427404\\
0.92125	0.275567425392491\\
0.92375	0.275493981209344\\
0.92625	0.275420974569675\\
0.92875	0.275348401239081\\
0.93125	0.275276257055684\\
0.93375	0.275204537928732\\
0.93625	0.27513323983715\\
0.93875	0.275062358828056\\
0.94125	0.274991891015242\\
0.94375	0.27492183257763\\
0.94625	0.274852179757701\\
0.94875	0.274782928859905\\
0.95125	0.274714076249059\\
0.95375	0.274645618348727\\
0.95625	0.274577551639592\\
0.95875	0.274509872657827\\
0.96125	0.274442577993462\\
0.96375	0.274375664288743\\
0.96625	0.27430912823651\\
0.96875	0.274242966578564\\
0.97125	0.27417717610405\\
0.97375	0.274111753647855\\
0.97625	0.274046696089\\
0.97875	0.273982000349069\\
0.98125	0.273917663390633\\
0.98375	0.273853682215703\\
0.98625	0.2737900538642\\
0.98875	0.273726775412439\\
0.99125	0.273663843971641\\
0.99375	0.273601256686464\\
0.99625	0.273539010733557\\
0.99875	0.273477103320142\\
};
\addlegendentry{Full soln.};
\addplot [color=black, dashed, very thick]
  table[row sep=crcr]{%
0	0.91698730536472\\
0.01	0.91698730536472\\
0.02	0.91698730536472\\
0.03	0.91698730536472\\
0.04	0.91698730536472\\
0.05	0.91698730536472\\
0.06	0.91698730536472\\
0.07	0.91698730536472\\
0.08	0.91698730536472\\
0.09	0.91698730536472\\
0.1	0.91698730536472\\
0.11	0.91698730536472\\
0.12	0.91698730536472\\
0.13	0.91698730536472\\
0.14	0.91698730536472\\
0.15	0.91698730536472\\
0.16	0.91698730536472\\
0.17	0.91698730536472\\
0.18	0.91698730536472\\
0.19	0.91698730536472\\
0.2	0.91698730536472\\
0.21	0.91698730536472\\
0.22	0.91698730536472\\
0.23	0.91698730536472\\
0.24	0.91698730536472\\
0.25	0.91698730536472\\
0.26	0.91698730536472\\
0.27	0.91698730536472\\
0.28	0.91698730536472\\
0.29	0.91698730536472\\
0.3	0.91698730536472\\
0.31	0.91698730536472\\
0.32	0.91698730536472\\
0.33	0.91698730536472\\
0.34	0.91698730536472\\
0.35	0.91698730536472\\
0.36	0.91698730536472\\
0.37	0.91698730536472\\
0.38	0.91698730536472\\
0.39	0.91698730536472\\
0.4	0.91698730536472\\
0.41	0.91698730536472\\
0.42	0.91698730536472\\
0.43	0.91698730536472\\
0.44	0.91698730536472\\
0.45	0.91698730536472\\
0.46	0.91698730536472\\
0.47	0.91698730536472\\
0.48	0.91698730536472\\
0.49	0.91698730536472\\
0.5	0.91698730536472\\
0.51	0.91698730536472\\
0.515	0.273766115912442\\
0.525	0.273766115912442\\
0.535	0.273766115912442\\
0.545	0.273766115912442\\
0.555	0.273766115912442\\
0.565	0.273766115912442\\
0.575	0.273766115912442\\
0.585	0.273766115912442\\
0.595	0.273766115912442\\
0.605	0.273766115912442\\
0.615	0.273766115912442\\
0.625	0.273766115912442\\
0.635	0.273766115912442\\
0.645	0.273766115912442\\
0.655	0.273766115912442\\
0.665	0.273766115912442\\
0.675	0.273766115912442\\
0.685	0.273766115912442\\
0.695	0.273766115912442\\
0.705	0.273766115912442\\
0.715	0.273766115912442\\
0.725	0.273766115912442\\
0.735	0.273766115912442\\
0.745	0.273766115912442\\
0.755	0.273766115912442\\
0.765	0.273766115912442\\
0.775	0.273766115912442\\
0.785	0.273766115912442\\
0.795	0.273766115912442\\
0.805	0.273766115912442\\
0.815	0.273766115912442\\
0.825	0.273766115912442\\
0.835	0.273766115912442\\
0.845	0.273766115912442\\
0.855	0.273766115912442\\
0.865	0.273766115912442\\
0.875	0.273766115912442\\
0.885	0.273766115912442\\
0.895	0.273766115912442\\
0.905	0.273766115912442\\
0.915	0.273766115912442\\
0.925	0.273766115912442\\
0.935	0.273766115912442\\
0.945	0.273766115912442\\
0.955	0.273766115912442\\
0.965	0.273766115912442\\
0.975	0.273766115912442\\
0.985	0.273766115912442\\
0.995	0.273766115912442\\
};
\addlegendentry{Step-fn.};
\end{axis}

\end{tikzpicture}%

%% file: Figs/Fig11b_Front_PorosityPert.tex
%
%
\begin{tikzpicture}

\definecolor{mycolor}{rgb}{0.8333,0.5208,0.3317}%

\begin{axis}[%
thick, 
width=0.45\textwidth, 
height=0.4\textwidth,
xmin=0, xmax=1, ymin=-0.01, ymax=0.025,
xlabel={$r/a$},
ylabel={$\porosity-\zerothorder{\porosity}_\pm$},
yticklabel style={
        /pgf/number format/fixed,
        /pgf/number format/precision=5
},
scaled y ticks=false
]
\addplot [color=mycolor, very thick, forget plot]
  table[row sep=crcr]{%
0.00125	-5.94876911053088e-06\\
0.00375	-5.95421352433956e-06\\
0.00625	-5.96538484953779e-06\\
0.00875	-5.98240619964496e-06\\
0.01125	-6.00526212779506e-06\\
0.01375	-6.03395712439614e-06\\
0.01625	-6.06851012119325e-06\\
0.01875	-6.10895039143688e-06\\
0.02125	-6.15531557734972e-06\\
0.02375	-6.20765075698415e-06\\
0.02625	-6.26600793585119e-06\\
0.02875	-6.3304457972313e-06\\
0.03125	-6.40102955806743e-06\\
0.03375	-6.47783090323983e-06\\
0.03625	-6.56092795503493e-06\\
0.03875	-6.65040527036975e-06\\
0.04125	-6.74635387021283e-06\\
0.04375	-6.84887124957623e-06\\
0.04625	-6.95806144990208e-06\\
0.04875	-7.07403507904658e-06\\
0.05125	-7.19690938710826e-06\\
0.05375	-7.32680831661003e-06\\
0.05625	-7.46386257743925e-06\\
0.05875	-7.60820970902021e-06\\
0.06125	-7.75999416302575e-06\\
0.06375	-7.91936738320231e-06\\
0.06625	-8.08648787842259e-06\\
0.06875	-8.26152132282765e-06\\
0.07125	-8.44464063232131e-06\\
0.07375	-8.63602606127056e-06\\
0.07625	-8.83586530353586e-06\\
0.07875	-9.04435358339839e-06\\
0.08125	-9.26169375270458e-06\\
0.08375	-9.4880964053301e-06\\
0.08625	-9.72377997021656e-06\\
0.08875	-9.9689708289441e-06\\
0.09125	-1.02239034196483e-05\\
0.09375	-1.04888203475983e-05\\
0.09625	-1.07639725006603e-05\\
0.09875	-1.10496191583209e-05\\
0.10125	-1.13460281186972e-05\\
0.10375	-1.16534757949038e-05\\
0.10625	-1.19722473522765e-05\\
0.10875	-1.23026368170631e-05\\
0.11125	-1.26449471965495e-05\\
0.11375	-1.29994905948561e-05\\
0.11625	-1.33665883313983e-05\\
0.11875	-1.37465710587925e-05\\
0.12125	-1.41397788832043e-05\\
0.12375	-1.45465614769247e-05\\
0.12625	-1.49672781906141e-05\\
0.12875	-1.54022981668778e-05\\
0.13125	-1.58520004530649e-05\\
0.13375	-1.63167741055181e-05\\
0.13625	-1.67970182971544e-05\\
0.13875	-1.72931424118339e-05\\
0.14125	-1.7805566149165e-05\\
0.14375	-1.83347196113237e-05\\
0.14625	-1.88810433902065e-05\\
0.14875	-1.94449886564696e-05\\
0.15125	-2.00270172295847e-05\\
0.15375	-2.06276016395668e-05\\
0.15625	-2.12472252015816e-05\\
0.15875	-2.18863820578008e-05\\
0.16125	-2.25455772367988e-05\\
0.16375	-2.32253266657656e-05\\
0.16625	-2.39261572130278e-05\\
0.16875	-2.46486066890483e-05\\
0.17125	-2.53932238551968e-05\\
0.17375	-2.61605684060973e-05\\
0.17625	-2.69512109456471e-05\\
0.17875	-2.77657329388337e-05\\
0.18125	-2.86047266596645e-05\\
0.18375	-2.94687951101213e-05\\
0.18625	-3.03585519264571e-05\\
0.18875	-3.12746212663972e-05\\
0.19125	-3.22176376678085e-05\\
0.19375	-3.31882458990407e-05\\
0.19625	-3.41871007601968e-05\\
0.19875	-3.52148668888441e-05\\
0.20125	-3.62722185147657e-05\\
0.20375	-3.73598391952834e-05\\
0.20625	-3.8478421522381e-05\\
0.20875	-3.96286667914136e-05\\
0.21125	-4.08112846308484e-05\\
0.21375	-4.20269926039163e-05\\
0.21625	-4.32765157755144e-05\\
0.21875	-4.45605862138265e-05\\
0.22125	-4.58799424796208e-05\\
0.22375	-4.72353290488225e-05\\
0.22625	-4.86274956902344e-05\\
0.22875	-5.00571967886332e-05\\
0.23125	-5.15251906241243e-05\\
0.23375	-5.3032238582551e-05\\
0.23625	-5.45791043194965e-05\\
0.23875	-5.61665528269195e-05\\
0.24125	-5.77953494750316e-05\\
0.24375	-5.94662589614714e-05\\
0.24625	-6.11800441776555e-05\\
0.24875	-6.29374650051862e-05\\
0.25125	-6.47392770360966e-05\\
0.25375	-6.6586230190846e-05\\
0.25625	-6.84790672516034e-05\\
0.25875	-7.04185223018294e-05\\
0.26125	-7.2405319051505e-05\\
0.26375	-7.44401690622176e-05\\
0.26625	-7.65237698612253e-05\\
0.26875	-7.86568029305101e-05\\
0.27125	-8.08399315738173e-05\\
0.27375	-8.30737986527996e-05\\
0.27625	-8.5359024180498e-05\\
0.27875	-8.76962027805961e-05\\
0.28125	-9.00859009677069e-05\\
0.28375	-9.25286543033144e-05\\
0.28625	-9.50249643519863e-05\\
0.28875	-9.75752954799347e-05\\
0.29125	-0.000100180071460509\\
0.29375	-0.000102839671880073\\
0.29625	-0.000105554428351473\\
0.29875	-0.000108324620498923\\
0.30125	-0.000111150471720833\\
0.30375	-0.00011403214473138\\
0.30625	-0.000116969736829842\\
0.30875	-0.000119963274923585\\
0.31125	-0.000123012710274151\\
0.31375	-0.000126117912953805\\
0.31625	-0.000129278666014754\\
0.31875	-0.0001324946593243\\
0.32125	-0.000135765483077144\\
0.32375	-0.000139090620965177\\
0.32625	-0.000142469442976467\\
0.32875	-0.000145901197813214\\
0.33125	-0.000149385004915015\\
0.33375	-0.000152919846057475\\
0.33625	-0.000156504556510173\\
0.33875	-0.000160137815740424\\
0.34125	-0.000163818137634442\\
0.34375	-0.000167543860218999\\
0.34625	-0.0001713131348553\\
0.34875	-0.000175123914895048\\
0.35125	-0.000178973943774863\\
0.35375	-0.000182860742494406\\
0.35625	-0.000186781596498209\\
0.35875	-0.000190733541877708\\
0.36125	-0.00019471335085719\\
0.36375	-0.000198717516508462\\
0.36625	-0.00020274223659511\\
0.36875	-0.000206783396415777\\
0.37125	-0.000210836550557647\\
0.37375	-0.00021489690338039\\
0.37625	-0.000218959288145526\\
0.37875	-0.000223018144723919\\
0.38125	-0.000227067495995881\\
0.38375	-0.000231100923272276\\
0.38625	-0.000235111541475042\\
0.38875	-0.000239091975273831\\
0.39125	-0.000243034337966241\\
0.39375	-0.000246930215487495\\
0.39625	-0.000250770658223431\\
0.39875	-0.000254546183054538\\
0.40125	-0.000258246786785765\\
0.40375	-0.000261861968917865\\
0.40625	-0.000265380756208966\\
0.40875	-0.000268791712458638\\
0.41125	-0.000272082904559268\\
0.41375	-0.000275241780024138\\
0.41625	-0.000278254894260366\\
0.41875	-0.000281107411868819\\
0.42125	-0.000283782303106994\\
0.42375	-0.000286259176567993\\
0.42625	-0.000288512749514136\\
0.42875	-0.000290511080418554\\
0.43125	-0.000292213898915983\\
0.43375	-0.00029357168861055\\
0.43625	-0.000294526616140933\\
0.43875	-0.000295016931362801\\
0.44125	-0.00029498700216779\\
0.44375	-0.000294405504606243\\
0.44625	-0.000293294120356502\\
0.44875	-0.000291767851192559\\
0.45125	-0.000290084968809889\\
0.45375	-0.000288698737386683\\
0.45625	-0.000288293492900715\\
0.45875	-0.000289774108231833\\
0.46125	-0.000294161399559179\\
0.46375	-0.000302330370958281\\
0.46625	-0.000314521189528016\\
0.46875	-0.000329567584112733\\
0.47125	-0.000343842444085141\\
0.47375	-0.000350036684037214\\
0.47625	-0.000336079671421796\\
0.47875	-0.000284768060288987\\
0.48125	-0.000174931534067557\\
0.48375	1.49172604777537e-05\\
0.48625	0.000299743890632809\\
0.48875	0.00067803865609295\\
0.49125	0.00112002573458059\\
0.49375	0.00155633011000911\\
0.49625	0.00186929318355389\\
0.49875	0.00188729780984209\\
0.50125	0.00137916817182193\\
0.50375	4.14984930493523e-05\\
0.50625	-0.00253469113735694\\
0.50875	-0.00694903155952697\\
0.51125	-0.0142312956998004\\
0.51375	-0.0267856965490005\\
};
\addplot [color=black, dashed, very thick, forget plot]
  table[row sep=crcr]{%
0	0\\
0.515	0\\
};
\addplot [color=black, dotted, forget plot]
  table[row sep=crcr]{%
0.515	-0.01\\
0.515	0.04\\
};
\addplot [color=mycolor, very thick, forget plot]
  table[row sep=crcr]{%
0.51625	0.0156852382352949\\
0.51875	0.0159714465858973\\
0.52125	0.0162163228133679\\
0.52375	0.0164257652851224\\
0.52625	0.0166044274120745\\
0.52875	0.0167560592975312\\
0.53125	0.0168837374012108\\
0.53375	0.0169900240882166\\
0.53625	0.0170770815956676\\
0.53875	0.0171467554039966\\
0.54125	0.0172006364998495\\
0.54375	0.0172401087230591\\
0.54625	0.0172663853494635\\
0.54875	0.0172805377595904\\
0.55125	0.0172835181910564\\
0.55375	0.0172761780016611\\
0.55625	0.0172592824797518\\
0.55875	0.0172335229664327\\
0.56125	0.0171995268614615\\
0.56375	0.0171578659459887\\
0.56625	0.0171090633541026\\
0.56875	0.0170535994503325\\
0.57125	0.0169919168142982\\
0.57375	0.016924424491366\\
0.57625	0.0168515016358323\\
0.57875	0.0167735006481902\\
0.58125	0.0166907498886255\\
0.58375	0.0166035560336336\\
0.58625	0.0165122061305925\\
0.58875	0.0164169693955201\\
0.59125	0.0163180987915259\\
0.59375	0.0162158324192463\\
0.59625	0.0161103947454822\\
0.59875	0.0160019976921267\\
0.60125	0.0158908416040615\\
0.60375	0.015777116111895\\
0.60625	0.0156610009030739\\
0.60875	0.0155426664129561\\
0.61125	0.0154222744457926\\
0.61375	0.0152999787341969\\
0.61625	0.0151759254445099\\
0.61875	0.0150502536344874\\
0.62125	0.0149230956688872\\
0.62375	0.0147945775978264\\
0.62625	0.0146648195021454\\
0.62875	0.0145339358095014\\
0.63125	0.0144020355844403\\
0.63375	0.0142692227953112\\
0.63625	0.0141355965605358\\
0.63875	0.0140012513764452\\
0.64125	0.0138662773286418\\
0.64375	0.013730760288608\\
0.64625	0.0135947820970857\\
0.64875	0.0134584207355771\\
0.65125	0.0133217504871599\\
0.65375	0.0131848420876766\\
0.65625	0.0130477628682327\\
0.65875	0.0129105768898413\\
0.66125	0.0127733450709486\\
0.66375	0.0126361253084992\\
0.66625	0.0124989725931202\\
0.66875	0.0123619391189478\\
0.67125	0.0122250743885486\\
0.67375	0.0120884253133497\\
0.67625	0.0119520363099389\\
0.67875	0.0118159493925576\\
0.68125	0.0116802042620728\\
0.68375	0.0115448383916841\\
0.68625	0.0114098871095913\\
0.68875	0.0112753836788258\\
0.69125	0.0111413593744203\\
0.69375	0.0110078435580835\\
0.69625	0.0108748637505128\\
0.69875	0.0107424457014785\\
0.70125	0.0106106134577865\\
0.70375	0.0104793894292224\\
0.70625	0.0103487944525644\\
0.70875	0.0102188478537462\\
0.71125	0.0100895675082392\\
0.71375	0.00996096989972028\\
0.71625	0.00983307017707929\\
0.71875	0.00970588220982094\\
0.72125	0.00957941864190565\\
0.72375	0.00945369094407306\\
0.72625	0.00932870946468467\\
0.72875	0.00920448347912378\\
0.73125	0.00908102123778276\\
0.73375	0.00895833001266788\\
0.73625	0.00883641614265157\\
0.73875	0.00871528507739566\\
0.74125	0.00859494141997202\\
0.74375	0.00847538896820371\\
0.74625	0.00835663075474841\\
0.74875	0.00823866908594684\\
0.75125	0.00812150557945512\\
0.75375	0.00800514120068474\\
0.75625	0.00788957629806536\\
0.75875	0.00777481063715502\\
0.76125	0.00766084343361334\\
0.76375	0.00754767338505907\\
0.76625	0.00743529870182991\\
0.76875	0.00732371713666496\\
0.77125	0.00721292601332718\\
0.77375	0.00710292225418718\\
0.77625	0.0069937024067856\\
0.77875	0.0068852626693946\\
0.78125	0.00677759891559743\\
0.78375	0.00667070671790532\\
0.78625	0.00656458137043303\\
0.78875	0.00645921791064935\\
0.79125	0.00635461114022745\\
0.79375	0.00625075564500888\\
0.79625	0.00614764581410765\\
0.79875	0.00604527585817022\\
0.80125	0.00594363982681279\\
0.80375	0.00584273162525728\\
0.80625	0.00574254503018407\\
0.80875	0.00564307370482547\\
0.81125	0.00554431121331506\\
0.81375	0.00544625103431939\\
0.81625	0.00534888657396698\\
0.81875	0.00525221117809754\\
0.82125	0.00515621814385214\\
0.82375	0.0050609007306221\\
0.82625	0.00496625217037883\\
0.82875	0.00487226567740284\\
0.83125	0.00477893445743288\\
0.83375	0.00468625171625325\\
0.83625	0.00459421066774052\\
0.83875	0.00450280454138585\\
0.84125	0.00441202658931616\\
0.84375	0.00432187009282764\\
0.84625	0.00423232836845366\\
0.84875	0.0041433947735845\\
0.85125	0.00405506271165468\\
0.85375	0.00396732563691804\\
0.85625	0.00388017705882698\\
0.85875	0.00379361054603061\\
0.86125	0.00370761973001138\\
0.86375	0.00362219830837501\\
0.86625	0.00353734004780859\\
0.86875	0.00345303878672404\\
0.87125	0.00336928843760059\\
0.87375	0.00328608298904243\\
0.87625	0.00320341650756339\\
0.87875	0.00312128313911536\\
0.88125	0.00303967711037334\\
0.88375	0.00295859272978705\\
0.88625	0.00287802438841778\\
0.88875	0.00279796656056663\\
0.89125	0.00271841380421045\\
0.89375	0.0026393607612557\\
0.89625	0.00256080215761972\\
0.89875	0.00248273280315342\\
0.90125	0.0024051475914128\\
0.90375	0.00232804149929067\\
0.90625	0.00225140958651832\\
0.90875	0.00217524699504573\\
0.91125	0.00209954894830933\\
0.91375	0.00202431075039622\\
0.91625	0.00194952778511198\\
0.91875	0.00187519551496174\\
0.92125	0.00180130948004881\\
0.92375	0.00172786529690172\\
0.92625	0.00165485865723286\\
0.92875	0.00158228532663823\\
0.93125	0.00151014114324116\\
0.93375	0.00143842201628958\\
0.93625	0.00136712392470806\\
0.93875	0.00129624291561403\\
0.94125	0.00122577510280009\\
0.94375	0.00115571666518771\\
0.94625	0.00108606384525817\\
0.94875	0.00101681294746236\\
0.95125	0.00094796033661676\\
0.95375	0.000879502436284207\\
0.95625	0.000811435727149146\\
0.95875	0.000743756745384627\\
0.96125	0.000676462081019125\\
0.96375	0.000609548376300839\\
0.96625	0.000543012324067726\\
0.96875	0.000476850666121187\\
0.97125	0.000411060191608081\\
0.97375	0.000345637735412241\\
0.97625	0.000280580176557743\\
0.97875	0.00021588443662629\\
0.98125	0.000151547478190184\\
0.98375	8.75663032608442e-05\\
0.98625	2.39379517579241e-05\\
0.98875	-3.93405000029268e-05\\
0.99125	-0.000102271940801202\\
0.99375	-0.000164859225978808\\
0.99625	-0.0002271051788858\\
0.99875	-0.000289012592299964\\
};
\addplot [color=black, dashed, very thick, forget plot]
  table[row sep=crcr]{%
0.515	0.0158302444733754\\
0.516215538847118	0.015921682052857\\
0.517431077694236	0.0160034443912618\\
0.518646616541353	0.0160764113465881\\
0.519862155388471	0.016141347809737\\
0.521077694235589	0.0161989234857191\\
0.522293233082707	0.016249728515231\\
0.523508771929825	0.0162942859577997\\
0.524724310776942	0.0163330618786002\\
0.52593984962406	0.0163664735708815\\
0.527155388471178	0.0163948963177659\\
0.528370927318296	0.0164186689996028\\
0.529586466165414	0.0164380987788208\\
0.530802005012531	0.0164534650365145\\
0.532017543859649	0.0164650227049658\\
0.533233082706767	0.0164730051081402\\
0.534448621553885	0.0164776263912187\\
0.535664160401003	0.0164790836098757\\
0.53687969924812	0.0164775585352956\\
0.538095238095238	0.0164732192262479\\
0.539310776942356	0.0164662213975792\\
0.540526315789474	0.0164567096179158\\
0.541741854636591	0.0164448183646647\\
0.542957393483709	0.0164306729518489\\
0.544172932330827	0.0164143903529271\\
0.545388471177945	0.0163960799332014\\
0.546604010025063	0.0163758440987887\\
0.54781954887218	0.0163537788773041\\
0.549035087719298	0.0163299744385108\\
0.550250626566416	0.0163045155612995\\
0.551466165413534	0.0162774820570113\\
0.552681704260652	0.0162489491463716\\
0.553897243107769	0.0162189878025171\\
0.555112781954887	0.0161876650648535\\
0.556328320802005	0.0161550443205294\\
0.557543859649123	0.0161211855613944\\
0.558759398496241	0.0160861456188762\\
0.559974937343358	0.016049978378577\\
0.561190476190476	0.0160127349766888\\
0.562406015037594	0.0159744639797789\\
0.563621553884712	0.0159352115495752\\
0.56483709273183	0.015895021595672\\
0.566052631578947	0.0158539359150662\\
0.567268170426065	0.0158119943208411\\
0.568483709273183	0.015769234762192\\
0.569699248120301	0.0157256934340291\\
0.570914786967418	0.0156814048796286\\
0.572130325814536	0.0156364020855685\\
0.573345864661654	0.015590716569304\\
0.574561403508772	0.0155443784611431\\
0.57577694235589	0.0154974165805557\\
0.576992481203008	0.0154498585072866\\
0.578208020050125	0.01540173064681\\
0.579423558897243	0.0153530582922871\\
0.580639097744361	0.0153038656830784\\
0.581854636591479	0.0152541760583534\\
0.583070175438597	0.0152040117075624\\
0.584285714285714	0.0151533940180351\\
0.585501253132832	0.015102343519645\\
0.58671679197995	0.0150508799264649\\
0.587932330827068	0.0149990221759043\\
0.589147869674185	0.0149467884661053\\
0.590363408521303	0.0148941962902818\\
0.591578947368421	0.0148412624696357\\
0.592794486215539	0.0147880031846921\\
0.594010025062657	0.0147344340040008\\
0.595225563909774	0.0146805699115041\\
0.596441102756892	0.0146264253329776\\
0.59765664160401	0.0145720141607553\\
0.598872180451128	0.0145173497762398\\
0.600087719298246	0.0144624450726525\\
0.601303258145363	0.0144073124760288\\
0.602518796992481	0.0143519639639418\\
0.603734335839599	0.0142964110855723\\
0.604949874686717	0.0142406649787502\\
0.606165413533835	0.0141847363866871\\
0.607380952380952	0.0141286356753451\\
0.60859649122807	0.0140723728473457\\
0.609812030075188	0.0140159575563153\\
0.611027568922306	0.0139593991220065\\
0.612243107769423	0.0139027065426837\\
0.613458646616541	0.0138458885077709\\
0.614674185463659	0.0137889534099091\\
0.615889724310777	0.0137319093561612\\
0.617105263157895	0.0136747641789856\\
0.618320802005012	0.013617525446417\\
0.61953634085213	0.0135602004723683\\
0.620751879699248	0.0135027963258956\\
0.621967418546366	0.013445319840042\\
0.623182957393484	0.0133877776208737\\
0.624398496240602	0.0133301760555461\\
0.625614035087719	0.0132725213200112\\
0.626829573934837	0.0132148193869768\\
0.628045112781955	0.0131570760329502\\
0.629260651629073	0.0130992968449756\\
0.630476190476191	0.0130414872273761\\
0.631691729323308	0.0129836524088103\\
0.632907268170426	0.0129257974475858\\
0.634122807017544	0.0128679272373143\\
0.635338345864662	0.0128100465135115\\
0.63655388471178	0.0127521598587863\\
0.637769423558897	0.0126942717069259\\
0.638984962406015	0.0126363863485502\\
0.640200501253133	0.0125785079368655\\
0.641416040100251	0.0125206404911596\\
0.642631578947368	0.0124627879010047\\
0.643847117794486	0.012404953930883\\
0.645062656641604	0.0123471422249434\\
0.646278195488722	0.0122893563101246\\
0.64749373433584	0.0122315995997193\\
0.648709273182957	0.0121738753982775\\
0.649924812030075	0.012116186904005\\
0.651140350877193	0.01205853721192\\
0.652355889724311	0.0120009293180694\\
0.653571428571429	0.0119433661221451\\
0.654786967418546	0.0118858504305786\\
0.656002506265664	0.0118283849595267\\
0.657218045112782	0.0117709723374553\\
0.6584335839599	0.0117136151082158\\
0.659649122807017	0.0116563157331361\\
0.660864661654135	0.0115990765936128\\
0.662080200501253	0.011541899994205\\
0.663295739348371	0.0114847881647561\\
0.664511278195489	0.011427743262138\\
0.665726817042607	0.0113707673728045\\
0.666942355889724	0.0113138625149735\\
0.668157894736842	0.0112570306407372\\
0.66937343358396	0.0112002736375103\\
0.670588972431078	0.0111435933302996\\
0.671804511278196	0.0110869914836135\\
0.673020050125313	0.0110304698036058\\
0.674235588972431	0.0109740299392712\\
0.675451127819549	0.010917673483585\\
0.676666666666667	0.0108614019766584\\
0.677882205513784	0.0108052169061778\\
0.679097744360902	0.0107491197086803\\
0.68031328320802	0.010693111772261\\
0.681528822055138	0.0106371944368635\\
0.682744360902256	0.0105813689957103\\
0.683959899749373	0.0105256366972805\\
0.685175438596491	0.0104699987460606\\
0.686390977443609	0.0104144563044394\\
0.687606516290727	0.0103590104930831\\
0.688822055137845	0.0103036623924559\\
0.690037593984962	0.0102484130443044\\
0.69125313283208	0.0101932634522186\\
0.692468671679198	0.0101382145830463\\
0.693684210526316	0.0100832673679789\\
0.694899749373434	0.0100284227030131\\
0.696115288220551	0.00997368145026984\\
0.697330827067669	0.00991904443928349\\
0.698546365914787	0.0098645124676704\\
0.699761904761905	0.00981008630147351\\
0.700977443609023	0.00975576667666422\\
0.70219298245614	0.00970155430032227\\
0.703408521303258	0.00964744985060654\\
0.704624060150376	0.00959345397788556\\
0.705839598997494	0.00953956730554853\\
0.707055137844612	0.00948579043020168\\
0.708270676691729	0.00943212392332256\\
0.709486215538847	0.00937856833200514\\
0.710701754385965	0.0093251241782043\\
0.711917293233083	0.00927179196062436\\
0.713132832080201	0.00921857215540431\\
0.714348370927318	0.00916546521619241\\
0.715563909774436	0.00911247157509063\\
0.716779448621554	0.00905959164328554\\
0.717994987468672	0.00900682581136628\\
0.719210526315789	0.00895417444962599\\
0.720426065162907	0.00890163790893938\\
0.721641604010025	0.00884921652192063\\
0.722857142857143	0.00879691060288257\\
0.724072681704261	0.00874472044748517\\
0.725288220551378	0.00869264633444189\\
0.726503759398496	0.0086406885257324\\
0.727719298245614	0.00858884726680216\\
0.728934837092732	0.00853712278704488\\
0.73015037593985	0.00848551529997679\\
0.731365914786967	0.00843402500428649\\
0.732581453634085	0.00838265208413331\\
0.733796992481203	0.00833139670913805\\
0.735012531328321	0.00828025903465883\\
0.736228070175439	0.00822923920264797\\
0.737443609022556	0.00817833734220263\\
0.738659147869674	0.00812755356921716\\
0.739874686716792	0.00807688798691363\\
0.74109022556391	0.00802634068680688\\
0.742305764411028	0.00797591174832812\\
0.743521303258145	0.0079256012393276\\
0.744736842105263	0.00787540921671648\\
0.745952380952381	0.00782533572650815\\
0.747167919799499	0.00777538080370327\\
0.748383458646617	0.00772554447320272\\
0.749598997493734	0.00767582674997277\\
0.750814536340852	0.00762622763905455\\
0.75203007518797	0.00757674713645419\\
0.753245614035088	0.00752738522854571\\
0.754461152882206	0.00747814189265086\\
0.755676691729323	0.00742901709761228\\
0.756892230576441	0.00738001080376781\\
0.758107769423559	0.00733112296291859\\
0.759323308270677	0.00728235351903067\\
0.760538847117794	0.00723370240848663\\
0.761754385964912	0.00718516955973926\\
0.76296992481203	0.00713675489399532\\
0.764185463659148	0.00708845832515385\\
0.765401002506266	0.00704027976033086\\
0.766616541353383	0.00699221909993467\\
0.767832080200501	0.00694427623758821\\
0.769047619047619	0.00689645106063803\\
0.770263157894737	0.00684874345021457\\
0.771478696741855	0.0068011532812875\\
0.772694235588972	0.00675368042316046\\
0.77390977443609	0.00670632473936916\\
0.775125313283208	0.00665908608757516\\
0.776340852130326	0.00661196432063917\\
0.777556390977444	0.00656495928591426\\
0.778771929824561	0.0065180708254229\\
0.779987468671679	0.00647129877662202\\
0.781203007518797	0.00642464297227613\\
0.782418546365915	0.00637810324032665\\
0.783634085213033	0.00633167940375759\\
0.78484962406015	0.00628537128193754\\
0.786065162907268	0.00623917869003409\\
0.787280701754386	0.00619310143857298\\
0.788496240601504	0.00614713933432574\\
0.789711779448622	0.00610129218001013\\
0.790927318295739	0.00605555977487536\\
0.792142857142857	0.00600994191439622\\
0.793358395989975	0.00596443839055617\\
0.794573934837093	0.00591904899212749\\
0.79578947368421	0.00587377350420847\\
0.797005012531328	0.0058286117090899\\
0.798220551378446	0.0057835633857867\\
0.799436090225564	0.0057386283101589\\
0.800651629072682	0.00569380625532603\\
0.8018671679198	0.00564909699178293\\
0.803082706766917	0.0056045002870691\\
0.804298245614035	0.00556001590602749\\
0.805513784461153	0.00551564361120913\\
0.806729323308271	0.00547138316268322\\
0.807944862155388	0.00542723431814108\\
0.809160401002506	0.00538319683270188\\
0.810375939849624	0.00533927045945615\\
0.811591478696742	0.00529545494956322\\
0.81280701754386	0.00525175005205063\\
0.814022556390977	0.00520815551405543\\
0.815238095238095	0.00516467108061965\\
0.816453634085213	0.00512129649492771\\
0.817669172932331	0.00507803149869006\\
0.818884711779449	0.00503487583163695\\
0.820100250626566	0.00499182923175038\\
0.821315789473684	0.00494889143564235\\
0.822531328320802	0.00490606217848737\\
0.82374686716792	0.00486334119410132\\
0.824962406015038	0.00482072821472265\\
0.826177944862155	0.00477822297123601\\
0.827393483709273	0.00473582519339432\\
0.828609022556391	0.0046935346095953\\
0.829824561403509	0.00465135094710055\\
0.831040100250627	0.00460927393225305\\
0.832255639097744	0.00456730329024946\\
0.833471177944862	0.00452543874520681\\
0.83468671679198	0.00448368002037593\\
0.835902255639098	0.00444202683776159\\
0.837117794486215	0.00440047891877722\\
0.838333333333333	0.00435903598415843\\
0.839548872180451	0.00431769775343137\\
0.840764411027569	0.00427646394585966\\
0.841979949874687	0.00423533427991035\\
0.843195488721804	0.00419430847316255\\
0.844411027568922	0.00415338624280707\\
0.84562656641604	0.00411256730540482\\
0.846842105263158	0.00407185137708815\\
0.848057644110276	0.00403123817346521\\
0.849273182957394	0.00399072740937516\\
0.850488721804511	0.00395031879938242\\
0.851704260651629	0.00391001205767789\\
0.852919799498747	0.0038698068982752\\
0.854135338345865	0.00382970303491007\\
0.855350877192982	0.0037897001807906\\
0.8565664160401	0.00374979804893878\\
0.857781954887218	0.00370999635193903\\
0.858997493734336	0.00367029480257389\\
0.860213032581454	0.00363069311327482\\
0.861428571428572	0.0035911909961642\\
0.862644110275689	0.00355178816339252\\
0.863859649122807	0.00351248432717867\\
0.865075187969925	0.00347327919970157\\
0.866290726817043	0.00343417249269495\\
0.86750626566416	0.0033951639179294\\
0.868721804511278	0.00335625318769365\\
0.869937343358396	0.00331744001394318\\
0.871152882205514	0.00327872410848408\\
0.872368421052632	0.00324010518374823\\
0.873583959899749	0.00320158295223578\\
0.874799498746867	0.00316315712595703\\
0.876015037593985	0.00312482741720557\\
0.877230576441103	0.00308659353903484\\
0.87844611528822	0.00304845520425402\\
0.879661654135338	0.0030104121257554\\
0.880877192982456	0.00297246401698902\\
0.882092731829574	0.00293461059154882\\
0.883308270676692	0.00289685156320219\\
0.884523809523809	0.00285918664606691\\
0.885739348370927	0.00282161555434361\\
0.886954887218045	0.00278413800263968\\
0.888170426065163	0.00274675370614456\\
0.889385964912281	0.00270946238021262\\
0.890601503759399	0.00267226374038946\\
0.891817042606516	0.00263515750273378\\
0.893032581453634	0.00259814338369467\\
0.894248120300752	0.00256122110043251\\
0.89546365914787	0.00252439036995532\\
0.896679197994988	0.00248765090958675\\
0.897894736842105	0.0024510024377295\\
0.899110275689223	0.00241444467314837\\
0.900325814536341	0.0023779773348449\\
0.901541353383459	0.00234160014207953\\
0.902756892230576	0.00230531281483735\\
0.903972431077694	0.00226911507384945\\
0.905187969924812	0.00223300663987396\\
0.90640350877193	0.00219698723430849\\
0.907619047619048	0.00216105657935839\\
0.908834586466165	0.00212521439731654\\
0.910050125313283	0.00208946041147078\\
0.911265664160401	0.00205379434567899\\
0.912481203007519	0.00201821592394379\\
0.913696741854637	0.00198272487102293\\
0.914912280701754	0.00194732091244732\\
0.916127819548872	0.00191200377424267\\
0.91734335839599	0.00187677318279896\\
0.918558897243108	0.00184162886533138\\
0.919774436090225	0.00180657054960114\\
0.920989974937343	0.0017715979640798\\
0.922205513784461	0.00173671083766948\\
0.923421052631579	0.00170190890001461\\
0.924636591478697	0.00166719188166549\\
0.925852130325814	0.00163255951335351\\
0.927067669172932	0.00159801152674612\\
0.92828320802005	0.00156354765416552\\
0.929498746867168	0.00152916762845505\\
0.930714285714286	0.00149487118314137\\
0.931929824561403	0.00146065805259629\\
0.933145363408521	0.00142652797190246\\
0.934360902255639	0.00139248067657073\\
0.935576441102757	0.00135851590270132\\
0.936791979949875	0.00132463338744062\\
0.938007518796993	0.00129083286854998\\
0.93922305764411	0.00125711408427002\\
0.940438596491228	0.0012234767739249\\
0.941654135338346	0.00118992067734229\\
0.942869674185464	0.00115644553501309\\
0.944085213032581	0.00112305108825103\\
0.945300751879699	0.00108973707905594\\
0.946516290726817	0.00105650325012491\\
0.947731829573935	0.00102334934471525\\
0.948947368421053	0.000990275106951115\\
0.95016290726817	0.000957280281834114\\
0.951378446115288	0.000924364615105617\\
0.952593984962406	0.000891527853108889\\
0.953809523809524	0.000858769742947083\\
0.955025062656642	0.00082609003249299\\
0.956240601503759	0.00079348847025067\\
0.957456140350877	0.000760964805660854\\
0.958671679197995	0.00072851878866621\\
0.959887218045113	0.000696150170164436\\
0.961102756892231	0.000663858702017161\\
0.962318295739348	0.000631644136318745\\
0.963533834586466	0.000599506226292839\\
0.964749373433584	0.000567444726004845\\
0.965964912280702	0.000535459389926222\\
0.967180451127819	0.000503549973386598\\
0.968395989974937	0.000471716232581768\\
0.969611528822055	0.000439957924285576\\
0.970827067669173	0.000408274806301571\\
0.972042606516291	0.000376666637322588\\
0.973258145363408	0.000345133176346192\\
0.974473684210526	0.000313674183273942\\
0.975689223057644	0.000282289418918586\\
0.976904761904762	0.00025097864486305\\
0.97812030075188	0.000219741623467356\\
0.979335839598998	0.000188578117727414\\
0.980551378446115	0.000157487891429654\\
0.981766917293233	0.000126470709157567\\
0.982982456140351	9.55263362981114e-05\\
0.984197994987469	6.46545390480122e-05\\
0.985413533834586	3.38550842719436e-05\\
0.986629072681704	3.1277396566153e-06\\
0.987844611528822	-2.75277264312977e-05\\
0.98906015037594	-5.8111544791293e-05\\
0.990275689223058	-8.86239452353016e-05\\
0.991491228070175	-0.000119065157026031\\
0.992706766917293	-0.000149435408723446\\
0.993922305764411	-0.000179734928179326\\
0.995137844611529	-0.000209963942235923\\
0.996353383458647	-0.000240122677016741\\
0.997568922305764	-0.000270211357921392\\
0.998784461152882	-0.000300230209620517\\
1	-0.000330179456050846\\
};
\end{axis}

\end{tikzpicture}%

%% file: Figs/Fig12a_Front_Strain.tex
%
%
\begin{tikzpicture}

\definecolor{mycolor1}{rgb}{0.2,0.6,1}%
\definecolor{mycolor2}{rgb}{1,0.5,0.2}%

\begin{axis}[%
thick, 
width=0.33\textwidth, 
height=0.4\textwidth,
y tick label style = overlay,
xmin=0,xmax=1,ymin=0,ymax=2.5, 
xlabel = {$r/a$},
ylabel = {$\lambda$}]
\addplot [color=mycolor1, very thick, forget plot]
  table[row sep=crcr]{%
0.00125	2.29231760192118\\
0.00375	2.29231757269003\\
0.00625	2.29231751201636\\
0.00875	2.29231741887795\\
0.01125	2.29231729340519\\
0.01375	2.29231713566342\\
0.01625	2.29231694562931\\
0.01875	2.29231672321972\\
0.02125	2.29231646830752\\
0.02375	2.29231618072981\\
0.02625	2.2923158602923\\
0.02875	2.29231550677182\\
0.03125	2.29231511991774\\
0.03375	2.29231469945286\\
0.03625	2.2923142450739\\
0.03875	2.29231375645176\\
0.04125	2.29231323323174\\
0.04375	2.29231267503351\\
0.04625	2.29231208145112\\
0.04875	2.29231145205294\\
0.05125	2.29231078638149\\
0.05375	2.29231008395334\\
0.05625	2.29230934425895\\
0.05875	2.29230856676243\\
0.06125	2.29230775090135\\
0.06375	2.29230689608652\\
0.06625	2.29230600170176\\
0.06875	2.29230506710361\\
0.07125	2.29230409162107\\
0.07375	2.29230307455535\\
0.07625	2.29230201517952\\
0.07875	2.2923009127383\\
0.08125	2.29229976644764\\
0.08375	2.29229857549449\\
0.08625	2.29229733903643\\
0.08875	2.29229605620135\\
0.09125	2.29229472608708\\
0.09375	2.29229334776106\\
0.09625	2.29229192025999\\
0.09875	2.29229044258943\\
0.10125	2.29228891372347\\
0.10375	2.29228733260434\\
0.10625	2.29228569814202\\
0.10875	2.29228400921385\\
0.11125	2.29228226466418\\
0.11375	2.29228046330394\\
0.11625	2.29227860391025\\
0.11875	2.29227668522605\\
0.12125	2.29227470595968\\
0.12375	2.29227266478449\\
0.12625	2.29227056033844\\
0.12875	2.29226839122373\\
0.13125	2.29226615600637\\
0.13375	2.2922638532158\\
0.13625	2.29226148134453\\
0.13875	2.29225903884772\\
0.14125	2.29225652414283\\
0.14375	2.29225393560927\\
0.14625	2.29225127158798\\
0.14875	2.29224853038114\\
0.15125	2.29224571025181\\
0.15375	2.29224280942359\\
0.15625	2.29223982608031\\
0.15875	2.29223675836577\\
0.16125	2.2922336043834\\
0.16375	2.29223036219602\\
0.16625	2.29222702982561\\
0.16875	2.29222360525309\\
0.17125	2.29222008641808\\
0.17375	2.29221647121877\\
0.17625	2.29221275751176\\
0.17875	2.29220894311195\\
0.18125	2.29220502579244\\
0.18375	2.29220100328452\\
0.18625	2.2921968732776\\
0.18875	2.29219263341929\\
0.19125	2.29218828131547\\
0.19375	2.29218381453036\\
0.19625	2.29217923058675\\
0.19875	2.29217452696616\\
0.20125	2.29216970110913\\
0.20375	2.2921647504156\\
0.20625	2.29215967224521\\
0.20875	2.29215446391785\\
0.21125	2.29214912271412\\
0.21375	2.29214364587598\\
0.21625	2.29213803060742\\
0.21875	2.2921322740752\\
0.22125	2.29212637340975\\
0.22375	2.29212032570608\\
0.22625	2.29211412802482\\
0.22875	2.29210777739341\\
0.23125	2.29210127080731\\
0.23375	2.29209460523143\\
0.23625	2.29208777760155\\
0.23875	2.29208078482601\\
0.24125	2.29207362378744\\
0.24375	2.29206629134467\\
0.24625	2.29205878433475\\
0.24875	2.29205109957521\\
0.25125	2.2920432338664\\
0.25375	2.29203518399405\\
0.25625	2.29202694673199\\
0.25875	2.29201851884508\\
0.26125	2.29200989709234\\
0.26375	2.29200107823027\\
0.26625	2.29199205901645\\
0.26875	2.29198283621329\\
0.27125	2.29197340659209\\
0.27375	2.29196376693734\\
0.27625	2.29195391405128\\
0.27875	2.29194384475875\\
0.28125	2.29193355591234\\
0.28375	2.29192304439785\\
0.28625	2.29191230714008\\
0.28875	2.29190134110892\\
0.29125	2.29189014332588\\
0.29375	2.29187871087092\\
0.29625	2.29186704088973\\
0.29875	2.29185513060133\\
0.30125	2.29184297730625\\
0.30375	2.291830578395\\
0.30625	2.29181793135712\\
0.30875	2.29180503379069\\
0.31125	2.29179188341233\\
0.31375	2.29177847806781\\
0.31625	2.29176481574312\\
0.31875	2.29175089457629\\
0.32125	2.29173671286961\\
0.32375	2.29172226910273\\
0.32625	2.29170756194627\\
0.32875	2.29169259027621\\
0.33125	2.29167735318903\\
0.33375	2.29166185001757\\
0.33625	2.29164608034774\\
0.33875	2.29163004403612\\
0.34125	2.29161374122833\\
0.34375	2.29159717237842\\
0.34625	2.29158033826921\\
0.34875	2.29156324003363\\
0.35125	2.29154587917707\\
0.35375	2.29152825760087\\
0.35625	2.291510377627\\
0.35875	2.29149224202381\\
0.36125	2.29147385403313\\
0.36375	2.29145521739868\\
0.36625	2.29143633639579\\
0.36875	2.29141721586268\\
0.37125	2.29139786123327\\
0.37375	2.2913782785717\\
0.37625	2.29135847460867\\
0.37875	2.29133845677981\\
0.38125	2.29131823326606\\
0.38375	2.29129781303636\\
0.38625	2.29127720589234\\
0.38875	2.29125642251519\\
0.39125	2.29123547451386\\
0.39375	2.29121437447419\\
0.39625	2.29119313600757\\
0.39875	2.29117177379766\\
0.40125	2.29115030364353\\
0.40375	2.29112874249751\\
0.40625	2.2911071084972\\
0.40875	2.29108542099307\\
0.41125	2.2910637005767\\
0.41375	2.29104196912054\\
0.41625	2.29102024984835\\
0.41875	2.29099856746527\\
0.42125	2.29097694838756\\
0.42375	2.2909554211207\\
0.42625	2.29093401683643\\
0.42875	2.29091277018603\\
0.43125	2.29089172034886\\
0.43375	2.29087091223677\\
0.43625	2.29085039764249\\
0.43875	2.29083023591988\\
0.44125	2.29081049351375\\
0.44375	2.29079124133369\\
0.44625	2.29077254864892\\
0.44875	2.29075447199103\\
0.45125	2.29073703770066\\
0.45375	2.29072021757232\\
0.45625	2.29070389897751\\
0.45875	2.29068785438171\\
0.46125	2.29067172072498\\
0.46375	2.29065500673297\\
0.46625	2.29063715503856\\
0.46875	2.29061769371859\\
0.47125	2.29059651404133\\
0.47375	2.290574300895\\
0.47625	2.29055311047653\\
0.47875	2.29053702759822\\
0.48125	2.29053273946238\\
0.48375	2.29054974595447\\
0.48625	2.29059982499362\\
0.48875	2.29069534642556\\
0.49125	2.29084614039815\\
0.49375	2.29105488372314\\
0.49625	2.29131127335046\\
0.49875	2.29158541725171\\
0.50125	2.29182068951315\\
0.50375	2.29192563764297\\
0.50625	2.2917632267263\\
0.50875	2.29113294968721\\
0.51125	2.28973302270312\\
0.51375	2.28704861388282\\
0.51625	2.24619550395849\\
0.51875	2.17493337452358\\
0.52125	2.11310436819727\\
0.52375	2.05878121538029\\
0.52625	2.0105522074958\\
0.52875	1.96735525185408\\
0.53125	1.92837314395143\\
0.53375	1.89296496193786\\
0.53625	1.8606196814046\\
0.53875	1.83092395123652\\
0.54125	1.80353917367073\\
0.54375	1.77818486191697\\
0.54625	1.75462633325769\\
0.54875	1.73266545893902\\
0.55125	1.71213360945711\\
0.55375	1.69288620295392\\
0.55625	1.67479844190039\\
0.55875	1.65776194265139\\
0.56125	1.64168204427764\\
0.56375	1.6264756400853\\
0.56625	1.61206941555745\\
0.56875	1.59839840537594\\
0.57125	1.58540480319846\\
0.57375	1.57303697331785\\
0.57625	1.56124862481869\\
0.57875	1.54999811747446\\
0.58125	1.53924787517155\\
0.58375	1.52896388765228\\
0.58625	1.5191152852317\\
0.58875	1.50967397414632\\
0.59125	1.50061432254599\\
0.59375	1.49191288899648\\
0.59625	1.48354818683408\\
0.59875	1.47550047889184\\
0.60125	1.46775159806334\\
0.60375	1.46028478993534\\
0.60625	1.45308457434248\\
0.60875	1.44613662320483\\
0.61125	1.43942765242613\\
0.61375	1.43294532597384\\
0.61625	1.42667817054668\\
0.61875	1.42061549947209\\
0.62125	1.41474734467309\\
0.62375	1.40906439570993\\
0.62625	1.40355794504075\\
0.62875	1.39821983876326\\
0.63125	1.39304243219865\\
0.63375	1.38801854976376\\
0.63625	1.38314144864929\\
0.63875	1.3784047858838\\
0.64125	1.37380258841564\\
0.64375	1.36932922589076\\
0.64625	1.3649793858431\\
0.64875	1.36074805104857\\
0.65125	1.3566304788223\\
0.65375	1.35262218206501\\
0.65625	1.34871891188591\\
0.65875	1.34491664164931\\
0.66125	1.34121155230877\\
0.66375	1.33760001890749\\
0.66625	1.33407859813684\\
0.66875	1.33064401685608\\
0.67125	1.32729316148672\\
0.67375	1.32402306820353\\
0.67625	1.32083091385258\\
0.67875	1.31771400753315\\
0.68125	1.31466978278699\\
0.68375	1.31169579034371\\
0.68625	1.30878969137599\\
0.68875	1.30594925122285\\
0.69125	1.30317233354301\\
0.69375	1.30045689486385\\
0.69625	1.2978009794948\\
0.69875	1.29520271477658\\
0.70125	1.29266030664051\\
0.70375	1.29017203545411\\
0.70625	1.2877362521315\\
0.70875	1.2853513744889\\
0.71125	1.28301588382712\\
0.71375	1.28072832172449\\
0.71625	1.27848728702528\\
0.71875	1.27629143300937\\
0.72125	1.27413946473074\\
0.72375	1.27203013651278\\
0.72625	1.26996224958978\\
0.72875	1.26793464988446\\
0.73125	1.26594622591261\\
0.73375	1.26399590680603\\
0.73625	1.26208266044632\\
0.73875	1.26020549170191\\
0.74125	1.25836344076192\\
0.74375	1.2565555815605\\
0.74625	1.25478102028594\\
0.74875	1.25303889396918\\
0.75125	1.25132836914694\\
0.75375	1.24964864059454\\
0.75625	1.24799893012456\\
0.75875	1.24637848544699\\
0.76125	1.24478657908744\\
0.76375	1.24322250735981\\
0.76625	1.24168558939022\\
0.76875	1.24017516618938\\
0.77125	1.2386905997703\\
0.77375	1.23723127230904\\
0.77625	1.23579658534585\\
0.77875	1.23438595902446\\
0.78125	1.23299883136754\\
0.78375	1.23163465758609\\
0.78625	1.23029290942102\\
0.78875	1.22897307451527\\
0.79125	1.22767465581453\\
0.79375	1.22639717099537\\
0.79625	1.22514015191905\\
0.79875	1.22390314410981\\
0.80125	1.22268570625628\\
0.80375	1.22148740973487\\
0.80625	1.22030783815396\\
0.80875	1.21914658691784\\
0.81125	1.21800326280944\\
0.81375	1.21687748359081\\
0.81625	1.21576887762055\\
0.81875	1.21467708348736\\
0.82125	1.2136017496588\\
0.82375	1.21254253414463\\
0.82625	1.21149910417404\\
0.82875	1.21047113588592\\
0.83125	1.20945831403185\\
0.83375	1.20846033169084\\
0.83625	1.20747688999563\\
0.83875	1.20650769786973\\
0.84125	1.20555247177483\\
0.84375	1.20461093546813\\
0.84625	1.20368281976903\\
0.84875	1.20276786233482\\
0.85125	1.20186580744499\\
0.85375	1.20097640579367\\
0.85625	1.20009941429001\\
0.85875	1.19923459586593\\
0.86125	1.19838171929109\\
0.86375	1.19754055899472\\
0.86625	1.19671089489393\\
0.86875	1.19589251222838\\
0.87125	1.19508520140084\\
0.87375	1.19428875782359\\
0.87625	1.19350298177024\\
0.87875	1.19272767823291\\
0.88125	1.19196265678434\\
0.88375	1.19120773144493\\
0.88625	1.19046272055441\\
0.88875	1.18972744664789\\
0.89125	1.18900173633631\\
0.89375	1.18828542019081\\
0.89625	1.18757833263123\\
0.89875	1.18688031181819\\
0.90125	1.18619119954895\\
0.90375	1.18551084115665\\
0.90625	1.18483908541298\\
0.90875	1.18417578443395\\
0.91125	1.18352079358888\\
0.91375	1.18287397141229\\
0.91625	1.18223517951861\\
0.91875	1.18160428251975\\
0.92125	1.18098114794521\\
0.92375	1.18036564616477\\
0.92625	1.17975765031369\\
0.92875	1.17915703622013\\
0.93125	1.17856368233501\\
0.93375	1.17797746966392\\
0.93625	1.17739828170121\\
0.93875	1.17682600436609\\
0.94125	1.17626052594069\\
0.94375	1.17570173701005\\
0.94625	1.17514953040381\\
0.94875	1.17460380113981\\
0.95125	1.17406444636929\\
0.95375	1.17353136532372\\
0.95625	1.17300445926326\\
0.95875	1.1724836314267\\
0.96125	1.17196878698289\\
0.96375	1.17145983298358\\
0.96625	1.17095667831763\\
0.96875	1.17045923366655\\
0.97125	1.16996741146133\\
0.97375	1.16948112584046\\
0.97625	1.16900029260921\\
0.97875	1.16852482920002\\
0.98125	1.16805465463401\\
0.98375	1.1675896894836\\
0.98625	1.16712985583613\\
0.98875	1.16667507725853\\
0.99125	1.16622527876294\\
0.99375	1.1657803867733\\
0.99625	1.16534032909284\\
0.99875	1.16490503487244\\
};
\addplot [color=black, dashed, very thick, forget plot]
  table[row sep=crcr]{%
0	2.29237235717068\\
0.01	2.29237235717068\\
0.02	2.29237235717068\\
0.03	2.29237235717068\\
0.04	2.29237235717068\\
0.05	2.29237235717068\\
0.06	2.29237235717068\\
0.07	2.29237235717068\\
0.08	2.29237235717068\\
0.09	2.29237235717068\\
0.1	2.29237235717068\\
0.11	2.29237235717068\\
0.12	2.29237235717068\\
0.13	2.29237235717068\\
0.14	2.29237235717068\\
0.15	2.29237235717068\\
0.16	2.29237235717068\\
0.17	2.29237235717068\\
0.18	2.29237235717068\\
0.19	2.29237235717068\\
0.2	2.29237235717068\\
0.21	2.29237235717068\\
0.22	2.29237235717068\\
0.23	2.29237235717068\\
0.24	2.29237235717068\\
0.25	2.29237235717068\\
0.26	2.29237235717068\\
0.27	2.29237235717068\\
0.28	2.29237235717068\\
0.29	2.29237235717068\\
0.3	2.29237235717068\\
0.31	2.29237235717068\\
0.32	2.29237235717068\\
0.33	2.29237235717068\\
0.34	2.29237235717068\\
0.35	2.29237235717068\\
0.36	2.29237235717068\\
0.37	2.29237235717068\\
0.38	2.29237235717068\\
0.39	2.29237235717068\\
0.4	2.29237235717068\\
0.41	2.29237235717068\\
0.42	2.29237235717068\\
0.43	2.29237235717068\\
0.44	2.29237235717068\\
0.45	2.29237235717068\\
0.46	2.29237235717068\\
0.47	2.29237235717068\\
0.48	2.29237235717068\\
0.49	2.29237235717068\\
0.5	2.29237235717068\\
0.51	2.29237235717068\\
0.515	2.29237235717068\\
0.525	2.03265093571979\\
0.535	1.8717858415478\\
0.545	1.75978718926734\\
0.555	1.676280279947\\
0.565	1.61113609678503\\
0.575	1.5586553617619\\
0.585	1.51534555504548\\
0.595	1.47892808550964\\
0.605	1.44784319020883\\
0.615	1.4209817530341\\
0.625	1.39753034983828\\
0.635	1.3768769439253\\
0.645	1.3585509877429\\
0.655	1.34218399113258\\
0.665	1.32748276012806\\
0.675	1.31421075202636\\
0.685	1.30217478498181\\
0.695	1.2912153722628\\
0.705	1.28119956648239\\
0.715	1.27201557729608\\
0.725	1.26356866496196\\
0.735	1.25577796678097\\
0.745	1.24857401571445\\
0.755	1.2418967794749\\
0.765	1.23569409577205\\
0.775	1.22992041247338\\
0.785	1.22453576487383\\
0.795	1.21950493910299\\
0.805	1.21479678294337\\
0.815	1.21038363434528\\
0.825	1.20624084462939\\
0.835	1.20234637840711\\
0.845	1.19868047607252\\
0.855	1.19522536764585\\
0.865	1.19196502900673\\
0.875	1.18888497331244\\
0.885	1.18597207177192\\
0.895	1.1832143990325\\
0.905	1.18060109929793\\
0.915	1.178122269985\\
0.925	1.17576886027976\\
0.935	1.17353258240091\\
0.945	1.17140583374185\\
0.955	1.16938162835877\\
0.965	1.16745353651583\\
0.975	1.16561563119862\\
0.985	1.163862440673\\
0.995	1.16218890630431\\
};
\addplot [color=mycolor2, very thick, forget plot]
  table[row sep=crcr]{%
0.00125	2.29231760192118\\
0.00375	2.29231751005187\\
0.00625	2.29231732293574\\
0.00875	2.29231703921784\\
0.01125	2.29231665906409\\
0.01375	2.29231618222001\\
0.01625	2.29231560820974\\
0.01875	2.29231493639157\\
0.02125	2.29231416598072\\
0.02375	2.29231329605869\\
0.02625	2.29231232557855\\
0.02875	2.2923112533668\\
0.03125	2.29231007812447\\
0.03375	2.29230879842725\\
0.03625	2.29230741272532\\
0.03875	2.29230591934281\\
0.04125	2.29230431647677\\
0.04375	2.29230260219671\\
0.04625	2.29230077444276\\
0.04875	2.29229883102522\\
0.05125	2.2922967696227\\
0.05375	2.29229458778106\\
0.05625	2.29229228291167\\
0.05875	2.29228985229007\\
0.06125	2.29228729305415\\
0.06375	2.2922846022024\\
0.06625	2.29228177659237\\
0.06875	2.29227881293845\\
0.07125	2.29227570781027\\
0.07375	2.29227245763064\\
0.07625	2.29226905867334\\
0.07875	2.29226550706119\\
0.08125	2.29226179876406\\
0.08375	2.2922579295963\\
0.08625	2.29225389521489\\
0.08875	2.29224969111684\\
0.09125	2.29224531263709\\
0.09375	2.29224075494609\\
0.09625	2.29223601304742\\
0.09875	2.29223108177546\\
0.10125	2.29222595579272\\
0.10375	2.29222062958788\\
0.10625	2.29221509747284\\
0.10875	2.29220935358044\\
0.11125	2.29220339186205\\
0.11375	2.29219720608507\\
0.11625	2.29219078983054\\
0.11875	2.2921841364907\\
0.12125	2.29217723926645\\
0.12375	2.29217009116512\\
0.12625	2.29216268499817\\
0.12875	2.29215501337887\\
0.13125	2.29214706872003\\
0.13375	2.29213884323191\\
0.13625	2.29213032892011\\
0.13875	2.29212151758374\\
0.14125	2.29211240081332\\
0.14375	2.29210296998916\\
0.14625	2.29209321627975\\
0.14875	2.29208313064003\\
0.15125	2.29207270381023\\
0.15375	2.29206192631479\\
0.15625	2.29205078846109\\
0.15875	2.29203928033889\\
0.16125	2.29202739181934\\
0.16375	2.29201511255524\\
0.16625	2.29200243198043\\
0.16875	2.29198933931026\\
0.17125	2.29197582354185\\
0.17375	2.291961873455\\
0.17625	2.29194747761319\\
0.17875	2.2919326243653\\
0.18125	2.29191730184724\\
0.18375	2.29190149798439\\
0.18625	2.29188520049435\\
0.18875	2.29186839689003\\
0.19125	2.29185107448358\\
0.19375	2.29183322039035\\
0.19625	2.29181482153419\\
0.19875	2.29179586465248\\
0.20125	2.29177633630246\\
0.20375	2.29175622286803\\
0.20625	2.29173551056716\\
0.20875	2.29171418546023\\
0.21125	2.29169223345936\\
0.21375	2.29166964033831\\
0.21625	2.2916463917432\\
0.21875	2.29162247320492\\
0.22125	2.29159787015161\\
0.22375	2.29157256792287\\
0.22625	2.29154655178503\\
0.22875	2.29151980694762\\
0.23125	2.29149231858095\\
0.23375	2.29146407183526\\
0.23625	2.29143505186099\\
0.23875	2.29140524383141\\
0.24125	2.29137463296571\\
0.24375	2.29134320455432\\
0.24625	2.29131094398629\\
0.24875	2.29127783677815\\
0.25125	2.29124386860464\\
0.25375	2.2912090253319\\
0.25625	2.29117329305256\\
0.25875	2.29113665812309\\
0.26125	2.29109910720382\\
0.26375	2.29106062730135\\
0.26625	2.29102120581353\\
0.26875	2.2909808305774\\
0.27125	2.29093948992001\\
0.27375	2.29089717271228\\
0.27625	2.29085386842624\\
0.27875	2.29080956719544\\
0.28125	2.29076425987957\\
0.28375	2.29071793813194\\
0.28625	2.2906705944717\\
0.28875	2.29062222235982\\
0.29125	2.29057281627942\\
0.29375	2.29052237182107\\
0.29625	2.29047088577232\\
0.29875	2.29041835621278\\
0.30125	2.29036478261442\\
0.30375	2.29031016594691\\
0.30625	2.29025450878952\\
0.30875	2.29019781544871\\
0.31125	2.2901400920823\\
0.31375	2.29008134683035\\
0.31625	2.2900215899528\\
0.31875	2.28996083397506\\
0.32125	2.28989909384105\\
0.32375	2.2898363870744\\
0.32625	2.28977273394827\\
0.32875	2.28970815766419\\
0.33125	2.28964268454015\\
0.33375	2.28957634420869\\
0.33625	2.28950916982535\\
0.33875	2.28944119828782\\
0.34125	2.28937247046635\\
0.34375	2.28930303144605\\
0.34625	2.2892329307815\\
0.34875	2.28916222276399\\
0.35125	2.28909096670201\\
0.35375	2.28901922721626\\
0.35625	2.28894707454862\\
0.35875	2.28887458488726\\
0.36125	2.28880184070878\\
0.36375	2.28872893113871\\
0.36625	2.28865595233312\\
0.36875	2.2885830078845\\
0.37125	2.28851020925443\\
0.37375	2.28843767623755\\
0.37625	2.28836553745899\\
0.37875	2.28829393090682\\
0.38125	2.28822300449617\\
0.38375	2.28815291665581\\
0.38625	2.28808383691701\\
0.38875	2.28801594647228\\
0.39125	2.28794943865565\\
0.39375	2.28788451928087\\
0.39625	2.28782140676632\\
0.39875	2.28776033198331\\
0.40125	2.28770153779966\\
0.40375	2.28764527837821\\
0.40625	2.287591818439\\
0.40875	2.28754143293768\\
0.41125	2.28749440794491\\
0.41375	2.28745104393488\\
0.41625	2.28741166313953\\
0.41875	2.28737662298859\\
0.42125	2.28734633772021\\
0.42375	2.28732130968261\\
0.42625	2.28730217018683\\
0.42875	2.28728972641859\\
0.43125	2.28728500521433\\
0.43375	2.28728927587519\\
0.43625	2.28730402243637\\
0.43875	2.28733082161417\\
0.44125	2.28737106838282\\
0.44375	2.28742548192004\\
0.44625	2.2874933298332\\
0.44875	2.28757134297091\\
0.45125	2.2876523777069\\
0.45375	2.28772404277184\\
0.45625	2.28776776715055\\
0.45875	2.28775915226979\\
0.46125	2.28767089390052\\
0.46375	2.28747997088837\\
0.46625	2.28718096038283\\
0.46875	2.28680689609029\\
0.47125	2.28645756118301\\
0.47375	2.28633201008365\\
0.47625	2.28675717353627\\
0.47875	2.28819795879656\\
0.48125	2.2912277527799\\
0.48375	2.2964345975188\\
0.48625	2.30424173944766\\
0.48875	2.31463575572964\\
0.49125	2.32682183944603\\
0.49375	2.33885872932692\\
0.49625	2.34735409550699\\
0.49875	2.34731333517743\\
0.50125	2.33222351936602\\
0.50375	2.2944130468905\\
0.50625	2.22563408804275\\
0.50875	2.11758883078436\\
0.51125	1.96140615265219\\
0.51375	1.74121833265626\\
0.51625	0.278940179895293\\
0.51875	0.297638589212138\\
0.52125	0.31541984494337\\
0.52375	0.33238287761825\\
0.52625	0.348608245954573\\
0.52875	0.364162813877483\\
0.53125	0.37910297108748\\
0.53375	0.39347692108782\\
0.53625	0.40732635014341\\
0.53875	0.420687672491448\\
0.54125	0.433592977769919\\
0.54375	0.446070764347607\\
0.54625	0.458146515604195\\
0.54875	0.469843158944776\\
0.55125	0.481181435857834\\
0.55375	0.492180203525954\\
0.55625	0.502856683089945\\
0.55875	0.513226665848552\\
0.56125	0.523304685935731\\
0.56375	0.533104166021712\\
0.56625	0.542637541110754\\
0.56875	0.551916364407157\\
0.57125	0.560951398388441\\
0.57375	0.56975269358832\\
0.57625	0.578329657101096\\
0.57875	0.586691112436769\\
0.58125	0.594845352055841\\
0.58375	0.60280018367508\\
0.58625	0.610562971245891\\
0.58875	0.618140671354652\\
0.59125	0.625539865671255\\
0.59375	0.632766789971917\\
0.59625	0.639827360180381\\
0.59875	0.646727195804152\\
0.60125	0.653471641086665\\
0.60375	0.660065784149855\\
0.60625	0.666514474362908\\
0.60875	0.672822338140429\\
0.61125	0.678993793345891\\
0.61375	0.685033062453061\\
0.61625	0.690944184598388\\
0.61875	0.696731026640586\\
0.62125	0.70239729332924\\
0.62375	0.707946536671992\\
0.62625	0.713382164579167\\
0.62875	0.71870744885559\\
0.63125	0.723925532601331\\
0.63375	0.729039437076248\\
0.63625	0.734052068077119\\
0.63875	0.738966221870934\\
0.64125	0.743784590723259\\
0.64375	0.748509768056559\\
0.64625	0.753144253269766\\
0.64875	0.75769045624726\\
0.65125	0.762150701582605\\
0.65375	0.766527232539953\\
0.65625	0.770822214773779\\
0.65875	0.775037739825722\\
0.66125	0.779175828415507\\
0.66375	0.783238433541401\\
0.66625	0.787227443404256\\
0.66875	0.791144684167929\\
0.67125	0.794991922567784\\
0.67375	0.79877086837792\\
0.67625	0.802483176746918\\
0.67875	0.806130450411035\\
0.68125	0.809714241793065\\
0.68375	0.813236054994397\\
0.68625	0.816697347687206\\
0.68875	0.820099532913156\\
0.69125	0.823443980794496\\
0.69375	0.826732020162986\\
0.69625	0.829964940111658\\
0.69875	0.833143991474048\\
0.70125	0.836270388235203\\
0.70375	0.839345308878439\\
0.70625	0.842369897671535\\
0.70875	0.845345265895794\\
0.71125	0.848272493021173\\
0.71375	0.851152627830424\\
0.71625	0.853986689495033\\
0.71875	0.856775668605511\\
0.72125	0.859520528158468\\
0.72375	0.862222204502685\\
0.72625	0.864881608246318\\
0.72875	0.867499625127165\\
0.73125	0.870077116847873\\
0.73375	0.87261492187778\\
0.73625	0.875113856223029\\
0.73875	0.877574714166477\\
0.74125	0.879998268978824\\
0.74375	0.8823852736023\\
0.74625	0.884736461308198\\
0.74875	0.88705254632943\\
0.75125	0.889334224469236\\
0.75375	0.891582173687116\\
0.75625	0.893797054662981\\
0.75875	0.895979511340477\\
0.76125	0.898130171450384\\
0.76375	0.900249647014933\\
0.76625	0.902338534833849\\
0.76875	0.904397416952899\\
0.77125	0.906426861115642\\
0.77375	0.908427421199101\\
0.77625	0.91039963763399\\
0.77875	0.912344037810123\\
0.78125	0.914261136467615\\
0.78375	0.91615143607441\\
0.78625	0.918015427190694\\
0.78875	0.919853588820706\\
0.79125	0.92166638875242\\
0.79375	0.923454283885584\\
0.79625	0.925217720548548\\
0.79875	0.926957134804312\\
0.80125	0.928672952746206\\
0.80375	0.930365590783581\\
0.80625	0.932035455917892\\
0.80875	0.933682946009531\\
0.81125	0.935308450035742\\
0.81375	0.936912348339964\\
0.81625	0.938495012872892\\
0.81875	0.940056807425587\\
0.82125	0.941598087854891\\
0.82375	0.94311920230146\\
0.82625	0.944620491400651\\
0.82875	0.946102288486539\\
0.83125	0.947564919789308\\
0.83375	0.94900870462624\\
0.83625	0.950433955586545\\
0.83875	0.951840978710245\\
0.84125	0.95323007366132\\
0.84375	0.954601533895327\\
0.84625	0.955955646821675\\
0.84875	0.957292693960756\\
0.85125	0.958612951096104\\
0.85375	0.959916688421765\\
0.85625	0.961204170685042\\
0.85875	0.962475657324765\\
0.86125	0.963731402605265\\
0.86375	0.964971655746186\\
0.86625	0.96619666104828\\
0.86875	0.967406658015337\\
0.87125	0.968601881472362\\
0.87375	0.969782561680161\\
0.87625	0.970948924446422\\
0.87875	0.972101191233446\\
0.88125	0.973239579262628\\
0.88375	0.974364301615794\\
0.88625	0.97547556733352\\
0.88875	0.976573581510521\\
0.89125	0.977658545388211\\
0.89375	0.978730656444556\\
0.89625	0.979790108481266\\
0.89875	0.980837091708466\\
0.90125	0.981871792826896\\
0.90375	0.982894395107746\\
0.90625	0.98390507847019\\
0.90875	0.984904019556714\\
0.91125	0.985891391806296\\
0.91375	0.986867365525521\\
0.91625	0.987832107957698\\
0.91875	0.988785783350042\\
0.92125	0.989728553018994\\
0.92375	0.990660575413734\\
0.92625	0.991582006177945\\
0.92875	0.992492998209897\\
0.93125	0.993393701720902\\
0.93375	0.994284264292176\\
0.93625	0.995164830930193\\
0.93875	0.996035544120558\\
0.94125	0.996896543880448\\
0.94375	0.997747967809674\\
0.94625	0.998589951140409\\
0.94875	0.999422626785618\\
0.95125	1.00024612538624\\
0.95375	1.00106057535717\\
0.95625	1.00186610293201\\
0.95875	1.00266283220677\\
0.96125	1.00345088518244\\
0.96375	1.00423038180641\\
0.96625	1.00500144001299\\
0.96875	1.00576417576285\\
0.97125	1.00651870308151\\
0.97375	1.00726513409688\\
0.97625	1.00800357907589\\
0.97875	1.00873414646027\\
0.98125	1.00945694290141\\
0.98375	1.01017207329447\\
0.98625	1.0108796408116\\
0.98875	1.01157974693443\\
0.99125	1.01227249148582\\
0.99375	1.0129579726608\\
0.99625	1.01363628705686\\
0.99875	1.01430752970355\\
};
\addplot [color=black, dashed, very thick, forget plot]
  table[row sep=crcr]{%
0	2.29237235717068\\
0.01	2.29237235717068\\
0.02	2.29237235717068\\
0.03	2.29237235717068\\
0.04	2.29237235717068\\
0.05	2.29237235717068\\
0.06	2.29237235717068\\
0.07	2.29237235717068\\
0.08	2.29237235717068\\
0.09	2.29237235717068\\
0.1	2.29237235717068\\
0.11	2.29237235717068\\
0.12	2.29237235717068\\
0.13	2.29237235717068\\
0.14	2.29237235717068\\
0.15	2.29237235717068\\
0.16	2.29237235717068\\
0.17	2.29237235717068\\
0.18	2.29237235717068\\
0.19	2.29237235717068\\
0.2	2.29237235717068\\
0.21	2.29237235717068\\
0.22	2.29237235717068\\
0.23	2.29237235717068\\
0.24	2.29237235717068\\
0.25	2.29237235717068\\
0.26	2.29237235717068\\
0.27	2.29237235717068\\
0.28	2.29237235717068\\
0.29	2.29237235717068\\
0.3	2.29237235717068\\
0.31	2.29237235717068\\
0.32	2.29237235717068\\
0.33	2.29237235717068\\
0.34	2.29237235717068\\
0.35	2.29237235717068\\
0.36	2.29237235717068\\
0.37	2.29237235717068\\
0.38	2.29237235717068\\
0.39	2.29237235717068\\
0.4	2.29237235717068\\
0.41	2.29237235717068\\
0.42	2.29237235717068\\
0.43	2.29237235717068\\
0.44	2.29237235717068\\
0.45	2.29237235717068\\
0.46	2.29237235717068\\
0.47	2.29237235717068\\
0.48	2.29237235717068\\
0.49	2.29237235717068\\
0.5	2.29237235717068\\
0.51	2.29237235717068\\
0.515	0.262031296866924\\
0.525	0.333271275349757\\
0.535	0.393016855150177\\
0.545	0.444634531644085\\
0.555	0.490038520621981\\
0.565	0.53046781126399\\
0.575	0.566791457735399\\
0.585	0.599653169019493\\
0.595	0.629548768941523\\
0.605	0.656871525737554\\
0.615	0.681940509407246\\
0.625	0.705019283227179\\
0.635	0.726328752747525\\
0.645	0.746056308189992\\
0.655	0.76436251901534\\
0.665	0.781386156333082\\
0.675	0.79724803956195\\
0.685	0.812054035525469\\
0.695	0.825897433165823\\
0.705	0.838860849457673\\
0.715	0.851017777376822\\
0.725	0.862433856461579\\
0.735	0.873167925505609\\
0.745	0.883272902090792\\
0.755	0.892796523009499\\
0.765	0.901781971841594\\
0.775	0.910268414183875\\
0.785	0.918291456699171\\
0.795	0.925883542861066\\
0.805	0.933074295740672\\
0.815	0.939890816217651\\
0.825	0.946357943457839\\
0.835	0.952498483281927\\
0.845	0.958333409078558\\
0.855	0.963882039134883\\
0.865	0.969162193626248\\
0.875	0.974190333992344\\
0.885	0.978981687005585\\
0.895	0.983550355489991\\
0.905	0.987909417360803\\
0.915	0.992071014415064\\
0.925	0.996046432102571\\
0.935	0.999846171337648\\
0.945	1.00348001326958\\
0.955	1.00695707780858\\
0.965	1.01028587660132\\
0.975	1.01347436106191\\
0.985	1.01652996598925\\
0.995	1.01945964923638\\
};
\end{axis}

\end{tikzpicture}%

%% file: Figs/Fig12b_Front_Stress.tex
%
%
\begin{tikzpicture}
\definecolor{mycolor1}{rgb}{0.2,0.6,1}%
\definecolor{mycolor2}{rgb}{1,0.5,0.2}%
\begin{axis}[%
thick, 
width=0.33\textwidth, 
height=0.4\textwidth,
y tick label style = overlay,
xmin=0,xmax=1,ymin=-1,ymax=2.5, 
xlabel = {$r/a$},
ylabel = {$\sigma'$}]
\addplot [color=mycolor1, very thick, forget plot]
  table[row sep=crcr]{%
0.00125	0.353221081475751\\
0.00375	0.353221093514535\\
0.00625	0.35322111795237\\
0.00875	0.353221154924001\\
0.01125	0.353221204413319\\
0.01375	0.353221266464274\\
0.01625	0.35322134114852\\
0.01875	0.353221428558932\\
0.02125	0.353221528807249\\
0.02375	0.353221642023224\\
0.02625	0.353221768354135\\
0.02875	0.353221907964671\\
0.03125	0.35322206103687\\
0.03375	0.353222227770163\\
0.03625	0.353222408381438\\
0.03875	0.353222603105137\\
0.04125	0.35322281219344\\
0.04375	0.353223035916319\\
0.04625	0.353223274561842\\
0.04875	0.353223528436221\\
0.05125	0.353223797864097\\
0.05375	0.353224083188696\\
0.05625	0.35322438477208\\
0.05875	0.353224702995337\\
0.06125	0.353225038258859\\
0.06375	0.353225390982587\\
0.06625	0.353225761606229\\
0.06875	0.353226150589592\\
0.07125	0.353226558412802\\
0.07375	0.353226985576611\\
0.07625	0.353227432602708\\
0.07875	0.353227900033998\\
0.08125	0.353228388434895\\
0.08375	0.353228898391684\\
0.08625	0.353229430512792\\
0.08875	0.35322998542916\\
0.09125	0.353230563794551\\
0.09375	0.353231166285885\\
0.09625	0.353231793603596\\
0.09875	0.353232446471954\\
0.10125	0.353233125639463\\
0.10375	0.353233831879125\\
0.10625	0.353234565988881\\
0.10875	0.353235328791923\\
0.11125	0.353236121137055\\
0.11375	0.353236943899036\\
0.11625	0.353237797978933\\
0.11875	0.353238684304471\\
0.12125	0.353239603830387\\
0.12375	0.353240557538762\\
0.12625	0.353241546439338\\
0.12875	0.353242571569857\\
0.13125	0.353243633996384\\
0.13375	0.353244734813598\\
0.13625	0.353245875145103\\
0.13875	0.353247056143679\\
0.14125	0.353248278991584\\
0.14375	0.353249544900781\\
0.14625	0.353250855113164\\
0.14875	0.353252210900804\\
0.15125	0.353253613566112\\
0.15375	0.353255064441975\\
0.15625	0.35325656489195\\
0.15875	0.353258116310321\\
0.16125	0.353259720122242\\
0.16375	0.353261377783686\\
0.16625	0.353263090781522\\
0.16875	0.353264860633437\\
0.17125	0.353266688887888\\
0.17375	0.353268577123956\\
0.17625	0.353270526951185\\
0.17875	0.353272540009324\\
0.18125	0.353274617968071\\
0.18375	0.353276762526702\\
0.18625	0.353278975413661\\
0.18875	0.353281258386087\\
0.19125	0.353283613229229\\
0.19375	0.353286041755849\\
0.19625	0.35328854580543\\
0.19875	0.353291127243426\\
0.20125	0.353293787960314\\
0.20375	0.353296529870584\\
0.20625	0.353299354911636\\
0.20875	0.353302265042538\\
0.21125	0.353305262242635\\
0.21375	0.353308348510085\\
0.21625	0.353311525860263\\
0.21875	0.353314796323918\\
0.22125	0.353318161945321\\
0.22375	0.353321624780159\\
0.22625	0.353325186893269\\
0.22875	0.35332885035619\\
0.23125	0.353332617244564\\
0.23375	0.353336489635292\\
0.23625	0.35334046960353\\
0.23875	0.353344559219329\\
0.24125	0.353348760544221\\
0.24375	0.353353075627454\\
0.24625	0.353357506501945\\
0.24875	0.353362055179988\\
0.25125	0.353366723648702\\
0.25375	0.353371513865115\\
0.25625	0.353376427750952\\
0.25875	0.353381467187096\\
0.26125	0.353386634007651\\
0.26375	0.353391929993653\\
0.26625	0.35339735686639\\
0.26875	0.353402916280287\\
0.27125	0.353408609815369\\
0.27375	0.35341443896925\\
0.27625	0.353420405148642\\
0.27875	0.353426509660381\\
0.28125	0.35343275370182\\
0.28375	0.353439138350809\\
0.28625	0.353445664554946\\
0.28875	0.353452333120298\\
0.29125	0.353459144699436\\
0.29375	0.35346609977876\\
0.29625	0.353473198665168\\
0.29875	0.353480441471905\\
0.30125	0.353487828103637\\
0.30375	0.353495358240757\\
0.30625	0.353503031322718\\
0.30875	0.353510846530513\\
0.31125	0.353518802768172\\
0.31375	0.353526898643239\\
0.31625	0.353535132446259\\
0.31875	0.353543502129056\\
0.32125	0.353552005281895\\
0.32375	0.353560639109431\\
0.32625	0.353569400405353\\
0.32875	0.353578285525681\\
0.33125	0.353587290360671\\
0.33375	0.353596410305225\\
0.33625	0.353605640227735\\
0.33875	0.353614974437342\\
0.34125	0.353624406649471\\
0.34375	0.353633929949621\\
0.34625	0.35364353675528\\
0.34875	0.353653218775951\\
0.35125	0.353662966971206\\
0.35375	0.353672771506543\\
0.35625	0.353682621707162\\
0.35875	0.353692506009317\\
0.36125	0.353702411909118\\
0.36375	0.353712325908563\\
0.36625	0.353722233458423\\
0.36875	0.353732118897434\\
0.37125	0.353741965387494\\
0.37375	0.353751754844115\\
0.37625	0.353761467861832\\
0.37875	0.353771083634352\\
0.38125	0.353780579869956\\
0.38375	0.353789932703598\\
0.38625	0.353799116608824\\
0.38875	0.353808104314518\\
0.39125	0.353816866733929\\
0.39375	0.353825372915781\\
0.39625	0.353833590028411\\
0.39875	0.353841483386652\\
0.40125	0.353849016525689\\
0.40375	0.353856151312607\\
0.40625	0.353862848063281\\
0.40875	0.353869065594743\\
0.41125	0.353874761091953\\
0.41375	0.353879889602865\\
0.41625	0.353884402906793\\
0.41875	0.35388824744562\\
0.42125	0.35389136099817\\
0.42375	0.353893667866111\\
0.42625	0.353895072597014\\
0.42875	0.353895452788219\\
0.43125	0.353894652395159\\
0.43375	0.353892478298112\\
0.43625	0.35388870469061\\
0.43875	0.353883092033119\\
0.44125	0.353875429499393\\
0.44375	0.353865611235443\\
0.44625	0.353853755912523\\
0.44875	0.353840373705296\\
0.45125	0.353826571754084\\
0.45375	0.353814264505538\\
0.45625	0.353806315488355\\
0.45875	0.353806480862417\\
0.46125	0.35381895723937\\
0.46375	0.353847272686641\\
0.46625	0.353892233497916\\
0.46875	0.353948705246938\\
0.47125	0.354001241598409\\
0.47375	0.354019064320859\\
0.47625	0.353951701481721\\
0.47875	0.353727662755524\\
0.48125	0.353259601212008\\
0.48375	0.35245986687096\\
0.48625	0.35126929290645\\
0.48875	0.349698763328626\\
0.49125	0.347878095258055\\
0.49375	0.346102592953012\\
0.49625	0.344868167026194\\
0.49875	0.344893569204255\\
0.50125	0.347141842490816\\
0.50375	0.35287011907137\\
0.50625	0.363762770543046\\
0.50875	0.382273480877756\\
0.51125	0.412594320637797\\
0.51375	0.464512125334624\\
0.51625	2.87444940059402\\
0.51875	2.64951696021368\\
0.52125	2.46036000383919\\
0.52375	2.29877044326334\\
0.52625	2.15892001570363\\
0.52875	2.03654665077994\\
0.53125	1.92845561128869\\
0.53375	1.8322012576339\\
0.53625	1.74587710250327\\
0.53875	1.66797325456389\\
0.54125	1.59727716625539\\
0.54375	1.53280299753662\\
0.54625	1.47374035986303\\
0.54875	1.41941647445629\\
0.55125	1.36926779773077\\
0.55375	1.32281844589269\\
0.55625	1.27966358025987\\
0.55875	1.23945646418165\\
0.56125	1.20189827318191\\
0.56375	1.16672999452289\\
0.56625	1.13372592997769\\
0.56875	1.10268844129583\\
0.57125	1.07344366801155\\
0.57375	1.04583801272547\\
0.57625	1.01973523709514\\
0.57875	0.995014047486626\\
0.58125	0.971566076025047\\
0.58375	0.949294183055744\\
0.58625	0.928111022508703\\
0.58875	0.907937823575718\\
0.59125	0.888703351354706\\
0.59375	0.870343016339669\\
0.59625	0.852798108318392\\
0.59875	0.836015134740864\\
0.60125	0.819945247207377\\
0.60375	0.804543742599159\\
0.60625	0.789769627690251\\
0.60875	0.77558523795533\\
0.61125	0.761955902815574\\
0.61375	0.748849650814078\\
0.61625	0.736236949239057\\
0.61875	0.724090473560469\\
0.62125	0.712384902748042\\
0.62375	0.701096737123013\\
0.62625	0.69020413588409\\
0.62875	0.679686771857261\\
0.63125	0.669525701363416\\
0.63375	0.659703247388272\\
0.63625	0.650202894485233\\
0.63875	0.641009194050838\\
0.64125	0.632107678790622\\
0.64375	0.623484785345394\\
0.64625	0.61512778417843\\
0.64875	0.607024715936099\\
0.65125	0.599164333591066\\
0.65375	0.591536049760544\\
0.65625	0.584129888664329\\
0.65875	0.576936442249921\\
0.66125	0.569946830066631\\
0.66375	0.563152662517959\\
0.66625	0.556546007163122\\
0.66875	0.550119357774871\\
0.67125	0.543865605892599\\
0.67375	0.537778014637786\\
0.67625	0.531850194583406\\
0.67875	0.526076081490735\\
0.68125	0.520449915746156\\
0.68375	0.514966223347614\\
0.68625	0.509619798305423\\
0.68875	0.504405686335531\\
0.69125	0.499319169735288\\
0.69375	0.494355753342323\\
0.69625	0.489511151486679\\
0.69875	0.484781275854758\\
0.70125	0.480162224191251\\
0.70375	0.475650269771972\\
0.70625	0.471241851586653\\
0.70875	0.466933565176195\\
0.71125	0.462722154073809\\
0.71375	0.458604501803917\\
0.71625	0.454577624396687\\
0.71875	0.450638663379694\\
0.72125	0.446784879211434\\
0.72375	0.44301364512444\\
0.72625	0.439322441348354\\
0.72875	0.43570884968581\\
0.73125	0.432170548416129\\
0.73375	0.428705307503893\\
0.73625	0.425310984091241\\
0.73875	0.421985518254432\\
0.74125	0.418726929006718\\
0.74375	0.415533310530964\\
0.74625	0.412402828626705\\
0.74875	0.409333717357515\\
0.75125	0.406324275885594\\
0.75375	0.403372865481456\\
0.75625	0.400477906697514\\
0.75875	0.39763787669514\\
0.76125	0.394851306715554\\
0.76375	0.392116779685581\\
0.76625	0.389432927949934\\
0.76875	0.386798431122293\\
0.77125	0.384212014047962\\
0.77375	0.3816724448714\\
0.77625	0.379178533202364\\
0.77875	0.37672912837486\\
0.78125	0.374323117793422\\
0.78375	0.371959425361694\\
0.78625	0.369637009988529\\
0.78875	0.367354864167204\\
0.79125	0.365112012623594\\
0.79375	0.362907511029432\\
0.79625	0.360740444777016\\
0.79875	0.358609927811989\\
0.80125	0.356515101520965\\
0.80375	0.354455133671058\\
0.80625	0.352429217398469\\
0.80875	0.35043657024353\\
0.81125	0.348476433229718\\
0.81375	0.346548069984315\\
0.81625	0.344650765898545\\
0.81875	0.342783827325118\\
0.82125	0.340946580811253\\
0.82375	0.339138372365369\\
0.82625	0.337358566755724\\
0.82875	0.335606546839377\\
0.83125	0.333881712919973\\
0.83375	0.332183482132901\\
0.83625	0.330511287856462\\
0.83875	0.328864579147795\\
0.84125	0.327242820202325\\
0.84375	0.325645489835607\\
0.84625	0.324072080986492\\
0.84875	0.322522100240581\\
0.85125	0.320995067373016\\
0.85375	0.319490514909692\\
0.85625	0.318007987706019\\
0.85875	0.31654704254243\\
0.86125	0.315107247735854\\
0.86375	0.313688182766415\\
0.86625	0.31228943791867\\
0.86875	0.310910613936722\\
0.87125	0.309551321692584\\
0.87375	0.308211181867199\\
0.87625	0.306889824643555\\
0.87875	0.305586889411357\\
0.88125	0.304302024482747\\
0.88375	0.303034886818599\\
0.88625	0.301785141764913\\
0.88875	0.300552462798887\\
0.89125	0.299336531284253\\
0.89375	0.298137036235457\\
0.89625	0.296953674090349\\
0.89875	0.29578614849099\\
0.90125	0.294634170072248\\
0.90375	0.293497456257871\\
0.90625	0.292375731063704\\
0.90875	0.291268724907777\\
0.91125	0.290176174426978\\
0.91375	0.289097822300028\\
0.91625	0.288033417076533\\
0.91875	0.286982713011844\\
0.92125	0.285945469907499\\
0.92375	0.284921452957037\\
0.92625	0.283910432596957\\
0.92875	0.282912184362632\\
0.93125	0.281926488748973\\
0.93375	0.280953131075676\\
0.93625	0.279991901356859\\
0.93875	0.279042594174921\\
0.94125	0.278105008558485\\
0.94375	0.277178947864239\\
0.94625	0.276264219662549\\
0.94875	0.275360635626699\\
0.95125	0.274468011425615\\
0.95375	0.273586166619956\\
0.95625	0.272714924561438\\
0.95875	0.27185411229528\\
0.96125	0.271003560465651\\
0.96375	0.270163103224022\\
0.96625	0.269332578140298\\
0.96875	0.268511826116653\\
0.97125	0.267700691303959\\
0.97375	0.266899021020715\\
0.97625	0.266106665674398\\
0.97875	0.265323478685145\\
0.98125	0.264549316411683\\
0.98375	0.263784038079438\\
0.98625	0.263027505710731\\
0.98875	0.262279584057013\\
0.99125	0.261540140533047\\
0.99375	0.260809045152989\\
0.99625	0.260086170468284\\
0.99875	0.259371391507344\\
};
\addplot [color=black, dashed, very thick, forget plot]
  table[row sep=crcr]{%
0	0.353216610290657\\
0.01	0.353216610290657\\
0.02	0.353216610290657\\
0.03	0.353216610290657\\
0.04	0.353216610290657\\
0.05	0.353216610290657\\
0.06	0.353216610290657\\
0.07	0.353216610290657\\
0.08	0.353216610290657\\
0.09	0.353216610290657\\
0.1	0.353216610290657\\
0.11	0.353216610290657\\
0.12	0.353216610290657\\
0.13	0.353216610290657\\
0.14	0.353216610290657\\
0.15	0.353216610290657\\
0.16	0.353216610290657\\
0.17	0.353216610290657\\
0.18	0.353216610290657\\
0.19	0.353216610290657\\
0.2	0.353216610290657\\
0.21	0.353216610290657\\
0.22	0.353216610290657\\
0.23	0.353216610290657\\
0.24	0.353216610290657\\
0.25	0.353216610290657\\
0.26	0.353216610290657\\
0.27	0.353216610290657\\
0.28	0.353216610290657\\
0.29	0.353216610290657\\
0.3	0.353216610290657\\
0.31	0.353216610290657\\
0.32	0.353216610290657\\
0.33	0.353216610290657\\
0.34	0.353216610290657\\
0.35	0.353216610290657\\
0.36	0.353216610290657\\
0.37	0.353216610290657\\
0.38	0.353216610290657\\
0.39	0.353216610290657\\
0.4	0.353216610290657\\
0.41	0.353216610290657\\
0.42	0.353216610290657\\
0.43	0.353216610290657\\
0.44	0.353216610290657\\
0.45	0.353216610290657\\
0.46	0.353216610290657\\
0.47	0.353216610290657\\
0.48	0.353216610290657\\
0.49	0.353216610290657\\
0.5	0.353216610290657\\
0.51	0.353216610290657\\
0.515	3.09010413338164\\
0.525	2.27432474176621\\
0.535	1.81818625183228\\
0.545	1.52280421086739\\
0.555	1.31442201931112\\
0.565	1.15889463599591\\
0.575	1.03808346112346\\
0.585	0.941396759197665\\
0.595	0.862205406614776\\
0.605	0.796133825301539\\
0.615	0.740169689545598\\
0.625	0.692167035987375\\
0.635	0.650553137240241\\
0.645	0.614147517453167\\
0.655	0.582045858995882\\
0.665	0.553543076247698\\
0.675	0.528080896729529\\
0.685	0.505211261496768\\
0.695	0.484570218023293\\
0.705	0.465858941136679\\
0.715	0.448829700462228\\
0.725	0.433275326395935\\
0.735	0.419021193188613\\
0.745	0.40591904128002\\
0.755	0.393842162612458\\
0.765	0.382681609041926\\
0.775	0.372343177818492\\
0.785	0.362744993701999\\
0.795	0.353815553782249\\
0.805	0.345492134477987\\
0.815	0.337719484479488\\
0.825	0.330448745263195\\
0.835	0.323636554083012\\
0.845	0.317244294307062\\
0.855	0.311237465516108\\
0.865	0.305585151546063\\
0.875	0.300259569098212\\
0.885	0.295235682987496\\
0.895	0.290490876793459\\
0.905	0.286004669798787\\
0.915	0.281758472779769\\
0.925	0.277735376551309\\
0.935	0.273919968241693\\
0.945	0.270298171136704\\
0.955	0.266857104632913\\
0.965	0.263584961409996\\
0.975	0.26047089939824\\
0.985	0.257504946500522\\
0.995	0.254677916344212\\
};
\addplot [color=mycolor2, very thick, forget plot]
  table[row sep=crcr]{%
0.00125	0.353221081475751\\
0.00375	0.353221069673812\\
0.00625	0.35322104598635\\
0.00875	0.35322101042146\\
0.01125	0.353220962976528\\
0.01375	0.35322090357377\\
0.01625	0.353220832112559\\
0.01875	0.353220748473137\\
0.02125	0.353220652516889\\
0.02375	0.353220544085935\\
0.02625	0.353220423002963\\
0.02875	0.353220289070893\\
0.03125	0.353220142072682\\
0.03375	0.353219981771075\\
0.03625	0.353219807908421\\
0.03875	0.353219620206464\\
0.04125	0.353219418366053\\
0.04375	0.353219202067037\\
0.04625	0.353218970967841\\
0.04875	0.353218724705399\\
0.05125	0.353218462894768\\
0.05375	0.353218185128918\\
0.05625	0.353217890978408\\
0.05875	0.353217579991135\\
0.06125	0.353217251691977\\
0.06375	0.35321690558248\\
0.06625	0.353216541140576\\
0.06875	0.353216157820159\\
0.07125	0.353215755050808\\
0.07375	0.3532153322374\\
0.07625	0.353214888759701\\
0.07875	0.353214423972031\\
0.08125	0.353213937202885\\
0.08375	0.35321342775447\\
0.08625	0.353212894902374\\
0.08875	0.353212337895085\\
0.09125	0.353211755953616\\
0.09375	0.353211148271074\\
0.09625	0.353210514012212\\
0.09875	0.353209852313035\\
0.10125	0.353209162280284\\
0.10375	0.353208442991127\\
0.10625	0.353207693492586\\
0.10875	0.353206912801164\\
0.11125	0.353206099902388\\
0.11375	0.353205253750382\\
0.11625	0.353204373267438\\
0.11875	0.353203457343578\\
0.12125	0.353202504836106\\
0.12375	0.353201514569223\\
0.12625	0.353200485333621\\
0.12875	0.353199415886076\\
0.13125	0.353198304949057\\
0.13375	0.35319715121037\\
0.13625	0.353195953322792\\
0.13875	0.35319470990377\\
0.14125	0.353193419535063\\
0.14375	0.353192080762482\\
0.14625	0.35319069209563\\
0.14875	0.35318925200762\\
0.15125	0.353187758934906\\
0.15375	0.353186211277143\\
0.15625	0.353184607396982\\
0.15875	0.353182945620037\\
0.16125	0.353181224234745\\
0.16375	0.353179441492496\\
0.16625	0.353177595607564\\
0.16875	0.353175684757286\\
0.17125	0.353173707082163\\
0.17375	0.353171660686122\\
0.17625	0.353169543636781\\
0.17875	0.353167353965865\\
0.18125	0.353165089669595\\
0.18375	0.35316274870925\\
0.18625	0.353160329011784\\
0.18875	0.353157828470514\\
0.19125	0.353155244945973\\
0.19375	0.353152576266774\\
0.19625	0.353149820230739\\
0.19875	0.353146974605974\\
0.20125	0.353144037132208\\
0.20375	0.353141005522217\\
0.20625	0.353137877463377\\
0.20875	0.353134650619403\\
0.21125	0.353131322632292\\
0.21375	0.353127891124361\\
0.21625	0.353124353700469\\
0.21875	0.353120707950568\\
0.22125	0.353116951452269\\
0.22375	0.353113081773755\\
0.22625	0.353109096476901\\
0.22875	0.353104993120657\\
0.23125	0.353100769264616\\
0.23375	0.353096422472928\\
0.23625	0.353091950318418\\
0.23875	0.353087350387205\\
0.24125	0.353082620283375\\
0.24375	0.353077757634131\\
0.24625	0.353072760095342\\
0.24875	0.353067625357403\\
0.25125	0.353062351151455\\
0.25375	0.3530569352561\\
0.25625	0.353051375504515\\
0.25875	0.353045669791999\\
0.26125	0.353039816084077\\
0.26375	0.353033812425065\\
0.26625	0.353027656947167\\
0.26875	0.353021347880161\\
0.27125	0.353014883561668\\
0.27375	0.353008262448031\\
0.27625	0.353001483125884\\
0.27875	0.352994544324335\\
0.28125	0.352987444928039\\
0.28375	0.352980183990806\\
0.28625	0.352972760750208\\
0.28875	0.352965174642893\\
0.29125	0.352957425320838\\
0.29375	0.352949512668557\\
0.29625	0.35294143682119\\
0.29875	0.352933198183718\\
0.30125	0.352924797451216\\
0.30375	0.352916235630136\\
0.30625	0.352907514060926\\
0.30875	0.352898634441799\\
0.31125	0.352889598853818\\
0.31375	0.352880409787382\\
0.31625	0.352871070170038\\
0.31875	0.352861583395937\\
0.32125	0.352851953356802\\
0.32375	0.352842184474539\\
0.32625	0.352832281735617\\
0.32875	0.352822250727272\\
0.33125	0.352812097675595\\
0.33375	0.352801829485663\\
0.33625	0.352791453783777\\
0.33875	0.352780978961862\\
0.34125	0.352770414224199\\
0.34375	0.352759769636531\\
0.34625	0.352749056177708\\
0.34875	0.352738285793906\\
0.35125	0.352727471455506\\
0.35375	0.352716627216978\\
0.35625	0.352705768279573\\
0.35875	0.35269491105734\\
0.36125	0.352684073246613\\
0.36375	0.352673273899302\\
0.36625	0.35266253350053\\
0.36875	0.352651874051373\\
0.37125	0.35264131915717\\
0.37375	0.352630894122453\\
0.37625	0.352620626052907\\
0.37875	0.352610543964686\\
0.38125	0.35260067890027\\
0.38375	0.352591064048774\\
0.38625	0.352581734866105\\
0.38875	0.352572729187693\\
0.39125	0.352564087322955\\
0.39375	0.352555852117273\\
0.39625	0.352548068965696\\
0.39875	0.352540785764434\\
0.40125	0.352534052794313\\
0.40375	0.352527922550134\\
0.40625	0.352522449563542\\
0.40875	0.352517690321611\\
0.41125	0.352513703457693\\
0.41375	0.352510550485355\\
0.41625	0.352508297445578\\
0.41875	0.352507017916843\\
0.42125	0.352506797849106\\
0.42375	0.352507742551967\\
0.42625	0.35250998578993\\
0.42875	0.35251370018074\\
0.43125	0.352519106810152\\
0.43375	0.352526480040878\\
0.43625	0.352536140865445\\
0.43875	0.352548428991545\\
0.44125	0.352563640692521\\
0.44375	0.352581917464503\\
0.44625	0.352603071806036\\
0.44875	0.352626344304647\\
0.45125	0.352650105368721\\
0.45375	0.352671551042296\\
0.45625	0.352686500606686\\
0.45875	0.352689485771373\\
0.46125	0.352674419895237\\
0.46375	0.352636226923823\\
0.46625	0.352573844412647\\
0.46875	0.352494912930461\\
0.47125	0.352422117767802\\
0.47375	0.352400446213923\\
0.47625	0.352503501425836\\
0.47875	0.352835548916055\\
0.48125	0.353524502604346\\
0.48375	0.354700285799141\\
0.48625	0.356453942110814\\
0.48875	0.358776448273944\\
0.49125	0.361482419355146\\
0.49375	0.364131140769463\\
0.49625	0.365962521989761\\
0.49875	0.365865795995397\\
0.50125	0.362392982921153\\
0.50375	0.353816662878308\\
0.50625	0.33820706366997\\
0.50875	0.313443068143054\\
0.51125	0.276864735601411\\
0.51375	0.223092903903015\\
0.51625	-0.655262544010489\\
0.51875	-0.647341190347123\\
0.52125	-0.639378142313954\\
0.52375	-0.631389669618793\\
0.52625	-0.623389814462109\\
0.52875	-0.615390738765207\\
0.53125	-0.607403008487447\\
0.53375	-0.599435826610499\\
0.53625	-0.591497224980286\\
0.53875	-0.583594223302673\\
0.54125	-0.575732961852474\\
0.54375	-0.567918813031824\\
0.54625	-0.560156475795405\\
0.54875	-0.55245005609598\\
0.55125	-0.544803135839283\\
0.55375	-0.537218832326181\\
0.55625	-0.52969984976497\\
0.55875	-0.522248524129782\\
0.56125	-0.514866862401004\\
0.56375	-0.507556577034534\\
0.56625	-0.500319116356735\\
0.56875	-0.493155691462217\\
0.57125	-0.486067300095251\\
0.57375	-0.47905474791773\\
0.57625	-0.472118667503069\\
0.57875	-0.465259535343494\\
0.58125	-0.458477687115256\\
0.58375	-0.451773331410854\\
0.58625	-0.44514656211771\\
0.58875	-0.43859736959805\\
0.59125	-0.43212565080385\\
0.59375	-0.425731218443182\\
0.59625	-0.419413809299298\\
0.59875	-0.413173091791148\\
0.60125	-0.407008672853083\\
0.60375	-0.400920104202197\\
0.60625	-0.394906888053733\\
0.60875	-0.388968482337971\\
0.61125	-0.383104305466055\\
0.61375	-0.377313740686931\\
0.61625	-0.37159614007301\\
0.61875	-0.365950828168168\\
0.62125	-0.360377105328204\\
0.62375	-0.354874250780745\\
0.62625	-0.349441525428939\\
0.62875	-0.344078174420799\\
0.63125	-0.338783429503997\\
0.63375	-0.333556511183972\\
0.63625	-0.328396630701561\\
0.63875	-0.323302991844858\\
0.64125	-0.318274792608658\\
0.64375	-0.313311226713671\\
0.64625	-0.308411484996583\\
0.64875	-0.303574756681117\\
0.65125	-0.298800230539333\\
0.65375	-0.294087095951668\\
0.65625	-0.289434543873491\\
0.65875	-0.284841767715295\\
0.66125	-0.2803079641431\\
0.66375	-0.275832333805104\\
0.66625	-0.271414081990112\\
0.66875	-0.267052419222906\\
0.67125	-0.262746561801249\\
0.67375	-0.258495732278907\\
0.67625	-0.254299159898723\\
0.67875	-0.250156080979479\\
0.68125	-0.246065739259992\\
0.68375	-0.242027386203674\\
0.68625	-0.238040281266493\\
0.68875	-0.234103692131126\\
0.69125	-0.230216894909838\\
0.69375	-0.226379174318476\\
0.69625	-0.222589823823785\\
0.69875	-0.218848145766102\\
0.70125	-0.215153451459338\\
0.70375	-0.211505061270011\\
0.70625	-0.207902304677024\\
0.70875	-0.204344520313679\\
0.71125	-0.200831055993396\\
0.71375	-0.197361268720467\\
0.71625	-0.193934524687079\\
0.71875	-0.190550199257777\\
0.72125	-0.187207676942444\\
0.72375	-0.183906351358799\\
0.72625	-0.180645625185362\\
0.72875	-0.177424910105757\\
0.73125	-0.174243626745153\\
0.73375	-0.17110120459962\\
0.73625	-0.167997081959099\\
0.73875	-0.164930705824636\\
0.74125	-0.1619015318205\\
0.74375	-0.158909024101749\\
0.74625	-0.155952655257774\\
0.74875	-0.153031906212306\\
0.75125	-0.150146266120365\\
0.75375	-0.147295232262527\\
0.75625	-0.144478309936959\\
0.75875	-0.141695012349545\\
0.76125	-0.138944860502447\\
0.76375	-0.136227383081424\\
0.76625	-0.133542116342194\\
0.76875	-0.130888603996084\\
0.77125	-0.128266397095245\\
0.77375	-0.125675053917631\\
0.77625	-0.123114139851955\\
0.77875	-0.120583227282826\\
0.78125	-0.118081895476214\\
0.78375	-0.115609730465423\\
0.78625	-0.113166324937713\\
0.78875	-0.110751278121685\\
0.79125	-0.108364195675576\\
0.79375	-0.106004689576544\\
0.79625	-0.103672378011054\\
0.79875	-0.101366885266451\\
0.80125	-0.0990878416237931\\
0.80375	-0.0968348832520104\\
0.80625	-0.0946076521034558\\
0.80875	-0.0924057958108956\\
0.81125	-0.0902289675859882\\
0.81375	-0.0880768261192806\\
0.81625	-0.0859490354817727\\
0.81875	-0.0838452650280543\\
0.82125	-0.081765189301052\\
0.82375	-0.079708487938395\\
0.82625	-0.0776748455804159\\
0.82875	-0.0756639517797896\\
0.83125	-0.0736755009128146\\
0.83375	-0.071709192092345\\
0.83625	-0.0697647290823575\\
0.83875	-0.0678418202141558\\
0.84125	-0.0659401783042037\\
0.84375	-0.0640595205735697\\
0.84625	-0.0621995685689772\\
0.84875	-0.0603600480854419\\
0.85125	-0.0585406890904743\\
0.85375	-0.0567412256498385\\
0.85625	-0.0549613958548333\\
0.85875	-0.053200941751092\\
0.86125	-0.051459609268863\\
0.86375	-0.0497371481547489\\
0.86625	-0.0480333119048911\\
0.86875	-0.0463478576995601\\
0.87125	-0.0446805463391396\\
0.87375	-0.0430311421814595\\
0.87625	-0.0413994130804769\\
0.87875	-0.0397851303262536\\
0.88125	-0.0381880685862114\\
0.88375	-0.0366080058476518\\
0.88625	-0.0350447233614836\\
0.88875	-0.0334980055871628\\
0.89125	-0.0319676401387958\\
0.89375	-0.0304534177323826\\
0.89625	-0.0289551321341831\\
0.89875	-0.0274725801101688\\
0.90125	-0.0260055613765409\\
0.90375	-0.0245538785512783\\
0.90625	-0.02311733710671\\
0.90875	-0.0216957453230609\\
0.91125	-0.0202889142429683\\
0.91375	-0.0188966576269304\\
0.91625	-0.0175187919096725\\
0.91875	-0.0161551361574004\\
0.92125	-0.0148055120259205\\
0.92375	-0.0134697437196082\\
0.92625	-0.0121476579511964\\
0.92875	-0.0108390839023655\\
0.93125	-0.0095438531851147\\
0.93375	-0.00826179980389491\\
0.93625	-0.00699276011848527\\
0.93875	-0.00573657280757905\\
0.94125	-0.00449307883308859\\
0.94375	-0.00326212140512229\\
0.94625	-0.00204354594763826\\
0.94875	-0.000837200064738681\\
0.95125	0.000357066492398051\\
0.95375	0.00153940185796294\\
0.95625	0.00270995208337569\\
0.95875	0.00386886116853775\\
0.96125	0.00501627109253614\\
0.96375	0.0061523218437674\\
0.96625	0.00727715144954059\\
0.96875	0.00839089600513291\\
0.97125	0.00949368970233951\\
0.97375	0.0105856648575171\\
0.97625	0.0116669519391324\\
0.97875	0.0127376795948365\\
0.98125	0.0137979746780713\\
0.98375	0.0148479622742075\\
0.98625	0.0158877657262614\\
0.98875	0.0169175066601502\\
0.99125	0.017937305009541\\
0.99375	0.0189472790402776\\
0.99625	0.0199475453744071\\
0.99875	0.0209382190138033\\
};
\addplot [color=black, dashed, very thick, forget plot]
  table[row sep=crcr]{%
0	0.353216610290657\\
0.01	0.353216610290657\\
0.02	0.353216610290657\\
0.03	0.353216610290657\\
0.04	0.353216610290657\\
0.05	0.353216610290657\\
0.06	0.353216610290657\\
0.07	0.353216610290657\\
0.08	0.353216610290657\\
0.09	0.353216610290657\\
0.1	0.353216610290657\\
0.11	0.353216610290657\\
0.12	0.353216610290657\\
0.13	0.353216610290657\\
0.14	0.353216610290657\\
0.15	0.353216610290657\\
0.16	0.353216610290657\\
0.17	0.353216610290657\\
0.18	0.353216610290657\\
0.19	0.353216610290657\\
0.2	0.353216610290657\\
0.21	0.353216610290657\\
0.22	0.353216610290657\\
0.23	0.353216610290657\\
0.24	0.353216610290657\\
0.25	0.353216610290657\\
0.26	0.353216610290657\\
0.27	0.353216610290657\\
0.28	0.353216610290657\\
0.29	0.353216610290657\\
0.3	0.353216610290657\\
0.31	0.353216610290657\\
0.32	0.353216610290657\\
0.33	0.353216610290657\\
0.34	0.353216610290657\\
0.35	0.353216610290657\\
0.36	0.353216610290657\\
0.37	0.353216610290657\\
0.38	0.353216610290657\\
0.39	0.353216610290657\\
0.4	0.353216610290657\\
0.41	0.353216610290657\\
0.42	0.353216610290657\\
0.43	0.353216610290657\\
0.44	0.353216610290657\\
0.45	0.353216610290657\\
0.46	0.353216610290657\\
0.47	0.353216610290657\\
0.48	0.353216610290657\\
0.49	0.353216610290657\\
0.5	0.353216610290657\\
0.51	0.353216610290657\\
0.515	-0.676370374722011\\
0.525	-0.645571273243485\\
0.535	-0.614058165463791\\
0.545	-0.58265754198838\\
0.555	-0.551837711959221\\
0.565	-0.521874502298237\\
0.575	-0.49292939216261\\
0.585	-0.465091854948222\\
0.595	-0.438404408723227\\
0.605	-0.41287834157258\\
0.615	-0.388504002785887\\
0.625	-0.365257741751404\\
0.635	-0.343106687897409\\
0.645	-0.322012093318119\\
0.655	-0.301931693390062\\
0.665	-0.282821382633113\\
0.675	-0.264636405362878\\
0.685	-0.247332198191673\\
0.695	-0.230864980315007\\
0.705	-0.215192159802337\\
0.715	-0.200272605040071\\
0.725	-0.186066817121753\\
0.735	-0.1725370294892\\
0.745	-0.159647254294219\\
0.755	-0.147363289974489\\
0.765	-0.135652700879049\\
0.775	-0.124484777067233\\
0.785	-0.113830480380827\\
0.795	-0.103662381369184\\
0.805	-0.0939545905000158\\
0.815	-0.0846826862194753\\
0.825	-0.0758236417645848\\
0.835	-0.0673557521280435\\
0.845	-0.0592585621917602\\
0.855	-0.0515127967528798\\
0.865	-0.0441002929430665\\
0.875	-0.0370039353722617\\
0.885	-0.0302075941996514\\
0.895	-0.0236960662377724\\
0.905	-0.0174550191233128\\
0.915	-0.0114709385346226\\
0.925	-0.00573107839697448\\
0.935	-0.000223413988829162\\
0.945	0.00506340215678348\\
0.955	0.0101400816742247\\
0.965	0.0150167392365168\\
0.975	0.0197029290055771\\
0.985	0.0242076787919346\\
0.995	0.0285395220531258\\
};
\end{axis}

\end{tikzpicture}%

%% file: Figs/Fig12c_Front_ChemGrad.tex
%
%
\begin{tikzpicture}

\definecolor{mycolor}{rgb}{0.8333,0.5208,0.3317}%

\begin{axis}[%
thick, 
width=0.33\textwidth, 
height=0.4\textwidth,
y tick label style = overlay,
xmin=0,xmax=1,ymin=-8,ymax=2, 
xlabel = {$r/a$},
ylabel = {$\mathrm{d}\mu/\mathrm{d}r$}]
\addplot [color=mycolor, very thick, forget plot]
  table[row sep=crcr]{%
0.00125	-1.04112290926542e-05\\
0.00375	-2.08324336953624e-05\\
0.00625	-3.1274790599721e-05\\
0.00875	-4.17492766126446e-05\\
0.01125	-5.22668178232391e-05\\
0.01375	-6.28383062415141e-05\\
0.01625	-7.34746830833559e-05\\
0.01875	-8.41868866052652e-05\\
0.02125	-9.49859362620501e-05\\
0.02375	-0.000105882869842015\\
0.02625	-0.000116888801541736\\
0.02875	-0.000128014907334573\\
0.03125	-0.000139272440881891\\
0.03375	-0.000150672729083805\\
0.03625	-0.000162227181462964\\
0.03875	-0.000173947321394995\\
0.04125	-0.000185844733958063\\
0.04375	-0.000197931166434561\\
0.04625	-0.000210218422293625\\
0.04875	-0.000222718460763356\\
0.05125	-0.000235443347710937\\
0.05375	-0.000248405289965039\\
0.05625	-0.000261616617560431\\
0.05875	-0.000275089816678638\\
0.06125	-0.000288837520783477\\
0.06375	-0.000302872499573\\
0.06625	-0.000317207709222406\\
0.06875	-0.00033185624463011\\
0.07125	-0.000346831378464515\\
0.07375	-0.000362146565845336\\
0.07625	-0.000377815429129395\\
0.07875	-0.000393851768403772\\
0.08125	-0.000410269589139668\\
0.08375	-0.000427083067162865\\
0.08625	-0.000444306590244739\\
0.08875	-0.000461954733764416\\
0.09125	-0.000480042273350835\\
0.09375	-0.000498584195184577\\
0.09625	-0.000517595679569151\\
0.09875	-0.000537092136950934\\
0.10125	-0.000557089152046527\\
0.10375	-0.000577602557154582\\
0.10625	-0.000598648378321368\\
0.10875	-0.000620242859799307\\
0.11125	-0.000642402454076272\\
0.11375	-0.000665143826149923\\
0.11625	-0.000688483853439544\\
0.11875	-0.000712439629307062\\
0.12125	-0.000737028447651001\\
0.12375	-0.000762267805162615\\
0.12625	-0.000788175404390692\\
0.12875	-0.000814769144760286\\
0.13125	-0.000842067116141157\\
0.13375	-0.000870087599238809\\
0.13625	-0.000898849041568228\\
0.13875	-0.000928370081998453\\
0.14125	-0.000958669507038151\\
0.14375	-0.000989766264970401\\
0.14625	-0.00102167945530843\\
0.14875	-0.0010544283051537\\
0.15125	-0.0010880321485721\\
0.15375	-0.00112251045058974\\
0.15625	-0.00115788274824846\\
0.15875	-0.00119416868141547\\
0.16125	-0.00123138790711208\\
0.16375	-0.0012695601560831\\
0.16625	-0.00130870515387613\\
0.16875	-0.0013488426391361\\
0.17125	-0.00138999231049778\\
0.17375	-0.00143217382010441\\
0.17625	-0.00147540672295181\\
0.17875	-0.00151971047865257\\
0.18125	-0.00156510439070927\\
0.18375	-0.00161160758490332\\
0.18625	-0.00165923897980403\\
0.18875	-0.00170801722783779\\
0.19125	-0.00175796070314825\\
0.19375	-0.00180908741412549\\
0.19625	-0.0018614150061369\\
0.19875	-0.00191496067522642\\
0.20125	-0.00196974112931752\\
0.20375	-0.00202577253553098\\
0.20625	-0.0020830704537182\\
0.20875	-0.00214164976226546\\
0.21125	-0.00220152461120988\\
0.21375	-0.0022627083548185\\
0.21625	-0.00232521343646587\\
0.21875	-0.00238905136273903\\
0.22125	-0.00245423258737722\\
0.22375	-0.00252076642407393\\
0.22625	-0.00258866095296252\\
0.22875	-0.0026579229359216\\
0.23125	-0.00272855769915287\\
0.23375	-0.00280056904024302\\
0.23625	-0.00287395905816503\\
0.23875	-0.00294872810761087\\
0.24125	-0.00302487462855918\\
0.24375	-0.0031023949996539\\
0.24625	-0.00318128341116768\\
0.24875	-0.00326153172648124\\
0.25125	-0.0033431292971773\\
0.25375	-0.00342606281353496\\
0.25625	-0.00351031612928797\\
0.25875	-0.00359587006117582\\
0.26125	-0.0036827022082394\\
0.26375	-0.00377078674841876\\
0.26625	-0.00386009421355569\\
0.26875	-0.00395059127558511\\
0.27125	-0.00404224050147996\\
0.27375	-0.00413500010808399\\
0.27625	-0.00422882371060196\\
0.27875	-0.00432366000610218\\
0.28125	-0.00441945255157197\\
0.28375	-0.00451613940064279\\
0.28625	-0.00461365282475374\\
0.28875	-0.00471191896615929\\
0.29125	-0.00481085747714673\\
0.29375	-0.00491038117916164\\
0.29625	-0.00501039565051548\\
0.29875	-0.0051107988311638\\
0.30125	-0.00521148062351781\\
0.30375	-0.0053123224066683\\
0.30625	-0.00541319659756206\\
0.30875	-0.00551396615360201\\
0.31125	-0.00561448405390209\\
0.31375	-0.00571459279008734\\
0.31625	-0.00581412376360171\\
0.31875	-0.00591289671520666\\
0.32125	-0.00601071910655308\\
0.32375	-0.00610738546305488\\
0.32625	-0.00620267669282227\\
0.32875	-0.00629635938170232\\
0.33125	-0.00638818503025215\\
0.33375	-0.00647788927672735\\
0.33625	-0.00656519107857044\\
0.33875	-0.00664979185457877\\
0.34125	-0.00673137459237107\\
0.34375	-0.00680960290658589\\
0.34625	-0.00688412007736908\\
0.34875	-0.00695454804995553\\
0.35125	-0.00702048633181666\\
0.35375	-0.00708151092858866\\
0.35625	-0.00713717312541546\\
0.35875	-0.00718699821126188\\
0.36125	-0.00723048409154984\\
0.36375	-0.00726709973144712\\
0.36625	-0.00729628335775907\\
0.36875	-0.00731744050156878\\
0.37125	-0.00732994171589542\\
0.37375	-0.00733312013870012\\
0.37625	-0.00732626893054685\\
0.37875	-0.00730863892249605\\
0.38125	-0.00727943684796367\\
0.38375	-0.00723782490966415\\
0.38625	-0.00718292248894991\\
0.38875	-0.00711381108782353\\
0.39125	-0.00702954354883867\\
0.39375	-0.00692915809572755\\
0.39625	-0.00681169672870478\\
0.39875	-0.00667622569378414\\
0.40125	-0.00652185224005436\\
0.40375	-0.00634772776417849\\
0.40625	-0.00615302106919346\\
0.40875	-0.00593683945187794\\
0.41125	-0.0056980690758397\\
0.41375	-0.00543510412295771\\
0.41625	-0.0051454395458982\\
0.41875	-0.00482512173310329\\
0.42125	-0.00446809295987262\\
0.42375	-0.00406553832890216\\
0.42625	-0.00360545689636984\\
0.42875	-0.00307283627791196\\
0.43125	-0.00245100752017293\\
0.43375	-0.00172496927816115\\
0.43625	-0.000887639439734995\\
0.43875	4.99931814488309e-05\\
0.44125	0.00104416909538697\\
0.44375	0.00199940683597872\\
0.44625	0.00274753114821143\\
0.44875	0.00303025213274363\\
0.45125	0.00249592133319176\\
0.45375	0.000727759010323942\\
0.45625	-0.00267189347741627\\
0.45875	-0.00791152739777453\\
0.46125	-0.0147266342287595\\
0.46375	-0.0219703839960206\\
0.46625	-0.0271071065500969\\
0.46875	-0.0257084495315447\\
0.47125	-0.0111560470833181\\
0.47375	0.0251238385292571\\
0.47625	0.0924682689190543\\
0.47875	0.198393436159611\\
0.48125	0.344356080357508\\
0.48375	0.520033542178896\\
0.48625	0.696998454781885\\
0.48875	0.823480698142214\\
0.49125	0.822497788936655\\
0.49375	0.595594543205851\\
0.49625	0.0335778187827021\\
0.49875	-0.965838405456012\\
0.50125	-2.4752484440847\\
0.50375	-4.5130495301744\\
0.50625	-7.022133685283\\
0.50875	-9.85726102930163\\
0.51125	-12.777537091815\\
0.51375	-5.16681118160744\\
0.51625	-3.30882162608969\\
0.51875	-3.31185667188262\\
0.52125	-3.31562906632923\\
0.52375	-3.32006406215189\\
0.52625	-3.32508920944414\\
0.52875	-3.33063426618789\\
0.53125	-3.3366311479412\\
0.53375	-3.34301390418298\\
0.53625	-3.34971871187988\\
0.53875	-3.35668388051182\\
0.54125	-3.36384986358927\\
0.54375	-3.37115927413515\\
0.54625	-3.37855690114468\\
0.54875	-3.38598972611132\\
0.55125	-3.39340693774815\\
0.55375	-3.40075994476229\\
0.55625	-3.40800238524793\\
0.55875	-3.41509013298511\\
0.56125	-3.42198129975818\\
0.56375	-3.42863623389861\\
0.56625	-3.43501751462739\\
0.56875	-3.4410899423746\\
0.57125	-3.44682052484414\\
0.57375	-3.45217845900935\\
0.57625	-3.45713510922056\\
0.57875	-3.46166398118298\\
0.58125	-3.46574069228154\\
0.58375	-3.46934293828137\\
0.58625	-3.47245045649724\\
0.58875	-3.47504498577237\\
0.59125	-3.47711022329808\\
0.59375	-3.47863177872827\\
0.59625	-3.47959712532334\\
0.59875	-3.47999554908247\\
0.60125	-3.47981809529201\\
0.60375	-3.47905751335381\\
0.60625	-3.47770819954706\\
0.60875	-3.47576613870608\\
0.61125	-3.4732288439959\\
0.61375	-3.47009529614185\\
0.61625	-3.46636588133678\\
0.61875	-3.46204232870107\\
0.62125	-3.457127647196\\
0.62375	-3.45162606219139\\
0.62625	-3.44554295195937\\
0.62875	-3.43888478432572\\
0.63125	-3.43165905333874\\
0.63375	-3.4238742166085\\
0.63625	-3.41553963298026\\
0.63875	-3.40666550106457\\
0.64125	-3.39726279841885\\
0.64375	-3.38734322191692\\
0.64625	-3.37691912896952\\
0.64875	-3.36600348005711\\
0.65125	-3.35460978246506\\
0.65375	-3.3427520354351\\
0.65625	-3.3304446767291\\
0.65875	-3.3177025306767\\
0.66125	-3.30454075786451\\
0.66375	-3.29097480643976\\
0.66625	-3.27702036496801\\
0.66875	-3.26269331724019\\
0.67125	-3.2480096985701\\
0.67375	-3.23298565404981\\
0.67625	-3.21763739855851\\
0.67875	-3.20198117841659\\
0.68125	-3.18603323508824\\
0.68375	-3.16980977052415\\
0.68625	-3.15332691436306\\
0.68875	-3.13660069289817\\
0.69125	-3.11964699987086\\
0.69375	-3.10248156898891\\
0.69625	-3.08511994813745\\
0.69875	-3.06757747541467\\
0.70125	-3.04986925676025\\
0.70375	-3.03201014525687\\
0.70625	-3.0140147220142\\
0.70875	-2.9958972787252\\
0.71125	-2.97767180165552\\
0.71375	-2.95935195713642\\
0.71625	-2.94095107863029\\
0.71875	-2.92248215506013\\
0.72125	-2.90395782049372\\
0.72375	-2.88539034534167\\
0.72625	-2.86679162856328\\
0.72875	-2.848173191281\\
0.73125	-2.82954617142243\\
0.73375	-2.81092131963642\\
0.73625	-2.79230899611815\\
0.73875	-2.7737191685724\\
0.74125	-2.7551614110913\\
0.74375	-2.73664490398447\\
0.74625	-2.71817843441623\\
0.74875	-2.69977039798093\\
0.75125	-2.68142880098979\\
0.75375	-2.66316126339792\\
0.75625	-2.64497502269058\\
0.75875	-2.62687693801171\\
0.76125	-2.60887349525611\\
0.76375	-2.59097081258101\\
0.76625	-2.57317464621201\\
0.76875	-2.55549039718646\\
0.77125	-2.53792311801914\\
0.77375	-2.52047752010337\\
0.77625	-2.50315798134657\\
0.77875	-2.48596855409151\\
0.78125	-2.46891297334183\\
0.78375	-2.45199466526555\\
0.78625	-2.43521675591209\\
0.78875	-2.41858207996432\\
0.79125	-2.40209318993566\\
0.79375	-2.38575236514074\\
0.79625	-2.36956162106197\\
0.79875	-2.35352271860511\\
0.80125	-2.33763717340929\\
0.80375	-2.32190626530831\\
0.80625	-2.3063310476313\\
0.80875	-2.29091235655253\\
0.81125	-2.27565082040823\\
0.81375	-2.26054686894599\\
0.81625	-2.24560074244766\\
0.81875	-2.23081250082602\\
0.82125	-2.21618203252532\\
0.82375	-2.20170906342359\\
0.82625	-2.18739316544279\\
0.82875	-2.17323376512332\\
0.83125	-2.15923015203929\\
0.83375	-2.14538148692862\\
0.83625	-2.13168680972724\\
0.83875	-2.11814504750174\\
0.84125	-2.10475502199571\\
0.84375	-2.09151545709704\\
0.84625	-2.07842498612426\\
0.84875	-2.06548215880032\\
0.85125	-2.05268544810513\\
0.85375	-2.04003325684599\\
0.85625	-2.02752392408024\\
0.85875	-2.01515573123775\\
0.86125	-2.00292690807696\\
0.86375	-1.99083563840968\\
0.86625	-1.97888006560125\\
0.86875	-1.96705829781714\\
0.87125	-1.9553684131139\\
0.87375	-1.94380846428033\\
0.87625	-1.93237648346789\\
0.87875	-1.92107048654224\\
0.88125	-1.9098884773796\\
0.88375	-1.8988284518063\\
0.88625	-1.88788840142877\\
0.88875	-1.87706631718547\\
0.89125	-1.86636019278574\\
0.89375	-1.85576802795521\\
0.89625	-1.8452878313366\\
0.89875	-1.83491762350027\\
0.90125	-1.82465543950918\\
0.90375	-1.81449933141972\\
0.90625	-1.80444737071629\\
0.90875	-1.79449765031459\\
0.91125	-1.7846482867656\\
0.91375	-1.77489742200053\\
0.91625	-1.76524322506777\\
0.91875	-1.7556838938244\\
0.92125	-1.74621765624386\\
0.92375	-1.73684277180172\\
0.92625	-1.72755753276749\\
0.92875	-1.71836026511392\\
0.93125	-1.70924932962029\\
0.93375	-1.70022312270285\\
0.93625	-1.69128007716786\\
0.93875	-1.68241866285405\\
0.94125	-1.67363738726837\\
0.94375	-1.66493479600306\\
0.94625	-1.65630947317338\\
0.94875	-1.64776004172249\\
0.95125	-1.63928516370694\\
0.95375	-1.63088354036257\\
0.95625	-1.62255391240043\\
0.95875	-1.6142950598212\\
0.96125	-1.60610580213015\\
0.96375	-1.59798499808222\\
0.96625	-1.58993154569567\\
0.96875	-1.58194438196309\\
0.97125	-1.5740224827447\\
0.97375	-1.56616486244315\\
0.97625	-1.55837057370371\\
0.97875	-1.55063870707422\\
0.98125	-1.5429683906989\\
0.98375	-1.53535878974191\\
0.98625	-1.52780910616243\\
0.98875	-1.52031857804463\\
0.99125	-1.51288647925338\\
0.99375	-1.50551211874868\\
0.99625	-1.49819484020904\\
0.99875	-1.49093402128551\\
};
\addplot [color=black, dashed, very thick, forget plot]
  table[row sep=crcr]{%
0	0\\
0.01	0\\
0.02	0\\
0.03	0\\
0.04	0\\
0.05	0\\
0.06	0\\
0.07	0\\
0.08	0\\
0.09	0\\
0.1	0\\
0.11	0\\
0.12	0\\
0.13	0\\
0.14	0\\
0.15	0\\
0.16	0\\
0.17	0\\
0.18	0\\
0.19	0\\
0.2	0\\
0.21	0\\
0.22	0\\
0.23	0\\
0.24	0\\
0.25	0\\
0.26	0\\
0.27	0\\
0.28	0\\
0.29	0\\
0.3	0\\
0.31	0\\
0.32	0\\
0.33	0\\
0.34	0\\
0.35	0\\
0.36	0\\
0.37	0\\
0.38	0\\
0.39	0\\
0.4	0\\
0.41	0\\
0.42	0\\
0.43	0\\
0.44	0\\
0.45	0\\
0.46	0\\
0.47	0\\
0.48	0\\
0.49	0\\
0.5	0\\
0.51	0\\
0.515	-5.41983448602831\\
0.525	-5.21533098070516\\
0.535	-5.02218744539037\\
0.545	-4.83957781855689\\
0.555	-4.66674978185816\\
0.565	-4.50301699916002\\
0.575	-4.34775229204343\\
0.585	-4.20038162482828\\
0.595	-4.06037879120644\\
0.605	-3.92726071048934\\
0.615	-3.80058325482678\\
0.625	-3.67993753998556\\
0.635	-3.56494662175425\\
0.645	-3.45526254806047\\
0.655	-3.35056372369176\\
0.665	-3.250552550301\\
0.675	-3.15495330931547\\
0.685	-3.06351025959158\\
0.695	-2.97598592527687\\
0.705	-2.8921595524508\\
0.715	-2.81182571579414\\
0.725	-2.73479305884777\\
0.735	-2.660883153421\\
0.745	-2.58992946544184\\
0.755	-2.52177641604642\\
0.765	-2.45627852801377\\
0.775	-2.39329964879394\\
0.785	-2.3327122423739\\
0.795	-2.27439674309855\\
0.805	-2.21824096532828\\
0.815	-2.16413956348656\\
0.825	-2.11199353764093\\
0.835	-2.06170978028163\\
0.845	-2.01320066042066\\
0.855	-1.96638364154011\\
0.865	-1.92118093027747\\
0.875	-1.87751915305386\\
0.885	-1.83532905813382\\
0.895	-1.79454524085623\\
0.905	-1.75510588999952\\
0.915	-1.71695255344365\\
0.925	-1.68002992146894\\
0.935	-1.64428562619104\\
0.945	-1.6096700557732\\
0.955	-1.57613618218454\\
0.965	-1.54363940138727\\
0.975	-1.51213738493818\\
0.985	-1.48158994208236\\
0.995	-1.45195889149957\\
};
\end{axis}

\end{tikzpicture}%

%% file: Figs/Fig13a_Front_Dynamics_CaiSuo.tex
%
%
\begin{tikzpicture}
\definecolor{mycolor1}{rgb}{0.2,0.6,1}%
\definecolor{mycolor2}{rgb}{1,0.5,0.2}%
\begin{axis}[%
thick, 
axis on top,
width=0.48\textwidth, 
height=0.4\textwidth,
xmin=0,xmax=0.3,ymin=0,ymax=2.5, 
xlabel = {$t$},
ylabel = {$r$},
scaled ticks=false, tick label style={/pgf/number format/fixed}],
\addplot [color=black, dashed, very thick,]
  table[row sep=crcr]{%
0.001	2.19503152096144\\
0.002	2.15194096875085\\
0.003	2.12154664441323\\
0.004	2.09621236494329\\
0.005	2.07388062956072\\
0.006	2.05360961739304\\
0.007	2.03487033431704\\
0.008	2.01732921359418\\
0.009	2.00075938625768\\
0.01	1.98499803445133\\
0.011	1.96992350738523\\
0.012	1.95544201197636\\
0.013	1.94147939360368\\
0.014	1.92797579792937\\
0.015	1.91488207072532\\
0.016	1.9021572508376\\
0.017	1.88976676844302\\
0.018	1.87768112610378\\
0.019	1.86587491376165\\
0.02	1.85432605137414\\
0.021	1.84301520951521\\
0.022	1.83192534919589\\
0.023	1.82104135823777\\
0.024	1.81034975869274\\
0.025	1.79983846899989\\
0.026	1.78949660736508\\
0.027	1.7793143330014\\
0.028	1.76928270994373\\
0.029	1.75939359410509\\
0.03	1.74963953954086\\
0.031	1.74001371505556\\
0.032	1.73050983585428\\
0.033	1.72112210512421\\
0.034	1.71184516068672\\
0.035	1.70267403043416\\
0.036	1.6936040948939\\
0.037	1.68463105184643\\
0.038	1.67575088677466\\
0.039	1.66695984619663\\
0.04	1.65825441427049\\
0.041	1.64963129297445\\
0.042	1.64108738188302\\
0.043	1.6326197630505\\
0.044	1.62422568560453\\
0.045	1.61590255262662\\
0.046	1.60764790913885\\
0.047	1.59945943141099\\
0.048	1.59133491703383\\
0.049	1.58327227619281\\
0.05	1.57526952382282\\
0.051	1.56732477127102\\
0.052	1.55943622115261\\
0.053	1.55160216019577\\
0.054	1.54382095400358\\
0.055	1.53609104184429\\
0.056	1.52841093197793\\
0.057	1.52077919726396\\
0.058	1.51319447132049\\
0.059	1.50565544469162\\
0.06	1.49816086149067\\
0.061	1.49070951630011\\
0.062	1.4833002512165\\
0.063	1.47593195315315\\
0.064	1.46860355107436\\
0.065	1.46131401455695\\
0.066	1.45406235033137\\
0.067	1.44684760109545\\
0.068	1.43966884358067\\
0.069	1.43252518585968\\
0.07	1.42541576752973\\
0.071	1.41833975637462\\
0.072	1.41129634771512\\
0.073	1.40428476299754\\
0.074	1.39730424842245\\
0.075	1.39035407376756\\
0.076	1.38343353125953\\
0.077	1.37654193451202\\
0.078	1.36967861749088\\
0.079	1.3628429336064\\
0.08	1.3560342548146\\
0.081	1.34925197073145\\
0.082	1.34249548786267\\
0.083	1.33576422903092\\
0.084	1.32905763199297\\
0.085	1.32237514969024\\
0.086	1.31571624903247\\
0.087	1.30908041037333\\
0.088	1.30246712699411\\
0.089	1.29587590454794\\
0.09	1.28930626045377\\
0.091	1.28275772349251\\
0.092	1.27622983328409\\
0.093	1.26972213983681\\
0.094	1.26323420311539\\
0.095	1.25676559264212\\
0.096	1.25031588711667\\
0.097	1.24388467393695\\
0.098	1.23747154898958\\
0.099	1.23107611617512\\
0.1	1.22469798701438\\
0.101	1.21833678082506\\
0.102	1.21199212365878\\
0.103	1.20566364849576\\
0.104	1.19935099485502\\
0.105	1.19305380851042\\
0.106	1.18677174124559\\
0.107	1.18050445047849\\
0.108	1.17425159957656\\
0.109	1.16801285650453\\
0.11	1.16178789473022\\
0.111	1.15557639241224\\
0.112	1.14937803235012\\
0.113	1.14319249178847\\
0.114	1.13701947184519\\
0.115	1.13085866793582\\
0.116	1.12470977930512\\
0.117	1.11857250884477\\
0.118	1.11244656291606\\
0.119	1.10633165117704\\
0.12	1.1002274864139\\
0.121	1.09413378437648\\
0.122	1.08805026361726\\
0.123	1.08197664533399\\
0.124	1.07591265321537\\
0.125	1.06985801328972\\
0.126	1.06381245377629\\
0.127	1.05777570493917\\
0.128	1.05174749894325\\
0.129	1.04572756971239\\
0.13	1.03971565278925\\
0.131	1.03371148519683\\
0.132	1.02771480530137\\
0.133	1.02172535267643\\
0.134	1.01574286796798\\
0.135	1.00976709276033\\
0.136	1.00379776944258\\
0.137	0.997834641075631\\
0.138	0.991877451259252\\
0.139	0.985925943999316\\
0.14	0.979979863574807\\
0.141	0.974038954404495\\
0.142	0.968102960913069\\
0.143	0.962171627396519\\
0.144	0.956244697886583\\
0.145	0.950321916014052\\
0.146	0.944403024870721\\
0.147	0.938487766869792\\
0.148	0.932575883604499\\
0.149	0.926667115704743\\
0.15	0.920761202691508\\
0.151	0.914857882828829\\
0.152	0.908956892973069\\
0.153	0.903057968419251\\
0.154	0.897160842744195\\
0.155	0.89126524764619\\
0.156	0.885370912780915\\
0.157	0.879477565593324\\
0.158	0.873584931145192\\
0.159	0.867692731937991\\
0.16	0.861800687730779\\
0.161	0.855908515352741\\
0.162	0.850015928510006\\
0.163	0.84412263758637\\
0.164	0.838228349437495\\
0.165	0.832332767178167\\
0.166	0.826435589962142\\
0.167	0.820536511641713\\
0.168	0.814635224980335\\
0.169	0.808731414734284\\
0.17	0.802824761845943\\
0.171	0.796914942066759\\
0.172	0.791001625681354\\
0.173	0.785084477220138\\
0.174	0.779163155159649\\
0.175	0.773237311609861\\
0.176	0.767306591987568\\
0.177	0.761370634674956\\
0.178	0.755429070662356\\
0.179	0.749481523174161\\
0.18	0.743527607276738\\
0.181	0.737566929467151\\
0.182	0.731599087241376\\
0.183	0.725623668640594\\
0.184	0.719640251774058\\
0.185	0.713648404316883\\
0.186	0.707647679466136\\
0.187	0.701637625677732\\
0.188	0.69561777601343\\
0.189	0.689587650801966\\
0.19	0.683546756981655\\
0.191	0.677494587404814\\
0.192	0.671430620101108\\
0.193	0.665354317496696\\
0.194	0.659265125585733\\
0.195	0.653162473050511\\
0.196	0.647045770326133\\
0.197	0.640914408605276\\
0.198	0.634767766877418\\
0.199	0.628605187133297\\
0.2	0.6224259962459\\
0.201	0.616229497165411\\
0.202	0.610014967541091\\
0.203	0.603781658245593\\
0.204	0.597528791793005\\
0.205	0.591255560640951\\
0.206	0.584961125366107\\
0.207	0.578644612701213\\
0.208	0.572305113420435\\
0.209	0.565941680058313\\
0.21	0.559553324445922\\
0.211	0.553139015045863\\
0.212	0.546697674065519\\
0.213	0.54022817432547\\
0.214	0.533729335857082\\
0.215	0.52719989193936\\
0.216	0.520638572181009\\
0.217	0.514044015145297\\
0.218	0.507414788819525\\
0.219	0.500749429830839\\
0.22	0.494046310772953\\
0.221	0.487303756882097\\
0.222	0.480519999146724\\
0.223	0.473693166537524\\
0.224	0.466821277403783\\
0.225	0.459902229924089\\
0.226	0.452933791481218\\
0.227	0.445913586809391\\
0.228	0.438839084736068\\
0.229	0.431707583309345\\
0.23	0.424516193064252\\
0.231	0.417261818135596\\
0.232	0.409941134869179\\
0.233	0.402550567514792\\
0.234	0.395086256066228\\
0.235	0.387544037530466\\
0.236	0.379919396652291\\
0.237	0.372207413313748\\
0.238	0.364402777278298\\
0.239	0.356499624763855\\
0.24	0.348491554618588\\
0.241	0.34037152035174\\
0.242	0.332131739795387\\
0.243	0.323763581885625\\
0.244	0.315257433614488\\
0.245	0.306602535253076\\
0.246	0.297786964029228\\
0.247	0.288796902714153\\
0.248	0.279616775468591\\
0.249	0.270228605001631\\
0.25	0.260611853248858\\
0.251	0.250742364275587\\
0.252	0.240591769586688\\
0.253	0.230126249814592\\
0.254	0.219305134042065\\
0.255	0.208078464899391\\
0.256	0.196384193962556\\
0.257	0.184143239103087\\
0.258	0.171252470703636\\
0.259	0.157573771931687\\
0.26	0.142913281345764\\
0.261	0.126986427135374\\
0.262	0.109345366269678\\
0.263	0.0892126390935905\\
0.264	0.0650172467930434\\
0.265	0.0325806441586002\\
0.266	-0.0302007850757924\\
};
\addplot [color=black, dashed, very thick,]
  table[row sep=crcr]{%
0.001	2.206061830112\\
0.002	2.16827747137478\\
0.003	2.14176151347204\\
0.004	2.11974465550974\\
0.005	2.10040228193962\\
0.006	2.08289890616324\\
0.007	2.0667651235204\\
0.008	2.05170486577497\\
0.009	2.03751672149015\\
0.01	2.02405611399767\\
0.011	2.01121500312577\\
0.012	1.99891007812514\\
0.013	1.98707546485282\\
0.014	1.97565798879955\\
0.015	1.96461397986837\\
0.016	1.95390704689138\\
0.017	1.94350647822832\\
0.018	1.93338607066429\\
0.019	1.92352325455364\\
0.02	1.91389842138273\\
0.021	1.90449440891546\\
0.022	1.89529609249127\\
0.023	1.88629006191511\\
0.024	1.8774643614481\\
0.025	1.86880827842052\\
0.026	1.860312168565\\
0.027	1.85196731471751\\
0.028	1.84376580584693\\
0.029	1.83570043663475\\
0.03	1.82776462399483\\
0.031	1.81995233306885\\
0.032	1.81225801635984\\
0.033	1.80467656166436\\
0.034	1.79720324475152\\
0.035	1.78983368970473\\
0.036	1.78256383547704\\
0.037	1.77538990450611\\
0.038	1.76830837637493\\
0.039	1.76131596409376\\
0.04	1.75440959327339\\
0.041	1.74758638438549\\
0.042	1.74084363492069\\
0.043	1.73417880579127\\
0.044	1.72758950766928\\
0.045	1.72107348928442\\
0.046	1.71462862672809\\
0.047	1.7082529139164\\
0.048	1.70194445376382\\
0.049	1.69570145039435\\
0.05	1.68952220212707\\
0.051	1.68340509416733\\
0.052	1.67734859384168\\
0.053	1.67135124434533\\
0.054	1.66541166003241\\
0.055	1.65952852178352\\
0.056	1.65370057283643\\
0.057	1.64792661487964\\
0.058	1.64220550460938\\
0.059	1.6365361503336\\
0.06	1.63091750897433\\
0.061	1.62534858329786\\
0.062	1.61982841928605\\
0.063	1.614356103731\\
0.064	1.60893076180796\\
0.065	1.60355155565833\\
0.066	1.5982176814988\\
0.067	1.59292836843849\\
0.068	1.58768287675949\\
0.069	1.58248049566581\\
0.07	1.57732054315234\\
0.071	1.57220236331716\\
0.072	1.56712532565519\\
0.073	1.56208882380811\\
0.074	1.55709227435857\\
0.075	1.55213511577803\\
0.076	1.54721680742197\\
0.077	1.54233682858422\\
0.078	1.53749467758189\\
0.079	1.53268987094096\\
0.08	1.52792194259788\\
0.081	1.52319044311985\\
0.082	1.51849493901546\\
0.083	1.51383501218997\\
0.084	1.50921025884739\\
0.085	1.50462028953311\\
0.086	1.5000647281652\\
0.087	1.49554321155356\\
0.088	1.49105538892988\\
0.089	1.48660092145613\\
0.09	1.48217948170534\\
0.091	1.47779075328509\\
0.092	1.47343443038523\\
0.093	1.46911021737981\\
0.094	1.46481782844619\\
0.095	1.46055698721082\\
0.096	1.45632742641147\\
0.097	1.452128887499\\
0.098	1.4479611204203\\
0.099	1.44382388323095\\
0.1	1.43971694176624\\
0.101	1.43564006968252\\
0.102	1.4315930477024\\
0.103	1.42757566367552\\
0.104	1.42358771226447\\
0.105	1.41962899470237\\
0.106	1.41569931857804\\
0.107	1.41179849754271\\
0.108	1.40792635144954\\
0.109	1.40408270545811\\
0.11	1.40026739054669\\
0.111	1.39648024295807\\
0.112	1.39272110412164\\
0.113	1.38898981445438\\
0.114	1.38528623113297\\
0.115	1.38161021008093\\
0.116	1.37796161173995\\
0.117	1.37434030092083\\
0.118	1.3707461466594\\
0.119	1.36717902207723\\
0.12	1.36363880424686\\
0.121	1.36012537406141\\
0.122	1.35663861610825\\
0.123	1.35317841854661\\
0.124	1.34974467298897\\
0.125	1.34633727438598\\
0.126	1.34295612091486\\
0.127	1.33960111387095\\
0.128	1.33627215756251\\
0.129	1.33296915920838\\
0.13	1.32969202883858\\
0.131	1.32644067919756\\
0.132	1.32321502565013\\
0.133	1.32001498608984\\
0.134	1.31684048084976\\
0.135	1.31369143261565\\
0.136	1.31056776634118\\
0.137	1.30746940916546\\
0.138	1.30439629033242\\
0.139	1.30134834111236\\
0.14	1.29832549472518\\
0.141	1.29532768626564\\
0.142	1.29235485263021\\
0.143	1.28940693244575\\
0.144	1.28648386599975\\
0.145	1.2835855951722\\
0.146	1.28071206336901\\
0.147	1.27786321545692\\
0.148	1.27503899769986\\
0.149	1.2722393576967\\
0.15	1.26946424432047\\
0.151	1.2667136076588\\
0.152	1.26398739895577\\
0.153	1.26128557055504\\
0.154	1.25860807584416\\
0.155	1.25595486920019\\
0.156	1.25332590593655\\
0.157	1.25072114225098\\
0.158	1.24814053517476\\
0.159	1.24558404252305\\
0.16	1.24305162284642\\
0.161	1.24054323538348\\
0.162	1.23805884001464\\
0.163	1.23559839721708\\
0.164	1.23316186802071\\
0.165	1.23074921396535\\
0.166	1.228360397059\\
0.167	1.22599537929123\\
0.168	1.22365412438203\\
0.169	1.22133659505423\\
0.17	1.21904275479353\\
0.171	1.21677256736194\\
0.172	1.21452599676283\\
0.173	1.21230300720709\\
0.174	1.21010356308053\\
0.175	1.2079276289124\\
0.176	1.20577516934512\\
0.177	1.20364614910524\\
0.178	1.20154053297563\\
0.179	1.19945828576893\\
0.18	1.19739937230221\\
0.181	1.19536375737307\\
0.182	1.19335140573691\\
0.183	1.19136228208567\\
0.184	1.18939635102788\\
0.185	1.18745357707019\\
0.186	1.18553392347589\\
0.187	1.18363735557421\\
0.188	1.18176383746704\\
0.189	1.17991333309507\\
0.19	1.17808580622695\\
0.191	1.17628122045007\\
0.192	1.17449953916312\\
0.193	1.17274072557031\\
0.194	1.1710047426775\\
0.195	1.16929155329018\\
0.196	1.16760112001343\\
0.197	1.16593340525392\\
0.198	1.16428837339153\\
0.199	1.16266598437866\\
0.2	1.16106620006583\\
0.201	1.15948898213533\\
0.202	1.15793429211469\\
0.203	1.15640209139264\\
0.204	1.15489234123794\\
0.205	1.15340500282093\\
0.206	1.15194003723817\\
0.207	1.1504974055402\\
0.208	1.14907706876264\\
0.209	1.14767898796084\\
0.21	1.14630312424824\\
0.211	1.14494943883869\\
0.212	1.14361789309304\\
0.213	1.14230844857018\\
0.214	1.14102106708286\\
0.215	1.13975570489448\\
0.216	1.13851232994398\\
0.217	1.13729090533066\\
0.218	1.13609139452936\\
0.219	1.13491376935575\\
0.22	1.1337579872543\\
0.221	1.13262401386556\\
0.222	1.13151181575435\\
0.223	1.1304213605301\\
0.224	1.12935261697786\\
0.225	1.12830555520117\\
0.226	1.12728014677795\\
0.227	1.12627636493074\\
0.228	1.12529418471301\\
0.229	1.12433358321329\\
0.23	1.12339453977922\\
0.231	1.12247703626397\\
0.232	1.12158105729789\\
0.233	1.1207065905885\\
0.234	1.11985362674626\\
0.235	1.11902216117846\\
0.236	1.1182121929715\\
0.237	1.11742372436876\\
0.238	1.11665676783491\\
0.239	1.11591133563381\\
0.24	1.1151874491311\\
0.241	1.11448513652088\\
0.242	1.11380443401745\\
0.243	1.1131453867289\\
0.244	1.11250804993088\\
0.245	1.11189249026646\\
0.246	1.1112987996529\\
0.247	1.11072706283236\\
0.248	1.1101773917484\\
0.249	1.10964991115019\\
0.25	1.10914478143992\\
0.251	1.10866217807049\\
0.252	1.1082023110226\\
0.253	1.1077654250837\\
0.254	1.1073518136017\\
0.255	1.10696181673862\\
0.256	1.10659584632635\\
0.257	1.10625439081983\\
0.258	1.10593804348353\\
0.259	1.10564754710888\\
0.26	1.10538381270497\\
0.261	1.1051480186667\\
0.262	1.10494172673498\\
0.263	1.10476710091376\\
0.264	1.10462735863821\\
0.265	1.10452783075794\\
0.266	1.10447944906735\\
};
\addplot [color=mycolor1, thick,]
  table[row sep=crcr]{%
0	2.29237235717068\\
0.001	2.206061830112\\
0.002	2.16835958767696\\
0.003	2.13663416749107\\
0.004	2.10868585064977\\
0.005	2.08428846122535\\
0.006	2.07307436921384\\
0.007	2.05104659005206\\
0.008	2.03171523270344\\
0.009	2.02178870854717\\
0.01	2.00340438778492\\
0.011	1.99485344440697\\
0.012	1.97790119749258\\
0.013	1.96494686746519\\
0.014	1.95427672700223\\
0.015	1.94075309415751\\
0.016	1.93214535777014\\
0.017	1.91897324141881\\
0.018	1.91126643426171\\
0.019	1.89861420974329\\
0.02	1.89146647575554\\
0.021	1.87935383043528\\
0.022	1.87261121317108\\
0.023	1.86101621335285\\
0.024	1.85459191753601\\
0.025	1.8434814283101\\
0.026	1.83731609929726\\
0.027	1.82665911332015\\
0.028	1.82068915120486\\
0.029	1.81047758280995\\
0.03	1.80395648577487\\
0.031	1.79487815782548\\
0.032	1.78749758026294\\
0.033	1.77981168174825\\
0.034	1.77199130929672\\
0.035	1.76523609304912\\
0.036	1.75722387460367\\
0.037	1.7511145075853\\
0.038	1.74305687407726\\
0.039	1.73741325644589\\
0.04	1.72939944917522\\
0.041	1.72409752395252\\
0.042	1.71618966173093\\
0.043	1.71086151229241\\
0.044	1.70338220010357\\
0.045	1.69693858224466\\
0.046	1.69094087219151\\
0.047	1.68408311704011\\
0.048	1.67883296213436\\
0.049	1.67185914875699\\
0.05	1.66700608259129\\
0.051	1.66007977591849\\
0.052	1.65418601953627\\
0.053	1.64865542371917\\
0.054	1.64232287033582\\
0.055	1.63752554587734\\
0.056	1.63110113048068\\
0.057	1.62569052291247\\
0.058	1.62027946412306\\
0.059	1.61433440055525\\
0.06	1.60974514395609\\
0.061	1.60370199234351\\
0.062	1.5982984107613\\
0.063	1.59346067517236\\
0.064	1.58773091305847\\
0.065	1.58286794322519\\
0.066	1.57773021253597\\
0.067	1.57228664667576\\
0.068	1.56782088036772\\
0.069	1.5625086406373\\
0.07	1.55729140560853\\
0.071	1.55297447250809\\
0.072	1.54775105577116\\
0.073	1.54269292970692\\
0.074	1.5383646225067\\
0.075	1.53342508276252\\
0.076	1.52847635525227\\
0.077	1.52402992219398\\
0.078	1.5195100682654\\
0.079	1.51465054645619\\
0.08	1.51009941912342\\
0.081	1.50598572919813\\
0.082	1.50122447600571\\
0.083	1.49663754756952\\
0.084	1.49238948694709\\
0.085	1.4881913930353\\
0.086	1.4836378102747\\
0.087	1.47924859718739\\
0.088	1.47513866813193\\
0.089	1.47105906111269\\
0.09	1.46667636889355\\
0.091	1.46241589303523\\
0.092	1.45833765570824\\
0.093	1.45448926253567\\
0.094	1.45030911305516\\
0.095	1.4461520192546\\
0.096	1.44209939006756\\
0.097	1.43818234787054\\
0.098	1.43441903360346\\
0.099	1.43048002403877\\
0.1	1.42649775511572\\
0.101	1.42258440388438\\
0.102	1.41874986772114\\
0.103	1.41500315749591\\
0.104	1.41134476677883\\
0.105	1.40767945639453\\
0.106	1.40390506071377\\
0.107	1.40017125124479\\
0.108	1.39648499420802\\
0.109	1.39284669678622\\
0.11	1.38925556535339\\
0.111	1.38570964512358\\
0.112	1.38220543179738\\
0.113	1.37873692108601\\
0.114	1.37529405657829\\
0.115	1.37186258336287\\
0.116	1.36843190678271\\
0.117	1.36501148785373\\
0.118	1.36161855214451\\
0.119	1.3582559580513\\
0.12	1.35492313846196\\
0.121	1.35161941170508\\
0.122	1.34834416294328\\
0.123	1.34509689225842\\
0.124	1.34187722810295\\
0.125	1.33868492012403\\
0.126	1.33551981290463\\
0.127	1.33238178132328\\
0.128	1.32927006193148\\
0.129	1.32617919855392\\
0.13	1.32310233477231\\
0.131	1.32004004824191\\
0.132	1.31699754202079\\
0.133	1.31398008828321\\
0.134	1.31099142114265\\
0.135	1.30803374565451\\
0.136	1.30510813704664\\
0.137	1.30221487196099\\
0.138	1.2993535460717\\
0.139	1.29651750249414\\
0.14	1.29367449624611\\
0.141	1.29084102934479\\
0.142	1.28803996601145\\
0.143	1.28527799727916\\
0.144	1.28255498764636\\
0.145	1.27985749474757\\
0.146	1.27714625910849\\
0.147	1.27445640636606\\
0.148	1.27180895351952\\
0.149	1.26920454948822\\
0.15	1.2666032224199\\
0.151	1.26400277279557\\
0.152	1.26144381740348\\
0.153	1.25892978769547\\
0.154	1.25641023544028\\
0.155	1.25390608083517\\
0.156	1.25145003362513\\
0.157	1.24901224420533\\
0.158	1.2465737582922\\
0.159	1.24418130559226\\
0.16	1.24181029077723\\
0.161	1.23944118281271\\
0.162	1.23711966950365\\
0.163	1.23480565317375\\
0.164	1.23251058321329\\
0.165	1.2302565651923\\
0.166	1.22800050674739\\
0.167	1.22578715838869\\
0.168	1.22358220385848\\
0.169	1.22140472898367\\
0.17	1.21925123159942\\
0.171	1.21711334578736\\
0.172	1.21500659291086\\
0.173	1.21291161871026\\
0.174	1.21084862587743\\
0.175	1.20879849665982\\
0.176	1.20677751190499\\
0.177	1.20477382961055\\
0.178	1.20279325924258\\
0.179	1.20083708380996\\
0.18	1.19889640621509\\
0.181	1.19698575855423\\
0.182	1.19508847763232\\
0.183	1.19321855797393\\
0.184	1.19136988170624\\
0.185	1.18953842825609\\
0.186	1.18773413492169\\
0.187	1.18594904488496\\
0.188	1.18418320500357\\
0.189	1.18244275219676\\
0.19	1.18072290184123\\
0.191	1.17902188909302\\
0.192	1.17734418637562\\
0.193	1.17568910268893\\
0.194	1.1740541498083\\
0.195	1.17243934144946\\
0.196	1.17084663779257\\
0.197	1.16927610714885\\
0.198	1.16772673560275\\
0.199	1.16619801274154\\
0.2	1.16469002895544\\
0.201	1.16320316172123\\
0.202	1.16173763774414\\
0.203	1.16029338758211\\
0.204	1.15887023538454\\
0.205	1.15746805468917\\
0.206	1.15608677889038\\
0.207	1.15472637333235\\
0.208	1.15338681386218\\
0.209	1.15206807580622\\
0.21	1.1507701307409\\
0.211	1.14949294798419\\
0.212	1.14823649890223\\
0.213	1.14700076302109\\
0.214	1.14578573283202\\
0.215	1.14459140353189\\
0.216	1.14341773723008\\
0.217	1.14226464259675\\
0.218	1.14113201819707\\
0.219	1.14001983120859\\
0.22	1.13892815962998\\
0.221	1.1378570636149\\
0.222	1.13680638239995\\
0.223	1.13577599663882\\
0.224	1.13476605078926\\
0.225	1.13377652068514\\
0.226	1.13280722239219\\
0.227	1.13185831290392\\
0.228	1.13092967874656\\
0.229	1.13002131358734\\
0.23	1.12913324344299\\
0.231	1.12826540795533\\
0.232	1.12741783466564\\
0.233	1.12659051454225\\
0.234	1.12578341942377\\
0.235	1.12499659709752\\
0.236	1.12423002605216\\
0.237	1.12348371799437\\
0.238	1.12275770387901\\
0.239	1.12205200762645\\
0.24	1.12136665737054\\
0.241	1.12070169157154\\
0.242	1.12005715836779\\
0.243	1.11943311561641\\
0.244	1.11882963223683\\
0.245	1.11824678909335\\
0.246	1.11768467923613\\
0.247	1.11714341336133\\
0.248	1.116623122076\\
0.249	1.1161239541918\\
0.25	1.11564608593054\\
0.251	1.11518972204233\\
0.252	1.11475510234909\\
0.253	1.11434250917901\\
0.254	1.11395227658997\\
0.255	1.11358480308726\\
0.256	1.11324056917746\\
0.257	1.11292016227803\\
0.258	1.11262431412556\\
0.259	1.11235396013641\\
0.26	1.1121103421294\\
0.261	1.11189521142014\\
0.262	1.11171135400774\\
0.263	1.11156440733988\\
0.264	1.11145800367408\\
0.265	1.11138513190084\\
0.266	1.1113361812287\\
0.267	1.11130351640919\\
0.268	1.11128177445327\\
0.269	1.11126731934839\\
0.27	1.11125771464769\\
0.271	1.11125133504193\\
0.272	1.11124709853061\\
0.273	1.11124428558588\\
0.274	1.11124241803022\\
0.275	1.11124117820866\\
0.276	1.11124035515697\\
0.277	1.1112398087914\\
0.278	1.11123944610461\\
0.279	1.11123920534979\\
0.28	1.11123904553579\\
0.281	1.11123893945117\\
0.282	1.11123886903239\\
0.283	1.11123882228862\\
0.284	1.11123879126027\\
0.285	1.11123877066378\\
0.286	1.11123875699192\\
0.287	1.11123874791655\\
0.288	1.11123874189229\\
0.289	1.11123873789339\\
0.29	1.11123873523883\\
0.291	1.11123873347665\\
0.292	1.11123873230673\\
0.293	1.11123873153002\\
0.294	1.1112387310143\\
0.295	1.11123873067127\\
0.296	1.11123873044385\\
0.297	1.11123873029167\\
0.298	1.11123873019121\\
0.299	1.11123873012213\\
0.3	1.11123873007772\\
};
\addplot [color=mycolor2, thick,]
  table[row sep=crcr]{%
0.261367354457691	0.00138977983450931\\
0.261167958196385	0.0236270691798647\\
0.2608	0.0421841895715607\\
0.2604	0.0585485402416478\\
0.2599	0.0727178411513048\\
0.2594	0.0863558004126307\\
0.2588	0.0978905585887129\\
0.258261988569915	0.109864411858796\\
0.257661213267281	0.121008492461052\\
0.257	0.13129579608744\\
0.2564	0.142295778556904\\
0.255759686631312	0.151690009537788\\
0.2551	0.161801925800228\\
0.2544	0.170319614161668\\
0.2537	0.178829089773747\\
0.253060609539333	0.188041203650656\\
0.252362142353041	0.196448805300873\\
0.251661031207062	0.204862874691424\\
0.2509	0.212543775100818\\
0.2502	0.220813328332543\\
0.2495	0.229178163540541\\
0.2488	0.237494637945432\\
0.248057659261209	0.244254891693818\\
0.2473	0.251764028103\\
0.2466	0.259842033596684\\
0.2458	0.265968242454586\\
0.2451	0.274275019355226\\
0.2443	0.280155150422489\\
0.2436	0.288374667795671\\
0.2428	0.294332852967066\\
0.242061156060432	0.301004952910255\\
0.2413	0.308511222755921\\
0.2405	0.314357002653406\\
0.239759978149377	0.321037768646933\\
0.239	0.328523447116009\\
0.2382	0.334397028545649\\
0.2374	0.340246172679642\\
0.2366	0.346285052130818\\
0.235864407756108	0.352759414584488\\
0.2351	0.360307917711951\\
0.2343	0.366192732889652\\
0.2335	0.372063543846062\\
0.2327	0.377937510537391\\
0.2319	0.383818200745221\\
0.2311	0.389706685480561\\
0.2303	0.395602810385068\\
0.2295	0.401505524820206\\
0.2287	0.407413626501921\\
0.2279	0.413327040148403\\
0.2271	0.419247197565736\\
0.2263	0.425175974892372\\
0.2255	0.431114684464231\\
0.2247	0.437063822073268\\
0.2239	0.44302304030795\\
0.2231	0.448990603221601\\
0.2223	0.454959701084119\\
0.2215	0.460831344194206\\
0.220661422835809	0.465246407990583\\
0.2198	0.470430505874603\\
0.219	0.476430411275734\\
0.2182	0.482461395217191\\
0.2174	0.488498114584076\\
0.216565308434668	0.492816723147642\\
0.2157	0.498091624754111\\
0.2149	0.504137361513927\\
0.2141	0.510227980322367\\
0.2133	0.516241082825974\\
0.2124	0.519911893282103\\
0.2116	0.526004062813818\\
0.2108	0.53214345985033\\
0.209962128265026	0.536569351749507\\
0.2091	0.541895598190569\\
0.2083	0.548077684380529\\
0.207463686922231	0.552526681929957\\
0.2066	0.557905883434866\\
0.2058	0.564130386931722\\
0.204961764290758	0.568632406579895\\
0.2041	0.574039953469118\\
0.2033	0.580303499221883\\
0.2024	0.584034058404246\\
0.2016	0.590309405088623\\
0.200763783714292	0.594866364733702\\
0.1999	0.600364096498513\\
0.1991	0.606698036917359\\
0.1982	0.610486401823009\\
0.1974	0.616863787207497\\
0.1965	0.620688946416276\\
0.1957	0.627078574289019\\
0.1948	0.631087364067894\\
0.194	0.637356248574903\\
0.193156667308729	0.642078809745421\\
0.1923	0.647700410405709\\
0.191450977052482	0.652462913308828\\
0.1906	0.65811291837285\\
0.1897	0.66208575227196\\
0.1889	0.668594565379628\\
0.188	0.67256009769217\\
0.1872	0.679142877526586\\
0.1863	0.683135834772866\\
0.1855	0.689724456065742\\
0.1846	0.693796817343204\\
0.183760137926296	0.698697828635129\\
0.1829	0.704534805708344\\
0.182	0.708614085226401\\
0.1812	0.715321095241012\\
0.1803	0.719459769091485\\
0.1794	0.723707850245846\\
0.1786	0.730386546005167\\
0.1777	0.734555864545157\\
0.176861634136741	0.739597817060272\\
0.176	0.745605280830826\\
0.1751	0.749829130120181\\
0.174261719371857	0.754928264579706\\
0.1734	0.760999781909102\\
0.1725	0.765283647218321\\
0.1716	0.769736892682382\\
0.1708	0.776570503102917\\
0.1699	0.780925121095995\\
0.169	0.785290672862679\\
0.168174862868097	0.790490673153823\\
0.1673	0.796752655132597\\
0.1664	0.80117967308684\\
0.1655	0.805633908140869\\
0.1647	0.812506195571809\\
0.1638	0.817264598912324\\
0.1629	0.821778049455778\\
0.162	0.826308000146856\\
0.161158357688698	0.831728503108763\\
0.1603	0.838108327419343\\
0.1594	0.842726325184551\\
0.1585	0.847346891244406\\
0.1576	0.851986458050229\\
0.1567	0.856696031753268\\
0.1559	0.86387803482513\\
0.155	0.868740395735645\\
0.1541	0.873477992655627\\
0.1532	0.878233622290129\\
0.1523	0.883011019341585\\
0.1514	0.887811654547835\\
0.1505	0.892712824780098\\
0.149665457726573	0.898307588832975\\
0.1488	0.905045534158316\\
0.1479	0.909954649570135\\
0.147	0.914874267184267\\
0.1461	0.919816209741608\\
0.1452	0.924782780602787\\
0.1443	0.929774273326289\\
0.1434	0.934790219542191\\
0.1425	0.93983000011084\\
0.1416	0.94489355067105\\
0.1407	0.949982178811307\\
0.1398	0.955098981195461\\
0.1389	0.960248768877579\\
0.138	0.965442331141359\\
0.1371	0.970816779323363\\
0.136252953997493	0.976676340654405\\
0.135356605711693	0.981904928134269\\
0.134456409525894	0.98717002967251\\
0.133551297479165	0.992474366668625\\
0.1326	0.997051510811175\\
0.1317	1.00228441639392\\
0.1308	1.00762720160846\\
0.1299	1.01301362810007\\
0.129	1.01843302737402\\
0.1281	1.02387822385748\\
0.1272	1.0293461304893\\
0.1263	1.03483805923661\\
0.1254	1.04035743246958\\
0.1245	1.04590775307584\\
0.1236	1.05149170269545\\
0.1227	1.05711084943139\\
0.1218	1.06276514599276\\
0.1209	1.06845019944312\\
0.12	1.07412151196159\\
0.11905913684356	1.0779756659546\\
0.1181	1.08286926196564\\
0.1172	1.08862268242617\\
0.1163	1.09440111179971\\
0.1154	1.10023454764748\\
0.1145	1.10612540486014\\
0.1136	1.11206442692547\\
0.1127	1.11789572477162\\
0.1117	1.12119595780843\\
0.1108	1.12712807042813\\
0.1099	1.1331035885969\\
0.109	1.1391706432757\\
0.1081	1.14531210613556\\
0.1072	1.15114002504969\\
0.1062	1.15479693554308\\
0.1053	1.16087329988868\\
0.1044	1.16707518583837\\
0.1035	1.17338967124344\\
0.1026	1.1795630724176\\
0.1016	1.18318128637939\\
0.1007	1.18940324635367\\
0.0998	1.19581153933712\\
0.0989	1.20233460224356\\
0.0979	1.2059792631957\\
0.097	1.21227552505117\\
0.0961	1.2187939429201\\
0.0952	1.22547875433587\\
0.0942	1.22927024577174\\
0.0933	1.2356654197549\\
0.0924	1.242357457898\\
0.0915	1.24922014239831\\
0.0905	1.25303169662354\\
0.0896	1.25961303032285\\
0.0887	1.2665548948278\\
0.0877604910979275	1.27151238433782\\
0.0868	1.27726984772907\\
0.0859	1.28423429189145\\
0.085	1.29141870137627\\
0.084	1.29531279375112\\
0.0831	1.30228563685686\\
0.0822	1.30962866768106\\
0.0812	1.31369969053403\\
0.0803	1.32071290779155\\
0.0794	1.3282075434276\\
0.0784	1.33238779897983\\
0.0775	1.3395258013675\\
0.0766	1.34719540820659\\
0.0756	1.35137301979092\\
0.0747	1.35875024163965\\
0.0738	1.36659263339025\\
0.0728	1.37071746151501\\
0.0719	1.37842787075817\\
0.0709577272427788	1.38427522681662\\
0.07	1.39052557037384\\
0.0691	1.39860320715423\\
0.0681	1.40325228209802\\
0.0672	1.41089123941629\\
0.0663	1.41924523822453\\
0.0653	1.42354773565149\\
0.0644	1.43187046023529\\
0.0634	1.436878874942\\
0.0625	1.44467368481924\\
0.0616	1.45337746948936\\
0.0606	1.45781828581244\\
0.0597	1.46657912267673\\
0.0587	1.47146885759868\\
0.0578	1.47992319482595\\
0.0568	1.48594701531912\\
0.0559	1.49355347292185\\
0.055	1.50288627717622\\
0.054	1.50752369120518\\
0.0531	1.51688972378893\\
0.0521	1.52190912456517\\
0.0512	1.53112657488612\\
0.0502	1.5368391183962\\
0.0493	1.54565309991757\\
0.0483	1.55248350343914\\
0.0474	1.56049178000302\\
0.0465	1.57052048872248\\
0.0455	1.57566065324206\\
0.0446	1.5861402545041\\
0.0436	1.59117604725652\\
0.0427	1.6018189527577\\
0.0417	1.60705283494424\\
0.0408	1.61783607091173\\
0.0398	1.62330427153405\\
0.0389	1.63422774242925\\
0.0379	1.6399422478472\\
0.037	1.65102190870125\\
0.036	1.65697866298258\\
0.0351	1.66824725764524\\
0.0341	1.67442826021857\\
0.0332	1.68593549317745\\
0.0322	1.69231249564681\\
0.0313	1.7041225182299\\
0.0303	1.71066323309197\\
0.0294	1.72284936717991\\
0.0284	1.7295249416397\\
0.0275	1.74216313862131\\
0.0265	1.74895486339089\\
0.0256	1.76211808598076\\
0.0246	1.76902184195542\\
0.0237	1.78277702681023\\
0.0227	1.78980534839705\\
0.0218	1.8042132646252\\
0.0208	1.81139628734931\\
0.0199	1.82651326737554\\
0.0189	1.83390079374738\\
0.018	1.8497804101183\\
0.017	1.85744852946352\\
0.0161	1.87414020047676\\
0.0151	1.88220999633137\\
0.0142	1.89974782330714\\
0.0132	1.90844114010913\\
0.0123	1.9268004473914\\
0.0113	1.93666063776407\\
0.0104	1.95556173002206\\
0.0094	1.96912312525448\\
0.0085	1.98642325156423\\
0.0076	2.00968330385655\\
0.0066	2.02013346355113\\
0.0057	2.04578993790589\\
0.0047	2.06227536695028\\
0.0038	2.08566704922946\\
0.0029	2.11726130405152\\
0.002	2.1526674172089\\
0.0011	2.1934562017214\\
0.0002	2.24610188955743\\
};
\addplot [color=gray, dotted]
  table[row sep=crcr]{%
0	1.47\\
0.09	1.47\\};
\addplot [color=gray, dotted]
  table[row sep=crcr]{%
0.09	1.47\\
0.09	0\\};
\end{axis}
\end{tikzpicture}%

%% file: Figs/Fig13b_Front_Dynamics_Hirotsu.tex
%
%
\begin{tikzpicture}
\definecolor{mycolor1}{rgb}{0.2,0.6,1}%
\definecolor{mycolor2}{rgb}{1,0.5,0.2}%
\begin{axis}[%
thick, 
axis on top,
width=0.48\textwidth, 
height=0.4\textwidth,
xmin=0,xmax=0.4,ymin=0,ymax=2.5, 
xlabel = {$t$},
ylabel = {$r$},
scaled ticks=false, tick label style={/pgf/number format/fixed}],
\addplot [color=black, dashed, very thick,]
  table[row sep=crcr]{%
0.01	2.49823196728886\\
0.011	2.47086955949376\\
0.012	2.44736831255255\\
0.013	2.42633618907868\\
0.014	2.40706616990322\\
0.015	2.38913826523061\\
0.016	2.37227801463414\\
0.017	2.35629404410125\\
0.018	2.34104645101186\\
0.019	2.32642920608559\\
0.02	2.31235962638047\\
0.021	2.29877173695679\\
0.022	2.28561187179956\\
0.023	2.27283567297134\\
0.024	2.26040596861744\\
0.025	2.24829123716309\\
0.026	2.23646446980565\\
0.027	2.22490231262743\\
0.028	2.21358440788966\\
0.029	2.20249288054177\\
0.03	2.19161193318589\\
0.031	2.18092752208151\\
0.032	2.17042709560292\\
0.033	2.16009938091997\\
0.034	2.14993420758507\\
0.035	2.13992236429672\\
0.036	2.13005547348953\\
0.037	2.12032589054147\\
0.038	2.11072661826746\\
0.039	2.10125122983276\\
0.04	2.09189380737157\\
0.041	2.08264888655721\\
0.042	2.07351140947733\\
0.043	2.06447668327254\\
0.044	2.05554034428455\\
0.045	2.04669832607996\\
0.046	2.03794683334076\\
0.047	2.02928231489288\\
0.048	2.02070144364623\\
0.049	2.01220109682614\\
0.05	2.00377833888027\\
0.051	1.99543040590704\\
0.052	1.98715469233812\\
0.053	1.97894873702585\\
0.054	1.97081021466446\\
0.055	1.96273692274204\\
0.056	1.95472677484902\\
0.057	1.94677779135148\\
0.058	1.93888809095551\\
0.059	1.9310558856286\\
0.06	1.9232794731434\\
0.061	1.9155572315628\\
0.062	1.90788761410192\\
0.063	1.90026914426784\\
0.064	1.89270041104166\\
0.065	1.88518006600915\\
0.066	1.8777068179914\\
0.067	1.87027943050709\\
0.068	1.86289671870873\\
0.069	1.85555754519749\\
0.07	1.84826081932875\\
0.071	1.84100549229493\\
0.072	1.83379055682264\\
0.073	1.82661504373644\\
0.074	1.81947802025648\\
0.075	1.81237858804719\\
0.076	1.80531588115033\\
0.077	1.79828906560939\\
0.078	1.79129733510755\\
0.079	1.7843399130216\\
0.08	1.77741604852792\\
0.081	1.77052501562682\\
0.082	1.7636661130539\\
0.083	1.75683866225848\\
0.084	1.75004200652726\\
0.085	1.74327550995913\\
0.086	1.7365385566086\\
0.087	1.72983054925766\\
0.088	1.72315090954438\\
0.089	1.71649907572092\\
0.09	1.70987450267916\\
0.091	1.70327666184506\\
0.092	1.69670503901021\\
0.093	1.69015913530361\\
0.094	1.68363846527587\\
0.095	1.67714255701746\\
0.096	1.67067095129316\\
0.097	1.66422320175741\\
0.098	1.65779887330559\\
0.099	1.65139754239461\\
0.1	1.64501879640437\\
0.101	1.6386622331856\\
0.102	1.63232746073053\\
0.103	1.62601409649026\\
0.104	1.61972176787794\\
0.105	1.6134501106771\\
0.106	1.60719876937113\\
0.107	1.60096739712624\\
0.108	1.59475565515936\\
0.109	1.58856321208017\\
0.11	1.58238974390579\\
0.111	1.57623493464761\\
0.112	1.57009847411941\\
0.113	1.5639800596118\\
0.114	1.55787939468139\\
0.115	1.55179618909564\\
0.116	1.54573015871332\\
0.117	1.53968102514811\\
0.118	1.53364851571312\\
0.119	1.52763236291967\\
0.12	1.52163230509319\\
0.121	1.51564808508376\\
0.122	1.50967945084188\\
0.123	1.50372615507742\\
0.124	1.49778795507656\\
0.125	1.49186461256414\\
0.126	1.48595589358361\\
0.127	1.48006156830689\\
0.128	1.47418141108044\\
0.129	1.46831519956782\\
0.13	1.46246271608463\\
0.131	1.45662374583549\\
0.132	1.45079807817141\\
0.133	1.44498550539646\\
0.134	1.43918582330375\\
0.135	1.43339883086003\\
0.136	1.4276243302854\\
0.137	1.42186212639273\\
0.138	1.41611202731174\\
0.139	1.410373843864\\
0.14	1.40464738959472\\
0.141	1.39893248081461\\
0.142	1.39322893597981\\
0.143	1.38753657634069\\
0.144	1.38185522577102\\
0.145	1.37618471007159\\
0.146	1.37052485743569\\
0.147	1.3648754982139\\
0.148	1.35923646485458\\
0.149	1.35360759199624\\
0.15	1.34798871585281\\
0.151	1.34237967490478\\
0.152	1.3367803093946\\
0.153	1.33119046131592\\
0.154	1.32560997435127\\
0.155	1.32003869391611\\
0.156	1.31447646695909\\
0.157	1.30892314212039\\
0.158	1.30337856919014\\
0.159	1.29784259976474\\
0.16	1.29231508657285\\
0.161	1.2867958840182\\
0.162	1.28128484747999\\
0.163	1.27578183367736\\
0.164	1.27028670050697\\
0.165	1.26479930691693\\
0.166	1.25931951327546\\
0.167	1.25384718066888\\
0.168	1.24838217130336\\
0.169	1.2429243483647\\
0.17	1.23747357595102\\
0.171	1.23202971901994\\
0.172	1.2265926433982\\
0.173	1.22116221567235\\
0.174	1.21573830320805\\
0.175	1.2103207740756\\
0.176	1.20490949702777\\
0.177	1.19950434134725\\
0.178	1.19410517718971\\
0.179	1.18871187501143\\
0.18	1.18332430584475\\
0.181	1.17794233199529\\
0.182	1.17256583448058\\
0.183	1.16719468567136\\
0.184	1.16182875830468\\
0.185	1.1564679254465\\
0.186	1.15111206045434\\
0.187	1.1457610369401\\
0.188	1.14041472873282\\
0.189	1.13507300984151\\
0.19	1.129735754418\\
0.191	1.12440283671965\\
0.192	1.11907413107202\\
0.193	1.11374951183145\\
0.194	1.10842885334743\\
0.195	1.10311202992483\\
0.196	1.09779891578587\\
0.197	1.09248938503189\\
0.198	1.08718331160478\\
0.199	1.08188056924811\\
0.2	1.07658103146786\\
0.201	1.07128457149286\\
0.202	1.06599106223464\\
0.203	1.06070037624691\\
0.204	1.05541238568453\\
0.205	1.05012696226183\\
0.206	1.04484397721046\\
0.207	1.03956330123653\\
0.208	1.03428480447704\\
0.209	1.02900835645566\\
0.21	1.0237338260377\\
0.211	1.01846108138423\\
0.212	1.01318998990541\\
0.213	1.0079204182128\\
0.214	1.00265223207077\\
0.215	0.997385296346857\\
0.216	0.992119474961054\\
0.217	0.98685463083392\\
0.218	0.981590625833525\\
0.219	0.976327320721113\\
0.22	0.971064575095434\\
0.221	0.965802247335679\\
0.222	0.960540194542944\\
0.223	0.955278272480141\\
0.224	0.950016335510302\\
0.225	0.944754236533165\\
0.226	0.939491826919982\\
0.227	0.934228956446447\\
0.228	0.928965473223658\\
0.229	0.923701223627015\\
0.23	0.918436052222954\\
0.231	0.913169801693405\\
0.232	0.907902312757874\\
0.233	0.902633424093021\\
0.234	0.897362972249621\\
0.235	0.892090791566767\\
0.236	0.886816714083198\\
0.237	0.88154056944559\\
0.238	0.876262184813672\\
0.239	0.870981384762007\\
0.24	0.86569799117827\\
0.241	0.860411823157842\\
0.242	0.855122696894548\\
0.243	0.849830425567329\\
0.244	0.844534819222652\\
0.245	0.839235684652429\\
0.246	0.833932825267222\\
0.247	0.828626040964485\\
0.248	0.823315127991583\\
0.249	0.817999878803312\\
0.25	0.812680081913637\\
0.251	0.80735552174132\\
0.252	0.802025978449127\\
0.253	0.796691227776255\\
0.254	0.791351040863606\\
0.255	0.786005184071526\\
0.256	0.78065341878956\\
0.257	0.775295501237812\\
0.258	0.769931182259389\\
0.259	0.76456019813599\\
0.26	0.759182306218876\\
0.261	0.753797230922201\\
0.262	0.748404699309159\\
0.263	0.74300443187569\\
0.264	0.737596142277527\\
0.265	0.73217953704397\\
0.266	0.726754315277523\\
0.267	0.721320168338547\\
0.268	0.715876779513958\\
0.269	0.71042382366898\\
0.27	0.704960966880841\\
0.271	0.699487866053242\\
0.272	0.694004168510332\\
0.273	0.688509493516629\\
0.274	0.683003503996262\\
0.275	0.677485797857318\\
0.276	0.671955979585372\\
0.277	0.666413641697028\\
0.278	0.660858364178031\\
0.279	0.655289713888625\\
0.28	0.649707243933794\\
0.281	0.644110492995768\\
0.282	0.638498984625997\\
0.283	0.632872226493503\\
0.284	0.627229709586271\\
0.285	0.621570907362\\
0.286	0.615895274844247\\
0.287	0.610202247659554\\
0.288	0.604491241010757\\
0.289	0.598761648581237\\
0.29	0.593012841364302\\
0.291	0.587244166411329\\
0.292	0.581454925498531\\
0.293	0.575644432506171\\
0.294	0.569811976388639\\
0.295	0.563956773021291\\
0.296	0.558078027589912\\
0.297	0.552174911671955\\
0.298	0.546246561171823\\
0.299	0.540292074092464\\
0.3	0.534310477444941\\
0.301	0.528300813989881\\
0.302	0.522262053387905\\
0.303	0.516193152894261\\
0.304	0.510092945635543\\
0.305	0.503960243940123\\
0.306	0.497793800416155\\
0.307	0.491592303631076\\
0.308	0.485354373383972\\
0.309	0.479078555522969\\
0.31	0.472763316252978\\
0.311	0.466407035871255\\
0.312	0.460008001858936\\
0.313	0.453564401245807\\
0.314	0.447074309426062\\
0.315	0.440535688833427\\
0.316	0.433946370530499\\
0.317	0.427304048183818\\
0.318	0.420606256081179\\
0.319	0.41385036328796\\
0.32	0.407033551920806\\
0.321	0.40015280546845\\
0.322	0.39320488084241\\
0.323	0.386186286652216\\
0.324	0.379093258460212\\
0.325	0.37192171965186\\
0.326	0.364667253019773\\
0.327	0.35732532522349\\
0.328	0.349890470829516\\
0.329	0.342356971157764\\
0.33	0.334718555812969\\
0.331	0.326968315465018\\
0.332	0.319098623224954\\
0.333	0.311100980184962\\
0.334	0.302966000205773\\
0.335	0.294683054877533\\
0.336	0.286240379644962\\
0.337	0.277624405977569\\
0.338	0.268819778283931\\
0.339	0.25980887238862\\
0.34	0.250571300223745\\
0.341	0.241083167559491\\
0.342	0.231316580289119\\
0.343	0.221238078847141\\
0.344	0.210807228853544\\
0.345	0.199974880516443\\
0.346	0.18867897376399\\
0.347	0.176840755015587\\
0.348	0.164357251595547\\
0.349	0.151089576723386\\
0.35	0.136842836510368\\
0.351	0.121328455221602\\
0.352	0.104088919494679\\
0.353	0.0843214732599378\\
0.354	0.060368680533554\\
0.355	0.0276460617237907\\
0.356	-0.041721310072315\\
};
\addplot [color=black, dashed, very thick]
  table[row sep=crcr]{%
0.01	2.51078589677272\\
0.011	2.48590407218367\\
0.012	2.46457835900244\\
0.013	2.44552909160179\\
0.014	2.42810620207219\\
0.015	2.41192346464728\\
0.016	2.39672844381342\\
0.017	2.38234509775055\\
0.018	2.3686447158938\\
0.019	2.35552974019696\\
0.02	2.34292408642499\\
0.021	2.33076703994329\\
0.022	2.31900921059367\\
0.023	2.30760977281181\\
0.024	2.29653451395363\\
0.025	2.28575442162436\\
0.026	2.27524463729648\\
0.027	2.26498366706484\\
0.028	2.25495277560115\\
0.029	2.24513551368467\\
0.03	2.23551734548003\\
0.031	2.22608535037689\\
0.032	2.21682798227716\\
0.033	2.20773487324784\\
0.034	2.19879667116501\\
0.035	2.1900049077525\\
0.036	2.18135188326933\\
0.037	2.1728305736384\\
0.038	2.16443455164851\\
0.039	2.15615791610291\\
0.04	2.14799523519842\\
0.041	2.13994149550837\\
0.042	2.13199205856759\\
0.043	2.12414262278303\\
0.044	2.11638919040272\\
0.045	2.1087280380825\\
0.046	2.10115569274216\\
0.047	2.09366890684914\\
0.048	2.08626463984541\\
0.049	2.07894003995654\\
0.05	2.07169242843286\\
0.051	2.06451928520963\\
0.052	2.05741823660296\\
0.053	2.050387042582\\
0.054	2.04342358869625\\
0.055	2.03652587423758\\
0.056	2.02969200591222\\
0.057	2.02292018930854\\
0.058	2.01620872118374\\
0.059	2.00955598465715\\
0.06	2.00296044241744\\
0.061	1.99642063162303\\
0.062	1.98993515915606\\
0.063	1.98350269714358\\
0.064	1.97712197854546\\
0.065	1.97079179441364\\
0.066	1.96451098906699\\
0.067	1.9582784576785\\
0.068	1.95209314343673\\
0.069	1.94595403378832\\
0.07	1.9398601596243\\
0.071	1.93381059094657\\
0.072	1.92780443641074\\
0.073	1.92184084026309\\
0.074	1.91591898074031\\
0.075	1.91003806827306\\
0.076	1.90419734360641\\
0.077	1.89839607733027\\
0.078	1.89263356612405\\
0.079	1.88690913431367\\
0.08	1.88122213052276\\
0.081	1.87557192674684\\
0.082	1.86995791815907\\
0.083	1.86437952133482\\
0.084	1.85883617342508\\
0.085	1.85332733121391\\
0.086	1.84785247031943\\
0.087	1.84241108409874\\
0.088	1.83700268365996\\
0.089	1.83162679595522\\
0.09	1.82628296372084\\
0.091	1.82097074530969\\
0.092	1.81568971286494\\
0.093	1.81043945303036\\
0.094	1.8052195653385\\
0.095	1.80002966223992\\
0.096	1.79486936834528\\
0.097	1.78973832053406\\
0.098	1.78463616657937\\
0.099	1.77956256534768\\
0.1	1.77451718623546\\
0.101	1.76949970875773\\
0.102	1.76450982223634\\
0.103	1.75954722521246\\
0.104	1.75461162579466\\
0.105	1.74970274036112\\
0.106	1.74482029377687\\
0.107	1.73996401933672\\
0.108	1.73513365822875\\
0.109	1.73032895898316\\
0.11	1.72554967744749\\
0.111	1.72079557720271\\
0.112	1.71606642783055\\
0.113	1.7113620061758\\
0.114	1.70668209537654\\
0.115	1.70202648479075\\
0.116	1.69739496987352\\
0.117	1.69278735188944\\
0.118	1.68820343784118\\
0.119	1.68364304006006\\
0.12	1.67910597664595\\
0.121	1.67459207046239\\
0.122	1.67010114954868\\
0.123	1.6656330468367\\
0.124	1.66118759998909\\
0.125	1.65676465127271\\
0.126	1.65236404744602\\
0.127	1.64798563959518\\
0.128	1.64362928314653\\
0.129	1.63929483721043\\
0.13	1.63498216554997\\
0.131	1.63069113525668\\
0.132	1.62642161766008\\
0.133	1.6221734874305\\
0.134	1.61794662295607\\
0.135	1.6137409060938\\
0.136	1.60955622221036\\
0.137	1.60539245968628\\
0.138	1.60124951042555\\
0.139	1.5971272693864\\
0.14	1.59302563458981\\
0.141	1.58894450713436\\
0.142	1.58488379073865\\
0.143	1.58084339219197\\
0.144	1.57682322121596\\
0.145	1.57282318995736\\
0.146	1.56884321330628\\
0.147	1.5648832087169\\
0.148	1.56094309615339\\
0.149	1.55702279814213\\
0.15	1.55312223933111\\
0.151	1.54924134696109\\
0.152	1.54538005050201\\
0.153	1.54153828163541\\
0.154	1.53771597420127\\
0.155	1.53391306421789\\
0.156	1.53012948973483\\
0.157	1.52636519093106\\
0.158	1.52262010973766\\
0.159	1.51889419027409\\
0.16	1.51518737838245\\
0.161	1.51149962198423\\
0.162	1.50783087060155\\
0.163	1.50418107559408\\
0.164	1.50055019004204\\
0.165	1.49693816865516\\
0.166	1.49334496800754\\
0.167	1.48977054606731\\
0.168	1.48621486245455\\
0.169	1.48267787834164\\
0.17	1.47915955640237\\
0.171	1.47565986077112\\
0.172	1.47217875704239\\
0.173	1.46871621219405\\
0.174	1.46527219459362\\
0.175	1.4618466739446\\
0.176	1.45843962126658\\
0.177	1.45505100879357\\
0.178	1.45168081018393\\
0.179	1.44832900015531\\
0.18	1.44499555465335\\
0.181	1.44168044511552\\
0.182	1.43838365541915\\
0.183	1.43510516492335\\
0.184	1.43184495406642\\
0.185	1.42860300433891\\
0.186	1.42537929825705\\
0.187	1.42217381933653\\
0.188	1.41898655206665\\
0.189	1.41581748188476\\
0.19	1.41266659515109\\
0.191	1.40953387912376\\
0.192	1.40641932193424\\
0.193	1.40332291256295\\
0.194	1.40024464081519\\
0.195	1.39718449729734\\
0.196	1.39414247339326\\
0.197	1.39111856124093\\
0.198	1.38811275370937\\
0.199	1.38512504437569\\
0.2	1.38215542750244\\
0.201	1.37920389801507\\
0.202	1.37627045147962\\
0.203	1.37335508408065\\
0.204	1.37045779259922\\
0.205	1.36757857439119\\
0.206	1.36471742736554\\
0.207	1.36187434996299\\
0.208	1.35904934113466\\
0.209	1.35624240032096\\
0.21	1.35345352743056\\
0.211	1.3506827228196\\
0.212	1.34792998727094\\
0.213	1.3451953219736\\
0.214	1.34247872850235\\
0.215	1.33978020879738\\
0.216	1.33709976514419\\
0.217	1.33443740015349\\
0.218	1.33179311674137\\
0.219	1.32916691810951\\
0.22	1.32655880772554\\
0.221	1.32396878930356\\
0.222	1.32139686678474\\
0.223	1.31884304431817\\
0.224	1.31630732624167\\
0.225	1.31378971706294\\
0.226	1.31129022144066\\
0.227	1.30880884416593\\
0.228	1.30634559014369\\
0.229	1.30390046437439\\
0.23	1.30147347193581\\
0.231	1.29906461796498\\
0.232	1.2966739076404\\
0.233	1.29430134616427\\
0.234	1.29194693874502\\
0.235	1.28961069057999\\
0.236	1.2872926068383\\
0.237	1.2849926926439\\
0.238	1.28271095305888\\
0.239	1.28044739306696\\
0.24	1.27820201755718\\
0.241	1.27597483130792\\
0.242	1.27376583897102\\
0.243	1.2715750450563\\
0.244	1.26940245391622\\
0.245	1.26724806973089\\
0.246	1.26511189649335\\
0.247	1.26299393799509\\
0.248	1.26089419781197\\
0.249	1.2588126792904\\
0.25	1.25674938553382\\
0.251	1.25470431938963\\
0.252	1.25267748343638\\
0.253	1.25066887997137\\
0.254	1.24867851099864\\
0.255	1.24670637821736\\
0.256	1.24475248301066\\
0.257	1.24281682643481\\
0.258	1.240899409209\\
0.259	1.23900022853451\\
0.26	1.23711929079934\\
0.261	1.23525659245302\\
0.262	1.2334121327728\\
0.263	1.23158591065465\\
0.264	1.22977792460611\\
0.265	1.22798817273959\\
0.266	1.22621665276643\\
0.267	1.22446336199147\\
0.268	1.22272829730833\\
0.269	1.22101145519541\\
0.27	1.21931283171253\\
0.271	1.2176324224983\\
0.272	1.21597022276839\\
0.273	1.21432622191319\\
0.274	1.21270042516194\\
0.275	1.21109282099774\\
0.276	1.20950340294235\\
0.277	1.20793216409797\\
0.278	1.2063790971507\\
0.279	1.20484419437503\\
0.28	1.20332744763933\\
0.281	1.20182884841249\\
0.282	1.20034838777166\\
0.283	1.19888605641128\\
0.284	1.19744184465326\\
0.285	1.19601574245854\\
0.286	1.19460773944001\\
0.287	1.19321782487689\\
0.288	1.1918459877306\\
0.289	1.19049221666222\\
0.29	1.1891565000517\\
0.291	1.18783882601872\\
0.292	1.18653917795723\\
0.293	1.1852575479252\\
0.294	1.18399392822131\\
0.295	1.18274830145736\\
0.296	1.18152065483491\\
0.297	1.18031097546397\\
0.298	1.1791192503996\\
0.299	1.17794546668127\\
0.3	1.17678960544626\\
0.301	1.17565165948943\\
0.302	1.17453161621728\\
0.303	1.17342946981122\\
0.304	1.1723452013155\\
0.305	1.1712787986876\\
0.306	1.17023025024295\\
0.307	1.16919954472572\\
0.308	1.16818667138517\\
0.309	1.16719162005791\\
0.31	1.16621438125666\\
0.311	1.16525494626618\\
0.312	1.16431330724692\\
0.313	1.16338945734726\\
0.314	1.16248339044446\\
0.315	1.16159510242374\\
0.316	1.1607245901701\\
0.317	1.15987185249798\\
0.318	1.15903688915239\\
0.319	1.15821970217033\\
0.32	1.15742029541695\\
0.321	1.15663867564118\\
0.322	1.15587485188457\\
0.323	1.15512883603533\\
0.324	1.15440064339949\\
0.325	1.15369029221869\\
0.326	1.15299780486784\\
0.327	1.15232323268105\\
0.328	1.15166658384569\\
0.329	1.15102789515938\\
0.33	1.15040720955266\\
0.331	1.14980457599602\\
0.332	1.14922005142149\\
0.333	1.14865369837784\\
0.334	1.14810559463427\\
0.335	1.1475758206024\\
0.336	1.14706448105405\\
0.337	1.14657167929551\\
0.338	1.14609754036929\\
0.339	1.14564220699696\\
0.34	1.14520584236694\\
0.341	1.1447886291643\\
0.342	1.14439078980171\\
0.343	1.1440125715068\\
0.344	1.14365426094114\\
0.345	1.14331620710964\\
0.346	1.14299879680472\\
0.347	1.14270250231384\\
0.348	1.1424278921545\\
0.349	1.14217566216777\\
0.35	1.141946680633\\
0.351	1.14174205092018\\
0.352	1.1415632389204\\
0.353	1.14141228605253\\
0.354	1.14129221752248\\
0.355	1.14120812459716\\
0.356	1.14117073301616\\
};
\addplot [color=mycolor1, thick]
  table[row sep=crcr]{%
0	2.67447742393749\\
0.001	2.65646086943707\\
0.002	2.59611851987564\\
0.003	2.58772404319215\\
0.004	2.58149656613345\\
0.005	2.5760963094366\\
0.006	2.53504466628969\\
0.007	2.52630803611367\\
0.008	2.52042088736167\\
0.009	2.515416648296\\
0.01	2.51078589677272\\
0.011	2.47487738941617\\
0.012	2.46780088428274\\
0.013	2.46255966629383\\
0.014	2.45798306813208\\
0.015	2.45367580587717\\
0.016	2.42181513478883\\
0.017	2.41541211413269\\
0.018	2.41052073450708\\
0.019	2.40619538549738\\
0.02	2.40207282912979\\
0.021	2.37323504819801\\
0.022	2.3674109716961\\
0.023	2.36280419116751\\
0.024	2.35867673553432\\
0.025	2.354621856217\\
0.026	2.32811706505726\\
0.027	2.32288071744436\\
0.028	2.31854965769234\\
0.029	2.31459800672907\\
0.03	2.294465963872\\
0.031	2.28595085463497\\
0.032	2.28125680347864\\
0.033	2.27719225059991\\
0.034	2.27337611436032\\
0.035	2.25176620424708\\
0.036	2.24639061535733\\
0.037	2.24215103777057\\
0.038	2.2383211316958\\
0.039	2.22174205328439\\
0.04	2.21360454067344\\
0.041	2.20914592772988\\
0.042	2.20526155406293\\
0.043	2.20153807260925\\
0.044	2.18258780341485\\
0.045	2.17788908359524\\
0.046	2.17395618989425\\
0.047	2.17029896284642\\
0.048	2.15303453954322\\
0.049	2.14813458883579\\
0.05	2.14418422173576\\
0.051	2.14055738185625\\
0.052	2.12463531023291\\
0.053	2.11968154518765\\
0.054	2.11576242783012\\
0.055	2.11217360597207\\
0.056	2.09716566634928\\
0.057	2.09237967595672\\
0.058	2.08854475491629\\
0.059	2.08500025249483\\
0.06	2.07059067784888\\
0.061	2.06612728132966\\
0.062	2.06241695186978\\
0.063	2.05879147017043\\
0.064	2.04496526289616\\
0.065	2.04085632912128\\
0.066	2.03728626293351\\
0.067	2.02556589623488\\
0.068	2.02030594464305\\
0.069	2.01651268173344\\
0.07	2.01305532974588\\
0.071	2.0007745046127\\
0.072	1.99658097852661\\
0.073	1.99303953245818\\
0.074	1.98261559380923\\
0.075	1.97742955142591\\
0.076	1.97373287788098\\
0.077	1.97031465399226\\
0.078	1.95899752735396\\
0.079	1.95509113047859\\
0.08	1.95167955928149\\
0.081	1.94123059512182\\
0.082	1.93705024504149\\
0.083	1.93359414358425\\
0.084	1.92408048906076\\
0.085	1.91954649900597\\
0.086	1.9160498846624\\
0.087	1.90747872321663\\
0.088	1.90252082957555\\
0.089	1.89900436243608\\
0.09	1.89118774982909\\
0.091	1.88592350864446\\
0.092	1.88241867058944\\
0.093	1.87483624358554\\
0.094	1.86971902282295\\
0.095	1.8662606551182\\
0.096	1.85851938858388\\
0.097	1.85388759547137\\
0.098	1.85050460264763\\
0.099	1.84256904080187\\
0.1	1.83842110613624\\
0.101	1.83512835151653\\
0.102	1.82708457127118\\
0.103	1.82331591529507\\
0.104	1.8200980087353\\
0.105	1.81204824624079\\
0.106	1.80856662088853\\
0.107	1.80212465928399\\
0.108	1.79742960928298\\
0.109	1.79416072470774\\
0.11	1.78687622632803\\
0.111	1.78319975339991\\
0.112	1.77940725280138\\
0.113	1.77261919897564\\
0.114	1.76932822499263\\
0.115	1.76257071459156\\
0.116	1.75886203685626\\
0.117	1.75499441108463\\
0.118	1.74870971978344\\
0.119	1.7454906126654\\
0.12	1.73893347998113\\
0.121	1.73544317018274\\
0.122	1.72987796602987\\
0.123	1.72563839148117\\
0.124	1.72254963980979\\
0.125	1.71608212768977\\
0.126	1.71288825523272\\
0.127	1.70679059694779\\
0.128	1.70340991553002\\
0.129	1.69782796659424\\
0.13	1.69411764896825\\
0.131	1.68946952224704\\
0.132	1.68500229768037\\
0.133	1.68197779624664\\
0.134	1.6760522253134\\
0.135	1.67300935495753\\
0.136	1.66725426022025\\
0.137	1.6641738223775\\
0.138	1.65859404636979\\
0.139	1.65548085356125\\
0.14	1.65005694288977\\
0.141	1.64692528429203\\
0.142	1.64163009326309\\
0.143	1.63850148711717\\
0.144	1.6333052576693\\
0.145	1.63020455172192\\
0.146	1.62508055492348\\
0.147	1.62203039236746\\
0.148	1.61695947804774\\
0.149	1.61397521506596\\
0.15	1.60894787303124\\
0.151	1.6060324689591\\
0.152	1.60105104759076\\
0.153	1.59756443293026\\
0.154	1.59327225547734\\
0.155	1.5890889323675\\
0.156	1.58561229635793\\
0.157	1.58115757213964\\
0.158	1.57806925572257\\
0.159	1.57350842276267\\
0.16	1.57063472843503\\
0.161	1.56604780223539\\
0.162	1.56226483727123\\
0.163	1.55873873399578\\
0.164	1.55450603697076\\
0.165	1.55155912793742\\
0.166	1.54720908165234\\
0.167	1.54374372066534\\
0.168	1.54012725545842\\
0.169	1.5360556241163\\
0.17	1.5331922162278\\
0.171	1.52900198124307\\
0.172	1.52529185745668\\
0.173	1.52217771472445\\
0.174	1.51815606034563\\
0.175	1.51488257488127\\
0.176	1.51142586283244\\
0.177	1.50755272248604\\
0.178	1.50460721523016\\
0.179	1.5009186958695\\
0.18	1.49715640643985\\
0.181	1.49424223079456\\
0.182	1.49064160329989\\
0.183	1.48695091148146\\
0.184	1.48384237130309\\
0.185	1.48058716146696\\
0.186	1.47694525800784\\
0.187	1.4736280202847\\
0.188	1.47075247390578\\
0.189	1.46715728218941\\
0.19	1.46373228310045\\
0.191	1.46089030341746\\
0.192	1.45759698504962\\
0.193	1.45415040590628\\
0.194	1.45089599907203\\
0.195	1.44822807597681\\
0.196	1.44484526000821\\
0.197	1.4415127032802\\
0.198	1.43833534221046\\
0.199	1.43557765681672\\
0.2	1.43247289912748\\
0.201	1.42922617454813\\
0.202	1.42607474970494\\
0.203	1.42310628929009\\
0.204	1.42046455405998\\
0.205	1.41730471460695\\
0.206	1.41419214443442\\
0.207	1.41115395322815\\
0.208	1.40823333073049\\
0.209	1.40553248276429\\
0.21	1.40271110344293\\
0.211	1.39970653522174\\
0.212	1.39674643415635\\
0.213	1.3938406201969\\
0.214	1.39100455629987\\
0.215	1.38826577228858\\
0.216	1.38565660124973\\
0.217	1.38291160036573\\
0.218	1.3800747180694\\
0.219	1.37726905261996\\
0.22	1.37449647202453\\
0.221	1.37175819063421\\
0.222	1.36905533830504\\
0.223	1.36638891498504\\
0.224	1.3637594772709\\
0.225	1.36116651963204\\
0.226	1.35860724866871\\
0.227	1.35607372393985\\
0.228	1.3535475055279\\
0.229	1.35100180730817\\
0.23	1.34844089456801\\
0.231	1.34589701182623\\
0.232	1.34337494650831\\
0.233	1.3408745225884\\
0.234	1.33839545852091\\
0.235	1.33593748922105\\
0.236	1.33350039475061\\
0.237	1.33108400183418\\
0.238	1.3286881598423\\
0.239	1.32631259349662\\
0.24	1.32395141170953\\
0.241	1.32158635757023\\
0.242	1.31921903581304\\
0.243	1.31686369626131\\
0.244	1.31453058360123\\
0.245	1.31222451939138\\
0.246	1.30994716039058\\
0.247	1.30769862270459\\
0.248	1.30547829543355\\
0.249	1.30328522057865\\
0.25	1.30111825407738\\
0.251	1.29897603445651\\
0.252	1.296845088631\\
0.253	1.29464990706237\\
0.254	1.29247431262165\\
0.255	1.29034828090592\\
0.256	1.28826693493493\\
0.257	1.28622312131274\\
0.258	1.28421048917286\\
0.259	1.28214920972271\\
0.26	1.28007039819844\\
0.261	1.27805258675549\\
0.262	1.27608726234617\\
0.263	1.27416170714294\\
0.264	1.27217759692984\\
0.265	1.27019434987036\\
0.266	1.26827958528948\\
0.267	1.26641713626481\\
0.268	1.26451851471561\\
0.269	1.2626030080956\\
0.27	1.26076134628565\\
0.271	1.25896935746127\\
0.272	1.25710004634195\\
0.273	1.25528265622366\\
0.274	1.2535346452327\\
0.275	1.25174124033571\\
0.276	1.24996436225839\\
0.277	1.24826370837693\\
0.278	1.24651867722733\\
0.279	1.24480003123179\\
0.28	1.24315147216681\\
0.281	1.24143902903775\\
0.282	1.23979387427478\\
0.283	1.23816492862308\\
0.284	1.23651780042938\\
0.285	1.23494642590888\\
0.286	1.23332161795335\\
0.287	1.23177117834468\\
0.288	1.23019654907429\\
0.289	1.22866383604102\\
0.29	1.2271362421288\\
0.291	1.22562825044443\\
0.292	1.22413827055332\\
0.293	1.2226622795901\\
0.294	1.2212028152367\\
0.295	1.21976534774982\\
0.296	1.21833130432518\\
0.297	1.21693716673115\\
0.298	1.21552648604141\\
0.299	1.2141682333639\\
0.3	1.21279229002095\\
0.301	1.2114545039582\\
0.302	1.2101321784318\\
0.303	1.2088093526976\\
0.304	1.2075264246338\\
0.305	1.20624380213546\\
0.306	1.20497579247912\\
0.307	1.2037398230246\\
0.308	1.20250903598459\\
0.309	1.20128865516365\\
0.31	1.20009675884909\\
0.311	1.1989224080997\\
0.312	1.19775250475001\\
0.313	1.19660152480852\\
0.314	1.19547321236864\\
0.315	1.19436271020248\\
0.316	1.19326436754841\\
0.317	1.19217811232717\\
0.318	1.1911099832125\\
0.319	1.19006058445387\\
0.32	1.18902867355196\\
0.321	1.18801324375195\\
0.322	1.1870137123378\\
0.323	1.18602993121861\\
0.324	1.18506203760454\\
0.325	1.18411024877162\\
0.326	1.18317471945453\\
0.327	1.1822554967941\\
0.328	1.18135253791536\\
0.329	1.18046575929238\\
0.33	1.17959511322642\\
0.331	1.17874067879889\\
0.332	1.17790272637203\\
0.333	1.17708171377926\\
0.334	1.17627809038696\\
0.335	1.17549086141257\\
0.336	1.17471862672326\\
0.337	1.17396244933339\\
0.338	1.17322424703757\\
0.339	1.1725025641877\\
0.34	1.1717959028235\\
0.341	1.17110735248955\\
0.342	1.17043466733868\\
0.343	1.16977864897318\\
0.344	1.16913965660259\\
0.345	1.16851720597942\\
0.346	1.16791146136405\\
0.347	1.16732313443697\\
0.348	1.16675118101556\\
0.349	1.16619647152515\\
0.35	1.16565915964972\\
0.351	1.16513906832404\\
0.352	1.16463631992091\\
0.353	1.16415110193847\\
0.354	1.1636835807335\\
0.355	1.16323394471758\\
0.356	1.16280249389441\\
0.357	1.16238941510412\\
0.358	1.16199488463304\\
0.359	1.16161935489386\\
0.36	1.16126314268736\\
0.361	1.16092667806687\\
0.362	1.16061049529476\\
0.363	1.16031524669585\\
0.364	1.16004169535436\\
0.365	1.15979083926015\\
0.366	1.15956396977948\\
0.367	1.1593628576461\\
0.368	1.15919011804879\\
0.369	1.15905019917184\\
0.37	1.15895454923723\\
0.371	1.15891129389365\\
0.372	1.1588948063818\\
0.373	1.15888860326965\\
0.374	1.15888627210047\\
0.375	1.15888539623036\\
0.376	1.15888506717175\\
0.377	1.15888494354999\\
0.378	1.1588848971078\\
0.379	1.1588848796604\\
0.38	1.15888487310565\\
0.381	1.15888487064291\\
0.382	1.15888486971725\\
0.383	1.1588848693689\\
0.384	1.15888486923652\\
0.385	1.15888486918507\\
0.386	1.15888486916287\\
0.387	1.15888486915327\\
0.388	1.15888486915327\\
0.389	1.15888486915327\\
0.39	1.15888486915327\\
0.391	1.15888486915327\\
0.392	1.15888486915327\\
0.393	1.15888486915327\\
0.394	1.15888486915327\\
0.395	1.15888486915327\\
0.396	1.15888486915327\\
0.397	1.15888486915327\\
0.398	1.15888486915327\\
0.399	1.15888486915327\\
0.4	1.15888486915327\\
};
\addplot [color=mycolor2, thick]
  table[row sep=crcr]{%
0.369176484161467	0.00144878693661371\\
0.368979745730316	0.0246298697044915\\
0.3686	0.0442934773656284\\
0.3682	0.06156982490678\\
0.3677	0.0761404271631527\\
0.3672	0.0905313592621332\\
0.3666	0.10227746274301\\
0.36607084301371	0.114505451562302\\
0.3654	0.125473196227909\\
0.3648	0.1369932731599\\
0.364173227475943	0.146449571464487\\
0.3635	0.157312647528215\\
0.3628	0.166108463575665\\
0.3621	0.174884569634357\\
0.3614	0.18364319758538\\
0.3607	0.192370981471455\\
0.36	0.201085554098261\\
0.3593	0.209798801275684\\
0.3586	0.218480454297092\\
0.357866848062852	0.225146483904263\\
0.3571	0.233209054048258\\
0.35637510190905	0.239795618095974\\
0.3556	0.247891178569286\\
0.35487524276878	0.254469497690702\\
0.3541	0.262560927199288\\
0.353371682396072	0.269169271067006\\
0.3526	0.277225054622318\\
0.3518	0.283204604411985\\
0.351	0.289257598024445\\
0.3503	0.297790686335014\\
0.3495	0.303853908095772\\
0.3487	0.309833411873812\\
0.3479	0.315819220982476\\
0.3471	0.321826612903278\\
0.346372900876409	0.328412819362637\\
0.345574397347237	0.334387885304641\\
0.3448	0.34087453726247\\
0.344	0.34846613226766\\
0.3432	0.354482070610929\\
0.3424	0.360489084901248\\
0.3416	0.366483676838711\\
0.340776669296482	0.370410846333984\\
0.339974114967954	0.376445292719036\\
0.339172319558356	0.382488594597513\\
0.3383	0.387916206974793\\
0.3375	0.39393942216804\\
0.3367	0.399983689523669\\
0.3359	0.406039910709374\\
0.3351	0.412101686929026\\
0.3343	0.417766595330748\\
0.3334	0.421567448611055\\
0.3326	0.427589300301797\\
0.3318	0.433681261630055\\
0.331	0.439778379506599\\
0.330172906807814	0.443766097428737\\
0.3293	0.449259128990439\\
0.3285	0.455384841799407\\
0.3277	0.461488097411916\\
0.3268	0.464920974340681\\
0.326	0.471061801025141\\
0.3252	0.477214977691741\\
0.3243	0.480652224836978\\
0.3235	0.486820541808708\\
0.3227	0.492971243784297\\
0.3218	0.496454704910478\\
0.321	0.50266526998539\\
0.320172237506711	0.506748211253766\\
0.3193	0.51235480428204\\
0.3185	0.51857358947716\\
0.3176	0.522086108400375\\
0.3168	0.528350126781175\\
0.3159	0.531865334761638\\
0.3151	0.538145642775861\\
0.3142	0.541698470694493\\
0.3134	0.547982668735852\\
0.3125	0.551604095777295\\
0.3117	0.557863948513743\\
0.3108	0.561492926334827\\
0.31	0.567790778800653\\
0.3091	0.571402335351937\\
0.3083	0.577764117924824\\
0.3074	0.581383079530629\\
0.3066	0.587783928292824\\
0.3057	0.591422883441617\\
0.3049	0.597834364836913\\
0.304	0.601515769446546\\
0.303172391566365	0.605806800411961\\
0.3023	0.611659763810288\\
0.3014	0.615359992627973\\
0.3006	0.62184815225104\\
0.2997	0.62557708480301\\
0.2988	0.629364025310483\\
0.298	0.635850433118508\\
0.2971	0.639611637158633\\
0.296273462474117	0.643991404513434\\
0.2954	0.64996359010785\\
0.2945	0.653764995050216\\
0.293673666223502	0.658180162451926\\
0.2928	0.664198364335637\\
0.2919	0.66803920468692\\
0.291070053108074	0.67251109775641\\
0.2902	0.678558651072406\\
0.2893	0.682436865560604\\
0.2884	0.686340278990419\\
0.2876	0.693022103062646\\
0.2867	0.696965651192208\\
0.2858	0.700904519130178\\
0.2849	0.70487324923356\\
0.2841	0.711625932969974\\
0.2832	0.715605336712516\\
0.2823	0.719601414831674\\
0.2814	0.723622257932936\\
0.2806	0.729758672627876\\
0.2797	0.734487978084383\\
0.2788	0.738534695544021\\
0.2779	0.742611955962953\\
0.277	0.746710435405071\\
0.2762	0.753516039570847\\
0.2753	0.757729186696995\\
0.2744	0.761848814256112\\
0.2735	0.765995519148472\\
0.2726	0.770171092018524\\
0.2717	0.774367529444657\\
0.270885098243459	0.779101885014789\\
0.27	0.785576246805644\\
0.2691	0.789788673275512\\
0.2682	0.794021077098374\\
0.2673	0.798279344762714\\
0.2664	0.802564729099642\\
0.2655	0.806869914988016\\
0.2646	0.811197397980615\\
0.263774088896613	0.816074923962027\\
0.2629	0.822673786575873\\
0.262	0.827028805558406\\
0.2611	0.831390822371806\\
0.2602	0.835770228644724\\
0.2593	0.840169022157956\\
0.2584	0.844587863486382\\
0.2575	0.849026907711393\\
0.2566	0.853486003376949\\
0.2557	0.857964763650963\\
0.2548	0.862462616374609\\
0.2539	0.866978870701889\\
0.253	0.871512811813799\\
0.2521	0.876063807959766\\
0.2512	0.880631388696593\\
0.2503	0.885215253945512\\
0.2494	0.889815213892896\\
0.2485	0.894431114172061\\
0.2476	0.899062823809309\\
0.2467	0.903710331108269\\
0.2458	0.908373914293686\\
0.2449	0.913054279921073\\
0.244	0.917752568729173\\
0.2431	0.922470222705449\\
0.2422	0.927208792291455\\
0.2413	0.931969764415345\\
0.2404	0.93675444198974\\
0.2395	0.941563866241835\\
0.2386	0.946398758072706\\
0.2377	0.951259435222626\\
0.2368	0.956145571932688\\
0.2359	0.961055104608135\\
0.235	0.965972759986857\\
0.234073389167816	0.968557526079065\\
0.2331	0.972927935196222\\
0.2322	0.977819872331258\\
0.2313	0.982737148635983\\
0.2304	0.987713914034059\\
0.2295	0.992743943031613\\
0.2286	0.99781938200921\\
0.2277	1.0029324794389\\
0.2268	1.00806039002479\\
0.2258	1.01031220431039\\
0.2249	1.01525580281424\\
0.224	1.0203082526205\\
0.2231	1.0254682006074\\
0.2222	1.03070235791801\\
0.2213	1.03598700317676\\
0.220374527827773	1.03884615428846\\
0.2194	1.04347101364078\\
0.2185	1.04859532463407\\
0.2176	1.05388955285596\\
0.2167	1.0592806675858\\
0.2158	1.06471939129419\\
0.2148	1.06709651609901\\
0.2139	1.07224026392194\\
0.213	1.07763081617453\\
0.2121	1.08315208059604\\
0.2112	1.08871231125912\\
0.2102	1.09105933829601\\
0.2093	1.09637170812196\\
0.2084	1.10195068614953\\
0.2075	1.10763667961927\\
0.2065	1.11019925212039\\
0.2056	1.11546880283469\\
0.2047	1.12112818968221\\
0.2038	1.1269280303028\\
0.2028	1.12953728880262\\
0.2019	1.1348976407642\\
0.201	1.14069578046372\\
0.2001	1.14661747770501\\
0.1991	1.14904060292166\\
0.1982	1.15470312526128\\
0.1973	1.16068500751243\\
0.1963	1.16358537089178\\
0.1954	1.16896450643609\\
0.1945	1.17496424216132\\
0.1936	1.18104374263173\\
0.1926	1.18346921398313\\
0.1917	1.18947249253007\\
0.1908	1.19571499575225\\
0.1898	1.19818009148076\\
0.1889	1.2042098076255\\
0.188	1.21055278183196\\
0.187	1.21305569691847\\
0.1861	1.2191793101712\\
0.1852	1.22562714817128\\
0.1842	1.22809792066784\\
0.1833	1.23439254666691\\
0.182375330932962	1.23830761420591\\
0.1814	1.24336134276499\\
0.1805	1.24986779025967\\
0.1795	1.25309347678399\\
0.1786	1.25890645102072\\
0.1777	1.26562127457084\\
0.1767	1.26827315796314\\
0.1758	1.27476809668539\\
0.174873354081519	1.27889254916642\\
0.1739	1.28408428566547\\
0.173	1.29095923487257\\
0.172	1.29372108594664\\
0.1711	1.30036240467509\\
0.17016789172385	1.30453110873361\\
0.1692	1.30991276903094\\
0.1683	1.31703226286889\\
0.1673	1.31966464671218\\
0.1664	1.32670088123735\\
0.1654	1.32976676208123\\
0.1645	1.33648548933054\\
0.1635	1.34103640300526\\
0.1626	1.34640189776359\\
0.1617	1.35384819479251\\
0.1607	1.35645964125722\\
0.1598	1.36393127051579\\
0.1588	1.36666819451035\\
0.1579	1.37411519858709\\
0.1569	1.37703706312768\\
0.156	1.3844171255982\\
0.155	1.38757059550368\\
0.1541	1.39484044541963\\
0.1531	1.39825415684134\\
0.1522	1.40538665371495\\
0.1512	1.40903972812645\\
0.1503	1.41605716106949\\
0.1493	1.41985575115252\\
0.1484	1.42685419306666\\
0.1474	1.43065098372372\\
0.1465	1.43778134278168\\
0.1455	1.44143827366923\\
0.1446	1.44884374640219\\
0.1436	1.45228998439652\\
0.1427	1.46004785893297\\
0.1417	1.46328552214509\\
0.1408	1.47140087857296\\
0.1398	1.47446788466467\\
0.1389	1.482909711827\\
0.1379	1.48584782886623\\
0.137	1.49457793190176\\
0.136	1.49742624224692\\
0.1351	1.50636171835554\\
0.1341	1.50920454043644\\
0.1331	1.51386306219113\\
0.1322	1.52118573383494\\
0.1312	1.52462236984611\\
0.1303	1.53337247023195\\
0.1293	1.5364372850313\\
0.1284	1.54575796090194\\
0.1274	1.54866011933625\\
0.1264	1.55446253521787\\
0.1255	1.56117979235435\\
0.1245	1.56465251361762\\
0.1236	1.5739639244945\\
0.1226	1.57701674929825\\
0.1217	1.58692700517403\\
0.1207	1.58988779337802\\
0.1197	1.59367081982948\\
0.1188	1.60310382888438\\
0.1178	1.60623346991533\\
0.1169	1.61653880511502\\
0.1159	1.61954445792151\\
0.1149	1.62321252256606\\
0.114	1.63326524486907\\
0.113	1.63636662338652\\
0.112	1.64283704957283\\
0.1111	1.65025168494628\\
0.1101	1.65362209689689\\
0.1092	1.66454687805209\\
0.1082	1.66759259463932\\
0.1072	1.67140140891466\\
0.1063	1.68216006709919\\
0.1053	1.68530222658903\\
0.1043	1.68999667934617\\
0.1034	1.70012257744752\\
0.1024	1.70338445808885\\
0.1014	1.71039193492376\\
0.1005	1.71846423430347\\
0.0995	1.72183648407103\\
0.0985745297352206	1.73002486859599\\
0.0976	1.73720277990051\\
0.0966	1.74065426862855\\
0.0957	1.7510193538668\\
0.0947	1.75635788700159\\
0.0937	1.75984308183925\\
0.0927755364129184	1.76893400778519\\
0.0918	1.77595485498923\\
0.0908	1.77942752467296\\
0.0898	1.78838785442581\\
0.0889	1.79602593200474\\
0.0879	1.79945213054117\\
0.0869	1.80499332573404\\
0.086	1.8166095486494\\
0.085	1.81997337709284\\
0.084	1.82436122198017\\
0.0831	1.83774629506608\\
0.0821	1.84105167069128\\
0.0811	1.84495986298434\\
0.0802	1.85945073490633\\
0.0792	1.86274732072779\\
0.0782	1.86637213153445\\
0.0772	1.87315670281265\\
0.0763	1.88511764204343\\
0.0753	1.88855987165686\\
0.0743	1.89279290390147\\
0.0734	1.90819664510541\\
0.0724	1.91154997905417\\
0.0714	1.91526899866926\\
0.0704	1.92158575380642\\
0.0695	1.93538391734926\\
0.0685	1.93885568447689\\
0.0675	1.94293953470705\\
0.0666	1.95996946067564\\
0.0656	1.96346306614228\\
0.0646	1.9670859550249\\
0.0636	1.9716128321439\\
0.0627	1.98907256318598\\
0.0617	1.99252456746748\\
0.0607	1.99628519608554\\
0.0597	2.00126003167876\\
0.0588	2.01915008689311\\
0.0578	2.02265042578379\\
0.0568	2.02650703476942\\
0.0558	2.03173206271676\\
0.0549	2.05039755424549\\
0.0539	2.05394038454527\\
0.0529	2.05783542503375\\
0.0519	2.06296322082348\\
0.051	2.08294686370226\\
0.05	2.08652213709106\\
0.049	2.09040146826849\\
0.048	2.09522947418327\\
0.0471	2.11693038169728\\
0.0461	2.12055622114787\\
0.0451	2.12438469107048\\
0.0441	2.12889296802592\\
0.0431815336762416	2.14653540645668\\
0.0422	2.15623587984241\\
0.0412	2.16000412309217\\
0.0402	2.16424248403751\\
0.0392	2.17019581677483\\
0.0383	2.19377542779959\\
0.0373	2.19751177975014\\
0.0363	2.20154279087425\\
0.0353	2.20635201591511\\
0.0344	2.23293762019361\\
0.0334	2.23718992039915\\
0.0324	2.24108104384386\\
0.0314	2.24541305663477\\
0.0304	2.25100324964257\\
0.0295	2.27930306439314\\
0.0285	2.28318873880148\\
0.0275	2.28726116235117\\
0.0265	2.29189635934573\\
0.0255	2.29849804084063\\
0.0246	2.32822615443449\\
0.0236	2.33220964711837\\
0.0226	2.33646311643264\\
0.0216	2.34137177074985\\
0.0206	2.34898387536425\\
0.0197	2.3807045481131\\
0.0187	2.38483179210483\\
0.0177	2.38926960253294\\
0.0167	2.39442211011925\\
0.0157	2.40246983989523\\
0.0148	2.4375785290847\\
0.0138	2.44190413556183\\
0.0128	2.44657030073616\\
0.0118	2.45201178920714\\
0.0108	2.46025800394549\\
0.0099	2.50017194928802\\
0.0089	2.50482699641623\\
0.0079	2.50989431114215\\
0.0069	2.51591947687609\\
0.0059	2.52572446968771\\
0.005	2.5711791509829\\
0.004	2.57659974433969\\
0.003	2.58284102745123\\
0.002	2.59125980488803\\
};
\addplot [color=gray, dotted]
  table[row sep=crcr]{%
0	1.67\\
0.137	1.67\\};
\addplot [color=gray, dotted]
  table[row sep=crcr]{%
0.123	1.67\\
0.123	0\\};
\addplot [color=gray, dotted]
  table[row sep=crcr]{%
0.137	1.67\\
0.137	0\\};
\end{axis}
\end{tikzpicture}%

%% file: Figs/expulsion_particles.tex
\begin{tikzpicture}[scale=0.6]

\definecolor{mycolor1}{rgb}{1,1,0.85}
\definecolor{mycolor2}{rgb}{0.2,1,0.85}

        \tikzset{test/.style n args={3}{
    draw = none,
    postaction={
    decorate,
    decoration={
    markings,
    mark=between positions 0 and \pgfdecoratedpathlength step 0.2pt with {
    \pgfmathsetmacro\myval{multiply(
        divide(
        \pgfkeysvalueof{/pgf/decoration/mark info/distance from start}, \pgfdecoratedpathlength
        ),
        100
    )};
    \pgfsetfillcolor{#3!\myval!#2};
    \pgfpathcircle{\pgfpointorigin}{#1};
    \pgfusepath{fill};}
}}}}

     \foreach \angle / \label in
    {0/3, 30/2, 60/1, 90/12, 120/11, 150/10, 180/9,
     210/8, 240/7, 270/6, 300/5, 330/4}
  {
  
    \draw[test={0.5ex}{mycolor1}{mycolor2},xshift = 1.7cm, yshift = 0.5cm] plot coordinates {(\angle:1.6cm) (\angle:2cm)};
    \draw[very thin,fill = mycolor2,xshift = 1.7cm, yshift = 0.5cm] (\angle:2cm) circle (.833ex);

  }
  
\end{tikzpicture}

%% file: Figs/FigA1a_ColourMap_CaiSuo_Shrink_DelayedFront.tex
\begin{tikzpicture}
\begin{axis}[axis on top,  
thick, 
/pgf/number format/.cd,
                fixed,
width=0.45\textwidth, 
height=0.4\textwidth,
enlargelimits=false, 
colorbar, 
point meta min = 0.2, 
point meta max = 1,
xmin=0,xmax=2.5,ymin=0,ymax=2.5, 
colorbar style = {
width = 0.3cm,
thick,
black,
title = {$\phi$},
title style = {overlay,yshift = -3pt},
at={(1.05,1)}},
xlabel = {$t$},
ylabel = {$r$},
legend style = {
draw=none,
fill=none,
font=\scriptsize},]
\addplot[forget plot] graphics [xmin=0,xmax=2.5,ymin=0,ymax=2.5] {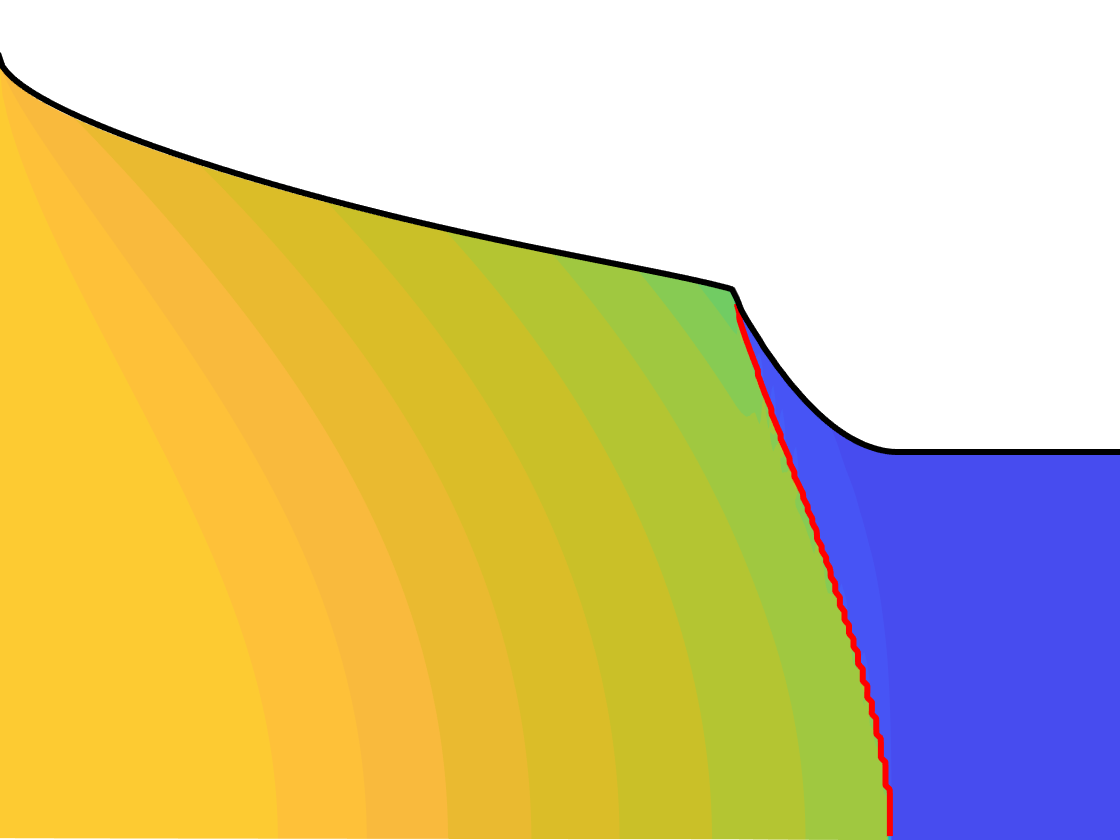};

\end{axis}

\end{tikzpicture}

%% file: Figs/FigA1b_ColourMap_CaiSuo_Shrink_Spinodal.tex
\begin{tikzpicture}
\begin{axis}[axis on top,  
thick, 
/pgf/number format/.cd,
                fixed,
width=0.45\textwidth, 
height=0.4\textwidth,
enlargelimits=false, 
colorbar, 
point meta min = 0.2, 
point meta max = 1,
xmin=0,xmax=0.2,ymin=0,ymax=2.5, 
colorbar style = {
width = 0.3cm,
thick,
black,
title = {$\phi$},
title style = {overlay,yshift = -3pt},
at={(1.05,1)}},
xlabel = {$t$},
ylabel = {$r$},
legend style = {
draw=none,
fill=none,
font=\scriptsize},]
\addplot[forget plot] graphics [xmin=0,xmax=0.2,ymin=0,ymax=2.5] {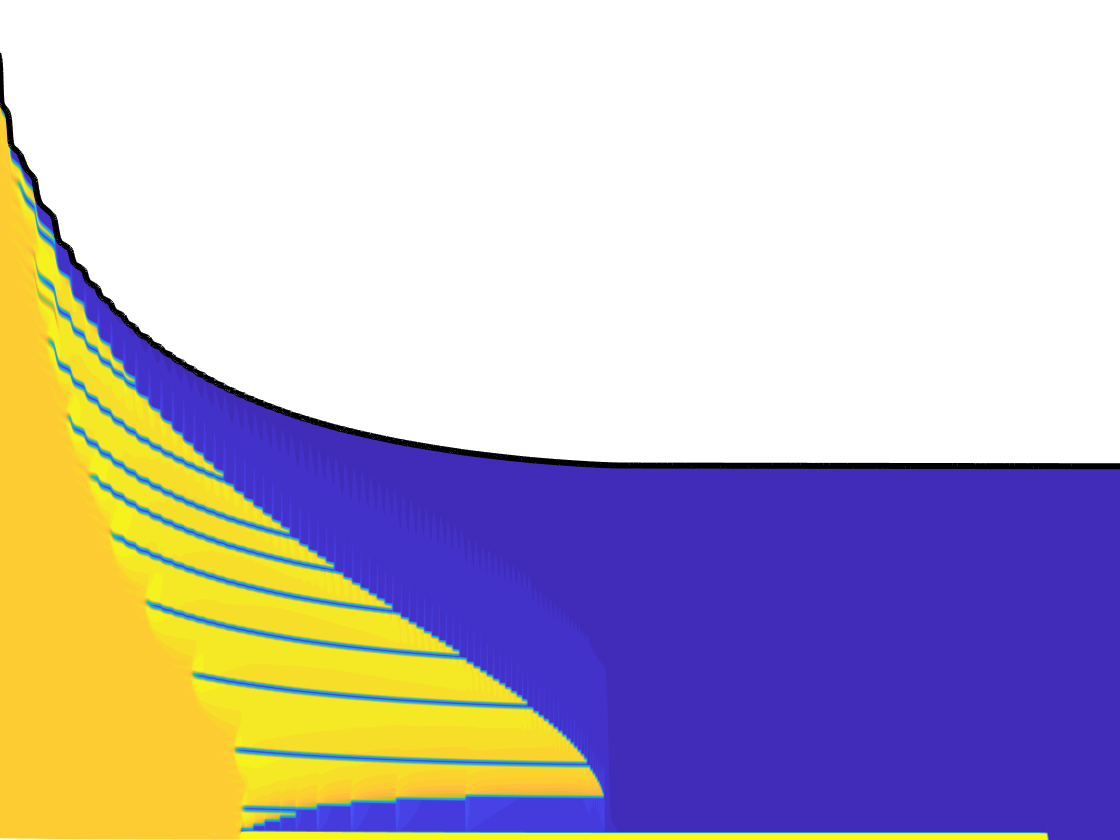};
\end{axis}
\end{tikzpicture}

%% file: Figs/FigA2a_ColourMap_Swell_VaryPerm.tex
\begin{tikzpicture}
\begin{axis}[axis on top,  
thick, 
/pgf/number format/.cd,
                fixed,
width=0.45\textwidth, 
height=0.4\textwidth,
enlargelimits=false, 
colorbar, 
point meta min = 0.25, 
point meta max = 1,
xmin=0,xmax=0.1,ymin=0,ymax=2.5, 
colorbar style = {
width = 0.3cm,
thick,
black,
title = {$\phi$},
title style = {overlay,yshift = -3pt},
at={(1.05,1)}},
xlabel = {$t$},
ylabel = {$r$},
legend style = {
draw=none,
fill=none,
font=\scriptsize},]
\addplot[forget plot] graphics [xmin=0,xmax=0.1,ymin=0,ymax=2.5] {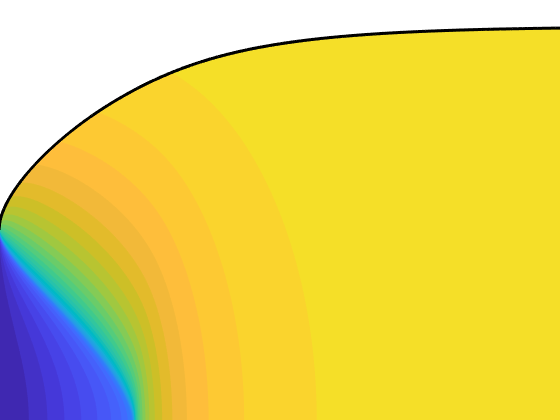};
%
%
\end{axis}
%
%
\end{tikzpicture}

%% file: Figs/FigA2b_ColourMap_ShrinkFront_VaryPerm.tex
\begin{tikzpicture}
\begin{axis}[axis on top,  
thick, 
/pgf/number format/.cd,
                fixed,
width=0.45\textwidth, 
height=0.4\textwidth,
enlargelimits=false, 
colorbar, 
point meta min = 0.25, 
point meta max = 1,
xmin=0,xmax=0.5,ymin=0,ymax=2.5, 
colorbar style = {
width = 0.3cm,
thick,
black,
title = {$\phi$},
title style = {overlay,yshift = -3pt},
at={(1.05,1)}},
xlabel = {$t$},
ylabel = {$r$},
legend style = {
draw=none,
fill=none,
font=\scriptsize},]
\addplot[forget plot] graphics [xmin=0,xmax=0.5,ymin=0,ymax=2.5] {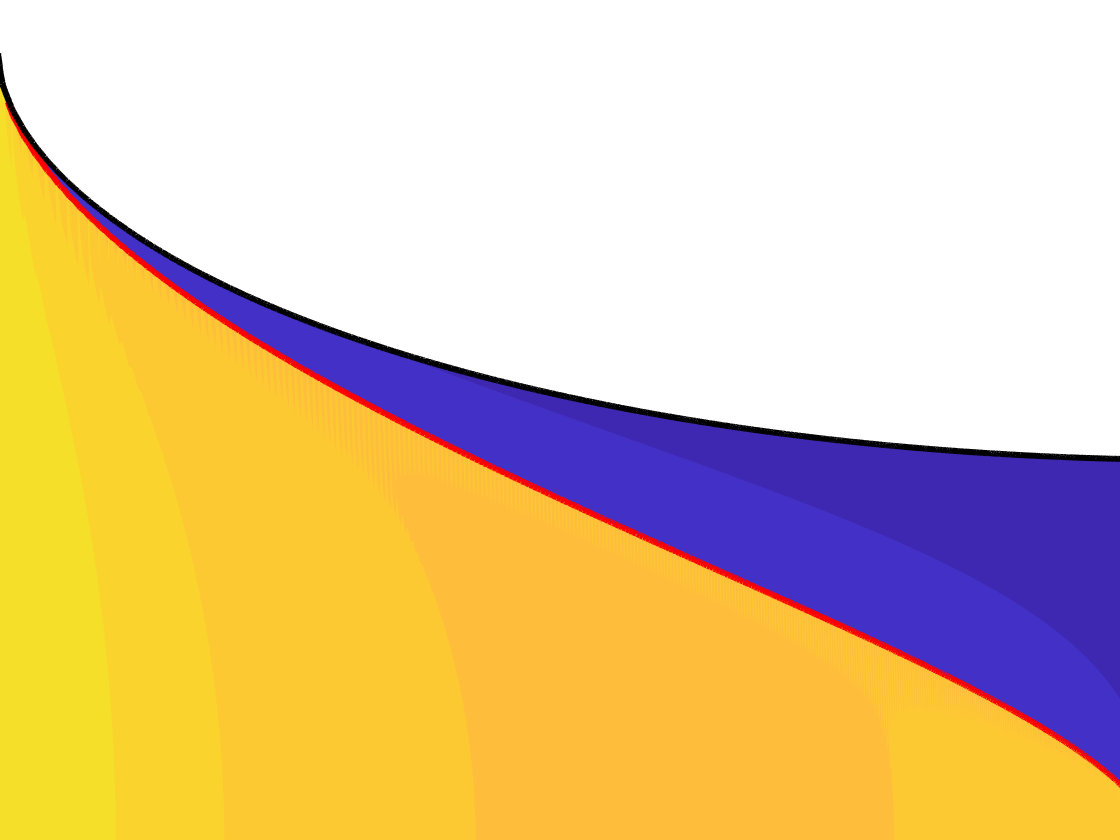};
%
%
\end{axis}
%
%
\end{tikzpicture}